\documentclass[copyright,creativecommons]{eptcs}
\providecommand{\event}{Peter Thiemann's Festschrift 2024} 

\usepackage{iftex}

\ifpdf
  \usepackage{underscore}         
  \usepackage[T1]{fontenc}        
\else
  \usepackage{breakurl}           
\fi

\usepackage{fancyvrb}
\usepackage{agda}
\usepackage{amssymb}
\usepackage{stmaryrd}
\usepackage[numbers]{natbib}
\usepackage{newunicodechar}
\newunicodechar{∷}{{::}}
\newunicodechar{⊤}{\ensuremath{\top}}
\newunicodechar{⊥}{\ensuremath{\bot}}
\newunicodechar{₁}{\ensuremath{_1}}
\newunicodechar{₂}{\ensuremath{_2}}
\newunicodechar{₃}{\ensuremath{_3}}

\newunicodechar{β}{\ensuremath{\beta}}
\newunicodechar{γ}{\ensuremath{\gamma}}
\newunicodechar{λ}{\ensuremath{\lambda}}
\newunicodechar{Λ}{\ensuremath{\Lambda}}
\newunicodechar{μ}{\ensuremath{\mu}}
\newunicodechar{ν}{\ensuremath{\nu}}
\newunicodechar{ρ}{\ensuremath{\rho}}
\newunicodechar{σ}{\ensuremath{\sigma}}
\newunicodechar{Π}{\ensuremath{\Pi}}
\newunicodechar{ξ}{\ensuremath{\xi}}
\newunicodechar{Γ}{\ensuremath{\Gamma}}
\newunicodechar{Δ}{\ensuremath{\Delta}}
\newunicodechar{Σ}{\ensuremath{\Sigma}}

\newunicodechar{∀}{\ensuremath{\forall}}
\newunicodechar{∃}{\ensuremath{\exists}}
\newunicodechar{⊎}{\ensuremath{\uplus}}
\newunicodechar{ᵖ}{\ensuremath{^p}}
\newunicodechar{≡}{\ensuremath{\equiv}}
\newunicodechar{≢}{\ensuremath{\not\equiv}}
\newunicodechar{⊑}{\ensuremath{\sqsubseteq}}
\newunicodechar{★}{\ensuremath{\star}}
\newunicodechar{∼}{\ensuremath{\sim}}
\newunicodechar{⌈}{\ensuremath{\lceil}}
\newunicodechar{⌉}{\ensuremath{\rceil}}
\newunicodechar{·}{\ensuremath{\cdot}}
\newunicodechar{•}{\ensuremath{\bullet}}

\newunicodechar{⇒}{\ensuremath{\Rightarrow}}
\newunicodechar{↠}{\ensuremath{\longrightarrow^{*}}}
\newunicodechar{⇓}{\ensuremath{\Downarrow}}
\newunicodechar{⇑}{\ensuremath{\Uparrow}}
\newunicodechar{⟶}{\ensuremath{\longrightarrow}}

\newunicodechar{ι}{\ensuremath{\iota}}
\newunicodechar{ₜ}{\ensuremath{_t}}
\newunicodechar{⊢}{\ensuremath{\vdash}}
\newunicodechar{⊩}{\ensuremath{\Vdash}}
\newunicodechar{⊨}{\ensuremath{\vDash}}
\newunicodechar{⦂}{\ensuremath{\mathop{:}}}
\newunicodechar{∈}{\ensuremath{\in}}
\newunicodechar{∋}{\ensuremath{\ni}}
\newunicodechar{′}{\ensuremath{'}}
\newunicodechar{″}{\ensuremath{''}}
\newunicodechar{ƛ}{\ensuremath{\lambda}}
\newunicodechar{≼}{\ensuremath{\preceq}}
\newunicodechar{≽}{\ensuremath{\succeq}}
\newunicodechar{ᵥ}{\ensuremath{_v}}
\newunicodechar{ˢ}{\ensuremath{^s}}
\newunicodechar{∅}{\ensuremath{\emptyset}}
\newunicodechar{∣}{\ensuremath{|}}
\newunicodechar{ᴸ}{\ensuremath{^L}}
\newunicodechar{ᴿ}{\ensuremath{^R}}
\newunicodechar{▷}{\ensuremath{\triangleright}}
\newunicodechar{ᵍ}{\ensuremath{^g}}
\newunicodechar{ᵒ}{\ensuremath{^{\circ}}}
\newunicodechar{⩦}{\ensuremath{\equiv}}
\newunicodechar{∎}{\ensuremath{\blacksquare}}
\newunicodechar{□}{\ensuremath{\square}}
\newunicodechar{■}{\ensuremath{\blacksquare}}
\newunicodechar{⇔}{\ensuremath{\iff}}
\newunicodechar{𝒱}{\ensuremath{\mathcal{V}}}
\newunicodechar{̬}{}
\newunicodechar{𝒫}{\ensuremath{\mathcal{P}}}
\newunicodechar{ℰ}{\ensuremath{\mathcal{E}}}
\newunicodechar{≤}{\ensuremath{\leq}}

\newunicodechar{ℕ}{\ensuremath{\mathbb{N}}}
\newunicodechar{𝔹}{\ensuremath{\mathbb{B}}}

\newunicodechar{⦉}{\ensuremath{\llfloor}}
\newunicodechar{⦊}{\ensuremath{\rrfloor}}
\newunicodechar{⟦}{\ensuremath{[}}
\newunicodechar{⟧}{\ensuremath{]}}
\newunicodechar{⦅}{\ensuremath{\llparenthesis}}
\newunicodechar{⦆}{\ensuremath{\rrparenthesis}}
\newunicodechar{⟪}{\ensuremath{\langle\!\!\langle}}
\newunicodechar{⟫}{\ensuremath{\rangle\!\!\rangle}}
\newunicodechar{⟨}{\ensuremath{\langle}}
\newunicodechar{⟩}{\ensuremath{\rangle}}
\newunicodechar{∶}{:}
\newunicodechar{⨟}{\ensuremath{\fatsemi}}
\newunicodechar{↑}{\ensuremath{\uparrow}}
\newunicodechar{≠}{\ensuremath{\neq}}

\title{Gradual Guarantee via \\ Step-Indexed Logical Relations in Agda}
\author{Jeremy G. Siek
\institute{School of Informatics, Computing, and Engineering\\
Indiana University \\
Bloomington, IN, USA}
\email{jsiek@iu.edu}
}
\def\titlerunning{Gradual Guarantee via Step-Indexed Logical Relations in Agda}
\def\authorrunning{Jeremy G. Siek}
\begin{document}
\maketitle

\begin{abstract}
  The gradual guarantee is an important litmus test for gradually
  typed languages, that is, languages that enable a mixture of static
  and dynamic typing. The gradual guarantee states that changing the
  precision of a type annotation does not change the behavior of the
  program, except perhaps to trigger an error if the type annotation
  is incorrect. Siek et al. (2015) proved that the Gradually Typed
  Lambda Calculus (GTLC) satisfies the gradual guarantee using a
  simulation-based proof and mechanized their proof in Isabelle. In
  the following decade, researchers have proved the gradual guarantee
  for more sophisticated calculi, using step-indexed logical
  relations.  However, given the complexity of that style of proof,
  there has not yet been a mechanized proof of the gradual guarantee
  using step-indexed logical relations. This paper reports on a
  mechanized proof of the gradual guarantee for the GTLC carried out
  in the Agda proof assistant.
\end{abstract}

\section{Introduction}

Gradually typed languages introduce the unknown type ★ as a way for
programmers to control the amount of type precision, and compile-time
type checking, in their program \cite{Siek:2006bh,Siek:2007qy}. If all
type annotations are just ★, then the program behaves like a
dynamically typed language. At the other end of the spectrum, if no
type annotation contains ★, then the program behaves like a statically
typed language.

Siek et al.~\cite{Siek:2015ac} introduce the \emph{gradual guarantee}
as a litmus test for gradually typed languages.  This property says
that the behavior of a program should not change (except for errors)
when the programmer changes type annotations to be more or less
precise.  Siek et al.~\cite{Siek:2015ac} prove that the Gradually
Typed Lambda Calculus (GTLC) satisfies the gradual guarantee using a
simulation-based proof and mechanize the result in the Isabelle proof
assistant~\cite{Nipkow:2002jl}. Using logical relations, New and
Ahmed~\cite{New:2018aa} prove the gradual guarantee for the GTLC and
New et al.~\cite{New:2019ab} prove the gradual guarantee for a
polymorphic calculus. Several researchers apply logical relations to
prove other properties of gradually typed languages, such as
noninterference\cite{Toro:2018aa},
parametricity~\cite{Ahmed:2011fk,New:2019ab,Labrada:2020tk}, and fully
abstract embedding~\cite{Jacobs:2021aa}. Of this later work, only the
abstract embedding was mechanized (in
Coq~\cite{The-Coq-Development-Team:2004kf} using the Iris
framework~\cite{JUNG:2018aa}.)

There are several technical challenges to overcome in developing a
mechanized proof of the gradual guarantee using step-indexed logical
relations. As in any programming language mechanization, one must
choose how to represent variables and perform substitution. Moreover,
proofs based on logical relations rely on the fact that substitutions
commute, and the proof of this standard result is quite technical and
lengthy. Repeating these proofs for each new programming language is
tedious, but this metatheory can be developed in a
language-independent way using the notion of \emph{abstract binding
  trees}~\cite{Harper:2012aa}. To this end we developed the Abstract
Binding Tree library in Agda~\cite{Siek:2021to}, representing
variables as de Bruijn indices and implementing substitution via
parallel renaming and
substitution~\cite{McBride:2005aa,Wadler:2020aa}.

The second technical challenge is that step-indexed logical relations
``involve tedious, error-prone, and proof-obscuring step-index
arithmetic''~\cite{Dreyer:2011wl}. Dreyer, Ahmed, and
Birkedal~\cite{Dreyer:2011wl} propose to abstract over the
step-indexing using a modal logic named LSLR. Dreyer and Birkedal,
with many colleagues, implemented this logic as part of the Iris
framework~\cite{JUNG:2018aa} in the Coq proof assistant. To make a
similar modal logic available in Agda, we developed the Step-Indexed
Logic (SIL)~\cite{Siek:2023aa}.  The proof of the gradual guarantee in
this paper is the first application of SIL and we report on the
experience. This paper is a literate Agda script, though we omit some
of the proofs. The full Agda development is available from the
following repository, in the \texttt{LogRel} directory.

\url{https://github.com/jsiek/gradual-typing-in-agda}

The semantics of the GTLC is defined by translation to a Cast
Calculus, so we present the Cast Calculus in
Section~\ref{sec:cast-calculus}. We define the precision relation on
types and terms in Section~\ref{sec:precision} and we review
Step-Indexed Logic in Section~\ref{sec:SIL}.  We define a logical
relation for precision in Section~\ref{sec:log-rel}. We prove the
Fundamental Theorem of the logical relation in
Section~\ref{sec:fundamental}. To finish the proof of the gradual
guarantee, in Section~\ref{sec:gradual-guarantee} we prove that the
logical relation implies the gradual
guarantee. Section~\ref{sec:conclusion} concludes this paper with a
comparison of using logical-relations versus simulation to prove the
gradual guarantee and it acknowledges Peter Thiemann and Philip Wadler
for their contributions to this work.

\begin{code}[hide]%
\>[0]\AgdaSymbol{\{-\#}\AgdaSpace{}%
\AgdaKeyword{OPTIONS}\AgdaSpace{}%
\AgdaPragma{--rewriting}\AgdaSpace{}%
\AgdaSymbol{\#-\}}\<%
\\
\>[0]\AgdaKeyword{module}\AgdaSpace{}%
\AgdaModule{LogRel.PeterCastCalculus}\AgdaSpace{}%
\AgdaKeyword{where}\<%
\\
\\[\AgdaEmptyExtraSkip]%
\>[0]\AgdaKeyword{open}\AgdaSpace{}%
\AgdaKeyword{import}\AgdaSpace{}%
\AgdaModule{Data.Empty}\AgdaSpace{}%
\AgdaKeyword{using}\AgdaSpace{}%
\AgdaSymbol{(}\AgdaDatatype{⊥}\AgdaSymbol{;}\AgdaSpace{}%
\AgdaFunction{⊥-elim}\AgdaSymbol{)}\<%
\\
\>[0]\AgdaKeyword{open}\AgdaSpace{}%
\AgdaKeyword{import}\AgdaSpace{}%
\AgdaModule{Data.List}\AgdaSpace{}%
\AgdaKeyword{using}\AgdaSpace{}%
\AgdaSymbol{(}\AgdaDatatype{List}\AgdaSymbol{;}\AgdaSpace{}%
\AgdaInductiveConstructor{[]}\AgdaSymbol{;}\AgdaSpace{}%
\AgdaOperator{\AgdaInductiveConstructor{\AgdaUnderscore{}∷\AgdaUnderscore{}}}\AgdaSymbol{;}\AgdaSpace{}%
\AgdaFunction{map}\AgdaSymbol{;}\AgdaSpace{}%
\AgdaFunction{length}\AgdaSymbol{)}\<%
\\
\>[0]\AgdaKeyword{open}\AgdaSpace{}%
\AgdaKeyword{import}\AgdaSpace{}%
\AgdaModule{Data.Nat}\<%
\\
\>[0]\AgdaKeyword{open}\AgdaSpace{}%
\AgdaKeyword{import}\AgdaSpace{}%
\AgdaModule{Data.Nat.Properties}\<%
\\
\>[0]\AgdaKeyword{open}\AgdaSpace{}%
\AgdaKeyword{import}\AgdaSpace{}%
\AgdaModule{Data.Bool}\AgdaSpace{}%
\AgdaKeyword{using}\AgdaSpace{}%
\AgdaSymbol{(}\AgdaInductiveConstructor{true}\AgdaSymbol{;}\AgdaSpace{}%
\AgdaInductiveConstructor{false}\AgdaSymbol{)}\AgdaSpace{}%
\AgdaKeyword{renaming}\AgdaSpace{}%
\AgdaSymbol{(}\AgdaDatatype{Bool}\AgdaSpace{}%
\AgdaSymbol{to}\AgdaSpace{}%
\AgdaDatatype{𝔹}\AgdaSymbol{)}\<%
\\
\>[0]\AgdaKeyword{open}\AgdaSpace{}%
\AgdaKeyword{import}\AgdaSpace{}%
\AgdaModule{Data.Product}\AgdaSpace{}%
\AgdaKeyword{using}\AgdaSpace{}%
\AgdaSymbol{(}\AgdaOperator{\AgdaInductiveConstructor{\AgdaUnderscore{},\AgdaUnderscore{}}}\AgdaSymbol{;}\AgdaOperator{\AgdaFunction{\AgdaUnderscore{}×\AgdaUnderscore{}}}\AgdaSymbol{;}\AgdaSpace{}%
\AgdaField{proj₁}\AgdaSymbol{;}\AgdaSpace{}%
\AgdaField{proj₂}\AgdaSymbol{;}\AgdaSpace{}%
\AgdaFunction{Σ-syntax}\AgdaSymbol{;}\AgdaSpace{}%
\AgdaFunction{∃-syntax}\AgdaSymbol{)}\<%
\\
\>[0]\AgdaKeyword{open}\AgdaSpace{}%
\AgdaKeyword{import}\AgdaSpace{}%
\AgdaModule{Data.Sum}\AgdaSpace{}%
\AgdaKeyword{using}\AgdaSpace{}%
\AgdaSymbol{(}\AgdaOperator{\AgdaDatatype{\AgdaUnderscore{}⊎\AgdaUnderscore{}}}\AgdaSymbol{;}\AgdaSpace{}%
\AgdaInductiveConstructor{inj₁}\AgdaSymbol{;}\AgdaSpace{}%
\AgdaInductiveConstructor{inj₂}\AgdaSymbol{)}\<%
\\
\>[0]\AgdaKeyword{open}\AgdaSpace{}%
\AgdaKeyword{import}\AgdaSpace{}%
\AgdaModule{Data.Unit}\AgdaSpace{}%
\AgdaKeyword{using}\AgdaSpace{}%
\AgdaSymbol{(}\AgdaRecord{⊤}\AgdaSymbol{;}\AgdaSpace{}%
\AgdaInductiveConstructor{tt}\AgdaSymbol{)}\<%
\\
\>[0]\AgdaKeyword{open}\AgdaSpace{}%
\AgdaKeyword{import}\AgdaSpace{}%
\AgdaModule{Data.Unit.Polymorphic}\AgdaSpace{}%
\AgdaKeyword{renaming}\AgdaSpace{}%
\AgdaSymbol{(}\AgdaFunction{⊤}\AgdaSpace{}%
\AgdaSymbol{to}\AgdaSpace{}%
\AgdaFunction{topᵖ}\AgdaSymbol{;}\AgdaSpace{}%
\AgdaFunction{tt}\AgdaSpace{}%
\AgdaSymbol{to}\AgdaSpace{}%
\AgdaFunction{ttᵖ}\AgdaSymbol{)}\<%
\\
\>[0]\AgdaKeyword{open}\AgdaSpace{}%
\AgdaKeyword{import}\AgdaSpace{}%
\AgdaModule{Relation.Binary.PropositionalEquality}\AgdaSpace{}%
\AgdaSymbol{as}\AgdaSpace{}%
\AgdaModule{Eq}\<%
\\
\>[0][@{}l@{\AgdaIndent{0}}]%
\>[2]\AgdaKeyword{using}\AgdaSpace{}%
\AgdaSymbol{(}\AgdaOperator{\AgdaDatatype{\AgdaUnderscore{}≡\AgdaUnderscore{}}}\AgdaSymbol{;}\AgdaSpace{}%
\AgdaOperator{\AgdaFunction{\AgdaUnderscore{}≢\AgdaUnderscore{}}}\AgdaSymbol{;}\AgdaSpace{}%
\AgdaInductiveConstructor{refl}\AgdaSymbol{;}\AgdaSpace{}%
\AgdaFunction{sym}\AgdaSymbol{;}\AgdaSpace{}%
\AgdaFunction{cong}\AgdaSymbol{;}\AgdaSpace{}%
\AgdaFunction{subst}\AgdaSymbol{;}\AgdaSpace{}%
\AgdaFunction{trans}\AgdaSymbol{)}\<%
\\
\>[0]\AgdaKeyword{open}\AgdaSpace{}%
\AgdaKeyword{import}\AgdaSpace{}%
\AgdaModule{Relation.Nullary}\AgdaSpace{}%
\AgdaKeyword{using}\AgdaSpace{}%
\AgdaSymbol{(}\AgdaOperator{\AgdaFunction{¬\AgdaUnderscore{}}}\AgdaSymbol{;}\AgdaSpace{}%
\AgdaRecord{Dec}\AgdaSymbol{;}\AgdaSpace{}%
\AgdaInductiveConstructor{yes}\AgdaSymbol{;}\AgdaSpace{}%
\AgdaInductiveConstructor{no}\AgdaSymbol{)}\<%
\\
\\[\AgdaEmptyExtraSkip]%
\>[0]\AgdaKeyword{open}\AgdaSpace{}%
\AgdaKeyword{import}\AgdaSpace{}%
\AgdaModule{Var}\<%
\\
\>[0]\AgdaKeyword{open}\AgdaSpace{}%
\AgdaKeyword{import}\AgdaSpace{}%
\AgdaModule{Sig}\<%
\end{code}

\section{Cast Calculus}
\label{sec:cast-calculus}

\begin{code}[hide]%
\>[0]\AgdaComment{\{-\ Base\ types\ -\}}\<%
\\
\\[\AgdaEmptyExtraSkip]%
\>[0]\AgdaKeyword{data}\AgdaSpace{}%
\AgdaDatatype{Base}\AgdaSpace{}%
\AgdaSymbol{:}\AgdaSpace{}%
\AgdaPrimitive{Set}\AgdaSpace{}%
\AgdaKeyword{where}\<%
\\
\>[0][@{}l@{\AgdaIndent{0}}]%
\>[2]\AgdaInductiveConstructor{′ℕ}\AgdaSpace{}%
\AgdaSymbol{:}\AgdaSpace{}%
\AgdaDatatype{Base}\<%
\\
\>[2]\AgdaInductiveConstructor{′𝔹}\AgdaSpace{}%
\AgdaSymbol{:}\AgdaSpace{}%
\AgdaDatatype{Base}\<%
\\
\\[\AgdaEmptyExtraSkip]%
\>[0]\AgdaOperator{\AgdaFunction{\AgdaUnderscore{}≡\$?\AgdaUnderscore{}}}\AgdaSpace{}%
\AgdaSymbol{:}\AgdaSpace{}%
\AgdaSymbol{(}\AgdaBound{ι}\AgdaSpace{}%
\AgdaSymbol{:}\AgdaSpace{}%
\AgdaDatatype{Base}\AgdaSymbol{)}\AgdaSpace{}%
\AgdaSymbol{→}\AgdaSpace{}%
\AgdaSymbol{(}\AgdaBound{ι′}\AgdaSpace{}%
\AgdaSymbol{:}\AgdaSpace{}%
\AgdaDatatype{Base}\AgdaSymbol{)}\AgdaSpace{}%
\AgdaSymbol{→}\AgdaSpace{}%
\AgdaRecord{Dec}\AgdaSpace{}%
\AgdaSymbol{(}\AgdaBound{ι}\AgdaSpace{}%
\AgdaOperator{\AgdaDatatype{≡}}\AgdaSpace{}%
\AgdaBound{ι′}\AgdaSymbol{)}\<%
\\
\>[0]\AgdaInductiveConstructor{′ℕ}%
\>[4]\AgdaOperator{\AgdaFunction{≡\$?}}\AgdaSpace{}%
\AgdaInductiveConstructor{′ℕ}%
\>[12]\AgdaSymbol{=}%
\>[15]\AgdaInductiveConstructor{yes}\AgdaSpace{}%
\AgdaInductiveConstructor{refl}\<%
\\
\>[0]\AgdaInductiveConstructor{′ℕ}%
\>[4]\AgdaOperator{\AgdaFunction{≡\$?}}\AgdaSpace{}%
\AgdaInductiveConstructor{′𝔹}%
\>[12]\AgdaSymbol{=}%
\>[15]\AgdaInductiveConstructor{no}\AgdaSpace{}%
\AgdaSymbol{(λ}\AgdaSpace{}%
\AgdaSymbol{())}\<%
\\
\>[0]\AgdaInductiveConstructor{′𝔹}%
\>[4]\AgdaOperator{\AgdaFunction{≡\$?}}\AgdaSpace{}%
\AgdaInductiveConstructor{′ℕ}%
\>[12]\AgdaSymbol{=}%
\>[15]\AgdaInductiveConstructor{no}\AgdaSpace{}%
\AgdaSymbol{(λ}\AgdaSpace{}%
\AgdaSymbol{())}\<%
\\
\>[0]\AgdaInductiveConstructor{′𝔹}%
\>[4]\AgdaOperator{\AgdaFunction{≡\$?}}\AgdaSpace{}%
\AgdaInductiveConstructor{′𝔹}%
\>[12]\AgdaSymbol{=}%
\>[15]\AgdaInductiveConstructor{yes}\AgdaSpace{}%
\AgdaInductiveConstructor{refl}\<%
\\
\\[\AgdaEmptyExtraSkip]%
\>[0]\AgdaKeyword{infixr}\AgdaSpace{}%
\AgdaNumber{7}\AgdaSpace{}%
\AgdaOperator{\AgdaInductiveConstructor{\AgdaUnderscore{}⇒\AgdaUnderscore{}}}\<%
\\
\>[0]\AgdaKeyword{infix}%
\>[7]\AgdaNumber{8}\AgdaSpace{}%
\AgdaOperator{\AgdaInductiveConstructor{\$ₜ\AgdaUnderscore{}}}\<%
\\
\>[0]\AgdaKeyword{infix}%
\>[7]\AgdaNumber{8}\AgdaSpace{}%
\AgdaOperator{\AgdaInductiveConstructor{\$ᵍ\AgdaUnderscore{}}}\<%
\end{code}

The type structure of the Cast Calculus includes base types (integer and Boolean),
function types, and the unknown type ★. The \emph{ground types} include
just the base types and function types from ★ to ★.

\begin{minipage}{0.5\textwidth}
\begin{code}%
\>[0]\AgdaKeyword{data}\AgdaSpace{}%
\AgdaDatatype{Type}\AgdaSpace{}%
\AgdaSymbol{:}\AgdaSpace{}%
\AgdaPrimitive{Set}\AgdaSpace{}%
\AgdaKeyword{where}\<%
\\
\>[0][@{}l@{\AgdaIndent{0}}]%
\>[2]\AgdaInductiveConstructor{★}\AgdaSpace{}%
\AgdaSymbol{:}\AgdaSpace{}%
\AgdaDatatype{Type}\<%
\\
\>[2]\AgdaOperator{\AgdaInductiveConstructor{\$ₜ\AgdaUnderscore{}}}\AgdaSpace{}%
\AgdaSymbol{:}\AgdaSpace{}%
\AgdaSymbol{(}\AgdaBound{ι}\AgdaSpace{}%
\AgdaSymbol{:}\AgdaSpace{}%
\AgdaDatatype{Base}\AgdaSymbol{)}\AgdaSpace{}%
\AgdaSymbol{→}\AgdaSpace{}%
\AgdaDatatype{Type}\<%
\\
\>[2]\AgdaOperator{\AgdaInductiveConstructor{\AgdaUnderscore{}⇒\AgdaUnderscore{}}}\AgdaSpace{}%
\AgdaSymbol{:}\AgdaSpace{}%
\AgdaSymbol{(}\AgdaBound{A}\AgdaSpace{}%
\AgdaSymbol{:}\AgdaSpace{}%
\AgdaDatatype{Type}\AgdaSymbol{)}\AgdaSpace{}%
\AgdaSymbol{→}\AgdaSpace{}%
\AgdaSymbol{(}\AgdaBound{B}\AgdaSpace{}%
\AgdaSymbol{:}\AgdaSpace{}%
\AgdaDatatype{Type}\AgdaSymbol{)}\AgdaSpace{}%
\AgdaSymbol{→}\AgdaSpace{}%
\AgdaDatatype{Type}\<%
\end{code}
\end{minipage}
\begin{minipage}{0.3\textwidth}
\begin{code}%
\>[0]\AgdaKeyword{data}\AgdaSpace{}%
\AgdaDatatype{Ground}\AgdaSpace{}%
\AgdaSymbol{:}\AgdaSpace{}%
\AgdaPrimitive{Set}\AgdaSpace{}%
\AgdaKeyword{where}\<%
\\
\>[0][@{}l@{\AgdaIndent{0}}]%
\>[2]\AgdaOperator{\AgdaInductiveConstructor{\$ᵍ\AgdaUnderscore{}}}%
\>[7]\AgdaSymbol{:}\AgdaSpace{}%
\AgdaSymbol{(}\AgdaBound{ι}\AgdaSpace{}%
\AgdaSymbol{:}\AgdaSpace{}%
\AgdaDatatype{Base}\AgdaSymbol{)}%
\>[21]\AgdaSymbol{→}\AgdaSpace{}%
\AgdaDatatype{Ground}\<%
\\
\>[2]\AgdaInductiveConstructor{★⇒★}\AgdaSpace{}%
\AgdaSymbol{:}\AgdaSpace{}%
\AgdaDatatype{Ground}\<%
\\
\\[\AgdaEmptyExtraSkip]%
\>[0]\AgdaOperator{\AgdaFunction{⌈\AgdaUnderscore{}⌉}}\AgdaSpace{}%
\AgdaSymbol{:}\AgdaSpace{}%
\AgdaDatatype{Ground}\AgdaSpace{}%
\AgdaSymbol{→}\AgdaSpace{}%
\AgdaDatatype{Type}\<%
\\
\>[0]\AgdaOperator{\AgdaFunction{⌈}}\AgdaSpace{}%
\AgdaOperator{\AgdaInductiveConstructor{\$ᵍ}}\AgdaSpace{}%
\AgdaBound{ι}\AgdaSpace{}%
\AgdaOperator{\AgdaFunction{⌉}}\AgdaSpace{}%
\AgdaSymbol{=}\AgdaSpace{}%
\AgdaOperator{\AgdaInductiveConstructor{\$ₜ}}\AgdaSpace{}%
\AgdaBound{ι}\<%
\\
\>[0]\AgdaOperator{\AgdaFunction{⌈}}\AgdaSpace{}%
\AgdaInductiveConstructor{★⇒★}\AgdaSpace{}%
\AgdaOperator{\AgdaFunction{⌉}}\AgdaSpace{}%
\AgdaSymbol{=}\AgdaSpace{}%
\AgdaInductiveConstructor{★}\AgdaSpace{}%
\AgdaOperator{\AgdaInductiveConstructor{⇒}}\AgdaSpace{}%
\AgdaInductiveConstructor{★}\<%
\end{code}
\end{minipage}

\begin{code}[hide]%
\>[0]\AgdaOperator{\AgdaFunction{\AgdaUnderscore{}≡ᵍ\AgdaUnderscore{}}}\AgdaSpace{}%
\AgdaSymbol{:}\AgdaSpace{}%
\AgdaSymbol{∀}\AgdaSpace{}%
\AgdaSymbol{(}\AgdaBound{G}\AgdaSpace{}%
\AgdaSymbol{:}\AgdaSpace{}%
\AgdaDatatype{Ground}\AgdaSymbol{)}\AgdaSpace{}%
\AgdaSymbol{(}\AgdaBound{H}\AgdaSpace{}%
\AgdaSymbol{:}\AgdaSpace{}%
\AgdaDatatype{Ground}\AgdaSymbol{)}\AgdaSpace{}%
\AgdaSymbol{→}\AgdaSpace{}%
\AgdaRecord{Dec}\AgdaSpace{}%
\AgdaSymbol{(}\AgdaBound{G}\AgdaSpace{}%
\AgdaOperator{\AgdaDatatype{≡}}\AgdaSpace{}%
\AgdaBound{H}\AgdaSymbol{)}\<%
\\
\>[0]\AgdaSymbol{(}\AgdaOperator{\AgdaInductiveConstructor{\$ᵍ}}%
\>[177I]\AgdaBound{ι}\AgdaSymbol{)}\AgdaSpace{}%
\AgdaOperator{\AgdaFunction{≡ᵍ}}\AgdaSpace{}%
\AgdaSymbol{(}\AgdaOperator{\AgdaInductiveConstructor{\$ᵍ}}\AgdaSpace{}%
\AgdaBound{ι′}\AgdaSymbol{)}\<%
\\
\>[.][@{}l@{}]\<[177I]%
\>[4]\AgdaKeyword{with}\AgdaSpace{}%
\AgdaBound{ι}\AgdaSpace{}%
\AgdaOperator{\AgdaFunction{≡\$?}}\AgdaSpace{}%
\AgdaBound{ι′}\<%
\\
\>[0]\AgdaSymbol{...}\AgdaSpace{}%
\AgdaSymbol{|}\AgdaSpace{}%
\AgdaInductiveConstructor{yes}\AgdaSpace{}%
\AgdaInductiveConstructor{refl}\AgdaSpace{}%
\AgdaSymbol{=}\AgdaSpace{}%
\AgdaInductiveConstructor{yes}\AgdaSpace{}%
\AgdaInductiveConstructor{refl}\<%
\\
\>[0]\AgdaSymbol{...}\AgdaSpace{}%
\AgdaSymbol{|}\AgdaSpace{}%
\AgdaInductiveConstructor{no}\AgdaSpace{}%
\AgdaBound{neq}\AgdaSpace{}%
\AgdaSymbol{=}\AgdaSpace{}%
\AgdaInductiveConstructor{no}\AgdaSpace{}%
\AgdaSymbol{λ}\AgdaSpace{}%
\AgdaSymbol{\{}\AgdaInductiveConstructor{refl}\AgdaSpace{}%
\AgdaSymbol{→}\AgdaSpace{}%
\AgdaBound{neq}\AgdaSpace{}%
\AgdaInductiveConstructor{refl}\AgdaSymbol{\}}\<%
\\
\>[0]\AgdaSymbol{(}\AgdaOperator{\AgdaInductiveConstructor{\$ᵍ}}\AgdaSpace{}%
\AgdaBound{ι}\AgdaSymbol{)}\AgdaSpace{}%
\AgdaOperator{\AgdaFunction{≡ᵍ}}\AgdaSpace{}%
\AgdaInductiveConstructor{★⇒★}\AgdaSpace{}%
\AgdaSymbol{=}\AgdaSpace{}%
\AgdaInductiveConstructor{no}\AgdaSpace{}%
\AgdaSymbol{λ}\AgdaSpace{}%
\AgdaSymbol{()}\<%
\\
\>[0]\AgdaInductiveConstructor{★⇒★}\AgdaSpace{}%
\AgdaOperator{\AgdaFunction{≡ᵍ}}\AgdaSpace{}%
\AgdaSymbol{(}\AgdaOperator{\AgdaInductiveConstructor{\$ᵍ}}\AgdaSpace{}%
\AgdaBound{ι}\AgdaSymbol{)}\AgdaSpace{}%
\AgdaSymbol{=}\AgdaSpace{}%
\AgdaInductiveConstructor{no}\AgdaSpace{}%
\AgdaSymbol{λ}\AgdaSpace{}%
\AgdaSymbol{()}\<%
\\
\>[0]\AgdaInductiveConstructor{★⇒★}\AgdaSpace{}%
\AgdaOperator{\AgdaFunction{≡ᵍ}}\AgdaSpace{}%
\AgdaInductiveConstructor{★⇒★}\AgdaSpace{}%
\AgdaSymbol{=}\AgdaSpace{}%
\AgdaInductiveConstructor{yes}\AgdaSpace{}%
\AgdaInductiveConstructor{refl}\<%
\end{code}

There are three special features in the Cast Calculus:
\begin{enumerate}
\item injection $M ⟨ G !⟩$, for casting from a ground type $G$
  to the unknown type ★,
\item projection $M ⟨ H ?⟩$, for casting from the unknown type ★
  to a ground type $H$, and
\item \textsf{blame} which represents a runtime exception if
  a projection fails.
\end{enumerate}
This Cast Calclulus differs from many of those in the
literature~\cite{Siek:2006bh,Siek:2021vf,Siek:2015ac} in that it does
not include casts from one function type to another, a choice that
reduces the number of reduction rules and simplifies the technical
development.  However, casts from one function type to another can be
compiled into a combination of lambda abstractions, injections, and
projections.

We define the terms of the Cast Calculus in Agda using the Abstract
Binding Tree (ABT) library by instantiating it with an appropriate set
of operators together with a description of their arity and
variable-binding structure. To that end, the following \texttt{Op}
data type includes one constructor for each term constructor that we
have in mind for the Cast Calculus, except for variables which are
always present. The \texttt{sig} function, shown below, specifies the
variable binding structure for each operator. It returns a list of
natural numbers where ■ represents zero and ν represents
successor. The list includes one number for each subterm; the number
specifices how many variables come into scope for that subterm. Lambda
abstraction (\texttt{op-lam}) has a single subterm and brings one
variable into scope, whereas application (\texttt{op-app}) has two
subterms but does not bind any variables.

\begin{minipage}{0.3\textwidth}
\begin{code}%
\>[0]\AgdaKeyword{data}\AgdaSpace{}%
\AgdaDatatype{Lit}\AgdaSpace{}%
\AgdaSymbol{:}\AgdaSpace{}%
\AgdaPrimitive{Set}\AgdaSpace{}%
\AgdaKeyword{where}\<%
\\
\>[0][@{}l@{\AgdaIndent{0}}]%
\>[2]\AgdaInductiveConstructor{Num}\AgdaSpace{}%
\AgdaSymbol{:}\AgdaSpace{}%
\AgdaDatatype{ℕ}\AgdaSpace{}%
\AgdaSymbol{→}\AgdaSpace{}%
\AgdaDatatype{Lit}\<%
\\
\>[2]\AgdaInductiveConstructor{Bool}\AgdaSpace{}%
\AgdaSymbol{:}\AgdaSpace{}%
\AgdaDatatype{𝔹}\AgdaSpace{}%
\AgdaSymbol{→}\AgdaSpace{}%
\AgdaDatatype{Lit}\<%
\end{code}
\end{minipage}
\begin{minipage}{0.3\textwidth}
\begin{code}%
\>[0]\AgdaKeyword{data}\AgdaSpace{}%
\AgdaDatatype{Op}\AgdaSpace{}%
\AgdaSymbol{:}\AgdaSpace{}%
\AgdaPrimitive{Set}\AgdaSpace{}%
\AgdaKeyword{where}\<%
\\
\>[0][@{}l@{\AgdaIndent{0}}]%
\>[2]\AgdaInductiveConstructor{op-lam}\AgdaSpace{}%
\AgdaSymbol{:}\AgdaSpace{}%
\AgdaDatatype{Op}\<%
\\
\>[2]\AgdaInductiveConstructor{op-app}\AgdaSpace{}%
\AgdaSymbol{:}\AgdaSpace{}%
\AgdaDatatype{Op}\<%
\\
\>[2]\AgdaInductiveConstructor{op-lit}\AgdaSpace{}%
\AgdaSymbol{:}\AgdaSpace{}%
\AgdaDatatype{Lit}\AgdaSpace{}%
\AgdaSymbol{→}\AgdaSpace{}%
\AgdaDatatype{Op}\<%
\\
\>[2]\AgdaInductiveConstructor{op-inject}\AgdaSpace{}%
\AgdaSymbol{:}\AgdaSpace{}%
\AgdaDatatype{Ground}\AgdaSpace{}%
\AgdaSymbol{→}\AgdaSpace{}%
\AgdaDatatype{Op}\<%
\\
\>[2]\AgdaInductiveConstructor{op-project}\AgdaSpace{}%
\AgdaSymbol{:}\AgdaSpace{}%
\AgdaDatatype{Ground}\AgdaSpace{}%
\AgdaSymbol{→}\AgdaSpace{}%
\AgdaDatatype{Op}\<%
\\
\>[2]\AgdaInductiveConstructor{op-blame}\AgdaSpace{}%
\AgdaSymbol{:}\AgdaSpace{}%
\AgdaDatatype{Op}\<%
\end{code}
\end{minipage}
\begin{minipage}{0.3\textwidth}
\begin{code}%
\>[0]\AgdaFunction{sig}\AgdaSpace{}%
\AgdaSymbol{:}\AgdaSpace{}%
\AgdaDatatype{Op}\AgdaSpace{}%
\AgdaSymbol{→}\AgdaSpace{}%
\AgdaDatatype{List}\AgdaSpace{}%
\AgdaDatatype{Sig}\<%
\\
\>[0]\AgdaFunction{sig}\AgdaSpace{}%
\AgdaInductiveConstructor{op-lam}\AgdaSpace{}%
\AgdaSymbol{=}\AgdaSpace{}%
\AgdaSymbol{(}\AgdaInductiveConstructor{ν}\AgdaSpace{}%
\AgdaInductiveConstructor{■}\AgdaSymbol{)}\AgdaSpace{}%
\AgdaOperator{\AgdaInductiveConstructor{∷}}\AgdaSpace{}%
\AgdaInductiveConstructor{[]}\<%
\\
\>[0]\AgdaFunction{sig}\AgdaSpace{}%
\AgdaInductiveConstructor{op-app}\AgdaSpace{}%
\AgdaSymbol{=}\AgdaSpace{}%
\AgdaInductiveConstructor{■}\AgdaSpace{}%
\AgdaOperator{\AgdaInductiveConstructor{∷}}\AgdaSpace{}%
\AgdaInductiveConstructor{■}\AgdaSpace{}%
\AgdaOperator{\AgdaInductiveConstructor{∷}}\AgdaSpace{}%
\AgdaInductiveConstructor{[]}\<%
\\
\>[0]\AgdaFunction{sig}\AgdaSpace{}%
\AgdaSymbol{(}\AgdaInductiveConstructor{op-lit}\AgdaSpace{}%
\AgdaBound{c}\AgdaSymbol{)}\AgdaSpace{}%
\AgdaSymbol{=}\AgdaSpace{}%
\AgdaInductiveConstructor{[]}\<%
\\
\>[0]\AgdaFunction{sig}\AgdaSpace{}%
\AgdaSymbol{(}\AgdaInductiveConstructor{op-inject}\AgdaSpace{}%
\AgdaBound{G}\AgdaSymbol{)}\AgdaSpace{}%
\AgdaSymbol{=}\AgdaSpace{}%
\AgdaInductiveConstructor{■}\AgdaSpace{}%
\AgdaOperator{\AgdaInductiveConstructor{∷}}\AgdaSpace{}%
\AgdaInductiveConstructor{[]}\<%
\\
\>[0]\AgdaFunction{sig}\AgdaSpace{}%
\AgdaSymbol{(}\AgdaInductiveConstructor{op-project}\AgdaSpace{}%
\AgdaBound{H}\AgdaSymbol{)}\AgdaSpace{}%
\AgdaSymbol{=}\AgdaSpace{}%
\AgdaInductiveConstructor{■}\AgdaSpace{}%
\AgdaOperator{\AgdaInductiveConstructor{∷}}\AgdaSpace{}%
\AgdaInductiveConstructor{[]}\<%
\\
\>[0]\AgdaFunction{sig}\AgdaSpace{}%
\AgdaSymbol{(}\AgdaInductiveConstructor{op-blame}\AgdaSymbol{)}\AgdaSpace{}%
\AgdaSymbol{=}\AgdaSpace{}%
\AgdaInductiveConstructor{[]}\<%
\end{code}
\end{minipage}

\noindent We instantiate the ABT library as follows, by applying it to
\texttt{Op} and \texttt{sig}. We rename the resulting \texttt{ABT}
type to \texttt{Term}.

\begin{code}%
\>[0]\AgdaKeyword{open}\AgdaSpace{}%
\AgdaKeyword{import}\AgdaSpace{}%
\AgdaModule{rewriting.AbstractBindingTree}\AgdaSpace{}%
\AgdaDatatype{Op}\AgdaSpace{}%
\AgdaFunction{sig}\AgdaSpace{}%
\AgdaKeyword{renaming}\AgdaSpace{}%
\AgdaSymbol{(}\AgdaDatatype{ABT}\AgdaSpace{}%
\AgdaSymbol{to}\AgdaSpace{}%
\AgdaDatatype{Term}\AgdaSymbol{)}\AgdaSpace{}%
\AgdaKeyword{public}\<%
\end{code}

\noindent We define Agda patterns to give succinct syntax to the
construction of abstract binding trees.

\begin{code}%
\>[0]\AgdaKeyword{pattern}\AgdaSpace{}%
\AgdaInductiveConstructor{ƛ}\AgdaSpace{}%
\AgdaBound{N}\AgdaSpace{}%
\AgdaSymbol{=}\AgdaSpace{}%
\AgdaInductiveConstructor{op-lam}\AgdaSpace{}%
\AgdaOperator{\AgdaInductiveConstructor{⦅}}\AgdaSpace{}%
\AgdaInductiveConstructor{cons}\AgdaSpace{}%
\AgdaSymbol{(}\AgdaInductiveConstructor{bind}\AgdaSpace{}%
\AgdaSymbol{(}\AgdaInductiveConstructor{ast}\AgdaSpace{}%
\AgdaBound{N}\AgdaSymbol{))}\AgdaSpace{}%
\AgdaInductiveConstructor{nil}\AgdaSpace{}%
\AgdaOperator{\AgdaInductiveConstructor{⦆}}\<%
\\
\>[0]\AgdaKeyword{infixl}\AgdaSpace{}%
\AgdaNumber{7}%
\>[10]\AgdaOperator{\AgdaInductiveConstructor{\AgdaUnderscore{}·\AgdaUnderscore{}}}\<%
\\
\>[0]\AgdaKeyword{pattern}\AgdaSpace{}%
\AgdaOperator{\AgdaInductiveConstructor{\AgdaUnderscore{}·\AgdaUnderscore{}}}\AgdaSpace{}%
\AgdaBound{L}\AgdaSpace{}%
\AgdaBound{M}\AgdaSpace{}%
\AgdaSymbol{=}\AgdaSpace{}%
\AgdaInductiveConstructor{op-app}\AgdaSpace{}%
\AgdaOperator{\AgdaInductiveConstructor{⦅}}\AgdaSpace{}%
\AgdaInductiveConstructor{cons}\AgdaSpace{}%
\AgdaSymbol{(}\AgdaInductiveConstructor{ast}\AgdaSpace{}%
\AgdaBound{L}\AgdaSymbol{)}\AgdaSpace{}%
\AgdaSymbol{(}\AgdaInductiveConstructor{cons}\AgdaSpace{}%
\AgdaSymbol{(}\AgdaInductiveConstructor{ast}\AgdaSpace{}%
\AgdaBound{M}\AgdaSymbol{)}\AgdaSpace{}%
\AgdaInductiveConstructor{nil}\AgdaSymbol{)}\AgdaSpace{}%
\AgdaOperator{\AgdaInductiveConstructor{⦆}}\<%
\\
\>[0]\AgdaKeyword{pattern}\AgdaSpace{}%
\AgdaInductiveConstructor{\$}\AgdaSpace{}%
\AgdaBound{c}\AgdaSpace{}%
\AgdaSymbol{=}\AgdaSpace{}%
\AgdaSymbol{(}\AgdaInductiveConstructor{op-lit}\AgdaSpace{}%
\AgdaBound{c}\AgdaSymbol{)}\AgdaSpace{}%
\AgdaOperator{\AgdaInductiveConstructor{⦅}}\AgdaSpace{}%
\AgdaInductiveConstructor{nil}\AgdaSpace{}%
\AgdaOperator{\AgdaInductiveConstructor{⦆}}\<%
\\
\>[0]\AgdaKeyword{pattern}\AgdaSpace{}%
\AgdaOperator{\AgdaInductiveConstructor{\AgdaUnderscore{}⟨\AgdaUnderscore{}!⟩}}\AgdaSpace{}%
\AgdaBound{M}\AgdaSpace{}%
\AgdaBound{G}\AgdaSpace{}%
\AgdaSymbol{=}\AgdaSpace{}%
\AgdaSymbol{(}\AgdaInductiveConstructor{op-inject}\AgdaSpace{}%
\AgdaBound{G}\AgdaSymbol{)}\AgdaSpace{}%
\AgdaOperator{\AgdaInductiveConstructor{⦅}}\AgdaSpace{}%
\AgdaInductiveConstructor{cons}\AgdaSpace{}%
\AgdaSymbol{(}\AgdaInductiveConstructor{ast}\AgdaSpace{}%
\AgdaBound{M}\AgdaSymbol{)}\AgdaSpace{}%
\AgdaInductiveConstructor{nil}\AgdaSpace{}%
\AgdaOperator{\AgdaInductiveConstructor{⦆}}\<%
\\
\>[0]\AgdaKeyword{pattern}\AgdaSpace{}%
\AgdaOperator{\AgdaInductiveConstructor{\AgdaUnderscore{}⟨\AgdaUnderscore{}?⟩}}\AgdaSpace{}%
\AgdaBound{M}\AgdaSpace{}%
\AgdaBound{H}\AgdaSpace{}%
\AgdaSymbol{=}\AgdaSpace{}%
\AgdaSymbol{(}\AgdaInductiveConstructor{op-project}\AgdaSpace{}%
\AgdaBound{H}\AgdaSymbol{)}\AgdaSpace{}%
\AgdaOperator{\AgdaInductiveConstructor{⦅}}\AgdaSpace{}%
\AgdaInductiveConstructor{cons}\AgdaSpace{}%
\AgdaSymbol{(}\AgdaInductiveConstructor{ast}\AgdaSpace{}%
\AgdaBound{M}\AgdaSymbol{)}\AgdaSpace{}%
\AgdaInductiveConstructor{nil}\AgdaSpace{}%
\AgdaOperator{\AgdaInductiveConstructor{⦆}}\<%
\\
\>[0]\AgdaKeyword{pattern}\AgdaSpace{}%
\AgdaInductiveConstructor{blame}\AgdaSpace{}%
\AgdaSymbol{=}\AgdaSpace{}%
\AgdaInductiveConstructor{op-blame}\AgdaSpace{}%
\AgdaOperator{\AgdaInductiveConstructor{⦅}}\AgdaSpace{}%
\AgdaInductiveConstructor{nil}\AgdaSpace{}%
\AgdaOperator{\AgdaInductiveConstructor{⦆}}\<%
\end{code}

\begin{code}[hide]%
\>[0]\AgdaSymbol{\{-\#}\AgdaSpace{}%
\AgdaKeyword{REWRITE}\AgdaSpace{}%
\AgdaFunction{sub-var}\AgdaSpace{}%
\AgdaSymbol{\#-\}}\<%
\end{code}

The ABT library represents variables with de Bruijn indices and
provides a definition of parallel substitution and many theorems about
substitution. The de Bruijn indices are represented directly as
natural numbers in Agda, with the constructors \texttt{0} and
\texttt{suc} for zero and successor. It is helpful to think of a
parallel substitution as an infinite stream, or equivalently, as a
function from natural numbers (de Bruijn indices) to terms. The •
operator is stream cons, that is, it adds a term to the front of the
stream.

\begin{code}%
\>[0]\AgdaFunction{sub-zero}\AgdaSpace{}%
\AgdaSymbol{:}\AgdaSpace{}%
\AgdaSymbol{∀}\AgdaSpace{}%
\AgdaSymbol{\{}\AgdaBound{V}\AgdaSpace{}%
\AgdaBound{σ}\AgdaSymbol{\}}\AgdaSpace{}%
\AgdaSymbol{→}\AgdaSpace{}%
\AgdaSymbol{(}\AgdaBound{V}\AgdaSpace{}%
\AgdaOperator{\AgdaFunction{•}}\AgdaSpace{}%
\AgdaBound{σ}\AgdaSymbol{)}\AgdaSpace{}%
\AgdaNumber{0}\AgdaSpace{}%
\AgdaOperator{\AgdaDatatype{≡}}\AgdaSpace{}%
\AgdaBound{V}\<%
\\
\>[0]\AgdaFunction{sub-zero}\AgdaSpace{}%
\AgdaSymbol{=}\AgdaSpace{}%
\AgdaInductiveConstructor{refl}\<%
\\
\>[0]\AgdaFunction{sub-suc}\AgdaSpace{}%
\AgdaSymbol{:}\AgdaSpace{}%
\AgdaSymbol{∀}\AgdaSpace{}%
\AgdaSymbol{\{}\AgdaBound{V}\AgdaSpace{}%
\AgdaBound{x}\AgdaSpace{}%
\AgdaBound{σ}\AgdaSymbol{\}}\AgdaSpace{}%
\AgdaSymbol{→}\AgdaSpace{}%
\AgdaSymbol{(}\AgdaBound{V}\AgdaSpace{}%
\AgdaOperator{\AgdaFunction{•}}\AgdaSpace{}%
\AgdaBound{σ}\AgdaSymbol{)}\AgdaSpace{}%
\AgdaSymbol{(}\AgdaInductiveConstructor{suc}\AgdaSpace{}%
\AgdaBound{x}\AgdaSymbol{)}\AgdaSpace{}%
\AgdaOperator{\AgdaDatatype{≡}}\AgdaSpace{}%
\AgdaBound{σ}\AgdaSpace{}%
\AgdaBound{x}\<%
\\
\>[0]\AgdaFunction{sub-suc}\AgdaSpace{}%
\AgdaSymbol{\{}\AgdaBound{V}\AgdaSymbol{\}\{}\AgdaBound{x}\AgdaSymbol{\}\{}\AgdaBound{σ}\AgdaSymbol{\}}\AgdaSpace{}%
\AgdaSymbol{=}\AgdaSpace{}%
\AgdaInductiveConstructor{refl}\<%
\end{code}

\noindent The ABT library provides the operator $⟪ σ ⟫ M$ for applying
a substitution to a term. Here are the equations for substitution
applied to variables, application, and lambda abstraction. The
\textsf{ext} operator transports a substitution over one variable
binder.

\begin{code}%
\>[0]\AgdaFunction{\AgdaUnderscore{}}\AgdaSpace{}%
\AgdaSymbol{:}\AgdaSpace{}%
\AgdaSymbol{∀}\AgdaSpace{}%
\AgdaSymbol{(}\AgdaBound{σ}\AgdaSpace{}%
\AgdaSymbol{:}\AgdaSpace{}%
\AgdaFunction{Subst}\AgdaSymbol{)}\AgdaSpace{}%
\AgdaSymbol{(}\AgdaBound{x}\AgdaSpace{}%
\AgdaSymbol{:}\AgdaSpace{}%
\AgdaDatatype{ℕ}\AgdaSymbol{)}\AgdaSpace{}%
\AgdaSymbol{→}\AgdaSpace{}%
\AgdaOperator{\AgdaFunction{⟪}}\AgdaSpace{}%
\AgdaBound{σ}\AgdaSpace{}%
\AgdaOperator{\AgdaFunction{⟫}}\AgdaSpace{}%
\AgdaSymbol{(}\AgdaOperator{\AgdaInductiveConstructor{`}}\AgdaSpace{}%
\AgdaBound{x}\AgdaSymbol{)}\AgdaSpace{}%
\AgdaOperator{\AgdaDatatype{≡}}\AgdaSpace{}%
\AgdaBound{σ}\AgdaSpace{}%
\AgdaBound{x}\<%
\\
\>[0]\AgdaSymbol{\AgdaUnderscore{}}\AgdaSpace{}%
\AgdaSymbol{=}\AgdaSpace{}%
\AgdaSymbol{λ}\AgdaSpace{}%
\AgdaBound{σ}\AgdaSpace{}%
\AgdaBound{x}\AgdaSpace{}%
\AgdaSymbol{→}\AgdaSpace{}%
\AgdaInductiveConstructor{refl}\<%
\\
\>[0]\AgdaFunction{\AgdaUnderscore{}}\AgdaSpace{}%
\AgdaSymbol{:}\AgdaSpace{}%
\AgdaSymbol{∀}\AgdaSpace{}%
\AgdaSymbol{(}\AgdaBound{σ}\AgdaSpace{}%
\AgdaSymbol{:}\AgdaSpace{}%
\AgdaFunction{Subst}\AgdaSymbol{)}\AgdaSpace{}%
\AgdaSymbol{(}\AgdaBound{L}\AgdaSpace{}%
\AgdaBound{M}\AgdaSpace{}%
\AgdaSymbol{:}\AgdaSpace{}%
\AgdaDatatype{Term}\AgdaSymbol{)}\AgdaSpace{}%
\AgdaSymbol{→}\AgdaSpace{}%
\AgdaOperator{\AgdaFunction{⟪}}\AgdaSpace{}%
\AgdaBound{σ}\AgdaSpace{}%
\AgdaOperator{\AgdaFunction{⟫}}\AgdaSpace{}%
\AgdaSymbol{(}\AgdaBound{L}\AgdaSpace{}%
\AgdaOperator{\AgdaInductiveConstructor{·}}\AgdaSpace{}%
\AgdaBound{M}\AgdaSymbol{)}\AgdaSpace{}%
\AgdaOperator{\AgdaDatatype{≡}}\AgdaSpace{}%
\AgdaOperator{\AgdaFunction{⟪}}\AgdaSpace{}%
\AgdaBound{σ}\AgdaSpace{}%
\AgdaOperator{\AgdaFunction{⟫}}\AgdaSpace{}%
\AgdaBound{L}\AgdaSpace{}%
\AgdaOperator{\AgdaInductiveConstructor{·}}\AgdaSpace{}%
\AgdaOperator{\AgdaFunction{⟪}}\AgdaSpace{}%
\AgdaBound{σ}\AgdaSpace{}%
\AgdaOperator{\AgdaFunction{⟫}}\AgdaSpace{}%
\AgdaBound{M}\<%
\\
\>[0]\AgdaSymbol{\AgdaUnderscore{}}\AgdaSpace{}%
\AgdaSymbol{=}\AgdaSpace{}%
\AgdaSymbol{λ}\AgdaSpace{}%
\AgdaBound{σ}\AgdaSpace{}%
\AgdaBound{L}\AgdaSpace{}%
\AgdaBound{M}\AgdaSpace{}%
\AgdaSymbol{→}\AgdaSpace{}%
\AgdaInductiveConstructor{refl}\<%
\\
\>[0]\AgdaFunction{\AgdaUnderscore{}}\AgdaSpace{}%
\AgdaSymbol{:}\AgdaSpace{}%
\AgdaSymbol{∀}\AgdaSpace{}%
\AgdaSymbol{(}\AgdaBound{σ}\AgdaSpace{}%
\AgdaSymbol{:}\AgdaSpace{}%
\AgdaFunction{Subst}\AgdaSymbol{)}\AgdaSpace{}%
\AgdaSymbol{(}\AgdaBound{N}\AgdaSpace{}%
\AgdaSymbol{:}\AgdaSpace{}%
\AgdaDatatype{Term}\AgdaSymbol{)}\AgdaSpace{}%
\AgdaSymbol{→}\AgdaSpace{}%
\AgdaOperator{\AgdaFunction{⟪}}\AgdaSpace{}%
\AgdaBound{σ}\AgdaSpace{}%
\AgdaOperator{\AgdaFunction{⟫}}\AgdaSpace{}%
\AgdaSymbol{(}\AgdaInductiveConstructor{ƛ}\AgdaSpace{}%
\AgdaBound{N}\AgdaSymbol{)}\AgdaSpace{}%
\AgdaOperator{\AgdaDatatype{≡}}\AgdaSpace{}%
\AgdaInductiveConstructor{ƛ}\AgdaSpace{}%
\AgdaSymbol{(}\AgdaOperator{\AgdaFunction{⟪}}\AgdaSpace{}%
\AgdaFunction{ext}\AgdaSpace{}%
\AgdaBound{σ}\AgdaSpace{}%
\AgdaOperator{\AgdaFunction{⟫}}\AgdaSpace{}%
\AgdaBound{N}\AgdaSymbol{)}\<%
\\
\>[0]\AgdaSymbol{\AgdaUnderscore{}}\AgdaSpace{}%
\AgdaSymbol{=}\AgdaSpace{}%
\AgdaSymbol{λ}\AgdaSpace{}%
\AgdaBound{σ}\AgdaSpace{}%
\AgdaBound{N}\AgdaSpace{}%
\AgdaSymbol{→}\AgdaSpace{}%
\AgdaInductiveConstructor{refl}\<%
\end{code}

\noindent The bracket notation $M [ N ]$ is defined to replace the
occurrences of variable 0 in $M$ with $N$ and decrement the other free
variables. For example,

\begin{code}%
\>[0]\AgdaFunction{\AgdaUnderscore{}}\AgdaSpace{}%
\AgdaSymbol{:}\AgdaSpace{}%
\AgdaSymbol{∀}\AgdaSpace{}%
\AgdaSymbol{(}\AgdaBound{N}\AgdaSpace{}%
\AgdaSymbol{:}\AgdaSpace{}%
\AgdaDatatype{Term}\AgdaSymbol{)}\AgdaSpace{}%
\AgdaSymbol{→}\AgdaSpace{}%
\AgdaSymbol{(}\AgdaOperator{\AgdaInductiveConstructor{`}}\AgdaSpace{}%
\AgdaNumber{1}\AgdaSpace{}%
\AgdaOperator{\AgdaInductiveConstructor{·}}\AgdaSpace{}%
\AgdaOperator{\AgdaInductiveConstructor{`}}\AgdaSpace{}%
\AgdaNumber{0}\AgdaSymbol{)}\AgdaSpace{}%
\AgdaOperator{\AgdaFunction{[}}\AgdaSpace{}%
\AgdaBound{N}\AgdaSpace{}%
\AgdaOperator{\AgdaFunction{]}}\AgdaSpace{}%
\AgdaOperator{\AgdaDatatype{≡}}\AgdaSpace{}%
\AgdaSymbol{(}\AgdaOperator{\AgdaInductiveConstructor{`}}\AgdaSpace{}%
\AgdaNumber{0}\AgdaSpace{}%
\AgdaOperator{\AgdaInductiveConstructor{·}}\AgdaSpace{}%
\AgdaBound{N}\AgdaSymbol{)}\<%
\\
\>[0]\AgdaSymbol{\AgdaUnderscore{}}\AgdaSpace{}%
\AgdaSymbol{=}\AgdaSpace{}%
\AgdaSymbol{λ}\AgdaSpace{}%
\AgdaBound{N}\AgdaSpace{}%
\AgdaSymbol{→}\AgdaSpace{}%
\AgdaInductiveConstructor{refl}\<%
\end{code}

\noindent Most importantly, the ABT library provides the following
theorem which is both difficult to prove and needed later for the
Compatibility Lemma for lambda abstraction.

\begin{code}%
\>[0]\AgdaFunction{ext-sub-cons}\AgdaSpace{}%
\AgdaSymbol{:}\AgdaSpace{}%
\AgdaSymbol{∀}\AgdaSpace{}%
\AgdaSymbol{\{}\AgdaBound{σ}\AgdaSpace{}%
\AgdaBound{N}\AgdaSpace{}%
\AgdaBound{V}\AgdaSymbol{\}}\AgdaSpace{}%
\AgdaSymbol{→}\AgdaSpace{}%
\AgdaSymbol{(}\AgdaOperator{\AgdaFunction{⟪}}\AgdaSpace{}%
\AgdaFunction{ext}\AgdaSpace{}%
\AgdaBound{σ}\AgdaSpace{}%
\AgdaOperator{\AgdaFunction{⟫}}\AgdaSpace{}%
\AgdaBound{N}\AgdaSymbol{)}\AgdaSpace{}%
\AgdaOperator{\AgdaFunction{[}}\AgdaSpace{}%
\AgdaBound{V}\AgdaSpace{}%
\AgdaOperator{\AgdaFunction{]}}\AgdaSpace{}%
\AgdaOperator{\AgdaDatatype{≡}}\AgdaSpace{}%
\AgdaOperator{\AgdaFunction{⟪}}\AgdaSpace{}%
\AgdaBound{V}\AgdaSpace{}%
\AgdaOperator{\AgdaFunction{•}}\AgdaSpace{}%
\AgdaBound{σ}\AgdaSpace{}%
\AgdaOperator{\AgdaFunction{⟫}}\AgdaSpace{}%
\AgdaBound{N}\<%
\\
\>[0]\AgdaFunction{ext-sub-cons}\AgdaSpace{}%
\AgdaSymbol{=}\AgdaSpace{}%
\AgdaInductiveConstructor{refl}\<%
\end{code}

\begin{code}[hide]%
\>[0]\AgdaComment{\{-\ Phil:\ consider\ ditching\ this\ and\ use\ M\ ≡\ blame\ -\}}\<%
\\
\>[0]\AgdaKeyword{data}\AgdaSpace{}%
\AgdaDatatype{Blame}\AgdaSpace{}%
\AgdaSymbol{:}\AgdaSpace{}%
\AgdaDatatype{Term}\AgdaSpace{}%
\AgdaSymbol{→}\AgdaSpace{}%
\AgdaPrimitive{Set}\AgdaSpace{}%
\AgdaKeyword{where}\<%
\\
\>[0][@{}l@{\AgdaIndent{0}}]%
\>[2]\AgdaInductiveConstructor{isBlame}\AgdaSpace{}%
\AgdaSymbol{:}\AgdaSpace{}%
\AgdaDatatype{Blame}\AgdaSpace{}%
\AgdaInductiveConstructor{blame}\<%
\\
\\[\AgdaEmptyExtraSkip]%
\>[0]\AgdaComment{\{---------------------\ Values\ -----------------------------\}}\<%
\\
\\[\AgdaEmptyExtraSkip]%
\>[0]\AgdaKeyword{data}\AgdaSpace{}%
\AgdaDatatype{Value}\AgdaSpace{}%
\AgdaSymbol{:}\AgdaSpace{}%
\AgdaDatatype{Term}\AgdaSpace{}%
\AgdaSymbol{→}\AgdaSpace{}%
\AgdaPrimitive{Set}\AgdaSpace{}%
\AgdaKeyword{where}\<%
\\
\>[0][@{}l@{\AgdaIndent{0}}]%
\>[2]\AgdaOperator{\AgdaInductiveConstructor{ƛ̬\AgdaUnderscore{}}}\AgdaSpace{}%
\AgdaSymbol{:}\AgdaSpace{}%
\AgdaSymbol{(}\AgdaBound{N}\AgdaSpace{}%
\AgdaSymbol{:}\AgdaSpace{}%
\AgdaDatatype{Term}\AgdaSymbol{)}\AgdaSpace{}%
\AgdaSymbol{→}\AgdaSpace{}%
\AgdaDatatype{Value}\AgdaSpace{}%
\AgdaSymbol{(}\AgdaInductiveConstructor{ƛ}\AgdaSpace{}%
\AgdaBound{N}\AgdaSymbol{)}\<%
\\
\>[2]\AgdaInductiveConstructor{\$̬}\AgdaSpace{}%
\AgdaSymbol{:}\AgdaSpace{}%
\AgdaSymbol{(}\AgdaBound{c}\AgdaSpace{}%
\AgdaSymbol{:}\AgdaSpace{}%
\AgdaDatatype{Lit}\AgdaSymbol{)}\AgdaSpace{}%
\AgdaSymbol{→}\AgdaSpace{}%
\AgdaDatatype{Value}\AgdaSpace{}%
\AgdaSymbol{(}\AgdaInductiveConstructor{\$}\AgdaSpace{}%
\AgdaBound{c}\AgdaSymbol{)}\<%
\\
\>[2]\AgdaOperator{\AgdaInductiveConstructor{\AgdaUnderscore{}〈\AgdaUnderscore{}〉}}\AgdaSpace{}%
\AgdaSymbol{:}\AgdaSpace{}%
\AgdaSymbol{∀\{}\AgdaBound{V}\AgdaSymbol{\}}\AgdaSpace{}%
\AgdaSymbol{→}\AgdaSpace{}%
\AgdaDatatype{Value}\AgdaSpace{}%
\AgdaBound{V}\AgdaSpace{}%
\AgdaSymbol{→}\AgdaSpace{}%
\AgdaSymbol{(}\AgdaBound{G}\AgdaSpace{}%
\AgdaSymbol{:}\AgdaSpace{}%
\AgdaDatatype{Ground}\AgdaSymbol{)}\AgdaSpace{}%
\AgdaSymbol{→}\AgdaSpace{}%
\AgdaDatatype{Value}\AgdaSpace{}%
\AgdaSymbol{(}\AgdaBound{V}\AgdaSpace{}%
\AgdaOperator{\AgdaInductiveConstructor{⟨}}\AgdaSpace{}%
\AgdaBound{G}\AgdaSpace{}%
\AgdaOperator{\AgdaInductiveConstructor{!⟩}}\AgdaSymbol{)}\<%
\\
\\[\AgdaEmptyExtraSkip]%
\>[0]\AgdaFunction{value}\AgdaSpace{}%
\AgdaSymbol{:}\AgdaSpace{}%
\AgdaSymbol{∀}\AgdaSpace{}%
\AgdaSymbol{\{}\AgdaBound{V}\AgdaSpace{}%
\AgdaSymbol{:}\AgdaSpace{}%
\AgdaDatatype{Term}\AgdaSymbol{\}}\<%
\\
\>[0][@{}l@{\AgdaIndent{0}}]%
\>[2]\AgdaSymbol{→}%
\>[577I]\AgdaSymbol{(}\AgdaBound{v}\AgdaSpace{}%
\AgdaSymbol{:}\AgdaSpace{}%
\AgdaDatatype{Value}\AgdaSpace{}%
\AgdaBound{V}\AgdaSymbol{)}\<%
\\
\>[.][@{}l@{}]\<[577I]%
\>[4]\AgdaComment{-------------}\<%
\\
\>[2]\AgdaSymbol{→}\AgdaSpace{}%
\AgdaDatatype{Term}\<%
\\
\>[0]\AgdaFunction{value}\AgdaSpace{}%
\AgdaSymbol{\{}\AgdaArgument{V}\AgdaSpace{}%
\AgdaSymbol{=}\AgdaSpace{}%
\AgdaBound{V}\AgdaSymbol{\}}\AgdaSpace{}%
\AgdaBound{v}%
\>[17]\AgdaSymbol{=}%
\>[20]\AgdaBound{V}\<%
\\
\\[\AgdaEmptyExtraSkip]%
\>[0]\AgdaFunction{rename-val}\AgdaSpace{}%
\AgdaSymbol{:}\AgdaSpace{}%
\AgdaSymbol{∀}\AgdaSpace{}%
\AgdaSymbol{\{}\AgdaBound{V}\AgdaSpace{}%
\AgdaSymbol{:}\AgdaSpace{}%
\AgdaDatatype{Term}\AgdaSymbol{\}}\<%
\\
\>[0][@{}l@{\AgdaIndent{0}}]%
\>[2]\AgdaSymbol{→}\AgdaSpace{}%
\AgdaSymbol{(}\AgdaBound{ρ}\AgdaSpace{}%
\AgdaSymbol{:}\AgdaSpace{}%
\AgdaFunction{Rename}\AgdaSymbol{)}\<%
\\
\>[2]\AgdaSymbol{→}%
\>[594I]\AgdaDatatype{Value}\AgdaSpace{}%
\AgdaBound{V}\<%
\\
\>[.][@{}l@{}]\<[594I]%
\>[4]\AgdaComment{------------------}\<%
\\
\>[2]\AgdaSymbol{→}\AgdaSpace{}%
\AgdaDatatype{Value}\AgdaSpace{}%
\AgdaSymbol{(}\AgdaFunction{rename}\AgdaSpace{}%
\AgdaBound{ρ}\AgdaSpace{}%
\AgdaBound{V}\AgdaSymbol{)}\<%
\\
\>[0]\AgdaFunction{rename-val}\AgdaSpace{}%
\AgdaBound{ρ}\AgdaSpace{}%
\AgdaSymbol{(}\AgdaOperator{\AgdaInductiveConstructor{ƛ̬}}\AgdaSpace{}%
\AgdaBound{N}\AgdaSymbol{)}%
\>[23]\AgdaSymbol{=}%
\>[26]\AgdaOperator{\AgdaInductiveConstructor{ƛ̬}}\AgdaSpace{}%
\AgdaSymbol{(}\AgdaFunction{rename}\AgdaSpace{}%
\AgdaSymbol{(}\AgdaFunction{extr}\AgdaSpace{}%
\AgdaBound{ρ}\AgdaSymbol{)}\AgdaSpace{}%
\AgdaBound{N}\AgdaSymbol{)}\<%
\\
\>[0]\AgdaFunction{rename-val}\AgdaSpace{}%
\AgdaBound{ρ}\AgdaSpace{}%
\AgdaSymbol{(}\AgdaInductiveConstructor{\$̬}\AgdaSpace{}%
\AgdaBound{c}\AgdaSymbol{)}%
\>[23]\AgdaSymbol{=}%
\>[26]\AgdaInductiveConstructor{\$̬}\AgdaSpace{}%
\AgdaBound{c}\<%
\\
\>[0]\AgdaFunction{rename-val}\AgdaSpace{}%
\AgdaBound{ρ}\AgdaSpace{}%
\AgdaSymbol{(}\AgdaBound{V}\AgdaSpace{}%
\AgdaOperator{\AgdaInductiveConstructor{〈}}\AgdaSpace{}%
\AgdaBound{g}\AgdaSpace{}%
\AgdaOperator{\AgdaInductiveConstructor{〉}}\AgdaSymbol{)}%
\>[24]\AgdaSymbol{=}%
\>[27]\AgdaSymbol{(}\AgdaFunction{rename-val}\AgdaSpace{}%
\AgdaBound{ρ}\AgdaSpace{}%
\AgdaBound{V}\AgdaSymbol{)}\AgdaSpace{}%
\AgdaOperator{\AgdaInductiveConstructor{〈}}\AgdaSpace{}%
\AgdaBound{g}\AgdaSpace{}%
\AgdaOperator{\AgdaInductiveConstructor{〉}}\<%
\\
\\[\AgdaEmptyExtraSkip]%
\>[0]\AgdaFunction{sub-val}\AgdaSpace{}%
\AgdaSymbol{:}\AgdaSpace{}%
\AgdaSymbol{∀}\AgdaSpace{}%
\AgdaSymbol{\{}\AgdaBound{V}\AgdaSymbol{\}}\<%
\\
\>[0][@{}l@{\AgdaIndent{0}}]%
\>[2]\AgdaSymbol{→}\AgdaSpace{}%
\AgdaSymbol{(}\AgdaBound{σ}\AgdaSpace{}%
\AgdaSymbol{:}\AgdaSpace{}%
\AgdaFunction{Subst}\AgdaSymbol{)}\<%
\\
\>[2]\AgdaSymbol{→}\AgdaSpace{}%
\AgdaDatatype{Value}\AgdaSpace{}%
\AgdaBound{V}\<%
\\
\>[2]\AgdaSymbol{→}\AgdaSpace{}%
\AgdaDatatype{Value}\AgdaSpace{}%
\AgdaSymbol{(}\AgdaOperator{\AgdaFunction{⟪}}\AgdaSpace{}%
\AgdaBound{σ}\AgdaSpace{}%
\AgdaOperator{\AgdaFunction{⟫}}\AgdaSpace{}%
\AgdaBound{V}\AgdaSymbol{)}\<%
\\
\>[0]\AgdaFunction{sub-val}\AgdaSpace{}%
\AgdaBound{σ}\AgdaSpace{}%
\AgdaSymbol{(}\AgdaOperator{\AgdaInductiveConstructor{ƛ̬}}\AgdaSpace{}%
\AgdaBound{N}\AgdaSymbol{)}\AgdaSpace{}%
\AgdaSymbol{=}\AgdaSpace{}%
\AgdaOperator{\AgdaInductiveConstructor{ƛ̬}}\AgdaSpace{}%
\AgdaOperator{\AgdaFunction{⟪}}\AgdaSpace{}%
\AgdaFunction{ext}\AgdaSpace{}%
\AgdaBound{σ}\AgdaSpace{}%
\AgdaOperator{\AgdaFunction{⟫}}\AgdaSpace{}%
\AgdaBound{N}\<%
\\
\>[0]\AgdaFunction{sub-val}\AgdaSpace{}%
\AgdaBound{σ}\AgdaSpace{}%
\AgdaSymbol{(}\AgdaInductiveConstructor{\$̬}\AgdaSpace{}%
\AgdaBound{c}\AgdaSymbol{)}\AgdaSpace{}%
\AgdaSymbol{=}\AgdaSpace{}%
\AgdaInductiveConstructor{\$̬}\AgdaSpace{}%
\AgdaBound{c}\<%
\\
\>[0]\AgdaFunction{sub-val}\AgdaSpace{}%
\AgdaBound{σ}\AgdaSpace{}%
\AgdaSymbol{(}\AgdaBound{V}\AgdaSpace{}%
\AgdaOperator{\AgdaInductiveConstructor{〈}}\AgdaSpace{}%
\AgdaBound{g}\AgdaSpace{}%
\AgdaOperator{\AgdaInductiveConstructor{〉}}\AgdaSymbol{)}%
\>[21]\AgdaSymbol{=}%
\>[24]\AgdaSymbol{(}\AgdaFunction{sub-val}\AgdaSpace{}%
\AgdaBound{σ}\AgdaSpace{}%
\AgdaBound{V}\AgdaSymbol{)}\AgdaSpace{}%
\AgdaOperator{\AgdaInductiveConstructor{〈}}\AgdaSpace{}%
\AgdaBound{g}\AgdaSpace{}%
\AgdaOperator{\AgdaInductiveConstructor{〉}}\<%
\\
\\[\AgdaEmptyExtraSkip]%
\>[0]\AgdaComment{\{-----------------\ Type\ System\ ------------------------\}}\<%
\\
\\[\AgdaEmptyExtraSkip]%
\>[0]\AgdaFunction{typeof}\AgdaSpace{}%
\AgdaSymbol{:}\AgdaSpace{}%
\AgdaDatatype{Lit}\AgdaSpace{}%
\AgdaSymbol{→}\AgdaSpace{}%
\AgdaDatatype{Base}\<%
\\
\>[0]\AgdaFunction{typeof}\AgdaSpace{}%
\AgdaSymbol{(}\AgdaInductiveConstructor{Num}\AgdaSpace{}%
\AgdaBound{n}\AgdaSymbol{)}\AgdaSpace{}%
\AgdaSymbol{=}\AgdaSpace{}%
\AgdaInductiveConstructor{′ℕ}\<%
\\
\>[0]\AgdaFunction{typeof}\AgdaSpace{}%
\AgdaSymbol{(}\AgdaInductiveConstructor{Bool}\AgdaSpace{}%
\AgdaBound{b}\AgdaSymbol{)}\AgdaSpace{}%
\AgdaSymbol{=}\AgdaSpace{}%
\AgdaInductiveConstructor{′𝔹}\<%
\\
\\[\AgdaEmptyExtraSkip]%
\>[0]\AgdaComment{\{-\ Consistency\ -\}}\<%
\\
\>[0]\AgdaKeyword{data}\AgdaSpace{}%
\AgdaOperator{\AgdaDatatype{\AgdaUnderscore{}∼\AgdaUnderscore{}}}\AgdaSpace{}%
\AgdaSymbol{:}\AgdaSpace{}%
\AgdaDatatype{Type}\AgdaSpace{}%
\AgdaSymbol{→}\AgdaSpace{}%
\AgdaDatatype{Type}\AgdaSpace{}%
\AgdaSymbol{→}\AgdaSpace{}%
\AgdaPrimitive{Set}\AgdaSpace{}%
\AgdaKeyword{where}\<%
\\
\>[0][@{}l@{\AgdaIndent{0}}]%
\>[2]\AgdaInductiveConstructor{★∼}%
\>[680I]\AgdaSymbol{:}%
\>[681I]\AgdaSymbol{∀\{}\AgdaBound{A}\AgdaSymbol{\}}\<%
\\
\>[.][@{}l@{}]\<[681I]%
\>[7]\AgdaComment{-----}\<%
\\
\>[.][@{}l@{}]\<[680I]%
\>[5]\AgdaSymbol{→}\AgdaSpace{}%
\AgdaInductiveConstructor{★}\AgdaSpace{}%
\AgdaOperator{\AgdaDatatype{∼}}\AgdaSpace{}%
\AgdaBound{A}\<%
\\
\\[\AgdaEmptyExtraSkip]%
\>[2]\AgdaInductiveConstructor{∼★}%
\>[685I]\AgdaSymbol{:}%
\>[686I]\AgdaSymbol{∀\{}\AgdaBound{A}\AgdaSymbol{\}}\<%
\\
\>[.][@{}l@{}]\<[686I]%
\>[7]\AgdaComment{-----}\<%
\\
\>[.][@{}l@{}]\<[685I]%
\>[5]\AgdaSymbol{→}\AgdaSpace{}%
\AgdaBound{A}\AgdaSpace{}%
\AgdaOperator{\AgdaDatatype{∼}}\AgdaSpace{}%
\AgdaInductiveConstructor{★}\<%
\\
\\[\AgdaEmptyExtraSkip]%
\>[2]\AgdaInductiveConstructor{∼-base}\AgdaSpace{}%
\AgdaSymbol{:}\AgdaSpace{}%
\AgdaSymbol{∀\{}\AgdaBound{ι}\AgdaSymbol{\}}\<%
\\
\>[2][@{}l@{\AgdaIndent{0}}]%
\>[7]\AgdaComment{--------------}\<%
\\
\>[2][@{}l@{\AgdaIndent{0}}]%
\>[5]\AgdaSymbol{→}\AgdaSpace{}%
\AgdaSymbol{(}\AgdaOperator{\AgdaInductiveConstructor{\$ₜ}}\AgdaSpace{}%
\AgdaBound{ι}\AgdaSymbol{)}\AgdaSpace{}%
\AgdaOperator{\AgdaDatatype{∼}}\AgdaSpace{}%
\AgdaSymbol{(}\AgdaOperator{\AgdaInductiveConstructor{\$ₜ}}\AgdaSpace{}%
\AgdaBound{ι}\AgdaSymbol{)}\<%
\\
\\[\AgdaEmptyExtraSkip]%
\>[2]\AgdaInductiveConstructor{∼-fun}\AgdaSpace{}%
\AgdaSymbol{:}\AgdaSpace{}%
\AgdaSymbol{∀\{}\AgdaBound{A}\AgdaSpace{}%
\AgdaBound{B}\AgdaSpace{}%
\AgdaBound{A′}\AgdaSpace{}%
\AgdaBound{B′}\AgdaSymbol{\}}\<%
\\
\>[2][@{}l@{\AgdaIndent{0}}]%
\>[5]\AgdaSymbol{→}\AgdaSpace{}%
\AgdaBound{A}\AgdaSpace{}%
\AgdaOperator{\AgdaDatatype{∼}}\AgdaSpace{}%
\AgdaBound{A′}\<%
\\
\>[5]\AgdaSymbol{→}%
\>[705I]\AgdaBound{B}\AgdaSpace{}%
\AgdaOperator{\AgdaDatatype{∼}}\AgdaSpace{}%
\AgdaBound{B′}\<%
\\
\>[.][@{}l@{}]\<[705I]%
\>[7]\AgdaComment{-------------------}\<%
\\
\>[5]\AgdaSymbol{→}\AgdaSpace{}%
\AgdaSymbol{(}\AgdaBound{A}\AgdaSpace{}%
\AgdaOperator{\AgdaInductiveConstructor{⇒}}\AgdaSpace{}%
\AgdaBound{B}\AgdaSymbol{)}\AgdaSpace{}%
\AgdaOperator{\AgdaDatatype{∼}}\AgdaSpace{}%
\AgdaSymbol{(}\AgdaBound{A′}\AgdaSpace{}%
\AgdaOperator{\AgdaInductiveConstructor{⇒}}\AgdaSpace{}%
\AgdaBound{B′}\AgdaSymbol{)}\<%
\end{code}
\begin{code}[hide]%
\>[0]\AgdaComment{\{-----------------------\ Frames\ ------------------------\}}\<%
\\
\\[\AgdaEmptyExtraSkip]%
\>[0]\AgdaKeyword{infix}%
\>[7]\AgdaNumber{6}\AgdaSpace{}%
\AgdaOperator{\AgdaInductiveConstructor{□·\AgdaUnderscore{}}}\<%
\\
\>[0]\AgdaKeyword{infix}%
\>[7]\AgdaNumber{6}\AgdaSpace{}%
\AgdaOperator{\AgdaInductiveConstructor{\AgdaUnderscore{}·□}}\<%
\\
\>[0]\AgdaKeyword{infix}%
\>[7]\AgdaNumber{6}\AgdaSpace{}%
\AgdaOperator{\AgdaInductiveConstructor{□⟨\AgdaUnderscore{}!⟩}}\<%
\\
\>[0]\AgdaKeyword{infix}%
\>[7]\AgdaNumber{6}\AgdaSpace{}%
\AgdaOperator{\AgdaInductiveConstructor{□⟨\AgdaUnderscore{}?⟩}}\<%
\\
\\[\AgdaEmptyExtraSkip]%
\>[0]\AgdaKeyword{data}\AgdaSpace{}%
\AgdaDatatype{Frame}\AgdaSpace{}%
\AgdaSymbol{:}\AgdaSpace{}%
\AgdaPrimitive{Set}\AgdaSpace{}%
\AgdaKeyword{where}\<%
\\
\\[\AgdaEmptyExtraSkip]%
\>[0][@{}l@{\AgdaIndent{0}}]%
\>[2]\AgdaOperator{\AgdaInductiveConstructor{□·\AgdaUnderscore{}}}%
\>[723I]\AgdaSymbol{:}\<%
\\
\>[.][@{}l@{}]\<[723I]%
\>[6]\AgdaSymbol{(}\AgdaBound{M}\AgdaSpace{}%
\AgdaSymbol{:}\AgdaSpace{}%
\AgdaDatatype{Term}\AgdaSymbol{)}\<%
\\
\>[6]\AgdaComment{----------}\<%
\\
\>[2][@{}l@{\AgdaIndent{0}}]%
\>[4]\AgdaSymbol{→}\AgdaSpace{}%
\AgdaDatatype{Frame}\<%
\\
\\[\AgdaEmptyExtraSkip]%
\>[2]\AgdaOperator{\AgdaInductiveConstructor{\AgdaUnderscore{}·□}}\AgdaSpace{}%
\AgdaSymbol{:}\AgdaSpace{}%
\AgdaSymbol{∀}\AgdaSpace{}%
\AgdaSymbol{\{}\AgdaBound{V}\AgdaSpace{}%
\AgdaSymbol{:}\AgdaSpace{}%
\AgdaDatatype{Term}\AgdaSymbol{\}}\<%
\\
\>[2][@{}l@{\AgdaIndent{0}}]%
\>[4]\AgdaSymbol{→}%
\>[732I]\AgdaSymbol{(}\AgdaBound{v}\AgdaSpace{}%
\AgdaSymbol{:}\AgdaSpace{}%
\AgdaDatatype{Value}\AgdaSpace{}%
\AgdaBound{V}\AgdaSymbol{)}\<%
\\
\>[.][@{}l@{}]\<[732I]%
\>[6]\AgdaComment{-------------}\<%
\\
\>[4]\AgdaSymbol{→}\AgdaSpace{}%
\AgdaDatatype{Frame}\<%
\\
\\[\AgdaEmptyExtraSkip]%
\>[2]\AgdaOperator{\AgdaInductiveConstructor{□⟨\AgdaUnderscore{}!⟩}}\AgdaSpace{}%
\AgdaSymbol{:}\<%
\\
\>[2][@{}l@{\AgdaIndent{0}}]%
\>[6]\AgdaSymbol{(}\AgdaBound{G}\AgdaSpace{}%
\AgdaSymbol{:}\AgdaSpace{}%
\AgdaDatatype{Ground}\AgdaSymbol{)}\<%
\\
\>[6]\AgdaComment{-----}\<%
\\
\>[2][@{}l@{\AgdaIndent{0}}]%
\>[4]\AgdaSymbol{→}\AgdaSpace{}%
\AgdaDatatype{Frame}\<%
\\
\\[\AgdaEmptyExtraSkip]%
\>[2]\AgdaOperator{\AgdaInductiveConstructor{□⟨\AgdaUnderscore{}?⟩}}\AgdaSpace{}%
\AgdaSymbol{:}\<%
\\
\>[2][@{}l@{\AgdaIndent{0}}]%
\>[6]\AgdaSymbol{(}\AgdaBound{H}\AgdaSpace{}%
\AgdaSymbol{:}\AgdaSpace{}%
\AgdaDatatype{Ground}\AgdaSymbol{)}\<%
\\
\>[6]\AgdaComment{-----}\<%
\\
\>[2][@{}l@{\AgdaIndent{0}}]%
\>[4]\AgdaSymbol{→}\AgdaSpace{}%
\AgdaDatatype{Frame}\<%
\\
\\[\AgdaEmptyExtraSkip]%
\>[0]\AgdaComment{\{-\ Plug\ an\ expression\ into\ a\ frame.\ -\}}\<%
\\
\\[\AgdaEmptyExtraSkip]%
\>[0]\AgdaOperator{\AgdaFunction{\AgdaUnderscore{}⟦\AgdaUnderscore{}⟧}}\AgdaSpace{}%
\AgdaSymbol{:}\AgdaSpace{}%
\AgdaDatatype{Frame}\AgdaSpace{}%
\AgdaSymbol{→}\AgdaSpace{}%
\AgdaDatatype{Term}\AgdaSpace{}%
\AgdaSymbol{→}\AgdaSpace{}%
\AgdaDatatype{Term}\<%
\\
\>[0]\AgdaSymbol{(}\AgdaOperator{\AgdaInductiveConstructor{□·}}\AgdaSpace{}%
\AgdaBound{M}\AgdaSymbol{)}\AgdaSpace{}%
\AgdaOperator{\AgdaFunction{⟦}}\AgdaSpace{}%
\AgdaBound{L}\AgdaSpace{}%
\AgdaOperator{\AgdaFunction{⟧}}%
\>[20]\AgdaSymbol{=}%
\>[23]\AgdaBound{L}\AgdaSpace{}%
\AgdaOperator{\AgdaInductiveConstructor{·}}\AgdaSpace{}%
\AgdaBound{M}\<%
\\
\>[0]\AgdaSymbol{(}\AgdaBound{v}\AgdaSpace{}%
\AgdaOperator{\AgdaInductiveConstructor{·□}}\AgdaSymbol{)}\AgdaSpace{}%
\AgdaOperator{\AgdaFunction{⟦}}\AgdaSpace{}%
\AgdaBound{M}\AgdaSpace{}%
\AgdaOperator{\AgdaFunction{⟧}}%
\>[20]\AgdaSymbol{=}%
\>[23]\AgdaFunction{value}\AgdaSpace{}%
\AgdaBound{v}\AgdaSpace{}%
\AgdaOperator{\AgdaInductiveConstructor{·}}\AgdaSpace{}%
\AgdaBound{M}\<%
\\
\>[0]\AgdaSymbol{(}\AgdaOperator{\AgdaInductiveConstructor{□⟨}}\AgdaSpace{}%
\AgdaBound{G}\AgdaSpace{}%
\AgdaOperator{\AgdaInductiveConstructor{!⟩}}\AgdaSymbol{)}\AgdaSpace{}%
\AgdaOperator{\AgdaFunction{⟦}}\AgdaSpace{}%
\AgdaBound{M}\AgdaSpace{}%
\AgdaOperator{\AgdaFunction{⟧}}%
\>[17]\AgdaSymbol{=}%
\>[20]\AgdaBound{M}\AgdaSpace{}%
\AgdaOperator{\AgdaInductiveConstructor{⟨}}\AgdaSpace{}%
\AgdaBound{G}\AgdaSpace{}%
\AgdaOperator{\AgdaInductiveConstructor{!⟩}}\<%
\\
\>[0]\AgdaSymbol{(}\AgdaOperator{\AgdaInductiveConstructor{□⟨}}\AgdaSpace{}%
\AgdaBound{H}\AgdaSpace{}%
\AgdaOperator{\AgdaInductiveConstructor{?⟩}}\AgdaSymbol{)}\AgdaSpace{}%
\AgdaOperator{\AgdaFunction{⟦}}\AgdaSpace{}%
\AgdaBound{M}\AgdaSpace{}%
\AgdaOperator{\AgdaFunction{⟧}}%
\>[17]\AgdaSymbol{=}%
\>[20]\AgdaBound{M}\AgdaSpace{}%
\AgdaOperator{\AgdaInductiveConstructor{⟨}}\AgdaSpace{}%
\AgdaBound{H}\AgdaSpace{}%
\AgdaOperator{\AgdaInductiveConstructor{?⟩}}\<%
\\
\\[\AgdaEmptyExtraSkip]%
\>[0]\AgdaComment{\{-\ Possibly-empty\ Frame\ -\}}\<%
\\
\\[\AgdaEmptyExtraSkip]%
\>[0]\AgdaKeyword{data}\AgdaSpace{}%
\AgdaDatatype{PEFrame}\AgdaSpace{}%
\AgdaSymbol{:}\AgdaSpace{}%
\AgdaPrimitive{Set}\AgdaSpace{}%
\AgdaKeyword{where}\<%
\\
\>[0][@{}l@{\AgdaIndent{0}}]%
\>[2]\AgdaOperator{\AgdaInductiveConstructor{`\AgdaUnderscore{}}}\AgdaSpace{}%
\AgdaSymbol{:}\AgdaSpace{}%
\AgdaDatatype{Frame}\AgdaSpace{}%
\AgdaSymbol{→}\AgdaSpace{}%
\AgdaDatatype{PEFrame}\<%
\\
\>[2]\AgdaInductiveConstructor{□}\AgdaSpace{}%
\AgdaSymbol{:}\AgdaSpace{}%
\AgdaDatatype{PEFrame}\<%
\\
\\[\AgdaEmptyExtraSkip]%
\>[0]\AgdaOperator{\AgdaFunction{\AgdaUnderscore{}⦉\AgdaUnderscore{}⦊}}\AgdaSpace{}%
\AgdaSymbol{:}\AgdaSpace{}%
\AgdaDatatype{PEFrame}\AgdaSpace{}%
\AgdaSymbol{→}\AgdaSpace{}%
\AgdaDatatype{Term}\AgdaSpace{}%
\AgdaSymbol{→}\AgdaSpace{}%
\AgdaDatatype{Term}\<%
\\
\>[0]\AgdaSymbol{(}\AgdaOperator{\AgdaInductiveConstructor{`}}\AgdaSpace{}%
\AgdaBound{F}\AgdaSymbol{)}\AgdaSpace{}%
\AgdaOperator{\AgdaFunction{⦉}}\AgdaSpace{}%
\AgdaBound{M}\AgdaSpace{}%
\AgdaOperator{\AgdaFunction{⦊}}\AgdaSpace{}%
\AgdaSymbol{=}\AgdaSpace{}%
\AgdaBound{F}\AgdaSpace{}%
\AgdaOperator{\AgdaFunction{⟦}}\AgdaSpace{}%
\AgdaBound{M}\AgdaSpace{}%
\AgdaOperator{\AgdaFunction{⟧}}\<%
\\
\>[0]\AgdaInductiveConstructor{□}\AgdaSpace{}%
\AgdaOperator{\AgdaFunction{⦉}}\AgdaSpace{}%
\AgdaBound{M}\AgdaSpace{}%
\AgdaOperator{\AgdaFunction{⦊}}\AgdaSpace{}%
\AgdaSymbol{=}\AgdaSpace{}%
\AgdaBound{M}\<%
\\
\\[\AgdaEmptyExtraSkip]%
\>[0]\AgdaComment{\{-\ Reduction\ -\}}\<%
\\
\\[\AgdaEmptyExtraSkip]%
\>[0]\AgdaKeyword{infixr}\AgdaSpace{}%
\AgdaNumber{1}\AgdaSpace{}%
\AgdaOperator{\AgdaFunction{\AgdaUnderscore{}++\AgdaUnderscore{}}}\<%
\\
\>[0]\AgdaComment{--infix\ \ 1\ begin\AgdaUnderscore{}}\<%
\\
\>[0]\AgdaKeyword{infix}%
\>[7]\AgdaNumber{2}\AgdaSpace{}%
\AgdaOperator{\AgdaDatatype{\AgdaUnderscore{}↠\AgdaUnderscore{}}}\<%
\\
\>[0]\AgdaKeyword{infixr}\AgdaSpace{}%
\AgdaNumber{2}\AgdaSpace{}%
\AgdaOperator{\AgdaInductiveConstructor{\AgdaUnderscore{}⟶⟨\AgdaUnderscore{}⟩\AgdaUnderscore{}}}\<%
\\
\>[0]\AgdaKeyword{infixr}\AgdaSpace{}%
\AgdaNumber{2}\AgdaSpace{}%
\AgdaOperator{\AgdaFunction{\AgdaUnderscore{}↠⟨\AgdaUnderscore{}⟩\AgdaUnderscore{}}}\<%
\\
\>[0]\AgdaKeyword{infix}%
\>[7]\AgdaNumber{3}\AgdaSpace{}%
\AgdaOperator{\AgdaInductiveConstructor{\AgdaUnderscore{}END}}\<%
\end{code}

Figure~\ref{fig:cast-calculus} defines the type system and reduction
for the Cast Calculus. The two rules specific to gradual typing are
\textsf{collapse} and \textsf{collide}. The \textsf{collapse} rule
states that when an injected value encounters a matching projection,
the result is the underlying value.  The \textsf{collide} says that if
the injection and projection do not match, the result is
\textsf{blame}. The reason we introduce the $M$ variable and the
equation $M ≡ V ⟨ G !⟩$ in those rules is that we ran into
difficulties with Agda when doing case analysis on reductions.  The
same is true for the ξξ and \textsf{ξξ-blame} rules.  The ξξ and
\textsf{ξξ-blame} rules handle the usual congruence rules that are
needed for reduction. The \textsf{Frame} type is a shallow
non-recursive evaluation context, representing just a single term
constructor with a hole. The notation $F[M]$ plugs term $M$ into the
hole inside $F$.

Figure~\ref{fig:cast-calculus} defines $M ⇓$ to mean that $M$
reduces to a value, $M ⇑$ to mean $M$ diverges, and $M ⇑⊎blame$
to mean that $M$ either diverges or reduces to \textsf{blame}.
(We ran into difficulties with the alternate formulation
of $(M ⇑) ⊎ (M ↠ \mathsf{blame})$ and could not prove them
equivalent.)

\begin{figure}[tbp]
\begin{code}%
\>[0]\AgdaKeyword{infix}\AgdaSpace{}%
\AgdaNumber{3}\AgdaSpace{}%
\AgdaOperator{\AgdaDatatype{\AgdaUnderscore{}⊢\AgdaUnderscore{}⦂\AgdaUnderscore{}}}\<%
\\
\>[0]\AgdaKeyword{data}\AgdaSpace{}%
\AgdaOperator{\AgdaDatatype{\AgdaUnderscore{}⊢\AgdaUnderscore{}⦂\AgdaUnderscore{}}}\AgdaSpace{}%
\AgdaSymbol{:}\AgdaSpace{}%
\AgdaDatatype{List}\AgdaSpace{}%
\AgdaDatatype{Type}\AgdaSpace{}%
\AgdaSymbol{→}\AgdaSpace{}%
\AgdaDatatype{Term}\AgdaSpace{}%
\AgdaSymbol{→}\AgdaSpace{}%
\AgdaDatatype{Type}\AgdaSpace{}%
\AgdaSymbol{→}\AgdaSpace{}%
\AgdaPrimitive{Set}\AgdaSpace{}%
\AgdaKeyword{where}\<%
\\
\>[0][@{}l@{\AgdaIndent{0}}]%
\>[2]\AgdaInductiveConstructor{⊢`}\AgdaSpace{}%
\AgdaSymbol{:}\AgdaSpace{}%
\AgdaSymbol{∀}\AgdaSpace{}%
\AgdaSymbol{\{}\AgdaBound{Γ}\AgdaSpace{}%
\AgdaBound{x}\AgdaSpace{}%
\AgdaBound{A}\AgdaSymbol{\}}\AgdaSpace{}%
\AgdaSymbol{→}\AgdaSpace{}%
\AgdaBound{Γ}\AgdaSpace{}%
\AgdaOperator{\AgdaFunction{∋}}\AgdaSpace{}%
\AgdaBound{x}\AgdaSpace{}%
\AgdaOperator{\AgdaFunction{⦂}}\AgdaSpace{}%
\AgdaBound{A}\AgdaSpace{}%
\AgdaSymbol{→}\AgdaSpace{}%
\AgdaBound{Γ}\AgdaSpace{}%
\AgdaOperator{\AgdaDatatype{⊢}}\AgdaSpace{}%
\AgdaOperator{\AgdaInductiveConstructor{`}}\AgdaSpace{}%
\AgdaBound{x}\AgdaSpace{}%
\AgdaOperator{\AgdaDatatype{⦂}}\AgdaSpace{}%
\AgdaBound{A}\<%
\\
\>[2]\AgdaInductiveConstructor{⊢\$}\AgdaSpace{}%
\AgdaSymbol{:}\AgdaSpace{}%
\AgdaSymbol{∀}\AgdaSpace{}%
\AgdaSymbol{\{}\AgdaBound{Γ}\AgdaSymbol{\}}\AgdaSpace{}%
\AgdaSymbol{(}\AgdaBound{l}\AgdaSpace{}%
\AgdaSymbol{:}\AgdaSpace{}%
\AgdaDatatype{Lit}\AgdaSymbol{)}\AgdaSpace{}%
\AgdaSymbol{→}\AgdaSpace{}%
\AgdaBound{Γ}\AgdaSpace{}%
\AgdaOperator{\AgdaDatatype{⊢}}\AgdaSpace{}%
\AgdaInductiveConstructor{\$}\AgdaSpace{}%
\AgdaBound{l}\AgdaSpace{}%
\AgdaOperator{\AgdaDatatype{⦂}}\AgdaSpace{}%
\AgdaOperator{\AgdaInductiveConstructor{\$ₜ}}\AgdaSpace{}%
\AgdaSymbol{(}\AgdaFunction{typeof}\AgdaSpace{}%
\AgdaBound{l}\AgdaSymbol{)}\<%
\\
\>[2]\AgdaInductiveConstructor{⊢·}\AgdaSpace{}%
\AgdaSymbol{:}\AgdaSpace{}%
\AgdaSymbol{∀}\AgdaSpace{}%
\AgdaSymbol{\{}\AgdaBound{Γ}\AgdaSpace{}%
\AgdaBound{L}\AgdaSpace{}%
\AgdaBound{M}\AgdaSpace{}%
\AgdaBound{A}\AgdaSpace{}%
\AgdaBound{B}\AgdaSymbol{\}}\AgdaSpace{}%
\AgdaSymbol{→}\AgdaSpace{}%
\AgdaBound{Γ}\AgdaSpace{}%
\AgdaOperator{\AgdaDatatype{⊢}}\AgdaSpace{}%
\AgdaBound{L}\AgdaSpace{}%
\AgdaOperator{\AgdaDatatype{⦂}}\AgdaSpace{}%
\AgdaSymbol{(}\AgdaBound{A}\AgdaSpace{}%
\AgdaOperator{\AgdaInductiveConstructor{⇒}}\AgdaSpace{}%
\AgdaBound{B}\AgdaSymbol{)}\AgdaSpace{}%
\AgdaSymbol{→}\AgdaSpace{}%
\AgdaBound{Γ}\AgdaSpace{}%
\AgdaOperator{\AgdaDatatype{⊢}}\AgdaSpace{}%
\AgdaBound{M}\AgdaSpace{}%
\AgdaOperator{\AgdaDatatype{⦂}}\AgdaSpace{}%
\AgdaBound{A}\AgdaSpace{}%
\AgdaSymbol{→}\AgdaSpace{}%
\AgdaBound{Γ}\AgdaSpace{}%
\AgdaOperator{\AgdaDatatype{⊢}}\AgdaSpace{}%
\AgdaBound{L}\AgdaSpace{}%
\AgdaOperator{\AgdaInductiveConstructor{·}}\AgdaSpace{}%
\AgdaBound{M}\AgdaSpace{}%
\AgdaOperator{\AgdaDatatype{⦂}}\AgdaSpace{}%
\AgdaBound{B}\<%
\\
\>[2]\AgdaInductiveConstructor{⊢ƛ}\AgdaSpace{}%
\AgdaSymbol{:}\AgdaSpace{}%
\AgdaSymbol{∀}\AgdaSpace{}%
\AgdaSymbol{\{}\AgdaBound{Γ}\AgdaSpace{}%
\AgdaBound{N}\AgdaSpace{}%
\AgdaBound{A}\AgdaSpace{}%
\AgdaBound{B}\AgdaSymbol{\}}\AgdaSpace{}%
\AgdaSymbol{→}\AgdaSpace{}%
\AgdaSymbol{(}\AgdaBound{A}\AgdaSpace{}%
\AgdaOperator{\AgdaInductiveConstructor{∷}}\AgdaSpace{}%
\AgdaBound{Γ}\AgdaSymbol{)}\AgdaSpace{}%
\AgdaOperator{\AgdaDatatype{⊢}}\AgdaSpace{}%
\AgdaBound{N}\AgdaSpace{}%
\AgdaOperator{\AgdaDatatype{⦂}}\AgdaSpace{}%
\AgdaBound{B}\AgdaSpace{}%
\AgdaSymbol{→}\AgdaSpace{}%
\AgdaBound{Γ}\AgdaSpace{}%
\AgdaOperator{\AgdaDatatype{⊢}}\AgdaSpace{}%
\AgdaInductiveConstructor{ƛ}\AgdaSpace{}%
\AgdaBound{N}\AgdaSpace{}%
\AgdaOperator{\AgdaDatatype{⦂}}\AgdaSpace{}%
\AgdaSymbol{(}\AgdaBound{A}\AgdaSpace{}%
\AgdaOperator{\AgdaInductiveConstructor{⇒}}\AgdaSpace{}%
\AgdaBound{B}\AgdaSymbol{)}\<%
\\
\>[2]\AgdaInductiveConstructor{⊢⟨!⟩}\AgdaSpace{}%
\AgdaSymbol{:}\AgdaSpace{}%
\AgdaSymbol{∀\{}\AgdaBound{Γ}\AgdaSpace{}%
\AgdaBound{M}\AgdaSpace{}%
\AgdaBound{G}\AgdaSymbol{\}}\AgdaSpace{}%
\AgdaSymbol{→}\AgdaSpace{}%
\AgdaBound{Γ}\AgdaSpace{}%
\AgdaOperator{\AgdaDatatype{⊢}}\AgdaSpace{}%
\AgdaBound{M}\AgdaSpace{}%
\AgdaOperator{\AgdaDatatype{⦂}}\AgdaSpace{}%
\AgdaOperator{\AgdaFunction{⌈}}\AgdaSpace{}%
\AgdaBound{G}\AgdaSpace{}%
\AgdaOperator{\AgdaFunction{⌉}}\AgdaSpace{}%
\AgdaSymbol{→}\AgdaSpace{}%
\AgdaBound{Γ}\AgdaSpace{}%
\AgdaOperator{\AgdaDatatype{⊢}}\AgdaSpace{}%
\AgdaBound{M}\AgdaSpace{}%
\AgdaOperator{\AgdaInductiveConstructor{⟨}}\AgdaSpace{}%
\AgdaBound{G}\AgdaSpace{}%
\AgdaOperator{\AgdaInductiveConstructor{!⟩}}\AgdaSpace{}%
\AgdaOperator{\AgdaDatatype{⦂}}\AgdaSpace{}%
\AgdaInductiveConstructor{★}\<%
\\
\>[2]\AgdaInductiveConstructor{⊢⟨?⟩}\AgdaSpace{}%
\AgdaSymbol{:}\AgdaSpace{}%
\AgdaSymbol{∀\{}\AgdaBound{Γ}\AgdaSpace{}%
\AgdaBound{M}\AgdaSymbol{\}}\AgdaSpace{}%
\AgdaSymbol{→}\AgdaSpace{}%
\AgdaBound{Γ}\AgdaSpace{}%
\AgdaOperator{\AgdaDatatype{⊢}}\AgdaSpace{}%
\AgdaBound{M}\AgdaSpace{}%
\AgdaOperator{\AgdaDatatype{⦂}}\AgdaSpace{}%
\AgdaInductiveConstructor{★}\AgdaSpace{}%
\AgdaSymbol{→}\AgdaSpace{}%
\AgdaSymbol{(}\AgdaBound{H}\AgdaSpace{}%
\AgdaSymbol{:}\AgdaSpace{}%
\AgdaDatatype{Ground}\AgdaSymbol{)}\AgdaSpace{}%
\AgdaSymbol{→}\AgdaSpace{}%
\AgdaBound{Γ}\AgdaSpace{}%
\AgdaOperator{\AgdaDatatype{⊢}}\AgdaSpace{}%
\AgdaBound{M}\AgdaSpace{}%
\AgdaOperator{\AgdaInductiveConstructor{⟨}}\AgdaSpace{}%
\AgdaBound{H}\AgdaSpace{}%
\AgdaOperator{\AgdaInductiveConstructor{?⟩}}\AgdaSpace{}%
\AgdaOperator{\AgdaDatatype{⦂}}\AgdaSpace{}%
\AgdaOperator{\AgdaFunction{⌈}}\AgdaSpace{}%
\AgdaBound{H}\AgdaSpace{}%
\AgdaOperator{\AgdaFunction{⌉}}\<%
\\
\>[2]\AgdaInductiveConstructor{⊢blame}\AgdaSpace{}%
\AgdaSymbol{:}\AgdaSpace{}%
\AgdaSymbol{∀\{}\AgdaBound{Γ}\AgdaSpace{}%
\AgdaBound{A}\AgdaSymbol{\}}\AgdaSpace{}%
\AgdaSymbol{→}\AgdaSpace{}%
\AgdaBound{Γ}\AgdaSpace{}%
\AgdaOperator{\AgdaDatatype{⊢}}\AgdaSpace{}%
\AgdaInductiveConstructor{blame}\AgdaSpace{}%
\AgdaOperator{\AgdaDatatype{⦂}}\AgdaSpace{}%
\AgdaBound{A}\<%
\\
\>[0]\<%
\\
\>[0]\AgdaKeyword{infix}\AgdaSpace{}%
\AgdaNumber{2}\AgdaSpace{}%
\AgdaOperator{\AgdaDatatype{\AgdaUnderscore{}⟶\AgdaUnderscore{}}}\<%
\\
\>[0]\AgdaKeyword{data}\AgdaSpace{}%
\AgdaOperator{\AgdaDatatype{\AgdaUnderscore{}⟶\AgdaUnderscore{}}}\AgdaSpace{}%
\AgdaSymbol{:}\AgdaSpace{}%
\AgdaDatatype{Term}\AgdaSpace{}%
\AgdaSymbol{→}\AgdaSpace{}%
\AgdaDatatype{Term}\AgdaSpace{}%
\AgdaSymbol{→}\AgdaSpace{}%
\AgdaPrimitive{Set}\AgdaSpace{}%
\AgdaKeyword{where}\<%
\\
\>[0][@{}l@{\AgdaIndent{0}}]%
\>[2]\AgdaInductiveConstructor{β}\AgdaSpace{}%
\AgdaSymbol{:}\AgdaSpace{}%
\AgdaSymbol{∀}\AgdaSpace{}%
\AgdaSymbol{\{}\AgdaBound{N}\AgdaSpace{}%
\AgdaBound{W}\AgdaSymbol{\}}\AgdaSpace{}%
\AgdaSymbol{→}\AgdaSpace{}%
\AgdaDatatype{Value}\AgdaSpace{}%
\AgdaBound{W}%
\>[25]\AgdaSymbol{→}%
\>[28]\AgdaSymbol{(}\AgdaInductiveConstructor{ƛ}\AgdaSpace{}%
\AgdaBound{N}\AgdaSymbol{)}\AgdaSpace{}%
\AgdaOperator{\AgdaInductiveConstructor{·}}\AgdaSpace{}%
\AgdaBound{W}\AgdaSpace{}%
\AgdaOperator{\AgdaDatatype{⟶}}\AgdaSpace{}%
\AgdaBound{N}\AgdaSpace{}%
\AgdaOperator{\AgdaFunction{[}}\AgdaSpace{}%
\AgdaBound{W}\AgdaSpace{}%
\AgdaOperator{\AgdaFunction{]}}\<%
\\
\>[2]\AgdaInductiveConstructor{collapse}\AgdaSpace{}%
\AgdaSymbol{:}\AgdaSpace{}%
\AgdaSymbol{∀\{}\AgdaBound{G}\AgdaSpace{}%
\AgdaBound{M}\AgdaSpace{}%
\AgdaBound{V}\AgdaSymbol{\}}\AgdaSpace{}%
\AgdaSymbol{→}\AgdaSpace{}%
\AgdaDatatype{Value}\AgdaSpace{}%
\AgdaBound{V}\AgdaSpace{}%
\AgdaSymbol{→}\AgdaSpace{}%
\AgdaBound{M}\AgdaSpace{}%
\AgdaOperator{\AgdaDatatype{≡}}\AgdaSpace{}%
\AgdaBound{V}\AgdaSpace{}%
\AgdaOperator{\AgdaInductiveConstructor{⟨}}\AgdaSpace{}%
\AgdaBound{G}\AgdaSpace{}%
\AgdaOperator{\AgdaInductiveConstructor{!⟩}}%
\>[48]\AgdaSymbol{→}%
\>[51]\AgdaBound{M}\AgdaSpace{}%
\AgdaOperator{\AgdaInductiveConstructor{⟨}}\AgdaSpace{}%
\AgdaBound{G}\AgdaSpace{}%
\AgdaOperator{\AgdaInductiveConstructor{?⟩}}\AgdaSpace{}%
\AgdaOperator{\AgdaDatatype{⟶}}\AgdaSpace{}%
\AgdaBound{V}\<%
\\
\>[2]\AgdaInductiveConstructor{collide}%
\>[11]\AgdaSymbol{:}\AgdaSpace{}%
\AgdaSymbol{∀\{}\AgdaBound{G}\AgdaSpace{}%
\AgdaBound{H}\AgdaSpace{}%
\AgdaBound{M}\AgdaSpace{}%
\AgdaBound{V}\AgdaSymbol{\}}\AgdaSpace{}%
\AgdaSymbol{→}\AgdaSpace{}%
\AgdaDatatype{Value}\AgdaSpace{}%
\AgdaBound{V}\AgdaSpace{}%
\AgdaSymbol{→}\AgdaSpace{}%
\AgdaBound{G}\AgdaSpace{}%
\AgdaOperator{\AgdaFunction{≢}}\AgdaSpace{}%
\AgdaBound{H}\AgdaSpace{}%
\AgdaSymbol{→}\AgdaSpace{}%
\AgdaBound{M}\AgdaSpace{}%
\AgdaOperator{\AgdaDatatype{≡}}\AgdaSpace{}%
\AgdaBound{V}\AgdaSpace{}%
\AgdaOperator{\AgdaInductiveConstructor{⟨}}\AgdaSpace{}%
\AgdaBound{G}\AgdaSpace{}%
\AgdaOperator{\AgdaInductiveConstructor{!⟩}}\AgdaSpace{}%
\AgdaSymbol{→}\AgdaSpace{}%
\AgdaBound{M}\AgdaSpace{}%
\AgdaOperator{\AgdaInductiveConstructor{⟨}}\AgdaSpace{}%
\AgdaBound{H}\AgdaSpace{}%
\AgdaOperator{\AgdaInductiveConstructor{?⟩}}\AgdaSpace{}%
\AgdaOperator{\AgdaDatatype{⟶}}\AgdaSpace{}%
\AgdaInductiveConstructor{blame}\<%
\\
\>[2]\AgdaInductiveConstructor{ξξ}\AgdaSpace{}%
\AgdaSymbol{:}\AgdaSpace{}%
\AgdaSymbol{∀}\AgdaSpace{}%
\AgdaSymbol{\{}\AgdaBound{M}\AgdaSpace{}%
\AgdaBound{N}\AgdaSpace{}%
\AgdaBound{M′}\AgdaSpace{}%
\AgdaBound{N′}\AgdaSymbol{\}}\AgdaSpace{}%
\AgdaSymbol{→}\AgdaSpace{}%
\AgdaSymbol{(}\AgdaBound{F}\AgdaSpace{}%
\AgdaSymbol{:}\AgdaSpace{}%
\AgdaDatatype{Frame}\AgdaSymbol{)}\<%
\\
\>[2][@{}l@{\AgdaIndent{0}}]%
\>[4]\AgdaSymbol{→}\AgdaSpace{}%
\AgdaBound{M′}\AgdaSpace{}%
\AgdaOperator{\AgdaDatatype{≡}}\AgdaSpace{}%
\AgdaBound{F}\AgdaSpace{}%
\AgdaOperator{\AgdaFunction{⟦}}\AgdaSpace{}%
\AgdaBound{M}\AgdaSpace{}%
\AgdaOperator{\AgdaFunction{⟧}}\AgdaSpace{}%
\AgdaSymbol{→}\AgdaSpace{}%
\AgdaBound{N′}\AgdaSpace{}%
\AgdaOperator{\AgdaDatatype{≡}}\AgdaSpace{}%
\AgdaBound{F}\AgdaSpace{}%
\AgdaOperator{\AgdaFunction{⟦}}\AgdaSpace{}%
\AgdaBound{N}\AgdaSpace{}%
\AgdaOperator{\AgdaFunction{⟧}}\AgdaSpace{}%
\AgdaSymbol{→}\AgdaSpace{}%
\AgdaBound{M}\AgdaSpace{}%
\AgdaOperator{\AgdaDatatype{⟶}}\AgdaSpace{}%
\AgdaBound{N}\AgdaSpace{}%
\AgdaSymbol{→}\AgdaSpace{}%
\AgdaBound{M′}\AgdaSpace{}%
\AgdaOperator{\AgdaDatatype{⟶}}\AgdaSpace{}%
\AgdaBound{N′}\<%
\\
\>[2]\AgdaInductiveConstructor{ξξ-blame}\AgdaSpace{}%
\AgdaSymbol{:}\AgdaSpace{}%
\AgdaSymbol{∀}\AgdaSpace{}%
\AgdaSymbol{\{}\AgdaBound{M′}\AgdaSymbol{\}}\AgdaSpace{}%
\AgdaSymbol{→}\AgdaSpace{}%
\AgdaSymbol{(}\AgdaBound{F}\AgdaSpace{}%
\AgdaSymbol{:}\AgdaSpace{}%
\AgdaDatatype{Frame}\AgdaSymbol{)}\AgdaSpace{}%
\AgdaSymbol{→}\AgdaSpace{}%
\AgdaBound{M′}\AgdaSpace{}%
\AgdaOperator{\AgdaDatatype{≡}}\AgdaSpace{}%
\AgdaBound{F}\AgdaSpace{}%
\AgdaOperator{\AgdaFunction{⟦}}\AgdaSpace{}%
\AgdaInductiveConstructor{blame}\AgdaSpace{}%
\AgdaOperator{\AgdaFunction{⟧}}\AgdaSpace{}%
\AgdaSymbol{→}\AgdaSpace{}%
\AgdaBound{M′}\AgdaSpace{}%
\AgdaOperator{\AgdaDatatype{⟶}}\AgdaSpace{}%
\AgdaInductiveConstructor{blame}\<%
\\
\\[\AgdaEmptyExtraSkip]%
\>[0]\AgdaKeyword{data}\AgdaSpace{}%
\AgdaOperator{\AgdaDatatype{\AgdaUnderscore{}↠\AgdaUnderscore{}}}\AgdaSpace{}%
\AgdaSymbol{:}\AgdaSpace{}%
\AgdaDatatype{Term}\AgdaSpace{}%
\AgdaSymbol{→}\AgdaSpace{}%
\AgdaDatatype{Term}\AgdaSpace{}%
\AgdaSymbol{→}\AgdaSpace{}%
\AgdaPrimitive{Set}\AgdaSpace{}%
\AgdaKeyword{where}\<%
\\
\>[0][@{}l@{\AgdaIndent{0}}]%
\>[2]\AgdaOperator{\AgdaInductiveConstructor{\AgdaUnderscore{}END}}\AgdaSpace{}%
\AgdaSymbol{:}\AgdaSpace{}%
\AgdaSymbol{(}\AgdaBound{M}\AgdaSpace{}%
\AgdaSymbol{:}\AgdaSpace{}%
\AgdaDatatype{Term}\AgdaSymbol{)}%
\>[21]\AgdaSymbol{→}%
\>[24]\AgdaBound{M}\AgdaSpace{}%
\AgdaOperator{\AgdaDatatype{↠}}\AgdaSpace{}%
\AgdaBound{M}\<%
\\
\>[2]\AgdaOperator{\AgdaInductiveConstructor{\AgdaUnderscore{}⟶⟨\AgdaUnderscore{}⟩\AgdaUnderscore{}}}\AgdaSpace{}%
\AgdaSymbol{:}\AgdaSpace{}%
\AgdaSymbol{(}\AgdaBound{L}\AgdaSpace{}%
\AgdaSymbol{:}\AgdaSpace{}%
\AgdaDatatype{Term}\AgdaSymbol{)}\AgdaSpace{}%
\AgdaSymbol{\{}\AgdaBound{M}\AgdaSpace{}%
\AgdaBound{N}\AgdaSpace{}%
\AgdaSymbol{:}\AgdaSpace{}%
\AgdaDatatype{Term}\AgdaSymbol{\}}%
\>[36]\AgdaSymbol{→}%
\>[39]\AgdaBound{L}\AgdaSpace{}%
\AgdaOperator{\AgdaDatatype{⟶}}\AgdaSpace{}%
\AgdaBound{M}%
\>[46]\AgdaSymbol{→}%
\>[49]\AgdaBound{M}\AgdaSpace{}%
\AgdaOperator{\AgdaDatatype{↠}}\AgdaSpace{}%
\AgdaBound{N}%
\>[56]\AgdaSymbol{→}%
\>[59]\AgdaBound{L}\AgdaSpace{}%
\AgdaOperator{\AgdaDatatype{↠}}\AgdaSpace{}%
\AgdaBound{N}\<%
\\
\\[\AgdaEmptyExtraSkip]%
\>[0]\AgdaFunction{len}\AgdaSpace{}%
\AgdaSymbol{:}\AgdaSpace{}%
\AgdaSymbol{∀\{}\AgdaBound{M}\AgdaSpace{}%
\AgdaBound{N}\AgdaSpace{}%
\AgdaSymbol{:}\AgdaSpace{}%
\AgdaDatatype{Term}\AgdaSymbol{\}}\AgdaSpace{}%
\AgdaSymbol{→}\AgdaSpace{}%
\AgdaSymbol{(}\AgdaBound{M→N}\AgdaSpace{}%
\AgdaSymbol{:}\AgdaSpace{}%
\AgdaBound{M}\AgdaSpace{}%
\AgdaOperator{\AgdaDatatype{↠}}\AgdaSpace{}%
\AgdaBound{N}\AgdaSymbol{)}\AgdaSpace{}%
\AgdaSymbol{→}\AgdaSpace{}%
\AgdaDatatype{ℕ}\<%
\\
\>[0]\AgdaFunction{len}\AgdaSpace{}%
\AgdaSymbol{(\AgdaUnderscore{}}\AgdaSpace{}%
\AgdaOperator{\AgdaInductiveConstructor{END}}\AgdaSymbol{)}\AgdaSpace{}%
\AgdaSymbol{=}\AgdaSpace{}%
\AgdaNumber{0}\<%
\\
\>[0]\AgdaFunction{len}\AgdaSpace{}%
\AgdaSymbol{(\AgdaUnderscore{}}\AgdaSpace{}%
\AgdaOperator{\AgdaInductiveConstructor{⟶⟨}}\AgdaSpace{}%
\AgdaBound{step}\AgdaSpace{}%
\AgdaOperator{\AgdaInductiveConstructor{⟩}}\AgdaSpace{}%
\AgdaBound{mstep}\AgdaSymbol{)}\AgdaSpace{}%
\AgdaSymbol{=}\AgdaSpace{}%
\AgdaInductiveConstructor{suc}\AgdaSpace{}%
\AgdaSymbol{(}\AgdaFunction{len}\AgdaSpace{}%
\AgdaBound{mstep}\AgdaSymbol{)}\<%
\\
\\[\AgdaEmptyExtraSkip]%
\>[0]\AgdaOperator{\AgdaFunction{\AgdaUnderscore{}⇓}}\AgdaSpace{}%
\AgdaSymbol{:}\AgdaSpace{}%
\AgdaDatatype{Term}\AgdaSpace{}%
\AgdaSymbol{→}\AgdaSpace{}%
\AgdaPrimitive{Set}\<%
\\
\>[0]\AgdaBound{M}\AgdaSpace{}%
\AgdaOperator{\AgdaFunction{⇓}}\AgdaSpace{}%
\AgdaSymbol{=}\AgdaSpace{}%
\AgdaFunction{∃[}\AgdaSpace{}%
\AgdaBound{V}\AgdaSpace{}%
\AgdaFunction{]}\AgdaSpace{}%
\AgdaSymbol{(}\AgdaBound{M}\AgdaSpace{}%
\AgdaOperator{\AgdaDatatype{↠}}\AgdaSpace{}%
\AgdaBound{V}\AgdaSymbol{)}\AgdaSpace{}%
\AgdaOperator{\AgdaFunction{×}}\AgdaSpace{}%
\AgdaDatatype{Value}\AgdaSpace{}%
\AgdaBound{V}\<%
\\
\>[0]\AgdaOperator{\AgdaFunction{\AgdaUnderscore{}⇑}}\AgdaSpace{}%
\AgdaSymbol{:}\AgdaSpace{}%
\AgdaDatatype{Term}\AgdaSpace{}%
\AgdaSymbol{→}\AgdaSpace{}%
\AgdaPrimitive{Set}\<%
\\
\>[0]\AgdaBound{M}\AgdaSpace{}%
\AgdaOperator{\AgdaFunction{⇑}}\AgdaSpace{}%
\AgdaSymbol{=}\AgdaSpace{}%
\AgdaSymbol{∀}\AgdaSpace{}%
\AgdaBound{k}\AgdaSpace{}%
\AgdaSymbol{→}\AgdaSpace{}%
\AgdaFunction{∃[}\AgdaSpace{}%
\AgdaBound{N}\AgdaSpace{}%
\AgdaFunction{]}\AgdaSpace{}%
\AgdaFunction{Σ[}\AgdaSpace{}%
\AgdaBound{r}\AgdaSpace{}%
\AgdaFunction{∈}\AgdaSpace{}%
\AgdaBound{M}\AgdaSpace{}%
\AgdaOperator{\AgdaDatatype{↠}}\AgdaSpace{}%
\AgdaBound{N}\AgdaSpace{}%
\AgdaFunction{]}\AgdaSpace{}%
\AgdaBound{k}\AgdaSpace{}%
\AgdaOperator{\AgdaDatatype{≡}}\AgdaSpace{}%
\AgdaFunction{len}\AgdaSpace{}%
\AgdaBound{r}\<%
\\
\>[0]\AgdaOperator{\AgdaFunction{\AgdaUnderscore{}⇑⊎blame}}\AgdaSpace{}%
\AgdaSymbol{:}\AgdaSpace{}%
\AgdaDatatype{Term}\AgdaSpace{}%
\AgdaSymbol{→}\AgdaSpace{}%
\AgdaPrimitive{Set}\<%
\\
\>[0]\AgdaBound{M}\AgdaSpace{}%
\AgdaOperator{\AgdaFunction{⇑⊎blame}}\AgdaSpace{}%
\AgdaSymbol{=}\AgdaSpace{}%
\AgdaSymbol{∀}\AgdaSpace{}%
\AgdaBound{k}\AgdaSpace{}%
\AgdaSymbol{→}\AgdaSpace{}%
\AgdaFunction{∃[}\AgdaSpace{}%
\AgdaBound{N}\AgdaSpace{}%
\AgdaFunction{]}\AgdaSpace{}%
\AgdaFunction{Σ[}\AgdaSpace{}%
\AgdaBound{r}\AgdaSpace{}%
\AgdaFunction{∈}\AgdaSpace{}%
\AgdaBound{M}\AgdaSpace{}%
\AgdaOperator{\AgdaDatatype{↠}}\AgdaSpace{}%
\AgdaBound{N}\AgdaSpace{}%
\AgdaFunction{]}\AgdaSpace{}%
\AgdaSymbol{((}\AgdaBound{k}\AgdaSpace{}%
\AgdaOperator{\AgdaDatatype{≡}}\AgdaSpace{}%
\AgdaFunction{len}\AgdaSpace{}%
\AgdaBound{r}\AgdaSymbol{)}\AgdaSpace{}%
\AgdaOperator{\AgdaDatatype{⊎}}\AgdaSpace{}%
\AgdaSymbol{(}\AgdaBound{N}\AgdaSpace{}%
\AgdaOperator{\AgdaDatatype{≡}}\AgdaSpace{}%
\AgdaInductiveConstructor{blame}\AgdaSymbol{))}\<%
\end{code}
\caption{Type System and Reduction for the Cast Calculus}
\label{fig:cast-calculus}
\end{figure}

\begin{code}[hide]%
\>[0]\AgdaKeyword{pattern}\AgdaSpace{}%
\AgdaInductiveConstructor{ξ}\AgdaSpace{}%
\AgdaBound{F}\AgdaSpace{}%
\AgdaBound{M⟶N}\AgdaSpace{}%
\AgdaSymbol{=}\AgdaSpace{}%
\AgdaInductiveConstructor{ξξ}\AgdaSpace{}%
\AgdaBound{F}\AgdaSpace{}%
\AgdaInductiveConstructor{refl}\AgdaSpace{}%
\AgdaInductiveConstructor{refl}\AgdaSpace{}%
\AgdaBound{M⟶N}\<%
\\
\>[0]\AgdaKeyword{pattern}\AgdaSpace{}%
\AgdaInductiveConstructor{ξ-blame}\AgdaSpace{}%
\AgdaBound{F}\AgdaSpace{}%
\AgdaSymbol{=}\AgdaSpace{}%
\AgdaInductiveConstructor{ξξ-blame}\AgdaSpace{}%
\AgdaBound{F}\AgdaSpace{}%
\AgdaInductiveConstructor{refl}\<%
\\
\\[\AgdaEmptyExtraSkip]%
\>[0]\AgdaFunction{ξ′}%
\>[1222I]\AgdaSymbol{:}\AgdaSpace{}%
\AgdaSymbol{∀}\AgdaSpace{}%
\AgdaSymbol{\{}\AgdaBound{M}\AgdaSpace{}%
\AgdaBound{N}\AgdaSpace{}%
\AgdaSymbol{:}\AgdaSpace{}%
\AgdaDatatype{Term}\AgdaSymbol{\}}\AgdaSpace{}%
\AgdaSymbol{\{}\AgdaBound{M′}\AgdaSpace{}%
\AgdaBound{N′}\AgdaSpace{}%
\AgdaSymbol{:}\AgdaSpace{}%
\AgdaDatatype{Term}\AgdaSymbol{\}}\<%
\\
\>[1222I][@{}l@{\AgdaIndent{0}}]%
\>[4]\AgdaSymbol{→}\AgdaSpace{}%
\AgdaSymbol{(}\AgdaBound{F}\AgdaSpace{}%
\AgdaSymbol{:}\AgdaSpace{}%
\AgdaDatatype{PEFrame}\AgdaSymbol{)}\<%
\\
\>[4]\AgdaSymbol{→}\AgdaSpace{}%
\AgdaBound{M′}\AgdaSpace{}%
\AgdaOperator{\AgdaDatatype{≡}}\AgdaSpace{}%
\AgdaBound{F}\AgdaSpace{}%
\AgdaOperator{\AgdaFunction{⦉}}\AgdaSpace{}%
\AgdaBound{M}\AgdaSpace{}%
\AgdaOperator{\AgdaFunction{⦊}}\<%
\\
\>[4]\AgdaSymbol{→}\AgdaSpace{}%
\AgdaBound{N′}\AgdaSpace{}%
\AgdaOperator{\AgdaDatatype{≡}}\AgdaSpace{}%
\AgdaBound{F}\AgdaSpace{}%
\AgdaOperator{\AgdaFunction{⦉}}\AgdaSpace{}%
\AgdaBound{N}\AgdaSpace{}%
\AgdaOperator{\AgdaFunction{⦊}}\<%
\\
\>[4]\AgdaSymbol{→}%
\>[1247I]\AgdaBound{M}\AgdaSpace{}%
\AgdaOperator{\AgdaDatatype{⟶}}\AgdaSpace{}%
\AgdaBound{N}\<%
\\
\>[.][@{}l@{}]\<[1247I]%
\>[6]\AgdaComment{--------}\<%
\\
\>[4]\AgdaSymbol{→}\AgdaSpace{}%
\AgdaBound{M′}\AgdaSpace{}%
\AgdaOperator{\AgdaDatatype{⟶}}\AgdaSpace{}%
\AgdaBound{N′}\<%
\\
\>[0]\AgdaFunction{ξ′}\AgdaSpace{}%
\AgdaSymbol{(}\AgdaOperator{\AgdaInductiveConstructor{`}}\AgdaSpace{}%
\AgdaBound{F}\AgdaSymbol{)}\AgdaSpace{}%
\AgdaInductiveConstructor{refl}\AgdaSpace{}%
\AgdaInductiveConstructor{refl}\AgdaSpace{}%
\AgdaBound{M→N}\AgdaSpace{}%
\AgdaSymbol{=}\AgdaSpace{}%
\AgdaInductiveConstructor{ξ}\AgdaSpace{}%
\AgdaBound{F}\AgdaSpace{}%
\AgdaBound{M→N}\<%
\\
\>[0]\AgdaFunction{ξ′}\AgdaSpace{}%
\AgdaInductiveConstructor{□}\AgdaSpace{}%
\AgdaInductiveConstructor{refl}\AgdaSpace{}%
\AgdaInductiveConstructor{refl}\AgdaSpace{}%
\AgdaBound{M→N}\AgdaSpace{}%
\AgdaSymbol{=}\AgdaSpace{}%
\AgdaBound{M→N}\<%
\\
\\[\AgdaEmptyExtraSkip]%
\>[0]\AgdaFunction{ξ′-blame}\AgdaSpace{}%
\AgdaSymbol{:}\AgdaSpace{}%
\AgdaSymbol{∀}\AgdaSpace{}%
\AgdaSymbol{\{}\AgdaBound{M′}\AgdaSpace{}%
\AgdaSymbol{:}\AgdaSpace{}%
\AgdaDatatype{Term}\AgdaSymbol{\}}\<%
\\
\>[0][@{}l@{\AgdaIndent{0}}]%
\>[3]\AgdaSymbol{→}\AgdaSpace{}%
\AgdaSymbol{(}\AgdaBound{F}\AgdaSpace{}%
\AgdaSymbol{:}\AgdaSpace{}%
\AgdaDatatype{PEFrame}\AgdaSymbol{)}\<%
\\
\>[3]\AgdaSymbol{→}%
\>[1276I]\AgdaBound{M′}\AgdaSpace{}%
\AgdaOperator{\AgdaDatatype{≡}}\AgdaSpace{}%
\AgdaBound{F}\AgdaSpace{}%
\AgdaOperator{\AgdaFunction{⦉}}\AgdaSpace{}%
\AgdaInductiveConstructor{blame}\AgdaSpace{}%
\AgdaOperator{\AgdaFunction{⦊}}\<%
\\
\>[.][@{}l@{}]\<[1276I]%
\>[5]\AgdaComment{------------------------}\<%
\\
\>[3]\AgdaSymbol{→}\AgdaSpace{}%
\AgdaBound{M′}\AgdaSpace{}%
\AgdaOperator{\AgdaDatatype{⟶}}\AgdaSpace{}%
\AgdaInductiveConstructor{blame}\AgdaSpace{}%
\AgdaOperator{\AgdaDatatype{⊎}}\AgdaSpace{}%
\AgdaBound{M′}\AgdaSpace{}%
\AgdaOperator{\AgdaDatatype{≡}}\AgdaSpace{}%
\AgdaInductiveConstructor{blame}\<%
\\
\>[0]\AgdaFunction{ξ′-blame}\AgdaSpace{}%
\AgdaSymbol{(}\AgdaOperator{\AgdaInductiveConstructor{`}}\AgdaSpace{}%
\AgdaBound{F}\AgdaSymbol{)}\AgdaSpace{}%
\AgdaInductiveConstructor{refl}\AgdaSpace{}%
\AgdaSymbol{=}\AgdaSpace{}%
\AgdaInductiveConstructor{inj₁}\AgdaSpace{}%
\AgdaSymbol{(}\AgdaInductiveConstructor{ξ-blame}\AgdaSpace{}%
\AgdaBound{F}\AgdaSymbol{)}\<%
\\
\>[0]\AgdaFunction{ξ′-blame}\AgdaSpace{}%
\AgdaInductiveConstructor{□}\AgdaSpace{}%
\AgdaInductiveConstructor{refl}\AgdaSpace{}%
\AgdaSymbol{=}\AgdaSpace{}%
\AgdaInductiveConstructor{inj₂}\AgdaSpace{}%
\AgdaInductiveConstructor{refl}\<%
\\
\\[\AgdaEmptyExtraSkip]%
\>[0]\AgdaFunction{ξ″-blame}\AgdaSpace{}%
\AgdaSymbol{:}\AgdaSpace{}%
\AgdaSymbol{∀}\AgdaSpace{}%
\AgdaSymbol{\{}\AgdaBound{M′}\AgdaSpace{}%
\AgdaSymbol{:}\AgdaSpace{}%
\AgdaDatatype{Term}\AgdaSymbol{\}}\<%
\\
\>[0][@{}l@{\AgdaIndent{0}}]%
\>[3]\AgdaSymbol{→}\AgdaSpace{}%
\AgdaSymbol{(}\AgdaBound{F}\AgdaSpace{}%
\AgdaSymbol{:}\AgdaSpace{}%
\AgdaDatatype{PEFrame}\AgdaSymbol{)}\<%
\\
\>[3]\AgdaSymbol{→}%
\>[1309I]\AgdaBound{M′}\AgdaSpace{}%
\AgdaOperator{\AgdaDatatype{≡}}\AgdaSpace{}%
\AgdaBound{F}\AgdaSpace{}%
\AgdaOperator{\AgdaFunction{⦉}}\AgdaSpace{}%
\AgdaInductiveConstructor{blame}\AgdaSpace{}%
\AgdaOperator{\AgdaFunction{⦊}}\<%
\\
\>[.][@{}l@{}]\<[1309I]%
\>[5]\AgdaComment{----------------------------------}\<%
\\
\>[3]\AgdaSymbol{→}\AgdaSpace{}%
\AgdaBound{M′}\AgdaSpace{}%
\AgdaOperator{\AgdaDatatype{⟶}}\AgdaSpace{}%
\AgdaInductiveConstructor{blame}\AgdaSpace{}%
\AgdaOperator{\AgdaDatatype{⊎}}\AgdaSpace{}%
\AgdaSymbol{(}\AgdaBound{M′}\AgdaSpace{}%
\AgdaOperator{\AgdaDatatype{≡}}\AgdaSpace{}%
\AgdaInductiveConstructor{blame}\AgdaSpace{}%
\AgdaOperator{\AgdaFunction{×}}\AgdaSpace{}%
\AgdaBound{F}\AgdaSpace{}%
\AgdaOperator{\AgdaDatatype{≡}}\AgdaSpace{}%
\AgdaInductiveConstructor{□}\AgdaSymbol{)}\<%
\\
\>[0]\AgdaFunction{ξ″-blame}\AgdaSpace{}%
\AgdaSymbol{(}\AgdaOperator{\AgdaInductiveConstructor{`}}\AgdaSpace{}%
\AgdaBound{F}\AgdaSymbol{)}\AgdaSpace{}%
\AgdaInductiveConstructor{refl}\AgdaSpace{}%
\AgdaSymbol{=}\AgdaSpace{}%
\AgdaInductiveConstructor{inj₁}\AgdaSpace{}%
\AgdaSymbol{(}\AgdaInductiveConstructor{ξ-blame}\AgdaSpace{}%
\AgdaBound{F}\AgdaSymbol{)}\<%
\\
\>[0]\AgdaFunction{ξ″-blame}\AgdaSpace{}%
\AgdaInductiveConstructor{□}\AgdaSpace{}%
\AgdaInductiveConstructor{refl}\AgdaSpace{}%
\AgdaSymbol{=}\AgdaSpace{}%
\AgdaInductiveConstructor{inj₂}\AgdaSpace{}%
\AgdaSymbol{(}\AgdaInductiveConstructor{refl}\AgdaSpace{}%
\AgdaOperator{\AgdaInductiveConstructor{,}}\AgdaSpace{}%
\AgdaInductiveConstructor{refl}\AgdaSymbol{)}\<%
\\
\\[\AgdaEmptyExtraSkip]%
\>[0]\AgdaComment{\{-\ Reflexive\ and\ transitive\ closure\ of\ reduction\ -\}}\<%
\\
\\[\AgdaEmptyExtraSkip]%
\\[\AgdaEmptyExtraSkip]%
\>[0]\AgdaComment{--begin\AgdaUnderscore{}\ :\ ∀\ \{M\ N\ :\ Term\}\ →\ (M\ ↠\ N)\ →\ (M\ ↠\ N)}\<%
\\
\>[0]\AgdaComment{--begin\ M↠N\ =\ M↠N}\<%
\\
\\[\AgdaEmptyExtraSkip]%
\>[0]\AgdaComment{\{-\ Convenience\ function\ to\ build\ a\ sequence\ of\ length\ one.\ -\}}\<%
\\
\\[\AgdaEmptyExtraSkip]%
\>[0]\AgdaFunction{unit}\AgdaSpace{}%
\AgdaSymbol{:}\AgdaSpace{}%
\AgdaSymbol{∀}\AgdaSpace{}%
\AgdaSymbol{\{}\AgdaBound{M}\AgdaSpace{}%
\AgdaBound{N}\AgdaSpace{}%
\AgdaSymbol{:}\AgdaSpace{}%
\AgdaDatatype{Term}\AgdaSymbol{\}}\AgdaSpace{}%
\AgdaSymbol{→}\AgdaSpace{}%
\AgdaSymbol{(}\AgdaBound{M}\AgdaSpace{}%
\AgdaOperator{\AgdaDatatype{⟶}}\AgdaSpace{}%
\AgdaBound{N}\AgdaSymbol{)}\AgdaSpace{}%
\AgdaSymbol{→}\AgdaSpace{}%
\AgdaSymbol{(}\AgdaBound{M}\AgdaSpace{}%
\AgdaOperator{\AgdaDatatype{↠}}\AgdaSpace{}%
\AgdaBound{N}\AgdaSymbol{)}\<%
\\
\>[0]\AgdaFunction{unit}\AgdaSpace{}%
\AgdaSymbol{\{}\AgdaArgument{M}\AgdaSpace{}%
\AgdaSymbol{=}\AgdaSpace{}%
\AgdaBound{M}\AgdaSymbol{\}}\AgdaSpace{}%
\AgdaSymbol{\{}\AgdaArgument{N}\AgdaSpace{}%
\AgdaSymbol{=}\AgdaSpace{}%
\AgdaBound{N}\AgdaSymbol{\}}\AgdaSpace{}%
\AgdaBound{M⟶N}%
\>[26]\AgdaSymbol{=}%
\>[29]\AgdaBound{M}\AgdaSpace{}%
\AgdaOperator{\AgdaInductiveConstructor{⟶⟨}}\AgdaSpace{}%
\AgdaBound{M⟶N}\AgdaSpace{}%
\AgdaOperator{\AgdaInductiveConstructor{⟩}}\AgdaSpace{}%
\AgdaSymbol{(}\AgdaBound{N}\AgdaSpace{}%
\AgdaOperator{\AgdaInductiveConstructor{END}}\AgdaSymbol{)}\<%
\\
\\[\AgdaEmptyExtraSkip]%
\\[\AgdaEmptyExtraSkip]%
\>[0]\AgdaComment{\{-\ Apply\ ξ\ to\ each\ element\ of\ a\ sequence\ -\}}\<%
\\
\\[\AgdaEmptyExtraSkip]%
\>[0]\AgdaFunction{ξ*}\AgdaSpace{}%
\AgdaSymbol{:}\AgdaSpace{}%
\AgdaSymbol{∀}\AgdaSpace{}%
\AgdaSymbol{\{}\AgdaBound{M}\AgdaSpace{}%
\AgdaBound{N}\AgdaSpace{}%
\AgdaSymbol{:}\AgdaSpace{}%
\AgdaDatatype{Term}\AgdaSymbol{\}}\AgdaSpace{}%
\AgdaSymbol{→}\AgdaSpace{}%
\AgdaSymbol{(}\AgdaBound{F}\AgdaSpace{}%
\AgdaSymbol{:}\AgdaSpace{}%
\AgdaDatatype{Frame}\AgdaSymbol{)}\AgdaSpace{}%
\AgdaSymbol{→}\AgdaSpace{}%
\AgdaBound{M}\AgdaSpace{}%
\AgdaOperator{\AgdaDatatype{↠}}\AgdaSpace{}%
\AgdaBound{N}\AgdaSpace{}%
\AgdaSymbol{→}\AgdaSpace{}%
\AgdaBound{F}\AgdaSpace{}%
\AgdaOperator{\AgdaFunction{⟦}}\AgdaSpace{}%
\AgdaBound{M}\AgdaSpace{}%
\AgdaOperator{\AgdaFunction{⟧}}\AgdaSpace{}%
\AgdaOperator{\AgdaDatatype{↠}}\AgdaSpace{}%
\AgdaBound{F}\AgdaSpace{}%
\AgdaOperator{\AgdaFunction{⟦}}\AgdaSpace{}%
\AgdaBound{N}\AgdaSpace{}%
\AgdaOperator{\AgdaFunction{⟧}}\<%
\\
\>[0]\AgdaFunction{ξ*}\AgdaSpace{}%
\AgdaBound{F}\AgdaSpace{}%
\AgdaSymbol{(}\AgdaBound{M}\AgdaSpace{}%
\AgdaOperator{\AgdaInductiveConstructor{END}}\AgdaSymbol{)}\AgdaSpace{}%
\AgdaSymbol{=}\AgdaSpace{}%
\AgdaBound{F}\AgdaSpace{}%
\AgdaOperator{\AgdaFunction{⟦}}\AgdaSpace{}%
\AgdaBound{M}\AgdaSpace{}%
\AgdaOperator{\AgdaFunction{⟧}}\AgdaSpace{}%
\AgdaOperator{\AgdaInductiveConstructor{END}}\<%
\\
\>[0]\AgdaFunction{ξ*}\AgdaSpace{}%
\AgdaBound{F}\AgdaSpace{}%
\AgdaSymbol{(}\AgdaBound{L}\AgdaSpace{}%
\AgdaOperator{\AgdaInductiveConstructor{⟶⟨}}\AgdaSpace{}%
\AgdaBound{L⟶M}\AgdaSpace{}%
\AgdaOperator{\AgdaInductiveConstructor{⟩}}\AgdaSpace{}%
\AgdaBound{M↠N}\AgdaSymbol{)}\AgdaSpace{}%
\AgdaSymbol{=}\AgdaSpace{}%
\AgdaSymbol{(}\AgdaBound{F}\AgdaSpace{}%
\AgdaOperator{\AgdaFunction{⟦}}\AgdaSpace{}%
\AgdaBound{L}\AgdaSpace{}%
\AgdaOperator{\AgdaFunction{⟧}}\AgdaSpace{}%
\AgdaOperator{\AgdaInductiveConstructor{⟶⟨}}\AgdaSpace{}%
\AgdaInductiveConstructor{ξ}\AgdaSpace{}%
\AgdaBound{F}\AgdaSpace{}%
\AgdaBound{L⟶M}\AgdaSpace{}%
\AgdaOperator{\AgdaInductiveConstructor{⟩}}\AgdaSpace{}%
\AgdaFunction{ξ*}\AgdaSpace{}%
\AgdaBound{F}\AgdaSpace{}%
\AgdaBound{M↠N}\AgdaSymbol{)}\<%
\\
\\[\AgdaEmptyExtraSkip]%
\>[0]\AgdaFunction{ξ′*}\AgdaSpace{}%
\AgdaSymbol{:}\AgdaSpace{}%
\AgdaSymbol{∀\{}\AgdaBound{M}\AgdaSpace{}%
\AgdaBound{N}\AgdaSymbol{\}}\AgdaSpace{}%
\AgdaSymbol{→}\AgdaSpace{}%
\AgdaSymbol{(}\AgdaBound{F}\AgdaSpace{}%
\AgdaSymbol{:}\AgdaSpace{}%
\AgdaDatatype{PEFrame}\AgdaSymbol{)}\AgdaSpace{}%
\AgdaSymbol{→}\AgdaSpace{}%
\AgdaBound{M}\AgdaSpace{}%
\AgdaOperator{\AgdaDatatype{↠}}\AgdaSpace{}%
\AgdaBound{N}\AgdaSpace{}%
\AgdaSymbol{→}\AgdaSpace{}%
\AgdaBound{F}\AgdaSpace{}%
\AgdaOperator{\AgdaFunction{⦉}}\AgdaSpace{}%
\AgdaBound{M}\AgdaSpace{}%
\AgdaOperator{\AgdaFunction{⦊}}\AgdaSpace{}%
\AgdaOperator{\AgdaDatatype{↠}}\AgdaSpace{}%
\AgdaBound{F}\AgdaSpace{}%
\AgdaOperator{\AgdaFunction{⦉}}\AgdaSpace{}%
\AgdaBound{N}\AgdaSpace{}%
\AgdaOperator{\AgdaFunction{⦊}}\<%
\\
\>[0]\AgdaFunction{ξ′*}\AgdaSpace{}%
\AgdaSymbol{\{}\AgdaBound{M}\AgdaSymbol{\}}\AgdaSpace{}%
\AgdaSymbol{\{}\AgdaBound{N}\AgdaSymbol{\}}\AgdaSpace{}%
\AgdaSymbol{(}\AgdaOperator{\AgdaInductiveConstructor{`}}\AgdaSpace{}%
\AgdaBound{F}\AgdaSymbol{)}\AgdaSpace{}%
\AgdaBound{M→N}\AgdaSpace{}%
\AgdaSymbol{=}\AgdaSpace{}%
\AgdaFunction{ξ*}\AgdaSpace{}%
\AgdaBound{F}\AgdaSpace{}%
\AgdaBound{M→N}\<%
\\
\>[0]\AgdaFunction{ξ′*}\AgdaSpace{}%
\AgdaSymbol{\{}\AgdaBound{M}\AgdaSymbol{\}}\AgdaSpace{}%
\AgdaSymbol{\{}\AgdaBound{N}\AgdaSymbol{\}}\AgdaSpace{}%
\AgdaInductiveConstructor{□}\AgdaSpace{}%
\AgdaBound{M→N}\AgdaSpace{}%
\AgdaSymbol{=}\AgdaSpace{}%
\AgdaBound{M→N}\<%
\\
\\[\AgdaEmptyExtraSkip]%
\>[0]\AgdaComment{\{-\ Concatenate\ two\ sequences.\ -\}}\<%
\\
\\[\AgdaEmptyExtraSkip]%
\>[0]\AgdaOperator{\AgdaFunction{\AgdaUnderscore{}++\AgdaUnderscore{}}}\AgdaSpace{}%
\AgdaSymbol{:}\AgdaSpace{}%
\AgdaSymbol{∀}\AgdaSpace{}%
\AgdaSymbol{\{}\AgdaBound{L}\AgdaSpace{}%
\AgdaBound{M}\AgdaSpace{}%
\AgdaBound{N}\AgdaSpace{}%
\AgdaSymbol{:}\AgdaSpace{}%
\AgdaDatatype{Term}\AgdaSymbol{\}}\AgdaSpace{}%
\AgdaSymbol{→}\AgdaSpace{}%
\AgdaBound{L}\AgdaSpace{}%
\AgdaOperator{\AgdaDatatype{↠}}\AgdaSpace{}%
\AgdaBound{M}\AgdaSpace{}%
\AgdaSymbol{→}\AgdaSpace{}%
\AgdaBound{M}\AgdaSpace{}%
\AgdaOperator{\AgdaDatatype{↠}}\AgdaSpace{}%
\AgdaBound{N}\AgdaSpace{}%
\AgdaSymbol{→}\AgdaSpace{}%
\AgdaBound{L}\AgdaSpace{}%
\AgdaOperator{\AgdaDatatype{↠}}\AgdaSpace{}%
\AgdaBound{N}\<%
\\
\>[0]\AgdaSymbol{(}\AgdaBound{M}\AgdaSpace{}%
\AgdaOperator{\AgdaInductiveConstructor{END}}\AgdaSymbol{)}\AgdaSpace{}%
\AgdaOperator{\AgdaFunction{++}}\AgdaSpace{}%
\AgdaBound{M↠N}%
\>[30]\AgdaSymbol{=}%
\>[33]\AgdaBound{M↠N}\<%
\\
\>[0]\AgdaSymbol{(}\AgdaBound{L}\AgdaSpace{}%
\AgdaOperator{\AgdaInductiveConstructor{⟶⟨}}\AgdaSpace{}%
\AgdaBound{L⟶M}\AgdaSpace{}%
\AgdaOperator{\AgdaInductiveConstructor{⟩}}\AgdaSpace{}%
\AgdaBound{M↠N}\AgdaSymbol{)}\AgdaSpace{}%
\AgdaOperator{\AgdaFunction{++}}\AgdaSpace{}%
\AgdaBound{N↠P}%
\>[25]\AgdaSymbol{=}%
\>[28]\AgdaBound{L}\AgdaSpace{}%
\AgdaOperator{\AgdaInductiveConstructor{⟶⟨}}\AgdaSpace{}%
\AgdaBound{L⟶M}\AgdaSpace{}%
\AgdaOperator{\AgdaInductiveConstructor{⟩}}\AgdaSpace{}%
\AgdaSymbol{(}\AgdaBound{M↠N}\AgdaSpace{}%
\AgdaOperator{\AgdaFunction{++}}\AgdaSpace{}%
\AgdaBound{N↠P}\AgdaSymbol{)}\<%
\\
\\[\AgdaEmptyExtraSkip]%
\>[0]\AgdaFunction{ξ-blame₃}\AgdaSpace{}%
\AgdaSymbol{:}\AgdaSpace{}%
\AgdaSymbol{∀}\AgdaSpace{}%
\AgdaSymbol{\{}\AgdaBound{M}\AgdaSymbol{\}\{}\AgdaBound{M′}\AgdaSpace{}%
\AgdaSymbol{:}\AgdaSpace{}%
\AgdaDatatype{Term}\AgdaSymbol{\}}\<%
\\
\>[0][@{}l@{\AgdaIndent{0}}]%
\>[3]\AgdaSymbol{→}\AgdaSpace{}%
\AgdaSymbol{(}\AgdaBound{F}\AgdaSpace{}%
\AgdaSymbol{:}\AgdaSpace{}%
\AgdaDatatype{PEFrame}\AgdaSymbol{)}\<%
\\
\>[3]\AgdaSymbol{→}\AgdaSpace{}%
\AgdaBound{M}\AgdaSpace{}%
\AgdaOperator{\AgdaDatatype{↠}}\AgdaSpace{}%
\AgdaInductiveConstructor{blame}\<%
\\
\>[3]\AgdaSymbol{→}%
\>[1499I]\AgdaBound{M′}\AgdaSpace{}%
\AgdaOperator{\AgdaDatatype{≡}}\AgdaSpace{}%
\AgdaBound{F}\AgdaSpace{}%
\AgdaOperator{\AgdaFunction{⦉}}\AgdaSpace{}%
\AgdaBound{M}\AgdaSpace{}%
\AgdaOperator{\AgdaFunction{⦊}}\<%
\\
\>[.][@{}l@{}]\<[1499I]%
\>[5]\AgdaComment{-----------}\<%
\\
\>[3]\AgdaSymbol{→}\AgdaSpace{}%
\AgdaBound{M′}\AgdaSpace{}%
\AgdaOperator{\AgdaDatatype{↠}}\AgdaSpace{}%
\AgdaInductiveConstructor{blame}\<%
\\
\>[0]\AgdaFunction{ξ-blame₃}\AgdaSpace{}%
\AgdaSymbol{(}\AgdaOperator{\AgdaInductiveConstructor{`}}\AgdaSpace{}%
\AgdaBound{F}\AgdaSymbol{)}\AgdaSpace{}%
\AgdaBound{M→b}\AgdaSpace{}%
\AgdaInductiveConstructor{refl}\AgdaSpace{}%
\AgdaSymbol{=}\AgdaSpace{}%
\AgdaSymbol{(}\AgdaFunction{ξ*}\AgdaSpace{}%
\AgdaBound{F}\AgdaSpace{}%
\AgdaBound{M→b}\AgdaSymbol{)}\AgdaSpace{}%
\AgdaOperator{\AgdaFunction{++}}\AgdaSpace{}%
\AgdaFunction{unit}\AgdaSpace{}%
\AgdaSymbol{(}\AgdaInductiveConstructor{ξ-blame}\AgdaSpace{}%
\AgdaBound{F}\AgdaSymbol{)}\<%
\\
\>[0]\AgdaFunction{ξ-blame₃}\AgdaSpace{}%
\AgdaInductiveConstructor{□}\AgdaSpace{}%
\AgdaBound{M→b}\AgdaSpace{}%
\AgdaInductiveConstructor{refl}\AgdaSpace{}%
\AgdaSymbol{=}\AgdaSpace{}%
\AgdaBound{M→b}\<%
\\
\\[\AgdaEmptyExtraSkip]%
\>[0]\AgdaComment{\{-\ Alternative\ notation\ for\ sequence\ concatenation.\ -\}}\<%
\\
\\[\AgdaEmptyExtraSkip]%
\>[0]\AgdaOperator{\AgdaFunction{\AgdaUnderscore{}↠⟨\AgdaUnderscore{}⟩\AgdaUnderscore{}}}\AgdaSpace{}%
\AgdaSymbol{:}\AgdaSpace{}%
\AgdaSymbol{(}\AgdaBound{L}\AgdaSpace{}%
\AgdaSymbol{:}\AgdaSpace{}%
\AgdaDatatype{Term}\AgdaSymbol{)}\AgdaSpace{}%
\AgdaSymbol{\{}\AgdaBound{M}\AgdaSpace{}%
\AgdaBound{N}\AgdaSpace{}%
\AgdaSymbol{:}\AgdaSpace{}%
\AgdaDatatype{Term}\AgdaSymbol{\}}\<%
\\
\>[0][@{}l@{\AgdaIndent{0}}]%
\>[2]\AgdaSymbol{→}\AgdaSpace{}%
\AgdaBound{L}\AgdaSpace{}%
\AgdaOperator{\AgdaDatatype{↠}}\AgdaSpace{}%
\AgdaBound{M}\<%
\\
\>[2]\AgdaSymbol{→}%
\>[1536I]\AgdaBound{M}\AgdaSpace{}%
\AgdaOperator{\AgdaDatatype{↠}}\AgdaSpace{}%
\AgdaBound{N}\<%
\\
\>[.][@{}l@{}]\<[1536I]%
\>[4]\AgdaComment{---------}\<%
\\
\>[2]\AgdaSymbol{→}\AgdaSpace{}%
\AgdaBound{L}\AgdaSpace{}%
\AgdaOperator{\AgdaDatatype{↠}}\AgdaSpace{}%
\AgdaBound{N}\<%
\\
\>[0]\AgdaBound{L}\AgdaSpace{}%
\AgdaOperator{\AgdaFunction{↠⟨}}\AgdaSpace{}%
\AgdaBound{L↠M}\AgdaSpace{}%
\AgdaOperator{\AgdaFunction{⟩}}\AgdaSpace{}%
\AgdaBound{M↠N}%
\>[16]\AgdaSymbol{=}%
\>[19]\AgdaBound{L↠M}\AgdaSpace{}%
\AgdaOperator{\AgdaFunction{++}}\AgdaSpace{}%
\AgdaBound{M↠N}\<%
\\
\\[\AgdaEmptyExtraSkip]%
\>[0]\AgdaFunction{reducible}\AgdaSpace{}%
\AgdaSymbol{:}\AgdaSpace{}%
\AgdaSymbol{(}\AgdaBound{M}\AgdaSpace{}%
\AgdaSymbol{:}\AgdaSpace{}%
\AgdaDatatype{Term}\AgdaSymbol{)}\AgdaSpace{}%
\AgdaSymbol{→}\AgdaSpace{}%
\AgdaPrimitive{Set}\<%
\\
\>[0]\AgdaFunction{reducible}\AgdaSpace{}%
\AgdaBound{M}\AgdaSpace{}%
\AgdaSymbol{=}\AgdaSpace{}%
\AgdaFunction{∃[}\AgdaSpace{}%
\AgdaBound{N}\AgdaSpace{}%
\AgdaFunction{]}\AgdaSpace{}%
\AgdaSymbol{(}\AgdaBound{M}\AgdaSpace{}%
\AgdaOperator{\AgdaDatatype{⟶}}\AgdaSpace{}%
\AgdaBound{N}\AgdaSymbol{)}\<%
\\
\\[\AgdaEmptyExtraSkip]%
\>[0]\AgdaFunction{irred}\AgdaSpace{}%
\AgdaSymbol{:}\AgdaSpace{}%
\AgdaSymbol{(}\AgdaBound{M}\AgdaSpace{}%
\AgdaSymbol{:}\AgdaSpace{}%
\AgdaDatatype{Term}\AgdaSymbol{)}\AgdaSpace{}%
\AgdaSymbol{→}\AgdaSpace{}%
\AgdaPrimitive{Set}\<%
\\
\>[0]\AgdaFunction{irred}\AgdaSpace{}%
\AgdaBound{M}\AgdaSpace{}%
\AgdaSymbol{=}\AgdaSpace{}%
\AgdaOperator{\AgdaFunction{¬}}\AgdaSpace{}%
\AgdaFunction{reducible}\AgdaSpace{}%
\AgdaBound{M}\<%
\\
\\[\AgdaEmptyExtraSkip]%
\>[0]\AgdaFunction{len-concat}\AgdaSpace{}%
\AgdaSymbol{:}\AgdaSpace{}%
\AgdaSymbol{∀}\AgdaSpace{}%
\AgdaSymbol{\{}\AgdaBound{L}\AgdaSymbol{\}\{}\AgdaBound{M}\AgdaSymbol{\}\{}\AgdaBound{N}\AgdaSymbol{\}}\AgdaSpace{}%
\AgdaSymbol{(}\AgdaBound{s}\AgdaSpace{}%
\AgdaSymbol{:}\AgdaSpace{}%
\AgdaBound{L}\AgdaSpace{}%
\AgdaOperator{\AgdaDatatype{↠}}\AgdaSpace{}%
\AgdaBound{M}\AgdaSymbol{)}\AgdaSpace{}%
\AgdaSymbol{(}\AgdaBound{r}\AgdaSpace{}%
\AgdaSymbol{:}\AgdaSpace{}%
\AgdaBound{M}\AgdaSpace{}%
\AgdaOperator{\AgdaDatatype{↠}}\AgdaSpace{}%
\AgdaBound{N}\AgdaSymbol{)}\<%
\\
\>[0][@{}l@{\AgdaIndent{0}}]%
\>[2]\AgdaSymbol{→}\AgdaSpace{}%
\AgdaFunction{len}\AgdaSpace{}%
\AgdaSymbol{(}\AgdaBound{s}\AgdaSpace{}%
\AgdaOperator{\AgdaFunction{++}}\AgdaSpace{}%
\AgdaBound{r}\AgdaSymbol{)}\AgdaSpace{}%
\AgdaOperator{\AgdaDatatype{≡}}\AgdaSpace{}%
\AgdaFunction{len}\AgdaSpace{}%
\AgdaBound{s}\AgdaSpace{}%
\AgdaOperator{\AgdaPrimitive{+}}\AgdaSpace{}%
\AgdaFunction{len}\AgdaSpace{}%
\AgdaBound{r}\<%
\\
\>[0]\AgdaFunction{len-concat}\AgdaSpace{}%
\AgdaSymbol{(\AgdaUnderscore{}}\AgdaSpace{}%
\AgdaOperator{\AgdaInductiveConstructor{END}}\AgdaSymbol{)}\AgdaSpace{}%
\AgdaBound{r}\AgdaSpace{}%
\AgdaSymbol{=}\AgdaSpace{}%
\AgdaInductiveConstructor{refl}\<%
\\
\>[0]\AgdaFunction{len-concat}\AgdaSpace{}%
\AgdaSymbol{(\AgdaUnderscore{}}\AgdaSpace{}%
\AgdaOperator{\AgdaInductiveConstructor{⟶⟨}}\AgdaSpace{}%
\AgdaBound{x}\AgdaSpace{}%
\AgdaOperator{\AgdaInductiveConstructor{⟩}}\AgdaSpace{}%
\AgdaBound{s}\AgdaSymbol{)}\AgdaSpace{}%
\AgdaBound{r}\AgdaSpace{}%
\AgdaKeyword{rewrite}\AgdaSpace{}%
\AgdaFunction{len-concat}\AgdaSpace{}%
\AgdaBound{s}\AgdaSpace{}%
\AgdaBound{r}\AgdaSpace{}%
\AgdaSymbol{=}\AgdaSpace{}%
\AgdaInductiveConstructor{refl}\<%
\\
\\[\AgdaEmptyExtraSkip]%
\\[\AgdaEmptyExtraSkip]%
\>[0]\AgdaComment{\{-}\<%
\\
\>[0]\AgdaComment{would\ prefer\ to\ say}\<%
\\
\>[0]\AgdaComment{(M\ ⇑)\ ⊎\ (M\ ↠\ blame)}\<%
\\
\>[0]\AgdaComment{but\ I'm\ having\ trouble\ showing}\<%
\\
\>[0]\<%
\\
\>[0]\AgdaComment{lemma\ :\ ∀\ M\ →\ M\ ⇑⊎blame\ →\ (M\ ⇑)\ ⊎\ (M\ ↠\ blame)}\<%
\\
\>[0]\AgdaComment{lemma\ M\ M⇑blame\ =\ \{!!\}}\<%
\\
\>[0]\<%
\\
\>[0]\AgdaComment{-\}}\<%
\\
\\[\AgdaEmptyExtraSkip]%
\>[0]\AgdaFunction{halt}\AgdaSpace{}%
\AgdaSymbol{:}\AgdaSpace{}%
\AgdaDatatype{Term}\AgdaSpace{}%
\AgdaSymbol{→}\AgdaSpace{}%
\AgdaPrimitive{Set}\<%
\\
\>[0]\AgdaFunction{halt}\AgdaSpace{}%
\AgdaBound{M}%
\>[8]\AgdaSymbol{=}\AgdaSpace{}%
\AgdaSymbol{(}\AgdaBound{M}\AgdaSpace{}%
\AgdaOperator{\AgdaDatatype{↠}}\AgdaSpace{}%
\AgdaInductiveConstructor{blame}\AgdaSymbol{)}\AgdaSpace{}%
\AgdaOperator{\AgdaDatatype{⊎}}\AgdaSpace{}%
\AgdaSymbol{(}\AgdaBound{M}\AgdaSpace{}%
\AgdaOperator{\AgdaFunction{⇓}}\AgdaSymbol{)}\<%
\end{code}

\begin{code}[hide]%
\>[0]\AgdaFunction{blame-not-value}\AgdaSpace{}%
\AgdaSymbol{:}\AgdaSpace{}%
\AgdaSymbol{∀\{}\AgdaBound{V}\AgdaSymbol{\}}\AgdaSpace{}%
\AgdaSymbol{→}\AgdaSpace{}%
\AgdaDatatype{Value}\AgdaSpace{}%
\AgdaBound{V}\AgdaSpace{}%
\AgdaSymbol{→}\AgdaSpace{}%
\AgdaBound{V}\AgdaSpace{}%
\AgdaOperator{\AgdaDatatype{≡}}\AgdaSpace{}%
\AgdaInductiveConstructor{blame}\AgdaSpace{}%
\AgdaSymbol{→}\AgdaSpace{}%
\AgdaDatatype{⊥}\<%
\\
\>[0]\AgdaFunction{blame-not-value}\AgdaSpace{}%
\AgdaSymbol{\{}\AgdaDottedPattern{\AgdaSymbol{.}}\AgdaDottedPattern{\AgdaInductiveConstructor{blame}}\AgdaSymbol{\}}\AgdaSpace{}%
\AgdaSymbol{()}\AgdaSpace{}%
\AgdaInductiveConstructor{refl}\<%
\\
\\[\AgdaEmptyExtraSkip]%
\>[0]\AgdaFunction{value-irreducible}\AgdaSpace{}%
\AgdaSymbol{:}\AgdaSpace{}%
\AgdaSymbol{∀}\AgdaSpace{}%
\AgdaSymbol{\{}\AgdaBound{V}\AgdaSpace{}%
\AgdaBound{M}\AgdaSpace{}%
\AgdaSymbol{:}\AgdaSpace{}%
\AgdaDatatype{Term}\AgdaSymbol{\}}\AgdaSpace{}%
\AgdaSymbol{→}\AgdaSpace{}%
\AgdaDatatype{Value}\AgdaSpace{}%
\AgdaBound{V}\AgdaSpace{}%
\AgdaSymbol{→}\AgdaSpace{}%
\AgdaOperator{\AgdaFunction{¬}}\AgdaSpace{}%
\AgdaSymbol{(}\AgdaBound{V}\AgdaSpace{}%
\AgdaOperator{\AgdaDatatype{⟶}}\AgdaSpace{}%
\AgdaBound{M}\AgdaSymbol{)}\<%
\\
\>[0]\AgdaFunction{value-irreducible}\AgdaSpace{}%
\AgdaBound{v}\AgdaSpace{}%
\AgdaBound{V⟶M}\AgdaSpace{}%
\AgdaSymbol{=}\AgdaSpace{}%
\AgdaFunction{nope}\AgdaSpace{}%
\AgdaBound{V⟶M}\AgdaSpace{}%
\AgdaBound{v}\<%
\\
\>[0][@{}l@{\AgdaIndent{0}}]%
\>[3]\AgdaKeyword{where}\<%
\\
\>[3]\AgdaFunction{nope}\AgdaSpace{}%
\AgdaSymbol{:}\AgdaSpace{}%
\AgdaSymbol{∀}\AgdaSpace{}%
\AgdaSymbol{\{}\AgdaBound{V}\AgdaSpace{}%
\AgdaBound{M}\AgdaSpace{}%
\AgdaSymbol{:}\AgdaSpace{}%
\AgdaDatatype{Term}\AgdaSymbol{\}}\AgdaSpace{}%
\AgdaSymbol{→}\AgdaSpace{}%
\AgdaBound{V}\AgdaSpace{}%
\AgdaOperator{\AgdaDatatype{⟶}}\AgdaSpace{}%
\AgdaBound{M}\AgdaSpace{}%
\AgdaSymbol{→}\AgdaSpace{}%
\AgdaDatatype{Value}\AgdaSpace{}%
\AgdaBound{V}\AgdaSpace{}%
\AgdaSymbol{→}\AgdaSpace{}%
\AgdaDatatype{⊥}\<%
\\
\>[3]\AgdaFunction{nope}\AgdaSpace{}%
\AgdaSymbol{(}\AgdaInductiveConstructor{ξξ}\AgdaSpace{}%
\AgdaSymbol{(}\AgdaOperator{\AgdaInductiveConstructor{□·}}\AgdaSpace{}%
\AgdaBound{M}\AgdaSymbol{)}\AgdaSpace{}%
\AgdaSymbol{()}\AgdaSpace{}%
\AgdaBound{x₁}\AgdaSpace{}%
\AgdaBound{V→M}\AgdaSymbol{)}\AgdaSpace{}%
\AgdaSymbol{(}\AgdaOperator{\AgdaInductiveConstructor{ƛ̬}}\AgdaSpace{}%
\AgdaBound{N}\AgdaSymbol{)}\<%
\\
\>[3]\AgdaFunction{nope}\AgdaSpace{}%
\AgdaSymbol{(}\AgdaInductiveConstructor{ξξ}\AgdaSpace{}%
\AgdaSymbol{(}\AgdaBound{v}\AgdaSpace{}%
\AgdaOperator{\AgdaInductiveConstructor{·□}}\AgdaSymbol{)}\AgdaSpace{}%
\AgdaSymbol{()}\AgdaSpace{}%
\AgdaBound{x₁}\AgdaSpace{}%
\AgdaBound{V→M}\AgdaSymbol{)}\AgdaSpace{}%
\AgdaSymbol{(}\AgdaOperator{\AgdaInductiveConstructor{ƛ̬}}\AgdaSpace{}%
\AgdaBound{N}\AgdaSymbol{)}\<%
\\
\>[3]\AgdaFunction{nope}\AgdaSpace{}%
\AgdaSymbol{(}\AgdaInductiveConstructor{ξξ}\AgdaSpace{}%
\AgdaOperator{\AgdaInductiveConstructor{□⟨}}\AgdaSpace{}%
\AgdaBound{G}\AgdaSpace{}%
\AgdaOperator{\AgdaInductiveConstructor{!⟩}}\AgdaSpace{}%
\AgdaSymbol{()}\AgdaSpace{}%
\AgdaBound{x₁}\AgdaSpace{}%
\AgdaBound{V→M}\AgdaSymbol{)}\AgdaSpace{}%
\AgdaSymbol{(}\AgdaOperator{\AgdaInductiveConstructor{ƛ̬}}\AgdaSpace{}%
\AgdaBound{N}\AgdaSymbol{)}\<%
\\
\>[3]\AgdaFunction{nope}\AgdaSpace{}%
\AgdaSymbol{(}\AgdaInductiveConstructor{ξξ}\AgdaSpace{}%
\AgdaOperator{\AgdaInductiveConstructor{□⟨}}\AgdaSpace{}%
\AgdaBound{H}\AgdaSpace{}%
\AgdaOperator{\AgdaInductiveConstructor{?⟩}}\AgdaSpace{}%
\AgdaSymbol{()}\AgdaSpace{}%
\AgdaBound{x₁}\AgdaSpace{}%
\AgdaBound{V→M}\AgdaSymbol{)}\AgdaSpace{}%
\AgdaSymbol{(}\AgdaOperator{\AgdaInductiveConstructor{ƛ̬}}\AgdaSpace{}%
\AgdaBound{N}\AgdaSymbol{)}\<%
\\
\>[3]\AgdaFunction{nope}\AgdaSpace{}%
\AgdaSymbol{(}\AgdaInductiveConstructor{ξξ-blame}\AgdaSpace{}%
\AgdaSymbol{(}\AgdaOperator{\AgdaInductiveConstructor{□·}}\AgdaSpace{}%
\AgdaBound{M}\AgdaSymbol{)}\AgdaSpace{}%
\AgdaSymbol{())}\AgdaSpace{}%
\AgdaSymbol{(}\AgdaOperator{\AgdaInductiveConstructor{ƛ̬}}\AgdaSpace{}%
\AgdaBound{N}\AgdaSymbol{)}\<%
\\
\>[3]\AgdaFunction{nope}\AgdaSpace{}%
\AgdaSymbol{(}\AgdaInductiveConstructor{ξξ-blame}\AgdaSpace{}%
\AgdaSymbol{(}\AgdaBound{v}\AgdaSpace{}%
\AgdaOperator{\AgdaInductiveConstructor{·□}}\AgdaSymbol{)}\AgdaSpace{}%
\AgdaSymbol{())}\AgdaSpace{}%
\AgdaSymbol{(}\AgdaOperator{\AgdaInductiveConstructor{ƛ̬}}\AgdaSpace{}%
\AgdaBound{N}\AgdaSymbol{)}\<%
\\
\>[3]\AgdaFunction{nope}\AgdaSpace{}%
\AgdaSymbol{(}\AgdaInductiveConstructor{ξξ-blame}\AgdaSpace{}%
\AgdaOperator{\AgdaInductiveConstructor{□⟨}}\AgdaSpace{}%
\AgdaBound{G}\AgdaSpace{}%
\AgdaOperator{\AgdaInductiveConstructor{!⟩}}\AgdaSpace{}%
\AgdaSymbol{())}\AgdaSpace{}%
\AgdaSymbol{(}\AgdaOperator{\AgdaInductiveConstructor{ƛ̬}}\AgdaSpace{}%
\AgdaBound{N}\AgdaSymbol{)}\<%
\\
\>[3]\AgdaFunction{nope}\AgdaSpace{}%
\AgdaSymbol{(}\AgdaInductiveConstructor{ξξ-blame}\AgdaSpace{}%
\AgdaOperator{\AgdaInductiveConstructor{□⟨}}\AgdaSpace{}%
\AgdaBound{H}\AgdaSpace{}%
\AgdaOperator{\AgdaInductiveConstructor{?⟩}}\AgdaSpace{}%
\AgdaSymbol{())}\AgdaSpace{}%
\AgdaSymbol{(}\AgdaOperator{\AgdaInductiveConstructor{ƛ̬}}\AgdaSpace{}%
\AgdaBound{N}\AgdaSymbol{)}\<%
\\
\>[3]\AgdaFunction{nope}\AgdaSpace{}%
\AgdaSymbol{(}\AgdaInductiveConstructor{ξξ}\AgdaSpace{}%
\AgdaSymbol{(}\AgdaOperator{\AgdaInductiveConstructor{□·}}\AgdaSpace{}%
\AgdaBound{M}\AgdaSymbol{)}\AgdaSpace{}%
\AgdaSymbol{()}\AgdaSpace{}%
\AgdaBound{x₁}\AgdaSpace{}%
\AgdaBound{V→M}\AgdaSymbol{)}\AgdaSpace{}%
\AgdaSymbol{(}\AgdaInductiveConstructor{\$̬}\AgdaSpace{}%
\AgdaBound{c}\AgdaSymbol{)}\<%
\\
\>[3]\AgdaFunction{nope}\AgdaSpace{}%
\AgdaSymbol{(}\AgdaInductiveConstructor{ξξ}\AgdaSpace{}%
\AgdaSymbol{(}\AgdaBound{v}\AgdaSpace{}%
\AgdaOperator{\AgdaInductiveConstructor{·□}}\AgdaSymbol{)}\AgdaSpace{}%
\AgdaSymbol{()}\AgdaSpace{}%
\AgdaBound{x₁}\AgdaSpace{}%
\AgdaBound{V→M}\AgdaSymbol{)}\AgdaSpace{}%
\AgdaSymbol{(}\AgdaInductiveConstructor{\$̬}\AgdaSpace{}%
\AgdaBound{c}\AgdaSymbol{)}\<%
\\
\>[3]\AgdaFunction{nope}\AgdaSpace{}%
\AgdaSymbol{(}\AgdaInductiveConstructor{ξξ}\AgdaSpace{}%
\AgdaOperator{\AgdaInductiveConstructor{□⟨}}\AgdaSpace{}%
\AgdaBound{G}\AgdaSpace{}%
\AgdaOperator{\AgdaInductiveConstructor{!⟩}}\AgdaSpace{}%
\AgdaSymbol{()}\AgdaSpace{}%
\AgdaBound{x₁}\AgdaSpace{}%
\AgdaBound{V→M}\AgdaSymbol{)}\AgdaSpace{}%
\AgdaSymbol{(}\AgdaInductiveConstructor{\$̬}\AgdaSpace{}%
\AgdaBound{c}\AgdaSymbol{)}\<%
\\
\>[3]\AgdaFunction{nope}\AgdaSpace{}%
\AgdaSymbol{(}\AgdaInductiveConstructor{ξξ}\AgdaSpace{}%
\AgdaOperator{\AgdaInductiveConstructor{□⟨}}\AgdaSpace{}%
\AgdaBound{H}\AgdaSpace{}%
\AgdaOperator{\AgdaInductiveConstructor{?⟩}}\AgdaSpace{}%
\AgdaSymbol{()}\AgdaSpace{}%
\AgdaBound{x₁}\AgdaSpace{}%
\AgdaBound{V→M}\AgdaSymbol{)}\AgdaSpace{}%
\AgdaSymbol{(}\AgdaInductiveConstructor{\$̬}\AgdaSpace{}%
\AgdaBound{c}\AgdaSymbol{)}\<%
\\
\>[3]\AgdaFunction{nope}\AgdaSpace{}%
\AgdaSymbol{(}\AgdaInductiveConstructor{ξξ-blame}\AgdaSpace{}%
\AgdaSymbol{(}\AgdaOperator{\AgdaInductiveConstructor{□·}}\AgdaSpace{}%
\AgdaBound{M}\AgdaSymbol{)}\AgdaSpace{}%
\AgdaSymbol{())}\AgdaSpace{}%
\AgdaSymbol{(}\AgdaInductiveConstructor{\$̬}\AgdaSpace{}%
\AgdaBound{c}\AgdaSymbol{)}\<%
\\
\>[3]\AgdaFunction{nope}\AgdaSpace{}%
\AgdaSymbol{(}\AgdaInductiveConstructor{ξξ-blame}\AgdaSpace{}%
\AgdaSymbol{(}\AgdaBound{v}\AgdaSpace{}%
\AgdaOperator{\AgdaInductiveConstructor{·□}}\AgdaSymbol{)}\AgdaSpace{}%
\AgdaSymbol{())}\AgdaSpace{}%
\AgdaSymbol{(}\AgdaInductiveConstructor{\$̬}\AgdaSpace{}%
\AgdaBound{c}\AgdaSymbol{)}\<%
\\
\>[3]\AgdaFunction{nope}\AgdaSpace{}%
\AgdaSymbol{(}\AgdaInductiveConstructor{ξξ-blame}\AgdaSpace{}%
\AgdaOperator{\AgdaInductiveConstructor{□⟨}}\AgdaSpace{}%
\AgdaBound{G}\AgdaSpace{}%
\AgdaOperator{\AgdaInductiveConstructor{!⟩}}\AgdaSpace{}%
\AgdaSymbol{())}\AgdaSpace{}%
\AgdaSymbol{(}\AgdaInductiveConstructor{\$̬}\AgdaSpace{}%
\AgdaBound{c}\AgdaSymbol{)}\<%
\\
\>[3]\AgdaFunction{nope}\AgdaSpace{}%
\AgdaSymbol{(}\AgdaInductiveConstructor{ξξ-blame}\AgdaSpace{}%
\AgdaOperator{\AgdaInductiveConstructor{□⟨}}\AgdaSpace{}%
\AgdaBound{H}\AgdaSpace{}%
\AgdaOperator{\AgdaInductiveConstructor{?⟩}}\AgdaSpace{}%
\AgdaSymbol{())}\AgdaSpace{}%
\AgdaSymbol{(}\AgdaInductiveConstructor{\$̬}\AgdaSpace{}%
\AgdaBound{c}\AgdaSymbol{)}\<%
\\
\>[3]\AgdaFunction{nope}\AgdaSpace{}%
\AgdaSymbol{(}\AgdaInductiveConstructor{ξ}\AgdaSpace{}%
\AgdaOperator{\AgdaInductiveConstructor{□⟨}}\AgdaSpace{}%
\AgdaBound{G}\AgdaSpace{}%
\AgdaOperator{\AgdaInductiveConstructor{!⟩}}\AgdaSpace{}%
\AgdaBound{V→M}\AgdaSymbol{)}\AgdaSpace{}%
\AgdaSymbol{(}\AgdaBound{v}\AgdaSpace{}%
\AgdaOperator{\AgdaInductiveConstructor{〈}}\AgdaSpace{}%
\AgdaBound{g}\AgdaSpace{}%
\AgdaOperator{\AgdaInductiveConstructor{〉}}\AgdaSymbol{)}\AgdaSpace{}%
\AgdaSymbol{=}\AgdaSpace{}%
\AgdaFunction{nope}\AgdaSpace{}%
\AgdaBound{V→M}\AgdaSpace{}%
\AgdaBound{v}\<%
\\
\>[3]\AgdaFunction{nope}\AgdaSpace{}%
\AgdaSymbol{(}\AgdaInductiveConstructor{ξ-blame}\AgdaSpace{}%
\AgdaOperator{\AgdaInductiveConstructor{□⟨}}\AgdaSpace{}%
\AgdaSymbol{\AgdaUnderscore{}}\AgdaSpace{}%
\AgdaOperator{\AgdaInductiveConstructor{!⟩}}\AgdaSymbol{)}\AgdaSpace{}%
\AgdaSymbol{(()}\AgdaSpace{}%
\AgdaOperator{\AgdaInductiveConstructor{〈}}\AgdaSpace{}%
\AgdaBound{g}\AgdaSpace{}%
\AgdaOperator{\AgdaInductiveConstructor{〉}}\AgdaSymbol{)}\<%
\\
\\[\AgdaEmptyExtraSkip]%
\>[0]\AgdaFunction{value-irred}\AgdaSpace{}%
\AgdaSymbol{:}\AgdaSpace{}%
\AgdaSymbol{∀\{}\AgdaBound{V}\AgdaSpace{}%
\AgdaSymbol{:}\AgdaSpace{}%
\AgdaDatatype{Term}\AgdaSymbol{\}}\AgdaSpace{}%
\AgdaSymbol{→}\AgdaSpace{}%
\AgdaDatatype{Value}\AgdaSpace{}%
\AgdaBound{V}\AgdaSpace{}%
\AgdaSymbol{→}\AgdaSpace{}%
\AgdaFunction{irred}\AgdaSpace{}%
\AgdaBound{V}\<%
\\
\>[0]\AgdaFunction{value-irred}\AgdaSpace{}%
\AgdaSymbol{\{}\AgdaBound{V}\AgdaSymbol{\}}\AgdaSpace{}%
\AgdaBound{v}\AgdaSpace{}%
\AgdaSymbol{(}\AgdaBound{N}\AgdaSpace{}%
\AgdaOperator{\AgdaInductiveConstructor{,}}\AgdaSpace{}%
\AgdaBound{V→N}\AgdaSymbol{)}\AgdaSpace{}%
\AgdaSymbol{=}\AgdaSpace{}%
\AgdaFunction{value-irreducible}\AgdaSpace{}%
\AgdaBound{v}\AgdaSpace{}%
\AgdaBound{V→N}\<%
\\
\\[\AgdaEmptyExtraSkip]%
\>[0]\AgdaFunction{value↠}\AgdaSpace{}%
\AgdaSymbol{:}\AgdaSpace{}%
\AgdaSymbol{∀\{}\AgdaBound{V}\AgdaSpace{}%
\AgdaBound{N}\AgdaSpace{}%
\AgdaSymbol{:}\AgdaSpace{}%
\AgdaDatatype{Term}\AgdaSymbol{\}}\<%
\\
\>[0][@{}l@{\AgdaIndent{0}}]%
\>[4]\AgdaSymbol{→}\AgdaSpace{}%
\AgdaDatatype{Value}\AgdaSpace{}%
\AgdaBound{V}\<%
\\
\>[4]\AgdaSymbol{→}\AgdaSpace{}%
\AgdaBound{V}\AgdaSpace{}%
\AgdaOperator{\AgdaDatatype{↠}}\AgdaSpace{}%
\AgdaBound{N}\<%
\\
\>[4]\AgdaSymbol{→}\AgdaSpace{}%
\AgdaBound{N}\AgdaSpace{}%
\AgdaOperator{\AgdaDatatype{≡}}\AgdaSpace{}%
\AgdaBound{V}\<%
\\
\>[0]\AgdaFunction{value↠}\AgdaSpace{}%
\AgdaBound{v}\AgdaSpace{}%
\AgdaSymbol{(\AgdaUnderscore{}}\AgdaSpace{}%
\AgdaOperator{\AgdaInductiveConstructor{END}}\AgdaSymbol{)}\AgdaSpace{}%
\AgdaSymbol{=}\AgdaSpace{}%
\AgdaInductiveConstructor{refl}\<%
\\
\>[0]\AgdaFunction{value↠}\AgdaSpace{}%
\AgdaBound{v}\AgdaSpace{}%
\AgdaSymbol{(\AgdaUnderscore{}}\AgdaSpace{}%
\AgdaOperator{\AgdaInductiveConstructor{⟶⟨}}\AgdaSpace{}%
\AgdaBound{V⟶M}\AgdaSpace{}%
\AgdaOperator{\AgdaInductiveConstructor{⟩}}\AgdaSpace{}%
\AgdaBound{M↠N}\AgdaSymbol{)}\AgdaSpace{}%
\AgdaSymbol{=}\AgdaSpace{}%
\AgdaFunction{⊥-elim}\AgdaSpace{}%
\AgdaSymbol{(}\AgdaFunction{value-irreducible}\AgdaSpace{}%
\AgdaBound{v}\AgdaSpace{}%
\AgdaBound{V⟶M}\AgdaSymbol{)}\<%
\\
\\[\AgdaEmptyExtraSkip]%
\>[0]\AgdaFunction{blame↠}\AgdaSpace{}%
\AgdaSymbol{:}\AgdaSpace{}%
\AgdaSymbol{∀\{}\AgdaBound{N}\AgdaSymbol{\}}\<%
\\
\>[0][@{}l@{\AgdaIndent{0}}]%
\>[3]\AgdaSymbol{→}\AgdaSpace{}%
\AgdaInductiveConstructor{blame}\AgdaSpace{}%
\AgdaOperator{\AgdaDatatype{↠}}\AgdaSpace{}%
\AgdaBound{N}\<%
\\
\>[3]\AgdaSymbol{→}\AgdaSpace{}%
\AgdaBound{N}\AgdaSpace{}%
\AgdaOperator{\AgdaDatatype{≡}}\AgdaSpace{}%
\AgdaInductiveConstructor{blame}\<%
\\
\>[0]\AgdaFunction{blame↠}\AgdaSpace{}%
\AgdaSymbol{\{}\AgdaDottedPattern{\AgdaSymbol{.}}\AgdaDottedPattern{\AgdaInductiveConstructor{blame}}\AgdaSymbol{\}}\AgdaSpace{}%
\AgdaSymbol{(}\AgdaDottedPattern{\AgdaSymbol{.}}\AgdaDottedPattern{\AgdaInductiveConstructor{blame}}\AgdaSpace{}%
\AgdaOperator{\AgdaInductiveConstructor{END}}\AgdaSymbol{)}\AgdaSpace{}%
\AgdaSymbol{=}\AgdaSpace{}%
\AgdaInductiveConstructor{refl}\<%
\\
\>[0]\AgdaFunction{blame↠}\AgdaSpace{}%
\AgdaSymbol{\{}\AgdaBound{N}\AgdaSymbol{\}}\AgdaSpace{}%
\AgdaSymbol{(}\AgdaDottedPattern{\AgdaSymbol{.}}\AgdaDottedPattern{\AgdaInductiveConstructor{blame}}\AgdaSpace{}%
\AgdaOperator{\AgdaInductiveConstructor{⟶⟨}}\AgdaSpace{}%
\AgdaInductiveConstructor{ξξ}\AgdaSpace{}%
\AgdaSymbol{(}\AgdaOperator{\AgdaInductiveConstructor{□·}}\AgdaSpace{}%
\AgdaBound{M}\AgdaSymbol{)}\AgdaSpace{}%
\AgdaSymbol{()}\AgdaSpace{}%
\AgdaBound{x₁}\AgdaSpace{}%
\AgdaBound{x₂}\AgdaSpace{}%
\AgdaOperator{\AgdaInductiveConstructor{⟩}}\AgdaSpace{}%
\AgdaBound{red}\AgdaSymbol{)}\<%
\\
\>[0]\AgdaFunction{blame↠}\AgdaSpace{}%
\AgdaSymbol{\{}\AgdaBound{N}\AgdaSymbol{\}}\AgdaSpace{}%
\AgdaSymbol{(}\AgdaDottedPattern{\AgdaSymbol{.}}\AgdaDottedPattern{\AgdaInductiveConstructor{blame}}\AgdaSpace{}%
\AgdaOperator{\AgdaInductiveConstructor{⟶⟨}}\AgdaSpace{}%
\AgdaInductiveConstructor{ξξ}\AgdaSpace{}%
\AgdaSymbol{(}\AgdaBound{v}\AgdaSpace{}%
\AgdaOperator{\AgdaInductiveConstructor{·□}}\AgdaSymbol{)}\AgdaSpace{}%
\AgdaSymbol{()}\AgdaSpace{}%
\AgdaBound{x₁}\AgdaSpace{}%
\AgdaBound{x₂}\AgdaSpace{}%
\AgdaOperator{\AgdaInductiveConstructor{⟩}}\AgdaSpace{}%
\AgdaBound{red}\AgdaSymbol{)}\<%
\\
\>[0]\AgdaFunction{blame↠}\AgdaSpace{}%
\AgdaSymbol{\{}\AgdaBound{N}\AgdaSymbol{\}}\AgdaSpace{}%
\AgdaSymbol{(}\AgdaDottedPattern{\AgdaSymbol{.}}\AgdaDottedPattern{\AgdaInductiveConstructor{blame}}\AgdaSpace{}%
\AgdaOperator{\AgdaInductiveConstructor{⟶⟨}}\AgdaSpace{}%
\AgdaInductiveConstructor{ξξ}\AgdaSpace{}%
\AgdaOperator{\AgdaInductiveConstructor{□⟨}}\AgdaSpace{}%
\AgdaBound{G}\AgdaSpace{}%
\AgdaOperator{\AgdaInductiveConstructor{!⟩}}\AgdaSpace{}%
\AgdaSymbol{()}\AgdaSpace{}%
\AgdaBound{x₁}\AgdaSpace{}%
\AgdaBound{x₂}\AgdaSpace{}%
\AgdaOperator{\AgdaInductiveConstructor{⟩}}\AgdaSpace{}%
\AgdaBound{red}\AgdaSymbol{)}\<%
\\
\>[0]\AgdaFunction{blame↠}\AgdaSpace{}%
\AgdaSymbol{\{}\AgdaBound{N}\AgdaSymbol{\}}\AgdaSpace{}%
\AgdaSymbol{(}\AgdaDottedPattern{\AgdaSymbol{.}}\AgdaDottedPattern{\AgdaInductiveConstructor{blame}}\AgdaSpace{}%
\AgdaOperator{\AgdaInductiveConstructor{⟶⟨}}\AgdaSpace{}%
\AgdaInductiveConstructor{ξξ}\AgdaSpace{}%
\AgdaOperator{\AgdaInductiveConstructor{□⟨}}\AgdaSpace{}%
\AgdaBound{H}\AgdaSpace{}%
\AgdaOperator{\AgdaInductiveConstructor{?⟩}}\AgdaSpace{}%
\AgdaSymbol{()}\AgdaSpace{}%
\AgdaBound{x₁}\AgdaSpace{}%
\AgdaBound{x₂}\AgdaSpace{}%
\AgdaOperator{\AgdaInductiveConstructor{⟩}}\AgdaSpace{}%
\AgdaBound{red}\AgdaSymbol{)}\<%
\\
\>[0]\AgdaFunction{blame↠}\AgdaSpace{}%
\AgdaSymbol{\{}\AgdaBound{N}\AgdaSymbol{\}}\AgdaSpace{}%
\AgdaSymbol{(}\AgdaDottedPattern{\AgdaSymbol{.}}\AgdaDottedPattern{\AgdaInductiveConstructor{blame}}\AgdaSpace{}%
\AgdaOperator{\AgdaInductiveConstructor{⟶⟨}}\AgdaSpace{}%
\AgdaInductiveConstructor{ξξ-blame}\AgdaSpace{}%
\AgdaSymbol{(}\AgdaOperator{\AgdaInductiveConstructor{□·}}\AgdaSpace{}%
\AgdaBound{M}\AgdaSymbol{)}\AgdaSpace{}%
\AgdaSymbol{()}\AgdaSpace{}%
\AgdaOperator{\AgdaInductiveConstructor{⟩}}\AgdaSpace{}%
\AgdaBound{red}\AgdaSymbol{)}\<%
\\
\>[0]\AgdaFunction{blame↠}\AgdaSpace{}%
\AgdaSymbol{\{}\AgdaBound{N}\AgdaSymbol{\}}\AgdaSpace{}%
\AgdaSymbol{(}\AgdaDottedPattern{\AgdaSymbol{.}}\AgdaDottedPattern{\AgdaInductiveConstructor{blame}}\AgdaSpace{}%
\AgdaOperator{\AgdaInductiveConstructor{⟶⟨}}\AgdaSpace{}%
\AgdaInductiveConstructor{ξξ-blame}\AgdaSpace{}%
\AgdaSymbol{(}\AgdaBound{v}\AgdaSpace{}%
\AgdaOperator{\AgdaInductiveConstructor{·□}}\AgdaSymbol{)}\AgdaSpace{}%
\AgdaSymbol{()}\AgdaSpace{}%
\AgdaOperator{\AgdaInductiveConstructor{⟩}}\AgdaSpace{}%
\AgdaBound{red}\AgdaSymbol{)}\<%
\\
\>[0]\AgdaFunction{blame↠}\AgdaSpace{}%
\AgdaSymbol{\{}\AgdaBound{N}\AgdaSymbol{\}}\AgdaSpace{}%
\AgdaSymbol{(}\AgdaDottedPattern{\AgdaSymbol{.}}\AgdaDottedPattern{\AgdaInductiveConstructor{blame}}\AgdaSpace{}%
\AgdaOperator{\AgdaInductiveConstructor{⟶⟨}}\AgdaSpace{}%
\AgdaInductiveConstructor{ξξ-blame}\AgdaSpace{}%
\AgdaOperator{\AgdaInductiveConstructor{□⟨}}\AgdaSpace{}%
\AgdaBound{G}\AgdaSpace{}%
\AgdaOperator{\AgdaInductiveConstructor{!⟩}}\AgdaSpace{}%
\AgdaSymbol{()}\AgdaSpace{}%
\AgdaOperator{\AgdaInductiveConstructor{⟩}}\AgdaSpace{}%
\AgdaBound{red}\AgdaSymbol{)}\<%
\\
\>[0]\AgdaFunction{blame↠}\AgdaSpace{}%
\AgdaSymbol{\{}\AgdaBound{N}\AgdaSymbol{\}}\AgdaSpace{}%
\AgdaSymbol{(}\AgdaDottedPattern{\AgdaSymbol{.}}\AgdaDottedPattern{\AgdaInductiveConstructor{blame}}\AgdaSpace{}%
\AgdaOperator{\AgdaInductiveConstructor{⟶⟨}}\AgdaSpace{}%
\AgdaInductiveConstructor{ξξ-blame}\AgdaSpace{}%
\AgdaOperator{\AgdaInductiveConstructor{□⟨}}\AgdaSpace{}%
\AgdaBound{H}\AgdaSpace{}%
\AgdaOperator{\AgdaInductiveConstructor{?⟩}}\AgdaSpace{}%
\AgdaSymbol{()}\AgdaSpace{}%
\AgdaOperator{\AgdaInductiveConstructor{⟩}}\AgdaSpace{}%
\AgdaBound{red}\AgdaSymbol{)}\<%
\\
\\[\AgdaEmptyExtraSkip]%
\>[0]\AgdaFunction{blame-irreducible}\AgdaSpace{}%
\AgdaSymbol{:}\AgdaSpace{}%
\AgdaSymbol{∀\{}\AgdaBound{M}\AgdaSymbol{\}}\AgdaSpace{}%
\AgdaSymbol{→}\AgdaSpace{}%
\AgdaOperator{\AgdaFunction{¬}}\AgdaSpace{}%
\AgdaSymbol{(}\AgdaInductiveConstructor{blame}\AgdaSpace{}%
\AgdaOperator{\AgdaDatatype{⟶}}\AgdaSpace{}%
\AgdaBound{M}\AgdaSymbol{)}\<%
\\
\>[0]\AgdaFunction{blame-irreducible}\AgdaSpace{}%
\AgdaSymbol{\{}\AgdaBound{M}\AgdaSymbol{\}}\AgdaSpace{}%
\AgdaSymbol{(}\AgdaInductiveConstructor{ξξ}\AgdaSpace{}%
\AgdaSymbol{(}\AgdaOperator{\AgdaInductiveConstructor{□·}}\AgdaSpace{}%
\AgdaBound{M₁}\AgdaSymbol{)}\AgdaSpace{}%
\AgdaSymbol{()}\AgdaSpace{}%
\AgdaBound{x₁}\AgdaSpace{}%
\AgdaBound{blame→M}\AgdaSymbol{)}\<%
\\
\>[0]\AgdaFunction{blame-irreducible}\AgdaSpace{}%
\AgdaSymbol{\{}\AgdaBound{M}\AgdaSymbol{\}}\AgdaSpace{}%
\AgdaSymbol{(}\AgdaInductiveConstructor{ξξ}\AgdaSpace{}%
\AgdaSymbol{(}\AgdaBound{v}\AgdaSpace{}%
\AgdaOperator{\AgdaInductiveConstructor{·□}}\AgdaSymbol{)}\AgdaSpace{}%
\AgdaSymbol{()}\AgdaSpace{}%
\AgdaBound{x₁}\AgdaSpace{}%
\AgdaBound{blame→M}\AgdaSymbol{)}\<%
\\
\>[0]\AgdaFunction{blame-irreducible}\AgdaSpace{}%
\AgdaSymbol{\{}\AgdaBound{M}\AgdaSymbol{\}}\AgdaSpace{}%
\AgdaSymbol{(}\AgdaInductiveConstructor{ξξ}\AgdaSpace{}%
\AgdaOperator{\AgdaInductiveConstructor{□⟨}}\AgdaSpace{}%
\AgdaBound{G}\AgdaSpace{}%
\AgdaOperator{\AgdaInductiveConstructor{!⟩}}\AgdaSpace{}%
\AgdaSymbol{()}\AgdaSpace{}%
\AgdaBound{x₁}\AgdaSpace{}%
\AgdaBound{blame→M}\AgdaSymbol{)}\<%
\\
\>[0]\AgdaFunction{blame-irreducible}\AgdaSpace{}%
\AgdaSymbol{\{}\AgdaBound{M}\AgdaSymbol{\}}\AgdaSpace{}%
\AgdaSymbol{(}\AgdaInductiveConstructor{ξξ}\AgdaSpace{}%
\AgdaOperator{\AgdaInductiveConstructor{□⟨}}\AgdaSpace{}%
\AgdaBound{H}\AgdaSpace{}%
\AgdaOperator{\AgdaInductiveConstructor{?⟩}}\AgdaSpace{}%
\AgdaSymbol{()}\AgdaSpace{}%
\AgdaBound{x₁}\AgdaSpace{}%
\AgdaBound{blame→M}\AgdaSymbol{)}\<%
\\
\>[0]\AgdaFunction{blame-irreducible}\AgdaSpace{}%
\AgdaSymbol{\{}\AgdaDottedPattern{\AgdaSymbol{.}}\AgdaDottedPattern{\AgdaInductiveConstructor{blame}}\AgdaSymbol{\}}\AgdaSpace{}%
\AgdaSymbol{(}\AgdaInductiveConstructor{ξξ-blame}\AgdaSpace{}%
\AgdaSymbol{(}\AgdaOperator{\AgdaInductiveConstructor{□·}}\AgdaSpace{}%
\AgdaBound{M}\AgdaSymbol{)}\AgdaSpace{}%
\AgdaSymbol{())}\<%
\\
\>[0]\AgdaFunction{blame-irreducible}\AgdaSpace{}%
\AgdaSymbol{\{}\AgdaDottedPattern{\AgdaSymbol{.}}\AgdaDottedPattern{\AgdaInductiveConstructor{blame}}\AgdaSymbol{\}}\AgdaSpace{}%
\AgdaSymbol{(}\AgdaInductiveConstructor{ξξ-blame}\AgdaSpace{}%
\AgdaSymbol{(}\AgdaBound{v}\AgdaSpace{}%
\AgdaOperator{\AgdaInductiveConstructor{·□}}\AgdaSymbol{)}\AgdaSpace{}%
\AgdaSymbol{())}\<%
\\
\>[0]\AgdaFunction{blame-irreducible}\AgdaSpace{}%
\AgdaSymbol{\{}\AgdaDottedPattern{\AgdaSymbol{.}}\AgdaDottedPattern{\AgdaInductiveConstructor{blame}}\AgdaSymbol{\}}\AgdaSpace{}%
\AgdaSymbol{(}\AgdaInductiveConstructor{ξξ-blame}\AgdaSpace{}%
\AgdaOperator{\AgdaInductiveConstructor{□⟨}}\AgdaSpace{}%
\AgdaBound{G}\AgdaSpace{}%
\AgdaOperator{\AgdaInductiveConstructor{!⟩}}\AgdaSpace{}%
\AgdaSymbol{())}\<%
\\
\>[0]\AgdaFunction{blame-irreducible}\AgdaSpace{}%
\AgdaSymbol{\{}\AgdaDottedPattern{\AgdaSymbol{.}}\AgdaDottedPattern{\AgdaInductiveConstructor{blame}}\AgdaSymbol{\}}\AgdaSpace{}%
\AgdaSymbol{(}\AgdaInductiveConstructor{ξξ-blame}\AgdaSpace{}%
\AgdaOperator{\AgdaInductiveConstructor{□⟨}}\AgdaSpace{}%
\AgdaBound{H}\AgdaSpace{}%
\AgdaOperator{\AgdaInductiveConstructor{?⟩}}\AgdaSpace{}%
\AgdaSymbol{())}\<%
\\
\\[\AgdaEmptyExtraSkip]%
\>[0]\AgdaFunction{blame-irred}\AgdaSpace{}%
\AgdaSymbol{:}\AgdaSpace{}%
\AgdaSymbol{∀\{}\AgdaBound{M}\AgdaSymbol{\}\{}\AgdaBound{N}\AgdaSymbol{\}}\<%
\\
\>[0][@{}l@{\AgdaIndent{0}}]%
\>[3]\AgdaSymbol{→}\AgdaSpace{}%
\AgdaDatatype{Blame}\AgdaSpace{}%
\AgdaBound{M}\<%
\\
\>[3]\AgdaSymbol{→}\AgdaSpace{}%
\AgdaBound{M}\AgdaSpace{}%
\AgdaOperator{\AgdaDatatype{⟶}}\AgdaSpace{}%
\AgdaBound{N}\<%
\\
\>[3]\AgdaSymbol{→}\AgdaSpace{}%
\AgdaDatatype{⊥}\<%
\\
\>[0]\AgdaFunction{blame-irred}\AgdaSpace{}%
\AgdaInductiveConstructor{isBlame}\AgdaSpace{}%
\AgdaBound{red}\AgdaSpace{}%
\AgdaSymbol{=}\AgdaSpace{}%
\AgdaFunction{blame-irreducible}\AgdaSpace{}%
\AgdaBound{red}\<%
\\
\\[\AgdaEmptyExtraSkip]%
\>[0]\AgdaFunction{app-multi-inv}\AgdaSpace{}%
\AgdaSymbol{:}\AgdaSpace{}%
\AgdaSymbol{∀\{}\AgdaBound{L}\AgdaSpace{}%
\AgdaBound{M}\AgdaSpace{}%
\AgdaBound{N}\AgdaSymbol{\}}\<%
\\
\>[0][@{}l@{\AgdaIndent{0}}]%
\>[2]\AgdaSymbol{→}\AgdaSpace{}%
\AgdaSymbol{(}\AgdaBound{r1}\AgdaSpace{}%
\AgdaSymbol{:}\AgdaSpace{}%
\AgdaBound{L}\AgdaSpace{}%
\AgdaOperator{\AgdaInductiveConstructor{·}}\AgdaSpace{}%
\AgdaBound{M}\AgdaSpace{}%
\AgdaOperator{\AgdaDatatype{↠}}\AgdaSpace{}%
\AgdaBound{N}\AgdaSymbol{)}\<%
\\
\>[2]\AgdaSymbol{→}%
\>[2042I]\AgdaSymbol{(}\AgdaFunction{∃[}\AgdaSpace{}%
\AgdaBound{L′}\AgdaSpace{}%
\AgdaFunction{]}\AgdaSpace{}%
\AgdaSymbol{(}\AgdaFunction{Σ[}\AgdaSpace{}%
\AgdaBound{r2}\AgdaSpace{}%
\AgdaFunction{∈}\AgdaSpace{}%
\AgdaSymbol{(}\AgdaBound{L}\AgdaSpace{}%
\AgdaOperator{\AgdaDatatype{↠}}\AgdaSpace{}%
\AgdaBound{L′}\AgdaSymbol{)}\AgdaSpace{}%
\AgdaFunction{]}\AgdaSpace{}%
\AgdaSymbol{(}\AgdaBound{N}\AgdaSpace{}%
\AgdaOperator{\AgdaDatatype{≡}}\AgdaSpace{}%
\AgdaBound{L′}\AgdaSpace{}%
\AgdaOperator{\AgdaInductiveConstructor{·}}\AgdaSpace{}%
\AgdaBound{M}\AgdaSpace{}%
\AgdaOperator{\AgdaFunction{×}}\AgdaSpace{}%
\AgdaFunction{len}\AgdaSpace{}%
\AgdaBound{r1}\AgdaSpace{}%
\AgdaOperator{\AgdaDatatype{≡}}\AgdaSpace{}%
\AgdaFunction{len}\AgdaSpace{}%
\AgdaBound{r2}\AgdaSymbol{)))}\<%
\\
\>[.][@{}l@{}]\<[2042I]%
\>[4]\AgdaOperator{\AgdaDatatype{⊎}}%
\>[2063I]\AgdaSymbol{(}\AgdaFunction{∃[}\AgdaSpace{}%
\AgdaBound{V}\AgdaSpace{}%
\AgdaFunction{]}\AgdaSpace{}%
\AgdaFunction{∃[}\AgdaSpace{}%
\AgdaBound{M′}\AgdaSpace{}%
\AgdaFunction{]}\AgdaSpace{}%
\AgdaFunction{Σ[}\AgdaSpace{}%
\AgdaBound{r2}\AgdaSpace{}%
\AgdaFunction{∈}\AgdaSpace{}%
\AgdaSymbol{(}\AgdaBound{L}\AgdaSpace{}%
\AgdaOperator{\AgdaDatatype{↠}}\AgdaSpace{}%
\AgdaBound{V}\AgdaSymbol{)}\AgdaSpace{}%
\AgdaFunction{]}\AgdaSpace{}%
\AgdaSymbol{(}\AgdaDatatype{Value}\AgdaSpace{}%
\AgdaBound{V}\AgdaSpace{}%
\AgdaOperator{\AgdaFunction{×}}\AgdaSpace{}%
\AgdaFunction{Σ[}\AgdaSpace{}%
\AgdaBound{r3}\AgdaSpace{}%
\AgdaFunction{∈}\AgdaSpace{}%
\AgdaSymbol{(}\AgdaBound{M}\AgdaSpace{}%
\AgdaOperator{\AgdaDatatype{↠}}\AgdaSpace{}%
\AgdaBound{M′}\AgdaSymbol{)}\AgdaSpace{}%
\AgdaFunction{]}\<%
\\
\>[2063I][@{}l@{\AgdaIndent{0}}]%
\>[8]\AgdaSymbol{(}\AgdaBound{N}\AgdaSpace{}%
\AgdaOperator{\AgdaDatatype{≡}}\AgdaSpace{}%
\AgdaBound{V}\AgdaSpace{}%
\AgdaOperator{\AgdaInductiveConstructor{·}}\AgdaSpace{}%
\AgdaBound{M′}\AgdaSpace{}%
\AgdaOperator{\AgdaFunction{×}}\AgdaSpace{}%
\AgdaFunction{len}\AgdaSpace{}%
\AgdaBound{r1}\AgdaSpace{}%
\AgdaOperator{\AgdaDatatype{≡}}\AgdaSpace{}%
\AgdaFunction{len}\AgdaSpace{}%
\AgdaBound{r2}\AgdaSpace{}%
\AgdaOperator{\AgdaPrimitive{+}}\AgdaSpace{}%
\AgdaFunction{len}\AgdaSpace{}%
\AgdaBound{r3}\AgdaSymbol{)))}\<%
\\
\>[4]\AgdaOperator{\AgdaDatatype{⊎}}%
\>[2099I]\AgdaSymbol{(}\AgdaFunction{∃[}\AgdaSpace{}%
\AgdaBound{V}\AgdaSpace{}%
\AgdaFunction{]}\AgdaSpace{}%
\AgdaFunction{∃[}\AgdaSpace{}%
\AgdaBound{W}\AgdaSpace{}%
\AgdaFunction{]}\AgdaSpace{}%
\AgdaFunction{Σ[}\AgdaSpace{}%
\AgdaBound{r2}\AgdaSpace{}%
\AgdaFunction{∈}\AgdaSpace{}%
\AgdaSymbol{(}\AgdaBound{L}\AgdaSpace{}%
\AgdaOperator{\AgdaDatatype{↠}}\AgdaSpace{}%
\AgdaBound{V}\AgdaSymbol{)}\AgdaSpace{}%
\AgdaFunction{]}\AgdaSpace{}%
\AgdaSymbol{(}\AgdaDatatype{Value}\AgdaSpace{}%
\AgdaBound{V}\AgdaSpace{}%
\AgdaOperator{\AgdaFunction{×}}\AgdaSpace{}%
\AgdaFunction{Σ[}\AgdaSpace{}%
\AgdaBound{r3}\AgdaSpace{}%
\AgdaFunction{∈}\AgdaSpace{}%
\AgdaSymbol{(}\AgdaBound{M}\AgdaSpace{}%
\AgdaOperator{\AgdaDatatype{↠}}\AgdaSpace{}%
\AgdaBound{W}\AgdaSymbol{)}\AgdaSpace{}%
\AgdaFunction{]}\<%
\\
\>[2099I][@{}l@{\AgdaIndent{0}}]%
\>[8]\AgdaSymbol{(}\AgdaDatatype{Value}\AgdaSpace{}%
\AgdaBound{W}\AgdaSpace{}%
\AgdaOperator{\AgdaFunction{×}}\AgdaSpace{}%
\AgdaFunction{Σ[}\AgdaSpace{}%
\AgdaBound{r4}\AgdaSpace{}%
\AgdaFunction{∈}\AgdaSpace{}%
\AgdaSymbol{(}\AgdaBound{V}\AgdaSpace{}%
\AgdaOperator{\AgdaInductiveConstructor{·}}\AgdaSpace{}%
\AgdaBound{W}\AgdaSpace{}%
\AgdaOperator{\AgdaDatatype{↠}}\AgdaSpace{}%
\AgdaBound{N}\AgdaSymbol{)}\AgdaSpace{}%
\AgdaFunction{]}\AgdaSpace{}%
\AgdaFunction{len}\AgdaSpace{}%
\AgdaBound{r1}\AgdaSpace{}%
\AgdaOperator{\AgdaDatatype{≡}}\AgdaSpace{}%
\AgdaFunction{len}\AgdaSpace{}%
\AgdaBound{r2}\AgdaSpace{}%
\AgdaOperator{\AgdaPrimitive{+}}\AgdaSpace{}%
\AgdaFunction{len}\AgdaSpace{}%
\AgdaBound{r3}\AgdaSpace{}%
\AgdaOperator{\AgdaPrimitive{+}}\AgdaSpace{}%
\AgdaFunction{len}\AgdaSpace{}%
\AgdaBound{r4}\AgdaSymbol{)))}\<%
\\
\>[4]\AgdaOperator{\AgdaDatatype{⊎}}\AgdaSpace{}%
\AgdaBound{N}\AgdaSpace{}%
\AgdaOperator{\AgdaDatatype{≡}}\AgdaSpace{}%
\AgdaInductiveConstructor{blame}\<%
\\
\>[0]\AgdaFunction{app-multi-inv}\AgdaSpace{}%
\AgdaSymbol{\{}\AgdaBound{L}\AgdaSymbol{\}\{}\AgdaBound{M}\AgdaSymbol{\}\{}\AgdaBound{N}\AgdaSymbol{\}}\AgdaSpace{}%
\AgdaSymbol{((}\AgdaBound{L}\AgdaSpace{}%
\AgdaOperator{\AgdaInductiveConstructor{·}}\AgdaSpace{}%
\AgdaBound{M}\AgdaSymbol{)}\AgdaSpace{}%
\AgdaOperator{\AgdaInductiveConstructor{END}}\AgdaSymbol{)}\AgdaSpace{}%
\AgdaSymbol{=}\AgdaSpace{}%
\AgdaInductiveConstructor{inj₁}\AgdaSpace{}%
\AgdaSymbol{(}\AgdaBound{L}\AgdaSpace{}%
\AgdaOperator{\AgdaInductiveConstructor{,}}\AgdaSpace{}%
\AgdaSymbol{(\AgdaUnderscore{}}\AgdaSpace{}%
\AgdaOperator{\AgdaInductiveConstructor{END}}\AgdaSymbol{)}\AgdaSpace{}%
\AgdaOperator{\AgdaInductiveConstructor{,}}\AgdaSpace{}%
\AgdaInductiveConstructor{refl}\AgdaSpace{}%
\AgdaOperator{\AgdaInductiveConstructor{,}}\AgdaSpace{}%
\AgdaInductiveConstructor{refl}\AgdaSymbol{)}\<%
\\
\>[0]\AgdaFunction{app-multi-inv}\AgdaSpace{}%
\AgdaSymbol{\{}\AgdaBound{L}\AgdaSymbol{\}}\AgdaSpace{}%
\AgdaSymbol{\{}\AgdaBound{M}\AgdaSymbol{\}}\AgdaSpace{}%
\AgdaSymbol{\{}\AgdaBound{N}\AgdaSymbol{\}}\AgdaSpace{}%
\AgdaSymbol{(}\AgdaDottedPattern{\AgdaSymbol{.(}}\AgdaDottedPattern{\AgdaBound{L}}\AgdaSpace{}%
\AgdaDottedPattern{\AgdaOperator{\AgdaInductiveConstructor{·}}}\AgdaSpace{}%
\AgdaDottedPattern{\AgdaBound{M}}\AgdaDottedPattern{\AgdaSymbol{)}}\AgdaSpace{}%
\AgdaOperator{\AgdaInductiveConstructor{⟶⟨}}\AgdaSpace{}%
\AgdaInductiveConstructor{ξξ}\AgdaSpace{}%
\AgdaSymbol{\{}\AgdaBound{L}\AgdaSymbol{\}\{}\AgdaBound{L′}\AgdaSymbol{\}}\AgdaSpace{}%
\AgdaSymbol{(}\AgdaOperator{\AgdaInductiveConstructor{□·}}\AgdaSpace{}%
\AgdaSymbol{\AgdaUnderscore{})}\AgdaSpace{}%
\AgdaInductiveConstructor{refl}\AgdaSpace{}%
\AgdaInductiveConstructor{refl}\AgdaSpace{}%
\AgdaBound{L⟶L′}\AgdaSpace{}%
\AgdaOperator{\AgdaInductiveConstructor{⟩}}\AgdaSpace{}%
\AgdaBound{rs}\AgdaSymbol{)}\<%
\\
\>[0][@{}l@{\AgdaIndent{0}}]%
\>[4]\AgdaKeyword{with}\AgdaSpace{}%
\AgdaFunction{app-multi-inv}\AgdaSpace{}%
\AgdaBound{rs}\<%
\\
\>[0]\AgdaSymbol{...}\AgdaSpace{}%
\AgdaSymbol{|}\AgdaSpace{}%
\AgdaInductiveConstructor{inj₁}\AgdaSpace{}%
\AgdaSymbol{(}\AgdaBound{L″}\AgdaSpace{}%
\AgdaOperator{\AgdaInductiveConstructor{,}}\AgdaSpace{}%
\AgdaBound{L′→L″}\AgdaSpace{}%
\AgdaOperator{\AgdaInductiveConstructor{,}}\AgdaSpace{}%
\AgdaInductiveConstructor{refl}\AgdaSpace{}%
\AgdaOperator{\AgdaInductiveConstructor{,}}\AgdaSpace{}%
\AgdaBound{eq}\AgdaSymbol{)}\AgdaSpace{}%
\AgdaSymbol{=}\AgdaSpace{}%
\AgdaInductiveConstructor{inj₁}\AgdaSpace{}%
\AgdaSymbol{(}\AgdaBound{L″}\AgdaSpace{}%
\AgdaOperator{\AgdaInductiveConstructor{,}}\AgdaSpace{}%
\AgdaSymbol{(}\AgdaBound{L}\AgdaSpace{}%
\AgdaOperator{\AgdaInductiveConstructor{⟶⟨}}\AgdaSpace{}%
\AgdaBound{L⟶L′}\AgdaSpace{}%
\AgdaOperator{\AgdaInductiveConstructor{⟩}}\AgdaSpace{}%
\AgdaBound{L′→L″}\AgdaSymbol{)}\AgdaSpace{}%
\AgdaOperator{\AgdaInductiveConstructor{,}}\AgdaSpace{}%
\AgdaInductiveConstructor{refl}\AgdaSpace{}%
\AgdaOperator{\AgdaInductiveConstructor{,}}\AgdaSpace{}%
\AgdaFunction{cong}\AgdaSpace{}%
\AgdaInductiveConstructor{suc}\AgdaSpace{}%
\AgdaBound{eq}\AgdaSymbol{)}\<%
\\
\>[0]\AgdaSymbol{...}\AgdaSpace{}%
\AgdaSymbol{|}%
\>[2205I]\AgdaInductiveConstructor{inj₂}\AgdaSpace{}%
\AgdaSymbol{(}\AgdaInductiveConstructor{inj₁}\AgdaSpace{}%
\AgdaSymbol{(}\AgdaBound{V}\AgdaSpace{}%
\AgdaOperator{\AgdaInductiveConstructor{,}}\AgdaSpace{}%
\AgdaBound{M′}\AgdaSpace{}%
\AgdaOperator{\AgdaInductiveConstructor{,}}\AgdaSpace{}%
\AgdaBound{L′→V}\AgdaSpace{}%
\AgdaOperator{\AgdaInductiveConstructor{,}}\AgdaSpace{}%
\AgdaBound{v}\AgdaSpace{}%
\AgdaOperator{\AgdaInductiveConstructor{,}}\AgdaSpace{}%
\AgdaBound{M→M′}\AgdaSpace{}%
\AgdaOperator{\AgdaInductiveConstructor{,}}\AgdaSpace{}%
\AgdaInductiveConstructor{refl}\AgdaSpace{}%
\AgdaOperator{\AgdaInductiveConstructor{,}}\AgdaSpace{}%
\AgdaBound{eq}\AgdaSymbol{))}\AgdaSpace{}%
\AgdaSymbol{=}\<%
\\
\>[.][@{}l@{}]\<[2205I]%
\>[6]\AgdaInductiveConstructor{inj₂}\AgdaSpace{}%
\AgdaSymbol{(}\AgdaInductiveConstructor{inj₁}\AgdaSpace{}%
\AgdaSymbol{(}\AgdaBound{V}\AgdaSpace{}%
\AgdaOperator{\AgdaInductiveConstructor{,}}\AgdaSpace{}%
\AgdaBound{M′}\AgdaSpace{}%
\AgdaOperator{\AgdaInductiveConstructor{,}}\AgdaSpace{}%
\AgdaSymbol{(}\AgdaBound{L}\AgdaSpace{}%
\AgdaOperator{\AgdaInductiveConstructor{⟶⟨}}\AgdaSpace{}%
\AgdaBound{L⟶L′}\AgdaSpace{}%
\AgdaOperator{\AgdaInductiveConstructor{⟩}}\AgdaSpace{}%
\AgdaBound{L′→V}\AgdaSymbol{)}\AgdaSpace{}%
\AgdaOperator{\AgdaInductiveConstructor{,}}\AgdaSpace{}%
\AgdaBound{v}\AgdaSpace{}%
\AgdaOperator{\AgdaInductiveConstructor{,}}\AgdaSpace{}%
\AgdaBound{M→M′}\AgdaSpace{}%
\AgdaOperator{\AgdaInductiveConstructor{,}}\AgdaSpace{}%
\AgdaInductiveConstructor{refl}\AgdaSpace{}%
\AgdaOperator{\AgdaInductiveConstructor{,}}\AgdaSpace{}%
\AgdaFunction{cong}\AgdaSpace{}%
\AgdaInductiveConstructor{suc}\AgdaSpace{}%
\AgdaBound{eq}\AgdaSymbol{))}\<%
\\
\>[0]\AgdaSymbol{...}\AgdaSpace{}%
\AgdaSymbol{|}%
\>[2242I]\AgdaInductiveConstructor{inj₂}\AgdaSpace{}%
\AgdaSymbol{(}\AgdaInductiveConstructor{inj₂}\AgdaSpace{}%
\AgdaSymbol{(}\AgdaInductiveConstructor{inj₁}\AgdaSpace{}%
\AgdaSymbol{(}\AgdaBound{V}\AgdaSpace{}%
\AgdaOperator{\AgdaInductiveConstructor{,}}\AgdaSpace{}%
\AgdaBound{W}\AgdaSpace{}%
\AgdaOperator{\AgdaInductiveConstructor{,}}\AgdaSpace{}%
\AgdaBound{L′→V}\AgdaSpace{}%
\AgdaOperator{\AgdaInductiveConstructor{,}}\AgdaSpace{}%
\AgdaBound{v}\AgdaSpace{}%
\AgdaOperator{\AgdaInductiveConstructor{,}}\AgdaSpace{}%
\AgdaBound{M→W}\AgdaSpace{}%
\AgdaOperator{\AgdaInductiveConstructor{,}}\AgdaSpace{}%
\AgdaBound{w}\AgdaSpace{}%
\AgdaOperator{\AgdaInductiveConstructor{,}}\AgdaSpace{}%
\AgdaBound{→N}\AgdaSpace{}%
\AgdaOperator{\AgdaInductiveConstructor{,}}\AgdaSpace{}%
\AgdaBound{eq}\AgdaSymbol{)))}\AgdaSpace{}%
\AgdaSymbol{=}\<%
\\
\>[.][@{}l@{}]\<[2242I]%
\>[6]\AgdaInductiveConstructor{inj₂}\AgdaSpace{}%
\AgdaSymbol{(}\AgdaInductiveConstructor{inj₂}\AgdaSpace{}%
\AgdaSymbol{(}\AgdaInductiveConstructor{inj₁}\AgdaSpace{}%
\AgdaSymbol{(}\AgdaBound{V}\AgdaSpace{}%
\AgdaOperator{\AgdaInductiveConstructor{,}}\AgdaSpace{}%
\AgdaBound{W}\AgdaSpace{}%
\AgdaOperator{\AgdaInductiveConstructor{,}}\AgdaSpace{}%
\AgdaSymbol{(}\AgdaBound{L}\AgdaSpace{}%
\AgdaOperator{\AgdaInductiveConstructor{⟶⟨}}\AgdaSpace{}%
\AgdaBound{L⟶L′}\AgdaSpace{}%
\AgdaOperator{\AgdaInductiveConstructor{⟩}}\AgdaSpace{}%
\AgdaBound{L′→V}\AgdaSymbol{)}\AgdaSpace{}%
\AgdaOperator{\AgdaInductiveConstructor{,}}\AgdaSpace{}%
\AgdaBound{v}\AgdaSpace{}%
\AgdaOperator{\AgdaInductiveConstructor{,}}\AgdaSpace{}%
\AgdaBound{M→W}\AgdaSpace{}%
\AgdaOperator{\AgdaInductiveConstructor{,}}\AgdaSpace{}%
\AgdaBound{w}\AgdaSpace{}%
\AgdaOperator{\AgdaInductiveConstructor{,}}\AgdaSpace{}%
\AgdaBound{→N}\AgdaSpace{}%
\AgdaOperator{\AgdaInductiveConstructor{,}}\AgdaSpace{}%
\AgdaFunction{cong}\AgdaSpace{}%
\AgdaInductiveConstructor{suc}\AgdaSpace{}%
\AgdaBound{eq}\AgdaSymbol{)))}\<%
\\
\>[0]\AgdaSymbol{...}\AgdaSpace{}%
\AgdaSymbol{|}\AgdaSpace{}%
\AgdaInductiveConstructor{inj₂}\AgdaSpace{}%
\AgdaSymbol{(}\AgdaInductiveConstructor{inj₂}\AgdaSpace{}%
\AgdaSymbol{(}\AgdaInductiveConstructor{inj₂}\AgdaSpace{}%
\AgdaInductiveConstructor{refl}\AgdaSymbol{))}\AgdaSpace{}%
\AgdaSymbol{=}\AgdaSpace{}%
\AgdaInductiveConstructor{inj₂}\AgdaSpace{}%
\AgdaSymbol{(}\AgdaInductiveConstructor{inj₂}\AgdaSpace{}%
\AgdaSymbol{(}\AgdaInductiveConstructor{inj₂}\AgdaSpace{}%
\AgdaInductiveConstructor{refl}\AgdaSymbol{))}\<%
\\
\>[0]\AgdaFunction{app-multi-inv}\AgdaSpace{}%
\AgdaSymbol{\{}\AgdaBound{V}\AgdaSymbol{\}}\AgdaSpace{}%
\AgdaSymbol{\{}\AgdaBound{M}\AgdaSymbol{\}}\AgdaSpace{}%
\AgdaSymbol{\{}\AgdaBound{N}\AgdaSymbol{\}}\AgdaSpace{}%
\AgdaSymbol{(}\AgdaDottedPattern{\AgdaSymbol{.(}}\AgdaDottedPattern{\AgdaBound{V}}\AgdaSpace{}%
\AgdaDottedPattern{\AgdaOperator{\AgdaInductiveConstructor{·}}}\AgdaSpace{}%
\AgdaDottedPattern{\AgdaBound{M}}\AgdaDottedPattern{\AgdaSymbol{)}}\AgdaSpace{}%
\AgdaOperator{\AgdaInductiveConstructor{⟶⟨}}\AgdaSpace{}%
\AgdaInductiveConstructor{ξξ}\AgdaSpace{}%
\AgdaSymbol{\{}\AgdaArgument{N}\AgdaSpace{}%
\AgdaSymbol{=}\AgdaSpace{}%
\AgdaBound{M′}\AgdaSymbol{\}}\AgdaSpace{}%
\AgdaSymbol{(}\AgdaBound{v}\AgdaSpace{}%
\AgdaOperator{\AgdaInductiveConstructor{·□}}\AgdaSymbol{)}\AgdaSpace{}%
\AgdaInductiveConstructor{refl}\AgdaSpace{}%
\AgdaInductiveConstructor{refl}\AgdaSpace{}%
\AgdaBound{M⟶M′}\AgdaSpace{}%
\AgdaOperator{\AgdaInductiveConstructor{⟩}}\AgdaSpace{}%
\AgdaBound{V·M′↠N}\AgdaSymbol{)}\<%
\\
\>[0][@{}l@{\AgdaIndent{0}}]%
\>[4]\AgdaKeyword{with}\AgdaSpace{}%
\AgdaFunction{app-multi-inv}\AgdaSpace{}%
\AgdaBound{V·M′↠N}\<%
\\
\>[0]\AgdaSymbol{...}%
\>[2314I]\AgdaSymbol{|}\AgdaSpace{}%
\AgdaInductiveConstructor{inj₁}\AgdaSpace{}%
\AgdaSymbol{(}\AgdaBound{L′}\AgdaSpace{}%
\AgdaOperator{\AgdaInductiveConstructor{,}}\AgdaSpace{}%
\AgdaBound{V→L′}\AgdaSpace{}%
\AgdaOperator{\AgdaInductiveConstructor{,}}\AgdaSpace{}%
\AgdaInductiveConstructor{refl}\AgdaSpace{}%
\AgdaOperator{\AgdaInductiveConstructor{,}}\AgdaSpace{}%
\AgdaBound{eq}\AgdaSymbol{)}\<%
\\
\>[.][@{}l@{}]\<[2314I]%
\>[4]\AgdaKeyword{with}\AgdaSpace{}%
\AgdaFunction{value↠}\AgdaSpace{}%
\AgdaBound{v}\AgdaSpace{}%
\AgdaBound{V→L′}\<%
\\
\>[0]\AgdaSymbol{...}%
\>[2326I]\AgdaSymbol{|}\AgdaSpace{}%
\AgdaInductiveConstructor{refl}\AgdaSpace{}%
\AgdaSymbol{=}\<%
\\
\>[.][@{}l@{}]\<[2326I]%
\>[4]\AgdaInductiveConstructor{inj₂}\AgdaSpace{}%
\AgdaSymbol{(}\AgdaInductiveConstructor{inj₁}\AgdaSpace{}%
\AgdaSymbol{(}\AgdaBound{V}\AgdaSpace{}%
\AgdaOperator{\AgdaInductiveConstructor{,}}\AgdaSpace{}%
\AgdaBound{M′}\AgdaSpace{}%
\AgdaOperator{\AgdaInductiveConstructor{,}}\AgdaSpace{}%
\AgdaBound{V→L′}\AgdaSpace{}%
\AgdaOperator{\AgdaInductiveConstructor{,}}\AgdaSpace{}%
\AgdaBound{v}\AgdaSpace{}%
\AgdaOperator{\AgdaInductiveConstructor{,}}\AgdaSpace{}%
\AgdaSymbol{(}\AgdaBound{M}\AgdaSpace{}%
\AgdaOperator{\AgdaInductiveConstructor{⟶⟨}}\AgdaSpace{}%
\AgdaBound{M⟶M′}\AgdaSpace{}%
\AgdaOperator{\AgdaInductiveConstructor{⟩}}\AgdaSpace{}%
\AgdaBound{M′}\AgdaSpace{}%
\AgdaOperator{\AgdaInductiveConstructor{END}}\AgdaSymbol{)}\AgdaSpace{}%
\AgdaOperator{\AgdaInductiveConstructor{,}}\AgdaSpace{}%
\AgdaInductiveConstructor{refl}\AgdaSpace{}%
\AgdaOperator{\AgdaInductiveConstructor{,}}\AgdaSpace{}%
\AgdaFunction{EQ}\AgdaSymbol{))}\<%
\\
\>[4]\AgdaKeyword{where}%
\>[2348I]\AgdaFunction{EQ}\AgdaSpace{}%
\AgdaSymbol{:}\AgdaSpace{}%
\AgdaInductiveConstructor{suc}\AgdaSpace{}%
\AgdaSymbol{(}\AgdaFunction{len}\AgdaSpace{}%
\AgdaBound{V·M′↠N}\AgdaSymbol{)}\AgdaSpace{}%
\AgdaOperator{\AgdaDatatype{≡}}\AgdaSpace{}%
\AgdaFunction{len}\AgdaSpace{}%
\AgdaBound{V→L′}\AgdaSpace{}%
\AgdaOperator{\AgdaPrimitive{+}}\AgdaSpace{}%
\AgdaNumber{1}\<%
\\
\>[.][@{}l@{}]\<[2348I]%
\>[10]\AgdaFunction{EQ}\AgdaSpace{}%
\AgdaSymbol{=}\AgdaSpace{}%
\AgdaFunction{trans}\AgdaSpace{}%
\AgdaSymbol{(}\AgdaFunction{cong}\AgdaSpace{}%
\AgdaInductiveConstructor{suc}\AgdaSpace{}%
\AgdaBound{eq}\AgdaSymbol{)}\AgdaSpace{}%
\AgdaSymbol{(}\AgdaFunction{sym}\AgdaSpace{}%
\AgdaSymbol{(}\AgdaFunction{trans}\AgdaSpace{}%
\AgdaSymbol{(}\AgdaFunction{+-suc}\AgdaSpace{}%
\AgdaSymbol{\AgdaUnderscore{}}\AgdaSpace{}%
\AgdaSymbol{\AgdaUnderscore{})}\AgdaSpace{}%
\AgdaSymbol{(}\AgdaFunction{+-identityʳ}\AgdaSpace{}%
\AgdaSymbol{\AgdaUnderscore{})))}\<%
\\
\>[0]\AgdaFunction{app-multi-inv}\AgdaSpace{}%
\AgdaSymbol{\{}\AgdaBound{V}\AgdaSymbol{\}}\AgdaSpace{}%
\AgdaSymbol{\{}\AgdaBound{M}\AgdaSymbol{\}}\AgdaSpace{}%
\AgdaSymbol{\{}\AgdaBound{N}\AgdaSymbol{\}}\AgdaSpace{}%
\AgdaSymbol{(}\AgdaDottedPattern{\AgdaSymbol{.(}}\AgdaDottedPattern{\AgdaBound{V}}\AgdaSpace{}%
\AgdaDottedPattern{\AgdaOperator{\AgdaInductiveConstructor{·}}}\AgdaSpace{}%
\AgdaDottedPattern{\AgdaBound{M}}\AgdaDottedPattern{\AgdaSymbol{)}}\AgdaSpace{}%
\AgdaOperator{\AgdaInductiveConstructor{⟶⟨}}\AgdaSpace{}%
\AgdaInductiveConstructor{ξξ}\AgdaSpace{}%
\AgdaSymbol{(}\AgdaBound{v}\AgdaSpace{}%
\AgdaOperator{\AgdaInductiveConstructor{·□}}\AgdaSymbol{)}\AgdaSpace{}%
\AgdaInductiveConstructor{refl}\AgdaSpace{}%
\AgdaInductiveConstructor{refl}\AgdaSpace{}%
\AgdaBound{M⟶M′}\AgdaSpace{}%
\AgdaOperator{\AgdaInductiveConstructor{⟩}}\AgdaSpace{}%
\AgdaBound{V·M′↠N}\AgdaSymbol{)}\<%
\\
\>[0][@{}l@{\AgdaIndent{0}}]%
\>[4]\AgdaSymbol{|}%
\>[2385I]\AgdaInductiveConstructor{inj₂}\AgdaSpace{}%
\AgdaSymbol{(}\AgdaInductiveConstructor{inj₁}\AgdaSpace{}%
\AgdaSymbol{(}\AgdaBound{V′}\AgdaSpace{}%
\AgdaOperator{\AgdaInductiveConstructor{,}}\AgdaSpace{}%
\AgdaBound{M″}\AgdaSpace{}%
\AgdaOperator{\AgdaInductiveConstructor{,}}\AgdaSpace{}%
\AgdaBound{V→V′}\AgdaSpace{}%
\AgdaOperator{\AgdaInductiveConstructor{,}}\AgdaSpace{}%
\AgdaBound{v′}\AgdaSpace{}%
\AgdaOperator{\AgdaInductiveConstructor{,}}\AgdaSpace{}%
\AgdaBound{M′→M″}\AgdaSpace{}%
\AgdaOperator{\AgdaInductiveConstructor{,}}\AgdaSpace{}%
\AgdaInductiveConstructor{refl}\AgdaSpace{}%
\AgdaOperator{\AgdaInductiveConstructor{,}}\AgdaSpace{}%
\AgdaBound{eq}\AgdaSymbol{))}\AgdaSpace{}%
\AgdaSymbol{=}\<%
\\
\>[.][@{}l@{}]\<[2385I]%
\>[6]\AgdaInductiveConstructor{inj₂}\AgdaSpace{}%
\AgdaSymbol{(}\AgdaInductiveConstructor{inj₁}\AgdaSpace{}%
\AgdaSymbol{(}\AgdaBound{V′}\AgdaSpace{}%
\AgdaOperator{\AgdaInductiveConstructor{,}}\AgdaSpace{}%
\AgdaBound{M″}\AgdaSpace{}%
\AgdaOperator{\AgdaInductiveConstructor{,}}\AgdaSpace{}%
\AgdaBound{V→V′}\AgdaSpace{}%
\AgdaOperator{\AgdaInductiveConstructor{,}}\AgdaSpace{}%
\AgdaBound{v′}\AgdaSpace{}%
\AgdaOperator{\AgdaInductiveConstructor{,}}\AgdaSpace{}%
\AgdaSymbol{(}\AgdaBound{M}\AgdaSpace{}%
\AgdaOperator{\AgdaInductiveConstructor{⟶⟨}}\AgdaSpace{}%
\AgdaBound{M⟶M′}\AgdaSpace{}%
\AgdaOperator{\AgdaInductiveConstructor{⟩}}\AgdaSpace{}%
\AgdaBound{M′→M″}\AgdaSymbol{)}\AgdaSpace{}%
\AgdaOperator{\AgdaInductiveConstructor{,}}\AgdaSpace{}%
\AgdaInductiveConstructor{refl}\AgdaSpace{}%
\AgdaOperator{\AgdaInductiveConstructor{,}}\AgdaSpace{}%
\AgdaFunction{EQ}\AgdaSymbol{))}\<%
\\
\>[4]\AgdaKeyword{where}%
\>[2419I]\AgdaFunction{EQ}\AgdaSpace{}%
\AgdaSymbol{:}\AgdaSpace{}%
\AgdaInductiveConstructor{suc}\AgdaSpace{}%
\AgdaSymbol{(}\AgdaFunction{len}\AgdaSpace{}%
\AgdaBound{V·M′↠N}\AgdaSymbol{)}\AgdaSpace{}%
\AgdaOperator{\AgdaDatatype{≡}}\AgdaSpace{}%
\AgdaFunction{len}\AgdaSpace{}%
\AgdaBound{V→V′}\AgdaSpace{}%
\AgdaOperator{\AgdaPrimitive{+}}\AgdaSpace{}%
\AgdaInductiveConstructor{suc}\AgdaSpace{}%
\AgdaSymbol{(}\AgdaFunction{len}\AgdaSpace{}%
\AgdaBound{M′→M″}\AgdaSymbol{)}\<%
\\
\>[.][@{}l@{}]\<[2419I]%
\>[10]\AgdaFunction{EQ}\AgdaSpace{}%
\AgdaKeyword{rewrite}\AgdaSpace{}%
\AgdaBound{eq}\AgdaSpace{}%
\AgdaSymbol{=}\AgdaSpace{}%
\AgdaFunction{sym}\AgdaSpace{}%
\AgdaSymbol{(}\AgdaFunction{+-suc}\AgdaSpace{}%
\AgdaSymbol{\AgdaUnderscore{}}\AgdaSpace{}%
\AgdaSymbol{\AgdaUnderscore{})}\<%
\\
\>[0]\AgdaFunction{app-multi-inv}\AgdaSpace{}%
\AgdaSymbol{\{}\AgdaBound{V}\AgdaSymbol{\}}\AgdaSpace{}%
\AgdaSymbol{\{}\AgdaBound{M}\AgdaSymbol{\}}\AgdaSpace{}%
\AgdaSymbol{\{}\AgdaBound{N}\AgdaSymbol{\}}\AgdaSpace{}%
\AgdaSymbol{(}\AgdaDottedPattern{\AgdaSymbol{.(}}\AgdaDottedPattern{\AgdaBound{V}}\AgdaSpace{}%
\AgdaDottedPattern{\AgdaOperator{\AgdaInductiveConstructor{·}}}\AgdaSpace{}%
\AgdaDottedPattern{\AgdaBound{M}}\AgdaDottedPattern{\AgdaSymbol{)}}\AgdaSpace{}%
\AgdaOperator{\AgdaInductiveConstructor{⟶⟨}}\AgdaSpace{}%
\AgdaInductiveConstructor{ξξ}\AgdaSpace{}%
\AgdaSymbol{(}\AgdaBound{v}\AgdaSpace{}%
\AgdaOperator{\AgdaInductiveConstructor{·□}}\AgdaSymbol{)}\AgdaSpace{}%
\AgdaInductiveConstructor{refl}\AgdaSpace{}%
\AgdaInductiveConstructor{refl}\AgdaSpace{}%
\AgdaBound{M⟶M′}\AgdaSpace{}%
\AgdaOperator{\AgdaInductiveConstructor{⟩}}\AgdaSpace{}%
\AgdaBound{V·M′↠N}\AgdaSymbol{)}\<%
\\
\>[0][@{}l@{\AgdaIndent{0}}]%
\>[4]\AgdaSymbol{|}%
\>[2453I]\AgdaInductiveConstructor{inj₂}\AgdaSpace{}%
\AgdaSymbol{(}\AgdaInductiveConstructor{inj₂}\AgdaSpace{}%
\AgdaSymbol{(}\AgdaInductiveConstructor{inj₁}\AgdaSpace{}%
\AgdaSymbol{(}\AgdaBound{V′}\AgdaSpace{}%
\AgdaOperator{\AgdaInductiveConstructor{,}}\AgdaSpace{}%
\AgdaBound{W}\AgdaSpace{}%
\AgdaOperator{\AgdaInductiveConstructor{,}}\AgdaSpace{}%
\AgdaBound{V→V′}\AgdaSpace{}%
\AgdaOperator{\AgdaInductiveConstructor{,}}\AgdaSpace{}%
\AgdaBound{v′}\AgdaSpace{}%
\AgdaOperator{\AgdaInductiveConstructor{,}}\AgdaSpace{}%
\AgdaBound{M′→W}\AgdaSpace{}%
\AgdaOperator{\AgdaInductiveConstructor{,}}\AgdaSpace{}%
\AgdaBound{w}\AgdaSpace{}%
\AgdaOperator{\AgdaInductiveConstructor{,}}\AgdaSpace{}%
\AgdaBound{V′·W→N}\AgdaSpace{}%
\AgdaOperator{\AgdaInductiveConstructor{,}}\AgdaSpace{}%
\AgdaBound{eq}\AgdaSpace{}%
\AgdaSymbol{)))}\AgdaSpace{}%
\AgdaSymbol{=}\<%
\\
\>[.][@{}l@{}]\<[2453I]%
\>[6]\AgdaInductiveConstructor{inj₂}\AgdaSpace{}%
\AgdaSymbol{(}\AgdaInductiveConstructor{inj₂}\AgdaSpace{}%
\AgdaSymbol{(}\AgdaInductiveConstructor{inj₁}\AgdaSpace{}%
\AgdaSymbol{(}\AgdaBound{V′}\AgdaSpace{}%
\AgdaOperator{\AgdaInductiveConstructor{,}}\AgdaSpace{}%
\AgdaBound{W}\AgdaSpace{}%
\AgdaOperator{\AgdaInductiveConstructor{,}}\AgdaSpace{}%
\AgdaBound{V→V′}\AgdaSpace{}%
\AgdaOperator{\AgdaInductiveConstructor{,}}\AgdaSpace{}%
\AgdaBound{v′}\AgdaSpace{}%
\AgdaOperator{\AgdaInductiveConstructor{,}}\AgdaSpace{}%
\AgdaSymbol{(}\AgdaBound{M}\AgdaSpace{}%
\AgdaOperator{\AgdaInductiveConstructor{⟶⟨}}\AgdaSpace{}%
\AgdaBound{M⟶M′}\AgdaSpace{}%
\AgdaOperator{\AgdaInductiveConstructor{⟩}}\AgdaSpace{}%
\AgdaBound{M′→W}\AgdaSymbol{)}\AgdaSpace{}%
\AgdaOperator{\AgdaInductiveConstructor{,}}\AgdaSpace{}%
\AgdaBound{w}\AgdaSpace{}%
\AgdaOperator{\AgdaInductiveConstructor{,}}\AgdaSpace{}%
\AgdaBound{V′·W→N}\AgdaSpace{}%
\AgdaOperator{\AgdaInductiveConstructor{,}}\AgdaSpace{}%
\AgdaFunction{EQ}\AgdaSymbol{)))}\<%
\\
\>[4]\AgdaKeyword{where}%
\>[2494I]\AgdaFunction{EQ}\AgdaSpace{}%
\AgdaSymbol{:}\AgdaSpace{}%
\AgdaInductiveConstructor{suc}\AgdaSpace{}%
\AgdaSymbol{(}\AgdaFunction{len}\AgdaSpace{}%
\AgdaBound{V·M′↠N}\AgdaSymbol{)}\AgdaSpace{}%
\AgdaOperator{\AgdaDatatype{≡}}\AgdaSpace{}%
\AgdaFunction{len}\AgdaSpace{}%
\AgdaBound{V→V′}\AgdaSpace{}%
\AgdaOperator{\AgdaPrimitive{+}}\AgdaSpace{}%
\AgdaInductiveConstructor{suc}\AgdaSpace{}%
\AgdaSymbol{(}\AgdaFunction{len}\AgdaSpace{}%
\AgdaBound{M′→W}\AgdaSymbol{)}\AgdaSpace{}%
\AgdaOperator{\AgdaPrimitive{+}}\AgdaSpace{}%
\AgdaFunction{len}\AgdaSpace{}%
\AgdaBound{V′·W→N}\<%
\\
\>[.][@{}l@{}]\<[2494I]%
\>[10]\AgdaFunction{EQ}\AgdaSpace{}%
\AgdaSymbol{=}\AgdaSpace{}%
\AgdaFunction{trans}\AgdaSpace{}%
\AgdaSymbol{(}\AgdaFunction{cong}\AgdaSpace{}%
\AgdaInductiveConstructor{suc}\AgdaSpace{}%
\AgdaBound{eq}\AgdaSymbol{)}%
\>[2514I]\AgdaSymbol{(}\AgdaFunction{sym}\AgdaSpace{}%
\AgdaSymbol{(}\AgdaFunction{cong}\AgdaSpace{}%
\AgdaSymbol{(λ}\AgdaSpace{}%
\AgdaBound{X}\AgdaSpace{}%
\AgdaSymbol{→}\AgdaSpace{}%
\AgdaBound{X}\AgdaSpace{}%
\AgdaOperator{\AgdaPrimitive{+}}\AgdaSpace{}%
\AgdaFunction{len}\AgdaSpace{}%
\AgdaBound{V′·W→N}\AgdaSymbol{)}\<%
\\
\>[2514I][@{}l@{\AgdaIndent{0}}]%
\>[39]\AgdaSymbol{(}\AgdaFunction{+-suc}\AgdaSpace{}%
\AgdaSymbol{(}\AgdaFunction{len}\AgdaSpace{}%
\AgdaBound{V→V′}\AgdaSymbol{)}\AgdaSpace{}%
\AgdaSymbol{(}\AgdaFunction{len}\AgdaSpace{}%
\AgdaBound{M′→W}\AgdaSymbol{))))}\<%
\\
\>[0]\AgdaFunction{app-multi-inv}\AgdaSpace{}%
\AgdaSymbol{\{}\AgdaBound{V}\AgdaSymbol{\}}\AgdaSpace{}%
\AgdaSymbol{\{}\AgdaBound{M}\AgdaSymbol{\}}\AgdaSpace{}%
\AgdaSymbol{\{}\AgdaBound{N}\AgdaSymbol{\}}\AgdaSpace{}%
\AgdaSymbol{(}\AgdaDottedPattern{\AgdaSymbol{.(}}\AgdaDottedPattern{\AgdaBound{V}}\AgdaSpace{}%
\AgdaDottedPattern{\AgdaOperator{\AgdaInductiveConstructor{·}}}\AgdaSpace{}%
\AgdaDottedPattern{\AgdaBound{M}}\AgdaDottedPattern{\AgdaSymbol{)}}\AgdaSpace{}%
\AgdaOperator{\AgdaInductiveConstructor{⟶⟨}}\AgdaSpace{}%
\AgdaInductiveConstructor{ξξ}\AgdaSpace{}%
\AgdaSymbol{(}\AgdaBound{v}\AgdaSpace{}%
\AgdaOperator{\AgdaInductiveConstructor{·□}}\AgdaSymbol{)}\AgdaSpace{}%
\AgdaInductiveConstructor{refl}\AgdaSpace{}%
\AgdaInductiveConstructor{refl}\AgdaSpace{}%
\AgdaBound{M⟶M′}\AgdaSpace{}%
\AgdaOperator{\AgdaInductiveConstructor{⟩}}\AgdaSpace{}%
\AgdaBound{V·M′↠N}\AgdaSymbol{)}\<%
\\
\>[0][@{}l@{\AgdaIndent{0}}]%
\>[4]\AgdaSymbol{|}\AgdaSpace{}%
\AgdaInductiveConstructor{inj₂}\AgdaSpace{}%
\AgdaSymbol{(}\AgdaInductiveConstructor{inj₂}\AgdaSpace{}%
\AgdaSymbol{(}\AgdaInductiveConstructor{inj₂}\AgdaSpace{}%
\AgdaInductiveConstructor{refl}\AgdaSymbol{))}\AgdaSpace{}%
\AgdaSymbol{=}\AgdaSpace{}%
\AgdaInductiveConstructor{inj₂}\AgdaSpace{}%
\AgdaSymbol{(}\AgdaInductiveConstructor{inj₂}\AgdaSpace{}%
\AgdaSymbol{(}\AgdaInductiveConstructor{inj₂}\AgdaSpace{}%
\AgdaInductiveConstructor{refl}\AgdaSymbol{))}\<%
\\
\>[0]\AgdaFunction{app-multi-inv}\AgdaSpace{}%
\AgdaSymbol{\{}\AgdaBound{L}\AgdaSymbol{\}}\AgdaSpace{}%
\AgdaSymbol{\{}\AgdaBound{M}\AgdaSymbol{\}}\AgdaSpace{}%
\AgdaSymbol{\{}\AgdaBound{N}\AgdaSymbol{\}}\AgdaSpace{}%
\AgdaSymbol{(}\AgdaDottedPattern{\AgdaSymbol{.(}}\AgdaDottedPattern{\AgdaBound{L}}\AgdaSpace{}%
\AgdaDottedPattern{\AgdaOperator{\AgdaInductiveConstructor{·}}}\AgdaSpace{}%
\AgdaDottedPattern{\AgdaBound{M}}\AgdaDottedPattern{\AgdaSymbol{)}}\AgdaSpace{}%
\AgdaOperator{\AgdaInductiveConstructor{⟶⟨}}\AgdaSpace{}%
\AgdaInductiveConstructor{ξξ-blame}\AgdaSpace{}%
\AgdaSymbol{(}\AgdaOperator{\AgdaInductiveConstructor{□·}}\AgdaSpace{}%
\AgdaSymbol{\AgdaUnderscore{})}\AgdaSpace{}%
\AgdaInductiveConstructor{refl}\AgdaSpace{}%
\AgdaOperator{\AgdaInductiveConstructor{⟩}}\AgdaSpace{}%
\AgdaBound{rs}\AgdaSymbol{)}\<%
\\
\>[0][@{}l@{\AgdaIndent{0}}]%
\>[4]\AgdaKeyword{with}\AgdaSpace{}%
\AgdaFunction{blame↠}\AgdaSpace{}%
\AgdaBound{rs}\<%
\\
\>[0]\AgdaSymbol{...}\AgdaSpace{}%
\AgdaSymbol{|}\AgdaSpace{}%
\AgdaInductiveConstructor{refl}\AgdaSpace{}%
\AgdaSymbol{=}\AgdaSpace{}%
\AgdaInductiveConstructor{inj₂}\AgdaSpace{}%
\AgdaSymbol{(}\AgdaInductiveConstructor{inj₂}\AgdaSpace{}%
\AgdaSymbol{(}\AgdaInductiveConstructor{inj₂}\AgdaSpace{}%
\AgdaInductiveConstructor{refl}\AgdaSymbol{))}\<%
\\
\>[0]\AgdaFunction{app-multi-inv}\AgdaSpace{}%
\AgdaSymbol{\{}\AgdaBound{L}\AgdaSymbol{\}}\AgdaSpace{}%
\AgdaSymbol{\{}\AgdaBound{M}\AgdaSymbol{\}}\AgdaSpace{}%
\AgdaSymbol{\{}\AgdaBound{N}\AgdaSymbol{\}}\AgdaSpace{}%
\AgdaSymbol{(}\AgdaDottedPattern{\AgdaSymbol{.(}}\AgdaDottedPattern{\AgdaBound{L}}\AgdaSpace{}%
\AgdaDottedPattern{\AgdaOperator{\AgdaInductiveConstructor{·}}}\AgdaSpace{}%
\AgdaDottedPattern{\AgdaBound{M}}\AgdaDottedPattern{\AgdaSymbol{)}}\AgdaSpace{}%
\AgdaOperator{\AgdaInductiveConstructor{⟶⟨}}\AgdaSpace{}%
\AgdaInductiveConstructor{ξξ-blame}\AgdaSpace{}%
\AgdaSymbol{(}\AgdaBound{v}\AgdaSpace{}%
\AgdaOperator{\AgdaInductiveConstructor{·□}}\AgdaSymbol{)}\AgdaSpace{}%
\AgdaInductiveConstructor{refl}\AgdaSpace{}%
\AgdaOperator{\AgdaInductiveConstructor{⟩}}\AgdaSpace{}%
\AgdaBound{rs}\AgdaSymbol{)}\<%
\\
\>[0][@{}l@{\AgdaIndent{0}}]%
\>[4]\AgdaKeyword{with}\AgdaSpace{}%
\AgdaFunction{blame↠}\AgdaSpace{}%
\AgdaBound{rs}\<%
\\
\>[0]\AgdaSymbol{...}\AgdaSpace{}%
\AgdaSymbol{|}\AgdaSpace{}%
\AgdaInductiveConstructor{refl}\AgdaSpace{}%
\AgdaSymbol{=}\AgdaSpace{}%
\AgdaInductiveConstructor{inj₂}\AgdaSpace{}%
\AgdaSymbol{(}\AgdaInductiveConstructor{inj₂}\AgdaSpace{}%
\AgdaSymbol{(}\AgdaInductiveConstructor{inj₂}\AgdaSpace{}%
\AgdaInductiveConstructor{refl}\AgdaSymbol{))}\<%
\\
\>[0]\AgdaFunction{app-multi-inv}\AgdaSpace{}%
\AgdaSymbol{\{}\AgdaDottedPattern{\AgdaSymbol{.(}}\AgdaDottedPattern{\AgdaInductiveConstructor{ƛ}}\AgdaSpace{}%
\AgdaDottedPattern{\AgdaSymbol{\AgdaUnderscore{})}}\AgdaSymbol{\}}\AgdaSpace{}%
\AgdaSymbol{\{}\AgdaBound{M}\AgdaSymbol{\}}\AgdaSpace{}%
\AgdaSymbol{\{}\AgdaBound{N}\AgdaSymbol{\}}\AgdaSpace{}%
\AgdaSymbol{(}\AgdaDottedPattern{\AgdaSymbol{.(}}\AgdaDottedPattern{\AgdaInductiveConstructor{ƛ}}\AgdaSpace{}%
\AgdaDottedPattern{\AgdaSymbol{\AgdaUnderscore{}}}\AgdaSpace{}%
\AgdaDottedPattern{\AgdaOperator{\AgdaInductiveConstructor{·}}}\AgdaSpace{}%
\AgdaDottedPattern{\AgdaBound{M}}\AgdaDottedPattern{\AgdaSymbol{)}}\AgdaSpace{}%
\AgdaOperator{\AgdaInductiveConstructor{⟶⟨}}\AgdaSpace{}%
\AgdaInductiveConstructor{β}\AgdaSpace{}%
\AgdaBound{v}\AgdaSpace{}%
\AgdaOperator{\AgdaInductiveConstructor{⟩}}\AgdaSpace{}%
\AgdaBound{M′↠N}\AgdaSymbol{)}\AgdaSpace{}%
\AgdaSymbol{=}\<%
\\
\>[0][@{}l@{\AgdaIndent{0}}]%
\>[2]\AgdaInductiveConstructor{inj₂}\AgdaSpace{}%
\AgdaSymbol{(}\AgdaInductiveConstructor{inj₂}\AgdaSpace{}%
\AgdaSymbol{(}\AgdaInductiveConstructor{inj₁}\AgdaSpace{}%
\AgdaSymbol{(}\AgdaInductiveConstructor{ƛ}\AgdaSpace{}%
\AgdaSymbol{\AgdaUnderscore{}}\AgdaSpace{}%
\AgdaOperator{\AgdaInductiveConstructor{,}}\AgdaSpace{}%
\AgdaBound{M}\AgdaSpace{}%
\AgdaOperator{\AgdaInductiveConstructor{,}}\AgdaSpace{}%
\AgdaSymbol{(\AgdaUnderscore{}}\AgdaSpace{}%
\AgdaOperator{\AgdaInductiveConstructor{END}}\AgdaSymbol{)}\AgdaSpace{}%
\AgdaOperator{\AgdaInductiveConstructor{,}}\AgdaSpace{}%
\AgdaOperator{\AgdaInductiveConstructor{ƛ̬}}\AgdaSpace{}%
\AgdaSymbol{\AgdaUnderscore{}}\AgdaSpace{}%
\AgdaOperator{\AgdaInductiveConstructor{,}}\AgdaSpace{}%
\AgdaSymbol{(}\AgdaBound{M}\AgdaSpace{}%
\AgdaOperator{\AgdaInductiveConstructor{END}}\AgdaSymbol{)}\AgdaSpace{}%
\AgdaOperator{\AgdaInductiveConstructor{,}}\AgdaSpace{}%
\AgdaBound{v}\AgdaSpace{}%
\AgdaOperator{\AgdaInductiveConstructor{,}}\AgdaSpace{}%
\AgdaSymbol{(\AgdaUnderscore{}}\AgdaSpace{}%
\AgdaOperator{\AgdaInductiveConstructor{⟶⟨}}\AgdaSpace{}%
\AgdaInductiveConstructor{β}\AgdaSpace{}%
\AgdaBound{v}\AgdaSpace{}%
\AgdaOperator{\AgdaInductiveConstructor{⟩}}\AgdaSpace{}%
\AgdaBound{M′↠N}\AgdaSymbol{)}\AgdaSpace{}%
\AgdaOperator{\AgdaInductiveConstructor{,}}\AgdaSpace{}%
\AgdaInductiveConstructor{refl}\AgdaSymbol{)))}\<%
\\
\\[\AgdaEmptyExtraSkip]%
\>[0]\AgdaFunction{inject-multi-inv}\AgdaSpace{}%
\AgdaSymbol{:}\AgdaSpace{}%
\AgdaSymbol{∀\{}\AgdaBound{M}\AgdaSpace{}%
\AgdaBound{N}\AgdaSymbol{\}\{}\AgdaBound{G}\AgdaSymbol{\}}\<%
\\
\>[0][@{}l@{\AgdaIndent{0}}]%
\>[2]\AgdaSymbol{→}\AgdaSpace{}%
\AgdaSymbol{(}\AgdaBound{red}\AgdaSpace{}%
\AgdaSymbol{:}\AgdaSpace{}%
\AgdaBound{M}\AgdaSpace{}%
\AgdaOperator{\AgdaInductiveConstructor{⟨}}\AgdaSpace{}%
\AgdaBound{G}\AgdaSpace{}%
\AgdaOperator{\AgdaInductiveConstructor{!⟩}}\AgdaSpace{}%
\AgdaOperator{\AgdaDatatype{↠}}\AgdaSpace{}%
\AgdaBound{N}\AgdaSymbol{)}\<%
\\
\>[2]\AgdaSymbol{→}%
\>[2646I]\AgdaSymbol{(}\AgdaFunction{∃[}\AgdaSpace{}%
\AgdaBound{M′}\AgdaSpace{}%
\AgdaFunction{]}\AgdaSpace{}%
\AgdaFunction{Σ[}\AgdaSpace{}%
\AgdaBound{r1}\AgdaSpace{}%
\AgdaFunction{∈}\AgdaSpace{}%
\AgdaBound{M}\AgdaSpace{}%
\AgdaOperator{\AgdaDatatype{↠}}\AgdaSpace{}%
\AgdaBound{M′}\AgdaSpace{}%
\AgdaFunction{]}\AgdaSpace{}%
\AgdaSymbol{(}\AgdaBound{N}\AgdaSpace{}%
\AgdaOperator{\AgdaDatatype{≡}}\AgdaSpace{}%
\AgdaBound{M′}\AgdaSpace{}%
\AgdaOperator{\AgdaInductiveConstructor{⟨}}\AgdaSpace{}%
\AgdaBound{G}\AgdaSpace{}%
\AgdaOperator{\AgdaInductiveConstructor{!⟩}}\AgdaSpace{}%
\AgdaOperator{\AgdaFunction{×}}\AgdaSpace{}%
\AgdaFunction{len}\AgdaSpace{}%
\AgdaBound{r1}\AgdaSpace{}%
\AgdaOperator{\AgdaDatatype{≡}}\AgdaSpace{}%
\AgdaFunction{len}\AgdaSpace{}%
\AgdaBound{red}\AgdaSymbol{))}\<%
\\
\>[.][@{}l@{}]\<[2646I]%
\>[4]\AgdaOperator{\AgdaDatatype{⊎}}\AgdaSpace{}%
\AgdaSymbol{(}\AgdaFunction{∃[}\AgdaSpace{}%
\AgdaBound{V}\AgdaSpace{}%
\AgdaFunction{]}\AgdaSpace{}%
\AgdaFunction{Σ[}\AgdaSpace{}%
\AgdaBound{r1}\AgdaSpace{}%
\AgdaFunction{∈}\AgdaSpace{}%
\AgdaBound{M}\AgdaSpace{}%
\AgdaOperator{\AgdaDatatype{↠}}\AgdaSpace{}%
\AgdaBound{V}\AgdaSpace{}%
\AgdaFunction{]}\AgdaSpace{}%
\AgdaDatatype{Value}\AgdaSpace{}%
\AgdaBound{V}\AgdaSpace{}%
\AgdaOperator{\AgdaFunction{×}}\AgdaSpace{}%
\AgdaFunction{Σ[}\AgdaSpace{}%
\AgdaBound{r2}\AgdaSpace{}%
\AgdaFunction{∈}\AgdaSpace{}%
\AgdaBound{V}\AgdaSpace{}%
\AgdaOperator{\AgdaInductiveConstructor{⟨}}\AgdaSpace{}%
\AgdaBound{G}\AgdaSpace{}%
\AgdaOperator{\AgdaInductiveConstructor{!⟩}}\AgdaSpace{}%
\AgdaOperator{\AgdaDatatype{↠}}\AgdaSpace{}%
\AgdaBound{N}\AgdaSpace{}%
\AgdaFunction{]}\AgdaSpace{}%
\AgdaFunction{len}\AgdaSpace{}%
\AgdaBound{red}\AgdaSpace{}%
\AgdaOperator{\AgdaDatatype{≡}}\AgdaSpace{}%
\AgdaFunction{len}\AgdaSpace{}%
\AgdaBound{r1}\AgdaSpace{}%
\AgdaOperator{\AgdaPrimitive{+}}\AgdaSpace{}%
\AgdaFunction{len}\AgdaSpace{}%
\AgdaBound{r2}\AgdaSymbol{)}\<%
\\
\>[4]\AgdaOperator{\AgdaDatatype{⊎}}\AgdaSpace{}%
\AgdaBound{N}\AgdaSpace{}%
\AgdaOperator{\AgdaDatatype{≡}}\AgdaSpace{}%
\AgdaInductiveConstructor{blame}\<%
\\
\>[0]\AgdaFunction{inject-multi-inv}\AgdaSpace{}%
\AgdaSymbol{\{}\AgdaBound{M}\AgdaSymbol{\}}\AgdaSpace{}%
\AgdaSymbol{(}\AgdaDottedPattern{\AgdaSymbol{.(\AgdaUnderscore{}}}\AgdaSpace{}%
\AgdaDottedPattern{\AgdaOperator{\AgdaInductiveConstructor{⟨}}}\AgdaSpace{}%
\AgdaDottedPattern{\AgdaSymbol{\AgdaUnderscore{}}}\AgdaSpace{}%
\AgdaDottedPattern{\AgdaOperator{\AgdaInductiveConstructor{!⟩}}}\AgdaDottedPattern{\AgdaSymbol{)}}\AgdaSpace{}%
\AgdaOperator{\AgdaInductiveConstructor{END}}\AgdaSymbol{)}\AgdaSpace{}%
\AgdaSymbol{=}\AgdaSpace{}%
\AgdaInductiveConstructor{inj₁}\AgdaSpace{}%
\AgdaSymbol{(}\AgdaBound{M}\AgdaSpace{}%
\AgdaOperator{\AgdaInductiveConstructor{,}}\AgdaSpace{}%
\AgdaSymbol{(\AgdaUnderscore{}}\AgdaSpace{}%
\AgdaOperator{\AgdaInductiveConstructor{END}}\AgdaSymbol{)}\AgdaSpace{}%
\AgdaOperator{\AgdaInductiveConstructor{,}}\AgdaSpace{}%
\AgdaInductiveConstructor{refl}\AgdaSpace{}%
\AgdaOperator{\AgdaInductiveConstructor{,}}\AgdaSpace{}%
\AgdaInductiveConstructor{refl}\AgdaSymbol{)}\<%
\\
\>[0]\AgdaFunction{inject-multi-inv}\AgdaSpace{}%
\AgdaSymbol{(}\AgdaDottedPattern{\AgdaSymbol{.(\AgdaUnderscore{}}}\AgdaSpace{}%
\AgdaDottedPattern{\AgdaOperator{\AgdaInductiveConstructor{⟨}}}\AgdaSpace{}%
\AgdaDottedPattern{\AgdaSymbol{\AgdaUnderscore{}}}\AgdaSpace{}%
\AgdaDottedPattern{\AgdaOperator{\AgdaInductiveConstructor{!⟩}}}\AgdaDottedPattern{\AgdaSymbol{)}}\AgdaSpace{}%
\AgdaOperator{\AgdaInductiveConstructor{⟶⟨}}\AgdaSpace{}%
\AgdaInductiveConstructor{ξξ}\AgdaSpace{}%
\AgdaOperator{\AgdaInductiveConstructor{□⟨}}\AgdaSpace{}%
\AgdaBound{G}\AgdaSpace{}%
\AgdaOperator{\AgdaInductiveConstructor{!⟩}}\AgdaSpace{}%
\AgdaInductiveConstructor{refl}\AgdaSpace{}%
\AgdaInductiveConstructor{refl}\AgdaSpace{}%
\AgdaBound{r1}\AgdaSpace{}%
\AgdaOperator{\AgdaInductiveConstructor{⟩}}\AgdaSpace{}%
\AgdaBound{r2}\AgdaSymbol{)}\<%
\\
\>[0][@{}l@{\AgdaIndent{0}}]%
\>[4]\AgdaKeyword{with}\AgdaSpace{}%
\AgdaFunction{inject-multi-inv}\AgdaSpace{}%
\AgdaBound{r2}\<%
\\
\>[0]\AgdaSymbol{...}\AgdaSpace{}%
\AgdaSymbol{|}\AgdaSpace{}%
\AgdaInductiveConstructor{inj₁}\AgdaSpace{}%
\AgdaSymbol{(}\AgdaBound{M′}\AgdaSpace{}%
\AgdaOperator{\AgdaInductiveConstructor{,}}\AgdaSpace{}%
\AgdaBound{→M′}\AgdaSpace{}%
\AgdaOperator{\AgdaInductiveConstructor{,}}\AgdaSpace{}%
\AgdaBound{eq}\AgdaSpace{}%
\AgdaOperator{\AgdaInductiveConstructor{,}}\AgdaSpace{}%
\AgdaBound{eqlen}\AgdaSymbol{)}\AgdaSpace{}%
\AgdaSymbol{=}\AgdaSpace{}%
\AgdaInductiveConstructor{inj₁}\AgdaSpace{}%
\AgdaSymbol{(\AgdaUnderscore{}}\AgdaSpace{}%
\AgdaOperator{\AgdaInductiveConstructor{,}}\AgdaSpace{}%
\AgdaSymbol{(\AgdaUnderscore{}}\AgdaSpace{}%
\AgdaOperator{\AgdaInductiveConstructor{⟶⟨}}\AgdaSpace{}%
\AgdaBound{r1}\AgdaSpace{}%
\AgdaOperator{\AgdaInductiveConstructor{⟩}}\AgdaSpace{}%
\AgdaBound{→M′}\AgdaSymbol{)}\AgdaSpace{}%
\AgdaOperator{\AgdaInductiveConstructor{,}}\AgdaSpace{}%
\AgdaBound{eq}\AgdaSpace{}%
\AgdaOperator{\AgdaInductiveConstructor{,}}\AgdaSpace{}%
\AgdaFunction{cong}\AgdaSpace{}%
\AgdaInductiveConstructor{suc}\AgdaSpace{}%
\AgdaBound{eqlen}\AgdaSymbol{)}\<%
\\
\>[0]\AgdaSymbol{...}\AgdaSpace{}%
\AgdaSymbol{|}\AgdaSpace{}%
\AgdaInductiveConstructor{inj₂}\AgdaSpace{}%
\AgdaSymbol{(}\AgdaInductiveConstructor{inj₁}\AgdaSpace{}%
\AgdaSymbol{(}\AgdaBound{V}\AgdaSpace{}%
\AgdaOperator{\AgdaInductiveConstructor{,}}\AgdaSpace{}%
\AgdaBound{→V}\AgdaSpace{}%
\AgdaOperator{\AgdaInductiveConstructor{,}}\AgdaSpace{}%
\AgdaBound{v}\AgdaSpace{}%
\AgdaOperator{\AgdaInductiveConstructor{,}}\AgdaSpace{}%
\AgdaBound{→N}\AgdaSpace{}%
\AgdaOperator{\AgdaInductiveConstructor{,}}\AgdaSpace{}%
\AgdaBound{eqlen}\AgdaSymbol{))}\AgdaSpace{}%
\AgdaSymbol{=}\AgdaSpace{}%
\AgdaInductiveConstructor{inj₂}\AgdaSpace{}%
\AgdaSymbol{(}\AgdaInductiveConstructor{inj₁}\AgdaSpace{}%
\AgdaSymbol{(\AgdaUnderscore{}}\AgdaSpace{}%
\AgdaOperator{\AgdaInductiveConstructor{,}}\AgdaSpace{}%
\AgdaSymbol{(\AgdaUnderscore{}}\AgdaSpace{}%
\AgdaOperator{\AgdaInductiveConstructor{⟶⟨}}\AgdaSpace{}%
\AgdaBound{r1}\AgdaSpace{}%
\AgdaOperator{\AgdaInductiveConstructor{⟩}}\AgdaSpace{}%
\AgdaBound{→V}\AgdaSymbol{)}\AgdaSpace{}%
\AgdaOperator{\AgdaInductiveConstructor{,}}\AgdaSpace{}%
\AgdaBound{v}\AgdaSpace{}%
\AgdaOperator{\AgdaInductiveConstructor{,}}\AgdaSpace{}%
\AgdaBound{→N}\AgdaSpace{}%
\AgdaOperator{\AgdaInductiveConstructor{,}}\AgdaSpace{}%
\AgdaFunction{cong}\AgdaSpace{}%
\AgdaInductiveConstructor{suc}\AgdaSpace{}%
\AgdaBound{eqlen}\AgdaSymbol{))}\<%
\\
\>[0]\AgdaSymbol{...}\AgdaSpace{}%
\AgdaSymbol{|}\AgdaSpace{}%
\AgdaInductiveConstructor{inj₂}\AgdaSpace{}%
\AgdaSymbol{(}\AgdaInductiveConstructor{inj₂}\AgdaSpace{}%
\AgdaInductiveConstructor{refl}\AgdaSymbol{)}\AgdaSpace{}%
\AgdaSymbol{=}\AgdaSpace{}%
\AgdaInductiveConstructor{inj₂}\AgdaSpace{}%
\AgdaSymbol{(}\AgdaInductiveConstructor{inj₂}\AgdaSpace{}%
\AgdaInductiveConstructor{refl}\AgdaSymbol{)}\<%
\\
\>[0]\AgdaFunction{inject-multi-inv}\AgdaSpace{}%
\AgdaSymbol{(}\AgdaDottedPattern{\AgdaSymbol{.(\AgdaUnderscore{}}}\AgdaSpace{}%
\AgdaDottedPattern{\AgdaOperator{\AgdaInductiveConstructor{⟨}}}\AgdaSpace{}%
\AgdaDottedPattern{\AgdaSymbol{\AgdaUnderscore{}}}\AgdaSpace{}%
\AgdaDottedPattern{\AgdaOperator{\AgdaInductiveConstructor{!⟩}}}\AgdaDottedPattern{\AgdaSymbol{)}}\AgdaSpace{}%
\AgdaOperator{\AgdaInductiveConstructor{⟶⟨}}\AgdaSpace{}%
\AgdaInductiveConstructor{ξξ-blame}\AgdaSpace{}%
\AgdaOperator{\AgdaInductiveConstructor{□⟨}}\AgdaSpace{}%
\AgdaBound{G}\AgdaSpace{}%
\AgdaOperator{\AgdaInductiveConstructor{!⟩}}\AgdaSpace{}%
\AgdaBound{x}\AgdaSpace{}%
\AgdaOperator{\AgdaInductiveConstructor{⟩}}\AgdaSpace{}%
\AgdaBound{red}\AgdaSymbol{)}\<%
\\
\>[0][@{}l@{\AgdaIndent{0}}]%
\>[4]\AgdaKeyword{with}\AgdaSpace{}%
\AgdaFunction{blame↠}\AgdaSpace{}%
\AgdaBound{red}\<%
\\
\>[0]\AgdaSymbol{...}\AgdaSpace{}%
\AgdaSymbol{|}\AgdaSpace{}%
\AgdaInductiveConstructor{refl}\AgdaSpace{}%
\AgdaSymbol{=}\AgdaSpace{}%
\AgdaInductiveConstructor{inj₂}\AgdaSpace{}%
\AgdaSymbol{(}\AgdaInductiveConstructor{inj₂}\AgdaSpace{}%
\AgdaInductiveConstructor{refl}\AgdaSymbol{)}\<%
\\
\\[\AgdaEmptyExtraSkip]%
\>[0]\AgdaFunction{project-multi-inv2}\AgdaSpace{}%
\AgdaSymbol{:}\AgdaSpace{}%
\AgdaSymbol{∀\{}\AgdaBound{M}\AgdaSpace{}%
\AgdaBound{N}\AgdaSymbol{\}\{}\AgdaBound{G}\AgdaSymbol{\}}\<%
\\
\>[0][@{}l@{\AgdaIndent{0}}]%
\>[2]\AgdaSymbol{→}\AgdaSpace{}%
\AgdaSymbol{(}\AgdaBound{red}\AgdaSpace{}%
\AgdaSymbol{:}\AgdaSpace{}%
\AgdaBound{M}\AgdaSpace{}%
\AgdaOperator{\AgdaInductiveConstructor{⟨}}\AgdaSpace{}%
\AgdaBound{G}\AgdaSpace{}%
\AgdaOperator{\AgdaInductiveConstructor{?⟩}}\AgdaSpace{}%
\AgdaOperator{\AgdaDatatype{↠}}\AgdaSpace{}%
\AgdaBound{N}\AgdaSymbol{)}\<%
\\
\>[2]\AgdaSymbol{→}%
\>[2827I]\AgdaSymbol{(}\AgdaFunction{∃[}\AgdaSpace{}%
\AgdaBound{M′}\AgdaSpace{}%
\AgdaFunction{]}\AgdaSpace{}%
\AgdaFunction{Σ[}\AgdaSpace{}%
\AgdaBound{r1}\AgdaSpace{}%
\AgdaFunction{∈}\AgdaSpace{}%
\AgdaBound{M}\AgdaSpace{}%
\AgdaOperator{\AgdaDatatype{↠}}\AgdaSpace{}%
\AgdaBound{M′}\AgdaSpace{}%
\AgdaFunction{]}\AgdaSpace{}%
\AgdaSymbol{(}\AgdaBound{N}\AgdaSpace{}%
\AgdaOperator{\AgdaDatatype{≡}}\AgdaSpace{}%
\AgdaBound{M′}\AgdaSpace{}%
\AgdaOperator{\AgdaInductiveConstructor{⟨}}\AgdaSpace{}%
\AgdaBound{G}\AgdaSpace{}%
\AgdaOperator{\AgdaInductiveConstructor{?⟩}}\AgdaSpace{}%
\AgdaOperator{\AgdaFunction{×}}\AgdaSpace{}%
\AgdaFunction{len}\AgdaSpace{}%
\AgdaBound{r1}\AgdaSpace{}%
\AgdaOperator{\AgdaDatatype{≡}}\AgdaSpace{}%
\AgdaFunction{len}\AgdaSpace{}%
\AgdaBound{red}\AgdaSymbol{))}\<%
\\
\>[.][@{}l@{}]\<[2827I]%
\>[4]\AgdaOperator{\AgdaDatatype{⊎}}\AgdaSpace{}%
\AgdaSymbol{(}\AgdaFunction{∃[}\AgdaSpace{}%
\AgdaBound{V}\AgdaSpace{}%
\AgdaFunction{]}\AgdaSpace{}%
\AgdaFunction{Σ[}\AgdaSpace{}%
\AgdaBound{r1}\AgdaSpace{}%
\AgdaFunction{∈}\AgdaSpace{}%
\AgdaBound{M}\AgdaSpace{}%
\AgdaOperator{\AgdaDatatype{↠}}\AgdaSpace{}%
\AgdaBound{V}\AgdaSpace{}%
\AgdaFunction{]}\AgdaSpace{}%
\AgdaDatatype{Value}\AgdaSpace{}%
\AgdaBound{V}\AgdaSpace{}%
\AgdaOperator{\AgdaFunction{×}}\AgdaSpace{}%
\AgdaFunction{Σ[}\AgdaSpace{}%
\AgdaBound{r2}\AgdaSpace{}%
\AgdaFunction{∈}\AgdaSpace{}%
\AgdaBound{V}\AgdaSpace{}%
\AgdaOperator{\AgdaInductiveConstructor{⟨}}\AgdaSpace{}%
\AgdaBound{G}\AgdaSpace{}%
\AgdaOperator{\AgdaInductiveConstructor{?⟩}}\AgdaSpace{}%
\AgdaOperator{\AgdaDatatype{↠}}\AgdaSpace{}%
\AgdaBound{N}\AgdaSpace{}%
\AgdaFunction{]}\AgdaSpace{}%
\AgdaFunction{len}\AgdaSpace{}%
\AgdaBound{red}\AgdaSpace{}%
\AgdaOperator{\AgdaDatatype{≡}}\AgdaSpace{}%
\AgdaFunction{len}\AgdaSpace{}%
\AgdaBound{r1}\AgdaSpace{}%
\AgdaOperator{\AgdaPrimitive{+}}\AgdaSpace{}%
\AgdaFunction{len}\AgdaSpace{}%
\AgdaBound{r2}\AgdaSymbol{)}\<%
\\
\>[4]\AgdaOperator{\AgdaDatatype{⊎}}\AgdaSpace{}%
\AgdaBound{N}\AgdaSpace{}%
\AgdaOperator{\AgdaDatatype{≡}}\AgdaSpace{}%
\AgdaInductiveConstructor{blame}\<%
\\
\>[0]\AgdaFunction{project-multi-inv2}\AgdaSpace{}%
\AgdaSymbol{(}\AgdaDottedPattern{\AgdaSymbol{.(\AgdaUnderscore{}}}\AgdaSpace{}%
\AgdaDottedPattern{\AgdaOperator{\AgdaInductiveConstructor{⟨}}}\AgdaSpace{}%
\AgdaDottedPattern{\AgdaSymbol{\AgdaUnderscore{}}}\AgdaSpace{}%
\AgdaDottedPattern{\AgdaOperator{\AgdaInductiveConstructor{?⟩}}}\AgdaDottedPattern{\AgdaSymbol{)}}\AgdaSpace{}%
\AgdaOperator{\AgdaInductiveConstructor{END}}\AgdaSymbol{)}\AgdaSpace{}%
\AgdaSymbol{=}\AgdaSpace{}%
\AgdaInductiveConstructor{inj₁}\AgdaSpace{}%
\AgdaSymbol{(\AgdaUnderscore{}}\AgdaSpace{}%
\AgdaOperator{\AgdaInductiveConstructor{,}}\AgdaSpace{}%
\AgdaSymbol{(\AgdaUnderscore{}}\AgdaSpace{}%
\AgdaOperator{\AgdaInductiveConstructor{END}}\AgdaSymbol{)}\AgdaSpace{}%
\AgdaOperator{\AgdaInductiveConstructor{,}}\AgdaSpace{}%
\AgdaInductiveConstructor{refl}\AgdaSpace{}%
\AgdaOperator{\AgdaInductiveConstructor{,}}\AgdaSpace{}%
\AgdaInductiveConstructor{refl}\AgdaSymbol{)}\<%
\\
\>[0]\AgdaFunction{project-multi-inv2}\AgdaSpace{}%
\AgdaSymbol{(}\AgdaDottedPattern{\AgdaSymbol{.(\AgdaUnderscore{}}}\AgdaSpace{}%
\AgdaDottedPattern{\AgdaOperator{\AgdaInductiveConstructor{⟨}}}\AgdaSpace{}%
\AgdaDottedPattern{\AgdaSymbol{\AgdaUnderscore{}}}\AgdaSpace{}%
\AgdaDottedPattern{\AgdaOperator{\AgdaInductiveConstructor{?⟩}}}\AgdaDottedPattern{\AgdaSymbol{)}}\AgdaSpace{}%
\AgdaOperator{\AgdaInductiveConstructor{⟶⟨}}\AgdaSpace{}%
\AgdaInductiveConstructor{ξξ}\AgdaSpace{}%
\AgdaOperator{\AgdaInductiveConstructor{□⟨}}\AgdaSpace{}%
\AgdaBound{H}\AgdaSpace{}%
\AgdaOperator{\AgdaInductiveConstructor{?⟩}}\AgdaSpace{}%
\AgdaInductiveConstructor{refl}\AgdaSpace{}%
\AgdaInductiveConstructor{refl}\AgdaSpace{}%
\AgdaBound{r}\AgdaSpace{}%
\AgdaOperator{\AgdaInductiveConstructor{⟩}}\AgdaSpace{}%
\AgdaBound{rs}\AgdaSymbol{)}\<%
\\
\>[0][@{}l@{\AgdaIndent{0}}]%
\>[4]\AgdaKeyword{with}\AgdaSpace{}%
\AgdaFunction{project-multi-inv2}\AgdaSpace{}%
\AgdaBound{rs}\<%
\\
\>[0]\AgdaSymbol{...}\AgdaSpace{}%
\AgdaSymbol{|}\AgdaSpace{}%
\AgdaInductiveConstructor{inj₁}\AgdaSpace{}%
\AgdaSymbol{(}\AgdaBound{M′}\AgdaSpace{}%
\AgdaOperator{\AgdaInductiveConstructor{,}}\AgdaSpace{}%
\AgdaBound{M→M′}\AgdaSpace{}%
\AgdaOperator{\AgdaInductiveConstructor{,}}\AgdaSpace{}%
\AgdaInductiveConstructor{refl}\AgdaSpace{}%
\AgdaOperator{\AgdaInductiveConstructor{,}}\AgdaSpace{}%
\AgdaBound{eq}\AgdaSymbol{)}\AgdaSpace{}%
\AgdaSymbol{=}\AgdaSpace{}%
\AgdaInductiveConstructor{inj₁}\AgdaSpace{}%
\AgdaSymbol{(}\AgdaBound{M′}\AgdaSpace{}%
\AgdaOperator{\AgdaInductiveConstructor{,}}\AgdaSpace{}%
\AgdaSymbol{(\AgdaUnderscore{}}\AgdaSpace{}%
\AgdaOperator{\AgdaInductiveConstructor{⟶⟨}}\AgdaSpace{}%
\AgdaBound{r}\AgdaSpace{}%
\AgdaOperator{\AgdaInductiveConstructor{⟩}}\AgdaSpace{}%
\AgdaBound{M→M′}\AgdaSymbol{)}\AgdaSpace{}%
\AgdaOperator{\AgdaInductiveConstructor{,}}\AgdaSpace{}%
\AgdaInductiveConstructor{refl}\AgdaSpace{}%
\AgdaOperator{\AgdaInductiveConstructor{,}}\AgdaSpace{}%
\AgdaFunction{cong}\AgdaSpace{}%
\AgdaInductiveConstructor{suc}\AgdaSpace{}%
\AgdaBound{eq}\AgdaSymbol{)}\<%
\\
\>[0]\AgdaSymbol{...}\AgdaSpace{}%
\AgdaSymbol{|}\AgdaSpace{}%
\AgdaInductiveConstructor{inj₂}\AgdaSpace{}%
\AgdaSymbol{(}\AgdaInductiveConstructor{inj₁}\AgdaSpace{}%
\AgdaSymbol{(}\AgdaBound{V}\AgdaSpace{}%
\AgdaOperator{\AgdaInductiveConstructor{,}}\AgdaSpace{}%
\AgdaBound{M→V}\AgdaSpace{}%
\AgdaOperator{\AgdaInductiveConstructor{,}}\AgdaSpace{}%
\AgdaBound{v}\AgdaSpace{}%
\AgdaOperator{\AgdaInductiveConstructor{,}}\AgdaSpace{}%
\AgdaBound{Vg→N}\AgdaSpace{}%
\AgdaOperator{\AgdaInductiveConstructor{,}}\AgdaSpace{}%
\AgdaBound{eq}\AgdaSymbol{))}\AgdaSpace{}%
\AgdaSymbol{=}\AgdaSpace{}%
\AgdaInductiveConstructor{inj₂}\AgdaSpace{}%
\AgdaSymbol{(}\AgdaInductiveConstructor{inj₁}\AgdaSpace{}%
\AgdaSymbol{(}\AgdaBound{V}\AgdaSpace{}%
\AgdaOperator{\AgdaInductiveConstructor{,}}\AgdaSpace{}%
\AgdaSymbol{(\AgdaUnderscore{}}\AgdaSpace{}%
\AgdaOperator{\AgdaInductiveConstructor{⟶⟨}}\AgdaSpace{}%
\AgdaBound{r}\AgdaSpace{}%
\AgdaOperator{\AgdaInductiveConstructor{⟩}}\AgdaSpace{}%
\AgdaBound{M→V}\AgdaSpace{}%
\AgdaSymbol{)}\AgdaSpace{}%
\AgdaOperator{\AgdaInductiveConstructor{,}}\AgdaSpace{}%
\AgdaBound{v}\AgdaSpace{}%
\AgdaOperator{\AgdaInductiveConstructor{,}}\AgdaSpace{}%
\AgdaBound{Vg→N}\AgdaSpace{}%
\AgdaOperator{\AgdaInductiveConstructor{,}}\AgdaSpace{}%
\AgdaFunction{cong}\AgdaSpace{}%
\AgdaInductiveConstructor{suc}\AgdaSpace{}%
\AgdaBound{eq}\AgdaSymbol{))}\<%
\\
\>[0]\AgdaSymbol{...}\AgdaSpace{}%
\AgdaSymbol{|}\AgdaSpace{}%
\AgdaInductiveConstructor{inj₂}\AgdaSpace{}%
\AgdaSymbol{(}\AgdaInductiveConstructor{inj₂}\AgdaSpace{}%
\AgdaInductiveConstructor{refl}\AgdaSymbol{)}\AgdaSpace{}%
\AgdaSymbol{=}\AgdaSpace{}%
\AgdaInductiveConstructor{inj₂}\AgdaSpace{}%
\AgdaSymbol{(}\AgdaInductiveConstructor{inj₂}\AgdaSpace{}%
\AgdaInductiveConstructor{refl}\AgdaSymbol{)}\<%
\\
\>[0]\AgdaFunction{project-multi-inv2}\AgdaSpace{}%
\AgdaSymbol{(}\AgdaDottedPattern{\AgdaSymbol{.(\AgdaUnderscore{}}}\AgdaSpace{}%
\AgdaDottedPattern{\AgdaOperator{\AgdaInductiveConstructor{⟨}}}\AgdaSpace{}%
\AgdaDottedPattern{\AgdaSymbol{\AgdaUnderscore{}}}\AgdaSpace{}%
\AgdaDottedPattern{\AgdaOperator{\AgdaInductiveConstructor{?⟩}}}\AgdaDottedPattern{\AgdaSymbol{)}}\AgdaSpace{}%
\AgdaOperator{\AgdaInductiveConstructor{⟶⟨}}\AgdaSpace{}%
\AgdaInductiveConstructor{ξξ-blame}\AgdaSpace{}%
\AgdaOperator{\AgdaInductiveConstructor{□⟨}}\AgdaSpace{}%
\AgdaBound{H}\AgdaSpace{}%
\AgdaOperator{\AgdaInductiveConstructor{?⟩}}\AgdaSpace{}%
\AgdaInductiveConstructor{refl}\AgdaSpace{}%
\AgdaOperator{\AgdaInductiveConstructor{⟩}}\AgdaSpace{}%
\AgdaBound{rs}\AgdaSymbol{)}\<%
\\
\>[0][@{}l@{\AgdaIndent{0}}]%
\>[4]\AgdaKeyword{with}\AgdaSpace{}%
\AgdaFunction{blame↠}\AgdaSpace{}%
\AgdaBound{rs}\<%
\\
\>[0]\AgdaSymbol{...}\AgdaSpace{}%
\AgdaSymbol{|}\AgdaSpace{}%
\AgdaInductiveConstructor{refl}\AgdaSpace{}%
\AgdaSymbol{=}\AgdaSpace{}%
\AgdaInductiveConstructor{inj₂}\AgdaSpace{}%
\AgdaSymbol{(}\AgdaInductiveConstructor{inj₂}\AgdaSpace{}%
\AgdaInductiveConstructor{refl}\AgdaSymbol{)}\<%
\\
\>[0]\AgdaFunction{project-multi-inv2}\AgdaSpace{}%
\AgdaSymbol{(}\AgdaDottedPattern{\AgdaSymbol{.(\AgdaUnderscore{}}}\AgdaSpace{}%
\AgdaDottedPattern{\AgdaOperator{\AgdaInductiveConstructor{⟨}}}\AgdaSpace{}%
\AgdaDottedPattern{\AgdaSymbol{\AgdaUnderscore{}}}\AgdaSpace{}%
\AgdaDottedPattern{\AgdaOperator{\AgdaInductiveConstructor{?⟩}}}\AgdaDottedPattern{\AgdaSymbol{)}}\AgdaSpace{}%
\AgdaOperator{\AgdaInductiveConstructor{⟶⟨}}\AgdaSpace{}%
\AgdaInductiveConstructor{collapse}\AgdaSpace{}%
\AgdaBound{v}\AgdaSpace{}%
\AgdaInductiveConstructor{refl}\AgdaSpace{}%
\AgdaOperator{\AgdaInductiveConstructor{⟩}}\AgdaSpace{}%
\AgdaBound{rs}\AgdaSymbol{)}\AgdaSpace{}%
\AgdaSymbol{=}\<%
\\
\>[0][@{}l@{\AgdaIndent{0}}]%
\>[4]\AgdaInductiveConstructor{inj₂}\AgdaSpace{}%
\AgdaSymbol{(}\AgdaInductiveConstructor{inj₁}\AgdaSpace{}%
\AgdaSymbol{((\AgdaUnderscore{}}\AgdaSpace{}%
\AgdaOperator{\AgdaInductiveConstructor{⟨}}\AgdaSpace{}%
\AgdaSymbol{\AgdaUnderscore{}}\AgdaSpace{}%
\AgdaOperator{\AgdaInductiveConstructor{!⟩}}\AgdaSymbol{)}\AgdaSpace{}%
\AgdaOperator{\AgdaInductiveConstructor{,}}\AgdaSpace{}%
\AgdaSymbol{(\AgdaUnderscore{}}\AgdaSpace{}%
\AgdaOperator{\AgdaInductiveConstructor{END}}\AgdaSymbol{)}\AgdaSpace{}%
\AgdaOperator{\AgdaInductiveConstructor{,}}\AgdaSpace{}%
\AgdaSymbol{(}\AgdaBound{v}\AgdaSpace{}%
\AgdaOperator{\AgdaInductiveConstructor{〈}}\AgdaSpace{}%
\AgdaSymbol{\AgdaUnderscore{}}\AgdaSpace{}%
\AgdaOperator{\AgdaInductiveConstructor{〉}}\AgdaSymbol{)}\AgdaSpace{}%
\AgdaOperator{\AgdaInductiveConstructor{,}}\AgdaSpace{}%
\AgdaSymbol{(\AgdaUnderscore{}}\AgdaSpace{}%
\AgdaOperator{\AgdaInductiveConstructor{⟶⟨}}\AgdaSpace{}%
\AgdaInductiveConstructor{collapse}\AgdaSpace{}%
\AgdaBound{v}\AgdaSpace{}%
\AgdaInductiveConstructor{refl}\AgdaSpace{}%
\AgdaOperator{\AgdaInductiveConstructor{⟩}}\AgdaSpace{}%
\AgdaBound{rs}\AgdaSymbol{)}\AgdaSpace{}%
\AgdaOperator{\AgdaInductiveConstructor{,}}\AgdaSpace{}%
\AgdaInductiveConstructor{refl}\AgdaSymbol{))}\<%
\\
\>[0]\AgdaFunction{project-multi-inv2}\AgdaSpace{}%
\AgdaSymbol{(}\AgdaDottedPattern{\AgdaSymbol{.(\AgdaUnderscore{}}}\AgdaSpace{}%
\AgdaDottedPattern{\AgdaOperator{\AgdaInductiveConstructor{⟨}}}\AgdaSpace{}%
\AgdaDottedPattern{\AgdaSymbol{\AgdaUnderscore{}}}\AgdaSpace{}%
\AgdaDottedPattern{\AgdaOperator{\AgdaInductiveConstructor{?⟩}}}\AgdaDottedPattern{\AgdaSymbol{)}}\AgdaSpace{}%
\AgdaOperator{\AgdaInductiveConstructor{⟶⟨}}\AgdaSpace{}%
\AgdaInductiveConstructor{collide}\AgdaSpace{}%
\AgdaBound{v}\AgdaSpace{}%
\AgdaBound{neq}\AgdaSpace{}%
\AgdaInductiveConstructor{refl}\AgdaSpace{}%
\AgdaOperator{\AgdaInductiveConstructor{⟩}}\AgdaSpace{}%
\AgdaBound{rs}\AgdaSymbol{)}\AgdaSpace{}%
\AgdaSymbol{=}\<%
\\
\>[0][@{}l@{\AgdaIndent{0}}]%
\>[4]\AgdaInductiveConstructor{inj₂}\AgdaSpace{}%
\AgdaSymbol{(}\AgdaInductiveConstructor{inj₁}\AgdaSpace{}%
\AgdaSymbol{((\AgdaUnderscore{}}\AgdaSpace{}%
\AgdaOperator{\AgdaInductiveConstructor{⟨}}\AgdaSpace{}%
\AgdaSymbol{\AgdaUnderscore{}}\AgdaSpace{}%
\AgdaOperator{\AgdaInductiveConstructor{!⟩}}\AgdaSymbol{)}\AgdaSpace{}%
\AgdaOperator{\AgdaInductiveConstructor{,}}\AgdaSpace{}%
\AgdaSymbol{(\AgdaUnderscore{}}\AgdaSpace{}%
\AgdaOperator{\AgdaInductiveConstructor{END}}\AgdaSymbol{)}\AgdaSpace{}%
\AgdaOperator{\AgdaInductiveConstructor{,}}\AgdaSpace{}%
\AgdaSymbol{(}\AgdaBound{v}\AgdaSpace{}%
\AgdaOperator{\AgdaInductiveConstructor{〈}}\AgdaSpace{}%
\AgdaSymbol{\AgdaUnderscore{}}\AgdaSpace{}%
\AgdaOperator{\AgdaInductiveConstructor{〉}}\AgdaSymbol{)}\AgdaSpace{}%
\AgdaOperator{\AgdaInductiveConstructor{,}}\AgdaSpace{}%
\AgdaSymbol{(\AgdaUnderscore{}}\AgdaSpace{}%
\AgdaOperator{\AgdaInductiveConstructor{⟶⟨}}\AgdaSpace{}%
\AgdaInductiveConstructor{collide}\AgdaSpace{}%
\AgdaBound{v}\AgdaSpace{}%
\AgdaBound{neq}\AgdaSpace{}%
\AgdaInductiveConstructor{refl}\AgdaSpace{}%
\AgdaOperator{\AgdaInductiveConstructor{⟩}}\AgdaSpace{}%
\AgdaBound{rs}\AgdaSymbol{)}\AgdaSpace{}%
\AgdaOperator{\AgdaInductiveConstructor{,}}\AgdaSpace{}%
\AgdaInductiveConstructor{refl}\AgdaSymbol{))}\<%
\\
\\[\AgdaEmptyExtraSkip]%
\>[0]\AgdaFunction{app-inv-left}\AgdaSpace{}%
\AgdaSymbol{:}\AgdaSpace{}%
\AgdaSymbol{∀\{}\AgdaBound{L}\AgdaSpace{}%
\AgdaBound{M}\AgdaSpace{}%
\AgdaBound{N}\AgdaSymbol{\}}\<%
\\
\>[0][@{}l@{\AgdaIndent{0}}]%
\>[2]\AgdaSymbol{→}\AgdaSpace{}%
\AgdaSymbol{(}\AgdaBound{r1}\AgdaSpace{}%
\AgdaSymbol{:}\AgdaSpace{}%
\AgdaBound{L}\AgdaSpace{}%
\AgdaOperator{\AgdaInductiveConstructor{·}}\AgdaSpace{}%
\AgdaBound{M}\AgdaSpace{}%
\AgdaOperator{\AgdaDatatype{↠}}\AgdaSpace{}%
\AgdaBound{N}\AgdaSymbol{)}\<%
\\
\>[2]\AgdaSymbol{→}%
\>[3078I]\AgdaFunction{irred}\AgdaSpace{}%
\AgdaBound{N}\<%
\\
\>[.][@{}l@{}]\<[3078I]%
\>[4]\AgdaComment{--------------------------}\<%
\\
\>[2]\AgdaSymbol{→}%
\>[3080I]\AgdaSymbol{(}\AgdaFunction{∃[}%
\>[3081I]\AgdaBound{L′}\AgdaSpace{}%
\AgdaFunction{]}\AgdaSpace{}%
\AgdaFunction{Σ[}\AgdaSpace{}%
\AgdaBound{r2}\AgdaSpace{}%
\AgdaFunction{∈}\AgdaSpace{}%
\AgdaSymbol{(}\AgdaBound{L}\AgdaSpace{}%
\AgdaOperator{\AgdaDatatype{↠}}\AgdaSpace{}%
\AgdaBound{L′}\AgdaSymbol{)}\AgdaSpace{}%
\AgdaFunction{]}\AgdaSpace{}%
\AgdaFunction{irred}\AgdaSpace{}%
\AgdaBound{L′}\<%
\\
\>[.][@{}l@{}]\<[3081I]%
\>[8]\AgdaOperator{\AgdaFunction{×}}\AgdaSpace{}%
\AgdaFunction{Σ[}\AgdaSpace{}%
\AgdaBound{r3}\AgdaSpace{}%
\AgdaFunction{∈}\AgdaSpace{}%
\AgdaSymbol{(}\AgdaBound{L′}\AgdaSpace{}%
\AgdaOperator{\AgdaInductiveConstructor{·}}\AgdaSpace{}%
\AgdaBound{M}\AgdaSpace{}%
\AgdaOperator{\AgdaDatatype{↠}}\AgdaSpace{}%
\AgdaBound{N}\AgdaSymbol{)}\AgdaSpace{}%
\AgdaFunction{]}\AgdaSpace{}%
\AgdaFunction{len}\AgdaSpace{}%
\AgdaBound{r1}\AgdaSpace{}%
\AgdaOperator{\AgdaDatatype{≡}}\AgdaSpace{}%
\AgdaFunction{len}\AgdaSpace{}%
\AgdaBound{r2}\AgdaSpace{}%
\AgdaOperator{\AgdaPrimitive{+}}\AgdaSpace{}%
\AgdaFunction{len}\AgdaSpace{}%
\AgdaBound{r3}\AgdaSymbol{)}\<%
\\
\>[.][@{}l@{}]\<[3080I]%
\>[4]\AgdaOperator{\AgdaDatatype{⊎}}\AgdaSpace{}%
\AgdaBound{N}\AgdaSpace{}%
\AgdaOperator{\AgdaDatatype{≡}}\AgdaSpace{}%
\AgdaInductiveConstructor{blame}\<%
\\
\>[0]\AgdaFunction{app-inv-left}\AgdaSpace{}%
\AgdaSymbol{\{}\AgdaBound{L}\AgdaSymbol{\}}\AgdaSpace{}%
\AgdaSymbol{\{}\AgdaBound{M}\AgdaSymbol{\}}\AgdaSpace{}%
\AgdaSymbol{\{}\AgdaDottedPattern{\AgdaSymbol{.(}}\AgdaDottedPattern{\AgdaBound{L}}\AgdaSpace{}%
\AgdaDottedPattern{\AgdaOperator{\AgdaInductiveConstructor{·}}}\AgdaSpace{}%
\AgdaDottedPattern{\AgdaBound{M}}\AgdaDottedPattern{\AgdaSymbol{)}}\AgdaSymbol{\}}\AgdaSpace{}%
\AgdaSymbol{(}\AgdaDottedPattern{\AgdaSymbol{.(}}\AgdaDottedPattern{\AgdaBound{L}}\AgdaSpace{}%
\AgdaDottedPattern{\AgdaOperator{\AgdaInductiveConstructor{·}}}\AgdaSpace{}%
\AgdaDottedPattern{\AgdaBound{M}}\AgdaDottedPattern{\AgdaSymbol{)}}\AgdaSpace{}%
\AgdaOperator{\AgdaInductiveConstructor{END}}\AgdaSymbol{)}\AgdaSpace{}%
\AgdaBound{irredN}\AgdaSpace{}%
\AgdaSymbol{=}\<%
\\
\>[0][@{}l@{\AgdaIndent{0}}]%
\>[4]\AgdaInductiveConstructor{inj₁}\AgdaSpace{}%
\AgdaSymbol{(}\AgdaBound{L}\AgdaSpace{}%
\AgdaOperator{\AgdaInductiveConstructor{,}}\AgdaSpace{}%
\AgdaSymbol{(\AgdaUnderscore{}}\AgdaSpace{}%
\AgdaOperator{\AgdaInductiveConstructor{END}}\AgdaSymbol{)}\AgdaSpace{}%
\AgdaOperator{\AgdaInductiveConstructor{,}}\AgdaSpace{}%
\AgdaFunction{IL}\AgdaSpace{}%
\AgdaOperator{\AgdaInductiveConstructor{,}}\AgdaSpace{}%
\AgdaSymbol{(\AgdaUnderscore{}}\AgdaSpace{}%
\AgdaOperator{\AgdaInductiveConstructor{END}}\AgdaSymbol{)}\AgdaSpace{}%
\AgdaOperator{\AgdaInductiveConstructor{,}}\AgdaSpace{}%
\AgdaInductiveConstructor{refl}\AgdaSymbol{)}\<%
\\
\>[4]\AgdaKeyword{where}%
\>[3134I]\AgdaFunction{IL}\AgdaSpace{}%
\AgdaSymbol{:}\AgdaSpace{}%
\AgdaFunction{irred}\AgdaSpace{}%
\AgdaBound{L}\<%
\\
\>[.][@{}l@{}]\<[3134I]%
\>[10]\AgdaFunction{IL}\AgdaSpace{}%
\AgdaSymbol{(}\AgdaBound{L′}\AgdaSpace{}%
\AgdaOperator{\AgdaInductiveConstructor{,}}\AgdaSpace{}%
\AgdaBound{L→L′}\AgdaSymbol{)}\AgdaSpace{}%
\AgdaSymbol{=}\AgdaSpace{}%
\AgdaFunction{⊥-elim}\AgdaSpace{}%
\AgdaSymbol{(}\AgdaBound{irredN}\AgdaSpace{}%
\AgdaSymbol{(\AgdaUnderscore{}}\AgdaSpace{}%
\AgdaOperator{\AgdaInductiveConstructor{,}}\AgdaSpace{}%
\AgdaSymbol{(}\AgdaInductiveConstructor{ξ}\AgdaSpace{}%
\AgdaSymbol{(}\AgdaOperator{\AgdaInductiveConstructor{□·}}\AgdaSpace{}%
\AgdaBound{M}\AgdaSymbol{)}\AgdaSpace{}%
\AgdaBound{L→L′}\AgdaSymbol{)))}\<%
\\
\>[0]\AgdaFunction{app-inv-left}\AgdaSpace{}%
\AgdaSymbol{\{}\AgdaBound{L}\AgdaSymbol{\}}\AgdaSpace{}%
\AgdaSymbol{\{}\AgdaBound{M}\AgdaSymbol{\}}\AgdaSpace{}%
\AgdaSymbol{\{}\AgdaBound{N}\AgdaSymbol{\}}\AgdaSpace{}%
\AgdaSymbol{(}\AgdaDottedPattern{\AgdaSymbol{.(}}\AgdaDottedPattern{\AgdaBound{L}}\AgdaSpace{}%
\AgdaDottedPattern{\AgdaOperator{\AgdaInductiveConstructor{·}}}\AgdaSpace{}%
\AgdaDottedPattern{\AgdaBound{M}}\AgdaDottedPattern{\AgdaSymbol{)}}\AgdaSpace{}%
\AgdaOperator{\AgdaInductiveConstructor{⟶⟨}}\AgdaSpace{}%
\AgdaInductiveConstructor{ξ}\AgdaSpace{}%
\AgdaSymbol{(}\AgdaOperator{\AgdaInductiveConstructor{□·}}\AgdaSpace{}%
\AgdaBound{M₁}\AgdaSymbol{)}\AgdaSpace{}%
\AgdaBound{r1}\AgdaSpace{}%
\AgdaOperator{\AgdaInductiveConstructor{⟩}}\AgdaSpace{}%
\AgdaBound{r2}\AgdaSymbol{)}\AgdaSpace{}%
\AgdaBound{irredN}\<%
\\
\>[0][@{}l@{\AgdaIndent{0}}]%
\>[4]\AgdaKeyword{with}\AgdaSpace{}%
\AgdaFunction{app-inv-left}\AgdaSpace{}%
\AgdaBound{r2}\AgdaSpace{}%
\AgdaBound{irredN}\<%
\\
\>[0]\AgdaSymbol{...}\AgdaSpace{}%
\AgdaSymbol{|}%
\>[3168I]\AgdaInductiveConstructor{inj₁}\AgdaSpace{}%
\AgdaSymbol{(}\AgdaBound{L′}\AgdaSpace{}%
\AgdaOperator{\AgdaInductiveConstructor{,}}\AgdaSpace{}%
\AgdaBound{L→L′}\AgdaSpace{}%
\AgdaOperator{\AgdaInductiveConstructor{,}}\AgdaSpace{}%
\AgdaBound{IL′}\AgdaSpace{}%
\AgdaOperator{\AgdaInductiveConstructor{,}}\AgdaSpace{}%
\AgdaBound{r3}\AgdaSpace{}%
\AgdaOperator{\AgdaInductiveConstructor{,}}\AgdaSpace{}%
\AgdaBound{eq}\AgdaSymbol{)}\AgdaSpace{}%
\AgdaSymbol{=}\<%
\\
\>[.][@{}l@{}]\<[3168I]%
\>[6]\AgdaInductiveConstructor{inj₁}\AgdaSpace{}%
\AgdaSymbol{(}\AgdaBound{L′}\AgdaSpace{}%
\AgdaOperator{\AgdaInductiveConstructor{,}}\AgdaSpace{}%
\AgdaSymbol{(}\AgdaBound{L}\AgdaSpace{}%
\AgdaOperator{\AgdaInductiveConstructor{⟶⟨}}\AgdaSpace{}%
\AgdaBound{r1}\AgdaSpace{}%
\AgdaOperator{\AgdaInductiveConstructor{⟩}}\AgdaSpace{}%
\AgdaBound{L→L′}\AgdaSymbol{)}\AgdaSpace{}%
\AgdaOperator{\AgdaInductiveConstructor{,}}\AgdaSpace{}%
\AgdaBound{IL′}\AgdaSpace{}%
\AgdaOperator{\AgdaInductiveConstructor{,}}\AgdaSpace{}%
\AgdaBound{r3}\AgdaSpace{}%
\AgdaOperator{\AgdaInductiveConstructor{,}}\AgdaSpace{}%
\AgdaFunction{cong}\AgdaSpace{}%
\AgdaInductiveConstructor{suc}\AgdaSpace{}%
\AgdaBound{eq}\AgdaSymbol{)}\<%
\\
\>[0]\AgdaSymbol{...}\AgdaSpace{}%
\AgdaSymbol{|}\AgdaSpace{}%
\AgdaInductiveConstructor{inj₂}\AgdaSpace{}%
\AgdaInductiveConstructor{refl}\AgdaSpace{}%
\AgdaSymbol{=}\AgdaSpace{}%
\AgdaInductiveConstructor{inj₂}\AgdaSpace{}%
\AgdaInductiveConstructor{refl}\<%
\\
\>[0]\AgdaFunction{app-inv-left}\AgdaSpace{}%
\AgdaSymbol{\{}\AgdaBound{L}\AgdaSymbol{\}}\AgdaSpace{}%
\AgdaSymbol{\{}\AgdaBound{M}\AgdaSymbol{\}}\AgdaSpace{}%
\AgdaSymbol{\{}\AgdaBound{N}\AgdaSymbol{\}}\AgdaSpace{}%
\AgdaSymbol{(}\AgdaDottedPattern{\AgdaSymbol{.(}}\AgdaDottedPattern{\AgdaBound{L}}\AgdaSpace{}%
\AgdaDottedPattern{\AgdaOperator{\AgdaInductiveConstructor{·}}}\AgdaSpace{}%
\AgdaDottedPattern{\AgdaBound{M}}\AgdaDottedPattern{\AgdaSymbol{)}}\AgdaSpace{}%
\AgdaOperator{\AgdaInductiveConstructor{⟶⟨}}\AgdaSpace{}%
\AgdaInductiveConstructor{ξ}\AgdaSpace{}%
\AgdaSymbol{(}\AgdaBound{v}\AgdaSpace{}%
\AgdaOperator{\AgdaInductiveConstructor{·□}}\AgdaSymbol{)}\AgdaSpace{}%
\AgdaBound{r1}\AgdaSpace{}%
\AgdaOperator{\AgdaInductiveConstructor{⟩}}\AgdaSpace{}%
\AgdaBound{r2}\AgdaSymbol{)}\AgdaSpace{}%
\AgdaBound{irredN}\AgdaSpace{}%
\AgdaSymbol{=}\<%
\\
\>[0][@{}l@{\AgdaIndent{0}}]%
\>[4]\AgdaInductiveConstructor{inj₁}%
\>[3215I]\AgdaSymbol{(}\AgdaFunction{value}\AgdaSpace{}%
\AgdaBound{v}\AgdaSpace{}%
\AgdaOperator{\AgdaInductiveConstructor{,}}\AgdaSpace{}%
\AgdaSymbol{(\AgdaUnderscore{}}\AgdaSpace{}%
\AgdaOperator{\AgdaInductiveConstructor{END}}\AgdaSymbol{)}\AgdaSpace{}%
\AgdaOperator{\AgdaInductiveConstructor{,}}\AgdaSpace{}%
\AgdaFunction{value-irred}\AgdaSpace{}%
\AgdaBound{v}\AgdaSpace{}%
\AgdaOperator{\AgdaInductiveConstructor{,}}\<%
\\
\>[3215I][@{}l@{\AgdaIndent{0}}]%
\>[10]\AgdaSymbol{((}\AgdaFunction{value}\AgdaSpace{}%
\AgdaBound{v}\AgdaSpace{}%
\AgdaOperator{\AgdaInductiveConstructor{·}}\AgdaSpace{}%
\AgdaBound{M}\AgdaSymbol{)}\AgdaSpace{}%
\AgdaOperator{\AgdaInductiveConstructor{⟶⟨}}\AgdaSpace{}%
\AgdaInductiveConstructor{ξ}\AgdaSpace{}%
\AgdaSymbol{(}\AgdaBound{v}\AgdaSpace{}%
\AgdaOperator{\AgdaInductiveConstructor{·□}}\AgdaSymbol{)}\AgdaSpace{}%
\AgdaBound{r1}\AgdaSpace{}%
\AgdaOperator{\AgdaInductiveConstructor{⟩}}\AgdaSpace{}%
\AgdaBound{r2}\AgdaSymbol{)}\AgdaSpace{}%
\AgdaOperator{\AgdaInductiveConstructor{,}}\AgdaSpace{}%
\AgdaInductiveConstructor{refl}\AgdaSymbol{)}\<%
\\
\>[0]\AgdaFunction{app-inv-left}\AgdaSpace{}%
\AgdaSymbol{\{}\AgdaBound{L}\AgdaSymbol{\}}\AgdaSpace{}%
\AgdaSymbol{\{}\AgdaBound{M}\AgdaSymbol{\}}\AgdaSpace{}%
\AgdaSymbol{\{}\AgdaBound{N}\AgdaSymbol{\}}\AgdaSpace{}%
\AgdaSymbol{(}\AgdaDottedPattern{\AgdaSymbol{.(}}\AgdaDottedPattern{\AgdaBound{L}}\AgdaSpace{}%
\AgdaDottedPattern{\AgdaOperator{\AgdaInductiveConstructor{·}}}\AgdaSpace{}%
\AgdaDottedPattern{\AgdaBound{M}}\AgdaDottedPattern{\AgdaSymbol{)}}\AgdaSpace{}%
\AgdaOperator{\AgdaInductiveConstructor{⟶⟨}}\AgdaSpace{}%
\AgdaInductiveConstructor{ξ-blame}\AgdaSpace{}%
\AgdaSymbol{(}\AgdaOperator{\AgdaInductiveConstructor{□·}}\AgdaSpace{}%
\AgdaBound{M₁}\AgdaSymbol{)}\AgdaSpace{}%
\AgdaOperator{\AgdaInductiveConstructor{⟩}}\AgdaSpace{}%
\AgdaBound{r2}\AgdaSymbol{)}\AgdaSpace{}%
\AgdaBound{irredN}\<%
\\
\>[0][@{}l@{\AgdaIndent{0}}]%
\>[4]\AgdaKeyword{with}\AgdaSpace{}%
\AgdaFunction{blame↠}\AgdaSpace{}%
\AgdaBound{r2}\<%
\\
\>[0]\AgdaSymbol{...}\AgdaSpace{}%
\AgdaSymbol{|}\AgdaSpace{}%
\AgdaInductiveConstructor{refl}\AgdaSpace{}%
\AgdaSymbol{=}\AgdaSpace{}%
\AgdaInductiveConstructor{inj₂}\AgdaSpace{}%
\AgdaInductiveConstructor{refl}\<%
\\
\>[0]\AgdaFunction{app-inv-left}\AgdaSpace{}%
\AgdaSymbol{\{}\AgdaBound{L}\AgdaSymbol{\}}\AgdaSpace{}%
\AgdaSymbol{\{}\AgdaBound{M}\AgdaSymbol{\}}\AgdaSpace{}%
\AgdaSymbol{\{}\AgdaBound{N}\AgdaSymbol{\}}\AgdaSpace{}%
\AgdaSymbol{(}\AgdaDottedPattern{\AgdaSymbol{.(}}\AgdaDottedPattern{\AgdaBound{L}}\AgdaSpace{}%
\AgdaDottedPattern{\AgdaOperator{\AgdaInductiveConstructor{·}}}\AgdaSpace{}%
\AgdaDottedPattern{\AgdaBound{M}}\AgdaDottedPattern{\AgdaSymbol{)}}\AgdaSpace{}%
\AgdaOperator{\AgdaInductiveConstructor{⟶⟨}}\AgdaSpace{}%
\AgdaInductiveConstructor{ξ-blame}\AgdaSpace{}%
\AgdaSymbol{(}\AgdaBound{v}\AgdaSpace{}%
\AgdaOperator{\AgdaInductiveConstructor{·□}}\AgdaSymbol{)}\AgdaSpace{}%
\AgdaOperator{\AgdaInductiveConstructor{⟩}}\AgdaSpace{}%
\AgdaBound{r2}\AgdaSymbol{)}\AgdaSpace{}%
\AgdaBound{irredN}\<%
\\
\>[0][@{}l@{\AgdaIndent{0}}]%
\>[4]\AgdaKeyword{with}\AgdaSpace{}%
\AgdaFunction{blame↠}\AgdaSpace{}%
\AgdaBound{r2}\<%
\\
\>[0]\AgdaSymbol{...}\AgdaSpace{}%
\AgdaSymbol{|}\AgdaSpace{}%
\AgdaInductiveConstructor{refl}\AgdaSpace{}%
\AgdaSymbol{=}\AgdaSpace{}%
\AgdaInductiveConstructor{inj₂}\AgdaSpace{}%
\AgdaInductiveConstructor{refl}\<%
\\
\>[0]\AgdaFunction{app-inv-left}\AgdaSpace{}%
\AgdaSymbol{\{}\AgdaDottedPattern{\AgdaSymbol{.(}}\AgdaDottedPattern{\AgdaInductiveConstructor{ƛ}}\AgdaSpace{}%
\AgdaDottedPattern{\AgdaSymbol{\AgdaUnderscore{})}}\AgdaSymbol{\}}\AgdaSpace{}%
\AgdaSymbol{\{}\AgdaBound{M}\AgdaSymbol{\}}\AgdaSpace{}%
\AgdaSymbol{\{}\AgdaBound{N}\AgdaSymbol{\}}\AgdaSpace{}%
\AgdaSymbol{(}\AgdaDottedPattern{\AgdaSymbol{.(}}\AgdaDottedPattern{\AgdaInductiveConstructor{ƛ}}\AgdaSpace{}%
\AgdaDottedPattern{\AgdaSymbol{\AgdaUnderscore{}}}\AgdaSpace{}%
\AgdaDottedPattern{\AgdaOperator{\AgdaInductiveConstructor{·}}}\AgdaSpace{}%
\AgdaDottedPattern{\AgdaBound{M}}\AgdaDottedPattern{\AgdaSymbol{)}}\AgdaSpace{}%
\AgdaOperator{\AgdaInductiveConstructor{⟶⟨}}\AgdaSpace{}%
\AgdaInductiveConstructor{β}\AgdaSpace{}%
\AgdaBound{v}\AgdaSpace{}%
\AgdaOperator{\AgdaInductiveConstructor{⟩}}\AgdaSpace{}%
\AgdaBound{r2}\AgdaSymbol{)}\AgdaSpace{}%
\AgdaBound{irredN}\AgdaSpace{}%
\AgdaSymbol{=}\<%
\\
\>[0][@{}l@{\AgdaIndent{0}}]%
\>[4]\AgdaInductiveConstructor{inj₁}\AgdaSpace{}%
\AgdaSymbol{(\AgdaUnderscore{}}\AgdaSpace{}%
\AgdaOperator{\AgdaInductiveConstructor{,}}\AgdaSpace{}%
\AgdaSymbol{(\AgdaUnderscore{}}\AgdaSpace{}%
\AgdaOperator{\AgdaInductiveConstructor{END}}\AgdaSymbol{)}\AgdaSpace{}%
\AgdaOperator{\AgdaInductiveConstructor{,}}\AgdaSpace{}%
\AgdaFunction{value-irred}\AgdaSpace{}%
\AgdaSymbol{(}\AgdaOperator{\AgdaInductiveConstructor{ƛ̬}}\AgdaSpace{}%
\AgdaSymbol{\AgdaUnderscore{})}\AgdaSpace{}%
\AgdaOperator{\AgdaInductiveConstructor{,}}\AgdaSpace{}%
\AgdaSymbol{(\AgdaUnderscore{}}\AgdaSpace{}%
\AgdaOperator{\AgdaInductiveConstructor{⟶⟨}}\AgdaSpace{}%
\AgdaInductiveConstructor{β}\AgdaSpace{}%
\AgdaBound{v}\AgdaSpace{}%
\AgdaOperator{\AgdaInductiveConstructor{⟩}}\AgdaSpace{}%
\AgdaBound{r2}\AgdaSymbol{)}\AgdaSpace{}%
\AgdaOperator{\AgdaInductiveConstructor{,}}\AgdaSpace{}%
\AgdaInductiveConstructor{refl}\AgdaSymbol{)}\<%
\\
\\[\AgdaEmptyExtraSkip]%
\>[0]\AgdaFunction{app-inv-right}\AgdaSpace{}%
\AgdaSymbol{:}\AgdaSpace{}%
\AgdaSymbol{∀\{}\AgdaBound{V}\AgdaSpace{}%
\AgdaBound{M}\AgdaSpace{}%
\AgdaBound{N}\AgdaSymbol{\}}\<%
\\
\>[0][@{}l@{\AgdaIndent{0}}]%
\>[2]\AgdaSymbol{→}\AgdaSpace{}%
\AgdaSymbol{(}\AgdaBound{r1}\AgdaSpace{}%
\AgdaSymbol{:}\AgdaSpace{}%
\AgdaBound{V}\AgdaSpace{}%
\AgdaOperator{\AgdaInductiveConstructor{·}}\AgdaSpace{}%
\AgdaBound{M}\AgdaSpace{}%
\AgdaOperator{\AgdaDatatype{↠}}\AgdaSpace{}%
\AgdaBound{N}\AgdaSymbol{)}\<%
\\
\>[2]\AgdaSymbol{→}\AgdaSpace{}%
\AgdaDatatype{Value}\AgdaSpace{}%
\AgdaBound{V}\<%
\\
\>[2]\AgdaSymbol{→}\AgdaSpace{}%
\AgdaFunction{irred}\AgdaSpace{}%
\AgdaBound{N}\<%
\\
\>[2]\AgdaSymbol{→}%
\>[3323I]\AgdaSymbol{(}\AgdaFunction{∃[}\AgdaSpace{}%
\AgdaBound{M′}\AgdaSpace{}%
\AgdaFunction{]}\AgdaSpace{}%
\AgdaFunction{Σ[}\AgdaSpace{}%
\AgdaBound{r2}\AgdaSpace{}%
\AgdaFunction{∈}\AgdaSpace{}%
\AgdaSymbol{(}\AgdaBound{M}\AgdaSpace{}%
\AgdaOperator{\AgdaDatatype{↠}}\AgdaSpace{}%
\AgdaBound{M′}\AgdaSymbol{)}\AgdaSpace{}%
\AgdaFunction{]}\AgdaSpace{}%
\AgdaFunction{irred}\AgdaSpace{}%
\AgdaBound{M′}\<%
\\
\>[3323I][@{}l@{\AgdaIndent{0}}]%
\>[7]\AgdaOperator{\AgdaFunction{×}}\AgdaSpace{}%
\AgdaFunction{Σ[}\AgdaSpace{}%
\AgdaBound{r3}\AgdaSpace{}%
\AgdaFunction{∈}\AgdaSpace{}%
\AgdaSymbol{(}\AgdaBound{V}\AgdaSpace{}%
\AgdaOperator{\AgdaInductiveConstructor{·}}\AgdaSpace{}%
\AgdaBound{M′}\AgdaSpace{}%
\AgdaOperator{\AgdaDatatype{↠}}\AgdaSpace{}%
\AgdaBound{N}\AgdaSymbol{)}\AgdaSpace{}%
\AgdaFunction{]}\AgdaSpace{}%
\AgdaFunction{len}\AgdaSpace{}%
\AgdaBound{r1}\AgdaSpace{}%
\AgdaOperator{\AgdaDatatype{≡}}\AgdaSpace{}%
\AgdaFunction{len}\AgdaSpace{}%
\AgdaBound{r2}\AgdaSpace{}%
\AgdaOperator{\AgdaPrimitive{+}}\AgdaSpace{}%
\AgdaFunction{len}\AgdaSpace{}%
\AgdaBound{r3}\AgdaSymbol{)}\<%
\\
\>[.][@{}l@{}]\<[3323I]%
\>[4]\AgdaOperator{\AgdaDatatype{⊎}}\AgdaSpace{}%
\AgdaBound{N}\AgdaSpace{}%
\AgdaOperator{\AgdaDatatype{≡}}\AgdaSpace{}%
\AgdaInductiveConstructor{blame}\<%
\\
\>[0]\AgdaFunction{app-inv-right}\AgdaSpace{}%
\AgdaSymbol{\{}\AgdaBound{V}\AgdaSymbol{\}\{}\AgdaBound{M}\AgdaSymbol{\}\{}\AgdaBound{N}\AgdaSymbol{\}}\AgdaSpace{}%
\AgdaSymbol{(}\AgdaDottedPattern{\AgdaSymbol{.(\AgdaUnderscore{}}}\AgdaSpace{}%
\AgdaDottedPattern{\AgdaOperator{\AgdaInductiveConstructor{·}}}\AgdaSpace{}%
\AgdaDottedPattern{\AgdaSymbol{\AgdaUnderscore{})}}\AgdaSpace{}%
\AgdaOperator{\AgdaInductiveConstructor{END}}\AgdaSymbol{)}\AgdaSpace{}%
\AgdaBound{v}\AgdaSpace{}%
\AgdaBound{irredN}\AgdaSpace{}%
\AgdaSymbol{=}\<%
\\
\>[0][@{}l@{\AgdaIndent{0}}]%
\>[4]\AgdaInductiveConstructor{inj₁}\AgdaSpace{}%
\AgdaSymbol{(}\AgdaBound{M}\AgdaSpace{}%
\AgdaOperator{\AgdaInductiveConstructor{,}}\AgdaSpace{}%
\AgdaSymbol{(\AgdaUnderscore{}}\AgdaSpace{}%
\AgdaOperator{\AgdaInductiveConstructor{END}}\AgdaSymbol{)}\AgdaSpace{}%
\AgdaOperator{\AgdaInductiveConstructor{,}}\AgdaSpace{}%
\AgdaFunction{irredM}\AgdaSpace{}%
\AgdaOperator{\AgdaInductiveConstructor{,}}\AgdaSpace{}%
\AgdaSymbol{(\AgdaUnderscore{}}\AgdaSpace{}%
\AgdaOperator{\AgdaInductiveConstructor{END}}\AgdaSymbol{)}\AgdaSpace{}%
\AgdaOperator{\AgdaInductiveConstructor{,}}\AgdaSpace{}%
\AgdaInductiveConstructor{refl}\AgdaSymbol{)}\<%
\\
\>[4]\AgdaKeyword{where}%
\>[3374I]\AgdaFunction{irredM}\AgdaSpace{}%
\AgdaSymbol{:}\AgdaSpace{}%
\AgdaFunction{irred}\AgdaSpace{}%
\AgdaBound{M}\<%
\\
\>[.][@{}l@{}]\<[3374I]%
\>[10]\AgdaFunction{irredM}\AgdaSpace{}%
\AgdaSymbol{(}\AgdaBound{M′}\AgdaSpace{}%
\AgdaOperator{\AgdaInductiveConstructor{,}}\AgdaSpace{}%
\AgdaBound{M→M′}\AgdaSymbol{)}\AgdaSpace{}%
\AgdaSymbol{=}\AgdaSpace{}%
\AgdaBound{irredN}\AgdaSpace{}%
\AgdaSymbol{((}\AgdaBound{V}\AgdaSpace{}%
\AgdaOperator{\AgdaInductiveConstructor{·}}\AgdaSpace{}%
\AgdaBound{M′}\AgdaSymbol{)}\AgdaSpace{}%
\AgdaOperator{\AgdaInductiveConstructor{,}}\AgdaSpace{}%
\AgdaSymbol{(}\AgdaInductiveConstructor{ξ}\AgdaSpace{}%
\AgdaSymbol{(}\AgdaBound{v}\AgdaSpace{}%
\AgdaOperator{\AgdaInductiveConstructor{·□}}\AgdaSymbol{)}\AgdaSpace{}%
\AgdaBound{M→M′}\AgdaSymbol{))}\<%
\\
\>[0]\AgdaFunction{app-inv-right}\AgdaSpace{}%
\AgdaSymbol{\{}\AgdaBound{V}\AgdaSymbol{\}}\AgdaSpace{}%
\AgdaSymbol{\{}\AgdaBound{M}\AgdaSymbol{\}}\AgdaSpace{}%
\AgdaSymbol{\{}\AgdaBound{N}\AgdaSymbol{\}}\AgdaSpace{}%
\AgdaSymbol{(}\AgdaDottedPattern{\AgdaSymbol{.(}}\AgdaDottedPattern{\AgdaBound{V}}\AgdaSpace{}%
\AgdaDottedPattern{\AgdaOperator{\AgdaInductiveConstructor{·}}}\AgdaSpace{}%
\AgdaDottedPattern{\AgdaBound{M}}\AgdaDottedPattern{\AgdaSymbol{)}}\AgdaSpace{}%
\AgdaOperator{\AgdaInductiveConstructor{⟶⟨}}\AgdaSpace{}%
\AgdaInductiveConstructor{ξ}\AgdaSpace{}%
\AgdaSymbol{(}\AgdaOperator{\AgdaInductiveConstructor{□·}}\AgdaSpace{}%
\AgdaBound{M}\AgdaSymbol{)}\AgdaSpace{}%
\AgdaBound{r1}\AgdaSpace{}%
\AgdaOperator{\AgdaInductiveConstructor{⟩}}\AgdaSpace{}%
\AgdaBound{r2}\AgdaSymbol{)}\AgdaSpace{}%
\AgdaBound{v}\AgdaSpace{}%
\AgdaBound{irredN}\AgdaSpace{}%
\AgdaSymbol{=}\<%
\\
\>[0][@{}l@{\AgdaIndent{0}}]%
\>[4]\AgdaFunction{⊥-elim}\AgdaSpace{}%
\AgdaSymbol{(}\AgdaFunction{value-irreducible}\AgdaSpace{}%
\AgdaBound{v}\AgdaSpace{}%
\AgdaBound{r1}\AgdaSymbol{)}\<%
\\
\>[0]\AgdaFunction{app-inv-right}\AgdaSpace{}%
\AgdaSymbol{\{}\AgdaBound{V}\AgdaSymbol{\}}\AgdaSpace{}%
\AgdaSymbol{\{}\AgdaBound{M}\AgdaSymbol{\}}\AgdaSpace{}%
\AgdaSymbol{\{}\AgdaBound{N}\AgdaSymbol{\}}\AgdaSpace{}%
\AgdaSymbol{(}\AgdaDottedPattern{\AgdaSymbol{.(}}\AgdaDottedPattern{\AgdaBound{V}}\AgdaSpace{}%
\AgdaDottedPattern{\AgdaOperator{\AgdaInductiveConstructor{·}}}\AgdaSpace{}%
\AgdaDottedPattern{\AgdaBound{M}}\AgdaDottedPattern{\AgdaSymbol{)}}\AgdaSpace{}%
\AgdaOperator{\AgdaInductiveConstructor{⟶⟨}}\AgdaSpace{}%
\AgdaInductiveConstructor{ξ}\AgdaSpace{}%
\AgdaSymbol{(}\AgdaBound{v′}\AgdaSpace{}%
\AgdaOperator{\AgdaInductiveConstructor{·□}}\AgdaSymbol{)}\AgdaSpace{}%
\AgdaBound{r1}\AgdaSpace{}%
\AgdaOperator{\AgdaInductiveConstructor{⟩}}\AgdaSpace{}%
\AgdaBound{r2}\AgdaSymbol{)}\AgdaSpace{}%
\AgdaBound{v}\AgdaSpace{}%
\AgdaBound{irredN}\<%
\\
\>[0][@{}l@{\AgdaIndent{0}}]%
\>[4]\AgdaKeyword{with}\AgdaSpace{}%
\AgdaFunction{app-inv-right}\AgdaSpace{}%
\AgdaBound{r2}\AgdaSpace{}%
\AgdaBound{v′}\AgdaSpace{}%
\AgdaBound{irredN}\<%
\\
\>[0]\AgdaSymbol{...}\AgdaSpace{}%
\AgdaSymbol{|}%
\>[3430I]\AgdaInductiveConstructor{inj₁}\AgdaSpace{}%
\AgdaSymbol{(}\AgdaBound{M′}\AgdaSpace{}%
\AgdaOperator{\AgdaInductiveConstructor{,}}\AgdaSpace{}%
\AgdaBound{M→M′}\AgdaSpace{}%
\AgdaOperator{\AgdaInductiveConstructor{,}}\AgdaSpace{}%
\AgdaBound{iM′}\AgdaSpace{}%
\AgdaOperator{\AgdaInductiveConstructor{,}}\AgdaSpace{}%
\AgdaBound{→N}\AgdaSpace{}%
\AgdaOperator{\AgdaInductiveConstructor{,}}\AgdaSpace{}%
\AgdaBound{eq}\AgdaSymbol{)}\AgdaSpace{}%
\AgdaSymbol{=}\<%
\\
\>[.][@{}l@{}]\<[3430I]%
\>[6]\AgdaInductiveConstructor{inj₁}\AgdaSpace{}%
\AgdaSymbol{(}\AgdaBound{M′}\AgdaSpace{}%
\AgdaOperator{\AgdaInductiveConstructor{,}}\AgdaSpace{}%
\AgdaSymbol{(}\AgdaBound{M}\AgdaSpace{}%
\AgdaOperator{\AgdaInductiveConstructor{⟶⟨}}\AgdaSpace{}%
\AgdaBound{r1}\AgdaSpace{}%
\AgdaOperator{\AgdaInductiveConstructor{⟩}}\AgdaSpace{}%
\AgdaBound{M→M′}\AgdaSymbol{)}\AgdaSpace{}%
\AgdaOperator{\AgdaInductiveConstructor{,}}\AgdaSpace{}%
\AgdaBound{iM′}\AgdaSpace{}%
\AgdaOperator{\AgdaInductiveConstructor{,}}\AgdaSpace{}%
\AgdaBound{→N}\AgdaSpace{}%
\AgdaOperator{\AgdaInductiveConstructor{,}}\AgdaSpace{}%
\AgdaFunction{cong}\AgdaSpace{}%
\AgdaInductiveConstructor{suc}\AgdaSpace{}%
\AgdaBound{eq}\AgdaSymbol{)}\<%
\\
\>[0]\AgdaSymbol{...}\AgdaSpace{}%
\AgdaSymbol{|}\AgdaSpace{}%
\AgdaInductiveConstructor{inj₂}\AgdaSpace{}%
\AgdaInductiveConstructor{refl}\AgdaSpace{}%
\AgdaSymbol{=}\AgdaSpace{}%
\AgdaInductiveConstructor{inj₂}\AgdaSpace{}%
\AgdaInductiveConstructor{refl}\<%
\\
\>[0]\AgdaFunction{app-inv-right}\AgdaSpace{}%
\AgdaSymbol{\{}\AgdaDottedPattern{\AgdaSymbol{.}}\AgdaDottedPattern{\AgdaInductiveConstructor{blame}}\AgdaSymbol{\}}\AgdaSpace{}%
\AgdaSymbol{\{}\AgdaBound{M}\AgdaSymbol{\}}\AgdaSpace{}%
\AgdaSymbol{\{}\AgdaBound{N}\AgdaSymbol{\}}\AgdaSpace{}%
\AgdaSymbol{(}\AgdaDottedPattern{\AgdaSymbol{.(}}\AgdaDottedPattern{\AgdaInductiveConstructor{blame}}\AgdaSpace{}%
\AgdaDottedPattern{\AgdaOperator{\AgdaInductiveConstructor{·}}}\AgdaSpace{}%
\AgdaDottedPattern{\AgdaBound{M}}\AgdaDottedPattern{\AgdaSymbol{)}}\AgdaSpace{}%
\AgdaOperator{\AgdaInductiveConstructor{⟶⟨}}\AgdaSpace{}%
\AgdaInductiveConstructor{ξ-blame}\AgdaSpace{}%
\AgdaSymbol{(}\AgdaOperator{\AgdaInductiveConstructor{□·}}\AgdaSpace{}%
\AgdaBound{M}\AgdaSymbol{)}\AgdaSpace{}%
\AgdaOperator{\AgdaInductiveConstructor{⟩}}\AgdaSpace{}%
\AgdaBound{r2}\AgdaSymbol{)}\AgdaSpace{}%
\AgdaSymbol{()}\AgdaSpace{}%
\AgdaBound{irredN}\<%
\\
\>[0]\AgdaFunction{app-inv-right}\AgdaSpace{}%
\AgdaSymbol{\{}\AgdaBound{V}\AgdaSymbol{\}}\AgdaSpace{}%
\AgdaSymbol{\{}\AgdaBound{M}\AgdaSymbol{\}}\AgdaSpace{}%
\AgdaSymbol{\{}\AgdaBound{N}\AgdaSymbol{\}}\AgdaSpace{}%
\AgdaSymbol{(}\AgdaDottedPattern{\AgdaSymbol{.(}}\AgdaDottedPattern{\AgdaBound{V}}\AgdaSpace{}%
\AgdaDottedPattern{\AgdaOperator{\AgdaInductiveConstructor{·}}}\AgdaSpace{}%
\AgdaDottedPattern{\AgdaBound{M}}\AgdaDottedPattern{\AgdaSymbol{)}}\AgdaSpace{}%
\AgdaOperator{\AgdaInductiveConstructor{⟶⟨}}\AgdaSpace{}%
\AgdaInductiveConstructor{ξ-blame}\AgdaSpace{}%
\AgdaSymbol{(}\AgdaBound{v₁}\AgdaSpace{}%
\AgdaOperator{\AgdaInductiveConstructor{·□}}\AgdaSymbol{)}\AgdaSpace{}%
\AgdaOperator{\AgdaInductiveConstructor{⟩}}\AgdaSpace{}%
\AgdaBound{r2}\AgdaSymbol{)}\AgdaSpace{}%
\AgdaBound{v}\AgdaSpace{}%
\AgdaBound{irredN}\<%
\\
\>[0][@{}l@{\AgdaIndent{0}}]%
\>[4]\AgdaKeyword{with}\AgdaSpace{}%
\AgdaFunction{blame↠}\AgdaSpace{}%
\AgdaBound{r2}\<%
\\
\>[0]\AgdaSymbol{...}\AgdaSpace{}%
\AgdaSymbol{|}\AgdaSpace{}%
\AgdaInductiveConstructor{refl}\AgdaSpace{}%
\AgdaSymbol{=}\AgdaSpace{}%
\AgdaInductiveConstructor{inj₂}\AgdaSpace{}%
\AgdaInductiveConstructor{refl}\<%
\\
\>[0]\AgdaFunction{app-inv-right}\AgdaSpace{}%
\AgdaSymbol{\{}\AgdaDottedPattern{\AgdaSymbol{.(}}\AgdaDottedPattern{\AgdaInductiveConstructor{ƛ}}\AgdaSpace{}%
\AgdaDottedPattern{\AgdaSymbol{\AgdaUnderscore{})}}\AgdaSymbol{\}}\AgdaSpace{}%
\AgdaSymbol{\{}\AgdaBound{M}\AgdaSymbol{\}}\AgdaSpace{}%
\AgdaSymbol{\{}\AgdaBound{N}\AgdaSymbol{\}}\AgdaSpace{}%
\AgdaSymbol{(}\AgdaDottedPattern{\AgdaSymbol{.(}}\AgdaDottedPattern{\AgdaInductiveConstructor{ƛ}}\AgdaSpace{}%
\AgdaDottedPattern{\AgdaSymbol{\AgdaUnderscore{}}}\AgdaSpace{}%
\AgdaDottedPattern{\AgdaOperator{\AgdaInductiveConstructor{·}}}\AgdaSpace{}%
\AgdaDottedPattern{\AgdaBound{M}}\AgdaDottedPattern{\AgdaSymbol{)}}\AgdaSpace{}%
\AgdaOperator{\AgdaInductiveConstructor{⟶⟨}}\AgdaSpace{}%
\AgdaInductiveConstructor{β}\AgdaSpace{}%
\AgdaBound{w}\AgdaSpace{}%
\AgdaOperator{\AgdaInductiveConstructor{⟩}}\AgdaSpace{}%
\AgdaBound{r2}\AgdaSymbol{)}\AgdaSpace{}%
\AgdaBound{v}\AgdaSpace{}%
\AgdaBound{irredN}\AgdaSpace{}%
\AgdaSymbol{=}\<%
\\
\>[0][@{}l@{\AgdaIndent{0}}]%
\>[4]\AgdaInductiveConstructor{inj₁}\AgdaSpace{}%
\AgdaSymbol{(}\AgdaBound{M}\AgdaSpace{}%
\AgdaOperator{\AgdaInductiveConstructor{,}}\AgdaSpace{}%
\AgdaSymbol{(\AgdaUnderscore{}}\AgdaSpace{}%
\AgdaOperator{\AgdaInductiveConstructor{END}}\AgdaSymbol{)}\AgdaSpace{}%
\AgdaOperator{\AgdaInductiveConstructor{,}}\AgdaSpace{}%
\AgdaFunction{value-irred}\AgdaSpace{}%
\AgdaBound{w}\AgdaSpace{}%
\AgdaOperator{\AgdaInductiveConstructor{,}}\AgdaSpace{}%
\AgdaSymbol{(\AgdaUnderscore{}}\AgdaSpace{}%
\AgdaOperator{\AgdaInductiveConstructor{⟶⟨}}\AgdaSpace{}%
\AgdaInductiveConstructor{β}\AgdaSpace{}%
\AgdaBound{w}\AgdaSpace{}%
\AgdaOperator{\AgdaInductiveConstructor{⟩}}\AgdaSpace{}%
\AgdaBound{r2}\AgdaSymbol{)}\AgdaSpace{}%
\AgdaOperator{\AgdaInductiveConstructor{,}}\AgdaSpace{}%
\AgdaInductiveConstructor{refl}\AgdaSymbol{)}\<%
\\
\\[\AgdaEmptyExtraSkip]%
\>[0]\AgdaFunction{frame-inv}\AgdaSpace{}%
\AgdaSymbol{:}\AgdaSpace{}%
\AgdaSymbol{∀\{}\AgdaBound{F}\AgdaSpace{}%
\AgdaBound{M}\AgdaSpace{}%
\AgdaBound{N}\AgdaSymbol{\}}\<%
\\
\>[0][@{}l@{\AgdaIndent{0}}]%
\>[2]\AgdaSymbol{→}\AgdaSpace{}%
\AgdaSymbol{(}\AgdaBound{r1}\AgdaSpace{}%
\AgdaSymbol{:}\AgdaSpace{}%
\AgdaBound{F}\AgdaSpace{}%
\AgdaOperator{\AgdaFunction{⟦}}\AgdaSpace{}%
\AgdaBound{M}\AgdaSpace{}%
\AgdaOperator{\AgdaFunction{⟧}}\AgdaSpace{}%
\AgdaOperator{\AgdaDatatype{↠}}\AgdaSpace{}%
\AgdaBound{N}\AgdaSymbol{)}\<%
\\
\>[2]\AgdaSymbol{→}\AgdaSpace{}%
\AgdaFunction{irred}\AgdaSpace{}%
\AgdaBound{N}\<%
\\
\>[2]\AgdaSymbol{→}%
\>[3543I]\AgdaSymbol{(}\AgdaFunction{∃[}%
\>[3544I]\AgdaBound{M′}\AgdaSpace{}%
\AgdaFunction{]}\AgdaSpace{}%
\AgdaFunction{Σ[}\AgdaSpace{}%
\AgdaBound{r2}\AgdaSpace{}%
\AgdaFunction{∈}\AgdaSpace{}%
\AgdaSymbol{(}\AgdaBound{M}\AgdaSpace{}%
\AgdaOperator{\AgdaDatatype{↠}}\AgdaSpace{}%
\AgdaBound{M′}\AgdaSymbol{)}\AgdaSpace{}%
\AgdaFunction{]}\AgdaSpace{}%
\AgdaFunction{irred}\AgdaSpace{}%
\AgdaBound{M′}\<%
\\
\>[.][@{}l@{}]\<[3544I]%
\>[8]\AgdaOperator{\AgdaFunction{×}}\AgdaSpace{}%
\AgdaFunction{Σ[}\AgdaSpace{}%
\AgdaBound{r3}\AgdaSpace{}%
\AgdaFunction{∈}\AgdaSpace{}%
\AgdaSymbol{(}\AgdaBound{F}\AgdaSpace{}%
\AgdaOperator{\AgdaFunction{⟦}}\AgdaSpace{}%
\AgdaBound{M′}\AgdaSpace{}%
\AgdaOperator{\AgdaFunction{⟧}}\AgdaSpace{}%
\AgdaOperator{\AgdaDatatype{↠}}\AgdaSpace{}%
\AgdaBound{N}\AgdaSymbol{)}\AgdaSpace{}%
\AgdaFunction{]}\AgdaSpace{}%
\AgdaFunction{len}\AgdaSpace{}%
\AgdaBound{r1}\AgdaSpace{}%
\AgdaOperator{\AgdaDatatype{≡}}\AgdaSpace{}%
\AgdaFunction{len}\AgdaSpace{}%
\AgdaBound{r2}\AgdaSpace{}%
\AgdaOperator{\AgdaPrimitive{+}}\AgdaSpace{}%
\AgdaFunction{len}\AgdaSpace{}%
\AgdaBound{r3}\AgdaSymbol{)}\<%
\\
\>[.][@{}l@{}]\<[3543I]%
\>[4]\AgdaOperator{\AgdaDatatype{⊎}}\AgdaSpace{}%
\AgdaBound{N}\AgdaSpace{}%
\AgdaOperator{\AgdaDatatype{≡}}\AgdaSpace{}%
\AgdaInductiveConstructor{blame}\<%
\\
\>[0]\AgdaFunction{frame-inv}\AgdaSpace{}%
\AgdaSymbol{\{}\AgdaOperator{\AgdaInductiveConstructor{□·}}\AgdaSpace{}%
\AgdaBound{M}\AgdaSymbol{\}}\AgdaSpace{}%
\AgdaSymbol{\{}\AgdaBound{L}\AgdaSymbol{\}}\AgdaSpace{}%
\AgdaSymbol{\{}\AgdaBound{N}\AgdaSymbol{\}}\AgdaSpace{}%
\AgdaBound{r1}\AgdaSpace{}%
\AgdaBound{irN}\AgdaSpace{}%
\AgdaSymbol{=}\AgdaSpace{}%
\AgdaFunction{app-inv-left}\AgdaSpace{}%
\AgdaBound{r1}\AgdaSpace{}%
\AgdaBound{irN}\<%
\\
\>[0]\AgdaFunction{frame-inv}\AgdaSpace{}%
\AgdaSymbol{\{}\AgdaBound{v}\AgdaSpace{}%
\AgdaOperator{\AgdaInductiveConstructor{·□}}\AgdaSymbol{\}}\AgdaSpace{}%
\AgdaSymbol{\{}\AgdaBound{M}\AgdaSymbol{\}}\AgdaSpace{}%
\AgdaSymbol{\{}\AgdaBound{N}\AgdaSymbol{\}}\AgdaSpace{}%
\AgdaBound{r1}\AgdaSpace{}%
\AgdaBound{irN}\AgdaSpace{}%
\AgdaSymbol{=}\AgdaSpace{}%
\AgdaFunction{app-inv-right}\AgdaSpace{}%
\AgdaBound{r1}\AgdaSpace{}%
\AgdaBound{v}\AgdaSpace{}%
\AgdaBound{irN}\<%
\\
\>[0]\AgdaFunction{frame-inv}\AgdaSpace{}%
\AgdaSymbol{\{}\AgdaOperator{\AgdaInductiveConstructor{□⟨}}\AgdaSpace{}%
\AgdaBound{G}\AgdaSpace{}%
\AgdaOperator{\AgdaInductiveConstructor{!⟩}}\AgdaSymbol{\}}\AgdaSpace{}%
\AgdaSymbol{\{}\AgdaBound{M}\AgdaSymbol{\}}\AgdaSpace{}%
\AgdaSymbol{(\AgdaUnderscore{}}\AgdaSpace{}%
\AgdaOperator{\AgdaInductiveConstructor{END}}\AgdaSymbol{)}\AgdaSpace{}%
\AgdaBound{irN}\AgdaSpace{}%
\AgdaSymbol{=}\AgdaSpace{}%
\AgdaInductiveConstructor{inj₁}\AgdaSpace{}%
\AgdaSymbol{(\AgdaUnderscore{}}\AgdaSpace{}%
\AgdaOperator{\AgdaInductiveConstructor{,}}\AgdaSpace{}%
\AgdaSymbol{(\AgdaUnderscore{}}\AgdaSpace{}%
\AgdaOperator{\AgdaInductiveConstructor{END}}\AgdaSymbol{)}\AgdaSpace{}%
\AgdaOperator{\AgdaInductiveConstructor{,}}\AgdaSpace{}%
\AgdaFunction{irM}\AgdaSpace{}%
\AgdaOperator{\AgdaInductiveConstructor{,}}\AgdaSpace{}%
\AgdaSymbol{(\AgdaUnderscore{}}\AgdaSpace{}%
\AgdaOperator{\AgdaInductiveConstructor{END}}\AgdaSymbol{)}\AgdaSpace{}%
\AgdaOperator{\AgdaInductiveConstructor{,}}\AgdaSpace{}%
\AgdaInductiveConstructor{refl}\AgdaSymbol{)}\<%
\\
\>[0][@{}l@{\AgdaIndent{0}}]%
\>[4]\AgdaKeyword{where}%
\>[3617I]\AgdaFunction{irM}\AgdaSpace{}%
\AgdaSymbol{:}\AgdaSpace{}%
\AgdaFunction{irred}\AgdaSpace{}%
\AgdaBound{M}\<%
\\
\>[.][@{}l@{}]\<[3617I]%
\>[10]\AgdaFunction{irM}\AgdaSpace{}%
\AgdaSymbol{(}\AgdaBound{M′}\AgdaSpace{}%
\AgdaOperator{\AgdaInductiveConstructor{,}}\AgdaSpace{}%
\AgdaBound{M→M′}\AgdaSymbol{)}\AgdaSpace{}%
\AgdaSymbol{=}\AgdaSpace{}%
\AgdaBound{irN}\AgdaSpace{}%
\AgdaSymbol{(\AgdaUnderscore{}}\AgdaSpace{}%
\AgdaOperator{\AgdaInductiveConstructor{,}}\AgdaSpace{}%
\AgdaSymbol{(}\AgdaInductiveConstructor{ξ}\AgdaSpace{}%
\AgdaOperator{\AgdaInductiveConstructor{□⟨}}\AgdaSpace{}%
\AgdaBound{G}\AgdaSpace{}%
\AgdaOperator{\AgdaInductiveConstructor{!⟩}}\AgdaSpace{}%
\AgdaBound{M→M′}\AgdaSymbol{))}\<%
\\
\>[0]\AgdaFunction{frame-inv}\AgdaSpace{}%
\AgdaSymbol{\{}\AgdaOperator{\AgdaInductiveConstructor{□⟨}}\AgdaSpace{}%
\AgdaBound{G}\AgdaSpace{}%
\AgdaOperator{\AgdaInductiveConstructor{!⟩}}\AgdaSymbol{\}}\AgdaSpace{}%
\AgdaSymbol{\{}\AgdaBound{M}\AgdaSymbol{\}}\AgdaSpace{}%
\AgdaSymbol{\{}\AgdaBound{N}\AgdaSymbol{\}}\AgdaSpace{}%
\AgdaSymbol{(}\AgdaDottedPattern{\AgdaSymbol{.(}}\AgdaDottedPattern{\AgdaOperator{\AgdaInductiveConstructor{□⟨}}}\AgdaSpace{}%
\AgdaDottedPattern{\AgdaBound{G}}\AgdaSpace{}%
\AgdaDottedPattern{\AgdaOperator{\AgdaInductiveConstructor{!⟩}}}\AgdaSpace{}%
\AgdaDottedPattern{\AgdaOperator{\AgdaFunction{⟦}}}\AgdaSpace{}%
\AgdaDottedPattern{\AgdaBound{M}}\AgdaSpace{}%
\AgdaDottedPattern{\AgdaOperator{\AgdaFunction{⟧}}}\AgdaDottedPattern{\AgdaSymbol{)}}\AgdaSpace{}%
\AgdaOperator{\AgdaInductiveConstructor{⟶⟨}}\AgdaSpace{}%
\AgdaInductiveConstructor{ξ}\AgdaSpace{}%
\AgdaOperator{\AgdaInductiveConstructor{□⟨}}\AgdaSpace{}%
\AgdaSymbol{\AgdaUnderscore{}}\AgdaSpace{}%
\AgdaOperator{\AgdaInductiveConstructor{!⟩}}\AgdaSpace{}%
\AgdaBound{r1}\AgdaSpace{}%
\AgdaOperator{\AgdaInductiveConstructor{⟩}}\AgdaSpace{}%
\AgdaBound{r2}\AgdaSymbol{)}\AgdaSpace{}%
\AgdaBound{irN}\<%
\\
\>[0][@{}l@{\AgdaIndent{0}}]%
\>[4]\AgdaKeyword{with}\AgdaSpace{}%
\AgdaFunction{frame-inv}\AgdaSymbol{\{}\AgdaOperator{\AgdaInductiveConstructor{□⟨}}\AgdaSpace{}%
\AgdaBound{G}\AgdaSpace{}%
\AgdaOperator{\AgdaInductiveConstructor{!⟩}}\AgdaSymbol{\}}\AgdaSpace{}%
\AgdaBound{r2}\AgdaSpace{}%
\AgdaBound{irN}\<%
\\
\>[0]\AgdaSymbol{...}\AgdaSpace{}%
\AgdaSymbol{|}\AgdaSpace{}%
\AgdaInductiveConstructor{inj₁}\AgdaSpace{}%
\AgdaSymbol{(}\AgdaBound{M′}\AgdaSpace{}%
\AgdaOperator{\AgdaInductiveConstructor{,}}\AgdaSpace{}%
\AgdaBound{r3}\AgdaSpace{}%
\AgdaOperator{\AgdaInductiveConstructor{,}}\AgdaSpace{}%
\AgdaBound{irM′}\AgdaSpace{}%
\AgdaOperator{\AgdaInductiveConstructor{,}}\AgdaSpace{}%
\AgdaBound{r4}\AgdaSpace{}%
\AgdaOperator{\AgdaInductiveConstructor{,}}\AgdaSpace{}%
\AgdaBound{eq}\AgdaSymbol{)}\AgdaSpace{}%
\AgdaSymbol{=}\AgdaSpace{}%
\AgdaInductiveConstructor{inj₁}\AgdaSpace{}%
\AgdaSymbol{(\AgdaUnderscore{}}\AgdaSpace{}%
\AgdaOperator{\AgdaInductiveConstructor{,}}\AgdaSpace{}%
\AgdaSymbol{(\AgdaUnderscore{}}\AgdaSpace{}%
\AgdaOperator{\AgdaInductiveConstructor{⟶⟨}}\AgdaSpace{}%
\AgdaBound{r1}\AgdaSpace{}%
\AgdaOperator{\AgdaInductiveConstructor{⟩}}\AgdaSpace{}%
\AgdaBound{r3}\AgdaSymbol{)}\AgdaSpace{}%
\AgdaOperator{\AgdaInductiveConstructor{,}}\AgdaSpace{}%
\AgdaBound{irM′}\AgdaSpace{}%
\AgdaOperator{\AgdaInductiveConstructor{,}}\AgdaSpace{}%
\AgdaBound{r4}\AgdaSpace{}%
\AgdaOperator{\AgdaInductiveConstructor{,}}\AgdaSpace{}%
\AgdaFunction{cong}\AgdaSpace{}%
\AgdaInductiveConstructor{suc}\AgdaSpace{}%
\AgdaBound{eq}\AgdaSymbol{)}\<%
\\
\>[0]\AgdaSymbol{...}\AgdaSpace{}%
\AgdaSymbol{|}\AgdaSpace{}%
\AgdaInductiveConstructor{inj₂}\AgdaSpace{}%
\AgdaInductiveConstructor{refl}\AgdaSpace{}%
\AgdaSymbol{=}\AgdaSpace{}%
\AgdaInductiveConstructor{inj₂}\AgdaSpace{}%
\AgdaInductiveConstructor{refl}\<%
\\
\>[0]\AgdaFunction{frame-inv}\AgdaSpace{}%
\AgdaSymbol{\{}\AgdaOperator{\AgdaInductiveConstructor{□⟨}}\AgdaSpace{}%
\AgdaBound{G}\AgdaSpace{}%
\AgdaOperator{\AgdaInductiveConstructor{!⟩}}\AgdaSymbol{\}}\AgdaSpace{}%
\AgdaSymbol{\{}\AgdaBound{M}\AgdaSymbol{\}}\AgdaSpace{}%
\AgdaSymbol{\{}\AgdaBound{N}\AgdaSymbol{\}}\AgdaSpace{}%
\AgdaSymbol{(}\AgdaDottedPattern{\AgdaSymbol{.(}}\AgdaDottedPattern{\AgdaOperator{\AgdaInductiveConstructor{□⟨}}}\AgdaSpace{}%
\AgdaDottedPattern{\AgdaBound{G}}\AgdaSpace{}%
\AgdaDottedPattern{\AgdaOperator{\AgdaInductiveConstructor{!⟩}}}\AgdaSpace{}%
\AgdaDottedPattern{\AgdaOperator{\AgdaFunction{⟦}}}\AgdaSpace{}%
\AgdaDottedPattern{\AgdaBound{M}}\AgdaSpace{}%
\AgdaDottedPattern{\AgdaOperator{\AgdaFunction{⟧}}}\AgdaDottedPattern{\AgdaSymbol{)}}\AgdaSpace{}%
\AgdaOperator{\AgdaInductiveConstructor{⟶⟨}}\AgdaSpace{}%
\AgdaInductiveConstructor{ξ-blame}\AgdaSpace{}%
\AgdaOperator{\AgdaInductiveConstructor{□⟨}}\AgdaSpace{}%
\AgdaSymbol{\AgdaUnderscore{}}\AgdaSpace{}%
\AgdaOperator{\AgdaInductiveConstructor{!⟩}}\AgdaSpace{}%
\AgdaOperator{\AgdaInductiveConstructor{⟩}}\AgdaSpace{}%
\AgdaBound{r2}\AgdaSymbol{)}\AgdaSpace{}%
\AgdaBound{irN}\<%
\\
\>[0][@{}l@{\AgdaIndent{0}}]%
\>[4]\AgdaKeyword{with}\AgdaSpace{}%
\AgdaFunction{blame↠}\AgdaSpace{}%
\AgdaBound{r2}\<%
\\
\>[0]\AgdaSymbol{...}\AgdaSpace{}%
\AgdaSymbol{|}\AgdaSpace{}%
\AgdaInductiveConstructor{refl}\AgdaSpace{}%
\AgdaSymbol{=}\AgdaSpace{}%
\AgdaInductiveConstructor{inj₂}\AgdaSpace{}%
\AgdaInductiveConstructor{refl}\<%
\\
\>[0]\AgdaFunction{frame-inv}\AgdaSpace{}%
\AgdaSymbol{\{}\AgdaOperator{\AgdaInductiveConstructor{□⟨}}\AgdaSpace{}%
\AgdaBound{H}\AgdaSpace{}%
\AgdaOperator{\AgdaInductiveConstructor{?⟩}}\AgdaSymbol{\}}\AgdaSpace{}%
\AgdaSymbol{\{}\AgdaBound{M}\AgdaSymbol{\}}\AgdaSpace{}%
\AgdaSymbol{(\AgdaUnderscore{}}\AgdaSpace{}%
\AgdaOperator{\AgdaInductiveConstructor{END}}\AgdaSymbol{)}\AgdaSpace{}%
\AgdaBound{irN}\AgdaSpace{}%
\AgdaSymbol{=}\AgdaSpace{}%
\AgdaInductiveConstructor{inj₁}\AgdaSpace{}%
\AgdaSymbol{(\AgdaUnderscore{}}\AgdaSpace{}%
\AgdaOperator{\AgdaInductiveConstructor{,}}\AgdaSpace{}%
\AgdaSymbol{(\AgdaUnderscore{}}\AgdaSpace{}%
\AgdaOperator{\AgdaInductiveConstructor{END}}\AgdaSymbol{)}\AgdaSpace{}%
\AgdaOperator{\AgdaInductiveConstructor{,}}\AgdaSpace{}%
\AgdaFunction{irM}\AgdaSpace{}%
\AgdaOperator{\AgdaInductiveConstructor{,}}\AgdaSpace{}%
\AgdaSymbol{(\AgdaUnderscore{}}\AgdaSpace{}%
\AgdaOperator{\AgdaInductiveConstructor{END}}\AgdaSymbol{)}\AgdaSpace{}%
\AgdaOperator{\AgdaInductiveConstructor{,}}\AgdaSpace{}%
\AgdaInductiveConstructor{refl}\AgdaSymbol{)}\<%
\\
\>[0][@{}l@{\AgdaIndent{0}}]%
\>[4]\AgdaKeyword{where}%
\>[3738I]\AgdaFunction{irM}\AgdaSpace{}%
\AgdaSymbol{:}\AgdaSpace{}%
\AgdaFunction{irred}\AgdaSpace{}%
\AgdaBound{M}\<%
\\
\>[.][@{}l@{}]\<[3738I]%
\>[10]\AgdaFunction{irM}\AgdaSpace{}%
\AgdaSymbol{(}\AgdaBound{M′}\AgdaSpace{}%
\AgdaOperator{\AgdaInductiveConstructor{,}}\AgdaSpace{}%
\AgdaBound{M→M′}\AgdaSymbol{)}\AgdaSpace{}%
\AgdaSymbol{=}\AgdaSpace{}%
\AgdaBound{irN}\AgdaSpace{}%
\AgdaSymbol{(\AgdaUnderscore{}}\AgdaSpace{}%
\AgdaOperator{\AgdaInductiveConstructor{,}}\AgdaSpace{}%
\AgdaSymbol{(}\AgdaInductiveConstructor{ξ}\AgdaSpace{}%
\AgdaOperator{\AgdaInductiveConstructor{□⟨}}\AgdaSpace{}%
\AgdaBound{H}\AgdaSpace{}%
\AgdaOperator{\AgdaInductiveConstructor{?⟩}}\AgdaSpace{}%
\AgdaBound{M→M′}\AgdaSymbol{))}\<%
\\
\>[0]\AgdaFunction{frame-inv}\AgdaSpace{}%
\AgdaSymbol{\{}\AgdaOperator{\AgdaInductiveConstructor{□⟨}}\AgdaSpace{}%
\AgdaBound{H}\AgdaSpace{}%
\AgdaOperator{\AgdaInductiveConstructor{?⟩}}\AgdaSymbol{\}}\AgdaSpace{}%
\AgdaSymbol{\{}\AgdaBound{M}\AgdaSymbol{\}}\AgdaSpace{}%
\AgdaSymbol{\{}\AgdaBound{N}\AgdaSymbol{\}}\AgdaSpace{}%
\AgdaSymbol{(}\AgdaDottedPattern{\AgdaSymbol{.(}}\AgdaDottedPattern{\AgdaOperator{\AgdaInductiveConstructor{□⟨}}}\AgdaSpace{}%
\AgdaDottedPattern{\AgdaBound{H}}\AgdaSpace{}%
\AgdaDottedPattern{\AgdaOperator{\AgdaInductiveConstructor{?⟩}}}\AgdaSpace{}%
\AgdaDottedPattern{\AgdaOperator{\AgdaFunction{⟦}}}\AgdaSpace{}%
\AgdaDottedPattern{\AgdaBound{M}}\AgdaSpace{}%
\AgdaDottedPattern{\AgdaOperator{\AgdaFunction{⟧}}}\AgdaDottedPattern{\AgdaSymbol{)}}\AgdaSpace{}%
\AgdaOperator{\AgdaInductiveConstructor{⟶⟨}}\AgdaSpace{}%
\AgdaInductiveConstructor{ξ}\AgdaSpace{}%
\AgdaOperator{\AgdaInductiveConstructor{□⟨}}\AgdaSpace{}%
\AgdaSymbol{\AgdaUnderscore{}}\AgdaSpace{}%
\AgdaOperator{\AgdaInductiveConstructor{?⟩}}\AgdaSpace{}%
\AgdaBound{r1}\AgdaSpace{}%
\AgdaOperator{\AgdaInductiveConstructor{⟩}}\AgdaSpace{}%
\AgdaBound{r2}\AgdaSymbol{)}\AgdaSpace{}%
\AgdaBound{irN}\<%
\\
\>[0][@{}l@{\AgdaIndent{0}}]%
\>[4]\AgdaKeyword{with}\AgdaSpace{}%
\AgdaFunction{frame-inv}\AgdaSymbol{\{}\AgdaOperator{\AgdaInductiveConstructor{□⟨}}\AgdaSpace{}%
\AgdaBound{H}\AgdaSpace{}%
\AgdaOperator{\AgdaInductiveConstructor{?⟩}}\AgdaSymbol{\}}\AgdaSpace{}%
\AgdaBound{r2}\AgdaSpace{}%
\AgdaBound{irN}\<%
\\
\>[0]\AgdaSymbol{...}\AgdaSpace{}%
\AgdaSymbol{|}\AgdaSpace{}%
\AgdaInductiveConstructor{inj₁}\AgdaSpace{}%
\AgdaSymbol{(}\AgdaBound{M′}\AgdaSpace{}%
\AgdaOperator{\AgdaInductiveConstructor{,}}\AgdaSpace{}%
\AgdaBound{r3}\AgdaSpace{}%
\AgdaOperator{\AgdaInductiveConstructor{,}}\AgdaSpace{}%
\AgdaBound{irM′}\AgdaSpace{}%
\AgdaOperator{\AgdaInductiveConstructor{,}}\AgdaSpace{}%
\AgdaBound{r4}\AgdaSpace{}%
\AgdaOperator{\AgdaInductiveConstructor{,}}\AgdaSpace{}%
\AgdaBound{eq}\AgdaSymbol{)}\AgdaSpace{}%
\AgdaSymbol{=}\AgdaSpace{}%
\AgdaInductiveConstructor{inj₁}\AgdaSpace{}%
\AgdaSymbol{(\AgdaUnderscore{}}\AgdaSpace{}%
\AgdaOperator{\AgdaInductiveConstructor{,}}\AgdaSpace{}%
\AgdaSymbol{(\AgdaUnderscore{}}\AgdaSpace{}%
\AgdaOperator{\AgdaInductiveConstructor{⟶⟨}}\AgdaSpace{}%
\AgdaBound{r1}\AgdaSpace{}%
\AgdaOperator{\AgdaInductiveConstructor{⟩}}\AgdaSpace{}%
\AgdaBound{r3}\AgdaSymbol{)}\AgdaSpace{}%
\AgdaOperator{\AgdaInductiveConstructor{,}}\AgdaSpace{}%
\AgdaBound{irM′}\AgdaSpace{}%
\AgdaOperator{\AgdaInductiveConstructor{,}}\AgdaSpace{}%
\AgdaBound{r4}\AgdaSpace{}%
\AgdaOperator{\AgdaInductiveConstructor{,}}\AgdaSpace{}%
\AgdaFunction{cong}\AgdaSpace{}%
\AgdaInductiveConstructor{suc}\AgdaSpace{}%
\AgdaBound{eq}\AgdaSymbol{)}\<%
\\
\>[0]\AgdaSymbol{...}\AgdaSpace{}%
\AgdaSymbol{|}\AgdaSpace{}%
\AgdaInductiveConstructor{inj₂}\AgdaSpace{}%
\AgdaInductiveConstructor{refl}\AgdaSpace{}%
\AgdaSymbol{=}\AgdaSpace{}%
\AgdaInductiveConstructor{inj₂}\AgdaSpace{}%
\AgdaInductiveConstructor{refl}\<%
\\
\>[0]\AgdaFunction{frame-inv}\AgdaSpace{}%
\AgdaSymbol{\{}\AgdaOperator{\AgdaInductiveConstructor{□⟨}}\AgdaSpace{}%
\AgdaBound{H}\AgdaSpace{}%
\AgdaOperator{\AgdaInductiveConstructor{?⟩}}\AgdaSymbol{\}}\AgdaSpace{}%
\AgdaSymbol{\{}\AgdaBound{M}\AgdaSymbol{\}}\AgdaSpace{}%
\AgdaSymbol{\{}\AgdaBound{N}\AgdaSymbol{\}}\AgdaSpace{}%
\AgdaSymbol{(}\AgdaDottedPattern{\AgdaSymbol{.(}}\AgdaDottedPattern{\AgdaOperator{\AgdaInductiveConstructor{□⟨}}}\AgdaSpace{}%
\AgdaDottedPattern{\AgdaBound{H}}\AgdaSpace{}%
\AgdaDottedPattern{\AgdaOperator{\AgdaInductiveConstructor{?⟩}}}\AgdaSpace{}%
\AgdaDottedPattern{\AgdaOperator{\AgdaFunction{⟦}}}\AgdaSpace{}%
\AgdaDottedPattern{\AgdaBound{M}}\AgdaSpace{}%
\AgdaDottedPattern{\AgdaOperator{\AgdaFunction{⟧}}}\AgdaDottedPattern{\AgdaSymbol{)}}\AgdaSpace{}%
\AgdaOperator{\AgdaInductiveConstructor{⟶⟨}}\AgdaSpace{}%
\AgdaInductiveConstructor{ξ-blame}\AgdaSpace{}%
\AgdaOperator{\AgdaInductiveConstructor{□⟨}}\AgdaSpace{}%
\AgdaSymbol{\AgdaUnderscore{}}\AgdaSpace{}%
\AgdaOperator{\AgdaInductiveConstructor{?⟩}}\AgdaSpace{}%
\AgdaOperator{\AgdaInductiveConstructor{⟩}}\AgdaSpace{}%
\AgdaBound{r2}\AgdaSymbol{)}\AgdaSpace{}%
\AgdaBound{irN}\<%
\\
\>[0][@{}l@{\AgdaIndent{0}}]%
\>[4]\AgdaKeyword{with}\AgdaSpace{}%
\AgdaFunction{blame↠}\AgdaSpace{}%
\AgdaBound{r2}\<%
\\
\>[0]\AgdaSymbol{...}\AgdaSpace{}%
\AgdaSymbol{|}\AgdaSpace{}%
\AgdaInductiveConstructor{refl}\AgdaSpace{}%
\AgdaSymbol{=}\AgdaSpace{}%
\AgdaInductiveConstructor{inj₂}\AgdaSpace{}%
\AgdaInductiveConstructor{refl}\<%
\\
\>[0]\AgdaFunction{frame-inv}\AgdaSpace{}%
\AgdaSymbol{\{}\AgdaOperator{\AgdaInductiveConstructor{□⟨}}\AgdaSpace{}%
\AgdaBound{H}\AgdaSpace{}%
\AgdaOperator{\AgdaInductiveConstructor{?⟩}}\AgdaSymbol{\}}\AgdaSpace{}%
\AgdaSymbol{\{}\AgdaBound{M}\AgdaSymbol{\}}\AgdaSpace{}%
\AgdaSymbol{\{}\AgdaBound{N}\AgdaSymbol{\}}\AgdaSpace{}%
\AgdaSymbol{(}\AgdaDottedPattern{\AgdaSymbol{.(}}\AgdaDottedPattern{\AgdaOperator{\AgdaInductiveConstructor{□⟨}}}\AgdaSpace{}%
\AgdaDottedPattern{\AgdaBound{H}}\AgdaSpace{}%
\AgdaDottedPattern{\AgdaOperator{\AgdaInductiveConstructor{?⟩}}}\AgdaSpace{}%
\AgdaDottedPattern{\AgdaOperator{\AgdaFunction{⟦}}}\AgdaSpace{}%
\AgdaDottedPattern{\AgdaBound{M}}\AgdaSpace{}%
\AgdaDottedPattern{\AgdaOperator{\AgdaFunction{⟧}}}\AgdaDottedPattern{\AgdaSymbol{)}}\AgdaSpace{}%
\AgdaOperator{\AgdaInductiveConstructor{⟶⟨}}\AgdaSpace{}%
\AgdaInductiveConstructor{collapse}\AgdaSpace{}%
\AgdaBound{v}\AgdaSpace{}%
\AgdaInductiveConstructor{refl}\AgdaSpace{}%
\AgdaOperator{\AgdaInductiveConstructor{⟩}}\AgdaSpace{}%
\AgdaBound{r2}\AgdaSymbol{)}\AgdaSpace{}%
\AgdaBound{irN}\AgdaSpace{}%
\AgdaSymbol{=}\<%
\\
\>[0][@{}l@{\AgdaIndent{0}}]%
\>[2]\AgdaInductiveConstructor{inj₁}\AgdaSpace{}%
\AgdaSymbol{(}\AgdaBound{M}\AgdaSpace{}%
\AgdaOperator{\AgdaInductiveConstructor{,}}\AgdaSpace{}%
\AgdaSymbol{(\AgdaUnderscore{}}\AgdaSpace{}%
\AgdaOperator{\AgdaInductiveConstructor{END}}\AgdaSymbol{)}\AgdaSpace{}%
\AgdaOperator{\AgdaInductiveConstructor{,}}\AgdaSpace{}%
\AgdaFunction{value-irred}\AgdaSpace{}%
\AgdaSymbol{(}\AgdaBound{v}\AgdaSpace{}%
\AgdaOperator{\AgdaInductiveConstructor{〈}}\AgdaSpace{}%
\AgdaSymbol{\AgdaUnderscore{}}\AgdaSpace{}%
\AgdaOperator{\AgdaInductiveConstructor{〉}}\AgdaSymbol{)}\AgdaSpace{}%
\AgdaOperator{\AgdaInductiveConstructor{,}}\AgdaSpace{}%
\AgdaSymbol{(\AgdaUnderscore{}}\AgdaSpace{}%
\AgdaOperator{\AgdaInductiveConstructor{⟶⟨}}\AgdaSpace{}%
\AgdaInductiveConstructor{collapse}\AgdaSpace{}%
\AgdaBound{v}\AgdaSpace{}%
\AgdaInductiveConstructor{refl}\AgdaSpace{}%
\AgdaOperator{\AgdaInductiveConstructor{⟩}}\AgdaSpace{}%
\AgdaBound{r2}\AgdaSymbol{)}\AgdaSpace{}%
\AgdaOperator{\AgdaInductiveConstructor{,}}\AgdaSpace{}%
\AgdaInductiveConstructor{refl}\AgdaSymbol{)}\<%
\\
\>[0]\AgdaFunction{frame-inv}\AgdaSpace{}%
\AgdaSymbol{\{}\AgdaOperator{\AgdaInductiveConstructor{□⟨}}\AgdaSpace{}%
\AgdaBound{H}\AgdaSpace{}%
\AgdaOperator{\AgdaInductiveConstructor{?⟩}}\AgdaSymbol{\}}\AgdaSpace{}%
\AgdaSymbol{\{}\AgdaBound{M}\AgdaSymbol{\}}\AgdaSpace{}%
\AgdaSymbol{\{}\AgdaBound{N}\AgdaSymbol{\}}\AgdaSpace{}%
\AgdaSymbol{(}\AgdaDottedPattern{\AgdaSymbol{.(}}\AgdaDottedPattern{\AgdaOperator{\AgdaInductiveConstructor{□⟨}}}\AgdaSpace{}%
\AgdaDottedPattern{\AgdaBound{H}}\AgdaSpace{}%
\AgdaDottedPattern{\AgdaOperator{\AgdaInductiveConstructor{?⟩}}}\AgdaSpace{}%
\AgdaDottedPattern{\AgdaOperator{\AgdaFunction{⟦}}}\AgdaSpace{}%
\AgdaDottedPattern{\AgdaBound{M}}\AgdaSpace{}%
\AgdaDottedPattern{\AgdaOperator{\AgdaFunction{⟧}}}\AgdaDottedPattern{\AgdaSymbol{)}}\AgdaSpace{}%
\AgdaOperator{\AgdaInductiveConstructor{⟶⟨}}\AgdaSpace{}%
\AgdaInductiveConstructor{collide}\AgdaSpace{}%
\AgdaBound{v}\AgdaSpace{}%
\AgdaBound{eq}\AgdaSpace{}%
\AgdaInductiveConstructor{refl}\AgdaSpace{}%
\AgdaOperator{\AgdaInductiveConstructor{⟩}}\AgdaSpace{}%
\AgdaBound{r2}\AgdaSymbol{)}\AgdaSpace{}%
\AgdaBound{irN}\AgdaSpace{}%
\AgdaSymbol{=}\<%
\\
\>[0][@{}l@{\AgdaIndent{0}}]%
\>[2]\AgdaInductiveConstructor{inj₁}\AgdaSpace{}%
\AgdaSymbol{(}\AgdaBound{M}\AgdaSpace{}%
\AgdaOperator{\AgdaInductiveConstructor{,}}\AgdaSpace{}%
\AgdaSymbol{(\AgdaUnderscore{}}\AgdaSpace{}%
\AgdaOperator{\AgdaInductiveConstructor{END}}\AgdaSymbol{)}\AgdaSpace{}%
\AgdaOperator{\AgdaInductiveConstructor{,}}\AgdaSpace{}%
\AgdaFunction{value-irred}\AgdaSpace{}%
\AgdaSymbol{(}\AgdaBound{v}\AgdaSpace{}%
\AgdaOperator{\AgdaInductiveConstructor{〈}}\AgdaSpace{}%
\AgdaSymbol{\AgdaUnderscore{}}\AgdaSpace{}%
\AgdaOperator{\AgdaInductiveConstructor{〉}}\AgdaSymbol{)}\AgdaSpace{}%
\AgdaOperator{\AgdaInductiveConstructor{,}}\AgdaSpace{}%
\AgdaSymbol{(\AgdaUnderscore{}}\AgdaSpace{}%
\AgdaOperator{\AgdaInductiveConstructor{⟶⟨}}\AgdaSpace{}%
\AgdaInductiveConstructor{collide}\AgdaSpace{}%
\AgdaBound{v}\AgdaSpace{}%
\AgdaBound{eq}\AgdaSpace{}%
\AgdaInductiveConstructor{refl}\AgdaSpace{}%
\AgdaOperator{\AgdaInductiveConstructor{⟩}}\AgdaSpace{}%
\AgdaBound{r2}\AgdaSymbol{)}\AgdaSpace{}%
\AgdaOperator{\AgdaInductiveConstructor{,}}\AgdaSpace{}%
\AgdaInductiveConstructor{refl}\AgdaSymbol{)}\<%
\\
\\[\AgdaEmptyExtraSkip]%
\>[0]\AgdaFunction{frame-blame}\AgdaSpace{}%
\AgdaSymbol{:}\AgdaSpace{}%
\AgdaSymbol{∀\{}\AgdaBound{F}\AgdaSymbol{\}\{}\AgdaBound{M}\AgdaSymbol{\}\{}\AgdaBound{N}\AgdaSymbol{\}}\<%
\\
\>[0][@{}l@{\AgdaIndent{0}}]%
\>[3]\AgdaSymbol{→}\AgdaSpace{}%
\AgdaBound{M}\AgdaSpace{}%
\AgdaOperator{\AgdaDatatype{↠}}\AgdaSpace{}%
\AgdaBound{N}\<%
\\
\>[3]\AgdaSymbol{→}\AgdaSpace{}%
\AgdaBound{M}\AgdaSpace{}%
\AgdaOperator{\AgdaDatatype{≡}}\AgdaSpace{}%
\AgdaBound{F}\AgdaSpace{}%
\AgdaOperator{\AgdaFunction{⟦}}\AgdaSpace{}%
\AgdaInductiveConstructor{blame}\AgdaSpace{}%
\AgdaOperator{\AgdaFunction{⟧}}\<%
\\
\>[3]\AgdaSymbol{→}\AgdaSpace{}%
\AgdaFunction{irred}\AgdaSpace{}%
\AgdaBound{N}\<%
\\
\>[3]\AgdaSymbol{→}\AgdaSpace{}%
\AgdaBound{N}\AgdaSpace{}%
\AgdaOperator{\AgdaDatatype{≡}}\AgdaSpace{}%
\AgdaInductiveConstructor{blame}\<%
\\
\>[0]\AgdaFunction{frame-blame}\AgdaSpace{}%
\AgdaSymbol{\{}\AgdaBound{F}\AgdaSymbol{\}}\AgdaSpace{}%
\AgdaSymbol{\{}\AgdaBound{N}\AgdaSymbol{\}}\AgdaSpace{}%
\AgdaSymbol{(}\AgdaDottedPattern{\AgdaSymbol{.}}\AgdaDottedPattern{\AgdaBound{N}}\AgdaSpace{}%
\AgdaOperator{\AgdaInductiveConstructor{END}}\AgdaSymbol{)}\AgdaSpace{}%
\AgdaInductiveConstructor{refl}\AgdaSpace{}%
\AgdaBound{irN}\AgdaSpace{}%
\AgdaSymbol{=}\AgdaSpace{}%
\AgdaFunction{⊥-elim}\AgdaSpace{}%
\AgdaSymbol{(}\AgdaBound{irN}\AgdaSpace{}%
\AgdaSymbol{(\AgdaUnderscore{}}\AgdaSpace{}%
\AgdaOperator{\AgdaInductiveConstructor{,}}\AgdaSpace{}%
\AgdaSymbol{(}\AgdaInductiveConstructor{ξ-blame}\AgdaSpace{}%
\AgdaBound{F}\AgdaSymbol{)))}\<%
\\
\>[0]\AgdaFunction{frame-blame}\AgdaSpace{}%
\AgdaSymbol{\{}\AgdaOperator{\AgdaInductiveConstructor{□·}}\AgdaSpace{}%
\AgdaBound{M}\AgdaSymbol{\}}\AgdaSpace{}%
\AgdaSymbol{\{}\AgdaDottedPattern{\AgdaSymbol{.((}}\AgdaDottedPattern{\AgdaOperator{\AgdaInductiveConstructor{□·}}}\AgdaSpace{}%
\AgdaDottedPattern{\AgdaBound{M}}\AgdaDottedPattern{\AgdaSymbol{)}}\AgdaSpace{}%
\AgdaDottedPattern{\AgdaOperator{\AgdaFunction{⟦}}}\AgdaSpace{}%
\AgdaDottedPattern{\AgdaInductiveConstructor{blame}}\AgdaSpace{}%
\AgdaDottedPattern{\AgdaOperator{\AgdaFunction{⟧}}}\AgdaDottedPattern{\AgdaSymbol{)}}\AgdaSymbol{\}}\AgdaSpace{}%
\AgdaSymbol{(}\AgdaDottedPattern{\AgdaSymbol{.((}}\AgdaDottedPattern{\AgdaOperator{\AgdaInductiveConstructor{□·}}}\AgdaSpace{}%
\AgdaDottedPattern{\AgdaBound{M}}\AgdaDottedPattern{\AgdaSymbol{)}}\AgdaSpace{}%
\AgdaDottedPattern{\AgdaOperator{\AgdaFunction{⟦}}}\AgdaSpace{}%
\AgdaDottedPattern{\AgdaInductiveConstructor{blame}}\AgdaSpace{}%
\AgdaDottedPattern{\AgdaOperator{\AgdaFunction{⟧}}}\AgdaDottedPattern{\AgdaSymbol{)}}\AgdaSpace{}%
\AgdaOperator{\AgdaInductiveConstructor{⟶⟨}}\AgdaSpace{}%
\AgdaInductiveConstructor{ξξ}\AgdaSpace{}%
\AgdaSymbol{(}\AgdaOperator{\AgdaInductiveConstructor{□·}}\AgdaSpace{}%
\AgdaBound{M₁}\AgdaSymbol{)}\AgdaSpace{}%
\AgdaInductiveConstructor{refl}\AgdaSpace{}%
\AgdaBound{x₁}\AgdaSpace{}%
\AgdaBound{r}\AgdaSpace{}%
\AgdaOperator{\AgdaInductiveConstructor{⟩}}\AgdaSpace{}%
\AgdaBound{M→N}\AgdaSymbol{)}\AgdaSpace{}%
\AgdaInductiveConstructor{refl}\AgdaSpace{}%
\AgdaBound{irN}\AgdaSpace{}%
\AgdaSymbol{=}\<%
\\
\>[0][@{}l@{\AgdaIndent{0}}]%
\>[2]\AgdaFunction{⊥-elim}\AgdaSpace{}%
\AgdaSymbol{(}\AgdaFunction{blame-irreducible}\AgdaSpace{}%
\AgdaBound{r}\AgdaSymbol{)}\<%
\\
\>[0]\AgdaFunction{frame-blame}\AgdaSpace{}%
\AgdaSymbol{\{}\AgdaOperator{\AgdaInductiveConstructor{□·}}\AgdaSpace{}%
\AgdaBound{M}\AgdaSymbol{\}}\AgdaSpace{}%
\AgdaSymbol{\{}\AgdaDottedPattern{\AgdaSymbol{.((}}\AgdaDottedPattern{\AgdaOperator{\AgdaInductiveConstructor{□·}}}\AgdaSpace{}%
\AgdaDottedPattern{\AgdaBound{M}}\AgdaDottedPattern{\AgdaSymbol{)}}\AgdaSpace{}%
\AgdaDottedPattern{\AgdaOperator{\AgdaFunction{⟦}}}\AgdaSpace{}%
\AgdaDottedPattern{\AgdaInductiveConstructor{blame}}\AgdaSpace{}%
\AgdaDottedPattern{\AgdaOperator{\AgdaFunction{⟧}}}\AgdaDottedPattern{\AgdaSymbol{)}}\AgdaSymbol{\}}\AgdaSpace{}%
\AgdaSymbol{(}\AgdaDottedPattern{\AgdaSymbol{.((}}\AgdaDottedPattern{\AgdaOperator{\AgdaInductiveConstructor{□·}}}\AgdaSpace{}%
\AgdaDottedPattern{\AgdaBound{M}}\AgdaDottedPattern{\AgdaSymbol{)}}\AgdaSpace{}%
\AgdaDottedPattern{\AgdaOperator{\AgdaFunction{⟦}}}\AgdaSpace{}%
\AgdaDottedPattern{\AgdaInductiveConstructor{blame}}\AgdaSpace{}%
\AgdaDottedPattern{\AgdaOperator{\AgdaFunction{⟧}}}\AgdaDottedPattern{\AgdaSymbol{)}}\AgdaSpace{}%
\AgdaOperator{\AgdaInductiveConstructor{⟶⟨}}\AgdaSpace{}%
\AgdaInductiveConstructor{ξξ}\AgdaSpace{}%
\AgdaSymbol{(()}\AgdaSpace{}%
\AgdaOperator{\AgdaInductiveConstructor{·□}}\AgdaSymbol{)}\AgdaSpace{}%
\AgdaInductiveConstructor{refl}\AgdaSpace{}%
\AgdaBound{x₁}\AgdaSpace{}%
\AgdaBound{r}\AgdaSpace{}%
\AgdaOperator{\AgdaInductiveConstructor{⟩}}\AgdaSpace{}%
\AgdaBound{M→N}\AgdaSymbol{)}\AgdaSpace{}%
\AgdaInductiveConstructor{refl}\AgdaSpace{}%
\AgdaBound{irN}\<%
\\
\>[0]\AgdaFunction{frame-blame}\AgdaSpace{}%
\AgdaSymbol{\{}\AgdaOperator{\AgdaInductiveConstructor{□·}}\AgdaSpace{}%
\AgdaBound{M}\AgdaSymbol{\}}\AgdaSpace{}%
\AgdaSymbol{\{}\AgdaDottedPattern{\AgdaSymbol{.((}}\AgdaDottedPattern{\AgdaOperator{\AgdaInductiveConstructor{□·}}}\AgdaSpace{}%
\AgdaDottedPattern{\AgdaBound{M}}\AgdaDottedPattern{\AgdaSymbol{)}}\AgdaSpace{}%
\AgdaDottedPattern{\AgdaOperator{\AgdaFunction{⟦}}}\AgdaSpace{}%
\AgdaDottedPattern{\AgdaInductiveConstructor{blame}}\AgdaSpace{}%
\AgdaDottedPattern{\AgdaOperator{\AgdaFunction{⟧}}}\AgdaDottedPattern{\AgdaSymbol{)}}\AgdaSymbol{\}}\AgdaSpace{}%
\AgdaSymbol{(}\AgdaDottedPattern{\AgdaSymbol{.((}}\AgdaDottedPattern{\AgdaOperator{\AgdaInductiveConstructor{□·}}}\AgdaSpace{}%
\AgdaDottedPattern{\AgdaBound{M}}\AgdaDottedPattern{\AgdaSymbol{)}}\AgdaSpace{}%
\AgdaDottedPattern{\AgdaOperator{\AgdaFunction{⟦}}}\AgdaSpace{}%
\AgdaDottedPattern{\AgdaInductiveConstructor{blame}}\AgdaSpace{}%
\AgdaDottedPattern{\AgdaOperator{\AgdaFunction{⟧}}}\AgdaDottedPattern{\AgdaSymbol{)}}\AgdaSpace{}%
\AgdaOperator{\AgdaInductiveConstructor{⟶⟨}}\AgdaSpace{}%
\AgdaInductiveConstructor{ξξ-blame}\AgdaSpace{}%
\AgdaBound{F}\AgdaSpace{}%
\AgdaBound{x}\AgdaSpace{}%
\AgdaOperator{\AgdaInductiveConstructor{⟩}}\AgdaSpace{}%
\AgdaBound{M→N}\AgdaSymbol{)}\AgdaSpace{}%
\AgdaInductiveConstructor{refl}\AgdaSpace{}%
\AgdaBound{irN}\<%
\\
\>[0][@{}l@{\AgdaIndent{0}}]%
\>[4]\AgdaKeyword{with}\AgdaSpace{}%
\AgdaFunction{blame↠}\AgdaSpace{}%
\AgdaBound{M→N}\<%
\\
\>[0]\AgdaSymbol{...}\AgdaSpace{}%
\AgdaSymbol{|}\AgdaSpace{}%
\AgdaInductiveConstructor{refl}\AgdaSpace{}%
\AgdaSymbol{=}\AgdaSpace{}%
\AgdaInductiveConstructor{refl}\<%
\\
\>[0]\AgdaFunction{frame-blame}\AgdaSpace{}%
\AgdaSymbol{\{}\AgdaBound{v}\AgdaSpace{}%
\AgdaOperator{\AgdaInductiveConstructor{·□}}\AgdaSymbol{\}}\AgdaSpace{}%
\AgdaSymbol{\{}\AgdaDottedPattern{\AgdaSymbol{.((}}\AgdaDottedPattern{\AgdaBound{v}}\AgdaSpace{}%
\AgdaDottedPattern{\AgdaOperator{\AgdaInductiveConstructor{·□}}}\AgdaDottedPattern{\AgdaSymbol{)}}\AgdaSpace{}%
\AgdaDottedPattern{\AgdaOperator{\AgdaFunction{⟦}}}\AgdaSpace{}%
\AgdaDottedPattern{\AgdaInductiveConstructor{blame}}\AgdaSpace{}%
\AgdaDottedPattern{\AgdaOperator{\AgdaFunction{⟧}}}\AgdaDottedPattern{\AgdaSymbol{)}}\AgdaSymbol{\}}\AgdaSpace{}%
\AgdaSymbol{(}\AgdaDottedPattern{\AgdaSymbol{.((}}\AgdaDottedPattern{\AgdaBound{v}}\AgdaSpace{}%
\AgdaDottedPattern{\AgdaOperator{\AgdaInductiveConstructor{·□}}}\AgdaDottedPattern{\AgdaSymbol{)}}\AgdaSpace{}%
\AgdaDottedPattern{\AgdaOperator{\AgdaFunction{⟦}}}\AgdaSpace{}%
\AgdaDottedPattern{\AgdaInductiveConstructor{blame}}\AgdaSpace{}%
\AgdaDottedPattern{\AgdaOperator{\AgdaFunction{⟧}}}\AgdaDottedPattern{\AgdaSymbol{)}}\AgdaSpace{}%
\AgdaOperator{\AgdaInductiveConstructor{⟶⟨}}\AgdaSpace{}%
\AgdaInductiveConstructor{ξξ}\AgdaSpace{}%
\AgdaSymbol{(}\AgdaOperator{\AgdaInductiveConstructor{□·}}\AgdaSpace{}%
\AgdaBound{M}\AgdaSymbol{)}\AgdaSpace{}%
\AgdaInductiveConstructor{refl}\AgdaSpace{}%
\AgdaInductiveConstructor{refl}\AgdaSpace{}%
\AgdaBound{r}\AgdaSpace{}%
\AgdaOperator{\AgdaInductiveConstructor{⟩}}\AgdaSpace{}%
\AgdaBound{M→N}\AgdaSymbol{)}\AgdaSpace{}%
\AgdaInductiveConstructor{refl}\AgdaSpace{}%
\AgdaBound{irN}\AgdaSpace{}%
\AgdaSymbol{=}\<%
\\
\>[0][@{}l@{\AgdaIndent{0}}]%
\>[4]\AgdaFunction{⊥-elim}\AgdaSpace{}%
\AgdaSymbol{(}\AgdaFunction{value-irreducible}\AgdaSpace{}%
\AgdaBound{v}\AgdaSpace{}%
\AgdaBound{r}\AgdaSymbol{)}\<%
\\
\>[0]\AgdaFunction{frame-blame}\AgdaSpace{}%
\AgdaSymbol{\{}\AgdaBound{v}\AgdaSpace{}%
\AgdaOperator{\AgdaInductiveConstructor{·□}}\AgdaSymbol{\}}\AgdaSpace{}%
\AgdaSymbol{\{}\AgdaDottedPattern{\AgdaSymbol{.((}}\AgdaDottedPattern{\AgdaBound{v}}\AgdaSpace{}%
\AgdaDottedPattern{\AgdaOperator{\AgdaInductiveConstructor{·□}}}\AgdaDottedPattern{\AgdaSymbol{)}}\AgdaSpace{}%
\AgdaDottedPattern{\AgdaOperator{\AgdaFunction{⟦}}}\AgdaSpace{}%
\AgdaDottedPattern{\AgdaInductiveConstructor{blame}}\AgdaSpace{}%
\AgdaDottedPattern{\AgdaOperator{\AgdaFunction{⟧}}}\AgdaDottedPattern{\AgdaSymbol{)}}\AgdaSymbol{\}}\AgdaSpace{}%
\AgdaSymbol{(}\AgdaDottedPattern{\AgdaSymbol{.((}}\AgdaDottedPattern{\AgdaBound{v}}\AgdaSpace{}%
\AgdaDottedPattern{\AgdaOperator{\AgdaInductiveConstructor{·□}}}\AgdaDottedPattern{\AgdaSymbol{)}}\AgdaSpace{}%
\AgdaDottedPattern{\AgdaOperator{\AgdaFunction{⟦}}}\AgdaSpace{}%
\AgdaDottedPattern{\AgdaInductiveConstructor{blame}}\AgdaSpace{}%
\AgdaDottedPattern{\AgdaOperator{\AgdaFunction{⟧}}}\AgdaDottedPattern{\AgdaSymbol{)}}\AgdaSpace{}%
\AgdaOperator{\AgdaInductiveConstructor{⟶⟨}}\AgdaSpace{}%
\AgdaInductiveConstructor{ξξ}\AgdaSpace{}%
\AgdaSymbol{(}\AgdaBound{v₁}\AgdaSpace{}%
\AgdaOperator{\AgdaInductiveConstructor{·□}}\AgdaSymbol{)}\AgdaSpace{}%
\AgdaInductiveConstructor{refl}\AgdaSpace{}%
\AgdaInductiveConstructor{refl}\AgdaSpace{}%
\AgdaBound{r}\AgdaSpace{}%
\AgdaOperator{\AgdaInductiveConstructor{⟩}}\AgdaSpace{}%
\AgdaBound{M→N}\AgdaSymbol{)}\AgdaSpace{}%
\AgdaInductiveConstructor{refl}\AgdaSpace{}%
\AgdaBound{irN}\AgdaSpace{}%
\AgdaSymbol{=}\<%
\\
\>[0][@{}l@{\AgdaIndent{0}}]%
\>[4]\AgdaFunction{⊥-elim}\AgdaSpace{}%
\AgdaSymbol{(}\AgdaFunction{blame-irreducible}\AgdaSpace{}%
\AgdaBound{r}\AgdaSymbol{)}\<%
\\
\>[0]\AgdaFunction{frame-blame}\AgdaSpace{}%
\AgdaSymbol{\{}\AgdaBound{v}\AgdaSpace{}%
\AgdaOperator{\AgdaInductiveConstructor{·□}}\AgdaSymbol{\}}\AgdaSpace{}%
\AgdaSymbol{\{}\AgdaDottedPattern{\AgdaSymbol{.((}}\AgdaDottedPattern{\AgdaBound{v}}\AgdaSpace{}%
\AgdaDottedPattern{\AgdaOperator{\AgdaInductiveConstructor{·□}}}\AgdaDottedPattern{\AgdaSymbol{)}}\AgdaSpace{}%
\AgdaDottedPattern{\AgdaOperator{\AgdaFunction{⟦}}}\AgdaSpace{}%
\AgdaDottedPattern{\AgdaInductiveConstructor{blame}}\AgdaSpace{}%
\AgdaDottedPattern{\AgdaOperator{\AgdaFunction{⟧}}}\AgdaDottedPattern{\AgdaSymbol{)}}\AgdaSymbol{\}}\AgdaSpace{}%
\AgdaSymbol{(}\AgdaDottedPattern{\AgdaSymbol{.((}}\AgdaDottedPattern{\AgdaBound{v}}\AgdaSpace{}%
\AgdaDottedPattern{\AgdaOperator{\AgdaInductiveConstructor{·□}}}\AgdaDottedPattern{\AgdaSymbol{)}}\AgdaSpace{}%
\AgdaDottedPattern{\AgdaOperator{\AgdaFunction{⟦}}}\AgdaSpace{}%
\AgdaDottedPattern{\AgdaInductiveConstructor{blame}}\AgdaSpace{}%
\AgdaDottedPattern{\AgdaOperator{\AgdaFunction{⟧}}}\AgdaDottedPattern{\AgdaSymbol{)}}\AgdaSpace{}%
\AgdaOperator{\AgdaInductiveConstructor{⟶⟨}}\AgdaSpace{}%
\AgdaInductiveConstructor{ξξ-blame}\AgdaSpace{}%
\AgdaBound{F}\AgdaSpace{}%
\AgdaBound{x}\AgdaSpace{}%
\AgdaOperator{\AgdaInductiveConstructor{⟩}}\AgdaSpace{}%
\AgdaBound{M→N}\AgdaSymbol{)}\AgdaSpace{}%
\AgdaInductiveConstructor{refl}\AgdaSpace{}%
\AgdaBound{irN}\<%
\\
\>[0][@{}l@{\AgdaIndent{0}}]%
\>[4]\AgdaKeyword{with}\AgdaSpace{}%
\AgdaFunction{blame↠}\AgdaSpace{}%
\AgdaBound{M→N}\<%
\\
\>[0]\AgdaSymbol{...}\AgdaSpace{}%
\AgdaSymbol{|}\AgdaSpace{}%
\AgdaInductiveConstructor{refl}\AgdaSpace{}%
\AgdaSymbol{=}\AgdaSpace{}%
\AgdaInductiveConstructor{refl}\<%
\\
\>[0]\AgdaFunction{frame-blame}\AgdaSpace{}%
\AgdaSymbol{\{}\AgdaOperator{\AgdaInductiveConstructor{□⟨}}\AgdaSpace{}%
\AgdaBound{G}\AgdaSpace{}%
\AgdaOperator{\AgdaInductiveConstructor{!⟩}}\AgdaSymbol{\}}\AgdaSpace{}%
\AgdaSymbol{\{}\AgdaDottedPattern{\AgdaSymbol{.(}}\AgdaDottedPattern{\AgdaOperator{\AgdaInductiveConstructor{□⟨}}}\AgdaSpace{}%
\AgdaDottedPattern{\AgdaBound{G}}\AgdaSpace{}%
\AgdaDottedPattern{\AgdaOperator{\AgdaInductiveConstructor{!⟩}}}\AgdaSpace{}%
\AgdaDottedPattern{\AgdaOperator{\AgdaFunction{⟦}}}\AgdaSpace{}%
\AgdaDottedPattern{\AgdaInductiveConstructor{blame}}\AgdaSpace{}%
\AgdaDottedPattern{\AgdaOperator{\AgdaFunction{⟧}}}\AgdaDottedPattern{\AgdaSymbol{)}}\AgdaSymbol{\}}\AgdaSpace{}%
\AgdaSymbol{(}\AgdaDottedPattern{\AgdaSymbol{.(}}\AgdaDottedPattern{\AgdaOperator{\AgdaInductiveConstructor{□⟨}}}\AgdaSpace{}%
\AgdaDottedPattern{\AgdaBound{G}}\AgdaSpace{}%
\AgdaDottedPattern{\AgdaOperator{\AgdaInductiveConstructor{!⟩}}}\AgdaSpace{}%
\AgdaDottedPattern{\AgdaOperator{\AgdaFunction{⟦}}}\AgdaSpace{}%
\AgdaDottedPattern{\AgdaInductiveConstructor{blame}}\AgdaSpace{}%
\AgdaDottedPattern{\AgdaOperator{\AgdaFunction{⟧}}}\AgdaDottedPattern{\AgdaSymbol{)}}\AgdaSpace{}%
\AgdaOperator{\AgdaInductiveConstructor{⟶⟨}}\AgdaSpace{}%
\AgdaInductiveConstructor{ξξ}\AgdaSpace{}%
\AgdaOperator{\AgdaInductiveConstructor{□⟨}}\AgdaSpace{}%
\AgdaSymbol{\AgdaUnderscore{}}\AgdaSpace{}%
\AgdaOperator{\AgdaInductiveConstructor{!⟩}}\AgdaSpace{}%
\AgdaInductiveConstructor{refl}\AgdaSpace{}%
\AgdaInductiveConstructor{refl}\AgdaSpace{}%
\AgdaBound{r}\AgdaSpace{}%
\AgdaOperator{\AgdaInductiveConstructor{⟩}}\AgdaSpace{}%
\AgdaBound{M→N}\AgdaSymbol{)}\AgdaSpace{}%
\AgdaInductiveConstructor{refl}\AgdaSpace{}%
\AgdaBound{irN}\AgdaSpace{}%
\AgdaSymbol{=}\<%
\\
\>[0][@{}l@{\AgdaIndent{0}}]%
\>[2]\AgdaFunction{⊥-elim}\AgdaSpace{}%
\AgdaSymbol{(}\AgdaFunction{blame-irreducible}\AgdaSpace{}%
\AgdaBound{r}\AgdaSymbol{)}\<%
\\
\>[0]\AgdaFunction{frame-blame}\AgdaSpace{}%
\AgdaSymbol{\{}\AgdaOperator{\AgdaInductiveConstructor{□⟨}}\AgdaSpace{}%
\AgdaBound{G}\AgdaSpace{}%
\AgdaOperator{\AgdaInductiveConstructor{!⟩}}\AgdaSymbol{\}}\AgdaSpace{}%
\AgdaSymbol{\{}\AgdaDottedPattern{\AgdaSymbol{.(}}\AgdaDottedPattern{\AgdaOperator{\AgdaInductiveConstructor{□⟨}}}\AgdaSpace{}%
\AgdaDottedPattern{\AgdaBound{G}}\AgdaSpace{}%
\AgdaDottedPattern{\AgdaOperator{\AgdaInductiveConstructor{!⟩}}}\AgdaSpace{}%
\AgdaDottedPattern{\AgdaOperator{\AgdaFunction{⟦}}}\AgdaSpace{}%
\AgdaDottedPattern{\AgdaInductiveConstructor{blame}}\AgdaSpace{}%
\AgdaDottedPattern{\AgdaOperator{\AgdaFunction{⟧}}}\AgdaDottedPattern{\AgdaSymbol{)}}\AgdaSymbol{\}}\AgdaSpace{}%
\AgdaSymbol{(}\AgdaDottedPattern{\AgdaSymbol{.(}}\AgdaDottedPattern{\AgdaOperator{\AgdaInductiveConstructor{□⟨}}}\AgdaSpace{}%
\AgdaDottedPattern{\AgdaBound{G}}\AgdaSpace{}%
\AgdaDottedPattern{\AgdaOperator{\AgdaInductiveConstructor{!⟩}}}\AgdaSpace{}%
\AgdaDottedPattern{\AgdaOperator{\AgdaFunction{⟦}}}\AgdaSpace{}%
\AgdaDottedPattern{\AgdaInductiveConstructor{blame}}\AgdaSpace{}%
\AgdaDottedPattern{\AgdaOperator{\AgdaFunction{⟧}}}\AgdaDottedPattern{\AgdaSymbol{)}}\AgdaSpace{}%
\AgdaOperator{\AgdaInductiveConstructor{⟶⟨}}\AgdaSpace{}%
\AgdaInductiveConstructor{ξξ-blame}\AgdaSpace{}%
\AgdaBound{F}\AgdaSpace{}%
\AgdaBound{x}\AgdaSpace{}%
\AgdaOperator{\AgdaInductiveConstructor{⟩}}\AgdaSpace{}%
\AgdaBound{M→N}\AgdaSymbol{)}\AgdaSpace{}%
\AgdaInductiveConstructor{refl}\AgdaSpace{}%
\AgdaBound{irN}\<%
\\
\>[0][@{}l@{\AgdaIndent{0}}]%
\>[4]\AgdaKeyword{with}\AgdaSpace{}%
\AgdaFunction{blame↠}\AgdaSpace{}%
\AgdaBound{M→N}\<%
\\
\>[0]\AgdaSymbol{...}\AgdaSpace{}%
\AgdaSymbol{|}\AgdaSpace{}%
\AgdaInductiveConstructor{refl}\AgdaSpace{}%
\AgdaSymbol{=}\AgdaSpace{}%
\AgdaInductiveConstructor{refl}\<%
\\
\>[0]\AgdaFunction{frame-blame}\AgdaSpace{}%
\AgdaSymbol{\{}\AgdaOperator{\AgdaInductiveConstructor{□⟨}}\AgdaSpace{}%
\AgdaBound{H}\AgdaSpace{}%
\AgdaOperator{\AgdaInductiveConstructor{?⟩}}\AgdaSymbol{\}}\AgdaSpace{}%
\AgdaSymbol{\{}\AgdaDottedPattern{\AgdaSymbol{.(}}\AgdaDottedPattern{\AgdaOperator{\AgdaInductiveConstructor{□⟨}}}\AgdaSpace{}%
\AgdaDottedPattern{\AgdaBound{H}}\AgdaSpace{}%
\AgdaDottedPattern{\AgdaOperator{\AgdaInductiveConstructor{?⟩}}}\AgdaSpace{}%
\AgdaDottedPattern{\AgdaOperator{\AgdaFunction{⟦}}}\AgdaSpace{}%
\AgdaDottedPattern{\AgdaInductiveConstructor{blame}}\AgdaSpace{}%
\AgdaDottedPattern{\AgdaOperator{\AgdaFunction{⟧}}}\AgdaDottedPattern{\AgdaSymbol{)}}\AgdaSymbol{\}}\AgdaSpace{}%
\AgdaSymbol{(}\AgdaDottedPattern{\AgdaSymbol{.(}}\AgdaDottedPattern{\AgdaOperator{\AgdaInductiveConstructor{□⟨}}}\AgdaSpace{}%
\AgdaDottedPattern{\AgdaBound{H}}\AgdaSpace{}%
\AgdaDottedPattern{\AgdaOperator{\AgdaInductiveConstructor{?⟩}}}\AgdaSpace{}%
\AgdaDottedPattern{\AgdaOperator{\AgdaFunction{⟦}}}\AgdaSpace{}%
\AgdaDottedPattern{\AgdaInductiveConstructor{blame}}\AgdaSpace{}%
\AgdaDottedPattern{\AgdaOperator{\AgdaFunction{⟧}}}\AgdaDottedPattern{\AgdaSymbol{)}}\AgdaSpace{}%
\AgdaOperator{\AgdaInductiveConstructor{⟶⟨}}\AgdaSpace{}%
\AgdaInductiveConstructor{ξξ}\AgdaSpace{}%
\AgdaOperator{\AgdaInductiveConstructor{□⟨}}\AgdaSpace{}%
\AgdaSymbol{\AgdaUnderscore{}}\AgdaSpace{}%
\AgdaOperator{\AgdaInductiveConstructor{?⟩}}\AgdaSpace{}%
\AgdaInductiveConstructor{refl}\AgdaSpace{}%
\AgdaInductiveConstructor{refl}\AgdaSpace{}%
\AgdaBound{r}\AgdaSpace{}%
\AgdaOperator{\AgdaInductiveConstructor{⟩}}\AgdaSpace{}%
\AgdaBound{M→N}\AgdaSymbol{)}\AgdaSpace{}%
\AgdaInductiveConstructor{refl}\AgdaSpace{}%
\AgdaBound{irN}\AgdaSpace{}%
\AgdaSymbol{=}\<%
\\
\>[0][@{}l@{\AgdaIndent{0}}]%
\>[2]\AgdaFunction{⊥-elim}\AgdaSpace{}%
\AgdaSymbol{(}\AgdaFunction{blame-irreducible}\AgdaSpace{}%
\AgdaBound{r}\AgdaSymbol{)}\<%
\\
\>[0]\AgdaFunction{frame-blame}\AgdaSpace{}%
\AgdaSymbol{\{}\AgdaOperator{\AgdaInductiveConstructor{□⟨}}\AgdaSpace{}%
\AgdaBound{H}\AgdaSpace{}%
\AgdaOperator{\AgdaInductiveConstructor{?⟩}}\AgdaSymbol{\}}\AgdaSpace{}%
\AgdaSymbol{\{}\AgdaDottedPattern{\AgdaSymbol{.(}}\AgdaDottedPattern{\AgdaOperator{\AgdaInductiveConstructor{□⟨}}}\AgdaSpace{}%
\AgdaDottedPattern{\AgdaBound{H}}\AgdaSpace{}%
\AgdaDottedPattern{\AgdaOperator{\AgdaInductiveConstructor{?⟩}}}\AgdaSpace{}%
\AgdaDottedPattern{\AgdaOperator{\AgdaFunction{⟦}}}\AgdaSpace{}%
\AgdaDottedPattern{\AgdaInductiveConstructor{blame}}\AgdaSpace{}%
\AgdaDottedPattern{\AgdaOperator{\AgdaFunction{⟧}}}\AgdaDottedPattern{\AgdaSymbol{)}}\AgdaSymbol{\}}\AgdaSpace{}%
\AgdaSymbol{(}\AgdaDottedPattern{\AgdaSymbol{.(}}\AgdaDottedPattern{\AgdaOperator{\AgdaInductiveConstructor{□⟨}}}\AgdaSpace{}%
\AgdaDottedPattern{\AgdaBound{H}}\AgdaSpace{}%
\AgdaDottedPattern{\AgdaOperator{\AgdaInductiveConstructor{?⟩}}}\AgdaSpace{}%
\AgdaDottedPattern{\AgdaOperator{\AgdaFunction{⟦}}}\AgdaSpace{}%
\AgdaDottedPattern{\AgdaInductiveConstructor{blame}}\AgdaSpace{}%
\AgdaDottedPattern{\AgdaOperator{\AgdaFunction{⟧}}}\AgdaDottedPattern{\AgdaSymbol{)}}\AgdaSpace{}%
\AgdaOperator{\AgdaInductiveConstructor{⟶⟨}}\AgdaSpace{}%
\AgdaInductiveConstructor{ξξ-blame}\AgdaSpace{}%
\AgdaOperator{\AgdaInductiveConstructor{□⟨}}\AgdaSpace{}%
\AgdaSymbol{\AgdaUnderscore{}}\AgdaSpace{}%
\AgdaOperator{\AgdaInductiveConstructor{?⟩}}\AgdaSpace{}%
\AgdaBound{x}\AgdaSpace{}%
\AgdaOperator{\AgdaInductiveConstructor{⟩}}\AgdaSpace{}%
\AgdaBound{M→N}\AgdaSymbol{)}\AgdaSpace{}%
\AgdaInductiveConstructor{refl}\AgdaSpace{}%
\AgdaBound{irN}\<%
\\
\>[0][@{}l@{\AgdaIndent{0}}]%
\>[4]\AgdaKeyword{with}\AgdaSpace{}%
\AgdaFunction{blame↠}\AgdaSpace{}%
\AgdaBound{M→N}\<%
\\
\>[0]\AgdaSymbol{...}\AgdaSpace{}%
\AgdaSymbol{|}\AgdaSpace{}%
\AgdaInductiveConstructor{refl}\AgdaSpace{}%
\AgdaSymbol{=}\AgdaSpace{}%
\AgdaInductiveConstructor{refl}\<%
\\
\\[\AgdaEmptyExtraSkip]%
\>[0]\AgdaFunction{app-invL}\AgdaSpace{}%
\AgdaSymbol{:}\AgdaSpace{}%
\AgdaSymbol{∀\{}\AgdaBound{L}\AgdaSpace{}%
\AgdaBound{M}\AgdaSpace{}%
\AgdaBound{N}\AgdaSpace{}%
\AgdaSymbol{:}\AgdaSpace{}%
\AgdaDatatype{Term}\AgdaSymbol{\}}\<%
\\
\>[0][@{}l@{\AgdaIndent{0}}]%
\>[3]\AgdaSymbol{→}\AgdaSpace{}%
\AgdaFunction{reducible}\AgdaSpace{}%
\AgdaBound{L}\<%
\\
\>[3]\AgdaSymbol{→}\AgdaSpace{}%
\AgdaBound{L}\AgdaSpace{}%
\AgdaOperator{\AgdaInductiveConstructor{·}}\AgdaSpace{}%
\AgdaBound{M}%
\>[12]\AgdaOperator{\AgdaDatatype{⟶}}\AgdaSpace{}%
\AgdaBound{N}\<%
\\
\>[3]\AgdaSymbol{→}\AgdaSpace{}%
\AgdaFunction{∃[}\AgdaSpace{}%
\AgdaBound{L′}\AgdaSpace{}%
\AgdaFunction{]}\AgdaSpace{}%
\AgdaSymbol{((}\AgdaBound{L}\AgdaSpace{}%
\AgdaOperator{\AgdaDatatype{⟶}}\AgdaSpace{}%
\AgdaBound{L′}\AgdaSymbol{)}\AgdaSpace{}%
\AgdaOperator{\AgdaFunction{×}}\AgdaSpace{}%
\AgdaSymbol{(}\AgdaBound{N}\AgdaSpace{}%
\AgdaOperator{\AgdaDatatype{≡}}\AgdaSpace{}%
\AgdaBound{L′}\AgdaSpace{}%
\AgdaOperator{\AgdaInductiveConstructor{·}}\AgdaSpace{}%
\AgdaBound{M}\AgdaSymbol{))}\<%
\\
\>[0]\AgdaFunction{app-invL}\AgdaSpace{}%
\AgdaBound{rl}\AgdaSpace{}%
\AgdaSymbol{(}\AgdaInductiveConstructor{ξ}\AgdaSpace{}%
\AgdaSymbol{(}\AgdaOperator{\AgdaInductiveConstructor{□·}}\AgdaSpace{}%
\AgdaBound{M}\AgdaSymbol{)}\AgdaSpace{}%
\AgdaBound{L→L′}\AgdaSymbol{)}\AgdaSpace{}%
\AgdaSymbol{=}\AgdaSpace{}%
\AgdaSymbol{\AgdaUnderscore{}}\AgdaSpace{}%
\AgdaOperator{\AgdaInductiveConstructor{,}}\AgdaSpace{}%
\AgdaSymbol{(}\AgdaBound{L→L′}\AgdaSpace{}%
\AgdaOperator{\AgdaInductiveConstructor{,}}\AgdaSpace{}%
\AgdaInductiveConstructor{refl}\AgdaSymbol{)}\<%
\\
\>[0]\AgdaFunction{app-invL}\AgdaSpace{}%
\AgdaSymbol{(}\AgdaBound{L′}\AgdaSpace{}%
\AgdaOperator{\AgdaInductiveConstructor{,}}\AgdaSpace{}%
\AgdaBound{L→L′}\AgdaSymbol{)}\AgdaSpace{}%
\AgdaSymbol{(}\AgdaInductiveConstructor{ξ}\AgdaSpace{}%
\AgdaSymbol{(}\AgdaBound{v}\AgdaSpace{}%
\AgdaOperator{\AgdaInductiveConstructor{·□}}\AgdaSymbol{)}\AgdaSpace{}%
\AgdaBound{M→M′}\AgdaSymbol{)}\AgdaSpace{}%
\AgdaSymbol{=}\AgdaSpace{}%
\AgdaFunction{⊥-elim}\AgdaSpace{}%
\AgdaSymbol{(}\AgdaFunction{value-irreducible}\AgdaSpace{}%
\AgdaBound{v}\AgdaSpace{}%
\AgdaBound{L→L′}\AgdaSymbol{)}\<%
\\
\>[0]\AgdaFunction{app-invL}\AgdaSpace{}%
\AgdaSymbol{(}\AgdaBound{L′}\AgdaSpace{}%
\AgdaOperator{\AgdaInductiveConstructor{,}}\AgdaSpace{}%
\AgdaBound{L→L′}\AgdaSymbol{)}\AgdaSpace{}%
\AgdaSymbol{(}\AgdaInductiveConstructor{ξ-blame}\AgdaSpace{}%
\AgdaSymbol{(}\AgdaOperator{\AgdaInductiveConstructor{□·}}\AgdaSpace{}%
\AgdaBound{M}\AgdaSymbol{))}\AgdaSpace{}%
\AgdaSymbol{=}\AgdaSpace{}%
\AgdaFunction{⊥-elim}\AgdaSpace{}%
\AgdaSymbol{(}\AgdaFunction{blame-irreducible}\AgdaSpace{}%
\AgdaBound{L→L′}\AgdaSymbol{)}\<%
\\
\>[0]\AgdaFunction{app-invL}\AgdaSpace{}%
\AgdaSymbol{(}\AgdaBound{L′}\AgdaSpace{}%
\AgdaOperator{\AgdaInductiveConstructor{,}}\AgdaSpace{}%
\AgdaBound{L→L′}\AgdaSymbol{)}\AgdaSpace{}%
\AgdaSymbol{(}\AgdaInductiveConstructor{ξ-blame}\AgdaSpace{}%
\AgdaSymbol{(}\AgdaBound{v}\AgdaSpace{}%
\AgdaOperator{\AgdaInductiveConstructor{·□}}\AgdaSymbol{))}\AgdaSpace{}%
\AgdaSymbol{=}\AgdaSpace{}%
\AgdaFunction{⊥-elim}\AgdaSpace{}%
\AgdaSymbol{(}\AgdaFunction{value-irreducible}\AgdaSpace{}%
\AgdaBound{v}\AgdaSpace{}%
\AgdaBound{L→L′}\AgdaSymbol{)}\<%
\\
\>[0]\AgdaFunction{app-invL}\AgdaSpace{}%
\AgdaSymbol{(}\AgdaBound{L′}\AgdaSpace{}%
\AgdaOperator{\AgdaInductiveConstructor{,}}\AgdaSpace{}%
\AgdaBound{L→L′}\AgdaSymbol{)}\AgdaSpace{}%
\AgdaSymbol{(}\AgdaInductiveConstructor{β}\AgdaSpace{}%
\AgdaBound{v}\AgdaSymbol{)}\AgdaSpace{}%
\AgdaSymbol{=}\AgdaSpace{}%
\AgdaFunction{⊥-elim}\AgdaSpace{}%
\AgdaSymbol{(}\AgdaFunction{value-irreducible}\AgdaSpace{}%
\AgdaSymbol{(}\AgdaOperator{\AgdaInductiveConstructor{ƛ̬}}\AgdaSpace{}%
\AgdaSymbol{\AgdaUnderscore{})}\AgdaSpace{}%
\AgdaBound{L→L′}\AgdaSymbol{)}\<%
\\
\\[\AgdaEmptyExtraSkip]%
\>[0]\AgdaFunction{blame-frame}\AgdaSpace{}%
\AgdaSymbol{:}\AgdaSpace{}%
\AgdaSymbol{∀\{}\AgdaBound{F}\AgdaSymbol{\}\{}\AgdaBound{N}\AgdaSymbol{\}}\<%
\\
\>[0][@{}l@{\AgdaIndent{0}}]%
\>[3]\AgdaSymbol{→}\AgdaSpace{}%
\AgdaSymbol{(}\AgdaBound{F}\AgdaSpace{}%
\AgdaOperator{\AgdaFunction{⟦}}\AgdaSpace{}%
\AgdaInductiveConstructor{blame}\AgdaSpace{}%
\AgdaOperator{\AgdaFunction{⟧}}\AgdaSymbol{)}\AgdaSpace{}%
\AgdaOperator{\AgdaDatatype{⟶}}\AgdaSpace{}%
\AgdaBound{N}\<%
\\
\>[3]\AgdaSymbol{→}\AgdaSpace{}%
\AgdaBound{N}\AgdaSpace{}%
\AgdaOperator{\AgdaDatatype{≡}}\AgdaSpace{}%
\AgdaInductiveConstructor{blame}\<%
\\
\>[0]\AgdaFunction{blame-frame}\AgdaSpace{}%
\AgdaSymbol{\{}\AgdaOperator{\AgdaInductiveConstructor{□·}}\AgdaSpace{}%
\AgdaBound{M}\AgdaSymbol{\}}\AgdaSpace{}%
\AgdaSymbol{\{}\AgdaDottedPattern{\AgdaSymbol{.((}}\AgdaDottedPattern{\AgdaOperator{\AgdaInductiveConstructor{□·}}}\AgdaSpace{}%
\AgdaDottedPattern{\AgdaBound{M₁}}\AgdaDottedPattern{\AgdaSymbol{)}}\AgdaSpace{}%
\AgdaDottedPattern{\AgdaOperator{\AgdaFunction{⟦}}}\AgdaSpace{}%
\AgdaDottedPattern{\AgdaSymbol{\AgdaUnderscore{}}}\AgdaSpace{}%
\AgdaDottedPattern{\AgdaOperator{\AgdaFunction{⟧}}}\AgdaDottedPattern{\AgdaSymbol{)}}\AgdaSymbol{\}}\AgdaSpace{}%
\AgdaSymbol{(}\AgdaInductiveConstructor{ξξ}\AgdaSpace{}%
\AgdaSymbol{(}\AgdaOperator{\AgdaInductiveConstructor{□·}}\AgdaSpace{}%
\AgdaBound{M₁}\AgdaSymbol{)}\AgdaSpace{}%
\AgdaInductiveConstructor{refl}\AgdaSpace{}%
\AgdaInductiveConstructor{refl}\AgdaSpace{}%
\AgdaBound{Fb→N}\AgdaSymbol{)}\AgdaSpace{}%
\AgdaSymbol{=}\<%
\\
\>[0][@{}l@{\AgdaIndent{0}}]%
\>[4]\AgdaFunction{⊥-elim}\AgdaSpace{}%
\AgdaSymbol{(}\AgdaFunction{blame-irreducible}\AgdaSpace{}%
\AgdaBound{Fb→N}\AgdaSymbol{)}\<%
\\
\>[0]\AgdaFunction{blame-frame}\AgdaSpace{}%
\AgdaSymbol{\{}\AgdaOperator{\AgdaInductiveConstructor{□·}}\AgdaSpace{}%
\AgdaBound{M}\AgdaSymbol{\}}\AgdaSpace{}%
\AgdaSymbol{(}\AgdaInductiveConstructor{ξξ}\AgdaSpace{}%
\AgdaSymbol{(()}\AgdaSpace{}%
\AgdaOperator{\AgdaInductiveConstructor{·□}}\AgdaSymbol{)}\AgdaSpace{}%
\AgdaInductiveConstructor{refl}\AgdaSpace{}%
\AgdaInductiveConstructor{refl}\AgdaSpace{}%
\AgdaBound{Fb→N}\AgdaSymbol{)}\<%
\\
\>[0]\AgdaFunction{blame-frame}\AgdaSpace{}%
\AgdaSymbol{\{}\AgdaOperator{\AgdaInductiveConstructor{□·}}\AgdaSpace{}%
\AgdaBound{M}\AgdaSymbol{\}}\AgdaSpace{}%
\AgdaSymbol{\{}\AgdaDottedPattern{\AgdaSymbol{.}}\AgdaDottedPattern{\AgdaInductiveConstructor{blame}}\AgdaSymbol{\}}\AgdaSpace{}%
\AgdaSymbol{(}\AgdaInductiveConstructor{ξξ-blame}\AgdaSpace{}%
\AgdaSymbol{(}\AgdaOperator{\AgdaInductiveConstructor{□·}}\AgdaSpace{}%
\AgdaBound{M₁}\AgdaSymbol{)}\AgdaSpace{}%
\AgdaInductiveConstructor{refl}\AgdaSymbol{)}\AgdaSpace{}%
\AgdaSymbol{=}\AgdaSpace{}%
\AgdaInductiveConstructor{refl}\<%
\\
\>[0]\AgdaFunction{blame-frame}\AgdaSpace{}%
\AgdaSymbol{\{}\AgdaOperator{\AgdaInductiveConstructor{□·}}\AgdaSpace{}%
\AgdaBound{M}\AgdaSymbol{\}}\AgdaSpace{}%
\AgdaSymbol{\{}\AgdaDottedPattern{\AgdaSymbol{.}}\AgdaDottedPattern{\AgdaInductiveConstructor{blame}}\AgdaSymbol{\}}\AgdaSpace{}%
\AgdaSymbol{(}\AgdaInductiveConstructor{ξξ-blame}\AgdaSpace{}%
\AgdaSymbol{(()}\AgdaSpace{}%
\AgdaOperator{\AgdaInductiveConstructor{·□}}\AgdaSymbol{)}\AgdaSpace{}%
\AgdaInductiveConstructor{refl}\AgdaSymbol{)}\<%
\\
\>[0]\AgdaFunction{blame-frame}\AgdaSpace{}%
\AgdaSymbol{\{}\AgdaBound{v}\AgdaSpace{}%
\AgdaOperator{\AgdaInductiveConstructor{·□}}\AgdaSymbol{\}}\AgdaSpace{}%
\AgdaSymbol{\{}\AgdaBound{N}\AgdaSymbol{\}}\AgdaSpace{}%
\AgdaSymbol{(}\AgdaInductiveConstructor{ξξ}\AgdaSpace{}%
\AgdaSymbol{(}\AgdaOperator{\AgdaInductiveConstructor{□·}}\AgdaSpace{}%
\AgdaBound{M}\AgdaSymbol{)}\AgdaSpace{}%
\AgdaInductiveConstructor{refl}\AgdaSpace{}%
\AgdaInductiveConstructor{refl}\AgdaSpace{}%
\AgdaBound{Fb→N}\AgdaSymbol{)}\AgdaSpace{}%
\AgdaSymbol{=}\<%
\\
\>[0][@{}l@{\AgdaIndent{0}}]%
\>[4]\AgdaFunction{⊥-elim}\AgdaSpace{}%
\AgdaSymbol{(}\AgdaFunction{value-irreducible}\AgdaSpace{}%
\AgdaBound{v}\AgdaSpace{}%
\AgdaBound{Fb→N}\AgdaSymbol{)}\<%
\\
\>[0]\AgdaFunction{blame-frame}\AgdaSpace{}%
\AgdaSymbol{\{}\AgdaBound{v}\AgdaSpace{}%
\AgdaOperator{\AgdaInductiveConstructor{·□}}\AgdaSymbol{\}}\AgdaSpace{}%
\AgdaSymbol{\{}\AgdaBound{N}\AgdaSymbol{\}}\AgdaSpace{}%
\AgdaSymbol{(}\AgdaInductiveConstructor{ξξ}\AgdaSpace{}%
\AgdaSymbol{(}\AgdaBound{v₁}\AgdaSpace{}%
\AgdaOperator{\AgdaInductiveConstructor{·□}}\AgdaSymbol{)}\AgdaSpace{}%
\AgdaInductiveConstructor{refl}\AgdaSpace{}%
\AgdaInductiveConstructor{refl}\AgdaSpace{}%
\AgdaBound{Fb→N}\AgdaSymbol{)}\AgdaSpace{}%
\AgdaSymbol{=}\<%
\\
\>[0][@{}l@{\AgdaIndent{0}}]%
\>[4]\AgdaFunction{⊥-elim}\AgdaSpace{}%
\AgdaSymbol{(}\AgdaFunction{blame-irreducible}\AgdaSpace{}%
\AgdaBound{Fb→N}\AgdaSymbol{)}\<%
\\
\>[0]\AgdaFunction{blame-frame}\AgdaSpace{}%
\AgdaSymbol{\{}\AgdaBound{v}\AgdaSpace{}%
\AgdaOperator{\AgdaInductiveConstructor{·□}}\AgdaSymbol{\}}\AgdaSpace{}%
\AgdaSymbol{\{}\AgdaDottedPattern{\AgdaSymbol{.}}\AgdaDottedPattern{\AgdaInductiveConstructor{blame}}\AgdaSymbol{\}}\AgdaSpace{}%
\AgdaSymbol{(}\AgdaInductiveConstructor{ξξ-blame}\AgdaSpace{}%
\AgdaBound{F}\AgdaSpace{}%
\AgdaBound{x}\AgdaSymbol{)}\AgdaSpace{}%
\AgdaSymbol{=}\AgdaSpace{}%
\AgdaInductiveConstructor{refl}\<%
\\
\>[0]\AgdaFunction{blame-frame}\AgdaSpace{}%
\AgdaSymbol{\{}\AgdaOperator{\AgdaInductiveConstructor{□⟨}}\AgdaSpace{}%
\AgdaBound{G}\AgdaSpace{}%
\AgdaOperator{\AgdaInductiveConstructor{!⟩}}\AgdaSymbol{\}}\AgdaSpace{}%
\AgdaSymbol{\{\AgdaUnderscore{}\}}\AgdaSpace{}%
\AgdaSymbol{(}\AgdaInductiveConstructor{ξξ}\AgdaSpace{}%
\AgdaOperator{\AgdaInductiveConstructor{□⟨}}\AgdaSpace{}%
\AgdaSymbol{\AgdaUnderscore{}}\AgdaSpace{}%
\AgdaOperator{\AgdaInductiveConstructor{!⟩}}\AgdaSpace{}%
\AgdaInductiveConstructor{refl}\AgdaSpace{}%
\AgdaInductiveConstructor{refl}\AgdaSpace{}%
\AgdaBound{Fb→N}\AgdaSymbol{)}\AgdaSpace{}%
\AgdaSymbol{=}\<%
\\
\>[0][@{}l@{\AgdaIndent{0}}]%
\>[4]\AgdaFunction{⊥-elim}\AgdaSpace{}%
\AgdaSymbol{(}\AgdaFunction{blame-irreducible}\AgdaSpace{}%
\AgdaBound{Fb→N}\AgdaSymbol{)}\<%
\\
\>[0]\AgdaFunction{blame-frame}\AgdaSpace{}%
\AgdaSymbol{\{}\AgdaOperator{\AgdaInductiveConstructor{□⟨}}\AgdaSpace{}%
\AgdaBound{G}\AgdaSpace{}%
\AgdaOperator{\AgdaInductiveConstructor{!⟩}}\AgdaSymbol{\}}\AgdaSpace{}%
\AgdaSymbol{\{}\AgdaDottedPattern{\AgdaSymbol{.}}\AgdaDottedPattern{\AgdaInductiveConstructor{blame}}\AgdaSymbol{\}}\AgdaSpace{}%
\AgdaSymbol{(}\AgdaInductiveConstructor{ξξ-blame}\AgdaSpace{}%
\AgdaBound{F}\AgdaSpace{}%
\AgdaBound{x}\AgdaSymbol{)}\AgdaSpace{}%
\AgdaSymbol{=}\AgdaSpace{}%
\AgdaInductiveConstructor{refl}\<%
\\
\>[0]\AgdaFunction{blame-frame}\AgdaSpace{}%
\AgdaSymbol{\{}\AgdaOperator{\AgdaInductiveConstructor{□⟨}}\AgdaSpace{}%
\AgdaBound{H}\AgdaSpace{}%
\AgdaOperator{\AgdaInductiveConstructor{?⟩}}\AgdaSymbol{\}}\AgdaSpace{}%
\AgdaSymbol{\{}\AgdaBound{N}\AgdaSymbol{\}}\AgdaSpace{}%
\AgdaSymbol{(}\AgdaInductiveConstructor{ξξ}\AgdaSpace{}%
\AgdaOperator{\AgdaInductiveConstructor{□⟨}}\AgdaSpace{}%
\AgdaSymbol{\AgdaUnderscore{}}\AgdaSpace{}%
\AgdaOperator{\AgdaInductiveConstructor{?⟩}}\AgdaSpace{}%
\AgdaInductiveConstructor{refl}\AgdaSpace{}%
\AgdaInductiveConstructor{refl}\AgdaSpace{}%
\AgdaBound{Fb→N}\AgdaSymbol{)}\AgdaSpace{}%
\AgdaSymbol{=}\<%
\\
\>[0][@{}l@{\AgdaIndent{0}}]%
\>[4]\AgdaFunction{⊥-elim}\AgdaSpace{}%
\AgdaSymbol{(}\AgdaFunction{blame-irreducible}\AgdaSpace{}%
\AgdaBound{Fb→N}\AgdaSymbol{)}\<%
\\
\>[0]\AgdaFunction{blame-frame}\AgdaSpace{}%
\AgdaSymbol{\{}\AgdaOperator{\AgdaInductiveConstructor{□⟨}}\AgdaSpace{}%
\AgdaBound{H}\AgdaSpace{}%
\AgdaOperator{\AgdaInductiveConstructor{?⟩}}\AgdaSymbol{\}}\AgdaSpace{}%
\AgdaSymbol{\{}\AgdaDottedPattern{\AgdaSymbol{.}}\AgdaDottedPattern{\AgdaInductiveConstructor{blame}}\AgdaSymbol{\}}\AgdaSpace{}%
\AgdaSymbol{(}\AgdaInductiveConstructor{ξξ-blame}\AgdaSpace{}%
\AgdaOperator{\AgdaInductiveConstructor{□⟨}}\AgdaSpace{}%
\AgdaSymbol{\AgdaUnderscore{}}\AgdaSpace{}%
\AgdaOperator{\AgdaInductiveConstructor{?⟩}}\AgdaSpace{}%
\AgdaBound{x}\AgdaSymbol{)}\AgdaSpace{}%
\AgdaSymbol{=}\AgdaSpace{}%
\AgdaInductiveConstructor{refl}\<%
\\
\\[\AgdaEmptyExtraSkip]%
\>[0]\AgdaFunction{collapse-inv}\AgdaSpace{}%
\AgdaSymbol{:}\AgdaSpace{}%
\AgdaSymbol{∀\{}\AgdaBound{V}\AgdaSymbol{\}\{}\AgdaBound{N}\AgdaSymbol{\}\{}\AgdaBound{G}\AgdaSymbol{\}}\<%
\\
\>[0][@{}l@{\AgdaIndent{0}}]%
\>[3]\AgdaSymbol{→}\AgdaSpace{}%
\AgdaDatatype{Value}\AgdaSpace{}%
\AgdaBound{V}\<%
\\
\>[3]\AgdaSymbol{→}\AgdaSpace{}%
\AgdaSymbol{((}\AgdaBound{V}\AgdaSpace{}%
\AgdaOperator{\AgdaInductiveConstructor{⟨}}\AgdaSpace{}%
\AgdaBound{G}\AgdaSpace{}%
\AgdaOperator{\AgdaInductiveConstructor{!⟩}}\AgdaSymbol{)}\AgdaSpace{}%
\AgdaOperator{\AgdaInductiveConstructor{⟨}}\AgdaSpace{}%
\AgdaBound{G}\AgdaSpace{}%
\AgdaOperator{\AgdaInductiveConstructor{?⟩}}\AgdaSymbol{)}\AgdaSpace{}%
\AgdaOperator{\AgdaDatatype{⟶}}\AgdaSpace{}%
\AgdaBound{N}\<%
\\
\>[3]\AgdaSymbol{→}\AgdaSpace{}%
\AgdaBound{N}\AgdaSpace{}%
\AgdaOperator{\AgdaDatatype{≡}}\AgdaSpace{}%
\AgdaBound{V}\<%
\\
\>[0]\AgdaFunction{collapse-inv}\AgdaSpace{}%
\AgdaSymbol{\{}\AgdaBound{V}\AgdaSymbol{\}}\AgdaSpace{}%
\AgdaSymbol{\{}\AgdaBound{N}\AgdaSymbol{\}}\AgdaSpace{}%
\AgdaBound{v}\AgdaSpace{}%
\AgdaSymbol{(}\AgdaInductiveConstructor{ξξ}\AgdaSpace{}%
\AgdaOperator{\AgdaInductiveConstructor{□⟨}}\AgdaSpace{}%
\AgdaBound{G}\AgdaSpace{}%
\AgdaOperator{\AgdaInductiveConstructor{?⟩}}\AgdaSpace{}%
\AgdaInductiveConstructor{refl}\AgdaSpace{}%
\AgdaBound{x₁}\AgdaSpace{}%
\AgdaBound{r}\AgdaSymbol{)}\AgdaSpace{}%
\AgdaSymbol{=}\<%
\\
\>[0][@{}l@{\AgdaIndent{0}}]%
\>[2]\AgdaFunction{⊥-elim}\AgdaSpace{}%
\AgdaSymbol{(}\AgdaFunction{value-irreducible}\AgdaSpace{}%
\AgdaSymbol{(}\AgdaBound{v}\AgdaSpace{}%
\AgdaOperator{\AgdaInductiveConstructor{〈}}\AgdaSpace{}%
\AgdaBound{G}\AgdaSpace{}%
\AgdaOperator{\AgdaInductiveConstructor{〉}}\AgdaSymbol{)}\AgdaSpace{}%
\AgdaBound{r}\AgdaSymbol{)}\<%
\\
\>[0]\AgdaFunction{collapse-inv}\AgdaSpace{}%
\AgdaSymbol{\{}\AgdaBound{V}\AgdaSymbol{\}}\AgdaSpace{}%
\AgdaSymbol{\{}\AgdaDottedPattern{\AgdaSymbol{.}}\AgdaDottedPattern{\AgdaInductiveConstructor{blame}}\AgdaSymbol{\}}\AgdaSpace{}%
\AgdaBound{v}\AgdaSpace{}%
\AgdaSymbol{(}\AgdaInductiveConstructor{ξξ-blame}\AgdaSpace{}%
\AgdaSymbol{(}\AgdaOperator{\AgdaInductiveConstructor{□·}}\AgdaSpace{}%
\AgdaBound{M}\AgdaSymbol{)}\AgdaSpace{}%
\AgdaSymbol{())}\<%
\\
\>[0]\AgdaFunction{collapse-inv}\AgdaSpace{}%
\AgdaSymbol{\{}\AgdaBound{V}\AgdaSymbol{\}}\AgdaSpace{}%
\AgdaSymbol{\{}\AgdaDottedPattern{\AgdaSymbol{.}}\AgdaDottedPattern{\AgdaInductiveConstructor{blame}}\AgdaSymbol{\}}\AgdaSpace{}%
\AgdaBound{v}\AgdaSpace{}%
\AgdaSymbol{(}\AgdaInductiveConstructor{ξξ-blame}\AgdaSpace{}%
\AgdaSymbol{(}\AgdaBound{v₁}\AgdaSpace{}%
\AgdaOperator{\AgdaInductiveConstructor{·□}}\AgdaSymbol{)}\AgdaSpace{}%
\AgdaSymbol{())}\<%
\\
\>[0]\AgdaFunction{collapse-inv}\AgdaSpace{}%
\AgdaSymbol{\{}\AgdaBound{V}\AgdaSymbol{\}}\AgdaSpace{}%
\AgdaSymbol{\{}\AgdaDottedPattern{\AgdaSymbol{.}}\AgdaDottedPattern{\AgdaInductiveConstructor{blame}}\AgdaSymbol{\}}\AgdaSpace{}%
\AgdaBound{v}\AgdaSpace{}%
\AgdaSymbol{(}\AgdaInductiveConstructor{ξξ-blame}\AgdaSpace{}%
\AgdaOperator{\AgdaInductiveConstructor{□⟨}}\AgdaSpace{}%
\AgdaBound{G}\AgdaSpace{}%
\AgdaOperator{\AgdaInductiveConstructor{!⟩}}\AgdaSpace{}%
\AgdaSymbol{())}\<%
\\
\>[0]\AgdaFunction{collapse-inv}\AgdaSpace{}%
\AgdaSymbol{\{}\AgdaBound{V}\AgdaSymbol{\}}\AgdaSpace{}%
\AgdaSymbol{\{}\AgdaDottedPattern{\AgdaSymbol{.}}\AgdaDottedPattern{\AgdaInductiveConstructor{blame}}\AgdaSymbol{\}}\AgdaSpace{}%
\AgdaBound{v}\AgdaSpace{}%
\AgdaSymbol{(}\AgdaInductiveConstructor{ξξ-blame}\AgdaSpace{}%
\AgdaOperator{\AgdaInductiveConstructor{□⟨}}\AgdaSpace{}%
\AgdaBound{H}\AgdaSpace{}%
\AgdaOperator{\AgdaInductiveConstructor{?⟩}}\AgdaSpace{}%
\AgdaSymbol{())}\<%
\\
\>[0]\AgdaFunction{collapse-inv}\AgdaSpace{}%
\AgdaSymbol{\{}\AgdaBound{V}\AgdaSymbol{\}}\AgdaSpace{}%
\AgdaSymbol{\{}\AgdaDottedPattern{\AgdaSymbol{.}}\AgdaDottedPattern{\AgdaBound{V}}\AgdaSymbol{\}}\AgdaSpace{}%
\AgdaBound{v}\AgdaSpace{}%
\AgdaSymbol{(}\AgdaInductiveConstructor{collapse}\AgdaSpace{}%
\AgdaBound{x}\AgdaSpace{}%
\AgdaInductiveConstructor{refl}\AgdaSymbol{)}\AgdaSpace{}%
\AgdaSymbol{=}\AgdaSpace{}%
\AgdaInductiveConstructor{refl}\<%
\\
\>[0]\AgdaFunction{collapse-inv}\AgdaSpace{}%
\AgdaSymbol{\{}\AgdaBound{V}\AgdaSymbol{\}}\AgdaSpace{}%
\AgdaSymbol{\{}\AgdaDottedPattern{\AgdaSymbol{.}}\AgdaDottedPattern{\AgdaInductiveConstructor{blame}}\AgdaSymbol{\}}\AgdaSpace{}%
\AgdaBound{v}\AgdaSpace{}%
\AgdaSymbol{(}\AgdaInductiveConstructor{collide}\AgdaSpace{}%
\AgdaBound{x}\AgdaSpace{}%
\AgdaBound{x₁}\AgdaSpace{}%
\AgdaInductiveConstructor{refl}\AgdaSymbol{)}\AgdaSpace{}%
\AgdaSymbol{=}\AgdaSpace{}%
\AgdaFunction{⊥-elim}\AgdaSpace{}%
\AgdaSymbol{(}\AgdaBound{x₁}\AgdaSpace{}%
\AgdaInductiveConstructor{refl}\AgdaSymbol{)}\<%
\\
\\[\AgdaEmptyExtraSkip]%
\>[0]\AgdaFunction{collide-inv}\AgdaSpace{}%
\AgdaSymbol{:}\AgdaSpace{}%
\AgdaSymbol{∀\{}\AgdaBound{V}\AgdaSymbol{\}\{}\AgdaBound{N}\AgdaSymbol{\}\{}\AgdaBound{G}\AgdaSymbol{\}\{}\AgdaBound{H}\AgdaSymbol{\}}\<%
\\
\>[0][@{}l@{\AgdaIndent{0}}]%
\>[3]\AgdaSymbol{→}\AgdaSpace{}%
\AgdaBound{G}\AgdaSpace{}%
\AgdaOperator{\AgdaFunction{≢}}\AgdaSpace{}%
\AgdaBound{H}\<%
\\
\>[3]\AgdaSymbol{→}\AgdaSpace{}%
\AgdaDatatype{Value}\AgdaSpace{}%
\AgdaBound{V}\<%
\\
\>[3]\AgdaSymbol{→}\AgdaSpace{}%
\AgdaSymbol{((}\AgdaBound{V}\AgdaSpace{}%
\AgdaOperator{\AgdaInductiveConstructor{⟨}}\AgdaSpace{}%
\AgdaBound{G}\AgdaSpace{}%
\AgdaOperator{\AgdaInductiveConstructor{!⟩}}\AgdaSymbol{)}\AgdaSpace{}%
\AgdaOperator{\AgdaInductiveConstructor{⟨}}\AgdaSpace{}%
\AgdaBound{H}\AgdaSpace{}%
\AgdaOperator{\AgdaInductiveConstructor{?⟩}}\AgdaSymbol{)}\AgdaSpace{}%
\AgdaOperator{\AgdaDatatype{⟶}}\AgdaSpace{}%
\AgdaBound{N}\<%
\\
\>[3]\AgdaSymbol{→}\AgdaSpace{}%
\AgdaBound{N}\AgdaSpace{}%
\AgdaOperator{\AgdaDatatype{≡}}\AgdaSpace{}%
\AgdaInductiveConstructor{blame}\<%
\\
\>[0]\AgdaFunction{collide-inv}\AgdaSpace{}%
\AgdaSymbol{\{}\AgdaBound{V}\AgdaSymbol{\}}\AgdaSpace{}%
\AgdaSymbol{\{}\AgdaBound{N}\AgdaSymbol{\}}\AgdaSpace{}%
\AgdaSymbol{\{}\AgdaBound{G}\AgdaSymbol{\}}\AgdaSpace{}%
\AgdaSymbol{\{}\AgdaBound{H}\AgdaSymbol{\}}\AgdaSpace{}%
\AgdaBound{neq}\AgdaSpace{}%
\AgdaBound{v}\AgdaSpace{}%
\AgdaSymbol{(}\AgdaInductiveConstructor{ξξ}\AgdaSpace{}%
\AgdaOperator{\AgdaInductiveConstructor{□⟨}}\AgdaSpace{}%
\AgdaBound{H₁}\AgdaSpace{}%
\AgdaOperator{\AgdaInductiveConstructor{?⟩}}\AgdaSpace{}%
\AgdaInductiveConstructor{refl}\AgdaSpace{}%
\AgdaBound{x₁}\AgdaSpace{}%
\AgdaBound{red}\AgdaSymbol{)}\AgdaSpace{}%
\AgdaSymbol{=}\<%
\\
\>[0][@{}l@{\AgdaIndent{0}}]%
\>[2]\AgdaFunction{⊥-elim}\AgdaSpace{}%
\AgdaSymbol{(}\AgdaFunction{value-irreducible}\AgdaSpace{}%
\AgdaSymbol{(}\AgdaBound{v}\AgdaSpace{}%
\AgdaOperator{\AgdaInductiveConstructor{〈}}\AgdaSpace{}%
\AgdaBound{G}\AgdaSpace{}%
\AgdaOperator{\AgdaInductiveConstructor{〉}}\AgdaSymbol{)}\AgdaSpace{}%
\AgdaBound{red}\AgdaSymbol{)}\<%
\\
\>[0]\AgdaFunction{collide-inv}\AgdaSpace{}%
\AgdaSymbol{\{}\AgdaBound{V}\AgdaSymbol{\}}\AgdaSpace{}%
\AgdaSymbol{\{}\AgdaDottedPattern{\AgdaSymbol{.}}\AgdaDottedPattern{\AgdaInductiveConstructor{blame}}\AgdaSymbol{\}}\AgdaSpace{}%
\AgdaSymbol{\{}\AgdaBound{G}\AgdaSymbol{\}}\AgdaSpace{}%
\AgdaSymbol{\{}\AgdaBound{H}\AgdaSymbol{\}}\AgdaSpace{}%
\AgdaBound{neq}\AgdaSpace{}%
\AgdaBound{v}\AgdaSpace{}%
\AgdaSymbol{(}\AgdaInductiveConstructor{ξξ-blame}\AgdaSpace{}%
\AgdaSymbol{(}\AgdaOperator{\AgdaInductiveConstructor{□·}}\AgdaSpace{}%
\AgdaBound{M}\AgdaSymbol{)}\AgdaSpace{}%
\AgdaSymbol{())}\<%
\\
\>[0]\AgdaFunction{collide-inv}\AgdaSpace{}%
\AgdaSymbol{\{}\AgdaBound{V}\AgdaSymbol{\}}\AgdaSpace{}%
\AgdaSymbol{\{}\AgdaDottedPattern{\AgdaSymbol{.}}\AgdaDottedPattern{\AgdaInductiveConstructor{blame}}\AgdaSymbol{\}}\AgdaSpace{}%
\AgdaSymbol{\{}\AgdaBound{G}\AgdaSymbol{\}}\AgdaSpace{}%
\AgdaSymbol{\{}\AgdaBound{H}\AgdaSymbol{\}}\AgdaSpace{}%
\AgdaBound{neq}\AgdaSpace{}%
\AgdaBound{v}\AgdaSpace{}%
\AgdaSymbol{(}\AgdaInductiveConstructor{ξξ-blame}\AgdaSpace{}%
\AgdaSymbol{(}\AgdaBound{v₁}\AgdaSpace{}%
\AgdaOperator{\AgdaInductiveConstructor{·□}}\AgdaSymbol{)}\AgdaSpace{}%
\AgdaSymbol{())}\<%
\\
\>[0]\AgdaFunction{collide-inv}\AgdaSpace{}%
\AgdaSymbol{\{}\AgdaBound{V}\AgdaSymbol{\}}\AgdaSpace{}%
\AgdaSymbol{\{}\AgdaDottedPattern{\AgdaSymbol{.}}\AgdaDottedPattern{\AgdaInductiveConstructor{blame}}\AgdaSymbol{\}}\AgdaSpace{}%
\AgdaSymbol{\{}\AgdaBound{G}\AgdaSymbol{\}}\AgdaSpace{}%
\AgdaSymbol{\{}\AgdaBound{H}\AgdaSymbol{\}}\AgdaSpace{}%
\AgdaBound{neq}\AgdaSpace{}%
\AgdaBound{v}\AgdaSpace{}%
\AgdaSymbol{(}\AgdaInductiveConstructor{ξξ-blame}\AgdaSpace{}%
\AgdaOperator{\AgdaInductiveConstructor{□⟨}}\AgdaSpace{}%
\AgdaBound{G₁}\AgdaSpace{}%
\AgdaOperator{\AgdaInductiveConstructor{!⟩}}\AgdaSpace{}%
\AgdaSymbol{())}\<%
\\
\>[0]\AgdaFunction{collide-inv}\AgdaSpace{}%
\AgdaSymbol{\{}\AgdaBound{V}\AgdaSymbol{\}}\AgdaSpace{}%
\AgdaSymbol{\{}\AgdaDottedPattern{\AgdaSymbol{.}}\AgdaDottedPattern{\AgdaInductiveConstructor{blame}}\AgdaSymbol{\}}\AgdaSpace{}%
\AgdaSymbol{\{}\AgdaBound{G}\AgdaSymbol{\}}\AgdaSpace{}%
\AgdaSymbol{\{}\AgdaBound{H}\AgdaSymbol{\}}\AgdaSpace{}%
\AgdaBound{neq}\AgdaSpace{}%
\AgdaBound{v}\AgdaSpace{}%
\AgdaSymbol{(}\AgdaInductiveConstructor{ξξ-blame}\AgdaSpace{}%
\AgdaOperator{\AgdaInductiveConstructor{□⟨}}\AgdaSpace{}%
\AgdaBound{H₁}\AgdaSpace{}%
\AgdaOperator{\AgdaInductiveConstructor{?⟩}}\AgdaSpace{}%
\AgdaSymbol{())}\<%
\\
\>[0]\AgdaFunction{collide-inv}\AgdaSpace{}%
\AgdaSymbol{\{}\AgdaBound{V}\AgdaSymbol{\}}\AgdaSpace{}%
\AgdaSymbol{\{}\AgdaBound{N}\AgdaSymbol{\}}\AgdaSpace{}%
\AgdaSymbol{\{}\AgdaBound{G}\AgdaSymbol{\}}\AgdaSpace{}%
\AgdaSymbol{\{}\AgdaBound{H}\AgdaSymbol{\}}\AgdaSpace{}%
\AgdaBound{neq}\AgdaSpace{}%
\AgdaBound{v}\AgdaSpace{}%
\AgdaSymbol{(}\AgdaInductiveConstructor{collapse}\AgdaSpace{}%
\AgdaBound{x}\AgdaSpace{}%
\AgdaInductiveConstructor{refl}\AgdaSymbol{)}\AgdaSpace{}%
\AgdaSymbol{=}\AgdaSpace{}%
\AgdaFunction{⊥-elim}\AgdaSpace{}%
\AgdaSymbol{(}\AgdaBound{neq}\AgdaSpace{}%
\AgdaInductiveConstructor{refl}\AgdaSymbol{)}\<%
\\
\>[0]\AgdaFunction{collide-inv}\AgdaSpace{}%
\AgdaSymbol{\{}\AgdaBound{V}\AgdaSymbol{\}}\AgdaSpace{}%
\AgdaSymbol{\{}\AgdaDottedPattern{\AgdaSymbol{.}}\AgdaDottedPattern{\AgdaInductiveConstructor{blame}}\AgdaSymbol{\}}\AgdaSpace{}%
\AgdaSymbol{\{}\AgdaBound{G}\AgdaSymbol{\}}\AgdaSpace{}%
\AgdaSymbol{\{}\AgdaBound{H}\AgdaSymbol{\}}\AgdaSpace{}%
\AgdaBound{neq}\AgdaSpace{}%
\AgdaBound{v}\AgdaSpace{}%
\AgdaSymbol{(}\AgdaInductiveConstructor{collide}\AgdaSpace{}%
\AgdaBound{x}\AgdaSpace{}%
\AgdaBound{x₁}\AgdaSpace{}%
\AgdaInductiveConstructor{refl}\AgdaSymbol{)}\AgdaSpace{}%
\AgdaSymbol{=}\AgdaSpace{}%
\AgdaInductiveConstructor{refl}\<%
\end{code}

\begin{code}[hide]%
\>[0]\AgdaFunction{inject-eq}\AgdaSpace{}%
\AgdaSymbol{:}\AgdaSpace{}%
\AgdaSymbol{∀\{}\AgdaBound{G}\AgdaSymbol{\}\{}\AgdaBound{N}\AgdaSpace{}%
\AgdaBound{N′}\AgdaSymbol{\}}\<%
\\
\>[0][@{}l@{\AgdaIndent{0}}]%
\>[3]\AgdaSymbol{→}\AgdaSpace{}%
\AgdaSymbol{(}\AgdaBound{N}\AgdaSpace{}%
\AgdaOperator{\AgdaInductiveConstructor{⟨}}\AgdaSpace{}%
\AgdaBound{G}\AgdaSpace{}%
\AgdaOperator{\AgdaInductiveConstructor{!⟩}}\AgdaSymbol{)}\AgdaSpace{}%
\AgdaOperator{\AgdaDatatype{≡}}\AgdaSpace{}%
\AgdaSymbol{(}\AgdaBound{N′}\AgdaSpace{}%
\AgdaOperator{\AgdaInductiveConstructor{⟨}}\AgdaSpace{}%
\AgdaBound{G}\AgdaSpace{}%
\AgdaOperator{\AgdaInductiveConstructor{!⟩}}\AgdaSymbol{)}\<%
\\
\>[3]\AgdaSymbol{→}\AgdaSpace{}%
\AgdaBound{N}\AgdaSpace{}%
\AgdaOperator{\AgdaDatatype{≡}}\AgdaSpace{}%
\AgdaBound{N′}\<%
\\
\>[0]\AgdaFunction{inject-eq}\AgdaSpace{}%
\AgdaSymbol{\{}\AgdaBound{G}\AgdaSymbol{\}}\AgdaSpace{}%
\AgdaSymbol{\{}\AgdaBound{N}\AgdaSymbol{\}}\AgdaSpace{}%
\AgdaSymbol{\{}\AgdaDottedPattern{\AgdaSymbol{.}}\AgdaDottedPattern{\AgdaBound{N}}\AgdaSymbol{\}}\AgdaSpace{}%
\AgdaInductiveConstructor{refl}\AgdaSpace{}%
\AgdaSymbol{=}\AgdaSpace{}%
\AgdaInductiveConstructor{refl}\<%
\\
\\[\AgdaEmptyExtraSkip]%
\>[0]\AgdaFunction{deterministic}\AgdaSpace{}%
\AgdaSymbol{:}\AgdaSpace{}%
\AgdaSymbol{∀\{}\AgdaBound{M}\AgdaSpace{}%
\AgdaBound{N}\AgdaSpace{}%
\AgdaBound{N′}\AgdaSymbol{\}}\<%
\\
\>[0][@{}l@{\AgdaIndent{0}}]%
\>[2]\AgdaSymbol{→}\AgdaSpace{}%
\AgdaBound{M}\AgdaSpace{}%
\AgdaOperator{\AgdaDatatype{⟶}}\AgdaSpace{}%
\AgdaBound{N}\<%
\\
\>[2]\AgdaSymbol{→}\AgdaSpace{}%
\AgdaBound{M}\AgdaSpace{}%
\AgdaOperator{\AgdaDatatype{⟶}}\AgdaSpace{}%
\AgdaBound{N′}\<%
\\
\>[2]\AgdaSymbol{→}\AgdaSpace{}%
\AgdaBound{N}\AgdaSpace{}%
\AgdaOperator{\AgdaDatatype{≡}}\AgdaSpace{}%
\AgdaBound{N′}\<%
\\
\>[0]\AgdaFunction{deterministic}\AgdaSpace{}%
\AgdaSymbol{(}\AgdaInductiveConstructor{ξ}\AgdaSpace{}%
\AgdaSymbol{(}\AgdaOperator{\AgdaInductiveConstructor{□·}}\AgdaSpace{}%
\AgdaBound{M}\AgdaSymbol{)}\AgdaSpace{}%
\AgdaBound{M⟶N}\AgdaSymbol{)}\AgdaSpace{}%
\AgdaSymbol{(}\AgdaInductiveConstructor{ξ}\AgdaSpace{}%
\AgdaSymbol{(}\AgdaOperator{\AgdaInductiveConstructor{□·}}\AgdaSpace{}%
\AgdaBound{M₁}\AgdaSymbol{)}\AgdaSpace{}%
\AgdaBound{M⟶N′}\AgdaSymbol{)}\<%
\\
\>[0][@{}l@{\AgdaIndent{0}}]%
\>[4]\AgdaKeyword{with}\AgdaSpace{}%
\AgdaFunction{deterministic}\AgdaSpace{}%
\AgdaBound{M⟶N}\AgdaSpace{}%
\AgdaBound{M⟶N′}\<%
\\
\>[0]\AgdaSymbol{...}\AgdaSpace{}%
\AgdaSymbol{|}\AgdaSpace{}%
\AgdaInductiveConstructor{refl}\AgdaSpace{}%
\AgdaSymbol{=}\AgdaSpace{}%
\AgdaInductiveConstructor{refl}\<%
\\
\>[0]\AgdaFunction{deterministic}\AgdaSpace{}%
\AgdaSymbol{(}\AgdaInductiveConstructor{ξ}\AgdaSpace{}%
\AgdaSymbol{(}\AgdaOperator{\AgdaInductiveConstructor{□·}}\AgdaSpace{}%
\AgdaBound{M}\AgdaSymbol{)}\AgdaSpace{}%
\AgdaBound{M⟶N}\AgdaSymbol{)}\AgdaSpace{}%
\AgdaSymbol{(}\AgdaInductiveConstructor{ξ}\AgdaSpace{}%
\AgdaSymbol{(}\AgdaBound{v}\AgdaSpace{}%
\AgdaOperator{\AgdaInductiveConstructor{·□}}\AgdaSymbol{)}\AgdaSpace{}%
\AgdaBound{M⟶N′}\AgdaSymbol{)}\AgdaSpace{}%
\AgdaSymbol{=}\<%
\\
\>[0][@{}l@{\AgdaIndent{0}}]%
\>[4]\AgdaFunction{⊥-elim}\AgdaSpace{}%
\AgdaSymbol{(}\AgdaFunction{value-irreducible}\AgdaSpace{}%
\AgdaBound{v}\AgdaSpace{}%
\AgdaBound{M⟶N}\AgdaSymbol{)}\<%
\\
\>[0]\AgdaFunction{deterministic}\AgdaSpace{}%
\AgdaSymbol{(}\AgdaInductiveConstructor{ξ}\AgdaSpace{}%
\AgdaSymbol{(}\AgdaOperator{\AgdaInductiveConstructor{□·}}\AgdaSpace{}%
\AgdaBound{M}\AgdaSymbol{)}\AgdaSpace{}%
\AgdaBound{M⟶N}\AgdaSymbol{)}\AgdaSpace{}%
\AgdaSymbol{(}\AgdaInductiveConstructor{ξ-blame}\AgdaSpace{}%
\AgdaSymbol{(}\AgdaOperator{\AgdaInductiveConstructor{□·}}\AgdaSpace{}%
\AgdaBound{M₁}\AgdaSymbol{))}\AgdaSpace{}%
\AgdaSymbol{=}\<%
\\
\>[0][@{}l@{\AgdaIndent{0}}]%
\>[4]\AgdaFunction{⊥-elim}\AgdaSpace{}%
\AgdaSymbol{(}\AgdaFunction{blame-irreducible}\AgdaSpace{}%
\AgdaBound{M⟶N}\AgdaSymbol{)}\<%
\\
\>[0]\AgdaFunction{deterministic}\AgdaSpace{}%
\AgdaSymbol{(}\AgdaInductiveConstructor{ξ}\AgdaSpace{}%
\AgdaSymbol{(}\AgdaOperator{\AgdaInductiveConstructor{□·}}\AgdaSpace{}%
\AgdaBound{M}\AgdaSymbol{)}\AgdaSpace{}%
\AgdaBound{M⟶N}\AgdaSymbol{)}\AgdaSpace{}%
\AgdaSymbol{(}\AgdaInductiveConstructor{ξ-blame}\AgdaSpace{}%
\AgdaSymbol{(}\AgdaBound{v}\AgdaSpace{}%
\AgdaOperator{\AgdaInductiveConstructor{·□}}\AgdaSymbol{))}\AgdaSpace{}%
\AgdaSymbol{=}\<%
\\
\>[0][@{}l@{\AgdaIndent{0}}]%
\>[4]\AgdaFunction{⊥-elim}\AgdaSpace{}%
\AgdaSymbol{(}\AgdaFunction{value-irreducible}\AgdaSpace{}%
\AgdaBound{v}\AgdaSpace{}%
\AgdaBound{M⟶N}\AgdaSymbol{)}\<%
\\
\>[0]\AgdaFunction{deterministic}\AgdaSpace{}%
\AgdaSymbol{(}\AgdaInductiveConstructor{ξ}\AgdaSpace{}%
\AgdaSymbol{(}\AgdaOperator{\AgdaInductiveConstructor{□·}}\AgdaSpace{}%
\AgdaBound{M}\AgdaSymbol{)}\AgdaSpace{}%
\AgdaBound{M⟶N}\AgdaSymbol{)}\AgdaSpace{}%
\AgdaSymbol{(}\AgdaInductiveConstructor{β}\AgdaSpace{}%
\AgdaBound{v}\AgdaSymbol{)}\AgdaSpace{}%
\AgdaSymbol{=}\<%
\\
\>[0][@{}l@{\AgdaIndent{0}}]%
\>[4]\AgdaFunction{⊥-elim}\AgdaSpace{}%
\AgdaSymbol{(}\AgdaFunction{value-irreducible}\AgdaSpace{}%
\AgdaSymbol{(}\AgdaOperator{\AgdaInductiveConstructor{ƛ̬}}\AgdaSpace{}%
\AgdaSymbol{\AgdaUnderscore{})}\AgdaSpace{}%
\AgdaBound{M⟶N}\AgdaSymbol{)}\<%
\\
\>[0]\AgdaFunction{deterministic}\AgdaSpace{}%
\AgdaSymbol{(}\AgdaInductiveConstructor{ξ}\AgdaSpace{}%
\AgdaSymbol{(}\AgdaBound{v}\AgdaSpace{}%
\AgdaOperator{\AgdaInductiveConstructor{·□}}\AgdaSymbol{)}\AgdaSpace{}%
\AgdaBound{M⟶N}\AgdaSymbol{)}\AgdaSpace{}%
\AgdaSymbol{(}\AgdaInductiveConstructor{ξ}\AgdaSpace{}%
\AgdaSymbol{(}\AgdaOperator{\AgdaInductiveConstructor{□·}}\AgdaSpace{}%
\AgdaBound{M}\AgdaSymbol{)}\AgdaSpace{}%
\AgdaBound{M⟶N′}\AgdaSymbol{)}\AgdaSpace{}%
\AgdaSymbol{=}\<%
\\
\>[0][@{}l@{\AgdaIndent{0}}]%
\>[4]\AgdaFunction{⊥-elim}\AgdaSpace{}%
\AgdaSymbol{(}\AgdaFunction{value-irreducible}\AgdaSpace{}%
\AgdaBound{v}\AgdaSpace{}%
\AgdaBound{M⟶N′}\AgdaSymbol{)}\<%
\\
\>[0]\AgdaFunction{deterministic}\AgdaSpace{}%
\AgdaSymbol{(}\AgdaInductiveConstructor{ξ}\AgdaSpace{}%
\AgdaSymbol{(}\AgdaBound{v}\AgdaSpace{}%
\AgdaOperator{\AgdaInductiveConstructor{·□}}\AgdaSymbol{)}\AgdaSpace{}%
\AgdaBound{M⟶N}\AgdaSymbol{)}\AgdaSpace{}%
\AgdaSymbol{(}\AgdaInductiveConstructor{ξ}\AgdaSpace{}%
\AgdaSymbol{(}\AgdaBound{v₁}\AgdaSpace{}%
\AgdaOperator{\AgdaInductiveConstructor{·□}}\AgdaSymbol{)}\AgdaSpace{}%
\AgdaBound{M⟶N′}\AgdaSymbol{)}\<%
\\
\>[0][@{}l@{\AgdaIndent{0}}]%
\>[4]\AgdaKeyword{with}\AgdaSpace{}%
\AgdaFunction{deterministic}\AgdaSpace{}%
\AgdaBound{M⟶N}\AgdaSpace{}%
\AgdaBound{M⟶N′}\<%
\\
\>[0]\AgdaSymbol{...}\AgdaSpace{}%
\AgdaSymbol{|}\AgdaSpace{}%
\AgdaInductiveConstructor{refl}\AgdaSpace{}%
\AgdaSymbol{=}\AgdaSpace{}%
\AgdaInductiveConstructor{refl}\<%
\\
\>[0]\AgdaFunction{deterministic}\AgdaSpace{}%
\AgdaSymbol{(}\AgdaInductiveConstructor{ξ}\AgdaSpace{}%
\AgdaSymbol{(()}\AgdaSpace{}%
\AgdaOperator{\AgdaInductiveConstructor{·□}}\AgdaSymbol{)}\AgdaSpace{}%
\AgdaBound{M⟶N}\AgdaSymbol{)}\AgdaSpace{}%
\AgdaSymbol{(}\AgdaInductiveConstructor{ξ-blame}\AgdaSpace{}%
\AgdaSymbol{(}\AgdaOperator{\AgdaInductiveConstructor{□·}}\AgdaSpace{}%
\AgdaBound{M}\AgdaSymbol{))}\<%
\\
\>[0]\AgdaFunction{deterministic}\AgdaSpace{}%
\AgdaSymbol{(}\AgdaInductiveConstructor{ξ}\AgdaSpace{}%
\AgdaSymbol{(}\AgdaBound{v}\AgdaSpace{}%
\AgdaOperator{\AgdaInductiveConstructor{·□}}\AgdaSymbol{)}\AgdaSpace{}%
\AgdaBound{M⟶N}\AgdaSymbol{)}\AgdaSpace{}%
\AgdaSymbol{(}\AgdaInductiveConstructor{ξ-blame}\AgdaSpace{}%
\AgdaSymbol{(}\AgdaBound{v₁}\AgdaSpace{}%
\AgdaOperator{\AgdaInductiveConstructor{·□}}\AgdaSymbol{))}\AgdaSpace{}%
\AgdaSymbol{=}\<%
\\
\>[0][@{}l@{\AgdaIndent{0}}]%
\>[4]\AgdaFunction{⊥-elim}\AgdaSpace{}%
\AgdaSymbol{(}\AgdaFunction{blame-irreducible}\AgdaSpace{}%
\AgdaBound{M⟶N}\AgdaSymbol{)}\<%
\\
\>[0]\AgdaFunction{deterministic}\AgdaSpace{}%
\AgdaSymbol{(}\AgdaInductiveConstructor{ξ}\AgdaSpace{}%
\AgdaSymbol{(}\AgdaBound{v}\AgdaSpace{}%
\AgdaOperator{\AgdaInductiveConstructor{·□}}\AgdaSymbol{)}\AgdaSpace{}%
\AgdaBound{M⟶N}\AgdaSymbol{)}\AgdaSpace{}%
\AgdaSymbol{(}\AgdaInductiveConstructor{β}\AgdaSpace{}%
\AgdaBound{x}\AgdaSymbol{)}\AgdaSpace{}%
\AgdaSymbol{=}\<%
\\
\>[0][@{}l@{\AgdaIndent{0}}]%
\>[4]\AgdaFunction{⊥-elim}\AgdaSpace{}%
\AgdaSymbol{(}\AgdaFunction{value-irreducible}\AgdaSpace{}%
\AgdaBound{x}\AgdaSpace{}%
\AgdaBound{M⟶N}\AgdaSymbol{)}\<%
\\
\>[0]\AgdaFunction{deterministic}\AgdaSpace{}%
\AgdaSymbol{(}\AgdaInductiveConstructor{ξ}\AgdaSpace{}%
\AgdaSymbol{(}\AgdaOperator{\AgdaInductiveConstructor{□⟨}}\AgdaSpace{}%
\AgdaBound{G}\AgdaSpace{}%
\AgdaOperator{\AgdaInductiveConstructor{!⟩}}\AgdaSymbol{)}\AgdaSpace{}%
\AgdaBound{M⟶N}\AgdaSymbol{)}\AgdaSpace{}%
\AgdaSymbol{(}\AgdaInductiveConstructor{ξ}\AgdaSpace{}%
\AgdaSymbol{(}\AgdaOperator{\AgdaInductiveConstructor{□⟨}}\AgdaSpace{}%
\AgdaSymbol{\AgdaUnderscore{}}\AgdaSpace{}%
\AgdaOperator{\AgdaInductiveConstructor{!⟩}}\AgdaSymbol{)}\AgdaSpace{}%
\AgdaBound{M⟶N′}\AgdaSymbol{)}\<%
\\
\>[0][@{}l@{\AgdaIndent{0}}]%
\>[4]\AgdaKeyword{with}\AgdaSpace{}%
\AgdaFunction{deterministic}\AgdaSpace{}%
\AgdaBound{M⟶N}\AgdaSpace{}%
\AgdaBound{M⟶N′}\<%
\\
\>[0]\AgdaSymbol{...}\AgdaSpace{}%
\AgdaSymbol{|}\AgdaSpace{}%
\AgdaInductiveConstructor{refl}\AgdaSpace{}%
\AgdaSymbol{=}\AgdaSpace{}%
\AgdaInductiveConstructor{refl}\<%
\\
\>[0]\AgdaFunction{deterministic}\AgdaSpace{}%
\AgdaSymbol{(}\AgdaInductiveConstructor{ξ}\AgdaSpace{}%
\AgdaSymbol{(}\AgdaOperator{\AgdaInductiveConstructor{□⟨}}\AgdaSpace{}%
\AgdaBound{G}\AgdaSpace{}%
\AgdaOperator{\AgdaInductiveConstructor{!⟩}}\AgdaSymbol{)}\AgdaSpace{}%
\AgdaBound{M⟶N}\AgdaSymbol{)}\AgdaSpace{}%
\AgdaSymbol{(}\AgdaInductiveConstructor{ξ-blame}\AgdaSpace{}%
\AgdaSymbol{(}\AgdaOperator{\AgdaInductiveConstructor{□⟨}}\AgdaSpace{}%
\AgdaSymbol{\AgdaUnderscore{}}\AgdaSpace{}%
\AgdaOperator{\AgdaInductiveConstructor{!⟩}}\AgdaSymbol{))}\AgdaSpace{}%
\AgdaSymbol{=}\<%
\\
\>[0][@{}l@{\AgdaIndent{0}}]%
\>[4]\AgdaFunction{⊥-elim}\AgdaSpace{}%
\AgdaSymbol{(}\AgdaFunction{blame-irreducible}\AgdaSpace{}%
\AgdaBound{M⟶N}\AgdaSymbol{)}\<%
\\
\>[0]\AgdaFunction{deterministic}\AgdaSpace{}%
\AgdaSymbol{(}\AgdaInductiveConstructor{ξ}\AgdaSpace{}%
\AgdaSymbol{(}\AgdaOperator{\AgdaInductiveConstructor{□⟨}}\AgdaSpace{}%
\AgdaBound{H}\AgdaSpace{}%
\AgdaOperator{\AgdaInductiveConstructor{?⟩}}\AgdaSymbol{)}\AgdaSpace{}%
\AgdaBound{M⟶N}\AgdaSymbol{)}\AgdaSpace{}%
\AgdaSymbol{(}\AgdaInductiveConstructor{ξ}\AgdaSpace{}%
\AgdaSymbol{(}\AgdaOperator{\AgdaInductiveConstructor{□⟨}}\AgdaSpace{}%
\AgdaSymbol{\AgdaUnderscore{}}\AgdaSpace{}%
\AgdaOperator{\AgdaInductiveConstructor{?⟩}}\AgdaSymbol{)}\AgdaSpace{}%
\AgdaBound{M⟶N′}\AgdaSymbol{)}\<%
\\
\>[0][@{}l@{\AgdaIndent{0}}]%
\>[4]\AgdaKeyword{with}\AgdaSpace{}%
\AgdaFunction{deterministic}\AgdaSpace{}%
\AgdaBound{M⟶N}\AgdaSpace{}%
\AgdaBound{M⟶N′}\<%
\\
\>[0]\AgdaSymbol{...}\AgdaSpace{}%
\AgdaSymbol{|}\AgdaSpace{}%
\AgdaInductiveConstructor{refl}\AgdaSpace{}%
\AgdaSymbol{=}\AgdaSpace{}%
\AgdaInductiveConstructor{refl}\<%
\\
\>[0]\AgdaFunction{deterministic}\AgdaSpace{}%
\AgdaSymbol{(}\AgdaInductiveConstructor{ξ}\AgdaSpace{}%
\AgdaSymbol{(}\AgdaOperator{\AgdaInductiveConstructor{□⟨}}\AgdaSpace{}%
\AgdaBound{H}\AgdaSpace{}%
\AgdaOperator{\AgdaInductiveConstructor{?⟩}}\AgdaSymbol{)}\AgdaSpace{}%
\AgdaBound{M⟶N}\AgdaSymbol{)}\AgdaSpace{}%
\AgdaSymbol{(}\AgdaInductiveConstructor{ξ-blame}\AgdaSpace{}%
\AgdaSymbol{(}\AgdaOperator{\AgdaInductiveConstructor{□⟨}}\AgdaSpace{}%
\AgdaSymbol{\AgdaUnderscore{}}\AgdaSpace{}%
\AgdaOperator{\AgdaInductiveConstructor{?⟩}}\AgdaSymbol{))}\AgdaSpace{}%
\AgdaSymbol{=}\<%
\\
\>[0][@{}l@{\AgdaIndent{0}}]%
\>[4]\AgdaFunction{⊥-elim}\AgdaSpace{}%
\AgdaSymbol{(}\AgdaFunction{blame-irreducible}\AgdaSpace{}%
\AgdaBound{M⟶N}\AgdaSymbol{)}\<%
\\
\>[0]\AgdaFunction{deterministic}\AgdaSpace{}%
\AgdaSymbol{(}\AgdaInductiveConstructor{ξ}\AgdaSpace{}%
\AgdaOperator{\AgdaInductiveConstructor{□⟨}}\AgdaSpace{}%
\AgdaBound{H}\AgdaSpace{}%
\AgdaOperator{\AgdaInductiveConstructor{?⟩}}\AgdaSpace{}%
\AgdaBound{r}\AgdaSymbol{)}\AgdaSpace{}%
\AgdaSymbol{(}\AgdaInductiveConstructor{collapse}\AgdaSpace{}%
\AgdaBound{v}\AgdaSpace{}%
\AgdaInductiveConstructor{refl}\AgdaSymbol{)}\AgdaSpace{}%
\AgdaSymbol{=}\<%
\\
\>[0][@{}l@{\AgdaIndent{0}}]%
\>[4]\AgdaFunction{⊥-elim}\AgdaSpace{}%
\AgdaSymbol{(}\AgdaFunction{value-irreducible}\AgdaSpace{}%
\AgdaSymbol{(}\AgdaBound{v}\AgdaSpace{}%
\AgdaOperator{\AgdaInductiveConstructor{〈}}\AgdaSpace{}%
\AgdaSymbol{\AgdaUnderscore{}}\AgdaSpace{}%
\AgdaOperator{\AgdaInductiveConstructor{〉}}\AgdaSymbol{)}\AgdaSpace{}%
\AgdaBound{r}\AgdaSymbol{)}\<%
\\
\>[0]\AgdaFunction{deterministic}\AgdaSpace{}%
\AgdaSymbol{(}\AgdaInductiveConstructor{ξ}\AgdaSpace{}%
\AgdaOperator{\AgdaInductiveConstructor{□⟨}}\AgdaSpace{}%
\AgdaBound{H}\AgdaSpace{}%
\AgdaOperator{\AgdaInductiveConstructor{?⟩}}\AgdaSpace{}%
\AgdaBound{r}\AgdaSymbol{)}\AgdaSpace{}%
\AgdaSymbol{(}\AgdaInductiveConstructor{collide}\AgdaSpace{}%
\AgdaBound{v}\AgdaSpace{}%
\AgdaBound{neq}\AgdaSpace{}%
\AgdaInductiveConstructor{refl}\AgdaSymbol{)}\AgdaSpace{}%
\AgdaSymbol{=}\<%
\\
\>[0][@{}l@{\AgdaIndent{0}}]%
\>[4]\AgdaFunction{⊥-elim}\AgdaSpace{}%
\AgdaSymbol{(}\AgdaFunction{value-irreducible}\AgdaSpace{}%
\AgdaSymbol{(}\AgdaBound{v}\AgdaSpace{}%
\AgdaOperator{\AgdaInductiveConstructor{〈}}\AgdaSpace{}%
\AgdaSymbol{\AgdaUnderscore{}}\AgdaSpace{}%
\AgdaOperator{\AgdaInductiveConstructor{〉}}\AgdaSymbol{)}\AgdaSpace{}%
\AgdaBound{r}\AgdaSymbol{)}\<%
\\
\>[0]\AgdaFunction{deterministic}\AgdaSpace{}%
\AgdaSymbol{(}\AgdaInductiveConstructor{ξ-blame}\AgdaSpace{}%
\AgdaSymbol{(}\AgdaOperator{\AgdaInductiveConstructor{□·}}\AgdaSpace{}%
\AgdaBound{M}\AgdaSymbol{))}\AgdaSpace{}%
\AgdaSymbol{(}\AgdaInductiveConstructor{ξ}\AgdaSpace{}%
\AgdaSymbol{(}\AgdaOperator{\AgdaInductiveConstructor{□·}}\AgdaSpace{}%
\AgdaBound{M₁}\AgdaSymbol{)}\AgdaSpace{}%
\AgdaBound{M⟶N′}\AgdaSymbol{)}\AgdaSpace{}%
\AgdaSymbol{=}\<%
\\
\>[0][@{}l@{\AgdaIndent{0}}]%
\>[4]\AgdaFunction{⊥-elim}\AgdaSpace{}%
\AgdaSymbol{(}\AgdaFunction{blame-irreducible}\AgdaSpace{}%
\AgdaBound{M⟶N′}\AgdaSymbol{)}\<%
\\
\>[0]\AgdaFunction{deterministic}\AgdaSpace{}%
\AgdaSymbol{(}\AgdaInductiveConstructor{ξ-blame}\AgdaSpace{}%
\AgdaSymbol{(}\AgdaOperator{\AgdaInductiveConstructor{□·}}\AgdaSpace{}%
\AgdaBound{M}\AgdaSymbol{))}\AgdaSpace{}%
\AgdaSymbol{(}\AgdaInductiveConstructor{ξ}\AgdaSpace{}%
\AgdaSymbol{(()}\AgdaSpace{}%
\AgdaOperator{\AgdaInductiveConstructor{·□}}\AgdaSymbol{)}\AgdaSpace{}%
\AgdaBound{M⟶N′}\AgdaSymbol{)}\<%
\\
\>[0]\AgdaFunction{deterministic}\AgdaSpace{}%
\AgdaSymbol{(}\AgdaInductiveConstructor{ξ-blame}\AgdaSpace{}%
\AgdaSymbol{(}\AgdaOperator{\AgdaInductiveConstructor{□·}}\AgdaSpace{}%
\AgdaBound{M}\AgdaSymbol{))}\AgdaSpace{}%
\AgdaSymbol{(}\AgdaInductiveConstructor{ξ-blame}\AgdaSpace{}%
\AgdaSymbol{(}\AgdaOperator{\AgdaInductiveConstructor{□·}}\AgdaSpace{}%
\AgdaBound{M₁}\AgdaSymbol{))}\AgdaSpace{}%
\AgdaSymbol{=}\AgdaSpace{}%
\AgdaInductiveConstructor{refl}\<%
\\
\>[0]\AgdaFunction{deterministic}\AgdaSpace{}%
\AgdaSymbol{(}\AgdaInductiveConstructor{ξ-blame}\AgdaSpace{}%
\AgdaSymbol{(}\AgdaOperator{\AgdaInductiveConstructor{□·}}\AgdaSpace{}%
\AgdaBound{M}\AgdaSymbol{))}\AgdaSpace{}%
\AgdaSymbol{(}\AgdaInductiveConstructor{ξ-blame}\AgdaSpace{}%
\AgdaSymbol{(}\AgdaBound{v}\AgdaSpace{}%
\AgdaOperator{\AgdaInductiveConstructor{·□}}\AgdaSymbol{))}\AgdaSpace{}%
\AgdaSymbol{=}\AgdaSpace{}%
\AgdaInductiveConstructor{refl}\<%
\\
\>[0]\AgdaFunction{deterministic}\AgdaSpace{}%
\AgdaSymbol{(}\AgdaInductiveConstructor{ξ-blame}\AgdaSpace{}%
\AgdaSymbol{(}\AgdaBound{v}\AgdaSpace{}%
\AgdaOperator{\AgdaInductiveConstructor{·□}}\AgdaSymbol{))}\AgdaSpace{}%
\AgdaSymbol{(}\AgdaInductiveConstructor{ξ}\AgdaSpace{}%
\AgdaSymbol{(}\AgdaOperator{\AgdaInductiveConstructor{□·}}\AgdaSpace{}%
\AgdaBound{M}\AgdaSymbol{)}\AgdaSpace{}%
\AgdaBound{M⟶N′}\AgdaSymbol{)}\AgdaSpace{}%
\AgdaSymbol{=}\<%
\\
\>[0][@{}l@{\AgdaIndent{0}}]%
\>[4]\AgdaFunction{⊥-elim}\AgdaSpace{}%
\AgdaSymbol{(}\AgdaFunction{value-irreducible}\AgdaSpace{}%
\AgdaBound{v}\AgdaSpace{}%
\AgdaBound{M⟶N′}\AgdaSymbol{)}\<%
\\
\>[0]\AgdaFunction{deterministic}\AgdaSpace{}%
\AgdaSymbol{(}\AgdaInductiveConstructor{ξ-blame}\AgdaSpace{}%
\AgdaSymbol{(}\AgdaBound{v}\AgdaSpace{}%
\AgdaOperator{\AgdaInductiveConstructor{·□}}\AgdaSymbol{))}\AgdaSpace{}%
\AgdaSymbol{(}\AgdaInductiveConstructor{ξ}\AgdaSpace{}%
\AgdaSymbol{(}\AgdaBound{v₁}\AgdaSpace{}%
\AgdaOperator{\AgdaInductiveConstructor{·□}}\AgdaSymbol{)}\AgdaSpace{}%
\AgdaBound{M⟶N′}\AgdaSymbol{)}\AgdaSpace{}%
\AgdaSymbol{=}\<%
\\
\>[0][@{}l@{\AgdaIndent{0}}]%
\>[4]\AgdaFunction{⊥-elim}\AgdaSpace{}%
\AgdaSymbol{(}\AgdaFunction{blame-irreducible}\AgdaSpace{}%
\AgdaBound{M⟶N′}\AgdaSymbol{)}\<%
\\
\>[0]\AgdaFunction{deterministic}\AgdaSpace{}%
\AgdaSymbol{(}\AgdaInductiveConstructor{ξ-blame}\AgdaSpace{}%
\AgdaSymbol{(()}\AgdaSpace{}%
\AgdaOperator{\AgdaInductiveConstructor{·□}}\AgdaSymbol{))}\AgdaSpace{}%
\AgdaSymbol{(}\AgdaInductiveConstructor{ξ-blame}\AgdaSpace{}%
\AgdaSymbol{(}\AgdaOperator{\AgdaInductiveConstructor{□·}}\AgdaSpace{}%
\AgdaDottedPattern{\AgdaSymbol{.}}\AgdaDottedPattern{\AgdaInductiveConstructor{blame}}\AgdaSymbol{))}\<%
\\
\>[0]\AgdaFunction{deterministic}\AgdaSpace{}%
\AgdaSymbol{(}\AgdaInductiveConstructor{ξ-blame}\AgdaSpace{}%
\AgdaSymbol{(}\AgdaBound{v}\AgdaSpace{}%
\AgdaOperator{\AgdaInductiveConstructor{·□}}\AgdaSymbol{))}\AgdaSpace{}%
\AgdaSymbol{(}\AgdaInductiveConstructor{ξ-blame}\AgdaSpace{}%
\AgdaSymbol{(}\AgdaBound{v₁}\AgdaSpace{}%
\AgdaOperator{\AgdaInductiveConstructor{·□}}\AgdaSymbol{))}\AgdaSpace{}%
\AgdaSymbol{=}\AgdaSpace{}%
\AgdaInductiveConstructor{refl}\<%
\\
\>[0]\AgdaFunction{deterministic}\AgdaSpace{}%
\AgdaSymbol{(}\AgdaInductiveConstructor{ξ-blame}\AgdaSpace{}%
\AgdaSymbol{(}\AgdaOperator{\AgdaInductiveConstructor{□⟨}}\AgdaSpace{}%
\AgdaBound{G}\AgdaSpace{}%
\AgdaOperator{\AgdaInductiveConstructor{!⟩}}\AgdaSymbol{))}\AgdaSpace{}%
\AgdaSymbol{(}\AgdaInductiveConstructor{ξ}\AgdaSpace{}%
\AgdaSymbol{(}\AgdaOperator{\AgdaInductiveConstructor{□⟨}}\AgdaSpace{}%
\AgdaSymbol{\AgdaUnderscore{}}\AgdaSpace{}%
\AgdaOperator{\AgdaInductiveConstructor{!⟩}}\AgdaSymbol{)}\AgdaSpace{}%
\AgdaBound{M⟶N′}\AgdaSymbol{)}\AgdaSpace{}%
\AgdaSymbol{=}\<%
\\
\>[0][@{}l@{\AgdaIndent{0}}]%
\>[4]\AgdaFunction{⊥-elim}\AgdaSpace{}%
\AgdaSymbol{(}\AgdaFunction{blame-irreducible}\AgdaSpace{}%
\AgdaBound{M⟶N′}\AgdaSymbol{)}\<%
\\
\>[0]\AgdaFunction{deterministic}\AgdaSpace{}%
\AgdaSymbol{(}\AgdaInductiveConstructor{ξ-blame}\AgdaSpace{}%
\AgdaSymbol{(}\AgdaOperator{\AgdaInductiveConstructor{□⟨}}\AgdaSpace{}%
\AgdaBound{G}\AgdaSpace{}%
\AgdaOperator{\AgdaInductiveConstructor{!⟩}}\AgdaSymbol{))}\AgdaSpace{}%
\AgdaSymbol{(}\AgdaInductiveConstructor{ξ-blame}\AgdaSpace{}%
\AgdaSymbol{(}\AgdaOperator{\AgdaInductiveConstructor{□⟨}}\AgdaSpace{}%
\AgdaSymbol{\AgdaUnderscore{}}\AgdaSpace{}%
\AgdaOperator{\AgdaInductiveConstructor{!⟩}}\AgdaSymbol{))}\AgdaSpace{}%
\AgdaSymbol{=}\AgdaSpace{}%
\AgdaInductiveConstructor{refl}\<%
\\
\>[0]\AgdaFunction{deterministic}\AgdaSpace{}%
\AgdaSymbol{(}\AgdaInductiveConstructor{ξ-blame}\AgdaSpace{}%
\AgdaSymbol{(}\AgdaOperator{\AgdaInductiveConstructor{□⟨}}\AgdaSpace{}%
\AgdaBound{H}\AgdaSpace{}%
\AgdaOperator{\AgdaInductiveConstructor{?⟩}}\AgdaSymbol{))}\AgdaSpace{}%
\AgdaSymbol{(}\AgdaInductiveConstructor{ξ}\AgdaSpace{}%
\AgdaSymbol{(}\AgdaOperator{\AgdaInductiveConstructor{□⟨}}\AgdaSpace{}%
\AgdaSymbol{\AgdaUnderscore{}}\AgdaSpace{}%
\AgdaOperator{\AgdaInductiveConstructor{?⟩}}\AgdaSymbol{)}\AgdaSpace{}%
\AgdaBound{M⟶N′}\AgdaSymbol{)}\AgdaSpace{}%
\AgdaSymbol{=}\<%
\\
\>[0][@{}l@{\AgdaIndent{0}}]%
\>[4]\AgdaFunction{⊥-elim}\AgdaSpace{}%
\AgdaSymbol{(}\AgdaFunction{blame-irreducible}\AgdaSpace{}%
\AgdaBound{M⟶N′}\AgdaSymbol{)}\<%
\\
\>[0]\AgdaFunction{deterministic}\AgdaSpace{}%
\AgdaSymbol{(}\AgdaInductiveConstructor{ξ-blame}\AgdaSpace{}%
\AgdaSymbol{(}\AgdaOperator{\AgdaInductiveConstructor{□⟨}}\AgdaSpace{}%
\AgdaBound{H}\AgdaSpace{}%
\AgdaOperator{\AgdaInductiveConstructor{?⟩}}\AgdaSymbol{))}\AgdaSpace{}%
\AgdaSymbol{(}\AgdaInductiveConstructor{ξ-blame}\AgdaSpace{}%
\AgdaSymbol{(}\AgdaOperator{\AgdaInductiveConstructor{□⟨}}\AgdaSpace{}%
\AgdaSymbol{\AgdaUnderscore{}}\AgdaSpace{}%
\AgdaOperator{\AgdaInductiveConstructor{?⟩}}\AgdaSymbol{))}\AgdaSpace{}%
\AgdaSymbol{=}\AgdaSpace{}%
\AgdaInductiveConstructor{refl}\<%
\\
\>[0]\AgdaFunction{deterministic}\AgdaSpace{}%
\AgdaSymbol{(}\AgdaInductiveConstructor{β}\AgdaSpace{}%
\AgdaBound{x}\AgdaSymbol{)}\AgdaSpace{}%
\AgdaSymbol{(}\AgdaInductiveConstructor{ξ}\AgdaSpace{}%
\AgdaSymbol{(}\AgdaOperator{\AgdaInductiveConstructor{□·}}\AgdaSpace{}%
\AgdaBound{M}\AgdaSymbol{)}\AgdaSpace{}%
\AgdaBound{M⟶N′}\AgdaSymbol{)}\AgdaSpace{}%
\AgdaSymbol{=}\AgdaSpace{}%
\AgdaFunction{⊥-elim}\AgdaSpace{}%
\AgdaSymbol{(}\AgdaFunction{value-irreducible}\AgdaSpace{}%
\AgdaSymbol{(}\AgdaOperator{\AgdaInductiveConstructor{ƛ̬}}\AgdaSpace{}%
\AgdaSymbol{\AgdaUnderscore{})}\AgdaSpace{}%
\AgdaBound{M⟶N′}\AgdaSymbol{)}\<%
\\
\>[0]\AgdaFunction{deterministic}\AgdaSpace{}%
\AgdaSymbol{(}\AgdaInductiveConstructor{β}\AgdaSpace{}%
\AgdaBound{x}\AgdaSymbol{)}\AgdaSpace{}%
\AgdaSymbol{(}\AgdaInductiveConstructor{ξ}\AgdaSpace{}%
\AgdaSymbol{(}\AgdaBound{v}\AgdaSpace{}%
\AgdaOperator{\AgdaInductiveConstructor{·□}}\AgdaSymbol{)}\AgdaSpace{}%
\AgdaBound{M⟶N′}\AgdaSymbol{)}\AgdaSpace{}%
\AgdaSymbol{=}\AgdaSpace{}%
\AgdaFunction{⊥-elim}\AgdaSpace{}%
\AgdaSymbol{(}\AgdaFunction{value-irreducible}\AgdaSpace{}%
\AgdaBound{x}\AgdaSpace{}%
\AgdaBound{M⟶N′}\AgdaSymbol{)}\<%
\\
\>[0]\AgdaFunction{deterministic}\AgdaSpace{}%
\AgdaSymbol{(}\AgdaInductiveConstructor{β}\AgdaSpace{}%
\AgdaSymbol{())}\AgdaSpace{}%
\AgdaSymbol{(}\AgdaInductiveConstructor{ξ-blame}\AgdaSpace{}%
\AgdaSymbol{(}\AgdaBound{v}\AgdaSpace{}%
\AgdaOperator{\AgdaInductiveConstructor{·□}}\AgdaSymbol{))}\<%
\\
\>[0]\AgdaFunction{deterministic}\AgdaSpace{}%
\AgdaSymbol{(}\AgdaInductiveConstructor{β}\AgdaSpace{}%
\AgdaBound{x}\AgdaSymbol{)}\AgdaSpace{}%
\AgdaSymbol{(}\AgdaInductiveConstructor{β}\AgdaSpace{}%
\AgdaBound{x₁}\AgdaSymbol{)}\AgdaSpace{}%
\AgdaSymbol{=}\AgdaSpace{}%
\AgdaInductiveConstructor{refl}\<%
\\
\>[0]\AgdaFunction{deterministic}\AgdaSpace{}%
\AgdaSymbol{(}\AgdaInductiveConstructor{collapse}\AgdaSpace{}%
\AgdaBound{v}\AgdaSpace{}%
\AgdaInductiveConstructor{refl}\AgdaSymbol{)}\AgdaSpace{}%
\AgdaSymbol{(}\AgdaInductiveConstructor{ξξ}\AgdaSpace{}%
\AgdaOperator{\AgdaInductiveConstructor{□⟨}}\AgdaSpace{}%
\AgdaSymbol{\AgdaUnderscore{}}\AgdaSpace{}%
\AgdaOperator{\AgdaInductiveConstructor{?⟩}}\AgdaSpace{}%
\AgdaInductiveConstructor{refl}\AgdaSpace{}%
\AgdaInductiveConstructor{refl}\AgdaSpace{}%
\AgdaBound{r}\AgdaSymbol{)}\AgdaSpace{}%
\AgdaSymbol{=}\<%
\\
\>[0][@{}l@{\AgdaIndent{0}}]%
\>[4]\AgdaFunction{⊥-elim}\AgdaSpace{}%
\AgdaSymbol{(}\AgdaFunction{value-irreducible}\AgdaSpace{}%
\AgdaSymbol{(}\AgdaBound{v}\AgdaSpace{}%
\AgdaOperator{\AgdaInductiveConstructor{〈}}\AgdaSpace{}%
\AgdaSymbol{\AgdaUnderscore{}}\AgdaSpace{}%
\AgdaOperator{\AgdaInductiveConstructor{〉}}\AgdaSymbol{)}\AgdaSpace{}%
\AgdaBound{r}\AgdaSymbol{)}\<%
\\
\>[0]\AgdaFunction{deterministic}\AgdaSpace{}%
\AgdaSymbol{(}\AgdaInductiveConstructor{collapse}\AgdaSpace{}%
\AgdaBound{v}\AgdaSpace{}%
\AgdaInductiveConstructor{refl}\AgdaSymbol{)}\AgdaSpace{}%
\AgdaSymbol{(}\AgdaInductiveConstructor{ξξ-blame}\AgdaSpace{}%
\AgdaSymbol{(}\AgdaOperator{\AgdaInductiveConstructor{□·}}\AgdaSpace{}%
\AgdaBound{M}\AgdaSymbol{)}\AgdaSpace{}%
\AgdaSymbol{())}\<%
\\
\>[0]\AgdaFunction{deterministic}\AgdaSpace{}%
\AgdaSymbol{(}\AgdaInductiveConstructor{collapse}\AgdaSpace{}%
\AgdaBound{v}\AgdaSpace{}%
\AgdaInductiveConstructor{refl}\AgdaSymbol{)}\AgdaSpace{}%
\AgdaSymbol{(}\AgdaInductiveConstructor{ξξ-blame}\AgdaSpace{}%
\AgdaSymbol{(}\AgdaBound{v₁}\AgdaSpace{}%
\AgdaOperator{\AgdaInductiveConstructor{·□}}\AgdaSymbol{)}\AgdaSpace{}%
\AgdaSymbol{())}\<%
\\
\>[0]\AgdaFunction{deterministic}\AgdaSpace{}%
\AgdaSymbol{(}\AgdaInductiveConstructor{collapse}\AgdaSpace{}%
\AgdaBound{v}\AgdaSpace{}%
\AgdaInductiveConstructor{refl}\AgdaSymbol{)}\AgdaSpace{}%
\AgdaSymbol{(}\AgdaInductiveConstructor{ξξ-blame}\AgdaSpace{}%
\AgdaOperator{\AgdaInductiveConstructor{□⟨}}\AgdaSpace{}%
\AgdaSymbol{\AgdaUnderscore{}}\AgdaSpace{}%
\AgdaOperator{\AgdaInductiveConstructor{!⟩}}\AgdaSpace{}%
\AgdaSymbol{())}\<%
\\
\>[0]\AgdaFunction{deterministic}\AgdaSpace{}%
\AgdaSymbol{(}\AgdaInductiveConstructor{collapse}\AgdaSpace{}%
\AgdaBound{v}\AgdaSpace{}%
\AgdaInductiveConstructor{refl}\AgdaSymbol{)}\AgdaSpace{}%
\AgdaSymbol{(}\AgdaInductiveConstructor{ξξ-blame}\AgdaSpace{}%
\AgdaOperator{\AgdaInductiveConstructor{□⟨}}\AgdaSpace{}%
\AgdaSymbol{\AgdaUnderscore{}}\AgdaSpace{}%
\AgdaOperator{\AgdaInductiveConstructor{?⟩}}\AgdaSpace{}%
\AgdaSymbol{())}\<%
\\
\>[0]\AgdaFunction{deterministic}\AgdaSpace{}%
\AgdaSymbol{(}\AgdaInductiveConstructor{collapse}\AgdaSpace{}%
\AgdaBound{v}\AgdaSpace{}%
\AgdaInductiveConstructor{refl}\AgdaSymbol{)}\AgdaSpace{}%
\AgdaSymbol{(}\AgdaInductiveConstructor{collapse}\AgdaSpace{}%
\AgdaBound{x}\AgdaSpace{}%
\AgdaInductiveConstructor{refl}\AgdaSymbol{)}\AgdaSpace{}%
\AgdaSymbol{=}\AgdaSpace{}%
\AgdaInductiveConstructor{refl}\<%
\\
\>[0]\AgdaFunction{deterministic}\AgdaSpace{}%
\AgdaSymbol{(}\AgdaInductiveConstructor{collapse}\AgdaSpace{}%
\AgdaBound{v}\AgdaSpace{}%
\AgdaInductiveConstructor{refl}\AgdaSymbol{)}\AgdaSpace{}%
\AgdaSymbol{(}\AgdaInductiveConstructor{collide}\AgdaSpace{}%
\AgdaBound{x}\AgdaSpace{}%
\AgdaBound{neq}\AgdaSpace{}%
\AgdaInductiveConstructor{refl}\AgdaSymbol{)}\AgdaSpace{}%
\AgdaSymbol{=}\<%
\\
\>[0][@{}l@{\AgdaIndent{0}}]%
\>[4]\AgdaFunction{⊥-elim}\AgdaSpace{}%
\AgdaSymbol{(}\AgdaBound{neq}\AgdaSpace{}%
\AgdaInductiveConstructor{refl}\AgdaSymbol{)}\<%
\\
\>[0]\AgdaFunction{deterministic}\AgdaSpace{}%
\AgdaSymbol{(}\AgdaInductiveConstructor{collide}\AgdaSpace{}%
\AgdaBound{v}\AgdaSpace{}%
\AgdaBound{neq}\AgdaSpace{}%
\AgdaInductiveConstructor{refl}\AgdaSymbol{)}\AgdaSpace{}%
\AgdaSymbol{(}\AgdaInductiveConstructor{ξξ}\AgdaSpace{}%
\AgdaOperator{\AgdaInductiveConstructor{□⟨}}\AgdaSpace{}%
\AgdaSymbol{\AgdaUnderscore{}}\AgdaSpace{}%
\AgdaOperator{\AgdaInductiveConstructor{?⟩}}\AgdaSpace{}%
\AgdaInductiveConstructor{refl}\AgdaSpace{}%
\AgdaInductiveConstructor{refl}\AgdaSpace{}%
\AgdaBound{r}\AgdaSymbol{)}\AgdaSpace{}%
\AgdaSymbol{=}\<%
\\
\>[0][@{}l@{\AgdaIndent{0}}]%
\>[4]\AgdaFunction{⊥-elim}\AgdaSpace{}%
\AgdaSymbol{(}\AgdaFunction{value-irreducible}\AgdaSpace{}%
\AgdaSymbol{(}\AgdaBound{v}\AgdaSpace{}%
\AgdaOperator{\AgdaInductiveConstructor{〈}}\AgdaSpace{}%
\AgdaSymbol{\AgdaUnderscore{}}\AgdaSpace{}%
\AgdaOperator{\AgdaInductiveConstructor{〉}}\AgdaSymbol{)}\AgdaSpace{}%
\AgdaBound{r}\AgdaSymbol{)}\<%
\\
\>[0]\AgdaFunction{deterministic}\AgdaSpace{}%
\AgdaSymbol{(}\AgdaInductiveConstructor{collide}\AgdaSpace{}%
\AgdaBound{v}\AgdaSpace{}%
\AgdaBound{neq}\AgdaSpace{}%
\AgdaInductiveConstructor{refl}\AgdaSymbol{)}\AgdaSpace{}%
\AgdaSymbol{(}\AgdaInductiveConstructor{ξξ-blame}\AgdaSpace{}%
\AgdaSymbol{(}\AgdaOperator{\AgdaInductiveConstructor{□·}}\AgdaSpace{}%
\AgdaBound{M}\AgdaSymbol{)}\AgdaSpace{}%
\AgdaSymbol{())}\<%
\\
\>[0]\AgdaFunction{deterministic}\AgdaSpace{}%
\AgdaSymbol{(}\AgdaInductiveConstructor{collide}\AgdaSpace{}%
\AgdaBound{v}\AgdaSpace{}%
\AgdaBound{neq}\AgdaSpace{}%
\AgdaInductiveConstructor{refl}\AgdaSymbol{)}\AgdaSpace{}%
\AgdaSymbol{(}\AgdaInductiveConstructor{ξξ-blame}\AgdaSpace{}%
\AgdaSymbol{(}\AgdaBound{v₁}\AgdaSpace{}%
\AgdaOperator{\AgdaInductiveConstructor{·□}}\AgdaSymbol{)}\AgdaSpace{}%
\AgdaSymbol{())}\<%
\\
\>[0]\AgdaFunction{deterministic}\AgdaSpace{}%
\AgdaSymbol{(}\AgdaInductiveConstructor{collide}\AgdaSpace{}%
\AgdaBound{v}\AgdaSpace{}%
\AgdaBound{neq}\AgdaSpace{}%
\AgdaInductiveConstructor{refl}\AgdaSymbol{)}\AgdaSpace{}%
\AgdaSymbol{(}\AgdaInductiveConstructor{ξξ-blame}\AgdaSpace{}%
\AgdaOperator{\AgdaInductiveConstructor{□⟨}}\AgdaSpace{}%
\AgdaSymbol{\AgdaUnderscore{}}\AgdaSpace{}%
\AgdaOperator{\AgdaInductiveConstructor{!⟩}}\AgdaSpace{}%
\AgdaSymbol{())}\<%
\\
\>[0]\AgdaFunction{deterministic}\AgdaSpace{}%
\AgdaSymbol{(}\AgdaInductiveConstructor{collide}\AgdaSpace{}%
\AgdaBound{v}\AgdaSpace{}%
\AgdaBound{neq}\AgdaSpace{}%
\AgdaInductiveConstructor{refl}\AgdaSymbol{)}\AgdaSpace{}%
\AgdaSymbol{(}\AgdaInductiveConstructor{ξξ-blame}\AgdaSpace{}%
\AgdaOperator{\AgdaInductiveConstructor{□⟨}}\AgdaSpace{}%
\AgdaSymbol{\AgdaUnderscore{}}\AgdaSpace{}%
\AgdaOperator{\AgdaInductiveConstructor{?⟩}}\AgdaSpace{}%
\AgdaSymbol{())}\<%
\\
\>[0]\AgdaFunction{deterministic}\AgdaSpace{}%
\AgdaSymbol{(}\AgdaInductiveConstructor{collide}\AgdaSpace{}%
\AgdaBound{v}\AgdaSpace{}%
\AgdaBound{neq}\AgdaSpace{}%
\AgdaInductiveConstructor{refl}\AgdaSymbol{)}\AgdaSpace{}%
\AgdaSymbol{(}\AgdaInductiveConstructor{collapse}\AgdaSpace{}%
\AgdaBound{x}\AgdaSpace{}%
\AgdaInductiveConstructor{refl}\AgdaSymbol{)}\AgdaSpace{}%
\AgdaSymbol{=}\<%
\\
\>[0][@{}l@{\AgdaIndent{0}}]%
\>[4]\AgdaFunction{⊥-elim}\AgdaSpace{}%
\AgdaSymbol{(}\AgdaBound{neq}\AgdaSpace{}%
\AgdaInductiveConstructor{refl}\AgdaSymbol{)}\<%
\\
\>[0]\AgdaFunction{deterministic}\AgdaSpace{}%
\AgdaSymbol{(}\AgdaInductiveConstructor{collide}\AgdaSpace{}%
\AgdaBound{v}\AgdaSpace{}%
\AgdaBound{neq}\AgdaSpace{}%
\AgdaInductiveConstructor{refl}\AgdaSymbol{)}\AgdaSpace{}%
\AgdaSymbol{(}\AgdaInductiveConstructor{collide}\AgdaSpace{}%
\AgdaBound{x}\AgdaSpace{}%
\AgdaBound{x₁}\AgdaSpace{}%
\AgdaBound{x₂}\AgdaSymbol{)}\AgdaSpace{}%
\AgdaSymbol{=}\AgdaSpace{}%
\AgdaInductiveConstructor{refl}\<%
\\
\\[\AgdaEmptyExtraSkip]%
\>[0]\AgdaFunction{frame-inv2}\AgdaSpace{}%
\AgdaSymbol{:}\AgdaSpace{}%
\AgdaSymbol{∀\{}\AgdaBound{L}\AgdaSpace{}%
\AgdaBound{N}\AgdaSpace{}%
\AgdaSymbol{:}\AgdaSpace{}%
\AgdaDatatype{Term}\AgdaSymbol{\}\{}\AgdaBound{F}\AgdaSymbol{\}}\<%
\\
\>[0][@{}l@{\AgdaIndent{0}}]%
\>[3]\AgdaSymbol{→}\AgdaSpace{}%
\AgdaFunction{reducible}\AgdaSpace{}%
\AgdaBound{L}\<%
\\
\>[3]\AgdaSymbol{→}\AgdaSpace{}%
\AgdaBound{F}\AgdaSpace{}%
\AgdaOperator{\AgdaFunction{⟦}}\AgdaSpace{}%
\AgdaBound{L}\AgdaSpace{}%
\AgdaOperator{\AgdaFunction{⟧}}\AgdaSpace{}%
\AgdaOperator{\AgdaDatatype{⟶}}\AgdaSpace{}%
\AgdaBound{N}\<%
\\
\>[3]\AgdaSymbol{→}\AgdaSpace{}%
\AgdaFunction{∃[}\AgdaSpace{}%
\AgdaBound{L′}\AgdaSpace{}%
\AgdaFunction{]}\AgdaSpace{}%
\AgdaSymbol{((}\AgdaBound{L}\AgdaSpace{}%
\AgdaOperator{\AgdaDatatype{⟶}}\AgdaSpace{}%
\AgdaBound{L′}\AgdaSymbol{)}\AgdaSpace{}%
\AgdaOperator{\AgdaFunction{×}}\AgdaSpace{}%
\AgdaSymbol{(}\AgdaBound{N}\AgdaSpace{}%
\AgdaOperator{\AgdaDatatype{≡}}\AgdaSpace{}%
\AgdaBound{F}\AgdaSpace{}%
\AgdaOperator{\AgdaFunction{⟦}}\AgdaSpace{}%
\AgdaBound{L′}\AgdaSpace{}%
\AgdaOperator{\AgdaFunction{⟧}}\AgdaSymbol{))}\<%
\\
\>[0]\AgdaFunction{frame-inv2}\AgdaSymbol{\{}\AgdaBound{L}\AgdaSymbol{\}\{}\AgdaDottedPattern{\AgdaSymbol{.((}}\AgdaDottedPattern{\AgdaOperator{\AgdaInductiveConstructor{□·}}}\AgdaSpace{}%
\AgdaDottedPattern{\AgdaBound{M₁}}\AgdaDottedPattern{\AgdaSymbol{)}}\AgdaSpace{}%
\AgdaDottedPattern{\AgdaOperator{\AgdaFunction{⟦}}}\AgdaSpace{}%
\AgdaDottedPattern{\AgdaSymbol{\AgdaUnderscore{}}}\AgdaSpace{}%
\AgdaDottedPattern{\AgdaOperator{\AgdaFunction{⟧}}}\AgdaDottedPattern{\AgdaSymbol{)}}\AgdaSymbol{\}\{}\AgdaOperator{\AgdaInductiveConstructor{□·}}\AgdaSpace{}%
\AgdaBound{M}\AgdaSymbol{\}}\AgdaSpace{}%
\AgdaSymbol{(}\AgdaBound{L′}\AgdaSpace{}%
\AgdaOperator{\AgdaInductiveConstructor{,}}\AgdaSpace{}%
\AgdaBound{L→L′}\AgdaSymbol{)}\AgdaSpace{}%
\AgdaSymbol{(}\AgdaInductiveConstructor{ξξ}\AgdaSpace{}%
\AgdaSymbol{(}\AgdaOperator{\AgdaInductiveConstructor{□·}}\AgdaSpace{}%
\AgdaBound{M₁}\AgdaSymbol{)}\AgdaSpace{}%
\AgdaInductiveConstructor{refl}\AgdaSpace{}%
\AgdaInductiveConstructor{refl}\AgdaSpace{}%
\AgdaBound{L→N}\AgdaSymbol{)}\AgdaSpace{}%
\AgdaSymbol{=}\<%
\\
\>[0][@{}l@{\AgdaIndent{0}}]%
\>[4]\AgdaBound{L′}\AgdaSpace{}%
\AgdaOperator{\AgdaInductiveConstructor{,}}\AgdaSpace{}%
\AgdaSymbol{(}\AgdaBound{L→L′}\AgdaSpace{}%
\AgdaOperator{\AgdaInductiveConstructor{,}}\AgdaSpace{}%
\AgdaFunction{cong}\AgdaSpace{}%
\AgdaSymbol{(λ}\AgdaSpace{}%
\AgdaBound{X}\AgdaSpace{}%
\AgdaSymbol{→}\AgdaSpace{}%
\AgdaBound{X}\AgdaSpace{}%
\AgdaOperator{\AgdaInductiveConstructor{·}}\AgdaSpace{}%
\AgdaBound{M}\AgdaSymbol{)}\AgdaSpace{}%
\AgdaSymbol{(}\AgdaFunction{deterministic}\AgdaSpace{}%
\AgdaBound{L→N}\AgdaSpace{}%
\AgdaBound{L→L′}\AgdaSymbol{))}\<%
\\
\>[0]\AgdaFunction{frame-inv2}\AgdaSpace{}%
\AgdaSymbol{\{}\AgdaBound{L}\AgdaSymbol{\}}\AgdaSpace{}%
\AgdaSymbol{\{}\AgdaDottedPattern{\AgdaSymbol{.((}}\AgdaDottedPattern{\AgdaBound{v}}\AgdaSpace{}%
\AgdaDottedPattern{\AgdaOperator{\AgdaInductiveConstructor{·□}}}\AgdaDottedPattern{\AgdaSymbol{)}}\AgdaSpace{}%
\AgdaDottedPattern{\AgdaOperator{\AgdaFunction{⟦}}}\AgdaSpace{}%
\AgdaDottedPattern{\AgdaSymbol{\AgdaUnderscore{}}}\AgdaSpace{}%
\AgdaDottedPattern{\AgdaOperator{\AgdaFunction{⟧}}}\AgdaDottedPattern{\AgdaSymbol{)}}\AgdaSymbol{\}}\AgdaSpace{}%
\AgdaSymbol{\{}\AgdaOperator{\AgdaInductiveConstructor{□·}}\AgdaSpace{}%
\AgdaBound{M}\AgdaSymbol{\}}\AgdaSpace{}%
\AgdaSymbol{(}\AgdaBound{L′}\AgdaSpace{}%
\AgdaOperator{\AgdaInductiveConstructor{,}}\AgdaSpace{}%
\AgdaBound{L→L′}\AgdaSymbol{)}\AgdaSpace{}%
\AgdaSymbol{(}\AgdaInductiveConstructor{ξξ}\AgdaSpace{}%
\AgdaSymbol{(}\AgdaBound{v}\AgdaSpace{}%
\AgdaOperator{\AgdaInductiveConstructor{·□}}\AgdaSymbol{)}\AgdaSpace{}%
\AgdaInductiveConstructor{refl}\AgdaSpace{}%
\AgdaInductiveConstructor{refl}\AgdaSpace{}%
\AgdaBound{FL→N}\AgdaSymbol{)}\AgdaSpace{}%
\AgdaSymbol{=}\<%
\\
\>[0][@{}l@{\AgdaIndent{0}}]%
\>[3]\AgdaFunction{⊥-elim}\AgdaSpace{}%
\AgdaSymbol{(}\AgdaFunction{value-irreducible}\AgdaSpace{}%
\AgdaBound{v}\AgdaSpace{}%
\AgdaBound{L→L′}\AgdaSymbol{)}\<%
\\
\>[0]\AgdaFunction{frame-inv2}\AgdaSpace{}%
\AgdaSymbol{\{}\AgdaBound{L}\AgdaSymbol{\}}\AgdaSpace{}%
\AgdaSymbol{\{}\AgdaDottedPattern{\AgdaSymbol{.}}\AgdaDottedPattern{\AgdaInductiveConstructor{blame}}\AgdaSymbol{\}}\AgdaSpace{}%
\AgdaSymbol{\{}\AgdaOperator{\AgdaInductiveConstructor{□·}}\AgdaSpace{}%
\AgdaBound{M}\AgdaSymbol{\}}\AgdaSpace{}%
\AgdaSymbol{(}\AgdaBound{L′}\AgdaSpace{}%
\AgdaOperator{\AgdaInductiveConstructor{,}}\AgdaSpace{}%
\AgdaBound{L→L′}\AgdaSymbol{)}\AgdaSpace{}%
\AgdaSymbol{(}\AgdaInductiveConstructor{ξξ-blame}\AgdaSpace{}%
\AgdaSymbol{(}\AgdaOperator{\AgdaInductiveConstructor{□·}}\AgdaSpace{}%
\AgdaBound{M₁}\AgdaSymbol{)}\AgdaSpace{}%
\AgdaInductiveConstructor{refl}\AgdaSymbol{)}\AgdaSpace{}%
\AgdaSymbol{=}\<%
\\
\>[0][@{}l@{\AgdaIndent{0}}]%
\>[4]\AgdaFunction{⊥-elim}\AgdaSpace{}%
\AgdaSymbol{(}\AgdaFunction{blame-irreducible}\AgdaSpace{}%
\AgdaBound{L→L′}\AgdaSymbol{)}\<%
\\
\>[0]\AgdaFunction{frame-inv2}\AgdaSpace{}%
\AgdaSymbol{\{}\AgdaBound{L}\AgdaSymbol{\}}\AgdaSpace{}%
\AgdaSymbol{\{}\AgdaDottedPattern{\AgdaSymbol{.}}\AgdaDottedPattern{\AgdaInductiveConstructor{blame}}\AgdaSymbol{\}}\AgdaSpace{}%
\AgdaSymbol{\{}\AgdaOperator{\AgdaInductiveConstructor{□·}}\AgdaSpace{}%
\AgdaBound{M}\AgdaSymbol{\}}\AgdaSpace{}%
\AgdaSymbol{(}\AgdaBound{L′}\AgdaSpace{}%
\AgdaOperator{\AgdaInductiveConstructor{,}}\AgdaSpace{}%
\AgdaBound{L→L′}\AgdaSymbol{)}\AgdaSpace{}%
\AgdaSymbol{(}\AgdaInductiveConstructor{ξξ-blame}\AgdaSpace{}%
\AgdaSymbol{(}\AgdaBound{v}\AgdaSpace{}%
\AgdaOperator{\AgdaInductiveConstructor{·□}}\AgdaSymbol{)}\AgdaSpace{}%
\AgdaInductiveConstructor{refl}\AgdaSymbol{)}\AgdaSpace{}%
\AgdaSymbol{=}\<%
\\
\>[0][@{}l@{\AgdaIndent{0}}]%
\>[4]\AgdaFunction{⊥-elim}\AgdaSpace{}%
\AgdaSymbol{(}\AgdaFunction{value-irreducible}\AgdaSpace{}%
\AgdaBound{v}\AgdaSpace{}%
\AgdaBound{L→L′}\AgdaSymbol{)}\<%
\\
\>[0]\AgdaFunction{frame-inv2}\AgdaSpace{}%
\AgdaSymbol{\{}\AgdaInductiveConstructor{ƛ}\AgdaSpace{}%
\AgdaBound{N}\AgdaSymbol{\}}\AgdaSpace{}%
\AgdaSymbol{\{\AgdaUnderscore{}\}}\AgdaSpace{}%
\AgdaSymbol{\{}\AgdaOperator{\AgdaInductiveConstructor{□·}}\AgdaSpace{}%
\AgdaBound{M}\AgdaSymbol{\}}\AgdaSpace{}%
\AgdaSymbol{(}\AgdaBound{L′}\AgdaSpace{}%
\AgdaOperator{\AgdaInductiveConstructor{,}}\AgdaSpace{}%
\AgdaBound{L→L′}\AgdaSymbol{)}\AgdaSpace{}%
\AgdaSymbol{(}\AgdaInductiveConstructor{β}\AgdaSpace{}%
\AgdaBound{x}\AgdaSymbol{)}\AgdaSpace{}%
\AgdaSymbol{=}\<%
\\
\>[0][@{}l@{\AgdaIndent{0}}]%
\>[4]\AgdaFunction{⊥-elim}\AgdaSpace{}%
\AgdaSymbol{(}\AgdaFunction{value-irreducible}\AgdaSpace{}%
\AgdaSymbol{(}\AgdaOperator{\AgdaInductiveConstructor{ƛ̬}}\AgdaSpace{}%
\AgdaBound{N}\AgdaSymbol{)}\AgdaSpace{}%
\AgdaBound{L→L′}\AgdaSymbol{)}\<%
\\
\>[0]\AgdaFunction{frame-inv2}\AgdaSpace{}%
\AgdaSymbol{\{}\AgdaBound{L}\AgdaSymbol{\}}\AgdaSpace{}%
\AgdaSymbol{\{}\AgdaBound{N}\AgdaSymbol{\}}\AgdaSpace{}%
\AgdaSymbol{\{}\AgdaBound{v}\AgdaSpace{}%
\AgdaOperator{\AgdaInductiveConstructor{·□}}\AgdaSymbol{\}}\AgdaSpace{}%
\AgdaSymbol{(}\AgdaBound{L′}\AgdaSpace{}%
\AgdaOperator{\AgdaInductiveConstructor{,}}\AgdaSpace{}%
\AgdaBound{L→L′}\AgdaSymbol{)}\AgdaSpace{}%
\AgdaSymbol{(}\AgdaInductiveConstructor{ξξ}\AgdaSpace{}%
\AgdaSymbol{(}\AgdaOperator{\AgdaInductiveConstructor{□·}}\AgdaSpace{}%
\AgdaBound{M}\AgdaSymbol{)}\AgdaSpace{}%
\AgdaInductiveConstructor{refl}\AgdaSpace{}%
\AgdaInductiveConstructor{refl}\AgdaSpace{}%
\AgdaBound{FL→N}\AgdaSymbol{)}\AgdaSpace{}%
\AgdaSymbol{=}\<%
\\
\>[0][@{}l@{\AgdaIndent{0}}]%
\>[4]\AgdaFunction{⊥-elim}\AgdaSpace{}%
\AgdaSymbol{(}\AgdaFunction{value-irreducible}\AgdaSpace{}%
\AgdaBound{v}\AgdaSpace{}%
\AgdaBound{FL→N}\AgdaSymbol{)}\<%
\\
\>[0]\AgdaFunction{frame-inv2}\AgdaSpace{}%
\AgdaSymbol{\{}\AgdaBound{L}\AgdaSymbol{\}}\AgdaSpace{}%
\AgdaSymbol{\{}\AgdaBound{N}\AgdaSymbol{\}}\AgdaSpace{}%
\AgdaSymbol{\{}\AgdaBound{v}\AgdaSpace{}%
\AgdaOperator{\AgdaInductiveConstructor{·□}}\AgdaSymbol{\}}\AgdaSpace{}%
\AgdaSymbol{(}\AgdaBound{L′}\AgdaSpace{}%
\AgdaOperator{\AgdaInductiveConstructor{,}}\AgdaSpace{}%
\AgdaBound{L→L′}\AgdaSymbol{)}\AgdaSpace{}%
\AgdaSymbol{(}\AgdaInductiveConstructor{ξξ}\AgdaSpace{}%
\AgdaSymbol{(}\AgdaBound{v₁}\AgdaSpace{}%
\AgdaOperator{\AgdaInductiveConstructor{·□}}\AgdaSymbol{)}\AgdaSpace{}%
\AgdaInductiveConstructor{refl}\AgdaSpace{}%
\AgdaInductiveConstructor{refl}\AgdaSpace{}%
\AgdaBound{FL→N}\AgdaSymbol{)}\<%
\\
\>[0][@{}l@{\AgdaIndent{0}}]%
\>[4]\AgdaKeyword{with}\AgdaSpace{}%
\AgdaFunction{deterministic}\AgdaSpace{}%
\AgdaBound{L→L′}\AgdaSpace{}%
\AgdaBound{FL→N}\<%
\\
\>[0]\AgdaSymbol{...}\AgdaSpace{}%
\AgdaSymbol{|}\AgdaSpace{}%
\AgdaInductiveConstructor{refl}\AgdaSpace{}%
\AgdaSymbol{=}\AgdaSpace{}%
\AgdaBound{L′}\AgdaSpace{}%
\AgdaOperator{\AgdaInductiveConstructor{,}}\AgdaSpace{}%
\AgdaSymbol{(}\AgdaBound{L→L′}\AgdaSpace{}%
\AgdaOperator{\AgdaInductiveConstructor{,}}\AgdaSpace{}%
\AgdaInductiveConstructor{refl}\AgdaSymbol{)}\<%
\\
\>[0]\AgdaFunction{frame-inv2}\AgdaSpace{}%
\AgdaSymbol{\{}\AgdaBound{L}\AgdaSymbol{\}}\AgdaSpace{}%
\AgdaSymbol{\{}\AgdaDottedPattern{\AgdaSymbol{.}}\AgdaDottedPattern{\AgdaInductiveConstructor{blame}}\AgdaSymbol{\}}\AgdaSpace{}%
\AgdaSymbol{\{()}\AgdaSpace{}%
\AgdaOperator{\AgdaInductiveConstructor{·□}}\AgdaSymbol{\}}\AgdaSpace{}%
\AgdaSymbol{(}\AgdaBound{L′}\AgdaSpace{}%
\AgdaOperator{\AgdaInductiveConstructor{,}}\AgdaSpace{}%
\AgdaBound{L→L′}\AgdaSymbol{)}\AgdaSpace{}%
\AgdaSymbol{(}\AgdaInductiveConstructor{ξξ-blame}\AgdaSpace{}%
\AgdaSymbol{(}\AgdaOperator{\AgdaInductiveConstructor{□·}}\AgdaSpace{}%
\AgdaBound{M}\AgdaSymbol{)}\AgdaSpace{}%
\AgdaInductiveConstructor{refl}\AgdaSymbol{)}\<%
\\
\>[0]\AgdaFunction{frame-inv2}\AgdaSpace{}%
\AgdaSymbol{\{}\AgdaBound{L}\AgdaSymbol{\}}\AgdaSpace{}%
\AgdaSymbol{\{}\AgdaDottedPattern{\AgdaSymbol{.}}\AgdaDottedPattern{\AgdaInductiveConstructor{blame}}\AgdaSymbol{\}}\AgdaSpace{}%
\AgdaSymbol{\{}\AgdaBound{v}\AgdaSpace{}%
\AgdaOperator{\AgdaInductiveConstructor{·□}}\AgdaSymbol{\}}\AgdaSpace{}%
\AgdaSymbol{(}\AgdaBound{L′}\AgdaSpace{}%
\AgdaOperator{\AgdaInductiveConstructor{,}}\AgdaSpace{}%
\AgdaBound{L→L′}\AgdaSymbol{)}\AgdaSpace{}%
\AgdaSymbol{(}\AgdaInductiveConstructor{ξξ-blame}\AgdaSpace{}%
\AgdaSymbol{(}\AgdaBound{v₁}\AgdaSpace{}%
\AgdaOperator{\AgdaInductiveConstructor{·□}}\AgdaSymbol{)}\AgdaSpace{}%
\AgdaInductiveConstructor{refl}\AgdaSymbol{)}\AgdaSpace{}%
\AgdaSymbol{=}\<%
\\
\>[0][@{}l@{\AgdaIndent{0}}]%
\>[4]\AgdaFunction{⊥-elim}\AgdaSpace{}%
\AgdaSymbol{(}\AgdaFunction{blame-irreducible}\AgdaSpace{}%
\AgdaBound{L→L′}\AgdaSymbol{)}\<%
\\
\>[0]\AgdaFunction{frame-inv2}\AgdaSpace{}%
\AgdaSymbol{\{}\AgdaBound{L}\AgdaSymbol{\}}\AgdaSpace{}%
\AgdaSymbol{\{\AgdaUnderscore{}\}}\AgdaSpace{}%
\AgdaSymbol{\{}\AgdaBound{v}\AgdaSpace{}%
\AgdaOperator{\AgdaInductiveConstructor{·□}}\AgdaSymbol{\}}\AgdaSpace{}%
\AgdaSymbol{(}\AgdaBound{L′}\AgdaSpace{}%
\AgdaOperator{\AgdaInductiveConstructor{,}}\AgdaSpace{}%
\AgdaBound{L→L′}\AgdaSymbol{)}\AgdaSpace{}%
\AgdaSymbol{(}\AgdaInductiveConstructor{β}\AgdaSpace{}%
\AgdaBound{w}\AgdaSymbol{)}\AgdaSpace{}%
\AgdaSymbol{=}\<%
\\
\>[0][@{}l@{\AgdaIndent{0}}]%
\>[4]\AgdaFunction{⊥-elim}\AgdaSpace{}%
\AgdaSymbol{(}\AgdaFunction{value-irreducible}\AgdaSpace{}%
\AgdaBound{w}\AgdaSpace{}%
\AgdaBound{L→L′}\AgdaSymbol{)}\<%
\\
\>[0]\AgdaFunction{frame-inv2}\AgdaSpace{}%
\AgdaSymbol{\{}\AgdaBound{L}\AgdaSymbol{\}}\AgdaSpace{}%
\AgdaSymbol{\{}\AgdaBound{N}\AgdaSymbol{\}}\AgdaSpace{}%
\AgdaSymbol{\{}\AgdaOperator{\AgdaInductiveConstructor{□⟨}}\AgdaSpace{}%
\AgdaBound{G}\AgdaSpace{}%
\AgdaOperator{\AgdaInductiveConstructor{!⟩}}\AgdaSymbol{\}}\AgdaSpace{}%
\AgdaSymbol{(}\AgdaBound{L′}\AgdaSpace{}%
\AgdaOperator{\AgdaInductiveConstructor{,}}\AgdaSpace{}%
\AgdaBound{L→L′}\AgdaSymbol{)}\AgdaSpace{}%
\AgdaSymbol{(}\AgdaInductiveConstructor{ξξ}\AgdaSpace{}%
\AgdaOperator{\AgdaInductiveConstructor{□⟨}}\AgdaSpace{}%
\AgdaSymbol{\AgdaUnderscore{}}\AgdaSpace{}%
\AgdaOperator{\AgdaInductiveConstructor{!⟩}}\AgdaSpace{}%
\AgdaInductiveConstructor{refl}\AgdaSpace{}%
\AgdaInductiveConstructor{refl}\AgdaSpace{}%
\AgdaBound{FL→N}\AgdaSymbol{)}\<%
\\
\>[0][@{}l@{\AgdaIndent{0}}]%
\>[4]\AgdaKeyword{with}\AgdaSpace{}%
\AgdaFunction{deterministic}\AgdaSpace{}%
\AgdaBound{L→L′}\AgdaSpace{}%
\AgdaBound{FL→N}\<%
\\
\>[0]\AgdaSymbol{...}\AgdaSpace{}%
\AgdaSymbol{|}\AgdaSpace{}%
\AgdaInductiveConstructor{refl}\AgdaSpace{}%
\AgdaSymbol{=}\AgdaSpace{}%
\AgdaBound{L′}\AgdaSpace{}%
\AgdaOperator{\AgdaInductiveConstructor{,}}\AgdaSpace{}%
\AgdaSymbol{(}\AgdaBound{L→L′}\AgdaSpace{}%
\AgdaOperator{\AgdaInductiveConstructor{,}}\AgdaSpace{}%
\AgdaInductiveConstructor{refl}\AgdaSymbol{)}\<%
\\
\>[0]\AgdaFunction{frame-inv2}\AgdaSpace{}%
\AgdaSymbol{\{}\AgdaBound{L}\AgdaSymbol{\}}\AgdaSpace{}%
\AgdaSymbol{\{}\AgdaDottedPattern{\AgdaSymbol{.}}\AgdaDottedPattern{\AgdaInductiveConstructor{blame}}\AgdaSymbol{\}}\AgdaSpace{}%
\AgdaSymbol{\{}\AgdaOperator{\AgdaInductiveConstructor{□⟨}}\AgdaSpace{}%
\AgdaBound{G}\AgdaSpace{}%
\AgdaOperator{\AgdaInductiveConstructor{!⟩}}\AgdaSymbol{\}}\AgdaSpace{}%
\AgdaSymbol{(}\AgdaBound{L′}\AgdaSpace{}%
\AgdaOperator{\AgdaInductiveConstructor{,}}\AgdaSpace{}%
\AgdaBound{L→L′}\AgdaSymbol{)}\AgdaSpace{}%
\AgdaSymbol{(}\AgdaInductiveConstructor{ξξ-blame}\AgdaSpace{}%
\AgdaOperator{\AgdaInductiveConstructor{□⟨}}\AgdaSpace{}%
\AgdaSymbol{\AgdaUnderscore{}}\AgdaSpace{}%
\AgdaOperator{\AgdaInductiveConstructor{!⟩}}\AgdaSpace{}%
\AgdaInductiveConstructor{refl}\AgdaSymbol{)}\AgdaSpace{}%
\AgdaSymbol{=}\<%
\\
\>[0][@{}l@{\AgdaIndent{0}}]%
\>[4]\AgdaFunction{⊥-elim}\AgdaSpace{}%
\AgdaSymbol{(}\AgdaFunction{blame-irreducible}\AgdaSpace{}%
\AgdaBound{L→L′}\AgdaSymbol{)}\<%
\\
\>[0]\AgdaFunction{frame-inv2}\AgdaSpace{}%
\AgdaSymbol{\{}\AgdaBound{L}\AgdaSymbol{\}}\AgdaSpace{}%
\AgdaSymbol{\{\AgdaUnderscore{}\}}\AgdaSpace{}%
\AgdaSymbol{\{}\AgdaOperator{\AgdaInductiveConstructor{□⟨}}\AgdaSpace{}%
\AgdaBound{H}\AgdaSpace{}%
\AgdaOperator{\AgdaInductiveConstructor{?⟩}}\AgdaSymbol{\}}\AgdaSpace{}%
\AgdaSymbol{(}\AgdaBound{L′}\AgdaSpace{}%
\AgdaOperator{\AgdaInductiveConstructor{,}}\AgdaSpace{}%
\AgdaBound{L→L′}\AgdaSymbol{)}\AgdaSpace{}%
\AgdaSymbol{(}\AgdaInductiveConstructor{ξξ}\AgdaSpace{}%
\AgdaOperator{\AgdaInductiveConstructor{□⟨}}\AgdaSpace{}%
\AgdaSymbol{\AgdaUnderscore{}}\AgdaSpace{}%
\AgdaOperator{\AgdaInductiveConstructor{?⟩}}\AgdaSpace{}%
\AgdaInductiveConstructor{refl}\AgdaSpace{}%
\AgdaInductiveConstructor{refl}\AgdaSpace{}%
\AgdaBound{FL→N}\AgdaSymbol{)}\<%
\\
\>[0][@{}l@{\AgdaIndent{0}}]%
\>[4]\AgdaKeyword{with}\AgdaSpace{}%
\AgdaFunction{deterministic}\AgdaSpace{}%
\AgdaBound{L→L′}\AgdaSpace{}%
\AgdaBound{FL→N}\<%
\\
\>[0]\AgdaSymbol{...}\AgdaSpace{}%
\AgdaSymbol{|}\AgdaSpace{}%
\AgdaInductiveConstructor{refl}\AgdaSpace{}%
\AgdaSymbol{=}\AgdaSpace{}%
\AgdaBound{L′}\AgdaSpace{}%
\AgdaOperator{\AgdaInductiveConstructor{,}}\AgdaSpace{}%
\AgdaSymbol{(}\AgdaBound{L→L′}\AgdaSpace{}%
\AgdaOperator{\AgdaInductiveConstructor{,}}\AgdaSpace{}%
\AgdaInductiveConstructor{refl}\AgdaSymbol{)}\<%
\\
\>[0]\AgdaFunction{frame-inv2}\AgdaSpace{}%
\AgdaSymbol{\{}\AgdaBound{L}\AgdaSymbol{\}}\AgdaSpace{}%
\AgdaSymbol{\{}\AgdaDottedPattern{\AgdaSymbol{.}}\AgdaDottedPattern{\AgdaInductiveConstructor{blame}}\AgdaSymbol{\}}\AgdaSpace{}%
\AgdaSymbol{\{}\AgdaOperator{\AgdaInductiveConstructor{□⟨}}\AgdaSpace{}%
\AgdaBound{H}\AgdaSpace{}%
\AgdaOperator{\AgdaInductiveConstructor{?⟩}}\AgdaSymbol{\}}\AgdaSpace{}%
\AgdaSymbol{(}\AgdaBound{L′}\AgdaSpace{}%
\AgdaOperator{\AgdaInductiveConstructor{,}}\AgdaSpace{}%
\AgdaBound{L→L′}\AgdaSymbol{)}\AgdaSpace{}%
\AgdaSymbol{(}\AgdaInductiveConstructor{ξξ-blame}\AgdaSpace{}%
\AgdaOperator{\AgdaInductiveConstructor{□⟨}}\AgdaSpace{}%
\AgdaSymbol{\AgdaUnderscore{}}\AgdaSpace{}%
\AgdaOperator{\AgdaInductiveConstructor{?⟩}}\AgdaSpace{}%
\AgdaInductiveConstructor{refl}\AgdaSymbol{)}\AgdaSpace{}%
\AgdaSymbol{=}\<%
\\
\>[0][@{}l@{\AgdaIndent{0}}]%
\>[4]\AgdaFunction{⊥-elim}\AgdaSpace{}%
\AgdaSymbol{(}\AgdaFunction{blame-irreducible}\AgdaSpace{}%
\AgdaBound{L→L′}\AgdaSymbol{)}\<%
\\
\>[0]\AgdaFunction{frame-inv2}\AgdaSpace{}%
\AgdaSymbol{\{}\AgdaBound{L}\AgdaSymbol{\}}\AgdaSpace{}%
\AgdaSymbol{\{}\AgdaBound{N}\AgdaSymbol{\}}\AgdaSpace{}%
\AgdaSymbol{\{}\AgdaOperator{\AgdaInductiveConstructor{□⟨}}\AgdaSpace{}%
\AgdaBound{H}\AgdaSpace{}%
\AgdaOperator{\AgdaInductiveConstructor{?⟩}}\AgdaSymbol{\}}\AgdaSpace{}%
\AgdaSymbol{(}\AgdaBound{L′}\AgdaSpace{}%
\AgdaOperator{\AgdaInductiveConstructor{,}}\AgdaSpace{}%
\AgdaBound{L→L′}\AgdaSymbol{)}\AgdaSpace{}%
\AgdaSymbol{(}\AgdaInductiveConstructor{collapse}\AgdaSpace{}%
\AgdaBound{v}\AgdaSpace{}%
\AgdaInductiveConstructor{refl}\AgdaSymbol{)}\AgdaSpace{}%
\AgdaSymbol{=}\<%
\\
\>[0][@{}l@{\AgdaIndent{0}}]%
\>[4]\AgdaFunction{⊥-elim}\AgdaSpace{}%
\AgdaSymbol{(}\AgdaFunction{value-irreducible}\AgdaSpace{}%
\AgdaSymbol{(}\AgdaBound{v}\AgdaSpace{}%
\AgdaOperator{\AgdaInductiveConstructor{〈}}\AgdaSpace{}%
\AgdaSymbol{\AgdaUnderscore{}}\AgdaSpace{}%
\AgdaOperator{\AgdaInductiveConstructor{〉}}\AgdaSymbol{)}\AgdaSpace{}%
\AgdaBound{L→L′}\AgdaSymbol{)}\<%
\\
\>[0]\AgdaFunction{frame-inv2}\AgdaSpace{}%
\AgdaSymbol{\{}\AgdaBound{L}\AgdaSymbol{\}}\AgdaSpace{}%
\AgdaSymbol{\{}\AgdaDottedPattern{\AgdaSymbol{.}}\AgdaDottedPattern{\AgdaInductiveConstructor{blame}}\AgdaSymbol{\}}\AgdaSpace{}%
\AgdaSymbol{\{}\AgdaOperator{\AgdaInductiveConstructor{□⟨}}\AgdaSpace{}%
\AgdaBound{H}\AgdaSpace{}%
\AgdaOperator{\AgdaInductiveConstructor{?⟩}}\AgdaSymbol{\}}\AgdaSpace{}%
\AgdaSymbol{(}\AgdaBound{L′}\AgdaSpace{}%
\AgdaOperator{\AgdaInductiveConstructor{,}}\AgdaSpace{}%
\AgdaBound{L→L′}\AgdaSymbol{)}\AgdaSpace{}%
\AgdaSymbol{(}\AgdaInductiveConstructor{collide}\AgdaSpace{}%
\AgdaBound{v}\AgdaSpace{}%
\AgdaBound{neq}\AgdaSpace{}%
\AgdaInductiveConstructor{refl}\AgdaSymbol{)}\AgdaSpace{}%
\AgdaSymbol{=}\<%
\\
\>[0][@{}l@{\AgdaIndent{0}}]%
\>[4]\AgdaFunction{⊥-elim}\AgdaSpace{}%
\AgdaSymbol{(}\AgdaFunction{value-irreducible}\AgdaSpace{}%
\AgdaSymbol{(}\AgdaBound{v}\AgdaSpace{}%
\AgdaOperator{\AgdaInductiveConstructor{〈}}\AgdaSpace{}%
\AgdaSymbol{\AgdaUnderscore{}}\AgdaSpace{}%
\AgdaOperator{\AgdaInductiveConstructor{〉}}\AgdaSymbol{)}\AgdaSpace{}%
\AgdaBound{L→L′}\AgdaSymbol{)}\<%
\\
\\[\AgdaEmptyExtraSkip]%
\>[0]\AgdaFunction{frame-inv3}\AgdaSpace{}%
\AgdaSymbol{:}\AgdaSpace{}%
\AgdaSymbol{∀\{}\AgdaBound{L}\AgdaSpace{}%
\AgdaBound{N}\AgdaSpace{}%
\AgdaSymbol{:}\AgdaSpace{}%
\AgdaDatatype{Term}\AgdaSymbol{\}\{}\AgdaBound{F}\AgdaSpace{}%
\AgdaSymbol{:}\AgdaSpace{}%
\AgdaDatatype{PEFrame}\AgdaSymbol{\}}\<%
\\
\>[0][@{}l@{\AgdaIndent{0}}]%
\>[3]\AgdaSymbol{→}\AgdaSpace{}%
\AgdaFunction{reducible}\AgdaSpace{}%
\AgdaBound{L}\<%
\\
\>[3]\AgdaSymbol{→}\AgdaSpace{}%
\AgdaBound{F}\AgdaSpace{}%
\AgdaOperator{\AgdaFunction{⦉}}\AgdaSpace{}%
\AgdaBound{L}\AgdaSpace{}%
\AgdaOperator{\AgdaFunction{⦊}}\AgdaSpace{}%
\AgdaOperator{\AgdaDatatype{⟶}}\AgdaSpace{}%
\AgdaBound{N}\<%
\\
\>[3]\AgdaSymbol{→}\AgdaSpace{}%
\AgdaFunction{∃[}\AgdaSpace{}%
\AgdaBound{L′}\AgdaSpace{}%
\AgdaFunction{]}\AgdaSpace{}%
\AgdaSymbol{((}\AgdaBound{L}\AgdaSpace{}%
\AgdaOperator{\AgdaDatatype{⟶}}\AgdaSpace{}%
\AgdaBound{L′}\AgdaSymbol{)}\AgdaSpace{}%
\AgdaOperator{\AgdaFunction{×}}\AgdaSpace{}%
\AgdaSymbol{(}\AgdaBound{N}\AgdaSpace{}%
\AgdaOperator{\AgdaDatatype{≡}}\AgdaSpace{}%
\AgdaBound{F}\AgdaSpace{}%
\AgdaOperator{\AgdaFunction{⦉}}\AgdaSpace{}%
\AgdaBound{L′}\AgdaSpace{}%
\AgdaOperator{\AgdaFunction{⦊}}\AgdaSymbol{))}\<%
\\
\>[0]\AgdaFunction{frame-inv3}\AgdaSpace{}%
\AgdaSymbol{\{}\AgdaBound{L}\AgdaSymbol{\}\{}\AgdaBound{N}\AgdaSymbol{\}\{}\AgdaInductiveConstructor{□}\AgdaSymbol{\}}\AgdaSpace{}%
\AgdaBound{rL}\AgdaSpace{}%
\AgdaBound{FL→N}\AgdaSpace{}%
\AgdaSymbol{=}\AgdaSpace{}%
\AgdaSymbol{\AgdaUnderscore{}}\AgdaSpace{}%
\AgdaOperator{\AgdaInductiveConstructor{,}}\AgdaSpace{}%
\AgdaSymbol{(}\AgdaBound{FL→N}\AgdaSpace{}%
\AgdaOperator{\AgdaInductiveConstructor{,}}\AgdaSpace{}%
\AgdaInductiveConstructor{refl}\AgdaSymbol{)}\<%
\\
\>[0]\AgdaFunction{frame-inv3}\AgdaSpace{}%
\AgdaSymbol{\{}\AgdaBound{L}\AgdaSymbol{\}\{}\AgdaBound{N}\AgdaSymbol{\}\{}\AgdaOperator{\AgdaInductiveConstructor{`}}\AgdaSpace{}%
\AgdaBound{F}\AgdaSymbol{\}}\AgdaSpace{}%
\AgdaBound{rL}\AgdaSpace{}%
\AgdaBound{FL→N}\AgdaSpace{}%
\AgdaSymbol{=}\AgdaSpace{}%
\AgdaFunction{frame-inv2}\AgdaSpace{}%
\AgdaBound{rL}\AgdaSpace{}%
\AgdaBound{FL→N}\<%
\\
\\[\AgdaEmptyExtraSkip]%
\>[0]\AgdaFunction{blame-frame2}\AgdaSpace{}%
\AgdaSymbol{:}\AgdaSpace{}%
\AgdaSymbol{∀\{}\AgdaBound{F}\AgdaSymbol{\}\{}\AgdaBound{N}\AgdaSymbol{\}}\<%
\\
\>[0][@{}l@{\AgdaIndent{0}}]%
\>[3]\AgdaSymbol{→}\AgdaSpace{}%
\AgdaSymbol{(}\AgdaBound{F}\AgdaSpace{}%
\AgdaOperator{\AgdaFunction{⦉}}\AgdaSpace{}%
\AgdaInductiveConstructor{blame}\AgdaSpace{}%
\AgdaOperator{\AgdaFunction{⦊}}\AgdaSymbol{)}\AgdaSpace{}%
\AgdaOperator{\AgdaDatatype{⟶}}\AgdaSpace{}%
\AgdaBound{N}\<%
\\
\>[3]\AgdaSymbol{→}\AgdaSpace{}%
\AgdaBound{N}\AgdaSpace{}%
\AgdaOperator{\AgdaDatatype{≡}}\AgdaSpace{}%
\AgdaInductiveConstructor{blame}\<%
\\
\>[0]\AgdaFunction{blame-frame2}\AgdaSpace{}%
\AgdaSymbol{\{}\AgdaInductiveConstructor{□}\AgdaSymbol{\}\{}\AgdaBound{N}\AgdaSymbol{\}}\AgdaSpace{}%
\AgdaBound{Fb→N}\AgdaSpace{}%
\AgdaSymbol{=}\AgdaSpace{}%
\AgdaFunction{⊥-elim}\AgdaSpace{}%
\AgdaSymbol{(}\AgdaFunction{blame-irreducible}\AgdaSpace{}%
\AgdaBound{Fb→N}\AgdaSymbol{)}\<%
\\
\>[0]\AgdaFunction{blame-frame2}\AgdaSpace{}%
\AgdaSymbol{\{}\AgdaOperator{\AgdaInductiveConstructor{`}}\AgdaSpace{}%
\AgdaBound{F}\AgdaSymbol{\}\{}\AgdaBound{N}\AgdaSymbol{\}}\AgdaSpace{}%
\AgdaBound{Fb→N}\AgdaSpace{}%
\AgdaSymbol{=}\AgdaSpace{}%
\AgdaFunction{blame-frame}\AgdaSpace{}%
\AgdaBound{Fb→N}\<%
\\
\\[\AgdaEmptyExtraSkip]%
\>[0]\AgdaFunction{step-value-plus-one}\AgdaSpace{}%
\AgdaSymbol{:}\AgdaSpace{}%
\AgdaSymbol{∀\{}\AgdaBound{M}\AgdaSpace{}%
\AgdaBound{N}\AgdaSpace{}%
\AgdaBound{V}\AgdaSymbol{\}}\<%
\\
\>[0][@{}l@{\AgdaIndent{0}}]%
\>[3]\AgdaSymbol{→}\AgdaSpace{}%
\AgdaSymbol{(}\AgdaBound{M→N}\AgdaSpace{}%
\AgdaSymbol{:}\AgdaSpace{}%
\AgdaBound{M}\AgdaSpace{}%
\AgdaOperator{\AgdaDatatype{↠}}\AgdaSpace{}%
\AgdaBound{N}\AgdaSymbol{)}\<%
\\
\>[3]\AgdaSymbol{→}\AgdaSpace{}%
\AgdaSymbol{(}\AgdaBound{M→V}\AgdaSpace{}%
\AgdaSymbol{:}\AgdaSpace{}%
\AgdaBound{M}\AgdaSpace{}%
\AgdaOperator{\AgdaDatatype{↠}}\AgdaSpace{}%
\AgdaBound{V}\AgdaSymbol{)}\<%
\\
\>[3]\AgdaSymbol{→}\AgdaSpace{}%
\AgdaDatatype{Value}\AgdaSpace{}%
\AgdaBound{V}\<%
\\
\>[3]\AgdaSymbol{→}\AgdaSpace{}%
\AgdaFunction{len}\AgdaSpace{}%
\AgdaBound{M→N}\AgdaSpace{}%
\AgdaOperator{\AgdaDatatype{≡}}\AgdaSpace{}%
\AgdaInductiveConstructor{suc}\AgdaSpace{}%
\AgdaSymbol{(}\AgdaFunction{len}\AgdaSpace{}%
\AgdaBound{M→V}\AgdaSymbol{)}\<%
\\
\>[3]\AgdaSymbol{→}\AgdaSpace{}%
\AgdaDatatype{⊥}\<%
\\
\>[0]\AgdaFunction{step-value-plus-one}\AgdaSpace{}%
\AgdaSymbol{(\AgdaUnderscore{}}\AgdaSpace{}%
\AgdaOperator{\AgdaInductiveConstructor{⟶⟨}}\AgdaSpace{}%
\AgdaBound{r}\AgdaSpace{}%
\AgdaOperator{\AgdaInductiveConstructor{⟩}}\AgdaSpace{}%
\AgdaSymbol{\AgdaUnderscore{}}\AgdaSpace{}%
\AgdaOperator{\AgdaInductiveConstructor{END}}\AgdaSymbol{)}\AgdaSpace{}%
\AgdaSymbol{(\AgdaUnderscore{}}\AgdaSpace{}%
\AgdaOperator{\AgdaInductiveConstructor{END}}\AgdaSymbol{)}\AgdaSpace{}%
\AgdaBound{v}\AgdaSpace{}%
\AgdaBound{eq}\AgdaSpace{}%
\AgdaSymbol{=}\AgdaSpace{}%
\AgdaFunction{value-irreducible}\AgdaSpace{}%
\AgdaBound{v}\AgdaSpace{}%
\AgdaBound{r}\<%
\\
\>[0]\AgdaCatchallClause{\AgdaFunction{step-value-plus-one}}\AgdaSpace{}%
\AgdaCatchallClause{\AgdaSymbol{(\AgdaUnderscore{}}}\AgdaSpace{}%
\AgdaCatchallClause{\AgdaOperator{\AgdaInductiveConstructor{⟶⟨}}}\AgdaSpace{}%
\AgdaCatchallClause{\AgdaBound{r1}}\AgdaSpace{}%
\AgdaCatchallClause{\AgdaOperator{\AgdaInductiveConstructor{⟩}}}\AgdaSpace{}%
\AgdaCatchallClause{\AgdaBound{M→N}}\AgdaCatchallClause{\AgdaSymbol{)}}\AgdaSpace{}%
\AgdaCatchallClause{\AgdaSymbol{(\AgdaUnderscore{}}}\AgdaSpace{}%
\AgdaCatchallClause{\AgdaOperator{\AgdaInductiveConstructor{⟶⟨}}}\AgdaSpace{}%
\AgdaCatchallClause{\AgdaBound{r2}}\AgdaSpace{}%
\AgdaCatchallClause{\AgdaOperator{\AgdaInductiveConstructor{⟩}}}\AgdaSpace{}%
\AgdaCatchallClause{\AgdaBound{M→V}}\AgdaCatchallClause{\AgdaSymbol{)}}\AgdaSpace{}%
\AgdaCatchallClause{\AgdaBound{v}}\AgdaSpace{}%
\AgdaCatchallClause{\AgdaBound{eq}}\<%
\\
\>[0][@{}l@{\AgdaIndent{0}}]%
\>[4]\AgdaKeyword{with}\AgdaSpace{}%
\AgdaFunction{deterministic}\AgdaSpace{}%
\AgdaBound{r1}\AgdaSpace{}%
\AgdaBound{r2}\<%
\\
\>[0]\AgdaSymbol{...}\AgdaSpace{}%
\AgdaSymbol{|}\AgdaSpace{}%
\AgdaInductiveConstructor{refl}\AgdaSpace{}%
\AgdaSymbol{=}\AgdaSpace{}%
\AgdaFunction{step-value-plus-one}\AgdaSpace{}%
\AgdaBound{M→N}\AgdaSpace{}%
\AgdaBound{M→V}\AgdaSpace{}%
\AgdaBound{v}\AgdaSpace{}%
\AgdaSymbol{(}\AgdaFunction{suc-injective}\AgdaSpace{}%
\AgdaBound{eq}\AgdaSymbol{)}\<%
\\
\\[\AgdaEmptyExtraSkip]%
\>[0]\AgdaFunction{step-blame-plus-one}\AgdaSpace{}%
\AgdaSymbol{:}\AgdaSpace{}%
\AgdaSymbol{∀\{}\AgdaBound{M}\AgdaSpace{}%
\AgdaBound{N}\AgdaSymbol{\}}\<%
\\
\>[0][@{}l@{\AgdaIndent{0}}]%
\>[3]\AgdaSymbol{→}\AgdaSpace{}%
\AgdaSymbol{(}\AgdaBound{M→N}\AgdaSpace{}%
\AgdaSymbol{:}\AgdaSpace{}%
\AgdaBound{M}\AgdaSpace{}%
\AgdaOperator{\AgdaDatatype{↠}}\AgdaSpace{}%
\AgdaBound{N}\AgdaSymbol{)}\<%
\\
\>[3]\AgdaSymbol{→}\AgdaSpace{}%
\AgdaSymbol{(}\AgdaBound{M→b}\AgdaSpace{}%
\AgdaSymbol{:}\AgdaSpace{}%
\AgdaBound{M}\AgdaSpace{}%
\AgdaOperator{\AgdaDatatype{↠}}\AgdaSpace{}%
\AgdaInductiveConstructor{blame}\AgdaSymbol{)}\<%
\\
\>[3]\AgdaSymbol{→}\AgdaSpace{}%
\AgdaFunction{len}\AgdaSpace{}%
\AgdaBound{M→N}\AgdaSpace{}%
\AgdaOperator{\AgdaDatatype{≡}}\AgdaSpace{}%
\AgdaInductiveConstructor{suc}\AgdaSpace{}%
\AgdaSymbol{(}\AgdaFunction{len}\AgdaSpace{}%
\AgdaBound{M→b}\AgdaSymbol{)}\<%
\\
\>[3]\AgdaSymbol{→}\AgdaSpace{}%
\AgdaDatatype{⊥}\<%
\\
\>[0]\AgdaFunction{step-blame-plus-one}\AgdaSpace{}%
\AgdaSymbol{(\AgdaUnderscore{}}\AgdaSpace{}%
\AgdaOperator{\AgdaInductiveConstructor{⟶⟨}}\AgdaSpace{}%
\AgdaBound{r}\AgdaSpace{}%
\AgdaOperator{\AgdaInductiveConstructor{⟩}}\AgdaSpace{}%
\AgdaSymbol{\AgdaUnderscore{}}\AgdaSpace{}%
\AgdaOperator{\AgdaInductiveConstructor{END}}\AgdaSymbol{)}\AgdaSpace{}%
\AgdaSymbol{(\AgdaUnderscore{}}\AgdaSpace{}%
\AgdaOperator{\AgdaInductiveConstructor{END}}\AgdaSymbol{)}\AgdaSpace{}%
\AgdaBound{eq}\AgdaSpace{}%
\AgdaSymbol{=}\AgdaSpace{}%
\AgdaFunction{blame-irreducible}\AgdaSpace{}%
\AgdaBound{r}\<%
\\
\>[0]\AgdaCatchallClause{\AgdaFunction{step-blame-plus-one}}\AgdaSpace{}%
\AgdaCatchallClause{\AgdaSymbol{(\AgdaUnderscore{}}}\AgdaSpace{}%
\AgdaCatchallClause{\AgdaOperator{\AgdaInductiveConstructor{⟶⟨}}}\AgdaSpace{}%
\AgdaCatchallClause{\AgdaBound{r1}}\AgdaSpace{}%
\AgdaCatchallClause{\AgdaOperator{\AgdaInductiveConstructor{⟩}}}\AgdaSpace{}%
\AgdaCatchallClause{\AgdaBound{M→N}}\AgdaCatchallClause{\AgdaSymbol{)}}\AgdaSpace{}%
\AgdaCatchallClause{\AgdaSymbol{(\AgdaUnderscore{}}}\AgdaSpace{}%
\AgdaCatchallClause{\AgdaOperator{\AgdaInductiveConstructor{⟶⟨}}}\AgdaSpace{}%
\AgdaCatchallClause{\AgdaBound{r2}}\AgdaSpace{}%
\AgdaCatchallClause{\AgdaOperator{\AgdaInductiveConstructor{⟩}}}\AgdaSpace{}%
\AgdaCatchallClause{\AgdaBound{M→b}}\AgdaCatchallClause{\AgdaSymbol{)}}\AgdaSpace{}%
\AgdaCatchallClause{\AgdaBound{eq}}\<%
\\
\>[0][@{}l@{\AgdaIndent{0}}]%
\>[4]\AgdaKeyword{with}\AgdaSpace{}%
\AgdaFunction{deterministic}\AgdaSpace{}%
\AgdaBound{r1}\AgdaSpace{}%
\AgdaBound{r2}\<%
\\
\>[0]\AgdaSymbol{...}\AgdaSpace{}%
\AgdaSymbol{|}\AgdaSpace{}%
\AgdaInductiveConstructor{refl}\AgdaSpace{}%
\AgdaSymbol{=}\AgdaSpace{}%
\AgdaFunction{step-blame-plus-one}\AgdaSpace{}%
\AgdaBound{M→N}\AgdaSpace{}%
\AgdaBound{M→b}\AgdaSpace{}%
\AgdaSymbol{(}\AgdaFunction{suc-injective}\AgdaSpace{}%
\AgdaBound{eq}\AgdaSymbol{)}\<%
\\
\\[\AgdaEmptyExtraSkip]%
\>[0]\AgdaFunction{diverge-not-halt}\AgdaSpace{}%
\AgdaSymbol{:}\AgdaSpace{}%
\AgdaSymbol{∀\{}\AgdaBound{M}\AgdaSymbol{\}}\<%
\\
\>[0][@{}l@{\AgdaIndent{0}}]%
\>[2]\AgdaSymbol{→}\AgdaSpace{}%
\AgdaBound{M}\AgdaSpace{}%
\AgdaOperator{\AgdaFunction{⇑}}\<%
\\
\>[2]\AgdaSymbol{→}\AgdaSpace{}%
\AgdaOperator{\AgdaFunction{¬}}\AgdaSpace{}%
\AgdaFunction{halt}\AgdaSpace{}%
\AgdaBound{M}\<%
\\
\>[0]\AgdaFunction{diverge-not-halt}\AgdaSpace{}%
\AgdaBound{divM}\AgdaSpace{}%
\AgdaSymbol{(}\AgdaInductiveConstructor{inj₁}\AgdaSpace{}%
\AgdaBound{M→blame}\AgdaSymbol{)}\<%
\\
\>[0][@{}l@{\AgdaIndent{0}}]%
\>[4]\AgdaKeyword{with}\AgdaSpace{}%
\AgdaBound{divM}\AgdaSpace{}%
\AgdaSymbol{(}\AgdaInductiveConstructor{suc}\AgdaSpace{}%
\AgdaSymbol{(}\AgdaFunction{len}\AgdaSpace{}%
\AgdaBound{M→blame}\AgdaSymbol{))}\<%
\\
\>[0]\AgdaSymbol{...}\AgdaSpace{}%
\AgdaSymbol{|}\AgdaSpace{}%
\AgdaBound{N}\AgdaSpace{}%
\AgdaOperator{\AgdaInductiveConstructor{,}}\AgdaSpace{}%
\AgdaBound{M→N}\AgdaSpace{}%
\AgdaOperator{\AgdaInductiveConstructor{,}}\AgdaSpace{}%
\AgdaBound{eq}\AgdaSpace{}%
\AgdaSymbol{=}\AgdaSpace{}%
\AgdaFunction{step-blame-plus-one}\AgdaSpace{}%
\AgdaBound{M→N}\AgdaSpace{}%
\AgdaBound{M→blame}\AgdaSpace{}%
\AgdaSymbol{(}\AgdaFunction{sym}\AgdaSpace{}%
\AgdaBound{eq}\AgdaSymbol{)}\<%
\\
\>[0]\AgdaFunction{diverge-not-halt}\AgdaSpace{}%
\AgdaBound{divM}\AgdaSpace{}%
\AgdaSymbol{(}\AgdaInductiveConstructor{inj₂}\AgdaSpace{}%
\AgdaSymbol{(}\AgdaBound{V}\AgdaSpace{}%
\AgdaOperator{\AgdaInductiveConstructor{,}}\AgdaSpace{}%
\AgdaBound{M→V}\AgdaSpace{}%
\AgdaOperator{\AgdaInductiveConstructor{,}}\AgdaSpace{}%
\AgdaBound{v}\AgdaSymbol{))}\<%
\\
\>[0][@{}l@{\AgdaIndent{0}}]%
\>[4]\AgdaKeyword{with}\AgdaSpace{}%
\AgdaBound{divM}\AgdaSpace{}%
\AgdaSymbol{(}\AgdaInductiveConstructor{suc}\AgdaSpace{}%
\AgdaSymbol{(}\AgdaFunction{len}\AgdaSpace{}%
\AgdaBound{M→V}\AgdaSymbol{))}\<%
\\
\>[0]\AgdaSymbol{...}\AgdaSpace{}%
\AgdaSymbol{|}\AgdaSpace{}%
\AgdaBound{N}\AgdaSpace{}%
\AgdaOperator{\AgdaInductiveConstructor{,}}\AgdaSpace{}%
\AgdaBound{M→N}\AgdaSpace{}%
\AgdaOperator{\AgdaInductiveConstructor{,}}\AgdaSpace{}%
\AgdaBound{eq}\AgdaSpace{}%
\AgdaSymbol{=}\AgdaSpace{}%
\AgdaFunction{step-value-plus-one}\AgdaSpace{}%
\AgdaBound{M→N}\AgdaSpace{}%
\AgdaBound{M→V}\AgdaSpace{}%
\AgdaBound{v}\AgdaSpace{}%
\AgdaSymbol{(}\AgdaFunction{sym}\AgdaSpace{}%
\AgdaBound{eq}\AgdaSymbol{)}\<%
\\
\>[0]\<%
\\
\>[0]\AgdaFunction{cant-reduce-value-and-blame}\AgdaSpace{}%
\AgdaSymbol{:}\AgdaSpace{}%
\AgdaSymbol{∀\{}\AgdaBound{M}\AgdaSymbol{\}\{}\AgdaBound{V}\AgdaSymbol{\}}\<%
\\
\>[0][@{}l@{\AgdaIndent{0}}]%
\>[3]\AgdaSymbol{→}\AgdaSpace{}%
\AgdaDatatype{Value}\AgdaSpace{}%
\AgdaBound{V}\<%
\\
\>[3]\AgdaSymbol{→}\AgdaSpace{}%
\AgdaBound{M}\AgdaSpace{}%
\AgdaOperator{\AgdaDatatype{↠}}\AgdaSpace{}%
\AgdaBound{V}\<%
\\
\>[3]\AgdaSymbol{→}\AgdaSpace{}%
\AgdaBound{M}\AgdaSpace{}%
\AgdaOperator{\AgdaDatatype{↠}}\AgdaSpace{}%
\AgdaInductiveConstructor{blame}\<%
\\
\>[3]\AgdaSymbol{→}\AgdaSpace{}%
\AgdaDatatype{⊥}\<%
\\
\>[0]\AgdaFunction{cant-reduce-value-and-blame}\AgdaSpace{}%
\AgdaBound{v}\AgdaSpace{}%
\AgdaSymbol{(}\AgdaBound{M}\AgdaSpace{}%
\AgdaOperator{\AgdaInductiveConstructor{END}}\AgdaSymbol{)}\AgdaSpace{}%
\AgdaSymbol{(}\AgdaBound{M}\AgdaSpace{}%
\AgdaOperator{\AgdaInductiveConstructor{⟶⟨}}\AgdaSpace{}%
\AgdaBound{M→N}\AgdaSpace{}%
\AgdaOperator{\AgdaInductiveConstructor{⟩}}\AgdaSpace{}%
\AgdaBound{N→b}\AgdaSymbol{)}\AgdaSpace{}%
\AgdaSymbol{=}\<%
\\
\>[0][@{}l@{\AgdaIndent{0}}]%
\>[2]\AgdaFunction{⊥-elim}\AgdaSpace{}%
\AgdaSymbol{(}\AgdaFunction{value-irreducible}\AgdaSpace{}%
\AgdaBound{v}\AgdaSpace{}%
\AgdaBound{M→N}\AgdaSymbol{)}\<%
\\
\>[0]\AgdaFunction{cant-reduce-value-and-blame}\AgdaSpace{}%
\AgdaBound{v}\AgdaSpace{}%
\AgdaSymbol{(}\AgdaDottedPattern{\AgdaSymbol{.}}\AgdaDottedPattern{\AgdaInductiveConstructor{blame}}\AgdaSpace{}%
\AgdaOperator{\AgdaInductiveConstructor{⟶⟨}}\AgdaSpace{}%
\AgdaBound{M→N}\AgdaSpace{}%
\AgdaOperator{\AgdaInductiveConstructor{⟩}}\AgdaSpace{}%
\AgdaBound{N→V}\AgdaSymbol{)}\AgdaSpace{}%
\AgdaSymbol{(}\AgdaDottedPattern{\AgdaSymbol{.}}\AgdaDottedPattern{\AgdaInductiveConstructor{blame}}\AgdaSpace{}%
\AgdaOperator{\AgdaInductiveConstructor{END}}\AgdaSymbol{)}\AgdaSpace{}%
\AgdaSymbol{=}\<%
\\
\>[0][@{}l@{\AgdaIndent{0}}]%
\>[2]\AgdaFunction{⊥-elim}\AgdaSpace{}%
\AgdaSymbol{(}\AgdaFunction{blame-irreducible}\AgdaSpace{}%
\AgdaBound{M→N}\AgdaSymbol{)}\<%
\\
\>[0]\AgdaFunction{cant-reduce-value-and-blame}\AgdaSpace{}%
\AgdaBound{v}\AgdaSpace{}%
\AgdaSymbol{(}\AgdaBound{M}\AgdaSpace{}%
\AgdaOperator{\AgdaInductiveConstructor{⟶⟨}}\AgdaSpace{}%
\AgdaBound{M→N}\AgdaSpace{}%
\AgdaOperator{\AgdaInductiveConstructor{⟩}}\AgdaSpace{}%
\AgdaBound{N→V}\AgdaSymbol{)}\AgdaSpace{}%
\AgdaSymbol{(}\AgdaDottedPattern{\AgdaSymbol{.}}\AgdaDottedPattern{\AgdaBound{M}}\AgdaSpace{}%
\AgdaOperator{\AgdaInductiveConstructor{⟶⟨}}\AgdaSpace{}%
\AgdaBound{M→N′}\AgdaSpace{}%
\AgdaOperator{\AgdaInductiveConstructor{⟩}}\AgdaSpace{}%
\AgdaBound{N′→b}\AgdaSymbol{)}\<%
\\
\>[0][@{}l@{\AgdaIndent{0}}]%
\>[2]\AgdaKeyword{rewrite}\AgdaSpace{}%
\AgdaFunction{deterministic}\AgdaSpace{}%
\AgdaBound{M→N}\AgdaSpace{}%
\AgdaBound{M→N′}\AgdaSpace{}%
\AgdaSymbol{=}\AgdaSpace{}%
\AgdaFunction{cant-reduce-value-and-blame}\AgdaSpace{}%
\AgdaBound{v}\AgdaSpace{}%
\AgdaBound{N→V}\AgdaSpace{}%
\AgdaBound{N′→b}\<%
\end{code}


\begin{code}[hide]%
\>[0]\AgdaSymbol{\{-\#}\AgdaSpace{}%
\AgdaKeyword{OPTIONS}\AgdaSpace{}%
\AgdaPragma{--rewriting}\AgdaSpace{}%
\AgdaSymbol{\#-\}}\<%
\\
\>[0]\AgdaKeyword{module}\AgdaSpace{}%
\AgdaModule{LogRel.PeterPrecision}\AgdaSpace{}%
\AgdaKeyword{where}\<%
\\
\\[\AgdaEmptyExtraSkip]%
\>[0]\AgdaKeyword{open}\AgdaSpace{}%
\AgdaKeyword{import}\AgdaSpace{}%
\AgdaModule{Data.Empty}\AgdaSpace{}%
\AgdaKeyword{using}\AgdaSpace{}%
\AgdaSymbol{(}\AgdaDatatype{⊥}\AgdaSymbol{;}\AgdaSpace{}%
\AgdaFunction{⊥-elim}\AgdaSymbol{)}\<%
\\
\>[0]\AgdaKeyword{open}\AgdaSpace{}%
\AgdaKeyword{import}\AgdaSpace{}%
\AgdaModule{Data.List}\AgdaSpace{}%
\AgdaKeyword{using}\AgdaSpace{}%
\AgdaSymbol{(}\AgdaDatatype{List}\AgdaSymbol{;}\AgdaSpace{}%
\AgdaInductiveConstructor{[]}\AgdaSymbol{;}\AgdaSpace{}%
\AgdaOperator{\AgdaInductiveConstructor{\AgdaUnderscore{}∷\AgdaUnderscore{}}}\AgdaSymbol{;}\AgdaSpace{}%
\AgdaFunction{map}\AgdaSymbol{;}\AgdaSpace{}%
\AgdaFunction{length}\AgdaSymbol{)}\<%
\\
\>[0]\AgdaKeyword{open}\AgdaSpace{}%
\AgdaKeyword{import}\AgdaSpace{}%
\AgdaModule{Data.Nat}\<%
\\
\>[0]\AgdaKeyword{open}\AgdaSpace{}%
\AgdaKeyword{import}\AgdaSpace{}%
\AgdaModule{Data.Nat.Properties}\<%
\\
\>[0]\AgdaKeyword{open}\AgdaSpace{}%
\AgdaKeyword{import}\AgdaSpace{}%
\AgdaModule{Data.Bool}\AgdaSpace{}%
\AgdaKeyword{using}\AgdaSpace{}%
\AgdaSymbol{(}\AgdaInductiveConstructor{true}\AgdaSymbol{;}\AgdaSpace{}%
\AgdaInductiveConstructor{false}\AgdaSymbol{)}\AgdaSpace{}%
\AgdaKeyword{renaming}\AgdaSpace{}%
\AgdaSymbol{(}\AgdaDatatype{Bool}\AgdaSpace{}%
\AgdaSymbol{to}\AgdaSpace{}%
\AgdaDatatype{𝔹}\AgdaSymbol{)}\<%
\\
\>[0]\AgdaKeyword{open}\AgdaSpace{}%
\AgdaKeyword{import}\AgdaSpace{}%
\AgdaModule{Data.Product}\AgdaSpace{}%
\AgdaKeyword{using}\AgdaSpace{}%
\AgdaSymbol{(}\AgdaOperator{\AgdaInductiveConstructor{\AgdaUnderscore{},\AgdaUnderscore{}}}\AgdaSymbol{;}\AgdaOperator{\AgdaFunction{\AgdaUnderscore{}×\AgdaUnderscore{}}}\AgdaSymbol{;}\AgdaSpace{}%
\AgdaField{proj₁}\AgdaSymbol{;}\AgdaSpace{}%
\AgdaField{proj₂}\AgdaSymbol{;}\AgdaSpace{}%
\AgdaFunction{Σ-syntax}\AgdaSymbol{;}\AgdaSpace{}%
\AgdaFunction{∃-syntax}\AgdaSymbol{)}\<%
\\
\>[0]\AgdaKeyword{open}\AgdaSpace{}%
\AgdaKeyword{import}\AgdaSpace{}%
\AgdaModule{Data.Sum}\AgdaSpace{}%
\AgdaKeyword{using}\AgdaSpace{}%
\AgdaSymbol{(}\AgdaOperator{\AgdaDatatype{\AgdaUnderscore{}⊎\AgdaUnderscore{}}}\AgdaSymbol{;}\AgdaSpace{}%
\AgdaInductiveConstructor{inj₁}\AgdaSymbol{;}\AgdaSpace{}%
\AgdaInductiveConstructor{inj₂}\AgdaSymbol{)}\<%
\\
\>[0]\AgdaKeyword{open}\AgdaSpace{}%
\AgdaKeyword{import}\AgdaSpace{}%
\AgdaModule{Data.Unit}\AgdaSpace{}%
\AgdaKeyword{using}\AgdaSpace{}%
\AgdaSymbol{(}\AgdaRecord{⊤}\AgdaSymbol{;}\AgdaSpace{}%
\AgdaInductiveConstructor{tt}\AgdaSymbol{)}\<%
\\
\>[0]\AgdaKeyword{open}\AgdaSpace{}%
\AgdaKeyword{import}\AgdaSpace{}%
\AgdaModule{Data.Unit.Polymorphic}\AgdaSpace{}%
\AgdaKeyword{renaming}\AgdaSpace{}%
\AgdaSymbol{(}\AgdaFunction{⊤}\AgdaSpace{}%
\AgdaSymbol{to}\AgdaSpace{}%
\AgdaFunction{topᵖ}\AgdaSymbol{;}\AgdaSpace{}%
\AgdaFunction{tt}\AgdaSpace{}%
\AgdaSymbol{to}\AgdaSpace{}%
\AgdaFunction{ttᵖ}\AgdaSymbol{)}\<%
\\
\>[0]\AgdaKeyword{open}\AgdaSpace{}%
\AgdaKeyword{import}\AgdaSpace{}%
\AgdaModule{Relation.Binary.PropositionalEquality}\AgdaSpace{}%
\AgdaSymbol{as}\AgdaSpace{}%
\AgdaModule{Eq}\<%
\\
\>[0][@{}l@{\AgdaIndent{0}}]%
\>[2]\AgdaKeyword{using}\AgdaSpace{}%
\AgdaSymbol{(}\AgdaOperator{\AgdaDatatype{\AgdaUnderscore{}≡\AgdaUnderscore{}}}\AgdaSymbol{;}\AgdaSpace{}%
\AgdaOperator{\AgdaFunction{\AgdaUnderscore{}≢\AgdaUnderscore{}}}\AgdaSymbol{;}\AgdaSpace{}%
\AgdaInductiveConstructor{refl}\AgdaSymbol{;}\AgdaSpace{}%
\AgdaFunction{sym}\AgdaSymbol{;}\AgdaSpace{}%
\AgdaFunction{cong}\AgdaSymbol{;}\AgdaSpace{}%
\AgdaFunction{subst}\AgdaSymbol{;}\AgdaSpace{}%
\AgdaFunction{trans}\AgdaSymbol{)}\<%
\\
\>[0]\AgdaKeyword{open}\AgdaSpace{}%
\AgdaKeyword{import}\AgdaSpace{}%
\AgdaModule{Relation.Nullary}\AgdaSpace{}%
\AgdaKeyword{using}\AgdaSpace{}%
\AgdaSymbol{(}\AgdaOperator{\AgdaFunction{¬\AgdaUnderscore{}}}\AgdaSymbol{;}\AgdaSpace{}%
\AgdaRecord{Dec}\AgdaSymbol{;}\AgdaSpace{}%
\AgdaInductiveConstructor{yes}\AgdaSymbol{;}\AgdaSpace{}%
\AgdaInductiveConstructor{no}\AgdaSymbol{)}\<%
\\
\\[\AgdaEmptyExtraSkip]%
\>[0]\AgdaKeyword{open}\AgdaSpace{}%
\AgdaKeyword{import}\AgdaSpace{}%
\AgdaModule{Var}\<%
\\
\>[0]\AgdaKeyword{open}\AgdaSpace{}%
\AgdaKeyword{import}\AgdaSpace{}%
\AgdaModule{Sig}\<%
\\
\>[0]\AgdaKeyword{open}\AgdaSpace{}%
\AgdaKeyword{import}\AgdaSpace{}%
\AgdaModule{LogRel.PeterCastCalculus}\<%
\\
\>[0]\AgdaKeyword{open}\AgdaSpace{}%
\AgdaKeyword{import}\AgdaSpace{}%
\AgdaModule{StepIndexedLogic}\<%
\end{code}

\section{Precision on Types and Terms}
\label{sec:precision}

To talk about the gradual guarantee, we first define when one type is
less precise than another one.\footnote{Some of the literature instead defines
a less imprecise relation, but here we stick to the original direction
of the relation~\cite{Siek:2006bh}.}

\begin{code}%
\>[0]\AgdaKeyword{infixr}\AgdaSpace{}%
\AgdaNumber{6}\AgdaSpace{}%
\AgdaOperator{\AgdaDatatype{\AgdaUnderscore{}⊑\AgdaUnderscore{}}}\<%
\\
\>[0]\AgdaKeyword{data}\AgdaSpace{}%
\AgdaOperator{\AgdaDatatype{\AgdaUnderscore{}⊑\AgdaUnderscore{}}}\AgdaSpace{}%
\AgdaSymbol{:}\AgdaSpace{}%
\AgdaDatatype{Type}\AgdaSpace{}%
\AgdaSymbol{→}\AgdaSpace{}%
\AgdaDatatype{Type}\AgdaSpace{}%
\AgdaSymbol{→}\AgdaSpace{}%
\AgdaPrimitive{Set}\AgdaSpace{}%
\AgdaKeyword{where}\<%
\\
\>[0][@{}l@{\AgdaIndent{0}}]%
\>[2]\AgdaInductiveConstructor{unk⊑unk}\AgdaSpace{}%
\AgdaSymbol{:}\AgdaSpace{}%
\AgdaInductiveConstructor{★}\AgdaSpace{}%
\AgdaOperator{\AgdaDatatype{⊑}}\AgdaSpace{}%
\AgdaInductiveConstructor{★}\<%
\\
\>[2]\AgdaInductiveConstructor{unk⊑}\AgdaSpace{}%
\AgdaSymbol{:}\AgdaSpace{}%
\AgdaSymbol{∀\{}\AgdaBound{G}\AgdaSymbol{\}\{}\AgdaBound{B}\AgdaSymbol{\}}\AgdaSpace{}%
\AgdaSymbol{→}\AgdaSpace{}%
\AgdaOperator{\AgdaFunction{⌈}}\AgdaSpace{}%
\AgdaBound{G}\AgdaSpace{}%
\AgdaOperator{\AgdaFunction{⌉}}\AgdaSpace{}%
\AgdaOperator{\AgdaDatatype{⊑}}\AgdaSpace{}%
\AgdaBound{B}\AgdaSpace{}%
\AgdaSymbol{→}\AgdaSpace{}%
\AgdaInductiveConstructor{★}\AgdaSpace{}%
\AgdaOperator{\AgdaDatatype{⊑}}\AgdaSpace{}%
\AgdaBound{B}\<%
\\
\>[2]\AgdaInductiveConstructor{base⊑}\AgdaSpace{}%
\AgdaSymbol{:}\AgdaSpace{}%
\AgdaSymbol{∀\{}\AgdaBound{ι}\AgdaSymbol{\}}\AgdaSpace{}%
\AgdaSymbol{→}\AgdaSpace{}%
\AgdaOperator{\AgdaInductiveConstructor{\$ₜ}}\AgdaSpace{}%
\AgdaBound{ι}\AgdaSpace{}%
\AgdaOperator{\AgdaDatatype{⊑}}\AgdaSpace{}%
\AgdaOperator{\AgdaInductiveConstructor{\$ₜ}}\AgdaSpace{}%
\AgdaBound{ι}\<%
\\
\>[2]\AgdaInductiveConstructor{fun⊑}\AgdaSpace{}%
\AgdaSymbol{:}\AgdaSpace{}%
\AgdaSymbol{∀\{}\AgdaBound{A}\AgdaSpace{}%
\AgdaBound{B}\AgdaSpace{}%
\AgdaBound{C}\AgdaSpace{}%
\AgdaBound{D}\AgdaSymbol{\}}%
\>[21]\AgdaSymbol{→}%
\>[24]\AgdaBound{A}\AgdaSpace{}%
\AgdaOperator{\AgdaDatatype{⊑}}\AgdaSpace{}%
\AgdaBound{C}%
\>[31]\AgdaSymbol{→}%
\>[34]\AgdaBound{B}\AgdaSpace{}%
\AgdaOperator{\AgdaDatatype{⊑}}\AgdaSpace{}%
\AgdaBound{D}%
\>[41]\AgdaSymbol{→}%
\>[44]\AgdaBound{A}\AgdaSpace{}%
\AgdaOperator{\AgdaInductiveConstructor{⇒}}\AgdaSpace{}%
\AgdaBound{B}\AgdaSpace{}%
\AgdaOperator{\AgdaDatatype{⊑}}\AgdaSpace{}%
\AgdaBound{C}\AgdaSpace{}%
\AgdaOperator{\AgdaInductiveConstructor{⇒}}\AgdaSpace{}%
\AgdaBound{D}\<%
\end{code}

The first two rules for precision are usually presented as a single
rule that says ★ is less precise than any type.  Instead we have
separated out the case for when both types are ★ from the case when
only the less-precise type is ★.  Also, for the rule \textsf{unk⊑},
instead of writing $B ≢ ★$ we have written $⌈ G ⌉ ⊑ B$, which turns
out to be important later when we define the logical relation and use
recursion on the precision relation.
The \textsf{Prec} type bundles two types in the precision relation.
Precision is reflexive and transitive, but we will not need transitivity.

\begin{code}%
\>[0]\AgdaFunction{Prec}\AgdaSpace{}%
\AgdaSymbol{:}\AgdaSpace{}%
\AgdaPrimitive{Set}\<%
\\
\>[0]\AgdaFunction{Prec}\AgdaSpace{}%
\AgdaSymbol{=}\AgdaSpace{}%
\AgdaSymbol{(}\AgdaFunction{∃[}\AgdaSpace{}%
\AgdaBound{A}\AgdaSpace{}%
\AgdaFunction{]}\AgdaSpace{}%
\AgdaFunction{∃[}\AgdaSpace{}%
\AgdaBound{B}\AgdaSpace{}%
\AgdaFunction{]}\AgdaSpace{}%
\AgdaBound{A}\AgdaSpace{}%
\AgdaOperator{\AgdaDatatype{⊑}}\AgdaSpace{}%
\AgdaBound{B}\AgdaSymbol{)}\<%
\\
\\[\AgdaEmptyExtraSkip]%
\>[0]\AgdaFunction{Refl⊑}\AgdaSpace{}%
\AgdaSymbol{:}\AgdaSpace{}%
\AgdaSymbol{∀\{}\AgdaBound{A}\AgdaSymbol{\}}\AgdaSpace{}%
\AgdaSymbol{→}\AgdaSpace{}%
\AgdaBound{A}\AgdaSpace{}%
\AgdaOperator{\AgdaDatatype{⊑}}\AgdaSpace{}%
\AgdaBound{A}\<%
\\
\>[0]\AgdaFunction{Refl⊑}\AgdaSpace{}%
\AgdaSymbol{\{}\AgdaInductiveConstructor{★}\AgdaSymbol{\}}\AgdaSpace{}%
\AgdaSymbol{=}\AgdaSpace{}%
\AgdaInductiveConstructor{unk⊑unk}\<%
\\
\>[0]\AgdaFunction{Refl⊑}\AgdaSpace{}%
\AgdaSymbol{\{}\AgdaOperator{\AgdaInductiveConstructor{\$ₜ}}\AgdaSpace{}%
\AgdaBound{ι}\AgdaSymbol{\}}\AgdaSpace{}%
\AgdaSymbol{=}\AgdaSpace{}%
\AgdaInductiveConstructor{base⊑}\<%
\\
\>[0]\AgdaFunction{Refl⊑}\AgdaSpace{}%
\AgdaSymbol{\{}\AgdaBound{A}\AgdaSpace{}%
\AgdaOperator{\AgdaInductiveConstructor{⇒}}\AgdaSpace{}%
\AgdaBound{B}\AgdaSymbol{\}}\AgdaSpace{}%
\AgdaSymbol{=}\AgdaSpace{}%
\AgdaInductiveConstructor{fun⊑}\AgdaSpace{}%
\AgdaFunction{Refl⊑}\AgdaSpace{}%
\AgdaFunction{Refl⊑}\<%
\end{code}

\begin{code}[hide]%
\>[0]\AgdaFunction{unk⊑gnd-inv}\AgdaSpace{}%
\AgdaSymbol{:}\AgdaSpace{}%
\AgdaSymbol{∀\{}\AgdaBound{G}\AgdaSymbol{\}}\<%
\\
\>[0][@{}l@{\AgdaIndent{0}}]%
\>[3]\AgdaSymbol{→}\AgdaSpace{}%
\AgdaSymbol{(}\AgdaBound{c}\AgdaSpace{}%
\AgdaSymbol{:}\AgdaSpace{}%
\AgdaInductiveConstructor{★}\AgdaSpace{}%
\AgdaOperator{\AgdaDatatype{⊑}}\AgdaSpace{}%
\AgdaOperator{\AgdaFunction{⌈}}\AgdaSpace{}%
\AgdaBound{G}\AgdaSpace{}%
\AgdaOperator{\AgdaFunction{⌉}}\AgdaSymbol{)}\<%
\\
\>[3]\AgdaSymbol{→}\AgdaSpace{}%
\AgdaFunction{∃[}\AgdaSpace{}%
\AgdaBound{d}\AgdaSpace{}%
\AgdaFunction{]}\AgdaSpace{}%
\AgdaBound{c}\AgdaSpace{}%
\AgdaOperator{\AgdaDatatype{≡}}\AgdaSpace{}%
\AgdaInductiveConstructor{unk⊑}\AgdaSymbol{\{}\AgdaBound{G}\AgdaSymbol{\}\{}\AgdaOperator{\AgdaFunction{⌈}}\AgdaSpace{}%
\AgdaBound{G}\AgdaSpace{}%
\AgdaOperator{\AgdaFunction{⌉}}\AgdaSymbol{\}}\AgdaSpace{}%
\AgdaBound{d}\<%
\\
\>[0]\AgdaFunction{unk⊑gnd-inv}\AgdaSpace{}%
\AgdaSymbol{\{}\AgdaOperator{\AgdaInductiveConstructor{\$ᵍ}}\AgdaSpace{}%
\AgdaBound{ι}\AgdaSymbol{\}}\AgdaSpace{}%
\AgdaSymbol{(}\AgdaInductiveConstructor{unk⊑}\AgdaSpace{}%
\AgdaSymbol{\{}\AgdaOperator{\AgdaInductiveConstructor{\$ᵍ}}\AgdaSpace{}%
\AgdaDottedPattern{\AgdaSymbol{.}}\AgdaDottedPattern{\AgdaBound{ι}}\AgdaSymbol{\}}\AgdaSpace{}%
\AgdaInductiveConstructor{base⊑}\AgdaSymbol{)}\AgdaSpace{}%
\AgdaSymbol{=}\AgdaSpace{}%
\AgdaInductiveConstructor{base⊑}\AgdaSpace{}%
\AgdaOperator{\AgdaInductiveConstructor{,}}\AgdaSpace{}%
\AgdaInductiveConstructor{refl}\<%
\\
\>[0]\AgdaFunction{unk⊑gnd-inv}\AgdaSpace{}%
\AgdaSymbol{\{}\AgdaInductiveConstructor{★⇒★}\AgdaSymbol{\}}\AgdaSpace{}%
\AgdaSymbol{(}\AgdaInductiveConstructor{unk⊑}\AgdaSpace{}%
\AgdaSymbol{\{}\AgdaInductiveConstructor{★⇒★}\AgdaSymbol{\}}\AgdaSpace{}%
\AgdaSymbol{(}\AgdaInductiveConstructor{fun⊑}\AgdaSpace{}%
\AgdaBound{c}\AgdaSpace{}%
\AgdaBound{d}\AgdaSymbol{))}\AgdaSpace{}%
\AgdaSymbol{=}\AgdaSpace{}%
\AgdaInductiveConstructor{fun⊑}\AgdaSpace{}%
\AgdaBound{c}\AgdaSpace{}%
\AgdaBound{d}\AgdaSpace{}%
\AgdaOperator{\AgdaInductiveConstructor{,}}\AgdaSpace{}%
\AgdaInductiveConstructor{refl}\<%
\\
\\[\AgdaEmptyExtraSkip]%
\>[0]\AgdaFunction{dyn-prec-unique}\AgdaSpace{}%
\AgdaSymbol{:}\AgdaSpace{}%
\AgdaSymbol{∀\{}\AgdaBound{A}\AgdaSymbol{\}}\<%
\\
\>[0][@{}l@{\AgdaIndent{0}}]%
\>[2]\AgdaSymbol{→}\AgdaSpace{}%
\AgdaSymbol{(}\AgdaBound{c}\AgdaSpace{}%
\AgdaSymbol{:}\AgdaSpace{}%
\AgdaInductiveConstructor{★}\AgdaSpace{}%
\AgdaOperator{\AgdaDatatype{⊑}}\AgdaSpace{}%
\AgdaBound{A}\AgdaSymbol{)}\<%
\\
\>[2]\AgdaSymbol{→}\AgdaSpace{}%
\AgdaSymbol{(}\AgdaBound{d}\AgdaSpace{}%
\AgdaSymbol{:}\AgdaSpace{}%
\AgdaInductiveConstructor{★}\AgdaSpace{}%
\AgdaOperator{\AgdaDatatype{⊑}}\AgdaSpace{}%
\AgdaBound{A}\AgdaSymbol{)}\<%
\\
\>[2]\AgdaSymbol{→}\AgdaSpace{}%
\AgdaBound{c}\AgdaSpace{}%
\AgdaOperator{\AgdaDatatype{≡}}\AgdaSpace{}%
\AgdaBound{d}\<%
\\
\>[0]\AgdaFunction{dyn-prec-unique}\AgdaSpace{}%
\AgdaSymbol{\{}\AgdaInductiveConstructor{★}\AgdaSymbol{\}}\AgdaSpace{}%
\AgdaInductiveConstructor{unk⊑unk}\AgdaSpace{}%
\AgdaInductiveConstructor{unk⊑unk}\AgdaSpace{}%
\AgdaSymbol{=}\AgdaSpace{}%
\AgdaInductiveConstructor{refl}\<%
\\
\>[0]\AgdaFunction{dyn-prec-unique}\AgdaSpace{}%
\AgdaSymbol{\{}\AgdaInductiveConstructor{★}\AgdaSymbol{\}}\AgdaSpace{}%
\AgdaInductiveConstructor{unk⊑unk}\AgdaSpace{}%
\AgdaSymbol{(}\AgdaInductiveConstructor{unk⊑}\AgdaSpace{}%
\AgdaSymbol{\{}\AgdaOperator{\AgdaInductiveConstructor{\$ᵍ}}\AgdaSpace{}%
\AgdaBound{ι}\AgdaSymbol{\}}\AgdaSpace{}%
\AgdaSymbol{())}\<%
\\
\>[0]\AgdaFunction{dyn-prec-unique}\AgdaSpace{}%
\AgdaSymbol{\{}\AgdaInductiveConstructor{★}\AgdaSymbol{\}}\AgdaSpace{}%
\AgdaInductiveConstructor{unk⊑unk}\AgdaSpace{}%
\AgdaSymbol{(}\AgdaInductiveConstructor{unk⊑}\AgdaSpace{}%
\AgdaSymbol{\{}\AgdaInductiveConstructor{★⇒★}\AgdaSymbol{\}}\AgdaSpace{}%
\AgdaSymbol{())}\<%
\\
\>[0]\AgdaFunction{dyn-prec-unique}\AgdaSpace{}%
\AgdaSymbol{\{}\AgdaInductiveConstructor{★}\AgdaSymbol{\}}\AgdaSpace{}%
\AgdaSymbol{(}\AgdaInductiveConstructor{unk⊑}\AgdaSpace{}%
\AgdaSymbol{\{}\AgdaOperator{\AgdaInductiveConstructor{\$ᵍ}}\AgdaSpace{}%
\AgdaBound{ι}\AgdaSymbol{\}}\AgdaSpace{}%
\AgdaSymbol{())}\AgdaSpace{}%
\AgdaBound{d}\<%
\\
\>[0]\AgdaFunction{dyn-prec-unique}\AgdaSpace{}%
\AgdaSymbol{\{}\AgdaInductiveConstructor{★}\AgdaSymbol{\}}\AgdaSpace{}%
\AgdaSymbol{(}\AgdaInductiveConstructor{unk⊑}\AgdaSpace{}%
\AgdaSymbol{\{}\AgdaInductiveConstructor{★⇒★}\AgdaSymbol{\}}\AgdaSpace{}%
\AgdaSymbol{())}\AgdaSpace{}%
\AgdaBound{d}\<%
\\
\>[0]\AgdaFunction{dyn-prec-unique}\AgdaSpace{}%
\AgdaSymbol{\{}\AgdaOperator{\AgdaInductiveConstructor{\$ₜ}}\AgdaSpace{}%
\AgdaBound{ι}\AgdaSymbol{\}}\AgdaSpace{}%
\AgdaSymbol{(}\AgdaInductiveConstructor{unk⊑}\AgdaSpace{}%
\AgdaSymbol{\{}\AgdaOperator{\AgdaInductiveConstructor{\$ᵍ}}\AgdaSpace{}%
\AgdaDottedPattern{\AgdaSymbol{.}}\AgdaDottedPattern{\AgdaBound{ι}}\AgdaSymbol{\}}\AgdaSpace{}%
\AgdaInductiveConstructor{base⊑}\AgdaSymbol{)}\AgdaSpace{}%
\AgdaSymbol{(}\AgdaInductiveConstructor{unk⊑}\AgdaSpace{}%
\AgdaSymbol{\{}\AgdaOperator{\AgdaInductiveConstructor{\$ᵍ}}\AgdaSpace{}%
\AgdaDottedPattern{\AgdaSymbol{.}}\AgdaDottedPattern{\AgdaBound{ι}}\AgdaSymbol{\}}\AgdaSpace{}%
\AgdaInductiveConstructor{base⊑}\AgdaSymbol{)}\AgdaSpace{}%
\AgdaSymbol{=}\AgdaSpace{}%
\AgdaInductiveConstructor{refl}\<%
\\
\>[0]\AgdaFunction{dyn-prec-unique}\AgdaSpace{}%
\AgdaSymbol{\{}\AgdaBound{A}\AgdaSpace{}%
\AgdaOperator{\AgdaInductiveConstructor{⇒}}\AgdaSpace{}%
\AgdaBound{A₁}\AgdaSymbol{\}}\AgdaSpace{}%
\AgdaSymbol{(}\AgdaInductiveConstructor{unk⊑}\AgdaSpace{}%
\AgdaSymbol{\{}\AgdaInductiveConstructor{★⇒★}\AgdaSymbol{\}}\AgdaSpace{}%
\AgdaSymbol{(}\AgdaInductiveConstructor{fun⊑}\AgdaSpace{}%
\AgdaBound{c}\AgdaSpace{}%
\AgdaBound{c₁}\AgdaSymbol{))}\AgdaSpace{}%
\AgdaSymbol{(}\AgdaInductiveConstructor{unk⊑}\AgdaSpace{}%
\AgdaSymbol{\{}\AgdaInductiveConstructor{★⇒★}\AgdaSymbol{\}}\AgdaSpace{}%
\AgdaSymbol{(}\AgdaInductiveConstructor{fun⊑}\AgdaSpace{}%
\AgdaBound{d}\AgdaSpace{}%
\AgdaBound{d₁}\AgdaSymbol{))}\<%
\\
\>[0][@{}l@{\AgdaIndent{0}}]%
\>[4]\AgdaKeyword{with}\AgdaSpace{}%
\AgdaFunction{dyn-prec-unique}\AgdaSpace{}%
\AgdaBound{c}\AgdaSpace{}%
\AgdaBound{d}\AgdaSpace{}%
\AgdaSymbol{|}\AgdaSpace{}%
\AgdaFunction{dyn-prec-unique}\AgdaSpace{}%
\AgdaBound{c₁}\AgdaSpace{}%
\AgdaBound{d₁}\<%
\\
\>[0]\AgdaSymbol{...}\AgdaSpace{}%
\AgdaSymbol{|}\AgdaSpace{}%
\AgdaInductiveConstructor{refl}\AgdaSpace{}%
\AgdaSymbol{|}\AgdaSpace{}%
\AgdaInductiveConstructor{refl}\AgdaSpace{}%
\AgdaSymbol{=}\AgdaSpace{}%
\AgdaInductiveConstructor{refl}\<%
\\
\\[\AgdaEmptyExtraSkip]%
\>[0]\AgdaFunction{gnd-prec-unique}\AgdaSpace{}%
\AgdaSymbol{:}\AgdaSpace{}%
\AgdaSymbol{∀\{}\AgdaBound{G}\AgdaSpace{}%
\AgdaBound{A}\AgdaSymbol{\}}\<%
\\
\>[0][@{}l@{\AgdaIndent{0}}]%
\>[3]\AgdaSymbol{→}\AgdaSpace{}%
\AgdaSymbol{(}\AgdaBound{c}\AgdaSpace{}%
\AgdaSymbol{:}\AgdaSpace{}%
\AgdaOperator{\AgdaFunction{⌈}}\AgdaSpace{}%
\AgdaBound{G}\AgdaSpace{}%
\AgdaOperator{\AgdaFunction{⌉}}\AgdaSpace{}%
\AgdaOperator{\AgdaDatatype{⊑}}\AgdaSpace{}%
\AgdaBound{A}\AgdaSymbol{)}\<%
\\
\>[3]\AgdaSymbol{→}\AgdaSpace{}%
\AgdaSymbol{(}\AgdaBound{d}\AgdaSpace{}%
\AgdaSymbol{:}\AgdaSpace{}%
\AgdaOperator{\AgdaFunction{⌈}}\AgdaSpace{}%
\AgdaBound{G}\AgdaSpace{}%
\AgdaOperator{\AgdaFunction{⌉}}\AgdaSpace{}%
\AgdaOperator{\AgdaDatatype{⊑}}\AgdaSpace{}%
\AgdaBound{A}\AgdaSymbol{)}\<%
\\
\>[3]\AgdaSymbol{→}\AgdaSpace{}%
\AgdaBound{c}\AgdaSpace{}%
\AgdaOperator{\AgdaDatatype{≡}}\AgdaSpace{}%
\AgdaBound{d}\<%
\\
\>[0]\AgdaFunction{gnd-prec-unique}\AgdaSpace{}%
\AgdaSymbol{\{}\AgdaOperator{\AgdaInductiveConstructor{\$ᵍ}}\AgdaSpace{}%
\AgdaBound{ι}\AgdaSymbol{\}}\AgdaSpace{}%
\AgdaSymbol{\{}\AgdaDottedPattern{\AgdaSymbol{.(}}\AgdaDottedPattern{\AgdaOperator{\AgdaInductiveConstructor{\$ₜ}}}\AgdaSpace{}%
\AgdaDottedPattern{\AgdaBound{ι}}\AgdaDottedPattern{\AgdaSymbol{)}}\AgdaSymbol{\}}\AgdaSpace{}%
\AgdaInductiveConstructor{base⊑}\AgdaSpace{}%
\AgdaInductiveConstructor{base⊑}\AgdaSpace{}%
\AgdaSymbol{=}\AgdaSpace{}%
\AgdaInductiveConstructor{refl}\<%
\\
\>[0]\AgdaFunction{gnd-prec-unique}\AgdaSpace{}%
\AgdaSymbol{\{}\AgdaInductiveConstructor{★⇒★}\AgdaSymbol{\}}\AgdaSpace{}%
\AgdaSymbol{\{}\AgdaDottedPattern{\AgdaSymbol{.(\AgdaUnderscore{}}}\AgdaSpace{}%
\AgdaDottedPattern{\AgdaOperator{\AgdaInductiveConstructor{⇒}}}\AgdaSpace{}%
\AgdaDottedPattern{\AgdaSymbol{\AgdaUnderscore{})}}\AgdaSymbol{\}}\AgdaSpace{}%
\AgdaSymbol{(}\AgdaInductiveConstructor{fun⊑}\AgdaSpace{}%
\AgdaBound{c}\AgdaSpace{}%
\AgdaBound{c₁}\AgdaSymbol{)}\AgdaSpace{}%
\AgdaSymbol{(}\AgdaInductiveConstructor{fun⊑}\AgdaSpace{}%
\AgdaBound{d}\AgdaSpace{}%
\AgdaBound{d₁}\AgdaSymbol{)}\<%
\\
\>[0][@{}l@{\AgdaIndent{0}}]%
\>[4]\AgdaKeyword{with}\AgdaSpace{}%
\AgdaFunction{dyn-prec-unique}\AgdaSpace{}%
\AgdaBound{c}\AgdaSpace{}%
\AgdaBound{d}\AgdaSpace{}%
\AgdaSymbol{|}\AgdaSpace{}%
\AgdaFunction{dyn-prec-unique}\AgdaSpace{}%
\AgdaBound{c₁}\AgdaSpace{}%
\AgdaBound{d₁}\<%
\\
\>[0]\AgdaSymbol{...}\AgdaSpace{}%
\AgdaSymbol{|}\AgdaSpace{}%
\AgdaInductiveConstructor{refl}\AgdaSpace{}%
\AgdaSymbol{|}\AgdaSpace{}%
\AgdaInductiveConstructor{refl}\AgdaSpace{}%
\AgdaSymbol{=}\AgdaSpace{}%
\AgdaInductiveConstructor{refl}\<%
\end{code}

Figure~\ref{fig:term-precision} defines the precision relation on
terms.  The judgment has the form $Γ ⊩ M ⊑ M′ ⦂ c$ where Γ is a list
of precision-related types and $c : A ⊑ A′$ is a precision derivation
for the types of $M$ and $M′$. There are two rules for injection and
also for projection, allowing either to appear on the left or right
across from an arbitrary term. However, when injection is on the
right, the term on the left must have type ★ (rule
\textsf{⊑-inj-R}).  Similarly, when projection is on the right, the
term on the left must have type ★ (rule \textsf{⊑-proj-R}).
The term \textsf{blame} is at least as precise as any term.

\begin{figure}[tbp]
\begin{code}%
\>[0]\AgdaKeyword{infix}\AgdaSpace{}%
\AgdaNumber{3}\AgdaSpace{}%
\AgdaOperator{\AgdaDatatype{\AgdaUnderscore{}⊩\AgdaUnderscore{}⊑\AgdaUnderscore{}⦂\AgdaUnderscore{}}}\<%
\\
\>[0]\AgdaKeyword{data}\AgdaSpace{}%
\AgdaOperator{\AgdaDatatype{\AgdaUnderscore{}⊩\AgdaUnderscore{}⊑\AgdaUnderscore{}⦂\AgdaUnderscore{}}}\AgdaSpace{}%
\AgdaSymbol{:}\AgdaSpace{}%
\AgdaDatatype{List}\AgdaSpace{}%
\AgdaFunction{Prec}\AgdaSpace{}%
\AgdaSymbol{→}\AgdaSpace{}%
\AgdaDatatype{Term}\AgdaSpace{}%
\AgdaSymbol{→}\AgdaSpace{}%
\AgdaDatatype{Term}\AgdaSpace{}%
\AgdaSymbol{→}\AgdaSpace{}%
\AgdaSymbol{∀\{}\AgdaBound{A}\AgdaSpace{}%
\AgdaBound{B}\AgdaSpace{}%
\AgdaSymbol{:}\AgdaSpace{}%
\AgdaDatatype{Type}\AgdaSymbol{\}}\AgdaSpace{}%
\AgdaSymbol{→}\AgdaSpace{}%
\AgdaBound{A}\AgdaSpace{}%
\AgdaOperator{\AgdaDatatype{⊑}}\AgdaSpace{}%
\AgdaBound{B}\AgdaSpace{}%
\AgdaSymbol{→}\AgdaSpace{}%
\AgdaPrimitive{Set}%
\>[70]\AgdaKeyword{where}\<%
\\
\>[0][@{}l@{\AgdaIndent{0}}]%
\>[2]\AgdaInductiveConstructor{⊑-var}\AgdaSpace{}%
\AgdaSymbol{:}\AgdaSpace{}%
\AgdaSymbol{∀}\AgdaSpace{}%
\AgdaSymbol{\{}\AgdaBound{Γ}\AgdaSpace{}%
\AgdaBound{x}\AgdaSpace{}%
\AgdaBound{A⊑B}\AgdaSymbol{\}}%
\>[23]\AgdaSymbol{→}%
\>[26]\AgdaBound{Γ}\AgdaSpace{}%
\AgdaOperator{\AgdaFunction{∋}}\AgdaSpace{}%
\AgdaBound{x}\AgdaSpace{}%
\AgdaOperator{\AgdaFunction{⦂}}\AgdaSpace{}%
\AgdaBound{A⊑B}%
\>[39]\AgdaSymbol{→}%
\>[42]\AgdaBound{Γ}\AgdaSpace{}%
\AgdaOperator{\AgdaDatatype{⊩}}\AgdaSpace{}%
\AgdaSymbol{(}\AgdaOperator{\AgdaInductiveConstructor{`}}\AgdaSpace{}%
\AgdaBound{x}\AgdaSymbol{)}\AgdaSpace{}%
\AgdaOperator{\AgdaDatatype{⊑}}\AgdaSpace{}%
\AgdaSymbol{(}\AgdaOperator{\AgdaInductiveConstructor{`}}\AgdaSpace{}%
\AgdaBound{x}\AgdaSymbol{)}\AgdaSpace{}%
\AgdaOperator{\AgdaDatatype{⦂}}\AgdaSpace{}%
\AgdaField{proj₂}\AgdaSpace{}%
\AgdaSymbol{(}\AgdaField{proj₂}\AgdaSpace{}%
\AgdaBound{A⊑B}\AgdaSymbol{)}\<%
\\
\>[2]\AgdaInductiveConstructor{⊑-lit}\AgdaSpace{}%
\AgdaSymbol{:}\AgdaSpace{}%
\AgdaSymbol{∀}\AgdaSpace{}%
\AgdaSymbol{\{}\AgdaBound{Γ}\AgdaSpace{}%
\AgdaBound{c}\AgdaSymbol{\}}\AgdaSpace{}%
\AgdaSymbol{→}%
\>[21]\AgdaBound{Γ}\AgdaSpace{}%
\AgdaOperator{\AgdaDatatype{⊩}}\AgdaSpace{}%
\AgdaSymbol{(}\AgdaInductiveConstructor{\$}\AgdaSpace{}%
\AgdaBound{c}\AgdaSymbol{)}\AgdaSpace{}%
\AgdaOperator{\AgdaDatatype{⊑}}\AgdaSpace{}%
\AgdaSymbol{(}\AgdaInductiveConstructor{\$}\AgdaSpace{}%
\AgdaBound{c}\AgdaSymbol{)}\AgdaSpace{}%
\AgdaOperator{\AgdaDatatype{⦂}}\AgdaSpace{}%
\AgdaInductiveConstructor{base⊑}\AgdaSymbol{\{}\AgdaFunction{typeof}\AgdaSpace{}%
\AgdaBound{c}\AgdaSymbol{\}}\<%
\\
\>[2]\AgdaInductiveConstructor{⊑-app}\AgdaSpace{}%
\AgdaSymbol{:}\AgdaSpace{}%
\AgdaSymbol{∀\{}\AgdaBound{Γ}\AgdaSpace{}%
\AgdaBound{L}\AgdaSpace{}%
\AgdaBound{M}\AgdaSpace{}%
\AgdaBound{L′}\AgdaSpace{}%
\AgdaBound{M′}\AgdaSpace{}%
\AgdaBound{A}\AgdaSpace{}%
\AgdaBound{B}\AgdaSpace{}%
\AgdaBound{C}\AgdaSpace{}%
\AgdaBound{D}\AgdaSymbol{\}\{}\AgdaBound{c}\AgdaSpace{}%
\AgdaSymbol{:}\AgdaSpace{}%
\AgdaBound{A}\AgdaSpace{}%
\AgdaOperator{\AgdaDatatype{⊑}}\AgdaSpace{}%
\AgdaBound{C}\AgdaSymbol{\}\{}\AgdaBound{d}\AgdaSpace{}%
\AgdaSymbol{:}\AgdaSpace{}%
\AgdaBound{B}\AgdaSpace{}%
\AgdaOperator{\AgdaDatatype{⊑}}\AgdaSpace{}%
\AgdaBound{D}\AgdaSymbol{\}}\<%
\\
\>[2][@{}l@{\AgdaIndent{0}}]%
\>[5]\AgdaSymbol{→}\AgdaSpace{}%
\AgdaBound{Γ}\AgdaSpace{}%
\AgdaOperator{\AgdaDatatype{⊩}}\AgdaSpace{}%
\AgdaBound{L}\AgdaSpace{}%
\AgdaOperator{\AgdaDatatype{⊑}}\AgdaSpace{}%
\AgdaBound{L′}\AgdaSpace{}%
\AgdaOperator{\AgdaDatatype{⦂}}\AgdaSpace{}%
\AgdaInductiveConstructor{fun⊑}\AgdaSpace{}%
\AgdaBound{c}\AgdaSpace{}%
\AgdaBound{d}%
\>[30]\AgdaSymbol{→}%
\>[33]\AgdaBound{Γ}\AgdaSpace{}%
\AgdaOperator{\AgdaDatatype{⊩}}\AgdaSpace{}%
\AgdaBound{M}\AgdaSpace{}%
\AgdaOperator{\AgdaDatatype{⊑}}\AgdaSpace{}%
\AgdaBound{M′}\AgdaSpace{}%
\AgdaOperator{\AgdaDatatype{⦂}}\AgdaSpace{}%
\AgdaBound{c}\<%
\\
\>[5]\AgdaSymbol{→}\AgdaSpace{}%
\AgdaBound{Γ}\AgdaSpace{}%
\AgdaOperator{\AgdaDatatype{⊩}}\AgdaSpace{}%
\AgdaBound{L}\AgdaSpace{}%
\AgdaOperator{\AgdaInductiveConstructor{·}}\AgdaSpace{}%
\AgdaBound{M}\AgdaSpace{}%
\AgdaOperator{\AgdaDatatype{⊑}}\AgdaSpace{}%
\AgdaBound{L′}\AgdaSpace{}%
\AgdaOperator{\AgdaInductiveConstructor{·}}\AgdaSpace{}%
\AgdaBound{M′}\AgdaSpace{}%
\AgdaOperator{\AgdaDatatype{⦂}}\AgdaSpace{}%
\AgdaBound{d}\<%
\\
\>[2]\AgdaInductiveConstructor{⊑-lam}\AgdaSpace{}%
\AgdaSymbol{:}\AgdaSpace{}%
\AgdaSymbol{∀\{}\AgdaBound{Γ}\AgdaSpace{}%
\AgdaBound{N}\AgdaSpace{}%
\AgdaBound{N′}\AgdaSpace{}%
\AgdaBound{A}\AgdaSpace{}%
\AgdaBound{B}\AgdaSpace{}%
\AgdaBound{C}\AgdaSpace{}%
\AgdaBound{D}\AgdaSymbol{\}\{}\AgdaBound{c}\AgdaSpace{}%
\AgdaSymbol{:}\AgdaSpace{}%
\AgdaBound{A}\AgdaSpace{}%
\AgdaOperator{\AgdaDatatype{⊑}}\AgdaSpace{}%
\AgdaBound{C}\AgdaSymbol{\}\{}\AgdaBound{d}\AgdaSpace{}%
\AgdaSymbol{:}\AgdaSpace{}%
\AgdaBound{B}\AgdaSpace{}%
\AgdaOperator{\AgdaDatatype{⊑}}\AgdaSpace{}%
\AgdaBound{D}\AgdaSymbol{\}}\<%
\\
\>[2][@{}l@{\AgdaIndent{0}}]%
\>[5]\AgdaSymbol{→}\AgdaSpace{}%
\AgdaSymbol{(}\AgdaBound{A}\AgdaSpace{}%
\AgdaOperator{\AgdaInductiveConstructor{,}}\AgdaSpace{}%
\AgdaBound{C}\AgdaSpace{}%
\AgdaOperator{\AgdaInductiveConstructor{,}}\AgdaSpace{}%
\AgdaBound{c}\AgdaSymbol{)}\AgdaSpace{}%
\AgdaOperator{\AgdaInductiveConstructor{∷}}\AgdaSpace{}%
\AgdaBound{Γ}\AgdaSpace{}%
\AgdaOperator{\AgdaDatatype{⊩}}\AgdaSpace{}%
\AgdaBound{N}\AgdaSpace{}%
\AgdaOperator{\AgdaDatatype{⊑}}\AgdaSpace{}%
\AgdaBound{N′}\AgdaSpace{}%
\AgdaOperator{\AgdaDatatype{⦂}}\AgdaSpace{}%
\AgdaBound{d}%
\>[37]\AgdaSymbol{→}%
\>[40]\AgdaBound{Γ}\AgdaSpace{}%
\AgdaOperator{\AgdaDatatype{⊩}}\AgdaSpace{}%
\AgdaInductiveConstructor{ƛ}\AgdaSpace{}%
\AgdaBound{N}\AgdaSpace{}%
\AgdaOperator{\AgdaDatatype{⊑}}\AgdaSpace{}%
\AgdaInductiveConstructor{ƛ}\AgdaSpace{}%
\AgdaBound{N′}\AgdaSpace{}%
\AgdaOperator{\AgdaDatatype{⦂}}\AgdaSpace{}%
\AgdaInductiveConstructor{fun⊑}\AgdaSpace{}%
\AgdaBound{c}\AgdaSpace{}%
\AgdaBound{d}\<%
\\
\>[2]\AgdaInductiveConstructor{⊑-inj-L}\AgdaSpace{}%
\AgdaSymbol{:}\AgdaSpace{}%
\AgdaSymbol{∀\{}\AgdaBound{Γ}\AgdaSpace{}%
\AgdaBound{M}\AgdaSpace{}%
\AgdaBound{M′}\AgdaSymbol{\}\{}\AgdaBound{G}\AgdaSpace{}%
\AgdaBound{B}\AgdaSymbol{\}\{}\AgdaBound{c}\AgdaSpace{}%
\AgdaSymbol{:}\AgdaSpace{}%
\AgdaOperator{\AgdaFunction{⌈}}\AgdaSpace{}%
\AgdaBound{G}\AgdaSpace{}%
\AgdaOperator{\AgdaFunction{⌉}}\AgdaSpace{}%
\AgdaOperator{\AgdaDatatype{⊑}}\AgdaSpace{}%
\AgdaBound{B}\AgdaSymbol{\}}\<%
\\
\>[2][@{}l@{\AgdaIndent{0}}]%
\>[5]\AgdaSymbol{→}\AgdaSpace{}%
\AgdaBound{Γ}\AgdaSpace{}%
\AgdaOperator{\AgdaDatatype{⊩}}\AgdaSpace{}%
\AgdaBound{M}\AgdaSpace{}%
\AgdaOperator{\AgdaDatatype{⊑}}\AgdaSpace{}%
\AgdaBound{M′}\AgdaSpace{}%
\AgdaOperator{\AgdaDatatype{⦂}}\AgdaSpace{}%
\AgdaBound{c}%
\>[23]\AgdaSymbol{→}%
\>[26]\AgdaBound{Γ}\AgdaSpace{}%
\AgdaOperator{\AgdaDatatype{⊩}}\AgdaSpace{}%
\AgdaBound{M}\AgdaSpace{}%
\AgdaOperator{\AgdaInductiveConstructor{⟨}}\AgdaSpace{}%
\AgdaBound{G}\AgdaSpace{}%
\AgdaOperator{\AgdaInductiveConstructor{!⟩}}\AgdaSpace{}%
\AgdaOperator{\AgdaDatatype{⊑}}\AgdaSpace{}%
\AgdaBound{M′}\AgdaSpace{}%
\AgdaOperator{\AgdaDatatype{⦂}}\AgdaSpace{}%
\AgdaInductiveConstructor{unk⊑}\AgdaSymbol{\{}\AgdaBound{G}\AgdaSymbol{\}\{}\AgdaBound{B}\AgdaSymbol{\}}\AgdaSpace{}%
\AgdaBound{c}\<%
\\
\>[2]\AgdaInductiveConstructor{⊑-inj-R}\AgdaSpace{}%
\AgdaSymbol{:}\AgdaSpace{}%
\AgdaSymbol{∀\{}\AgdaBound{Γ}\AgdaSpace{}%
\AgdaBound{M}\AgdaSpace{}%
\AgdaBound{M′}\AgdaSymbol{\}\{}\AgdaBound{G}\AgdaSymbol{\}\{}\AgdaBound{c}\AgdaSpace{}%
\AgdaSymbol{:}\AgdaSpace{}%
\AgdaInductiveConstructor{★}\AgdaSpace{}%
\AgdaOperator{\AgdaDatatype{⊑}}\AgdaSpace{}%
\AgdaOperator{\AgdaFunction{⌈}}\AgdaSpace{}%
\AgdaBound{G}\AgdaSpace{}%
\AgdaOperator{\AgdaFunction{⌉}}\AgdaSymbol{\}}\<%
\\
\>[2][@{}l@{\AgdaIndent{0}}]%
\>[5]\AgdaSymbol{→}\AgdaSpace{}%
\AgdaBound{Γ}\AgdaSpace{}%
\AgdaOperator{\AgdaDatatype{⊩}}\AgdaSpace{}%
\AgdaBound{M}\AgdaSpace{}%
\AgdaOperator{\AgdaDatatype{⊑}}\AgdaSpace{}%
\AgdaBound{M′}\AgdaSpace{}%
\AgdaOperator{\AgdaDatatype{⦂}}\AgdaSpace{}%
\AgdaBound{c}%
\>[23]\AgdaSymbol{→}%
\>[26]\AgdaBound{Γ}\AgdaSpace{}%
\AgdaOperator{\AgdaDatatype{⊩}}\AgdaSpace{}%
\AgdaBound{M}\AgdaSpace{}%
\AgdaOperator{\AgdaDatatype{⊑}}\AgdaSpace{}%
\AgdaBound{M′}\AgdaSpace{}%
\AgdaOperator{\AgdaInductiveConstructor{⟨}}\AgdaSpace{}%
\AgdaBound{G}\AgdaSpace{}%
\AgdaOperator{\AgdaInductiveConstructor{!⟩}}\AgdaSpace{}%
\AgdaOperator{\AgdaDatatype{⦂}}\AgdaSpace{}%
\AgdaInductiveConstructor{unk⊑unk}\<%
\\
\>[2]\AgdaInductiveConstructor{⊑-proj-L}\AgdaSpace{}%
\AgdaSymbol{:}\AgdaSpace{}%
\AgdaSymbol{∀\{}\AgdaBound{Γ}\AgdaSpace{}%
\AgdaBound{M}\AgdaSpace{}%
\AgdaBound{M′}\AgdaSpace{}%
\AgdaBound{H}\AgdaSpace{}%
\AgdaBound{B}\AgdaSymbol{\}\{}\AgdaBound{c}\AgdaSpace{}%
\AgdaSymbol{:}\AgdaSpace{}%
\AgdaOperator{\AgdaFunction{⌈}}\AgdaSpace{}%
\AgdaBound{H}\AgdaSpace{}%
\AgdaOperator{\AgdaFunction{⌉}}\AgdaSpace{}%
\AgdaOperator{\AgdaDatatype{⊑}}\AgdaSpace{}%
\AgdaBound{B}\AgdaSymbol{\}}\<%
\\
\>[2][@{}l@{\AgdaIndent{0}}]%
\>[5]\AgdaSymbol{→}\AgdaSpace{}%
\AgdaBound{Γ}\AgdaSpace{}%
\AgdaOperator{\AgdaDatatype{⊩}}\AgdaSpace{}%
\AgdaBound{M}\AgdaSpace{}%
\AgdaOperator{\AgdaDatatype{⊑}}\AgdaSpace{}%
\AgdaBound{M′}\AgdaSpace{}%
\AgdaOperator{\AgdaDatatype{⦂}}\AgdaSpace{}%
\AgdaInductiveConstructor{unk⊑}\AgdaSpace{}%
\AgdaBound{c}%
\>[28]\AgdaSymbol{→}%
\>[31]\AgdaBound{Γ}\AgdaSpace{}%
\AgdaOperator{\AgdaDatatype{⊩}}\AgdaSpace{}%
\AgdaBound{M}\AgdaSpace{}%
\AgdaOperator{\AgdaInductiveConstructor{⟨}}\AgdaSpace{}%
\AgdaBound{H}\AgdaSpace{}%
\AgdaOperator{\AgdaInductiveConstructor{?⟩}}\AgdaSpace{}%
\AgdaOperator{\AgdaDatatype{⊑}}\AgdaSpace{}%
\AgdaBound{M′}\AgdaSpace{}%
\AgdaOperator{\AgdaDatatype{⦂}}\AgdaSpace{}%
\AgdaBound{c}\<%
\\
\>[2]\AgdaInductiveConstructor{⊑-proj-R}\AgdaSpace{}%
\AgdaSymbol{:}\AgdaSpace{}%
\AgdaSymbol{∀\{}\AgdaBound{Γ}\AgdaSpace{}%
\AgdaBound{M}\AgdaSpace{}%
\AgdaBound{M′}\AgdaSpace{}%
\AgdaBound{H}\AgdaSymbol{\}\{}\AgdaBound{c}\AgdaSpace{}%
\AgdaSymbol{:}\AgdaSpace{}%
\AgdaInductiveConstructor{★}\AgdaSpace{}%
\AgdaOperator{\AgdaDatatype{⊑}}\AgdaSpace{}%
\AgdaOperator{\AgdaFunction{⌈}}\AgdaSpace{}%
\AgdaBound{H}\AgdaSpace{}%
\AgdaOperator{\AgdaFunction{⌉}}\AgdaSymbol{\}}\<%
\\
\>[2][@{}l@{\AgdaIndent{0}}]%
\>[5]\AgdaSymbol{→}\AgdaSpace{}%
\AgdaBound{Γ}\AgdaSpace{}%
\AgdaOperator{\AgdaDatatype{⊩}}\AgdaSpace{}%
\AgdaBound{M}\AgdaSpace{}%
\AgdaOperator{\AgdaDatatype{⊑}}\AgdaSpace{}%
\AgdaBound{M′}\AgdaSpace{}%
\AgdaOperator{\AgdaDatatype{⦂}}\AgdaSpace{}%
\AgdaInductiveConstructor{unk⊑unk}%
\>[29]\AgdaSymbol{→}%
\>[32]\AgdaBound{Γ}\AgdaSpace{}%
\AgdaOperator{\AgdaDatatype{⊩}}\AgdaSpace{}%
\AgdaBound{M}\AgdaSpace{}%
\AgdaOperator{\AgdaDatatype{⊑}}\AgdaSpace{}%
\AgdaBound{M′}\AgdaSpace{}%
\AgdaOperator{\AgdaInductiveConstructor{⟨}}\AgdaSpace{}%
\AgdaBound{H}\AgdaSpace{}%
\AgdaOperator{\AgdaInductiveConstructor{?⟩}}%
\>[51]\AgdaOperator{\AgdaDatatype{⦂}}\AgdaSpace{}%
\AgdaBound{c}\<%
\\
\>[2]\AgdaInductiveConstructor{⊑-blame}\AgdaSpace{}%
\AgdaSymbol{:}\AgdaSpace{}%
\AgdaSymbol{∀\{}\AgdaBound{Γ}\AgdaSpace{}%
\AgdaBound{M}\AgdaSpace{}%
\AgdaBound{A}\AgdaSymbol{\}}%
\>[22]\AgdaSymbol{→}%
\>[25]\AgdaFunction{map}\AgdaSpace{}%
\AgdaField{proj₁}\AgdaSpace{}%
\AgdaBound{Γ}\AgdaSpace{}%
\AgdaOperator{\AgdaDatatype{⊢}}\AgdaSpace{}%
\AgdaBound{M}\AgdaSpace{}%
\AgdaOperator{\AgdaDatatype{⦂}}\AgdaSpace{}%
\AgdaBound{A}%
\>[46]\AgdaSymbol{→}%
\>[49]\AgdaBound{Γ}\AgdaSpace{}%
\AgdaOperator{\AgdaDatatype{⊩}}\AgdaSpace{}%
\AgdaBound{M}\AgdaSpace{}%
\AgdaOperator{\AgdaDatatype{⊑}}\AgdaSpace{}%
\AgdaInductiveConstructor{blame}\AgdaSpace{}%
\AgdaOperator{\AgdaDatatype{⦂}}\AgdaSpace{}%
\AgdaFunction{Refl⊑}\AgdaSymbol{\{}\AgdaBound{A}\AgdaSymbol{\}}\<%
\end{code}
\caption{Precision Relation on Terms}
\label{fig:term-precision}
\end{figure}

With precision defined, we are ready to discuss the gradual guarantee.
It states that if $M$ is less precise than $M′$, then $M$ and $M′$
behave in a similar way, as defined below by the predicate
$\mathsf{gradual}\,M\,M′$. In particular, it says that if the
more-precise term terminates or diverges, then the less-precise term
does likewise.  On the other hand the more-precise term may reduce to
\textsf{blame} even though the less-precise term does not.

\begin{code}%
\>[0]\AgdaFunction{gradual}\AgdaSpace{}%
\AgdaSymbol{:}\AgdaSpace{}%
\AgdaSymbol{(}\AgdaBound{M}\AgdaSpace{}%
\AgdaBound{M′}\AgdaSpace{}%
\AgdaSymbol{:}\AgdaSpace{}%
\AgdaDatatype{Term}\AgdaSymbol{)}\AgdaSpace{}%
\AgdaSymbol{→}\AgdaSpace{}%
\AgdaPrimitive{Set}\<%
\\
\>[0]\AgdaFunction{gradual}\AgdaSpace{}%
\AgdaBound{M}\AgdaSpace{}%
\AgdaBound{M′}\AgdaSpace{}%
\AgdaSymbol{=}\AgdaSpace{}%
\AgdaSymbol{(}\AgdaBound{M′}\AgdaSpace{}%
\AgdaOperator{\AgdaFunction{⇓}}\AgdaSpace{}%
\AgdaSymbol{→}\AgdaSpace{}%
\AgdaBound{M}\AgdaSpace{}%
\AgdaOperator{\AgdaFunction{⇓}}\AgdaSymbol{)}\AgdaSpace{}%
\AgdaOperator{\AgdaFunction{×}}\AgdaSpace{}%
\AgdaSymbol{(}\AgdaBound{M′}\AgdaSpace{}%
\AgdaOperator{\AgdaFunction{⇑}}\AgdaSpace{}%
\AgdaSymbol{→}\AgdaSpace{}%
\AgdaBound{M}\AgdaSpace{}%
\AgdaOperator{\AgdaFunction{⇑}}\AgdaSymbol{)}\AgdaSpace{}%
\AgdaOperator{\AgdaFunction{×}}\AgdaSpace{}%
\AgdaSymbol{(}\AgdaBound{M}\AgdaSpace{}%
\AgdaOperator{\AgdaFunction{⇓}}\AgdaSpace{}%
\AgdaSymbol{→}\AgdaSpace{}%
\AgdaBound{M′}\AgdaSpace{}%
\AgdaOperator{\AgdaFunction{⇓}}\AgdaSpace{}%
\AgdaOperator{\AgdaDatatype{⊎}}\AgdaSpace{}%
\AgdaBound{M′}\AgdaSpace{}%
\AgdaOperator{\AgdaDatatype{↠}}\AgdaSpace{}%
\AgdaInductiveConstructor{blame}\AgdaSymbol{)}\<%
\\
\>[0][@{}l@{\AgdaIndent{0}}]%
\>[4]\AgdaOperator{\AgdaFunction{×}}\AgdaSpace{}%
\AgdaSymbol{(}\AgdaBound{M}\AgdaSpace{}%
\AgdaOperator{\AgdaFunction{⇑}}\AgdaSpace{}%
\AgdaSymbol{→}\AgdaSpace{}%
\AgdaBound{M′}\AgdaSpace{}%
\AgdaOperator{\AgdaFunction{⇑⊎blame}}\AgdaSymbol{)}\AgdaSpace{}%
\AgdaOperator{\AgdaFunction{×}}\AgdaSpace{}%
\AgdaSymbol{(}\AgdaBound{M}\AgdaSpace{}%
\AgdaOperator{\AgdaDatatype{↠}}\AgdaSpace{}%
\AgdaInductiveConstructor{blame}\AgdaSpace{}%
\AgdaSymbol{→}\AgdaSpace{}%
\AgdaBound{M′}\AgdaSpace{}%
\AgdaOperator{\AgdaDatatype{↠}}\AgdaSpace{}%
\AgdaInductiveConstructor{blame}\AgdaSymbol{)}\<%
\end{code}

\begin{code}[hide]%
\>[0]\AgdaSymbol{\{-\#}\AgdaSpace{}%
\AgdaKeyword{OPTIONS}\AgdaSpace{}%
\AgdaPragma{--rewriting}\AgdaSpace{}%
\AgdaSymbol{\#-\}}\<%
\\
\>[0]\AgdaKeyword{module}\AgdaSpace{}%
\AgdaModule{LogRel.PeterLogRel}\AgdaSpace{}%
\AgdaKeyword{where}\<%
\\
\\[\AgdaEmptyExtraSkip]%
\>[0]\AgdaKeyword{open}\AgdaSpace{}%
\AgdaKeyword{import}\AgdaSpace{}%
\AgdaModule{Data.Empty}\AgdaSpace{}%
\AgdaKeyword{using}\AgdaSpace{}%
\AgdaSymbol{(}\AgdaDatatype{⊥}\AgdaSymbol{;}\AgdaSpace{}%
\AgdaFunction{⊥-elim}\AgdaSymbol{)}\<%
\\
\>[0]\AgdaKeyword{open}\AgdaSpace{}%
\AgdaKeyword{import}\AgdaSpace{}%
\AgdaModule{Data.List}\AgdaSpace{}%
\AgdaKeyword{using}\AgdaSpace{}%
\AgdaSymbol{(}\AgdaDatatype{List}\AgdaSymbol{;}\AgdaSpace{}%
\AgdaInductiveConstructor{[]}\AgdaSymbol{;}\AgdaSpace{}%
\AgdaOperator{\AgdaInductiveConstructor{\AgdaUnderscore{}∷\AgdaUnderscore{}}}\AgdaSymbol{;}\AgdaSpace{}%
\AgdaFunction{map}\AgdaSymbol{;}\AgdaSpace{}%
\AgdaFunction{length}\AgdaSymbol{)}\<%
\\
\>[0]\AgdaKeyword{open}\AgdaSpace{}%
\AgdaKeyword{import}\AgdaSpace{}%
\AgdaModule{Data.Nat}\<%
\\
\>[0]\AgdaKeyword{open}\AgdaSpace{}%
\AgdaKeyword{import}\AgdaSpace{}%
\AgdaModule{Data.Nat.Properties}\<%
\\
\>[0]\AgdaKeyword{open}\AgdaSpace{}%
\AgdaKeyword{import}\AgdaSpace{}%
\AgdaModule{Data.Bool}\AgdaSpace{}%
\AgdaKeyword{using}\AgdaSpace{}%
\AgdaSymbol{(}\AgdaInductiveConstructor{true}\AgdaSymbol{;}\AgdaSpace{}%
\AgdaInductiveConstructor{false}\AgdaSymbol{)}\AgdaSpace{}%
\AgdaKeyword{renaming}\AgdaSpace{}%
\AgdaSymbol{(}\AgdaDatatype{Bool}\AgdaSpace{}%
\AgdaSymbol{to}\AgdaSpace{}%
\AgdaDatatype{𝔹}\AgdaSymbol{)}\<%
\\
\>[0]\AgdaKeyword{open}\AgdaSpace{}%
\AgdaKeyword{import}\AgdaSpace{}%
\AgdaModule{Data.Product}\AgdaSpace{}%
\AgdaKeyword{using}\AgdaSpace{}%
\AgdaSymbol{(}\AgdaOperator{\AgdaInductiveConstructor{\AgdaUnderscore{},\AgdaUnderscore{}}}\AgdaSymbol{;}\AgdaOperator{\AgdaFunction{\AgdaUnderscore{}×\AgdaUnderscore{}}}\AgdaSymbol{;}\AgdaSpace{}%
\AgdaField{proj₁}\AgdaSymbol{;}\AgdaSpace{}%
\AgdaField{proj₂}\AgdaSymbol{;}\AgdaSpace{}%
\AgdaFunction{Σ-syntax}\AgdaSymbol{;}\AgdaSpace{}%
\AgdaFunction{∃-syntax}\AgdaSymbol{)}\<%
\\
\>[0]\AgdaKeyword{open}\AgdaSpace{}%
\AgdaKeyword{import}\AgdaSpace{}%
\AgdaModule{Data.Sum}\AgdaSpace{}%
\AgdaKeyword{using}\AgdaSpace{}%
\AgdaSymbol{(}\AgdaOperator{\AgdaDatatype{\AgdaUnderscore{}⊎\AgdaUnderscore{}}}\AgdaSymbol{;}\AgdaSpace{}%
\AgdaInductiveConstructor{inj₁}\AgdaSymbol{;}\AgdaSpace{}%
\AgdaInductiveConstructor{inj₂}\AgdaSymbol{)}\<%
\\
\>[0]\AgdaKeyword{open}\AgdaSpace{}%
\AgdaKeyword{import}\AgdaSpace{}%
\AgdaModule{Data.Unit}\AgdaSpace{}%
\AgdaKeyword{using}\AgdaSpace{}%
\AgdaSymbol{(}\AgdaRecord{⊤}\AgdaSymbol{;}\AgdaSpace{}%
\AgdaInductiveConstructor{tt}\AgdaSymbol{)}\<%
\\
\>[0]\AgdaKeyword{open}\AgdaSpace{}%
\AgdaKeyword{import}\AgdaSpace{}%
\AgdaModule{Data.Unit.Polymorphic}\AgdaSpace{}%
\AgdaKeyword{renaming}\AgdaSpace{}%
\AgdaSymbol{(}\AgdaFunction{⊤}\AgdaSpace{}%
\AgdaSymbol{to}\AgdaSpace{}%
\AgdaFunction{topᵖ}\AgdaSymbol{;}\AgdaSpace{}%
\AgdaFunction{tt}\AgdaSpace{}%
\AgdaSymbol{to}\AgdaSpace{}%
\AgdaFunction{ttᵖ}\AgdaSymbol{)}\<%
\\
\>[0]\AgdaKeyword{open}\AgdaSpace{}%
\AgdaKeyword{import}\AgdaSpace{}%
\AgdaModule{Relation.Binary.PropositionalEquality}\AgdaSpace{}%
\AgdaSymbol{as}\AgdaSpace{}%
\AgdaModule{Eq}\<%
\\
\>[0][@{}l@{\AgdaIndent{0}}]%
\>[2]\AgdaKeyword{using}\AgdaSpace{}%
\AgdaSymbol{(}\AgdaOperator{\AgdaDatatype{\AgdaUnderscore{}≡\AgdaUnderscore{}}}\AgdaSymbol{;}\AgdaSpace{}%
\AgdaOperator{\AgdaFunction{\AgdaUnderscore{}≢\AgdaUnderscore{}}}\AgdaSymbol{;}\AgdaSpace{}%
\AgdaInductiveConstructor{refl}\AgdaSymbol{;}\AgdaSpace{}%
\AgdaFunction{sym}\AgdaSymbol{;}\AgdaSpace{}%
\AgdaFunction{cong}\AgdaSymbol{;}\AgdaSpace{}%
\AgdaFunction{subst}\AgdaSymbol{;}\AgdaSpace{}%
\AgdaFunction{trans}\AgdaSymbol{)}\<%
\\
\>[0]\AgdaKeyword{open}\AgdaSpace{}%
\AgdaKeyword{import}\AgdaSpace{}%
\AgdaModule{Relation.Nullary}\AgdaSpace{}%
\AgdaKeyword{using}\AgdaSpace{}%
\AgdaSymbol{(}\AgdaOperator{\AgdaFunction{¬\AgdaUnderscore{}}}\AgdaSymbol{;}\AgdaSpace{}%
\AgdaRecord{Dec}\AgdaSymbol{;}\AgdaSpace{}%
\AgdaInductiveConstructor{yes}\AgdaSymbol{;}\AgdaSpace{}%
\AgdaInductiveConstructor{no}\AgdaSymbol{)}\<%
\\
\\[\AgdaEmptyExtraSkip]%
\>[0]\AgdaKeyword{open}\AgdaSpace{}%
\AgdaKeyword{import}\AgdaSpace{}%
\AgdaModule{Var}\<%
\\
\>[0]\AgdaKeyword{open}\AgdaSpace{}%
\AgdaKeyword{import}\AgdaSpace{}%
\AgdaModule{Sig}\<%
\\
\>[0]\AgdaKeyword{open}\AgdaSpace{}%
\AgdaKeyword{import}\AgdaSpace{}%
\AgdaModule{LogRel.PeterCastCalculus}\<%
\\
\>[0]\AgdaKeyword{open}\AgdaSpace{}%
\AgdaKeyword{import}\AgdaSpace{}%
\AgdaModule{LogRel.PeterPrecision}\<%
\\
\>[0]\AgdaKeyword{open}\AgdaSpace{}%
\AgdaKeyword{import}\AgdaSpace{}%
\AgdaModule{StepIndexedLogic}\<%
\end{code}

\section{Step-Indexed Logic}
\label{sec:SIL}

The Step-Indexed Logic (SIL) library~\cite{Siek:2023aa} is a shallow
embedding of a modal logic into Agda. The formulas of this logic have
type \textsf{Setᵒ}, which is a record with three fields, the most
important of which is named \textsf{\#} and is a function from ℕ to
\textsf{Set} which expresses the meaning of the formula in Agda.
Think of the ℕ as a count-down clock, with smaller numbers
representing later points in time. The other two fields of the record
contain proofs of the LSLR invariants: (1) that the formula is true at
0, and (2) if the formula is true at some number, then it is true at
all smaller numbers (monotonicity). Each of the constructors for SIL
formulas proves these two properties, thereby saving the client of SIL
from these tedious proofs.

SIL includes the connectives of first-order logic (conjunction,
disjunction, existential and universal quantification, etc.).  Each
connective comes in two versions, one with a superscript ``o'' and
another with superscript ``s''. The ``o'' version has type
\textsf{Setᵒ} whereas the ``s'' version has type
$\mathsf{Set}ˢ\,Γ\,Δ$, which we explain next. What makes SIL special
is that it includes an operator μᵒ for defining recursive
predicates. In the body of the μᵒ, de Bruijn index 0 refers to itself,
that is, the entire μᵒ. However, variable 0 may only be used
``later'', that is, underneath at least one use of the modal operator
▷ˢ.  The formula in the body of a μᵒ has type $\mathsf{Set}ˢ\,Γ\,Δ$,
where $Γ$ is a list of types, one for each recursive predicate in scope
(one can nest μˢ an arbitrary number of times).

The $Δ$ records when each recursive predicate is used (now or
later). It is represented by a list-like data structured indexed
by Γ to ensure they have the same length, with $∅$ as the empty list
and $\textsf{cons}$ to add an element to the front of a list.
\textsf{Setˢ} is a record whose field \textsf{\#} is a
function from a tuple of step-indexed predicates to \textsf{Setᵒ}.
(These tuples are structured like cons-lists with the
always-true predicate $\mathsf{tt}ᵖ$ playing the role of nil.)
From the client's perspective, use the ``s'' connectives when
writing formulas under a μᵒ and use the ``o'' connectives
otherwise. During this work we found that the ``s'' versus ``o''
distinction created unnecessary complexity for the client and have
developed a new version of the SIL (file \texttt{StepIndexedLogic2.lagda})
that has one version of each logical connective.

The majority of the lines of code in the SIL library are dedicated to
proving the \textsf{fixpointᵒ} theorem, which states that a recursive
predicate is equivalent to one unrolling of itself. The proof of
\textsf{fixpointᵒ} is an adaptation of the fixpoint theorem of Appel
and McAllester~\cite{Appel:2001aa}.

\begin{code}%
\>[0]\AgdaFunction{\AgdaUnderscore{}}\AgdaSpace{}%
\AgdaSymbol{:}%
\>[88I]\AgdaSymbol{∀(}\AgdaBound{A}\AgdaSpace{}%
\AgdaSymbol{:}\AgdaSpace{}%
\AgdaPrimitive{Set}\AgdaSymbol{)}\AgdaSpace{}%
\AgdaSymbol{(}\AgdaBound{P}\AgdaSpace{}%
\AgdaSymbol{:}\AgdaSpace{}%
\AgdaBound{A}\AgdaSpace{}%
\AgdaSymbol{→}\AgdaSpace{}%
\AgdaRecord{Setˢ}\AgdaSpace{}%
\AgdaSymbol{(}\AgdaBound{A}\AgdaSpace{}%
\AgdaOperator{\AgdaInductiveConstructor{∷}}\AgdaSpace{}%
\AgdaInductiveConstructor{[]}\AgdaSymbol{)}\AgdaSpace{}%
\AgdaSymbol{(}\AgdaInductiveConstructor{cons}\AgdaSpace{}%
\AgdaInductiveConstructor{Later}\AgdaSpace{}%
\AgdaInductiveConstructor{∅}\AgdaSymbol{))}\AgdaSpace{}%
\AgdaSymbol{(}\AgdaBound{a}\AgdaSpace{}%
\AgdaSymbol{:}\AgdaSpace{}%
\AgdaBound{A}\AgdaSymbol{)}\<%
\\
\>[.][@{}l@{}]\<[88I]%
\>[4]\AgdaSymbol{→}\AgdaSpace{}%
\AgdaFunction{μᵒ}\AgdaSpace{}%
\AgdaBound{P}\AgdaSpace{}%
\AgdaBound{a}\AgdaSpace{}%
\AgdaOperator{\AgdaFunction{≡ᵒ}}\AgdaSpace{}%
\AgdaField{\#}\AgdaSpace{}%
\AgdaSymbol{(}\AgdaBound{P}\AgdaSpace{}%
\AgdaBound{a}\AgdaSymbol{)}\AgdaSpace{}%
\AgdaSymbol{(}\AgdaFunction{μᵒ}\AgdaSpace{}%
\AgdaBound{P}\AgdaSpace{}%
\AgdaOperator{\AgdaInductiveConstructor{,}}\AgdaSpace{}%
\AgdaFunction{ttᵖ}\AgdaSymbol{)}\<%
\\
\>[0]\AgdaSymbol{\AgdaUnderscore{}}\AgdaSpace{}%
\AgdaSymbol{=}\AgdaSpace{}%
\AgdaSymbol{λ}\AgdaSpace{}%
\AgdaBound{A}\AgdaSpace{}%
\AgdaBound{P}\AgdaSpace{}%
\AgdaBound{a}\AgdaSpace{}%
\AgdaSymbol{→}\AgdaSpace{}%
\AgdaFunction{fixpointᵒ}\AgdaSpace{}%
\AgdaBound{P}\AgdaSpace{}%
\AgdaBound{a}\<%
\end{code}

\section{A Logical Relation for Precision}
\label{sec:log-rel}

\begin{code}[hide]%
\>[0]\AgdaKeyword{data}\AgdaSpace{}%
\AgdaDatatype{Dir}\AgdaSpace{}%
\AgdaSymbol{:}\AgdaSpace{}%
\AgdaPrimitive{Set}\AgdaSpace{}%
\AgdaKeyword{where}\<%
\\
\>[0][@{}l@{\AgdaIndent{0}}]%
\>[2]\AgdaInductiveConstructor{≼}\AgdaSpace{}%
\AgdaSymbol{:}\AgdaSpace{}%
\AgdaDatatype{Dir}\<%
\\
\>[2]\AgdaInductiveConstructor{≽}\AgdaSpace{}%
\AgdaSymbol{:}\AgdaSpace{}%
\AgdaDatatype{Dir}\<%
\end{code}

To define a logical relation for precision, we adapt the logical
relation of New~\cite{New:2020ab}, which used explicit step indexing,
into the Step-Indexed Logic. The logical relation has two directions
(of type \textsf{Dir}): the ≼ direction requires the more-precise term
to simulate the less-precise term whereas the ≽ direction requires the
less-precise term to simulate the more-precise.  
logical relation consists of mutually-recursive relations on both
terms and values. SIL does not directly support mutual recursion, but
it can be expressed by combining the two relations into a single
relation whose input is a disjoint sum.  The formula for expressing
membership in these recursive relations is verbose, so we define the
below shorthands. Note that these shorthands are only intended for use
inside the definition of the logical relation.

\begin{code}%
\>[0]\AgdaFunction{LR-type}\AgdaSpace{}%
\AgdaSymbol{:}\AgdaSpace{}%
\AgdaPrimitive{Set}\<%
\\
\>[0]\AgdaFunction{LR-type}\AgdaSpace{}%
\AgdaSymbol{=}\AgdaSpace{}%
\AgdaSymbol{(}\AgdaFunction{Prec}\AgdaSpace{}%
\AgdaOperator{\AgdaFunction{×}}\AgdaSpace{}%
\AgdaDatatype{Dir}\AgdaSpace{}%
\AgdaOperator{\AgdaFunction{×}}\AgdaSpace{}%
\AgdaDatatype{Term}\AgdaSpace{}%
\AgdaOperator{\AgdaFunction{×}}\AgdaSpace{}%
\AgdaDatatype{Term}\AgdaSymbol{)}\AgdaSpace{}%
\AgdaOperator{\AgdaDatatype{⊎}}\AgdaSpace{}%
\AgdaSymbol{(}\AgdaFunction{Prec}\AgdaSpace{}%
\AgdaOperator{\AgdaFunction{×}}\AgdaSpace{}%
\AgdaDatatype{Dir}\AgdaSpace{}%
\AgdaOperator{\AgdaFunction{×}}\AgdaSpace{}%
\AgdaDatatype{Term}\AgdaSpace{}%
\AgdaOperator{\AgdaFunction{×}}\AgdaSpace{}%
\AgdaDatatype{Term}\AgdaSymbol{)}\<%
\\
\\[\AgdaEmptyExtraSkip]%
\>[0]\AgdaFunction{LR-ctx}\AgdaSpace{}%
\AgdaSymbol{:}\AgdaSpace{}%
\AgdaDatatype{List}\AgdaSpace{}%
\AgdaPrimitive{Set}\<%
\\
\>[0]\AgdaFunction{LR-ctx}\AgdaSpace{}%
\AgdaSymbol{=}\AgdaSpace{}%
\AgdaFunction{LR-type}\AgdaSpace{}%
\AgdaOperator{\AgdaInductiveConstructor{∷}}\AgdaSpace{}%
\AgdaInductiveConstructor{[]}\<%
\\
\\[\AgdaEmptyExtraSkip]%
\>[0]\AgdaOperator{\AgdaFunction{\AgdaUnderscore{}∣\AgdaUnderscore{}ˢ⊑ᴸᴿₜ\AgdaUnderscore{}⦂\AgdaUnderscore{}}}\AgdaSpace{}%
\AgdaSymbol{:}\AgdaSpace{}%
\AgdaDatatype{Dir}\AgdaSpace{}%
\AgdaSymbol{→}\AgdaSpace{}%
\AgdaDatatype{Term}\AgdaSpace{}%
\AgdaSymbol{→}\AgdaSpace{}%
\AgdaDatatype{Term}\AgdaSpace{}%
\AgdaSymbol{→}\AgdaSpace{}%
\AgdaSymbol{∀\{}\AgdaBound{A}\AgdaSymbol{\}\{}\AgdaBound{A′}\AgdaSymbol{\}}\AgdaSpace{}%
\AgdaSymbol{(}\AgdaBound{c}\AgdaSpace{}%
\AgdaSymbol{:}\AgdaSpace{}%
\AgdaBound{A}\AgdaSpace{}%
\AgdaOperator{\AgdaDatatype{⊑}}\AgdaSpace{}%
\AgdaBound{A′}\AgdaSymbol{)}\AgdaSpace{}%
\AgdaSymbol{→}\AgdaSpace{}%
\AgdaRecord{Setˢ}\AgdaSpace{}%
\AgdaFunction{LR-ctx}\AgdaSpace{}%
\AgdaSymbol{(}\AgdaInductiveConstructor{cons}\AgdaSpace{}%
\AgdaInductiveConstructor{Now}\AgdaSpace{}%
\AgdaInductiveConstructor{∅}\AgdaSymbol{)}\<%
\\
\>[0]\AgdaBound{dir}\AgdaSpace{}%
\AgdaOperator{\AgdaFunction{∣}}\AgdaSpace{}%
\AgdaBound{M}\AgdaSpace{}%
\AgdaOperator{\AgdaFunction{ˢ⊑ᴸᴿₜ}}\AgdaSpace{}%
\AgdaBound{M′}\AgdaSpace{}%
\AgdaOperator{\AgdaFunction{⦂}}\AgdaSpace{}%
\AgdaBound{c}\AgdaSpace{}%
\AgdaSymbol{=}\AgdaSpace{}%
\AgdaSymbol{(}\AgdaInductiveConstructor{inj₂}\AgdaSpace{}%
\AgdaSymbol{((\AgdaUnderscore{}}\AgdaSpace{}%
\AgdaOperator{\AgdaInductiveConstructor{,}}\AgdaSpace{}%
\AgdaSymbol{\AgdaUnderscore{}}\AgdaSpace{}%
\AgdaOperator{\AgdaInductiveConstructor{,}}\AgdaSpace{}%
\AgdaBound{c}\AgdaSymbol{)}\AgdaSpace{}%
\AgdaOperator{\AgdaInductiveConstructor{,}}\AgdaSpace{}%
\AgdaBound{dir}\AgdaSpace{}%
\AgdaOperator{\AgdaInductiveConstructor{,}}\AgdaSpace{}%
\AgdaBound{M}\AgdaSpace{}%
\AgdaOperator{\AgdaInductiveConstructor{,}}\AgdaSpace{}%
\AgdaBound{M′}\AgdaSymbol{))}\AgdaSpace{}%
\AgdaOperator{\AgdaFunction{∈}}\AgdaSpace{}%
\AgdaInductiveConstructor{zeroˢ}\<%
\\
\\[\AgdaEmptyExtraSkip]%
\>[0]\AgdaOperator{\AgdaFunction{\AgdaUnderscore{}∣\AgdaUnderscore{}ˢ⊑ᴸᴿᵥ\AgdaUnderscore{}⦂\AgdaUnderscore{}}}\AgdaSpace{}%
\AgdaSymbol{:}\AgdaSpace{}%
\AgdaDatatype{Dir}\AgdaSpace{}%
\AgdaSymbol{→}\AgdaSpace{}%
\AgdaDatatype{Term}\AgdaSpace{}%
\AgdaSymbol{→}\AgdaSpace{}%
\AgdaDatatype{Term}\AgdaSpace{}%
\AgdaSymbol{→}\AgdaSpace{}%
\AgdaSymbol{∀\{}\AgdaBound{A}\AgdaSymbol{\}\{}\AgdaBound{A′}\AgdaSymbol{\}}\AgdaSpace{}%
\AgdaSymbol{(}\AgdaBound{c}\AgdaSpace{}%
\AgdaSymbol{:}\AgdaSpace{}%
\AgdaBound{A}\AgdaSpace{}%
\AgdaOperator{\AgdaDatatype{⊑}}\AgdaSpace{}%
\AgdaBound{A′}\AgdaSymbol{)}\AgdaSpace{}%
\AgdaSymbol{→}\AgdaSpace{}%
\AgdaRecord{Setˢ}\AgdaSpace{}%
\AgdaFunction{LR-ctx}\AgdaSpace{}%
\AgdaSymbol{(}\AgdaInductiveConstructor{cons}\AgdaSpace{}%
\AgdaInductiveConstructor{Now}\AgdaSpace{}%
\AgdaInductiveConstructor{∅}\AgdaSymbol{)}\<%
\\
\>[0]\AgdaBound{dir}\AgdaSpace{}%
\AgdaOperator{\AgdaFunction{∣}}\AgdaSpace{}%
\AgdaBound{V}\AgdaSpace{}%
\AgdaOperator{\AgdaFunction{ˢ⊑ᴸᴿᵥ}}\AgdaSpace{}%
\AgdaBound{V′}\AgdaSpace{}%
\AgdaOperator{\AgdaFunction{⦂}}\AgdaSpace{}%
\AgdaBound{c}\AgdaSpace{}%
\AgdaSymbol{=}\AgdaSpace{}%
\AgdaSymbol{(}\AgdaInductiveConstructor{inj₁}\AgdaSpace{}%
\AgdaSymbol{((\AgdaUnderscore{}}\AgdaSpace{}%
\AgdaOperator{\AgdaInductiveConstructor{,}}\AgdaSpace{}%
\AgdaSymbol{\AgdaUnderscore{}}\AgdaSpace{}%
\AgdaOperator{\AgdaInductiveConstructor{,}}\AgdaSpace{}%
\AgdaBound{c}\AgdaSymbol{)}\AgdaSpace{}%
\AgdaOperator{\AgdaInductiveConstructor{,}}\AgdaSpace{}%
\AgdaBound{dir}\AgdaSpace{}%
\AgdaOperator{\AgdaInductiveConstructor{,}}\AgdaSpace{}%
\AgdaBound{V}\AgdaSpace{}%
\AgdaOperator{\AgdaInductiveConstructor{,}}\AgdaSpace{}%
\AgdaBound{V′}\AgdaSymbol{))}\AgdaSpace{}%
\AgdaOperator{\AgdaFunction{∈}}\AgdaSpace{}%
\AgdaInductiveConstructor{zeroˢ}\<%
\end{code}
\begin{code}[hide]%
\>[0]\AgdaKeyword{instance}\<%
\\
\>[0][@{}l@{\AgdaIndent{0}}]%
\>[2]\AgdaFunction{TermInhabited}\AgdaSpace{}%
\AgdaSymbol{:}\AgdaSpace{}%
\AgdaRecord{Inhabited}\AgdaSpace{}%
\AgdaDatatype{Term}\<%
\\
\>[2]\AgdaFunction{TermInhabited}\AgdaSpace{}%
\AgdaSymbol{=}\AgdaSpace{}%
\AgdaKeyword{record}\AgdaSpace{}%
\AgdaSymbol{\{}\AgdaSpace{}%
\AgdaField{elt}\AgdaSpace{}%
\AgdaSymbol{=}\AgdaSpace{}%
\AgdaOperator{\AgdaInductiveConstructor{`}}\AgdaSpace{}%
\AgdaNumber{0}\AgdaSpace{}%
\AgdaSymbol{\}}\<%
\end{code}

\begin{figure}[tbp]
\begin{code}%
\>[0]\AgdaFunction{LRₜ}\AgdaSpace{}%
\AgdaSymbol{:}\AgdaSpace{}%
\AgdaFunction{Prec}\AgdaSpace{}%
\AgdaSymbol{→}\AgdaSpace{}%
\AgdaDatatype{Dir}\AgdaSpace{}%
\AgdaSymbol{→}\AgdaSpace{}%
\AgdaDatatype{Term}\AgdaSpace{}%
\AgdaSymbol{→}\AgdaSpace{}%
\AgdaDatatype{Term}\AgdaSpace{}%
\AgdaSymbol{→}\AgdaSpace{}%
\AgdaRecord{Setˢ}\AgdaSpace{}%
\AgdaFunction{LR-ctx}\AgdaSpace{}%
\AgdaSymbol{(}\AgdaInductiveConstructor{cons}\AgdaSpace{}%
\AgdaInductiveConstructor{Later}\AgdaSpace{}%
\AgdaInductiveConstructor{∅}\AgdaSymbol{)}\<%
\\
\>[0]\AgdaFunction{LRᵥ}\AgdaSpace{}%
\AgdaSymbol{:}\AgdaSpace{}%
\AgdaFunction{Prec}\AgdaSpace{}%
\AgdaSymbol{→}\AgdaSpace{}%
\AgdaDatatype{Dir}\AgdaSpace{}%
\AgdaSymbol{→}\AgdaSpace{}%
\AgdaDatatype{Term}\AgdaSpace{}%
\AgdaSymbol{→}\AgdaSpace{}%
\AgdaDatatype{Term}\AgdaSpace{}%
\AgdaSymbol{→}\AgdaSpace{}%
\AgdaRecord{Setˢ}\AgdaSpace{}%
\AgdaFunction{LR-ctx}\AgdaSpace{}%
\AgdaSymbol{(}\AgdaInductiveConstructor{cons}\AgdaSpace{}%
\AgdaInductiveConstructor{Later}\AgdaSpace{}%
\AgdaInductiveConstructor{∅}\AgdaSymbol{)}\<%
\\
\\[\AgdaEmptyExtraSkip]%
\>[0]\AgdaFunction{LRₜ}\AgdaSpace{}%
\AgdaSymbol{(}\AgdaBound{A}\AgdaSpace{}%
\AgdaOperator{\AgdaInductiveConstructor{,}}\AgdaSpace{}%
\AgdaBound{A′}\AgdaSpace{}%
\AgdaOperator{\AgdaInductiveConstructor{,}}\AgdaSpace{}%
\AgdaBound{c}\AgdaSymbol{)}\AgdaSpace{}%
\AgdaInductiveConstructor{≼}\AgdaSpace{}%
\AgdaBound{M}\AgdaSpace{}%
\AgdaBound{M′}\AgdaSpace{}%
\AgdaSymbol{=}\<%
\\
\>[0][@{}l@{\AgdaIndent{0}}]%
\>[3]\AgdaSymbol{(}\AgdaFunction{∃ˢ[}\AgdaSpace{}%
\AgdaBound{N}\AgdaSpace{}%
\AgdaFunction{]}\AgdaSpace{}%
\AgdaSymbol{(}\AgdaBound{M}\AgdaSpace{}%
\AgdaOperator{\AgdaDatatype{⟶}}\AgdaSpace{}%
\AgdaBound{N}\AgdaSymbol{)}\AgdaOperator{\AgdaFunction{ˢ}}\AgdaSpace{}%
\AgdaOperator{\AgdaFunction{×ˢ}}\AgdaSpace{}%
\AgdaFunction{▷ˢ}\AgdaSpace{}%
\AgdaSymbol{(}\AgdaInductiveConstructor{≼}\AgdaSpace{}%
\AgdaOperator{\AgdaFunction{∣}}\AgdaSpace{}%
\AgdaBound{N}\AgdaSpace{}%
\AgdaOperator{\AgdaFunction{ˢ⊑ᴸᴿₜ}}\AgdaSpace{}%
\AgdaBound{M′}\AgdaSpace{}%
\AgdaOperator{\AgdaFunction{⦂}}\AgdaSpace{}%
\AgdaBound{c}\AgdaSymbol{))}\<%
\\
\>[3]\AgdaOperator{\AgdaFunction{⊎ˢ}}\AgdaSpace{}%
\AgdaSymbol{(}\AgdaBound{M′}\AgdaSpace{}%
\AgdaOperator{\AgdaDatatype{↠}}\AgdaSpace{}%
\AgdaInductiveConstructor{blame}\AgdaSymbol{)}\AgdaOperator{\AgdaFunction{ˢ}}\<%
\\
\>[3]\AgdaOperator{\AgdaFunction{⊎ˢ}}\AgdaSpace{}%
\AgdaSymbol{((}\AgdaDatatype{Value}\AgdaSpace{}%
\AgdaBound{M}\AgdaSymbol{)}\AgdaOperator{\AgdaFunction{ˢ}}\AgdaSpace{}%
\AgdaOperator{\AgdaFunction{×ˢ}}\AgdaSpace{}%
\AgdaSymbol{(}\AgdaFunction{∃ˢ[}\AgdaSpace{}%
\AgdaBound{V′}\AgdaSpace{}%
\AgdaFunction{]}\AgdaSpace{}%
\AgdaSymbol{(}\AgdaBound{M′}\AgdaSpace{}%
\AgdaOperator{\AgdaDatatype{↠}}\AgdaSpace{}%
\AgdaBound{V′}\AgdaSymbol{)}\AgdaOperator{\AgdaFunction{ˢ}}\AgdaSpace{}%
\AgdaOperator{\AgdaFunction{×ˢ}}\AgdaSpace{}%
\AgdaSymbol{(}\AgdaDatatype{Value}\AgdaSpace{}%
\AgdaBound{V′}\AgdaSymbol{)}\AgdaOperator{\AgdaFunction{ˢ}}\AgdaSpace{}%
\AgdaOperator{\AgdaFunction{×ˢ}}\AgdaSpace{}%
\AgdaSymbol{(}\AgdaFunction{LRᵥ}\AgdaSpace{}%
\AgdaSymbol{(\AgdaUnderscore{}}\AgdaSpace{}%
\AgdaOperator{\AgdaInductiveConstructor{,}}\AgdaSpace{}%
\AgdaSymbol{\AgdaUnderscore{}}\AgdaSpace{}%
\AgdaOperator{\AgdaInductiveConstructor{,}}\AgdaSpace{}%
\AgdaBound{c}\AgdaSymbol{)}\AgdaSpace{}%
\AgdaInductiveConstructor{≼}\AgdaSpace{}%
\AgdaBound{M}\AgdaSpace{}%
\AgdaBound{V′}\AgdaSymbol{)))}\<%
\\
\>[0]\AgdaFunction{LRₜ}\AgdaSpace{}%
\AgdaSymbol{(}\AgdaBound{A}\AgdaSpace{}%
\AgdaOperator{\AgdaInductiveConstructor{,}}\AgdaSpace{}%
\AgdaBound{A′}\AgdaSpace{}%
\AgdaOperator{\AgdaInductiveConstructor{,}}\AgdaSpace{}%
\AgdaBound{c}\AgdaSymbol{)}\AgdaSpace{}%
\AgdaInductiveConstructor{≽}\AgdaSpace{}%
\AgdaBound{M}\AgdaSpace{}%
\AgdaBound{M′}\AgdaSpace{}%
\AgdaSymbol{=}\<%
\\
\>[0][@{}l@{\AgdaIndent{0}}]%
\>[3]\AgdaSymbol{(}\AgdaFunction{∃ˢ[}\AgdaSpace{}%
\AgdaBound{N′}\AgdaSpace{}%
\AgdaFunction{]}\AgdaSpace{}%
\AgdaSymbol{(}\AgdaBound{M′}\AgdaSpace{}%
\AgdaOperator{\AgdaDatatype{⟶}}\AgdaSpace{}%
\AgdaBound{N′}\AgdaSymbol{)}\AgdaOperator{\AgdaFunction{ˢ}}\AgdaSpace{}%
\AgdaOperator{\AgdaFunction{×ˢ}}\AgdaSpace{}%
\AgdaFunction{▷ˢ}\AgdaSpace{}%
\AgdaSymbol{(}\AgdaInductiveConstructor{≽}\AgdaSpace{}%
\AgdaOperator{\AgdaFunction{∣}}\AgdaSpace{}%
\AgdaBound{M}\AgdaSpace{}%
\AgdaOperator{\AgdaFunction{ˢ⊑ᴸᴿₜ}}\AgdaSpace{}%
\AgdaBound{N′}\AgdaSpace{}%
\AgdaOperator{\AgdaFunction{⦂}}\AgdaSpace{}%
\AgdaBound{c}\AgdaSymbol{))}\<%
\\
\>[3]\AgdaOperator{\AgdaFunction{⊎ˢ}}\AgdaSpace{}%
\AgdaSymbol{(}\AgdaDatatype{Blame}\AgdaSpace{}%
\AgdaBound{M′}\AgdaSymbol{)}\AgdaOperator{\AgdaFunction{ˢ}}\<%
\\
\>[3]\AgdaOperator{\AgdaFunction{⊎ˢ}}\AgdaSpace{}%
\AgdaSymbol{((}\AgdaDatatype{Value}\AgdaSpace{}%
\AgdaBound{M′}\AgdaSymbol{)}\AgdaOperator{\AgdaFunction{ˢ}}\AgdaSpace{}%
\AgdaOperator{\AgdaFunction{×ˢ}}\AgdaSpace{}%
\AgdaSymbol{(}\AgdaFunction{∃ˢ[}\AgdaSpace{}%
\AgdaBound{V}\AgdaSpace{}%
\AgdaFunction{]}\AgdaSpace{}%
\AgdaSymbol{(}\AgdaBound{M}\AgdaSpace{}%
\AgdaOperator{\AgdaDatatype{↠}}\AgdaSpace{}%
\AgdaBound{V}\AgdaSymbol{)}\AgdaOperator{\AgdaFunction{ˢ}}\AgdaSpace{}%
\AgdaOperator{\AgdaFunction{×ˢ}}\AgdaSpace{}%
\AgdaSymbol{(}\AgdaDatatype{Value}\AgdaSpace{}%
\AgdaBound{V}\AgdaSymbol{)}\AgdaOperator{\AgdaFunction{ˢ}}\AgdaSpace{}%
\AgdaOperator{\AgdaFunction{×ˢ}}\AgdaSpace{}%
\AgdaSymbol{(}\AgdaFunction{LRᵥ}\AgdaSpace{}%
\AgdaSymbol{(\AgdaUnderscore{}}\AgdaSpace{}%
\AgdaOperator{\AgdaInductiveConstructor{,}}\AgdaSpace{}%
\AgdaSymbol{\AgdaUnderscore{}}\AgdaSpace{}%
\AgdaOperator{\AgdaInductiveConstructor{,}}\AgdaSpace{}%
\AgdaBound{c}\AgdaSymbol{)}\AgdaSpace{}%
\AgdaInductiveConstructor{≽}\AgdaSpace{}%
\AgdaBound{V}\AgdaSpace{}%
\AgdaBound{M′}\AgdaSymbol{)))}\<%
\\
\\[\AgdaEmptyExtraSkip]%
\>[0]\AgdaFunction{LRᵥ}\AgdaSpace{}%
\AgdaSymbol{(}\AgdaDottedPattern{\AgdaSymbol{.(}}\AgdaDottedPattern{\AgdaOperator{\AgdaInductiveConstructor{\$ₜ}}}\AgdaSpace{}%
\AgdaDottedPattern{\AgdaBound{ι}}\AgdaDottedPattern{\AgdaSymbol{)}}\AgdaSpace{}%
\AgdaOperator{\AgdaInductiveConstructor{,}}\AgdaSpace{}%
\AgdaDottedPattern{\AgdaSymbol{.(}}\AgdaDottedPattern{\AgdaOperator{\AgdaInductiveConstructor{\$ₜ}}}\AgdaSpace{}%
\AgdaDottedPattern{\AgdaBound{ι}}\AgdaDottedPattern{\AgdaSymbol{)}}\AgdaSpace{}%
\AgdaOperator{\AgdaInductiveConstructor{,}}\AgdaSpace{}%
\AgdaInductiveConstructor{base⊑}\AgdaSymbol{\{}\AgdaBound{ι}\AgdaSymbol{\})}\AgdaSpace{}%
\AgdaBound{dir}\AgdaSpace{}%
\AgdaSymbol{(}\AgdaInductiveConstructor{\$}\AgdaSpace{}%
\AgdaBound{c}\AgdaSymbol{)}\AgdaSpace{}%
\AgdaSymbol{(}\AgdaInductiveConstructor{\$}\AgdaSpace{}%
\AgdaBound{c′}\AgdaSymbol{)}\AgdaSpace{}%
\AgdaSymbol{=}\AgdaSpace{}%
\AgdaSymbol{(}\AgdaBound{c}\AgdaSpace{}%
\AgdaOperator{\AgdaDatatype{≡}}\AgdaSpace{}%
\AgdaBound{c′}\AgdaSymbol{)}\AgdaSpace{}%
\AgdaOperator{\AgdaFunction{ˢ}}\<%
\\
\>[0]\AgdaCatchallClause{\AgdaFunction{LRᵥ}}\AgdaSpace{}%
\AgdaCatchallClause{\AgdaSymbol{(}}\AgdaDottedPattern{\AgdaCatchallClause{\AgdaSymbol{.(}}}\AgdaDottedPattern{\AgdaCatchallClause{\AgdaOperator{\AgdaInductiveConstructor{\$ₜ}}}}\AgdaSpace{}%
\AgdaDottedPattern{\AgdaCatchallClause{\AgdaBound{ι}}}\AgdaDottedPattern{\AgdaCatchallClause{\AgdaSymbol{)}}}\AgdaSpace{}%
\AgdaCatchallClause{\AgdaOperator{\AgdaInductiveConstructor{,}}}\AgdaSpace{}%
\AgdaDottedPattern{\AgdaCatchallClause{\AgdaSymbol{.(}}}\AgdaDottedPattern{\AgdaCatchallClause{\AgdaOperator{\AgdaInductiveConstructor{\$ₜ}}}}\AgdaSpace{}%
\AgdaDottedPattern{\AgdaCatchallClause{\AgdaBound{ι}}}\AgdaDottedPattern{\AgdaCatchallClause{\AgdaSymbol{)}}}\AgdaSpace{}%
\AgdaCatchallClause{\AgdaOperator{\AgdaInductiveConstructor{,}}}\AgdaSpace{}%
\AgdaCatchallClause{\AgdaInductiveConstructor{base⊑}}\AgdaCatchallClause{\AgdaSymbol{\{}}\AgdaCatchallClause{\AgdaBound{ι}}\AgdaCatchallClause{\AgdaSymbol{\})}}\AgdaSpace{}%
\AgdaCatchallClause{\AgdaBound{dir}}\AgdaSpace{}%
\AgdaCatchallClause{\AgdaBound{V}}\AgdaSpace{}%
\AgdaCatchallClause{\AgdaBound{V′}}\AgdaSpace{}%
\AgdaSymbol{=}\AgdaSpace{}%
\AgdaDatatype{⊥}\AgdaSpace{}%
\AgdaOperator{\AgdaFunction{ˢ}}\<%
\\
\>[0]\AgdaFunction{LRᵥ}%
\>[402I]\AgdaSymbol{(}\AgdaDottedPattern{\AgdaSymbol{.(}}\AgdaDottedPattern{\AgdaBound{A}}\AgdaSpace{}%
\AgdaDottedPattern{\AgdaOperator{\AgdaInductiveConstructor{⇒}}}\AgdaSpace{}%
\AgdaDottedPattern{\AgdaBound{B}}\AgdaDottedPattern{\AgdaSymbol{)}}\AgdaSpace{}%
\AgdaOperator{\AgdaInductiveConstructor{,}}\AgdaSpace{}%
\AgdaDottedPattern{\AgdaSymbol{.(}}\AgdaDottedPattern{\AgdaBound{A′}}\AgdaSpace{}%
\AgdaDottedPattern{\AgdaOperator{\AgdaInductiveConstructor{⇒}}}\AgdaSpace{}%
\AgdaDottedPattern{\AgdaBound{B′}}\AgdaDottedPattern{\AgdaSymbol{)}}\AgdaSpace{}%
\AgdaOperator{\AgdaInductiveConstructor{,}}\AgdaSpace{}%
\AgdaInductiveConstructor{fun⊑}\AgdaSymbol{\{}\AgdaBound{A}\AgdaSymbol{\}\{}\AgdaBound{B}\AgdaSymbol{\}\{}\AgdaBound{A′}\AgdaSymbol{\}\{}\AgdaBound{B′}\AgdaSymbol{\}}\AgdaSpace{}%
\AgdaBound{A⊑A′}\AgdaSpace{}%
\AgdaBound{B⊑B′}\AgdaSymbol{)}\AgdaSpace{}%
\AgdaBound{dir}\AgdaSpace{}%
\AgdaSymbol{(}\AgdaInductiveConstructor{ƛ}\AgdaSpace{}%
\AgdaBound{N}\AgdaSymbol{)(}\AgdaInductiveConstructor{ƛ}\AgdaSpace{}%
\AgdaBound{N′}\AgdaSymbol{)}\AgdaSpace{}%
\AgdaSymbol{=}\<%
\\
\>[.][@{}l@{}]\<[402I]%
\>[4]\AgdaFunction{∀ˢ[}\AgdaSpace{}%
\AgdaBound{W}\AgdaSpace{}%
\AgdaFunction{]}\AgdaSpace{}%
\AgdaFunction{∀ˢ[}%
\>[421I]\AgdaBound{W′}\AgdaSpace{}%
\AgdaFunction{]}\AgdaSpace{}%
\AgdaFunction{▷ˢ}\AgdaSpace{}%
\AgdaSymbol{(}\AgdaBound{dir}\AgdaSpace{}%
\AgdaOperator{\AgdaFunction{∣}}\AgdaSpace{}%
\AgdaBound{W}\AgdaSpace{}%
\AgdaOperator{\AgdaFunction{ˢ⊑ᴸᴿᵥ}}\AgdaSpace{}%
\AgdaBound{W′}\AgdaSpace{}%
\AgdaOperator{\AgdaFunction{⦂}}\AgdaSpace{}%
\AgdaBound{A⊑A′}\AgdaSymbol{)}\<%
\\
\>[421I][@{}l@{\AgdaIndent{0}}]%
\>[18]\AgdaOperator{\AgdaFunction{→ˢ}}\AgdaSpace{}%
\AgdaFunction{▷ˢ}\AgdaSpace{}%
\AgdaSymbol{(}\AgdaBound{dir}\AgdaSpace{}%
\AgdaOperator{\AgdaFunction{∣}}\AgdaSpace{}%
\AgdaSymbol{(}\AgdaBound{N}\AgdaSpace{}%
\AgdaOperator{\AgdaFunction{[}}\AgdaSpace{}%
\AgdaBound{W}\AgdaSpace{}%
\AgdaOperator{\AgdaFunction{]}}\AgdaSymbol{)}\AgdaSpace{}%
\AgdaOperator{\AgdaFunction{ˢ⊑ᴸᴿₜ}}\AgdaSpace{}%
\AgdaSymbol{(}\AgdaBound{N′}\AgdaSpace{}%
\AgdaOperator{\AgdaFunction{[}}\AgdaSpace{}%
\AgdaBound{W′}\AgdaSpace{}%
\AgdaOperator{\AgdaFunction{]}}\AgdaSymbol{)}\AgdaSpace{}%
\AgdaOperator{\AgdaFunction{⦂}}\AgdaSpace{}%
\AgdaBound{B⊑B′}\AgdaSymbol{)}\<%
\\
\>[0]\AgdaCatchallClause{\AgdaFunction{LRᵥ}}\AgdaSpace{}%
\AgdaCatchallClause{\AgdaSymbol{(}}\AgdaDottedPattern{\AgdaCatchallClause{\AgdaSymbol{.(}}}\AgdaDottedPattern{\AgdaCatchallClause{\AgdaBound{A}}}\AgdaSpace{}%
\AgdaDottedPattern{\AgdaCatchallClause{\AgdaOperator{\AgdaInductiveConstructor{⇒}}}}\AgdaSpace{}%
\AgdaDottedPattern{\AgdaCatchallClause{\AgdaBound{B}}}\AgdaDottedPattern{\AgdaCatchallClause{\AgdaSymbol{)}}}\AgdaSpace{}%
\AgdaCatchallClause{\AgdaOperator{\AgdaInductiveConstructor{,}}}\AgdaSpace{}%
\AgdaDottedPattern{\AgdaCatchallClause{\AgdaSymbol{.(}}}\AgdaDottedPattern{\AgdaCatchallClause{\AgdaBound{A′}}}\AgdaSpace{}%
\AgdaDottedPattern{\AgdaCatchallClause{\AgdaOperator{\AgdaInductiveConstructor{⇒}}}}\AgdaSpace{}%
\AgdaDottedPattern{\AgdaCatchallClause{\AgdaBound{B′}}}\AgdaDottedPattern{\AgdaCatchallClause{\AgdaSymbol{)}}}\AgdaSpace{}%
\AgdaCatchallClause{\AgdaOperator{\AgdaInductiveConstructor{,}}}\AgdaSpace{}%
\AgdaCatchallClause{\AgdaInductiveConstructor{fun⊑}}\AgdaCatchallClause{\AgdaSymbol{\{}}\AgdaCatchallClause{\AgdaBound{A}}\AgdaCatchallClause{\AgdaSymbol{\}\{}}\AgdaCatchallClause{\AgdaBound{B}}\AgdaCatchallClause{\AgdaSymbol{\}\{}}\AgdaCatchallClause{\AgdaBound{A′}}\AgdaCatchallClause{\AgdaSymbol{\}\{}}\AgdaCatchallClause{\AgdaBound{B′}}\AgdaCatchallClause{\AgdaSymbol{\}}}\AgdaSpace{}%
\AgdaCatchallClause{\AgdaBound{A⊑A′}}\AgdaSpace{}%
\AgdaCatchallClause{\AgdaBound{B⊑B′}}\AgdaCatchallClause{\AgdaSymbol{)}}\AgdaSpace{}%
\AgdaCatchallClause{\AgdaBound{dir}}\AgdaSpace{}%
\AgdaCatchallClause{\AgdaBound{V}}\AgdaSpace{}%
\AgdaCatchallClause{\AgdaBound{V′}}\AgdaSpace{}%
\AgdaSymbol{=}\AgdaSpace{}%
\AgdaDatatype{⊥}\AgdaSpace{}%
\AgdaOperator{\AgdaFunction{ˢ}}\<%
\\
\>[0]\AgdaFunction{LRᵥ}%
\>[462I]\AgdaSymbol{(}\AgdaDottedPattern{\AgdaSymbol{.}}\AgdaDottedPattern{\AgdaInductiveConstructor{★}}\AgdaSpace{}%
\AgdaOperator{\AgdaInductiveConstructor{,}}\AgdaSpace{}%
\AgdaDottedPattern{\AgdaSymbol{.}}\AgdaDottedPattern{\AgdaInductiveConstructor{★}}\AgdaSpace{}%
\AgdaOperator{\AgdaInductiveConstructor{,}}\AgdaSpace{}%
\AgdaInductiveConstructor{unk⊑unk}\AgdaSymbol{)}\AgdaSpace{}%
\AgdaBound{dir}\AgdaSpace{}%
\AgdaSymbol{(}\AgdaBound{V}\AgdaSpace{}%
\AgdaOperator{\AgdaInductiveConstructor{⟨}}\AgdaSpace{}%
\AgdaBound{G}\AgdaSpace{}%
\AgdaOperator{\AgdaInductiveConstructor{!⟩}}\AgdaSymbol{)}\AgdaSpace{}%
\AgdaSymbol{(}\AgdaBound{V′}\AgdaSpace{}%
\AgdaOperator{\AgdaInductiveConstructor{⟨}}\AgdaSpace{}%
\AgdaBound{H}\AgdaSpace{}%
\AgdaOperator{\AgdaInductiveConstructor{!⟩}}\AgdaSymbol{)}\<%
\\
\>[.][@{}l@{}]\<[462I]%
\>[4]\AgdaKeyword{with}\AgdaSpace{}%
\AgdaBound{G}\AgdaSpace{}%
\AgdaOperator{\AgdaFunction{≡ᵍ}}\AgdaSpace{}%
\AgdaBound{H}\<%
\\
\>[0]\AgdaSymbol{...}\AgdaSpace{}%
\AgdaSymbol{|}\AgdaSpace{}%
\AgdaInductiveConstructor{yes}\AgdaSpace{}%
\AgdaInductiveConstructor{refl}\AgdaSpace{}%
\AgdaSymbol{=}\AgdaSpace{}%
\AgdaSymbol{(}\AgdaDatatype{Value}\AgdaSpace{}%
\AgdaBound{V}\AgdaSymbol{)}\AgdaOperator{\AgdaFunction{ˢ}}\AgdaSpace{}%
\AgdaOperator{\AgdaFunction{×ˢ}}\AgdaSpace{}%
\AgdaSymbol{(}\AgdaDatatype{Value}\AgdaSpace{}%
\AgdaBound{V′}\AgdaSymbol{)}\AgdaOperator{\AgdaFunction{ˢ}}\AgdaSpace{}%
\AgdaOperator{\AgdaFunction{×ˢ}}\AgdaSpace{}%
\AgdaSymbol{(}\AgdaFunction{▷ˢ}\AgdaSpace{}%
\AgdaSymbol{(}\AgdaBound{dir}\AgdaSpace{}%
\AgdaOperator{\AgdaFunction{∣}}\AgdaSpace{}%
\AgdaBound{V}\AgdaSpace{}%
\AgdaOperator{\AgdaFunction{ˢ⊑ᴸᴿᵥ}}\AgdaSpace{}%
\AgdaBound{V′}\AgdaSpace{}%
\AgdaOperator{\AgdaFunction{⦂}}\AgdaSpace{}%
\AgdaFunction{Refl⊑}\AgdaSymbol{\{}\AgdaOperator{\AgdaFunction{⌈}}\AgdaSpace{}%
\AgdaBound{G}\AgdaSpace{}%
\AgdaOperator{\AgdaFunction{⌉}}\AgdaSymbol{\}))}\<%
\\
\>[0]\AgdaSymbol{...}\AgdaSpace{}%
\AgdaSymbol{|}\AgdaSpace{}%
\AgdaInductiveConstructor{no}\AgdaSpace{}%
\AgdaBound{neq}\AgdaSpace{}%
\AgdaSymbol{=}\AgdaSpace{}%
\AgdaDatatype{⊥}\AgdaSpace{}%
\AgdaOperator{\AgdaFunction{ˢ}}\<%
\\
\>[0]\AgdaCatchallClause{\AgdaFunction{LRᵥ}}\AgdaSpace{}%
\AgdaCatchallClause{\AgdaSymbol{(}}\AgdaDottedPattern{\AgdaCatchallClause{\AgdaSymbol{.}}}\AgdaDottedPattern{\AgdaCatchallClause{\AgdaInductiveConstructor{★}}}\AgdaSpace{}%
\AgdaCatchallClause{\AgdaOperator{\AgdaInductiveConstructor{,}}}\AgdaSpace{}%
\AgdaDottedPattern{\AgdaCatchallClause{\AgdaSymbol{.}}}\AgdaDottedPattern{\AgdaCatchallClause{\AgdaInductiveConstructor{★}}}\AgdaSpace{}%
\AgdaCatchallClause{\AgdaOperator{\AgdaInductiveConstructor{,}}}\AgdaSpace{}%
\AgdaCatchallClause{\AgdaInductiveConstructor{unk⊑unk}}\AgdaCatchallClause{\AgdaSymbol{)}}\AgdaSpace{}%
\AgdaCatchallClause{\AgdaBound{dir}}\AgdaSpace{}%
\AgdaCatchallClause{\AgdaBound{V}}\AgdaSpace{}%
\AgdaCatchallClause{\AgdaBound{V′}}\AgdaSpace{}%
\AgdaSymbol{=}\AgdaSpace{}%
\AgdaDatatype{⊥}\AgdaSpace{}%
\AgdaOperator{\AgdaFunction{ˢ}}\<%
\\
\>[0]\AgdaFunction{LRᵥ}%
\>[516I]\AgdaSymbol{(}\AgdaDottedPattern{\AgdaSymbol{.}}\AgdaDottedPattern{\AgdaInductiveConstructor{★}}\AgdaSpace{}%
\AgdaOperator{\AgdaInductiveConstructor{,}}\AgdaSpace{}%
\AgdaDottedPattern{\AgdaSymbol{.}}\AgdaDottedPattern{\AgdaBound{A′}}\AgdaSpace{}%
\AgdaOperator{\AgdaInductiveConstructor{,}}\AgdaSpace{}%
\AgdaInductiveConstructor{unk⊑}\AgdaSymbol{\{}\AgdaBound{H}\AgdaSymbol{\}\{}\AgdaBound{A′}\AgdaSymbol{\}}\AgdaSpace{}%
\AgdaBound{d}\AgdaSymbol{)}\AgdaSpace{}%
\AgdaInductiveConstructor{≼}\AgdaSpace{}%
\AgdaSymbol{(}\AgdaBound{V}\AgdaSpace{}%
\AgdaOperator{\AgdaInductiveConstructor{⟨}}\AgdaSpace{}%
\AgdaBound{G}\AgdaSpace{}%
\AgdaOperator{\AgdaInductiveConstructor{!⟩}}\AgdaSymbol{)}\AgdaSpace{}%
\AgdaBound{V′}\<%
\\
\>[.][@{}l@{}]\<[516I]%
\>[4]\AgdaKeyword{with}\AgdaSpace{}%
\AgdaBound{G}\AgdaSpace{}%
\AgdaOperator{\AgdaFunction{≡ᵍ}}\AgdaSpace{}%
\AgdaBound{H}\<%
\\
\>[0]\AgdaSymbol{...}\AgdaSpace{}%
\AgdaSymbol{|}\AgdaSpace{}%
\AgdaInductiveConstructor{yes}\AgdaSpace{}%
\AgdaInductiveConstructor{refl}\AgdaSpace{}%
\AgdaSymbol{=}\AgdaSpace{}%
\AgdaSymbol{(}\AgdaDatatype{Value}\AgdaSpace{}%
\AgdaBound{V}\AgdaSymbol{)}\AgdaOperator{\AgdaFunction{ˢ}}\AgdaSpace{}%
\AgdaOperator{\AgdaFunction{×ˢ}}\AgdaSpace{}%
\AgdaSymbol{(}\AgdaDatatype{Value}\AgdaSpace{}%
\AgdaBound{V′}\AgdaSymbol{)}\AgdaOperator{\AgdaFunction{ˢ}}\AgdaSpace{}%
\AgdaOperator{\AgdaFunction{×ˢ}}\AgdaSpace{}%
\AgdaFunction{▷ˢ}\AgdaSpace{}%
\AgdaSymbol{(}\AgdaInductiveConstructor{≼}\AgdaSpace{}%
\AgdaOperator{\AgdaFunction{∣}}\AgdaSpace{}%
\AgdaBound{V}\AgdaSpace{}%
\AgdaOperator{\AgdaFunction{ˢ⊑ᴸᴿᵥ}}\AgdaSpace{}%
\AgdaBound{V′}\AgdaSpace{}%
\AgdaOperator{\AgdaFunction{⦂}}\AgdaSpace{}%
\AgdaBound{d}\AgdaSymbol{)}\<%
\\
\>[0]\AgdaSymbol{...}\AgdaSpace{}%
\AgdaSymbol{|}\AgdaSpace{}%
\AgdaInductiveConstructor{no}\AgdaSpace{}%
\AgdaBound{neq}\AgdaSpace{}%
\AgdaSymbol{=}\AgdaSpace{}%
\AgdaDatatype{⊥}\AgdaSpace{}%
\AgdaOperator{\AgdaFunction{ˢ}}\<%
\\
\>[0]\AgdaFunction{LRᵥ}%
\>[555I]\AgdaSymbol{(}\AgdaDottedPattern{\AgdaSymbol{.}}\AgdaDottedPattern{\AgdaInductiveConstructor{★}}\AgdaSpace{}%
\AgdaOperator{\AgdaInductiveConstructor{,}}\AgdaSpace{}%
\AgdaDottedPattern{\AgdaSymbol{.}}\AgdaDottedPattern{\AgdaBound{A′}}\AgdaSpace{}%
\AgdaOperator{\AgdaInductiveConstructor{,}}\AgdaSpace{}%
\AgdaInductiveConstructor{unk⊑}\AgdaSymbol{\{}\AgdaBound{H}\AgdaSymbol{\}\{}\AgdaBound{A′}\AgdaSymbol{\}}\AgdaSpace{}%
\AgdaBound{d}\AgdaSymbol{)}\AgdaSpace{}%
\AgdaInductiveConstructor{≽}\AgdaSpace{}%
\AgdaSymbol{(}\AgdaBound{V}\AgdaSpace{}%
\AgdaOperator{\AgdaInductiveConstructor{⟨}}\AgdaSpace{}%
\AgdaBound{G}\AgdaSpace{}%
\AgdaOperator{\AgdaInductiveConstructor{!⟩}}\AgdaSymbol{)}\AgdaSpace{}%
\AgdaBound{V′}\<%
\\
\>[.][@{}l@{}]\<[555I]%
\>[4]\AgdaKeyword{with}\AgdaSpace{}%
\AgdaBound{G}\AgdaSpace{}%
\AgdaOperator{\AgdaFunction{≡ᵍ}}\AgdaSpace{}%
\AgdaBound{H}\<%
\\
\>[0]\AgdaSymbol{...}\AgdaSpace{}%
\AgdaSymbol{|}\AgdaSpace{}%
\AgdaInductiveConstructor{yes}\AgdaSpace{}%
\AgdaInductiveConstructor{refl}\AgdaSpace{}%
\AgdaSymbol{=}\AgdaSpace{}%
\AgdaSymbol{(}\AgdaDatatype{Value}\AgdaSpace{}%
\AgdaBound{V}\AgdaSymbol{)}\AgdaOperator{\AgdaFunction{ˢ}}\AgdaSpace{}%
\AgdaOperator{\AgdaFunction{×ˢ}}\AgdaSpace{}%
\AgdaSymbol{(}\AgdaDatatype{Value}\AgdaSpace{}%
\AgdaBound{V′}\AgdaSymbol{)}\AgdaOperator{\AgdaFunction{ˢ}}\AgdaSpace{}%
\AgdaOperator{\AgdaFunction{×ˢ}}\AgdaSpace{}%
\AgdaSymbol{(}\AgdaFunction{LRᵥ}\AgdaSpace{}%
\AgdaSymbol{(}\AgdaOperator{\AgdaFunction{⌈}}\AgdaSpace{}%
\AgdaBound{G}\AgdaSpace{}%
\AgdaOperator{\AgdaFunction{⌉}}\AgdaSpace{}%
\AgdaOperator{\AgdaInductiveConstructor{,}}\AgdaSpace{}%
\AgdaBound{A′}\AgdaSpace{}%
\AgdaOperator{\AgdaInductiveConstructor{,}}\AgdaSpace{}%
\AgdaBound{d}\AgdaSymbol{)}\AgdaSpace{}%
\AgdaInductiveConstructor{≽}\AgdaSpace{}%
\AgdaBound{V}\AgdaSpace{}%
\AgdaBound{V′}\AgdaSymbol{)}\<%
\\
\>[0]\AgdaSymbol{...}\AgdaSpace{}%
\AgdaSymbol{|}\AgdaSpace{}%
\AgdaInductiveConstructor{no}\AgdaSpace{}%
\AgdaBound{neq}\AgdaSpace{}%
\AgdaSymbol{=}\AgdaSpace{}%
\AgdaDatatype{⊥}\AgdaSpace{}%
\AgdaOperator{\AgdaFunction{ˢ}}\<%
\\
\>[0]\AgdaCatchallClause{\AgdaFunction{LRᵥ}}\AgdaSpace{}%
\AgdaCatchallClause{\AgdaSymbol{(}}\AgdaCatchallClause{\AgdaInductiveConstructor{★}}\AgdaSpace{}%
\AgdaCatchallClause{\AgdaOperator{\AgdaInductiveConstructor{,}}}\AgdaSpace{}%
\AgdaDottedPattern{\AgdaCatchallClause{\AgdaSymbol{.}}}\AgdaDottedPattern{\AgdaCatchallClause{\AgdaBound{A′}}}\AgdaSpace{}%
\AgdaCatchallClause{\AgdaOperator{\AgdaInductiveConstructor{,}}}\AgdaSpace{}%
\AgdaCatchallClause{\AgdaInductiveConstructor{unk⊑}}\AgdaCatchallClause{\AgdaSymbol{\{}}\AgdaCatchallClause{\AgdaBound{H}}\AgdaCatchallClause{\AgdaSymbol{\}\{}}\AgdaCatchallClause{\AgdaBound{A′}}\AgdaCatchallClause{\AgdaSymbol{\}}}\AgdaSpace{}%
\AgdaCatchallClause{\AgdaBound{d}}\AgdaCatchallClause{\AgdaSymbol{)}}\AgdaSpace{}%
\AgdaCatchallClause{\AgdaBound{dir}}\AgdaSpace{}%
\AgdaCatchallClause{\AgdaBound{V}}\AgdaSpace{}%
\AgdaCatchallClause{\AgdaBound{V′}}\AgdaSpace{}%
\AgdaSymbol{=}\AgdaSpace{}%
\AgdaDatatype{⊥}\AgdaSpace{}%
\AgdaOperator{\AgdaFunction{ˢ}}\<%
\end{code}

\caption{Logical Relation for Precision on Terms $\mathsf{LR}_t$
  and Values $\mathsf{LR}_v$}
\label{fig:log-rel-precision}
\end{figure}

The logical relation is defined in Figure~\ref{fig:log-rel-precision}
and explained in the following paragraphs.  The definition of the
logical relation for terms is based on the requirements of the gradual
guarantee, but it only talks about one step at a time of the term
begin simulated. In the ≼ direction, the first clause says that the
less-precise $M$ takes a step to $N$ and that $N$ is related to $M′$
at one tick later in time. The second clause allows the more-precise
$M′$ to reduce to an error. The third clause says that the
less-precise $M$ is already a value, and requires $M′$ to reduce to a
value that is related at the current time to $M$. The other direction
≽ is defined in a similar way, but with the more precise term $M′$
taking one step at a time.

The definition of the logical relation for values is by recursion on
the precision relation and by cases on the values and their types.
When both values are of the same base type ($\mathsf{base}⊑$), they
are related if they are identical constants.  When the values are of
function type ($\mathsf{fun}⊑$), then they are related if they are
both lambda abstractions that, when later applied to related
arguments, behave in a related way. When the values are both of unkown
type ($\mathsf{unk⊑unk}$), then they are related if they are both
injections from the same ground type and the underlying values are
related one step later. If the less-precise value is of unknown type
but the more-precise value is not ($\mathsf{unk⊑}$), then they are
related if (1) the less-precise value is an injection and (2) the
ground type of the injection is less-precise than the type of the
more-precise value. Furthermore, for direction ≼, (3a) the underlying
value of the injection is related one step later to the more-precise
value. For direction ≽, (3b) the underlying value of the injection is
related now to the more-precise value. Note that the recursive call
to $\mathsf{LRᵥ}$ is fine from a termination perspective because
argument $d$ is a subterm of $\mathsf{unk⊑}\,d$. This is why
the $\mathsf{unk⊑}$ rule needs to be recursive, with the
premise $⌈ G ⌉ ⊑ B$.

The following definitions combine the LRᵥ and LRₜ functions into a
single function, pre-LRₜ⊎LRᵥ, and than applies the μᵒ operator to
produce the recursive relation LRₜ⊎LRᵥ. Finally, we define some
shorthand for the logical relation on values, written ⊑ᴸᴿᵥ, and the
logical relation on terms, ⊑ᴸᴿₜ.

\begin{code}%
\>[0]\AgdaFunction{pre-LRₜ⊎LRᵥ}\AgdaSpace{}%
\AgdaSymbol{:}\AgdaSpace{}%
\AgdaFunction{LR-type}\AgdaSpace{}%
\AgdaSymbol{→}\AgdaSpace{}%
\AgdaRecord{Setˢ}\AgdaSpace{}%
\AgdaFunction{LR-ctx}\AgdaSpace{}%
\AgdaSymbol{(}\AgdaInductiveConstructor{cons}\AgdaSpace{}%
\AgdaInductiveConstructor{Later}\AgdaSpace{}%
\AgdaInductiveConstructor{∅}\AgdaSymbol{)}\<%
\\
\>[0]\AgdaFunction{pre-LRₜ⊎LRᵥ}\AgdaSpace{}%
\AgdaSymbol{(}\AgdaInductiveConstructor{inj₁}\AgdaSpace{}%
\AgdaSymbol{(}\AgdaBound{c}\AgdaSpace{}%
\AgdaOperator{\AgdaInductiveConstructor{,}}\AgdaSpace{}%
\AgdaBound{dir}\AgdaSpace{}%
\AgdaOperator{\AgdaInductiveConstructor{,}}\AgdaSpace{}%
\AgdaBound{V}\AgdaSpace{}%
\AgdaOperator{\AgdaInductiveConstructor{,}}\AgdaSpace{}%
\AgdaBound{V′}\AgdaSymbol{))}\AgdaSpace{}%
\AgdaSymbol{=}\AgdaSpace{}%
\AgdaFunction{LRᵥ}\AgdaSpace{}%
\AgdaBound{c}\AgdaSpace{}%
\AgdaBound{dir}\AgdaSpace{}%
\AgdaBound{V}\AgdaSpace{}%
\AgdaBound{V′}\<%
\\
\>[0]\AgdaFunction{pre-LRₜ⊎LRᵥ}\AgdaSpace{}%
\AgdaSymbol{(}\AgdaInductiveConstructor{inj₂}\AgdaSpace{}%
\AgdaSymbol{(}\AgdaBound{c}\AgdaSpace{}%
\AgdaOperator{\AgdaInductiveConstructor{,}}\AgdaSpace{}%
\AgdaBound{dir}\AgdaSpace{}%
\AgdaOperator{\AgdaInductiveConstructor{,}}\AgdaSpace{}%
\AgdaBound{M}\AgdaSpace{}%
\AgdaOperator{\AgdaInductiveConstructor{,}}\AgdaSpace{}%
\AgdaBound{M′}\AgdaSymbol{))}\AgdaSpace{}%
\AgdaSymbol{=}\AgdaSpace{}%
\AgdaFunction{LRₜ}\AgdaSpace{}%
\AgdaBound{c}\AgdaSpace{}%
\AgdaBound{dir}\AgdaSpace{}%
\AgdaBound{M}\AgdaSpace{}%
\AgdaBound{M′}\<%
\\
\\[\AgdaEmptyExtraSkip]%
\>[0]\AgdaFunction{LRₜ⊎LRᵥ}\AgdaSpace{}%
\AgdaSymbol{:}\AgdaSpace{}%
\AgdaFunction{LR-type}\AgdaSpace{}%
\AgdaSymbol{→}\AgdaSpace{}%
\AgdaRecord{Setᵒ}\<%
\\
\>[0]\AgdaFunction{LRₜ⊎LRᵥ}\AgdaSpace{}%
\AgdaBound{X}\AgdaSpace{}%
\AgdaSymbol{=}\AgdaSpace{}%
\AgdaFunction{μᵒ}\AgdaSpace{}%
\AgdaFunction{pre-LRₜ⊎LRᵥ}\AgdaSpace{}%
\AgdaBound{X}\<%
\\
\\[\AgdaEmptyExtraSkip]%
\>[0]\AgdaOperator{\AgdaFunction{\AgdaUnderscore{}∣\AgdaUnderscore{}⊑ᴸᴿᵥ\AgdaUnderscore{}⦂\AgdaUnderscore{}}}\AgdaSpace{}%
\AgdaSymbol{:}\AgdaSpace{}%
\AgdaDatatype{Dir}\AgdaSpace{}%
\AgdaSymbol{→}\AgdaSpace{}%
\AgdaDatatype{Term}\AgdaSpace{}%
\AgdaSymbol{→}\AgdaSpace{}%
\AgdaDatatype{Term}\AgdaSpace{}%
\AgdaSymbol{→}\AgdaSpace{}%
\AgdaSymbol{∀\{}\AgdaBound{A}\AgdaSpace{}%
\AgdaBound{A′}\AgdaSymbol{\}}\AgdaSpace{}%
\AgdaSymbol{→}\AgdaSpace{}%
\AgdaBound{A}\AgdaSpace{}%
\AgdaOperator{\AgdaDatatype{⊑}}\AgdaSpace{}%
\AgdaBound{A′}\AgdaSpace{}%
\AgdaSymbol{→}\AgdaSpace{}%
\AgdaRecord{Setᵒ}\<%
\\
\>[0]\AgdaBound{dir}\AgdaSpace{}%
\AgdaOperator{\AgdaFunction{∣}}\AgdaSpace{}%
\AgdaBound{V}\AgdaSpace{}%
\AgdaOperator{\AgdaFunction{⊑ᴸᴿᵥ}}\AgdaSpace{}%
\AgdaBound{V′}\AgdaSpace{}%
\AgdaOperator{\AgdaFunction{⦂}}\AgdaSpace{}%
\AgdaBound{A⊑A′}\AgdaSpace{}%
\AgdaSymbol{=}\AgdaSpace{}%
\AgdaFunction{LRₜ⊎LRᵥ}\AgdaSpace{}%
\AgdaSymbol{(}\AgdaInductiveConstructor{inj₁}\AgdaSpace{}%
\AgdaSymbol{((\AgdaUnderscore{}}\AgdaSpace{}%
\AgdaOperator{\AgdaInductiveConstructor{,}}\AgdaSpace{}%
\AgdaSymbol{\AgdaUnderscore{}}\AgdaSpace{}%
\AgdaOperator{\AgdaInductiveConstructor{,}}\AgdaSpace{}%
\AgdaBound{A⊑A′}\AgdaSymbol{)}\AgdaSpace{}%
\AgdaOperator{\AgdaInductiveConstructor{,}}\AgdaSpace{}%
\AgdaBound{dir}\AgdaSpace{}%
\AgdaOperator{\AgdaInductiveConstructor{,}}\AgdaSpace{}%
\AgdaBound{V}\AgdaSpace{}%
\AgdaOperator{\AgdaInductiveConstructor{,}}\AgdaSpace{}%
\AgdaBound{V′}\AgdaSymbol{))}\<%
\\
\\[\AgdaEmptyExtraSkip]%
\>[0]\AgdaOperator{\AgdaFunction{\AgdaUnderscore{}∣\AgdaUnderscore{}⊑ᴸᴿₜ\AgdaUnderscore{}⦂\AgdaUnderscore{}}}\AgdaSpace{}%
\AgdaSymbol{:}\AgdaSpace{}%
\AgdaDatatype{Dir}\AgdaSpace{}%
\AgdaSymbol{→}\AgdaSpace{}%
\AgdaDatatype{Term}\AgdaSpace{}%
\AgdaSymbol{→}\AgdaSpace{}%
\AgdaDatatype{Term}\AgdaSpace{}%
\AgdaSymbol{→}\AgdaSpace{}%
\AgdaSymbol{∀\{}\AgdaBound{A}\AgdaSpace{}%
\AgdaBound{A′}\AgdaSymbol{\}}\AgdaSpace{}%
\AgdaSymbol{→}\AgdaSpace{}%
\AgdaBound{A}\AgdaSpace{}%
\AgdaOperator{\AgdaDatatype{⊑}}\AgdaSpace{}%
\AgdaBound{A′}\AgdaSpace{}%
\AgdaSymbol{→}\AgdaSpace{}%
\AgdaRecord{Setᵒ}\<%
\\
\>[0]\AgdaBound{dir}\AgdaSpace{}%
\AgdaOperator{\AgdaFunction{∣}}\AgdaSpace{}%
\AgdaBound{M}\AgdaSpace{}%
\AgdaOperator{\AgdaFunction{⊑ᴸᴿₜ}}\AgdaSpace{}%
\AgdaBound{M′}\AgdaSpace{}%
\AgdaOperator{\AgdaFunction{⦂}}\AgdaSpace{}%
\AgdaBound{A⊑A′}\AgdaSpace{}%
\AgdaSymbol{=}\AgdaSpace{}%
\AgdaFunction{LRₜ⊎LRᵥ}\AgdaSpace{}%
\AgdaSymbol{(}\AgdaInductiveConstructor{inj₂}\AgdaSpace{}%
\AgdaSymbol{((\AgdaUnderscore{}}\AgdaSpace{}%
\AgdaOperator{\AgdaInductiveConstructor{,}}\AgdaSpace{}%
\AgdaSymbol{\AgdaUnderscore{}}\AgdaSpace{}%
\AgdaOperator{\AgdaInductiveConstructor{,}}\AgdaSpace{}%
\AgdaBound{A⊑A′}\AgdaSymbol{)}\AgdaSpace{}%
\AgdaOperator{\AgdaInductiveConstructor{,}}\AgdaSpace{}%
\AgdaBound{dir}\AgdaSpace{}%
\AgdaOperator{\AgdaInductiveConstructor{,}}\AgdaSpace{}%
\AgdaBound{M}\AgdaSpace{}%
\AgdaOperator{\AgdaInductiveConstructor{,}}\AgdaSpace{}%
\AgdaBound{M′}\AgdaSymbol{))}\<%
\\
\\[\AgdaEmptyExtraSkip]%
\>[0]\AgdaOperator{\AgdaFunction{\AgdaUnderscore{}⊑ᴸᴿₜ\AgdaUnderscore{}⦂\AgdaUnderscore{}}}\AgdaSpace{}%
\AgdaSymbol{:}\AgdaSpace{}%
\AgdaDatatype{Term}\AgdaSpace{}%
\AgdaSymbol{→}\AgdaSpace{}%
\AgdaDatatype{Term}\AgdaSpace{}%
\AgdaSymbol{→}\AgdaSpace{}%
\AgdaSymbol{∀\{}\AgdaBound{A}\AgdaSpace{}%
\AgdaBound{A′}\AgdaSymbol{\}}\AgdaSpace{}%
\AgdaSymbol{→}\AgdaSpace{}%
\AgdaBound{A}\AgdaSpace{}%
\AgdaOperator{\AgdaDatatype{⊑}}\AgdaSpace{}%
\AgdaBound{A′}\AgdaSpace{}%
\AgdaSymbol{→}\AgdaSpace{}%
\AgdaRecord{Setᵒ}\<%
\\
\>[0]\AgdaBound{M}\AgdaSpace{}%
\AgdaOperator{\AgdaFunction{⊑ᴸᴿₜ}}\AgdaSpace{}%
\AgdaBound{M′}\AgdaSpace{}%
\AgdaOperator{\AgdaFunction{⦂}}\AgdaSpace{}%
\AgdaBound{A⊑A′}\AgdaSpace{}%
\AgdaSymbol{=}\AgdaSpace{}%
\AgdaSymbol{(}\AgdaInductiveConstructor{≼}\AgdaSpace{}%
\AgdaOperator{\AgdaFunction{∣}}\AgdaSpace{}%
\AgdaBound{M}\AgdaSpace{}%
\AgdaOperator{\AgdaFunction{⊑ᴸᴿₜ}}\AgdaSpace{}%
\AgdaBound{M′}\AgdaSpace{}%
\AgdaOperator{\AgdaFunction{⦂}}\AgdaSpace{}%
\AgdaBound{A⊑A′}\AgdaSymbol{)}\AgdaSpace{}%
\AgdaOperator{\AgdaFunction{×ᵒ}}\AgdaSpace{}%
\AgdaSymbol{(}\AgdaInductiveConstructor{≽}\AgdaSpace{}%
\AgdaOperator{\AgdaFunction{∣}}\AgdaSpace{}%
\AgdaBound{M}\AgdaSpace{}%
\AgdaOperator{\AgdaFunction{⊑ᴸᴿₜ}}\AgdaSpace{}%
\AgdaBound{M′}\AgdaSpace{}%
\AgdaOperator{\AgdaFunction{⦂}}\AgdaSpace{}%
\AgdaBound{A⊑A′}\AgdaSymbol{)}\<%
\end{code}

The relations that we have defined so far, ⊑ᴸᴿᵥ and ⊑ᴸᴿₜ, only apply
to closed terms, that is, terms with no free variables.  We also need
to relate open terms. The standard way to do that is to apply two
substitutions to the two terms, replacing each free variable with
related values. We relate a pair of substitutions γ and γ′ with the
following definition, which says that the substitutions must be
point-wise related using the logical relation for values.

\begin{code}%
\>[0]\AgdaOperator{\AgdaFunction{\AgdaUnderscore{}∣\AgdaUnderscore{}⊨\AgdaUnderscore{}⊑ᴸᴿ\AgdaUnderscore{}}}\AgdaSpace{}%
\AgdaSymbol{:}\AgdaSpace{}%
\AgdaSymbol{(}\AgdaBound{Γ}\AgdaSpace{}%
\AgdaSymbol{:}\AgdaSpace{}%
\AgdaDatatype{List}\AgdaSpace{}%
\AgdaFunction{Prec}\AgdaSymbol{)}\AgdaSpace{}%
\AgdaSymbol{→}\AgdaSpace{}%
\AgdaDatatype{Dir}\AgdaSpace{}%
\AgdaSymbol{→}\AgdaSpace{}%
\AgdaFunction{Subst}\AgdaSpace{}%
\AgdaSymbol{→}\AgdaSpace{}%
\AgdaFunction{Subst}\AgdaSpace{}%
\AgdaSymbol{→}\AgdaSpace{}%
\AgdaDatatype{List}\AgdaSpace{}%
\AgdaRecord{Setᵒ}\<%
\\
\>[0]\AgdaInductiveConstructor{[]}\AgdaSpace{}%
\AgdaOperator{\AgdaFunction{∣}}\AgdaSpace{}%
\AgdaBound{dir}\AgdaSpace{}%
\AgdaOperator{\AgdaFunction{⊨}}\AgdaSpace{}%
\AgdaBound{γ}\AgdaSpace{}%
\AgdaOperator{\AgdaFunction{⊑ᴸᴿ}}\AgdaSpace{}%
\AgdaBound{γ′}\AgdaSpace{}%
\AgdaSymbol{=}\AgdaSpace{}%
\AgdaInductiveConstructor{[]}\<%
\\
\>[0]\AgdaSymbol{((\AgdaUnderscore{}}%
\>[779I]\AgdaOperator{\AgdaInductiveConstructor{,}}\AgdaSpace{}%
\AgdaSymbol{\AgdaUnderscore{}}\AgdaSpace{}%
\AgdaOperator{\AgdaInductiveConstructor{,}}\AgdaSpace{}%
\AgdaBound{A⊑A′}\AgdaSymbol{)}\AgdaSpace{}%
\AgdaOperator{\AgdaInductiveConstructor{∷}}\AgdaSpace{}%
\AgdaBound{Γ}\AgdaSymbol{)}\AgdaSpace{}%
\AgdaOperator{\AgdaFunction{∣}}\AgdaSpace{}%
\AgdaBound{dir}\AgdaSpace{}%
\AgdaOperator{\AgdaFunction{⊨}}\AgdaSpace{}%
\AgdaBound{γ}\AgdaSpace{}%
\AgdaOperator{\AgdaFunction{⊑ᴸᴿ}}\AgdaSpace{}%
\AgdaBound{γ′}\AgdaSpace{}%
\AgdaSymbol{=}\<%
\\
\>[.][@{}l@{}]\<[779I]%
\>[4]\AgdaSymbol{(}\AgdaBound{dir}\AgdaSpace{}%
\AgdaOperator{\AgdaFunction{∣}}\AgdaSpace{}%
\AgdaSymbol{(}\AgdaBound{γ}\AgdaSpace{}%
\AgdaNumber{0}\AgdaSymbol{)}\AgdaSpace{}%
\AgdaOperator{\AgdaFunction{⊑ᴸᴿᵥ}}\AgdaSpace{}%
\AgdaSymbol{(}\AgdaBound{γ′}\AgdaSpace{}%
\AgdaNumber{0}\AgdaSymbol{)}\AgdaSpace{}%
\AgdaOperator{\AgdaFunction{⦂}}\AgdaSpace{}%
\AgdaBound{A⊑A′}\AgdaSymbol{)}\AgdaSpace{}%
\AgdaOperator{\AgdaInductiveConstructor{∷}}\AgdaSpace{}%
\AgdaSymbol{(}\AgdaBound{Γ}\AgdaSpace{}%
\AgdaOperator{\AgdaFunction{∣}}\AgdaSpace{}%
\AgdaBound{dir}\AgdaSpace{}%
\AgdaOperator{\AgdaFunction{⊨}}\AgdaSpace{}%
\AgdaSymbol{(λ}\AgdaSpace{}%
\AgdaBound{x}\AgdaSpace{}%
\AgdaSymbol{→}\AgdaSpace{}%
\AgdaBound{γ}\AgdaSpace{}%
\AgdaSymbol{(}\AgdaInductiveConstructor{suc}\AgdaSpace{}%
\AgdaBound{x}\AgdaSymbol{))}\AgdaSpace{}%
\AgdaOperator{\AgdaFunction{⊑ᴸᴿ}}\AgdaSpace{}%
\AgdaSymbol{(λ}\AgdaSpace{}%
\AgdaBound{x}\AgdaSpace{}%
\AgdaSymbol{→}\AgdaSpace{}%
\AgdaBound{γ′}\AgdaSpace{}%
\AgdaSymbol{(}\AgdaInductiveConstructor{suc}\AgdaSpace{}%
\AgdaBound{x}\AgdaSymbol{)))}\<%
\end{code}

We then define two open terms $M$ and $M′$ to be logically related
if there are a pair of related substitutions $γ$ and $γ′$ such that
applying them to $M$ and $M′$ produces related terms.

\begin{code}%
\>[0]\AgdaOperator{\AgdaFunction{\AgdaUnderscore{}∣\AgdaUnderscore{}⊨\AgdaUnderscore{}⊑ᴸᴿ\AgdaUnderscore{}⦂\AgdaUnderscore{}}}\AgdaSpace{}%
\AgdaSymbol{:}\AgdaSpace{}%
\AgdaDatatype{List}\AgdaSpace{}%
\AgdaFunction{Prec}\AgdaSpace{}%
\AgdaSymbol{→}\AgdaSpace{}%
\AgdaDatatype{Dir}\AgdaSpace{}%
\AgdaSymbol{→}\AgdaSpace{}%
\AgdaDatatype{Term}\AgdaSpace{}%
\AgdaSymbol{→}\AgdaSpace{}%
\AgdaDatatype{Term}\AgdaSpace{}%
\AgdaSymbol{→}\AgdaSpace{}%
\AgdaFunction{Prec}\AgdaSpace{}%
\AgdaSymbol{→}\AgdaSpace{}%
\AgdaPrimitive{Set}\<%
\\
\>[0]\AgdaBound{Γ}%
\>[831I]\AgdaOperator{\AgdaFunction{∣}}\AgdaSpace{}%
\AgdaBound{dir}\AgdaSpace{}%
\AgdaOperator{\AgdaFunction{⊨}}\AgdaSpace{}%
\AgdaBound{M}\AgdaSpace{}%
\AgdaOperator{\AgdaFunction{⊑ᴸᴿ}}\AgdaSpace{}%
\AgdaBound{M′}\AgdaSpace{}%
\AgdaOperator{\AgdaFunction{⦂}}\AgdaSpace{}%
\AgdaSymbol{(\AgdaUnderscore{}}\AgdaSpace{}%
\AgdaOperator{\AgdaInductiveConstructor{,}}\AgdaSpace{}%
\AgdaSymbol{\AgdaUnderscore{}}\AgdaSpace{}%
\AgdaOperator{\AgdaInductiveConstructor{,}}\AgdaSpace{}%
\AgdaBound{A⊑A′}\AgdaSymbol{)}\AgdaSpace{}%
\AgdaSymbol{=}\AgdaSpace{}%
\AgdaSymbol{∀}\AgdaSpace{}%
\AgdaSymbol{(}\AgdaBound{γ}\AgdaSpace{}%
\AgdaBound{γ′}\AgdaSpace{}%
\AgdaSymbol{:}\AgdaSpace{}%
\AgdaFunction{Subst}\AgdaSymbol{)}\<%
\\
\>[831I][@{}l@{\AgdaIndent{0}}]%
\>[3]\AgdaSymbol{→}\AgdaSpace{}%
\AgdaSymbol{(}\AgdaBound{Γ}\AgdaSpace{}%
\AgdaOperator{\AgdaFunction{∣}}\AgdaSpace{}%
\AgdaBound{dir}\AgdaSpace{}%
\AgdaOperator{\AgdaFunction{⊨}}\AgdaSpace{}%
\AgdaBound{γ}\AgdaSpace{}%
\AgdaOperator{\AgdaFunction{⊑ᴸᴿ}}\AgdaSpace{}%
\AgdaBound{γ′}\AgdaSymbol{)}\AgdaSpace{}%
\AgdaOperator{\AgdaFunction{⊢ᵒ}}\AgdaSpace{}%
\AgdaBound{dir}\AgdaSpace{}%
\AgdaOperator{\AgdaFunction{∣}}\AgdaSpace{}%
\AgdaSymbol{(}\AgdaOperator{\AgdaFunction{⟪}}\AgdaSpace{}%
\AgdaBound{γ}\AgdaSpace{}%
\AgdaOperator{\AgdaFunction{⟫}}\AgdaSpace{}%
\AgdaBound{M}\AgdaSymbol{)}\AgdaSpace{}%
\AgdaOperator{\AgdaFunction{⊑ᴸᴿₜ}}\AgdaSpace{}%
\AgdaSymbol{(}\AgdaOperator{\AgdaFunction{⟪}}\AgdaSpace{}%
\AgdaBound{γ′}\AgdaSpace{}%
\AgdaOperator{\AgdaFunction{⟫}}\AgdaSpace{}%
\AgdaBound{M′}\AgdaSymbol{)}\AgdaSpace{}%
\AgdaOperator{\AgdaFunction{⦂}}\AgdaSpace{}%
\AgdaBound{A⊑A′}\<%
\end{code}

\noindent We use the following notation for the conjunction of the two
directions and define the \textsf{proj} function for accessing each
direction.

\begin{code}%
\>[0]\AgdaOperator{\AgdaFunction{\AgdaUnderscore{}⊨\AgdaUnderscore{}⊑ᴸᴿ\AgdaUnderscore{}⦂\AgdaUnderscore{}}}\AgdaSpace{}%
\AgdaSymbol{:}\AgdaSpace{}%
\AgdaDatatype{List}\AgdaSpace{}%
\AgdaFunction{Prec}\AgdaSpace{}%
\AgdaSymbol{→}\AgdaSpace{}%
\AgdaDatatype{Term}\AgdaSpace{}%
\AgdaSymbol{→}\AgdaSpace{}%
\AgdaDatatype{Term}\AgdaSpace{}%
\AgdaSymbol{→}\AgdaSpace{}%
\AgdaFunction{Prec}\AgdaSpace{}%
\AgdaSymbol{→}\AgdaSpace{}%
\AgdaPrimitive{Set}\<%
\\
\>[0]\AgdaBound{Γ}\AgdaSpace{}%
\AgdaOperator{\AgdaFunction{⊨}}\AgdaSpace{}%
\AgdaBound{M}\AgdaSpace{}%
\AgdaOperator{\AgdaFunction{⊑ᴸᴿ}}\AgdaSpace{}%
\AgdaBound{M′}\AgdaSpace{}%
\AgdaOperator{\AgdaFunction{⦂}}\AgdaSpace{}%
\AgdaBound{c}\AgdaSpace{}%
\AgdaSymbol{=}\AgdaSpace{}%
\AgdaSymbol{(}\AgdaBound{Γ}\AgdaSpace{}%
\AgdaOperator{\AgdaFunction{∣}}\AgdaSpace{}%
\AgdaInductiveConstructor{≼}\AgdaSpace{}%
\AgdaOperator{\AgdaFunction{⊨}}\AgdaSpace{}%
\AgdaBound{M}\AgdaSpace{}%
\AgdaOperator{\AgdaFunction{⊑ᴸᴿ}}\AgdaSpace{}%
\AgdaBound{M′}\AgdaSpace{}%
\AgdaOperator{\AgdaFunction{⦂}}\AgdaSpace{}%
\AgdaBound{c}\AgdaSymbol{)}\AgdaSpace{}%
\AgdaOperator{\AgdaFunction{×}}\AgdaSpace{}%
\AgdaSymbol{(}\AgdaBound{Γ}\AgdaSpace{}%
\AgdaOperator{\AgdaFunction{∣}}\AgdaSpace{}%
\AgdaInductiveConstructor{≽}\AgdaSpace{}%
\AgdaOperator{\AgdaFunction{⊨}}\AgdaSpace{}%
\AgdaBound{M}\AgdaSpace{}%
\AgdaOperator{\AgdaFunction{⊑ᴸᴿ}}\AgdaSpace{}%
\AgdaBound{M′}\AgdaSpace{}%
\AgdaOperator{\AgdaFunction{⦂}}\AgdaSpace{}%
\AgdaBound{c}\AgdaSymbol{)}\<%
\\
\\[\AgdaEmptyExtraSkip]%
\>[0]\AgdaFunction{proj}\AgdaSpace{}%
\AgdaSymbol{:}\AgdaSpace{}%
\AgdaSymbol{∀}\AgdaSpace{}%
\AgdaSymbol{\{}\AgdaBound{Γ}\AgdaSymbol{\}\{}\AgdaBound{c}\AgdaSymbol{\}}\AgdaSpace{}%
\AgdaSymbol{→}\AgdaSpace{}%
\AgdaSymbol{(}\AgdaBound{dir}\AgdaSpace{}%
\AgdaSymbol{:}\AgdaSpace{}%
\AgdaDatatype{Dir}\AgdaSymbol{)}\AgdaSpace{}%
\AgdaSymbol{→}\AgdaSpace{}%
\AgdaSymbol{(}\AgdaBound{M}\AgdaSpace{}%
\AgdaBound{M′}\AgdaSpace{}%
\AgdaSymbol{:}\AgdaSpace{}%
\AgdaDatatype{Term}\AgdaSymbol{)}\AgdaSpace{}%
\AgdaSymbol{→}\AgdaSpace{}%
\AgdaBound{Γ}\AgdaSpace{}%
\AgdaOperator{\AgdaFunction{⊨}}\AgdaSpace{}%
\AgdaBound{M}\AgdaSpace{}%
\AgdaOperator{\AgdaFunction{⊑ᴸᴿ}}\AgdaSpace{}%
\AgdaBound{M′}\AgdaSpace{}%
\AgdaOperator{\AgdaFunction{⦂}}\AgdaSpace{}%
\AgdaBound{c}\AgdaSpace{}%
\AgdaSymbol{→}\AgdaSpace{}%
\AgdaBound{Γ}\AgdaSpace{}%
\AgdaOperator{\AgdaFunction{∣}}\AgdaSpace{}%
\AgdaBound{dir}\AgdaSpace{}%
\AgdaOperator{\AgdaFunction{⊨}}\AgdaSpace{}%
\AgdaBound{M}\AgdaSpace{}%
\AgdaOperator{\AgdaFunction{⊑ᴸᴿ}}\AgdaSpace{}%
\AgdaBound{M′}\AgdaSpace{}%
\AgdaOperator{\AgdaFunction{⦂}}\AgdaSpace{}%
\AgdaBound{c}\<%
\\
\>[0]\AgdaFunction{proj}\AgdaSpace{}%
\AgdaInductiveConstructor{≼}\AgdaSpace{}%
\AgdaBound{M}\AgdaSpace{}%
\AgdaBound{M′}\AgdaSpace{}%
\AgdaBound{M⊑M′}\AgdaSpace{}%
\AgdaSymbol{=}\AgdaSpace{}%
\AgdaField{proj₁}\AgdaSpace{}%
\AgdaBound{M⊑M′}\<%
\\
\>[0]\AgdaFunction{proj}\AgdaSpace{}%
\AgdaInductiveConstructor{≽}\AgdaSpace{}%
\AgdaBound{M}\AgdaSpace{}%
\AgdaBound{M′}\AgdaSpace{}%
\AgdaBound{M⊑M′}\AgdaSpace{}%
\AgdaSymbol{=}\AgdaSpace{}%
\AgdaField{proj₂}\AgdaSpace{}%
\AgdaBound{M⊑M′}\<%
\end{code}

\begin{code}[hide]%
\>[0]\AgdaFunction{LRₜ-def}\AgdaSpace{}%
\AgdaSymbol{:}\AgdaSpace{}%
\AgdaSymbol{∀\{}\AgdaBound{A}\AgdaSymbol{\}\{}\AgdaBound{A′}\AgdaSymbol{\}}\AgdaSpace{}%
\AgdaSymbol{→}\AgdaSpace{}%
\AgdaSymbol{(}\AgdaBound{A⊑A′}\AgdaSpace{}%
\AgdaSymbol{:}\AgdaSpace{}%
\AgdaBound{A}\AgdaSpace{}%
\AgdaOperator{\AgdaDatatype{⊑}}\AgdaSpace{}%
\AgdaBound{A′}\AgdaSymbol{)}\AgdaSpace{}%
\AgdaSymbol{→}\AgdaSpace{}%
\AgdaDatatype{Dir}\AgdaSpace{}%
\AgdaSymbol{→}\AgdaSpace{}%
\AgdaDatatype{Term}\AgdaSpace{}%
\AgdaSymbol{→}\AgdaSpace{}%
\AgdaDatatype{Term}\AgdaSpace{}%
\AgdaSymbol{→}\AgdaSpace{}%
\AgdaRecord{Setᵒ}\<%
\\
\>[0]\AgdaFunction{LRₜ-def}\AgdaSpace{}%
\AgdaBound{A⊑A′}\AgdaSpace{}%
\AgdaInductiveConstructor{≼}\AgdaSpace{}%
\AgdaBound{M}\AgdaSpace{}%
\AgdaBound{M′}\AgdaSpace{}%
\AgdaSymbol{=}\<%
\\
\>[0][@{}l@{\AgdaIndent{0}}]%
\>[3]\AgdaSymbol{(}\AgdaFunction{∃ᵒ[}\AgdaSpace{}%
\AgdaBound{N}\AgdaSpace{}%
\AgdaFunction{]}\AgdaSpace{}%
\AgdaSymbol{(}\AgdaBound{M}\AgdaSpace{}%
\AgdaOperator{\AgdaDatatype{⟶}}\AgdaSpace{}%
\AgdaBound{N}\AgdaSymbol{)}\AgdaOperator{\AgdaFunction{ᵒ}}\AgdaSpace{}%
\AgdaOperator{\AgdaFunction{×ᵒ}}\AgdaSpace{}%
\AgdaOperator{\AgdaFunction{▷ᵒ}}\AgdaSpace{}%
\AgdaSymbol{(}\AgdaInductiveConstructor{≼}\AgdaSpace{}%
\AgdaOperator{\AgdaFunction{∣}}\AgdaSpace{}%
\AgdaBound{N}\AgdaSpace{}%
\AgdaOperator{\AgdaFunction{⊑ᴸᴿₜ}}\AgdaSpace{}%
\AgdaBound{M′}\AgdaSpace{}%
\AgdaOperator{\AgdaFunction{⦂}}\AgdaSpace{}%
\AgdaBound{A⊑A′}\AgdaSymbol{))}\<%
\\
\>[3]\AgdaOperator{\AgdaFunction{⊎ᵒ}}\AgdaSpace{}%
\AgdaSymbol{(}\AgdaBound{M′}\AgdaSpace{}%
\AgdaOperator{\AgdaDatatype{↠}}\AgdaSpace{}%
\AgdaInductiveConstructor{blame}\AgdaSymbol{)}\AgdaOperator{\AgdaFunction{ᵒ}}\<%
\\
\>[3]\AgdaOperator{\AgdaFunction{⊎ᵒ}}\AgdaSpace{}%
\AgdaSymbol{((}\AgdaDatatype{Value}\AgdaSpace{}%
\AgdaBound{M}\AgdaSymbol{)}\AgdaOperator{\AgdaFunction{ᵒ}}\AgdaSpace{}%
\AgdaOperator{\AgdaFunction{×ᵒ}}\AgdaSpace{}%
\AgdaSymbol{(}\AgdaFunction{∃ᵒ[}\AgdaSpace{}%
\AgdaBound{V′}\AgdaSpace{}%
\AgdaFunction{]}\AgdaSpace{}%
\AgdaSymbol{(}\AgdaBound{M′}\AgdaSpace{}%
\AgdaOperator{\AgdaDatatype{↠}}\AgdaSpace{}%
\AgdaBound{V′}\AgdaSymbol{)}\AgdaOperator{\AgdaFunction{ᵒ}}\AgdaSpace{}%
\AgdaOperator{\AgdaFunction{×ᵒ}}\AgdaSpace{}%
\AgdaSymbol{(}\AgdaDatatype{Value}\AgdaSpace{}%
\AgdaBound{V′}\AgdaSymbol{)}\AgdaOperator{\AgdaFunction{ᵒ}}\AgdaSpace{}%
\AgdaOperator{\AgdaFunction{×ᵒ}}\AgdaSpace{}%
\AgdaSymbol{(}\AgdaInductiveConstructor{≼}\AgdaSpace{}%
\AgdaOperator{\AgdaFunction{∣}}\AgdaSpace{}%
\AgdaBound{M}\AgdaSpace{}%
\AgdaOperator{\AgdaFunction{⊑ᴸᴿᵥ}}\AgdaSpace{}%
\AgdaBound{V′}\AgdaSpace{}%
\AgdaOperator{\AgdaFunction{⦂}}\AgdaSpace{}%
\AgdaBound{A⊑A′}\AgdaSymbol{)))}\<%
\\
\>[0]\AgdaFunction{LRₜ-def}\AgdaSpace{}%
\AgdaBound{A⊑A′}\AgdaSpace{}%
\AgdaInductiveConstructor{≽}\AgdaSpace{}%
\AgdaBound{M}\AgdaSpace{}%
\AgdaBound{M′}\AgdaSpace{}%
\AgdaSymbol{=}\<%
\\
\>[0][@{}l@{\AgdaIndent{0}}]%
\>[3]\AgdaSymbol{(}\AgdaFunction{∃ᵒ[}\AgdaSpace{}%
\AgdaBound{N′}\AgdaSpace{}%
\AgdaFunction{]}\AgdaSpace{}%
\AgdaSymbol{(}\AgdaBound{M′}\AgdaSpace{}%
\AgdaOperator{\AgdaDatatype{⟶}}\AgdaSpace{}%
\AgdaBound{N′}\AgdaSymbol{)}\AgdaOperator{\AgdaFunction{ᵒ}}\AgdaSpace{}%
\AgdaOperator{\AgdaFunction{×ᵒ}}\AgdaSpace{}%
\AgdaOperator{\AgdaFunction{▷ᵒ}}\AgdaSpace{}%
\AgdaSymbol{(}\AgdaInductiveConstructor{≽}\AgdaSpace{}%
\AgdaOperator{\AgdaFunction{∣}}\AgdaSpace{}%
\AgdaBound{M}\AgdaSpace{}%
\AgdaOperator{\AgdaFunction{⊑ᴸᴿₜ}}\AgdaSpace{}%
\AgdaBound{N′}\AgdaSpace{}%
\AgdaOperator{\AgdaFunction{⦂}}\AgdaSpace{}%
\AgdaBound{A⊑A′}\AgdaSymbol{))}\<%
\\
\>[3]\AgdaOperator{\AgdaFunction{⊎ᵒ}}\AgdaSpace{}%
\AgdaSymbol{(}\AgdaDatatype{Blame}\AgdaSpace{}%
\AgdaBound{M′}\AgdaSymbol{)}\AgdaOperator{\AgdaFunction{ᵒ}}\<%
\\
\>[3]\AgdaOperator{\AgdaFunction{⊎ᵒ}}\AgdaSpace{}%
\AgdaSymbol{((}\AgdaDatatype{Value}\AgdaSpace{}%
\AgdaBound{M′}\AgdaSymbol{)}\AgdaOperator{\AgdaFunction{ᵒ}}\AgdaSpace{}%
\AgdaOperator{\AgdaFunction{×ᵒ}}\AgdaSpace{}%
\AgdaSymbol{(}\AgdaFunction{∃ᵒ[}\AgdaSpace{}%
\AgdaBound{V}\AgdaSpace{}%
\AgdaFunction{]}\AgdaSpace{}%
\AgdaSymbol{(}\AgdaBound{M}\AgdaSpace{}%
\AgdaOperator{\AgdaDatatype{↠}}\AgdaSpace{}%
\AgdaBound{V}\AgdaSymbol{)}\AgdaOperator{\AgdaFunction{ᵒ}}\AgdaSpace{}%
\AgdaOperator{\AgdaFunction{×ᵒ}}\AgdaSpace{}%
\AgdaSymbol{(}\AgdaDatatype{Value}\AgdaSpace{}%
\AgdaBound{V}\AgdaSymbol{)}\AgdaOperator{\AgdaFunction{ᵒ}}\AgdaSpace{}%
\AgdaOperator{\AgdaFunction{×ᵒ}}\AgdaSpace{}%
\AgdaSymbol{(}\AgdaInductiveConstructor{≽}\AgdaSpace{}%
\AgdaOperator{\AgdaFunction{∣}}\AgdaSpace{}%
\AgdaBound{V}\AgdaSpace{}%
\AgdaOperator{\AgdaFunction{⊑ᴸᴿᵥ}}\AgdaSpace{}%
\AgdaBound{M′}\AgdaSpace{}%
\AgdaOperator{\AgdaFunction{⦂}}\AgdaSpace{}%
\AgdaBound{A⊑A′}\AgdaSymbol{)))}\<%
\end{code}
\begin{code}[hide]%
\>[0]\AgdaFunction{LRₜ-stmt}\AgdaSpace{}%
\AgdaSymbol{:}\AgdaSpace{}%
\AgdaSymbol{∀\{}\AgdaBound{A}\AgdaSymbol{\}\{}\AgdaBound{A′}\AgdaSymbol{\}\{}\AgdaBound{A⊑A′}\AgdaSpace{}%
\AgdaSymbol{:}\AgdaSpace{}%
\AgdaBound{A}\AgdaSpace{}%
\AgdaOperator{\AgdaDatatype{⊑}}\AgdaSpace{}%
\AgdaBound{A′}\AgdaSymbol{\}\{}\AgdaBound{dir}\AgdaSymbol{\}\{}\AgdaBound{M}\AgdaSymbol{\}\{}\AgdaBound{M′}\AgdaSymbol{\}}\<%
\\
\>[0][@{}l@{\AgdaIndent{0}}]%
\>[3]\AgdaSymbol{→}\AgdaSpace{}%
\AgdaBound{dir}\AgdaSpace{}%
\AgdaOperator{\AgdaFunction{∣}}\AgdaSpace{}%
\AgdaBound{M}\AgdaSpace{}%
\AgdaOperator{\AgdaFunction{⊑ᴸᴿₜ}}\AgdaSpace{}%
\AgdaBound{M′}\AgdaSpace{}%
\AgdaOperator{\AgdaFunction{⦂}}\AgdaSpace{}%
\AgdaBound{A⊑A′}\AgdaSpace{}%
\AgdaOperator{\AgdaFunction{≡ᵒ}}\AgdaSpace{}%
\AgdaFunction{LRₜ-def}\AgdaSpace{}%
\AgdaBound{A⊑A′}\AgdaSpace{}%
\AgdaBound{dir}\AgdaSpace{}%
\AgdaBound{M}\AgdaSpace{}%
\AgdaBound{M′}\<%
\end{code}
\begin{code}[hide]%
\>[0]\AgdaFunction{LRₜ-stmt}\AgdaSpace{}%
\AgdaSymbol{\{}\AgdaBound{A}\AgdaSymbol{\}\{}\AgdaBound{A′}\AgdaSymbol{\}\{}\AgdaBound{A⊑A′}\AgdaSymbol{\}\{}\AgdaBound{dir}\AgdaSymbol{\}\{}\AgdaBound{M}\AgdaSymbol{\}\{}\AgdaBound{M′}\AgdaSymbol{\}}\AgdaSpace{}%
\AgdaSymbol{=}\<%
\\
\>[0][@{}l@{\AgdaIndent{0}}]%
\>[2]\AgdaBound{dir}\AgdaSpace{}%
\AgdaOperator{\AgdaFunction{∣}}\AgdaSpace{}%
\AgdaBound{M}\AgdaSpace{}%
\AgdaOperator{\AgdaFunction{⊑ᴸᴿₜ}}\AgdaSpace{}%
\AgdaBound{M′}\AgdaSpace{}%
\AgdaOperator{\AgdaFunction{⦂}}\AgdaSpace{}%
\AgdaBound{A⊑A′}%
\>[43]\AgdaOperator{\AgdaFunction{⩦⟨}}\AgdaSpace{}%
\AgdaFunction{≡ᵒ-refl}\AgdaSpace{}%
\AgdaInductiveConstructor{refl}\AgdaSpace{}%
\AgdaOperator{\AgdaFunction{⟩}}\<%
\\
\>[2]\AgdaFunction{μᵒ}\AgdaSpace{}%
\AgdaFunction{pre-LRₜ⊎LRᵥ}\AgdaSpace{}%
\AgdaSymbol{(}\AgdaFunction{X₂}\AgdaSpace{}%
\AgdaBound{dir}\AgdaSymbol{)}%
\>[43]\AgdaOperator{\AgdaFunction{⩦⟨}}\AgdaSpace{}%
\AgdaFunction{fixpointᵒ}\AgdaSpace{}%
\AgdaFunction{pre-LRₜ⊎LRᵥ}\AgdaSpace{}%
\AgdaSymbol{(}\AgdaFunction{X₂}\AgdaSpace{}%
\AgdaBound{dir}\AgdaSymbol{)}\AgdaSpace{}%
\AgdaOperator{\AgdaFunction{⟩}}\<%
\\
\>[2]\AgdaField{\#}\AgdaSpace{}%
\AgdaSymbol{(}\AgdaFunction{pre-LRₜ⊎LRᵥ}\AgdaSpace{}%
\AgdaSymbol{(}\AgdaFunction{X₂}\AgdaSpace{}%
\AgdaBound{dir}\AgdaSymbol{))}\AgdaSpace{}%
\AgdaSymbol{(}\AgdaFunction{LRₜ⊎LRᵥ}\AgdaSpace{}%
\AgdaOperator{\AgdaInductiveConstructor{,}}\AgdaSpace{}%
\AgdaFunction{ttᵖ}\AgdaSymbol{)}\AgdaSpace{}%
\AgdaOperator{\AgdaFunction{⩦⟨}}\AgdaSpace{}%
\AgdaFunction{EQ}\AgdaSymbol{\{}\AgdaBound{dir}\AgdaSymbol{\}}\AgdaSpace{}%
\AgdaOperator{\AgdaFunction{⟩}}\<%
\\
\>[2]\AgdaFunction{LRₜ-def}\AgdaSpace{}%
\AgdaBound{A⊑A′}\AgdaSpace{}%
\AgdaBound{dir}\AgdaSpace{}%
\AgdaBound{M}\AgdaSpace{}%
\AgdaBound{M′}%
\>[43]\AgdaOperator{\AgdaFunction{∎}}\<%
\\
\>[2]\AgdaKeyword{where}\<%
\\
\>[2]\AgdaFunction{c}\AgdaSpace{}%
\AgdaSymbol{=}\AgdaSpace{}%
\AgdaSymbol{(}\AgdaBound{A}\AgdaSpace{}%
\AgdaOperator{\AgdaInductiveConstructor{,}}\AgdaSpace{}%
\AgdaBound{A′}\AgdaSpace{}%
\AgdaOperator{\AgdaInductiveConstructor{,}}\AgdaSpace{}%
\AgdaBound{A⊑A′}\AgdaSymbol{)}\<%
\\
\>[2]\AgdaFunction{X₁}\AgdaSpace{}%
\AgdaSymbol{:}\AgdaSpace{}%
\AgdaDatatype{Dir}\AgdaSpace{}%
\AgdaSymbol{→}\AgdaSpace{}%
\AgdaFunction{LR-type}\<%
\\
\>[2]\AgdaFunction{X₁}\AgdaSpace{}%
\AgdaSymbol{=}\AgdaSpace{}%
\AgdaSymbol{λ}\AgdaSpace{}%
\AgdaBound{dir}\AgdaSpace{}%
\AgdaSymbol{→}\AgdaSpace{}%
\AgdaInductiveConstructor{inj₁}\AgdaSpace{}%
\AgdaSymbol{(}\AgdaFunction{c}\AgdaSpace{}%
\AgdaOperator{\AgdaInductiveConstructor{,}}\AgdaSpace{}%
\AgdaBound{dir}\AgdaSpace{}%
\AgdaOperator{\AgdaInductiveConstructor{,}}\AgdaSpace{}%
\AgdaBound{M}\AgdaSpace{}%
\AgdaOperator{\AgdaInductiveConstructor{,}}\AgdaSpace{}%
\AgdaBound{M′}\AgdaSymbol{)}\<%
\\
\>[2]\AgdaFunction{X₂}\AgdaSpace{}%
\AgdaSymbol{=}\AgdaSpace{}%
\AgdaSymbol{λ}\AgdaSpace{}%
\AgdaBound{dir}\AgdaSpace{}%
\AgdaSymbol{→}\AgdaSpace{}%
\AgdaInductiveConstructor{inj₂}\AgdaSpace{}%
\AgdaSymbol{(}\AgdaFunction{c}\AgdaSpace{}%
\AgdaOperator{\AgdaInductiveConstructor{,}}\AgdaSpace{}%
\AgdaBound{dir}\AgdaSpace{}%
\AgdaOperator{\AgdaInductiveConstructor{,}}\AgdaSpace{}%
\AgdaBound{M}\AgdaSpace{}%
\AgdaOperator{\AgdaInductiveConstructor{,}}\AgdaSpace{}%
\AgdaBound{M′}\AgdaSymbol{)}\<%
\\
\>[2]\AgdaFunction{EQ}\AgdaSpace{}%
\AgdaSymbol{:}\AgdaSpace{}%
\AgdaSymbol{∀\{}\AgdaBound{dir}\AgdaSymbol{\}}\AgdaSpace{}%
\AgdaSymbol{→}\AgdaSpace{}%
\AgdaField{\#}\AgdaSpace{}%
\AgdaSymbol{(}\AgdaFunction{pre-LRₜ⊎LRᵥ}\AgdaSpace{}%
\AgdaSymbol{(}\AgdaFunction{X₂}\AgdaSpace{}%
\AgdaBound{dir}\AgdaSymbol{))}\AgdaSpace{}%
\AgdaSymbol{(}\AgdaFunction{LRₜ⊎LRᵥ}\AgdaSpace{}%
\AgdaOperator{\AgdaInductiveConstructor{,}}\AgdaSpace{}%
\AgdaFunction{ttᵖ}\AgdaSymbol{)}\AgdaSpace{}%
\AgdaOperator{\AgdaFunction{≡ᵒ}}\AgdaSpace{}%
\AgdaFunction{LRₜ-def}\AgdaSpace{}%
\AgdaBound{A⊑A′}\AgdaSpace{}%
\AgdaBound{dir}\AgdaSpace{}%
\AgdaBound{M}\AgdaSpace{}%
\AgdaBound{M′}\<%
\\
\>[2]\AgdaFunction{EQ}\AgdaSpace{}%
\AgdaSymbol{\{}\AgdaInductiveConstructor{≼}\AgdaSymbol{\}}\AgdaSpace{}%
\AgdaSymbol{=}%
\>[1153I]\AgdaFunction{cong-⊎ᵒ}\AgdaSpace{}%
\AgdaSymbol{(}\AgdaFunction{≡ᵒ-refl}\AgdaSpace{}%
\AgdaInductiveConstructor{refl}\AgdaSymbol{)}\AgdaSpace{}%
\AgdaSymbol{(}\AgdaFunction{cong-⊎ᵒ}\AgdaSpace{}%
\AgdaSymbol{(}\AgdaFunction{≡ᵒ-refl}\AgdaSpace{}%
\AgdaInductiveConstructor{refl}\AgdaSymbol{)}\AgdaSpace{}%
\AgdaSymbol{(}\AgdaFunction{cong-×ᵒ}\AgdaSpace{}%
\AgdaSymbol{(}\AgdaFunction{≡ᵒ-refl}\AgdaSpace{}%
\AgdaInductiveConstructor{refl}\AgdaSymbol{)}\<%
\\
\>[1153I][@{}l@{\AgdaIndent{0}}]%
\>[13]\AgdaSymbol{(}\AgdaFunction{cong-∃}\AgdaSpace{}%
\AgdaSymbol{λ}\AgdaSpace{}%
\AgdaBound{V′}\AgdaSpace{}%
\AgdaSymbol{→}\AgdaSpace{}%
\AgdaFunction{cong-×ᵒ}\AgdaSpace{}%
\AgdaSymbol{(}\AgdaFunction{≡ᵒ-refl}\AgdaSpace{}%
\AgdaInductiveConstructor{refl}\AgdaSymbol{)}\AgdaSpace{}%
\AgdaSymbol{(}\AgdaFunction{cong-×ᵒ}\AgdaSpace{}%
\AgdaSymbol{(}\AgdaFunction{≡ᵒ-refl}\AgdaSpace{}%
\AgdaInductiveConstructor{refl}\AgdaSymbol{)}\<%
\\
\>[13][@{}l@{\AgdaIndent{0}}]%
\>[14]\AgdaSymbol{((}\AgdaFunction{≡ᵒ-sym}\AgdaSpace{}%
\AgdaSymbol{(}\AgdaFunction{fixpointᵒ}\AgdaSpace{}%
\AgdaFunction{pre-LRₜ⊎LRᵥ}\AgdaSpace{}%
\AgdaSymbol{(}\AgdaInductiveConstructor{inj₁}\AgdaSpace{}%
\AgdaSymbol{(}\AgdaFunction{c}\AgdaSpace{}%
\AgdaOperator{\AgdaInductiveConstructor{,}}\AgdaSpace{}%
\AgdaInductiveConstructor{≼}\AgdaSpace{}%
\AgdaOperator{\AgdaInductiveConstructor{,}}\AgdaSpace{}%
\AgdaBound{M}\AgdaSpace{}%
\AgdaOperator{\AgdaInductiveConstructor{,}}\AgdaSpace{}%
\AgdaBound{V′}\AgdaSymbol{)))))))))}\<%
\\
\>[2]\AgdaFunction{EQ}\AgdaSpace{}%
\AgdaSymbol{\{}\AgdaInductiveConstructor{≽}\AgdaSymbol{\}}\AgdaSpace{}%
\AgdaSymbol{=}%
\>[1183I]\AgdaFunction{cong-⊎ᵒ}\AgdaSpace{}%
\AgdaSymbol{(}\AgdaFunction{≡ᵒ-refl}\AgdaSpace{}%
\AgdaInductiveConstructor{refl}\AgdaSymbol{)}\AgdaSpace{}%
\AgdaSymbol{(}\AgdaFunction{cong-⊎ᵒ}\AgdaSpace{}%
\AgdaSymbol{(}\AgdaFunction{≡ᵒ-refl}\AgdaSpace{}%
\AgdaInductiveConstructor{refl}\AgdaSymbol{)}\<%
\\
\>[1183I][@{}l@{\AgdaIndent{0}}]%
\>[12]\AgdaSymbol{(}\AgdaFunction{cong-×ᵒ}\AgdaSpace{}%
\AgdaSymbol{(}\AgdaFunction{≡ᵒ-refl}\AgdaSpace{}%
\AgdaInductiveConstructor{refl}\AgdaSymbol{)}\AgdaSpace{}%
\AgdaSymbol{(}\AgdaFunction{cong-∃}\AgdaSpace{}%
\AgdaSymbol{λ}\AgdaSpace{}%
\AgdaBound{V}\AgdaSpace{}%
\AgdaSymbol{→}\AgdaSpace{}%
\AgdaFunction{cong-×ᵒ}\AgdaSpace{}%
\AgdaSymbol{(}\AgdaFunction{≡ᵒ-refl}\AgdaSpace{}%
\AgdaInductiveConstructor{refl}\AgdaSymbol{)}\<%
\\
\>[12][@{}l@{\AgdaIndent{0}}]%
\>[14]\AgdaSymbol{(}\AgdaFunction{cong-×ᵒ}\AgdaSpace{}%
\AgdaSymbol{(}\AgdaFunction{≡ᵒ-refl}\AgdaSpace{}%
\AgdaInductiveConstructor{refl}\AgdaSymbol{)}\<%
\\
\>[14][@{}l@{\AgdaIndent{0}}]%
\>[15]\AgdaSymbol{(}\AgdaFunction{≡ᵒ-sym}\AgdaSpace{}%
\AgdaSymbol{(}\AgdaFunction{fixpointᵒ}\AgdaSpace{}%
\AgdaFunction{pre-LRₜ⊎LRᵥ}\AgdaSpace{}%
\AgdaSymbol{(}\AgdaInductiveConstructor{inj₁}\AgdaSpace{}%
\AgdaSymbol{(}\AgdaFunction{c}\AgdaSpace{}%
\AgdaOperator{\AgdaInductiveConstructor{,}}\AgdaSpace{}%
\AgdaInductiveConstructor{≽}\AgdaSpace{}%
\AgdaOperator{\AgdaInductiveConstructor{,}}\AgdaSpace{}%
\AgdaBound{V}\AgdaSpace{}%
\AgdaOperator{\AgdaInductiveConstructor{,}}\AgdaSpace{}%
\AgdaBound{M′}\AgdaSymbol{))))))))}\<%
\end{code}
\begin{code}[hide]%
\>[0]\AgdaFunction{LRₜ-suc}\AgdaSpace{}%
\AgdaSymbol{:}\AgdaSpace{}%
\AgdaSymbol{∀\{}\AgdaBound{A}\AgdaSymbol{\}\{}\AgdaBound{A′}\AgdaSymbol{\}\{}\AgdaBound{A⊑A′}\AgdaSpace{}%
\AgdaSymbol{:}\AgdaSpace{}%
\AgdaBound{A}\AgdaSpace{}%
\AgdaOperator{\AgdaDatatype{⊑}}\AgdaSpace{}%
\AgdaBound{A′}\AgdaSymbol{\}\{}\AgdaBound{dir}\AgdaSymbol{\}\{}\AgdaBound{M}\AgdaSymbol{\}\{}\AgdaBound{M′}\AgdaSymbol{\}\{}\AgdaBound{k}\AgdaSymbol{\}}\<%
\\
\>[0][@{}l@{\AgdaIndent{0}}]%
\>[2]\AgdaSymbol{→}\AgdaSpace{}%
\AgdaField{\#}\AgdaSymbol{(}\AgdaBound{dir}\AgdaSpace{}%
\AgdaOperator{\AgdaFunction{∣}}\AgdaSpace{}%
\AgdaBound{M}\AgdaSpace{}%
\AgdaOperator{\AgdaFunction{⊑ᴸᴿₜ}}\AgdaSpace{}%
\AgdaBound{M′}\AgdaSpace{}%
\AgdaOperator{\AgdaFunction{⦂}}\AgdaSpace{}%
\AgdaBound{A⊑A′}\AgdaSymbol{)}\AgdaSpace{}%
\AgdaSymbol{(}\AgdaInductiveConstructor{suc}\AgdaSpace{}%
\AgdaBound{k}\AgdaSymbol{)}\AgdaSpace{}%
\AgdaOperator{\AgdaFunction{⇔}}\AgdaSpace{}%
\AgdaField{\#}\AgdaSymbol{(}\AgdaFunction{LRₜ-def}\AgdaSpace{}%
\AgdaBound{A⊑A′}\AgdaSpace{}%
\AgdaBound{dir}\AgdaSpace{}%
\AgdaBound{M}\AgdaSpace{}%
\AgdaBound{M′}\AgdaSymbol{)}\AgdaSpace{}%
\AgdaSymbol{(}\AgdaInductiveConstructor{suc}\AgdaSpace{}%
\AgdaBound{k}\AgdaSymbol{)}\<%
\end{code}
\begin{code}[hide]%
\>[0]\AgdaFunction{LRₜ-suc}\AgdaSpace{}%
\AgdaSymbol{\{}\AgdaBound{A}\AgdaSymbol{\}\{}\AgdaBound{A′}\AgdaSymbol{\}\{}\AgdaBound{A⊑A′}\AgdaSymbol{\}\{}\AgdaBound{dir}\AgdaSymbol{\}\{}\AgdaBound{M}\AgdaSymbol{\}\{}\AgdaBound{M′}\AgdaSymbol{\}\{}\AgdaBound{k}\AgdaSymbol{\}}\AgdaSpace{}%
\AgdaSymbol{=}\<%
\\
\>[0][@{}l@{\AgdaIndent{0}}]%
\>[3]\AgdaFunction{≡ᵒ⇒⇔}\AgdaSymbol{\{}\AgdaArgument{k}\AgdaSpace{}%
\AgdaSymbol{=}\AgdaSpace{}%
\AgdaInductiveConstructor{suc}\AgdaSpace{}%
\AgdaBound{k}\AgdaSymbol{\}}\AgdaSpace{}%
\AgdaSymbol{(}\AgdaFunction{LRₜ-stmt}\AgdaSymbol{\{}\AgdaBound{A}\AgdaSymbol{\}\{}\AgdaBound{A′}\AgdaSymbol{\}\{}\AgdaBound{A⊑A′}\AgdaSymbol{\}\{}\AgdaBound{dir}\AgdaSymbol{\}\{}\AgdaBound{M}\AgdaSymbol{\}\{}\AgdaBound{M′}\AgdaSymbol{\})}\<%
\end{code}

The definition of ⊑ᴸᴿᵥ includes several clauses that ensured that the
related values are indeed syntactic values. Here we make use of that
to prove that indeed, logically related values are syntactic values.

\begin{code}[hide]%
\>[0]\AgdaFunction{LRᵥ⇒Value}\AgdaSpace{}%
\AgdaSymbol{:}\AgdaSpace{}%
\AgdaSymbol{∀}\AgdaSpace{}%
\AgdaSymbol{\{}\AgdaBound{k}\AgdaSymbol{\}\{}\AgdaBound{dir}\AgdaSymbol{\}\{}\AgdaBound{A}\AgdaSymbol{\}\{}\AgdaBound{A′}\AgdaSymbol{\}}\AgdaSpace{}%
\AgdaSymbol{(}\AgdaBound{A⊑A′}\AgdaSpace{}%
\AgdaSymbol{:}\AgdaSpace{}%
\AgdaBound{A}\AgdaSpace{}%
\AgdaOperator{\AgdaDatatype{⊑}}\AgdaSpace{}%
\AgdaBound{A′}\AgdaSymbol{)}\AgdaSpace{}%
\AgdaBound{M}\AgdaSpace{}%
\AgdaBound{M′}\<%
\\
\>[0][@{}l@{\AgdaIndent{0}}]%
\>[3]\AgdaSymbol{→}\AgdaSpace{}%
\AgdaField{\#}\AgdaSpace{}%
\AgdaSymbol{(}\AgdaBound{dir}\AgdaSpace{}%
\AgdaOperator{\AgdaFunction{∣}}\AgdaSpace{}%
\AgdaBound{M}\AgdaSpace{}%
\AgdaOperator{\AgdaFunction{⊑ᴸᴿᵥ}}\AgdaSpace{}%
\AgdaBound{M′}\AgdaSpace{}%
\AgdaOperator{\AgdaFunction{⦂}}\AgdaSpace{}%
\AgdaBound{A⊑A′}\AgdaSymbol{)}\AgdaSpace{}%
\AgdaSymbol{(}\AgdaInductiveConstructor{suc}\AgdaSpace{}%
\AgdaBound{k}\AgdaSymbol{)}%
\>[41]\AgdaSymbol{→}%
\>[44]\AgdaDatatype{Value}\AgdaSpace{}%
\AgdaBound{M}\AgdaSpace{}%
\AgdaOperator{\AgdaFunction{×}}\AgdaSpace{}%
\AgdaDatatype{Value}\AgdaSpace{}%
\AgdaBound{M′}\<%
\\
\>[0]\AgdaFunction{LRᵥ⇒Value}\AgdaSpace{}%
\AgdaSymbol{\{}\AgdaBound{k}\AgdaSymbol{\}\{}\AgdaBound{dir}\AgdaSymbol{\}}\AgdaSpace{}%
\AgdaInductiveConstructor{unk⊑unk}\AgdaSpace{}%
\AgdaSymbol{(}\AgdaBound{V}\AgdaSpace{}%
\AgdaOperator{\AgdaInductiveConstructor{⟨}}\AgdaSpace{}%
\AgdaBound{G}\AgdaSpace{}%
\AgdaOperator{\AgdaInductiveConstructor{!⟩}}\AgdaSymbol{)}\AgdaSpace{}%
\AgdaSymbol{(}\AgdaBound{V′}\AgdaSpace{}%
\AgdaOperator{\AgdaInductiveConstructor{⟨}}\AgdaSpace{}%
\AgdaBound{H}\AgdaSpace{}%
\AgdaOperator{\AgdaInductiveConstructor{!⟩}}\AgdaSymbol{)}\AgdaSpace{}%
\AgdaBound{𝒱MM′}\<%
\\
\>[0][@{}l@{\AgdaIndent{0}}]%
\>[4]\AgdaKeyword{with}\AgdaSpace{}%
\AgdaBound{G}\AgdaSpace{}%
\AgdaOperator{\AgdaFunction{≡ᵍ}}\AgdaSpace{}%
\AgdaBound{H}\<%
\\
\>[0]\AgdaSymbol{...}\AgdaSpace{}%
\AgdaSymbol{|}\AgdaSpace{}%
\AgdaInductiveConstructor{no}\AgdaSpace{}%
\AgdaBound{neq}\AgdaSpace{}%
\AgdaSymbol{=}\AgdaSpace{}%
\AgdaFunction{⊥-elim}\AgdaSpace{}%
\AgdaBound{𝒱MM′}\<%
\\
\>[0]\AgdaSymbol{...}%
\>[1283I]\AgdaSymbol{|}\AgdaSpace{}%
\AgdaInductiveConstructor{yes}\AgdaSpace{}%
\AgdaInductiveConstructor{refl}\<%
\\
\>[.][@{}l@{}]\<[1283I]%
\>[4]\AgdaKeyword{with}\AgdaSpace{}%
\AgdaBound{𝒱MM′}\<%
\\
\>[0]\AgdaSymbol{...}\AgdaSpace{}%
\AgdaSymbol{|}\AgdaSpace{}%
\AgdaBound{v}\AgdaSpace{}%
\AgdaOperator{\AgdaInductiveConstructor{,}}\AgdaSpace{}%
\AgdaBound{v′}\AgdaSpace{}%
\AgdaOperator{\AgdaInductiveConstructor{,}}\AgdaSpace{}%
\AgdaSymbol{\AgdaUnderscore{}}\AgdaSpace{}%
\AgdaSymbol{=}\AgdaSpace{}%
\AgdaSymbol{(}\AgdaBound{v}\AgdaSpace{}%
\AgdaOperator{\AgdaInductiveConstructor{〈}}\AgdaSpace{}%
\AgdaBound{G}\AgdaSpace{}%
\AgdaOperator{\AgdaInductiveConstructor{〉}}\AgdaSymbol{)}\AgdaSpace{}%
\AgdaOperator{\AgdaInductiveConstructor{,}}\AgdaSpace{}%
\AgdaSymbol{(}\AgdaBound{v′}\AgdaSpace{}%
\AgdaOperator{\AgdaInductiveConstructor{〈}}\AgdaSpace{}%
\AgdaBound{G}\AgdaSpace{}%
\AgdaOperator{\AgdaInductiveConstructor{〉}}\AgdaSymbol{)}\<%
\\
\>[0]\AgdaFunction{LRᵥ⇒Value}\AgdaSpace{}%
\AgdaSymbol{\{}\AgdaBound{k}\AgdaSymbol{\}\{}\AgdaInductiveConstructor{≼}\AgdaSymbol{\}}\AgdaSpace{}%
\AgdaSymbol{(}\AgdaInductiveConstructor{unk⊑}\AgdaSymbol{\{}\AgdaBound{H}\AgdaSymbol{\}\{}\AgdaBound{A′}\AgdaSymbol{\}}\AgdaSpace{}%
\AgdaBound{d}\AgdaSymbol{)}\AgdaSpace{}%
\AgdaSymbol{(}\AgdaBound{V}\AgdaSpace{}%
\AgdaOperator{\AgdaInductiveConstructor{⟨}}\AgdaSpace{}%
\AgdaBound{G}\AgdaSpace{}%
\AgdaOperator{\AgdaInductiveConstructor{!⟩}}\AgdaSymbol{)}\AgdaSpace{}%
\AgdaBound{V′}\AgdaSpace{}%
\AgdaBound{𝒱VGV′}\<%
\\
\>[0][@{}l@{\AgdaIndent{0}}]%
\>[4]\AgdaKeyword{with}\AgdaSpace{}%
\AgdaBound{G}\AgdaSpace{}%
\AgdaOperator{\AgdaFunction{≡ᵍ}}\AgdaSpace{}%
\AgdaBound{H}\<%
\\
\>[0]\AgdaSymbol{...}%
\>[1315I]\AgdaSymbol{|}\AgdaSpace{}%
\AgdaInductiveConstructor{yes}\AgdaSpace{}%
\AgdaInductiveConstructor{refl}\<%
\\
\>[.][@{}l@{}]\<[1315I]%
\>[4]\AgdaKeyword{with}\AgdaSpace{}%
\AgdaBound{𝒱VGV′}\<%
\\
\>[0]\AgdaSymbol{...}\AgdaSpace{}%
\AgdaSymbol{|}\AgdaSpace{}%
\AgdaBound{v}\AgdaSpace{}%
\AgdaOperator{\AgdaInductiveConstructor{,}}\AgdaSpace{}%
\AgdaBound{v′}\AgdaSpace{}%
\AgdaOperator{\AgdaInductiveConstructor{,}}\AgdaSpace{}%
\AgdaSymbol{\AgdaUnderscore{}}\AgdaSpace{}%
\AgdaSymbol{=}\AgdaSpace{}%
\AgdaSymbol{(}\AgdaBound{v}\AgdaSpace{}%
\AgdaOperator{\AgdaInductiveConstructor{〈}}\AgdaSpace{}%
\AgdaSymbol{\AgdaUnderscore{}}\AgdaSpace{}%
\AgdaOperator{\AgdaInductiveConstructor{〉}}\AgdaSymbol{)}\AgdaSpace{}%
\AgdaOperator{\AgdaInductiveConstructor{,}}\AgdaSpace{}%
\AgdaBound{v′}\<%
\\
\>[0]\AgdaFunction{LRᵥ⇒Value}\AgdaSpace{}%
\AgdaSymbol{\{}\AgdaBound{k}\AgdaSymbol{\}\{}\AgdaInductiveConstructor{≽}\AgdaSymbol{\}}\AgdaSpace{}%
\AgdaSymbol{(}\AgdaInductiveConstructor{unk⊑}\AgdaSymbol{\{}\AgdaBound{H}\AgdaSymbol{\}\{}\AgdaBound{A′}\AgdaSymbol{\}}\AgdaSpace{}%
\AgdaBound{d}\AgdaSymbol{)}\AgdaSpace{}%
\AgdaSymbol{(}\AgdaBound{V}\AgdaSpace{}%
\AgdaOperator{\AgdaInductiveConstructor{⟨}}\AgdaSpace{}%
\AgdaBound{G}\AgdaSpace{}%
\AgdaOperator{\AgdaInductiveConstructor{!⟩}}\AgdaSymbol{)}\AgdaSpace{}%
\AgdaBound{V′}\AgdaSpace{}%
\AgdaBound{𝒱VGV′}\<%
\\
\>[0][@{}l@{\AgdaIndent{0}}]%
\>[4]\AgdaKeyword{with}\AgdaSpace{}%
\AgdaBound{G}\AgdaSpace{}%
\AgdaOperator{\AgdaFunction{≡ᵍ}}\AgdaSpace{}%
\AgdaBound{H}\<%
\\
\>[0]\AgdaSymbol{...}%
\>[1344I]\AgdaSymbol{|}\AgdaSpace{}%
\AgdaInductiveConstructor{yes}\AgdaSpace{}%
\AgdaInductiveConstructor{refl}\<%
\\
\>[.][@{}l@{}]\<[1344I]%
\>[4]\AgdaKeyword{with}\AgdaSpace{}%
\AgdaBound{𝒱VGV′}\<%
\\
\>[0]\AgdaSymbol{...}\AgdaSpace{}%
\AgdaSymbol{|}\AgdaSpace{}%
\AgdaBound{v}\AgdaSpace{}%
\AgdaOperator{\AgdaInductiveConstructor{,}}\AgdaSpace{}%
\AgdaBound{v′}\AgdaSpace{}%
\AgdaOperator{\AgdaInductiveConstructor{,}}\AgdaSpace{}%
\AgdaSymbol{\AgdaUnderscore{}}\AgdaSpace{}%
\AgdaSymbol{=}\AgdaSpace{}%
\AgdaSymbol{(}\AgdaBound{v}\AgdaSpace{}%
\AgdaOperator{\AgdaInductiveConstructor{〈}}\AgdaSpace{}%
\AgdaSymbol{\AgdaUnderscore{}}\AgdaSpace{}%
\AgdaOperator{\AgdaInductiveConstructor{〉}}\AgdaSymbol{)}\AgdaSpace{}%
\AgdaOperator{\AgdaInductiveConstructor{,}}\AgdaSpace{}%
\AgdaBound{v′}\<%
\\
\>[0]\AgdaFunction{LRᵥ⇒Value}\AgdaSpace{}%
\AgdaSymbol{\{}\AgdaBound{k}\AgdaSymbol{\}\{}\AgdaBound{dir}\AgdaSymbol{\}}\AgdaSpace{}%
\AgdaSymbol{(}\AgdaInductiveConstructor{unk⊑}\AgdaSymbol{\{}\AgdaBound{H}\AgdaSymbol{\}\{}\AgdaBound{A′}\AgdaSymbol{\}}\AgdaSpace{}%
\AgdaBound{d}\AgdaSymbol{)}\AgdaSpace{}%
\AgdaSymbol{(}\AgdaBound{V}\AgdaSpace{}%
\AgdaOperator{\AgdaInductiveConstructor{⟨}}\AgdaSpace{}%
\AgdaBound{G}\AgdaSpace{}%
\AgdaOperator{\AgdaInductiveConstructor{!⟩}}\AgdaSymbol{)}\AgdaSpace{}%
\AgdaBound{V′}\AgdaSpace{}%
\AgdaBound{𝒱VGV′}\<%
\\
\>[0][@{}l@{\AgdaIndent{0}}]%
\>[4]\AgdaSymbol{|}\AgdaSpace{}%
\AgdaInductiveConstructor{no}\AgdaSpace{}%
\AgdaBound{neq}\AgdaSpace{}%
\AgdaSymbol{=}\AgdaSpace{}%
\AgdaFunction{⊥-elim}\AgdaSpace{}%
\AgdaBound{𝒱VGV′}\<%
\\
\>[0]\AgdaFunction{LRᵥ⇒Value}\AgdaSpace{}%
\AgdaSymbol{\{}\AgdaBound{k}\AgdaSymbol{\}\{}\AgdaBound{dir}\AgdaSymbol{\}}\AgdaSpace{}%
\AgdaSymbol{(}\AgdaInductiveConstructor{base⊑}\AgdaSymbol{\{}\AgdaBound{ι}\AgdaSymbol{\})}\AgdaSpace{}%
\AgdaSymbol{(}\AgdaInductiveConstructor{\$}\AgdaSpace{}%
\AgdaBound{c}\AgdaSymbol{)}\AgdaSpace{}%
\AgdaSymbol{(}\AgdaInductiveConstructor{\$}\AgdaSpace{}%
\AgdaBound{c′}\AgdaSymbol{)}\AgdaSpace{}%
\AgdaInductiveConstructor{refl}\AgdaSpace{}%
\AgdaSymbol{=}\AgdaSpace{}%
\AgdaSymbol{(}\AgdaInductiveConstructor{\$̬}\AgdaSpace{}%
\AgdaBound{c}\AgdaSymbol{)}\AgdaSpace{}%
\AgdaOperator{\AgdaInductiveConstructor{,}}\AgdaSpace{}%
\AgdaSymbol{(}\AgdaInductiveConstructor{\$̬}\AgdaSpace{}%
\AgdaBound{c}\AgdaSymbol{)}\<%
\\
\>[0]\AgdaFunction{LRᵥ⇒Value}\AgdaSpace{}%
\AgdaSymbol{\{}\AgdaBound{k}\AgdaSymbol{\}\{}\AgdaBound{dir}\AgdaSymbol{\}}\AgdaSpace{}%
\AgdaSymbol{(}\AgdaInductiveConstructor{fun⊑}\AgdaSpace{}%
\AgdaBound{A⊑A′}\AgdaSpace{}%
\AgdaBound{B⊑B′}\AgdaSymbol{)}\AgdaSpace{}%
\AgdaSymbol{(}\AgdaInductiveConstructor{ƛ}\AgdaSpace{}%
\AgdaBound{N}\AgdaSymbol{)}\AgdaSpace{}%
\AgdaSymbol{(}\AgdaInductiveConstructor{ƛ}\AgdaSpace{}%
\AgdaBound{N′}\AgdaSymbol{)}\AgdaSpace{}%
\AgdaBound{𝒱VV′}\AgdaSpace{}%
\AgdaSymbol{=}\<%
\\
\>[0][@{}l@{\AgdaIndent{0}}]%
\>[4]\AgdaSymbol{(}\AgdaOperator{\AgdaInductiveConstructor{ƛ̬}}\AgdaSpace{}%
\AgdaBound{N}\AgdaSymbol{)}\AgdaSpace{}%
\AgdaOperator{\AgdaInductiveConstructor{,}}\AgdaSpace{}%
\AgdaSymbol{(}\AgdaOperator{\AgdaInductiveConstructor{ƛ̬}}\AgdaSpace{}%
\AgdaBound{N′}\AgdaSymbol{)}\<%
\end{code}

\begin{code}%
\>[0]\AgdaFunction{LRᵥ⇒Valueᵒ}\AgdaSpace{}%
\AgdaSymbol{:}\AgdaSpace{}%
\AgdaSymbol{∀}\AgdaSpace{}%
\AgdaSymbol{\{}\AgdaBound{dir}\AgdaSymbol{\}\{}\AgdaBound{A}\AgdaSymbol{\}\{}\AgdaBound{A′}\AgdaSymbol{\}\{}\AgdaBound{𝒫}\AgdaSymbol{\}}\AgdaSpace{}%
\AgdaSymbol{(}\AgdaBound{A⊑A′}\AgdaSpace{}%
\AgdaSymbol{:}\AgdaSpace{}%
\AgdaBound{A}\AgdaSpace{}%
\AgdaOperator{\AgdaDatatype{⊑}}\AgdaSpace{}%
\AgdaBound{A′}\AgdaSymbol{)}\AgdaSpace{}%
\AgdaBound{M}\AgdaSpace{}%
\AgdaBound{M′}\<%
\\
\>[0][@{}l@{\AgdaIndent{0}}]%
\>[3]\AgdaSymbol{→}\AgdaSpace{}%
\AgdaBound{𝒫}\AgdaSpace{}%
\AgdaOperator{\AgdaFunction{⊢ᵒ}}\AgdaSpace{}%
\AgdaSymbol{(}\AgdaBound{dir}\AgdaSpace{}%
\AgdaOperator{\AgdaFunction{∣}}\AgdaSpace{}%
\AgdaBound{M}\AgdaSpace{}%
\AgdaOperator{\AgdaFunction{⊑ᴸᴿᵥ}}\AgdaSpace{}%
\AgdaBound{M′}\AgdaSpace{}%
\AgdaOperator{\AgdaFunction{⦂}}\AgdaSpace{}%
\AgdaBound{A⊑A′}\AgdaSymbol{)}\AgdaSpace{}%
\AgdaSymbol{→}\AgdaSpace{}%
\AgdaBound{𝒫}\AgdaSpace{}%
\AgdaOperator{\AgdaFunction{⊢ᵒ}}\AgdaSpace{}%
\AgdaSymbol{(}\AgdaDatatype{Value}\AgdaSpace{}%
\AgdaBound{M}\AgdaSymbol{)}\AgdaOperator{\AgdaFunction{ᵒ}}\AgdaSpace{}%
\AgdaOperator{\AgdaFunction{×ᵒ}}\AgdaSpace{}%
\AgdaSymbol{(}\AgdaDatatype{Value}\AgdaSpace{}%
\AgdaBound{M′}\AgdaSymbol{)}\AgdaOperator{\AgdaFunction{ᵒ}}\<%
\end{code}
\begin{code}[hide]%
\>[0]\AgdaFunction{LRᵥ⇒Valueᵒ}\AgdaSpace{}%
\AgdaBound{A⊑A′}\AgdaSpace{}%
\AgdaBound{M}\AgdaSpace{}%
\AgdaBound{M′}\AgdaSpace{}%
\AgdaBound{M⊑M′}\AgdaSpace{}%
\AgdaSymbol{=}\<%
\\
\>[0][@{}l@{\AgdaIndent{0}}]%
\>[4]\AgdaFunction{⊢ᵒ-intro}\AgdaSpace{}%
\AgdaSymbol{λ}\AgdaSpace{}%
\AgdaSymbol{\{}%
\>[1436I]\AgdaInductiveConstructor{zero}\AgdaSpace{}%
\AgdaBound{𝒫k}\AgdaSpace{}%
\AgdaSymbol{→}\AgdaSpace{}%
\AgdaInductiveConstructor{tt}\AgdaSpace{}%
\AgdaOperator{\AgdaInductiveConstructor{,}}\AgdaSpace{}%
\AgdaInductiveConstructor{tt}\AgdaSpace{}%
\AgdaSymbol{;}\<%
\\
\>[.][@{}l@{}]\<[1436I]%
\>[17]\AgdaSymbol{(}\AgdaInductiveConstructor{suc}\AgdaSpace{}%
\AgdaBound{k}\AgdaSymbol{)}\AgdaSpace{}%
\AgdaBound{𝒫k}\AgdaSpace{}%
\AgdaSymbol{→}\AgdaSpace{}%
\AgdaFunction{LRᵥ⇒Value}\AgdaSpace{}%
\AgdaBound{A⊑A′}\AgdaSpace{}%
\AgdaBound{M}\AgdaSpace{}%
\AgdaBound{M′}\AgdaSpace{}%
\AgdaSymbol{(}\AgdaFunction{⊢ᵒ-elim}\AgdaSpace{}%
\AgdaBound{M⊑M′}\AgdaSpace{}%
\AgdaSymbol{(}\AgdaInductiveConstructor{suc}\AgdaSpace{}%
\AgdaBound{k}\AgdaSymbol{)}\AgdaSpace{}%
\AgdaBound{𝒫k}\AgdaSymbol{)\}}\<%
\end{code}

If two values are related via ⊑ᴸᴿᵥ, then they are also related via
⊑ᴸᴿₜ.

\begin{code}[hide]%
\>[0]\AgdaFunction{LRᵥ⇒LRₜ-step}\AgdaSpace{}%
\AgdaSymbol{:}\AgdaSpace{}%
\AgdaSymbol{∀\{}\AgdaBound{A}\AgdaSymbol{\}\{}\AgdaBound{A′}\AgdaSymbol{\}\{}\AgdaBound{A⊑A′}\AgdaSpace{}%
\AgdaSymbol{:}\AgdaSpace{}%
\AgdaBound{A}\AgdaSpace{}%
\AgdaOperator{\AgdaDatatype{⊑}}\AgdaSpace{}%
\AgdaBound{A′}\AgdaSymbol{\}\{}\AgdaBound{V}\AgdaSpace{}%
\AgdaBound{V′}\AgdaSymbol{\}\{}\AgdaBound{dir}\AgdaSymbol{\}\{}\AgdaBound{k}\AgdaSymbol{\}}\<%
\\
\>[0][@{}l@{\AgdaIndent{0}}]%
\>[3]\AgdaSymbol{→}\AgdaSpace{}%
\AgdaField{\#}\AgdaSymbol{(}\AgdaBound{dir}\AgdaSpace{}%
\AgdaOperator{\AgdaFunction{∣}}\AgdaSpace{}%
\AgdaBound{V}\AgdaSpace{}%
\AgdaOperator{\AgdaFunction{⊑ᴸᴿᵥ}}\AgdaSpace{}%
\AgdaBound{V′}\AgdaSpace{}%
\AgdaOperator{\AgdaFunction{⦂}}\AgdaSpace{}%
\AgdaBound{A⊑A′}\AgdaSymbol{)}\AgdaSpace{}%
\AgdaBound{k}%
\>[34]\AgdaSymbol{→}%
\>[37]\AgdaField{\#}\AgdaSymbol{(}\AgdaBound{dir}\AgdaSpace{}%
\AgdaOperator{\AgdaFunction{∣}}\AgdaSpace{}%
\AgdaBound{V}\AgdaSpace{}%
\AgdaOperator{\AgdaFunction{⊑ᴸᴿₜ}}\AgdaSpace{}%
\AgdaBound{V′}\AgdaSpace{}%
\AgdaOperator{\AgdaFunction{⦂}}\AgdaSpace{}%
\AgdaBound{A⊑A′}\AgdaSymbol{)}\AgdaSpace{}%
\AgdaBound{k}\<%
\\
\>[0]\AgdaFunction{LRᵥ⇒LRₜ-step}\AgdaSpace{}%
\AgdaSymbol{\{}\AgdaBound{A}\AgdaSymbol{\}\{}\AgdaBound{A′}\AgdaSymbol{\}\{}\AgdaBound{A⊑A′}\AgdaSymbol{\}\{}\AgdaBound{V}\AgdaSymbol{\}}\AgdaSpace{}%
\AgdaSymbol{\{}\AgdaBound{V′}\AgdaSymbol{\}}\AgdaSpace{}%
\AgdaSymbol{\{}\AgdaBound{dir}\AgdaSymbol{\}}\AgdaSpace{}%
\AgdaSymbol{\{}\AgdaInductiveConstructor{zero}\AgdaSymbol{\}}\AgdaSpace{}%
\AgdaBound{𝒱VV′k}\AgdaSpace{}%
\AgdaSymbol{=}\<%
\\
\>[0][@{}l@{\AgdaIndent{0}}]%
\>[3]\AgdaField{tz}\AgdaSpace{}%
\AgdaSymbol{(}\AgdaBound{dir}\AgdaSpace{}%
\AgdaOperator{\AgdaFunction{∣}}\AgdaSpace{}%
\AgdaBound{V}\AgdaSpace{}%
\AgdaOperator{\AgdaFunction{⊑ᴸᴿₜ}}\AgdaSpace{}%
\AgdaBound{V′}\AgdaSpace{}%
\AgdaOperator{\AgdaFunction{⦂}}\AgdaSpace{}%
\AgdaBound{A⊑A′}\AgdaSymbol{)}\<%
\\
\>[0]\AgdaFunction{LRᵥ⇒LRₜ-step}\AgdaSpace{}%
\AgdaSymbol{\{}\AgdaBound{A}\AgdaSymbol{\}\{}\AgdaBound{A′}\AgdaSymbol{\}\{}\AgdaBound{A⊑A′}\AgdaSymbol{\}\{}\AgdaBound{V}\AgdaSymbol{\}}\AgdaSpace{}%
\AgdaSymbol{\{}\AgdaBound{V′}\AgdaSymbol{\}}\AgdaSpace{}%
\AgdaSymbol{\{}\AgdaInductiveConstructor{≼}\AgdaSymbol{\}}\AgdaSpace{}%
\AgdaSymbol{\{}\AgdaInductiveConstructor{suc}\AgdaSpace{}%
\AgdaBound{k}\AgdaSymbol{\}}\AgdaSpace{}%
\AgdaBound{𝒱VV′sk}\AgdaSpace{}%
\AgdaSymbol{=}\<%
\\
\>[0][@{}l@{\AgdaIndent{0}}]%
\>[2]\AgdaFunction{⇔-fro}\AgdaSpace{}%
\AgdaSymbol{(}\AgdaFunction{LRₜ-suc}\AgdaSymbol{\{}\AgdaArgument{dir}\AgdaSpace{}%
\AgdaSymbol{=}\AgdaSpace{}%
\AgdaInductiveConstructor{≼}\AgdaSymbol{\})}\<%
\\
\>[2]\AgdaSymbol{(}\AgdaKeyword{let}\AgdaSpace{}%
\AgdaSymbol{(}\AgdaBound{v}\AgdaSpace{}%
\AgdaOperator{\AgdaInductiveConstructor{,}}\AgdaSpace{}%
\AgdaBound{v′}\AgdaSymbol{)}\AgdaSpace{}%
\AgdaSymbol{=}\AgdaSpace{}%
\AgdaFunction{LRᵥ⇒Value}\AgdaSpace{}%
\AgdaBound{A⊑A′}\AgdaSpace{}%
\AgdaBound{V}\AgdaSpace{}%
\AgdaBound{V′}\AgdaSpace{}%
\AgdaBound{𝒱VV′sk}\AgdaSpace{}%
\AgdaKeyword{in}\<%
\\
\>[2]\AgdaSymbol{(}\AgdaInductiveConstructor{inj₂}\AgdaSpace{}%
\AgdaSymbol{(}\AgdaInductiveConstructor{inj₂}\AgdaSpace{}%
\AgdaSymbol{(}\AgdaBound{v}\AgdaSpace{}%
\AgdaOperator{\AgdaInductiveConstructor{,}}\AgdaSpace{}%
\AgdaSymbol{(}\AgdaBound{V′}\AgdaSpace{}%
\AgdaOperator{\AgdaInductiveConstructor{,}}\AgdaSpace{}%
\AgdaSymbol{(}\AgdaBound{V′}\AgdaSpace{}%
\AgdaOperator{\AgdaInductiveConstructor{END}}\AgdaSymbol{)}\AgdaSpace{}%
\AgdaOperator{\AgdaInductiveConstructor{,}}\AgdaSpace{}%
\AgdaBound{v′}\AgdaSpace{}%
\AgdaOperator{\AgdaInductiveConstructor{,}}\AgdaSpace{}%
\AgdaBound{𝒱VV′sk}\AgdaSymbol{)))))}\<%
\\
\>[0]\AgdaFunction{LRᵥ⇒LRₜ-step}\AgdaSpace{}%
\AgdaSymbol{\{}\AgdaBound{A}\AgdaSymbol{\}\{}\AgdaBound{A′}\AgdaSymbol{\}\{}\AgdaBound{A⊑A′}\AgdaSymbol{\}\{}\AgdaBound{V}\AgdaSymbol{\}}\AgdaSpace{}%
\AgdaSymbol{\{}\AgdaBound{V′}\AgdaSymbol{\}}\AgdaSpace{}%
\AgdaSymbol{\{}\AgdaInductiveConstructor{≽}\AgdaSymbol{\}}\AgdaSpace{}%
\AgdaSymbol{\{}\AgdaInductiveConstructor{suc}\AgdaSpace{}%
\AgdaBound{k}\AgdaSymbol{\}}\AgdaSpace{}%
\AgdaBound{𝒱VV′sk}\AgdaSpace{}%
\AgdaSymbol{=}\<%
\\
\>[0][@{}l@{\AgdaIndent{0}}]%
\>[2]\AgdaFunction{⇔-fro}\AgdaSpace{}%
\AgdaSymbol{(}\AgdaFunction{LRₜ-suc}\AgdaSymbol{\{}\AgdaArgument{dir}\AgdaSpace{}%
\AgdaSymbol{=}\AgdaSpace{}%
\AgdaInductiveConstructor{≽}\AgdaSymbol{\})}\<%
\\
\>[2]\AgdaSymbol{(}\AgdaKeyword{let}\AgdaSpace{}%
\AgdaSymbol{(}\AgdaBound{v}\AgdaSpace{}%
\AgdaOperator{\AgdaInductiveConstructor{,}}\AgdaSpace{}%
\AgdaBound{v′}\AgdaSymbol{)}\AgdaSpace{}%
\AgdaSymbol{=}\AgdaSpace{}%
\AgdaFunction{LRᵥ⇒Value}\AgdaSpace{}%
\AgdaBound{A⊑A′}\AgdaSpace{}%
\AgdaBound{V}\AgdaSpace{}%
\AgdaBound{V′}\AgdaSpace{}%
\AgdaBound{𝒱VV′sk}\AgdaSpace{}%
\AgdaKeyword{in}\<%
\\
\>[2]\AgdaInductiveConstructor{inj₂}\AgdaSpace{}%
\AgdaSymbol{(}\AgdaInductiveConstructor{inj₂}\AgdaSpace{}%
\AgdaSymbol{(}\AgdaBound{v′}\AgdaSpace{}%
\AgdaOperator{\AgdaInductiveConstructor{,}}\AgdaSpace{}%
\AgdaBound{V}\AgdaSpace{}%
\AgdaOperator{\AgdaInductiveConstructor{,}}\AgdaSpace{}%
\AgdaSymbol{(}\AgdaBound{V}\AgdaSpace{}%
\AgdaOperator{\AgdaInductiveConstructor{END}}\AgdaSymbol{)}\AgdaSpace{}%
\AgdaOperator{\AgdaInductiveConstructor{,}}\AgdaSpace{}%
\AgdaBound{v}\AgdaSpace{}%
\AgdaOperator{\AgdaInductiveConstructor{,}}\AgdaSpace{}%
\AgdaBound{𝒱VV′sk}\AgdaSymbol{)))}\<%
\end{code}
\begin{code}%
\>[0]\AgdaFunction{LRᵥ⇒LRₜ}\AgdaSpace{}%
\AgdaSymbol{:}\AgdaSpace{}%
\AgdaSymbol{∀\{}\AgdaBound{A}\AgdaSymbol{\}\{}\AgdaBound{A′}\AgdaSymbol{\}\{}\AgdaBound{A⊑A′}\AgdaSpace{}%
\AgdaSymbol{:}\AgdaSpace{}%
\AgdaBound{A}\AgdaSpace{}%
\AgdaOperator{\AgdaDatatype{⊑}}\AgdaSpace{}%
\AgdaBound{A′}\AgdaSymbol{\}\{}\AgdaBound{𝒫}\AgdaSymbol{\}\{}\AgdaBound{V}\AgdaSpace{}%
\AgdaBound{V′}\AgdaSymbol{\}\{}\AgdaBound{dir}\AgdaSymbol{\}}\<%
\\
\>[0][@{}l@{\AgdaIndent{0}}]%
\>[3]\AgdaSymbol{→}\AgdaSpace{}%
\AgdaBound{𝒫}\AgdaSpace{}%
\AgdaOperator{\AgdaFunction{⊢ᵒ}}\AgdaSpace{}%
\AgdaBound{dir}\AgdaSpace{}%
\AgdaOperator{\AgdaFunction{∣}}\AgdaSpace{}%
\AgdaBound{V}\AgdaSpace{}%
\AgdaOperator{\AgdaFunction{⊑ᴸᴿᵥ}}\AgdaSpace{}%
\AgdaBound{V′}\AgdaSpace{}%
\AgdaOperator{\AgdaFunction{⦂}}\AgdaSpace{}%
\AgdaBound{A⊑A′}%
\>[34]\AgdaSymbol{→}%
\>[37]\AgdaBound{𝒫}\AgdaSpace{}%
\AgdaOperator{\AgdaFunction{⊢ᵒ}}\AgdaSpace{}%
\AgdaBound{dir}\AgdaSpace{}%
\AgdaOperator{\AgdaFunction{∣}}\AgdaSpace{}%
\AgdaBound{V}\AgdaSpace{}%
\AgdaOperator{\AgdaFunction{⊑ᴸᴿₜ}}\AgdaSpace{}%
\AgdaBound{V′}\AgdaSpace{}%
\AgdaOperator{\AgdaFunction{⦂}}\AgdaSpace{}%
\AgdaBound{A⊑A′}\<%
\end{code}
\begin{code}[hide]%
\>[0]\AgdaFunction{LRᵥ⇒LRₜ}\AgdaSpace{}%
\AgdaSymbol{\{}\AgdaBound{A}\AgdaSymbol{\}\{}\AgdaBound{A′}\AgdaSymbol{\}\{}\AgdaBound{A⊑A′}\AgdaSymbol{\}\{}\AgdaBound{𝒫}\AgdaSymbol{\}\{}\AgdaBound{V}\AgdaSymbol{\}\{}\AgdaBound{V′}\AgdaSymbol{\}\{}\AgdaBound{dir}\AgdaSymbol{\}}\AgdaSpace{}%
\AgdaBound{⊢𝒱VV′}\AgdaSpace{}%
\AgdaSymbol{=}\AgdaSpace{}%
\AgdaFunction{⊢ᵒ-intro}\AgdaSpace{}%
\AgdaSymbol{λ}\AgdaSpace{}%
\AgdaBound{k}\AgdaSpace{}%
\AgdaBound{𝒫k}\AgdaSpace{}%
\AgdaSymbol{→}\<%
\\
\>[0][@{}l@{\AgdaIndent{0}}]%
\>[2]\AgdaFunction{LRᵥ⇒LRₜ-step}\AgdaSymbol{\{}\AgdaArgument{V}\AgdaSpace{}%
\AgdaSymbol{=}\AgdaSpace{}%
\AgdaBound{V}\AgdaSymbol{\}\{}\AgdaBound{V′}\AgdaSymbol{\}\{}\AgdaBound{dir}\AgdaSymbol{\}\{}\AgdaBound{k}\AgdaSymbol{\}}\AgdaSpace{}%
\AgdaSymbol{(}\AgdaFunction{⊢ᵒ-elim}\AgdaSpace{}%
\AgdaBound{⊢𝒱VV′}\AgdaSpace{}%
\AgdaBound{k}\AgdaSpace{}%
\AgdaBound{𝒫k}\AgdaSymbol{)}\<%
\end{code}


\begin{code}[hide]%
\>[0]\AgdaSymbol{\{-\#}\AgdaSpace{}%
\AgdaKeyword{OPTIONS}\AgdaSpace{}%
\AgdaPragma{--rewriting}\AgdaSpace{}%
\AgdaSymbol{\#-\}}\<%
\\
\>[0]\AgdaKeyword{module}\AgdaSpace{}%
\AgdaModule{LogRel.PeterFundamental}\AgdaSpace{}%
\AgdaKeyword{where}\<%
\\
\\[\AgdaEmptyExtraSkip]%
\>[0]\AgdaKeyword{open}\AgdaSpace{}%
\AgdaKeyword{import}\AgdaSpace{}%
\AgdaModule{Data.Empty}\AgdaSpace{}%
\AgdaKeyword{using}\AgdaSpace{}%
\AgdaSymbol{(}\AgdaDatatype{⊥}\AgdaSymbol{;}\AgdaSpace{}%
\AgdaFunction{⊥-elim}\AgdaSymbol{)}\<%
\\
\>[0]\AgdaKeyword{open}\AgdaSpace{}%
\AgdaKeyword{import}\AgdaSpace{}%
\AgdaModule{Data.List}\AgdaSpace{}%
\AgdaKeyword{using}\AgdaSpace{}%
\AgdaSymbol{(}\AgdaDatatype{List}\AgdaSymbol{;}\AgdaSpace{}%
\AgdaInductiveConstructor{[]}\AgdaSymbol{;}\AgdaSpace{}%
\AgdaOperator{\AgdaInductiveConstructor{\AgdaUnderscore{}∷\AgdaUnderscore{}}}\AgdaSymbol{;}\AgdaSpace{}%
\AgdaFunction{map}\AgdaSymbol{;}\AgdaSpace{}%
\AgdaFunction{length}\AgdaSymbol{)}\<%
\\
\>[0]\AgdaKeyword{open}\AgdaSpace{}%
\AgdaKeyword{import}\AgdaSpace{}%
\AgdaModule{Data.Nat}\<%
\\
\>[0]\AgdaKeyword{open}\AgdaSpace{}%
\AgdaKeyword{import}\AgdaSpace{}%
\AgdaModule{Data.Nat.Properties}\<%
\\
\>[0]\AgdaKeyword{open}\AgdaSpace{}%
\AgdaKeyword{import}\AgdaSpace{}%
\AgdaModule{Data.Bool}\AgdaSpace{}%
\AgdaKeyword{using}\AgdaSpace{}%
\AgdaSymbol{(}\AgdaInductiveConstructor{true}\AgdaSymbol{;}\AgdaSpace{}%
\AgdaInductiveConstructor{false}\AgdaSymbol{)}\AgdaSpace{}%
\AgdaKeyword{renaming}\AgdaSpace{}%
\AgdaSymbol{(}\AgdaDatatype{Bool}\AgdaSpace{}%
\AgdaSymbol{to}\AgdaSpace{}%
\AgdaDatatype{𝔹}\AgdaSymbol{)}\<%
\\
\>[0]\AgdaKeyword{open}\AgdaSpace{}%
\AgdaKeyword{import}\AgdaSpace{}%
\AgdaModule{Data.Product}\AgdaSpace{}%
\AgdaKeyword{using}\AgdaSpace{}%
\AgdaSymbol{(}\AgdaOperator{\AgdaInductiveConstructor{\AgdaUnderscore{},\AgdaUnderscore{}}}\AgdaSymbol{;}\AgdaOperator{\AgdaFunction{\AgdaUnderscore{}×\AgdaUnderscore{}}}\AgdaSymbol{;}\AgdaSpace{}%
\AgdaField{proj₁}\AgdaSymbol{;}\AgdaSpace{}%
\AgdaField{proj₂}\AgdaSymbol{;}\AgdaSpace{}%
\AgdaFunction{Σ-syntax}\AgdaSymbol{;}\AgdaSpace{}%
\AgdaFunction{∃-syntax}\AgdaSymbol{)}\<%
\\
\>[0]\AgdaKeyword{open}\AgdaSpace{}%
\AgdaKeyword{import}\AgdaSpace{}%
\AgdaModule{Data.Sum}\AgdaSpace{}%
\AgdaKeyword{using}\AgdaSpace{}%
\AgdaSymbol{(}\AgdaOperator{\AgdaDatatype{\AgdaUnderscore{}⊎\AgdaUnderscore{}}}\AgdaSymbol{;}\AgdaSpace{}%
\AgdaInductiveConstructor{inj₁}\AgdaSymbol{;}\AgdaSpace{}%
\AgdaInductiveConstructor{inj₂}\AgdaSymbol{)}\<%
\\
\>[0]\AgdaKeyword{open}\AgdaSpace{}%
\AgdaKeyword{import}\AgdaSpace{}%
\AgdaModule{Data.Unit}\AgdaSpace{}%
\AgdaKeyword{using}\AgdaSpace{}%
\AgdaSymbol{(}\AgdaRecord{⊤}\AgdaSymbol{;}\AgdaSpace{}%
\AgdaInductiveConstructor{tt}\AgdaSymbol{)}\<%
\\
\>[0]\AgdaKeyword{open}\AgdaSpace{}%
\AgdaKeyword{import}\AgdaSpace{}%
\AgdaModule{Data.Unit.Polymorphic}\AgdaSpace{}%
\AgdaKeyword{renaming}\AgdaSpace{}%
\AgdaSymbol{(}\AgdaFunction{⊤}\AgdaSpace{}%
\AgdaSymbol{to}\AgdaSpace{}%
\AgdaFunction{topᵖ}\AgdaSymbol{;}\AgdaSpace{}%
\AgdaFunction{tt}\AgdaSpace{}%
\AgdaSymbol{to}\AgdaSpace{}%
\AgdaFunction{ttᵖ}\AgdaSymbol{)}\<%
\\
\>[0]\AgdaKeyword{open}\AgdaSpace{}%
\AgdaKeyword{import}\AgdaSpace{}%
\AgdaModule{Relation.Binary.PropositionalEquality}\AgdaSpace{}%
\AgdaSymbol{as}\AgdaSpace{}%
\AgdaModule{Eq}\<%
\\
\>[0][@{}l@{\AgdaIndent{0}}]%
\>[2]\AgdaKeyword{using}\AgdaSpace{}%
\AgdaSymbol{(}\AgdaOperator{\AgdaDatatype{\AgdaUnderscore{}≡\AgdaUnderscore{}}}\AgdaSymbol{;}\AgdaSpace{}%
\AgdaOperator{\AgdaFunction{\AgdaUnderscore{}≢\AgdaUnderscore{}}}\AgdaSymbol{;}\AgdaSpace{}%
\AgdaInductiveConstructor{refl}\AgdaSymbol{;}\AgdaSpace{}%
\AgdaFunction{sym}\AgdaSymbol{;}\AgdaSpace{}%
\AgdaFunction{cong}\AgdaSymbol{;}\AgdaSpace{}%
\AgdaFunction{subst}\AgdaSymbol{;}\AgdaSpace{}%
\AgdaFunction{trans}\AgdaSymbol{)}\<%
\\
\>[0]\AgdaKeyword{open}\AgdaSpace{}%
\AgdaKeyword{import}\AgdaSpace{}%
\AgdaModule{Relation.Nullary}\AgdaSpace{}%
\AgdaKeyword{using}\AgdaSpace{}%
\AgdaSymbol{(}\AgdaOperator{\AgdaFunction{¬\AgdaUnderscore{}}}\AgdaSymbol{;}\AgdaSpace{}%
\AgdaRecord{Dec}\AgdaSymbol{;}\AgdaSpace{}%
\AgdaInductiveConstructor{yes}\AgdaSymbol{;}\AgdaSpace{}%
\AgdaInductiveConstructor{no}\AgdaSymbol{)}\<%
\\
\\[\AgdaEmptyExtraSkip]%
\>[0]\AgdaKeyword{open}\AgdaSpace{}%
\AgdaKeyword{import}\AgdaSpace{}%
\AgdaModule{Var}\<%
\\
\>[0]\AgdaKeyword{open}\AgdaSpace{}%
\AgdaKeyword{import}\AgdaSpace{}%
\AgdaModule{Sig}\<%
\\
\>[0]\AgdaKeyword{open}\AgdaSpace{}%
\AgdaKeyword{import}\AgdaSpace{}%
\AgdaModule{LogRel.PeterCastCalculus}\<%
\\
\>[0]\AgdaKeyword{open}\AgdaSpace{}%
\AgdaKeyword{import}\AgdaSpace{}%
\AgdaModule{LogRel.PeterPrecision}\<%
\\
\>[0]\AgdaKeyword{open}\AgdaSpace{}%
\AgdaKeyword{import}\AgdaSpace{}%
\AgdaModule{LogRel.PeterLogRel}\<%
\\
\>[0]\AgdaKeyword{open}\AgdaSpace{}%
\AgdaKeyword{import}\AgdaSpace{}%
\AgdaModule{StepIndexedLogic}\<%
\end{code}

\section{Fundamental Theorem of the Logical Relation}
\label{sec:fundamental}

The fundamental theorem of the logical relation states that if two
terms are related by precision, then they are in the logical relation.
The fundamental theorem is proved by induction on the term precision
relation. Each case of that proof is split out into a separate lemma,
which by tradition are called Compatibility Lemmas.

\paragraph{Compatibility for Literals, Blame, and Variables}

The proof of compatibility for literals uses the LRᵥ⇒LRₜ lemma.

\begin{code}[hide]%
\>[0]\AgdaFunction{LRᵥ-base-intro}\AgdaSpace{}%
\AgdaSymbol{:}\AgdaSpace{}%
\AgdaSymbol{∀\{}\AgdaBound{𝒫}\AgdaSymbol{\}\{}\AgdaBound{ι}\AgdaSymbol{\}\{}\AgdaBound{c}\AgdaSymbol{\}\{}\AgdaBound{dir}\AgdaSymbol{\}}\AgdaSpace{}%
\AgdaSymbol{→}\AgdaSpace{}%
\AgdaBound{𝒫}\AgdaSpace{}%
\AgdaOperator{\AgdaFunction{⊢ᵒ}}\AgdaSpace{}%
\AgdaBound{dir}\AgdaSpace{}%
\AgdaOperator{\AgdaFunction{∣}}\AgdaSpace{}%
\AgdaSymbol{(}\AgdaInductiveConstructor{\$}\AgdaSpace{}%
\AgdaBound{c}\AgdaSymbol{)}\AgdaSpace{}%
\AgdaOperator{\AgdaFunction{⊑ᴸᴿᵥ}}\AgdaSpace{}%
\AgdaSymbol{(}\AgdaInductiveConstructor{\$}\AgdaSpace{}%
\AgdaBound{c}\AgdaSymbol{)}\AgdaSpace{}%
\AgdaOperator{\AgdaFunction{⦂}}\AgdaSpace{}%
\AgdaInductiveConstructor{base⊑}\AgdaSymbol{\{}\AgdaBound{ι}\AgdaSymbol{\}}\<%
\\
\>[0]\AgdaFunction{LRᵥ-base-intro}\AgdaSymbol{\{}\AgdaBound{𝒫}\AgdaSymbol{\}\{}\AgdaBound{ι}\AgdaSymbol{\}\{}\AgdaBound{c}\AgdaSymbol{\}\{}\AgdaBound{dir}\AgdaSymbol{\}}\AgdaSpace{}%
\AgdaSymbol{=}\AgdaSpace{}%
\AgdaFunction{⊢ᵒ-intro}\AgdaSpace{}%
\AgdaSymbol{λ}\AgdaSpace{}%
\AgdaBound{k}\AgdaSpace{}%
\AgdaBound{𝒫k}\AgdaSpace{}%
\AgdaSymbol{→}\AgdaSpace{}%
\AgdaFunction{step}\AgdaSpace{}%
\AgdaSymbol{\{}\AgdaBound{ι}\AgdaSymbol{\}\{}\AgdaBound{dir}\AgdaSymbol{\}\{}\AgdaBound{c}\AgdaSymbol{\}\{}\AgdaBound{k}\AgdaSymbol{\}}\<%
\\
\>[0][@{}l@{\AgdaIndent{0}}]%
\>[2]\AgdaKeyword{where}\<%
\\
\>[2]\AgdaFunction{step}\AgdaSpace{}%
\AgdaSymbol{:}\AgdaSpace{}%
\AgdaSymbol{∀\{}\AgdaBound{ι}\AgdaSymbol{\}\{}\AgdaBound{dir}\AgdaSymbol{\}\{}\AgdaBound{c}\AgdaSymbol{\}\{}\AgdaBound{k}\AgdaSymbol{\}}\AgdaSpace{}%
\AgdaSymbol{→}\AgdaSpace{}%
\AgdaField{\#}\AgdaSpace{}%
\AgdaSymbol{(}\AgdaBound{dir}\AgdaSpace{}%
\AgdaOperator{\AgdaFunction{∣}}\AgdaSpace{}%
\AgdaSymbol{(}\AgdaInductiveConstructor{\$}\AgdaSpace{}%
\AgdaBound{c}\AgdaSymbol{)}\AgdaSpace{}%
\AgdaOperator{\AgdaFunction{⊑ᴸᴿᵥ}}\AgdaSpace{}%
\AgdaSymbol{(}\AgdaInductiveConstructor{\$}\AgdaSpace{}%
\AgdaBound{c}\AgdaSymbol{)}\AgdaSpace{}%
\AgdaOperator{\AgdaFunction{⦂}}\AgdaSpace{}%
\AgdaInductiveConstructor{base⊑}\AgdaSymbol{\{}\AgdaBound{ι}\AgdaSymbol{\})}\AgdaSpace{}%
\AgdaBound{k}\<%
\\
\>[2]\AgdaFunction{step}\AgdaSpace{}%
\AgdaSymbol{\{}\AgdaBound{ι}\AgdaSymbol{\}}\AgdaSpace{}%
\AgdaSymbol{\{}\AgdaBound{dir}\AgdaSymbol{\}}\AgdaSpace{}%
\AgdaSymbol{\{}\AgdaBound{c}\AgdaSymbol{\}}\AgdaSpace{}%
\AgdaSymbol{\{}\AgdaInductiveConstructor{zero}\AgdaSymbol{\}}\AgdaSpace{}%
\AgdaSymbol{=}\AgdaSpace{}%
\AgdaInductiveConstructor{tt}\<%
\\
\>[2]\AgdaFunction{step}\AgdaSpace{}%
\AgdaSymbol{\{}\AgdaBound{ι}\AgdaSymbol{\}}\AgdaSpace{}%
\AgdaSymbol{\{}\AgdaBound{dir}\AgdaSymbol{\}}\AgdaSpace{}%
\AgdaSymbol{\{}\AgdaBound{c}\AgdaSymbol{\}}\AgdaSpace{}%
\AgdaSymbol{\{}\AgdaInductiveConstructor{suc}\AgdaSpace{}%
\AgdaBound{k}\AgdaSymbol{\}}\AgdaSpace{}%
\AgdaSymbol{=}\AgdaSpace{}%
\AgdaInductiveConstructor{refl}\<%
\end{code}

\begin{code}%
\>[0]\AgdaFunction{compatible-literal}\AgdaSpace{}%
\AgdaSymbol{:}\AgdaSpace{}%
\AgdaSymbol{∀\{}\AgdaBound{Γ}\AgdaSymbol{\}\{}\AgdaBound{c}\AgdaSymbol{\}\{}\AgdaBound{ι}\AgdaSymbol{\}}\AgdaSpace{}%
\AgdaSymbol{→}\AgdaSpace{}%
\AgdaBound{Γ}\AgdaSpace{}%
\AgdaOperator{\AgdaFunction{⊨}}\AgdaSpace{}%
\AgdaInductiveConstructor{\$}\AgdaSpace{}%
\AgdaBound{c}\AgdaSpace{}%
\AgdaOperator{\AgdaFunction{⊑ᴸᴿ}}\AgdaSpace{}%
\AgdaInductiveConstructor{\$}\AgdaSpace{}%
\AgdaBound{c}\AgdaSpace{}%
\AgdaOperator{\AgdaFunction{⦂}}\AgdaSpace{}%
\AgdaSymbol{(}\AgdaOperator{\AgdaInductiveConstructor{\$ₜ}}\AgdaSpace{}%
\AgdaBound{ι}\AgdaSpace{}%
\AgdaOperator{\AgdaInductiveConstructor{,}}\AgdaSpace{}%
\AgdaOperator{\AgdaInductiveConstructor{\$ₜ}}\AgdaSpace{}%
\AgdaBound{ι}\AgdaSpace{}%
\AgdaOperator{\AgdaInductiveConstructor{,}}\AgdaSpace{}%
\AgdaInductiveConstructor{base⊑}\AgdaSymbol{)}\<%
\end{code}
\begin{code}[hide]%
\>[0]\AgdaFunction{compatible-literal}\AgdaSpace{}%
\AgdaSymbol{\{}\AgdaBound{Γ}\AgdaSymbol{\}\{}\AgdaBound{c}\AgdaSymbol{\}\{}\AgdaBound{ι}\AgdaSymbol{\}}\AgdaSpace{}%
\AgdaSymbol{=}\<%
\\
\>[0][@{}l@{\AgdaIndent{0}}]%
\>[2]\AgdaSymbol{(λ}\AgdaSpace{}%
\AgdaBound{γ}\AgdaSpace{}%
\AgdaBound{γ′}\AgdaSpace{}%
\AgdaSymbol{→}\AgdaSpace{}%
\AgdaFunction{LRᵥ⇒LRₜ}\AgdaSpace{}%
\AgdaFunction{LRᵥ-base-intro}\AgdaSymbol{)}\AgdaSpace{}%
\AgdaOperator{\AgdaInductiveConstructor{,}}\AgdaSpace{}%
\AgdaSymbol{(λ}\AgdaSpace{}%
\AgdaBound{γ}\AgdaSpace{}%
\AgdaBound{γ′}\AgdaSpace{}%
\AgdaSymbol{→}\AgdaSpace{}%
\AgdaFunction{LRᵥ⇒LRₜ}\AgdaSpace{}%
\AgdaFunction{LRᵥ-base-intro}\AgdaSymbol{)}\<%
\end{code}

\noindent The proof of compatibility for blame is direct from the definitions.

\begin{code}[hide]%
\>[0]\AgdaFunction{LRₜ-blame-step}\AgdaSpace{}%
\AgdaSymbol{:}\AgdaSpace{}%
\AgdaSymbol{∀\{}\AgdaBound{A}\AgdaSymbol{\}\{}\AgdaBound{A′}\AgdaSymbol{\}\{}\AgdaBound{A⊑A′}\AgdaSpace{}%
\AgdaSymbol{:}\AgdaSpace{}%
\AgdaBound{A}\AgdaSpace{}%
\AgdaOperator{\AgdaDatatype{⊑}}\AgdaSpace{}%
\AgdaBound{A′}\AgdaSymbol{\}\{}\AgdaBound{dir}\AgdaSymbol{\}\{}\AgdaBound{M}\AgdaSymbol{\}\{}\AgdaBound{k}\AgdaSymbol{\}}\<%
\\
\>[0][@{}l@{\AgdaIndent{0}}]%
\>[3]\AgdaSymbol{→}\AgdaSpace{}%
\AgdaField{\#}\AgdaSymbol{(}\AgdaBound{dir}\AgdaSpace{}%
\AgdaOperator{\AgdaFunction{∣}}\AgdaSpace{}%
\AgdaBound{M}\AgdaSpace{}%
\AgdaOperator{\AgdaFunction{⊑ᴸᴿₜ}}\AgdaSpace{}%
\AgdaInductiveConstructor{blame}\AgdaSpace{}%
\AgdaOperator{\AgdaFunction{⦂}}\AgdaSpace{}%
\AgdaBound{A⊑A′}\AgdaSymbol{)}\AgdaSpace{}%
\AgdaBound{k}\<%
\\
\>[0]\AgdaFunction{LRₜ-blame-step}\AgdaSpace{}%
\AgdaSymbol{\{}\AgdaBound{A}\AgdaSymbol{\}\{}\AgdaBound{A′}\AgdaSymbol{\}\{}\AgdaBound{A⊑A′}\AgdaSymbol{\}\{}\AgdaBound{dir}\AgdaSymbol{\}}\AgdaSpace{}%
\AgdaSymbol{\{}\AgdaBound{M}\AgdaSymbol{\}}\AgdaSpace{}%
\AgdaSymbol{\{}\AgdaInductiveConstructor{zero}\AgdaSymbol{\}}\AgdaSpace{}%
\AgdaSymbol{=}\AgdaSpace{}%
\AgdaField{tz}\AgdaSpace{}%
\AgdaSymbol{(}\AgdaBound{dir}\AgdaSpace{}%
\AgdaOperator{\AgdaFunction{∣}}\AgdaSpace{}%
\AgdaBound{M}\AgdaSpace{}%
\AgdaOperator{\AgdaFunction{⊑ᴸᴿₜ}}\AgdaSpace{}%
\AgdaInductiveConstructor{blame}\AgdaSpace{}%
\AgdaOperator{\AgdaFunction{⦂}}\AgdaSpace{}%
\AgdaBound{A⊑A′}\AgdaSymbol{)}\<%
\\
\>[0]\AgdaFunction{LRₜ-blame-step}\AgdaSpace{}%
\AgdaSymbol{\{}\AgdaBound{A}\AgdaSymbol{\}\{}\AgdaBound{A′}\AgdaSymbol{\}\{}\AgdaBound{A⊑A′}\AgdaSymbol{\}\{}\AgdaInductiveConstructor{≼}\AgdaSymbol{\}}\AgdaSpace{}%
\AgdaSymbol{\{}\AgdaBound{M}\AgdaSymbol{\}}\AgdaSpace{}%
\AgdaSymbol{\{}\AgdaInductiveConstructor{suc}\AgdaSpace{}%
\AgdaBound{k}\AgdaSymbol{\}}\AgdaSpace{}%
\AgdaSymbol{=}\AgdaSpace{}%
\AgdaInductiveConstructor{inj₂}\AgdaSpace{}%
\AgdaSymbol{(}\AgdaInductiveConstructor{inj₁}\AgdaSpace{}%
\AgdaSymbol{(}\AgdaInductiveConstructor{blame}\AgdaSpace{}%
\AgdaOperator{\AgdaInductiveConstructor{END}}\AgdaSymbol{))}\<%
\\
\>[0]\AgdaFunction{LRₜ-blame-step}\AgdaSpace{}%
\AgdaSymbol{\{}\AgdaBound{A}\AgdaSymbol{\}\{}\AgdaBound{A′}\AgdaSymbol{\}\{}\AgdaBound{A⊑A′}\AgdaSymbol{\}\{}\AgdaInductiveConstructor{≽}\AgdaSymbol{\}}\AgdaSpace{}%
\AgdaSymbol{\{}\AgdaBound{M}\AgdaSymbol{\}}\AgdaSpace{}%
\AgdaSymbol{\{}\AgdaInductiveConstructor{suc}\AgdaSpace{}%
\AgdaBound{k}\AgdaSymbol{\}}\AgdaSpace{}%
\AgdaSymbol{=}\AgdaSpace{}%
\AgdaInductiveConstructor{inj₂}\AgdaSpace{}%
\AgdaSymbol{(}\AgdaInductiveConstructor{inj₁}\AgdaSpace{}%
\AgdaInductiveConstructor{isBlame}\AgdaSymbol{)}\<%
\end{code}
\begin{code}[hide]%
\>[0]\AgdaFunction{LRₜ-blame}\AgdaSpace{}%
\AgdaSymbol{:}\AgdaSpace{}%
\AgdaSymbol{∀\{}\AgdaBound{𝒫}\AgdaSymbol{\}\{}\AgdaBound{A}\AgdaSymbol{\}\{}\AgdaBound{A′}\AgdaSymbol{\}\{}\AgdaBound{A⊑A′}\AgdaSpace{}%
\AgdaSymbol{:}\AgdaSpace{}%
\AgdaBound{A}\AgdaSpace{}%
\AgdaOperator{\AgdaDatatype{⊑}}\AgdaSpace{}%
\AgdaBound{A′}\AgdaSymbol{\}\{}\AgdaBound{M}\AgdaSymbol{\}\{}\AgdaBound{dir}\AgdaSymbol{\}}\AgdaSpace{}%
\AgdaSymbol{→}\AgdaSpace{}%
\AgdaBound{𝒫}\AgdaSpace{}%
\AgdaOperator{\AgdaFunction{⊢ᵒ}}\AgdaSpace{}%
\AgdaBound{dir}\AgdaSpace{}%
\AgdaOperator{\AgdaFunction{∣}}\AgdaSpace{}%
\AgdaBound{M}\AgdaSpace{}%
\AgdaOperator{\AgdaFunction{⊑ᴸᴿₜ}}\AgdaSpace{}%
\AgdaInductiveConstructor{blame}\AgdaSpace{}%
\AgdaOperator{\AgdaFunction{⦂}}\AgdaSpace{}%
\AgdaBound{A⊑A′}\<%
\\
\>[0]\AgdaFunction{LRₜ-blame}\AgdaSpace{}%
\AgdaSymbol{\{}\AgdaBound{𝒫}\AgdaSymbol{\}\{}\AgdaBound{A}\AgdaSymbol{\}\{}\AgdaBound{A′}\AgdaSymbol{\}\{}\AgdaBound{A⊑A′}\AgdaSymbol{\}\{}\AgdaBound{M}\AgdaSymbol{\}\{}\AgdaBound{dir}\AgdaSymbol{\}}\AgdaSpace{}%
\AgdaSymbol{=}\AgdaSpace{}%
\AgdaFunction{⊢ᵒ-intro}\AgdaSpace{}%
\AgdaSymbol{λ}\AgdaSpace{}%
\AgdaBound{n}\AgdaSpace{}%
\AgdaBound{x}\AgdaSpace{}%
\AgdaSymbol{→}\AgdaSpace{}%
\AgdaFunction{LRₜ-blame-step}\AgdaSymbol{\{}\AgdaArgument{dir}\AgdaSpace{}%
\AgdaSymbol{=}\AgdaSpace{}%
\AgdaBound{dir}\AgdaSymbol{\}}\<%
\end{code}
\begin{code}%
\>[0]\AgdaFunction{compatible-blame}\AgdaSpace{}%
\AgdaSymbol{:}\AgdaSpace{}%
\AgdaSymbol{∀\{}\AgdaBound{Γ}\AgdaSymbol{\}\{}\AgdaBound{A}\AgdaSymbol{\}\{}\AgdaBound{M}\AgdaSymbol{\}}\AgdaSpace{}%
\AgdaSymbol{→}\AgdaSpace{}%
\AgdaFunction{map}\AgdaSpace{}%
\AgdaField{proj₁}\AgdaSpace{}%
\AgdaBound{Γ}\AgdaSpace{}%
\AgdaOperator{\AgdaDatatype{⊢}}\AgdaSpace{}%
\AgdaBound{M}\AgdaSpace{}%
\AgdaOperator{\AgdaDatatype{⦂}}\AgdaSpace{}%
\AgdaBound{A}\AgdaSpace{}%
\AgdaSymbol{→}\AgdaSpace{}%
\AgdaBound{Γ}\AgdaSpace{}%
\AgdaOperator{\AgdaFunction{⊨}}\AgdaSpace{}%
\AgdaBound{M}\AgdaSpace{}%
\AgdaOperator{\AgdaFunction{⊑ᴸᴿ}}\AgdaSpace{}%
\AgdaInductiveConstructor{blame}\AgdaSpace{}%
\AgdaOperator{\AgdaFunction{⦂}}\AgdaSpace{}%
\AgdaSymbol{(}\AgdaBound{A}\AgdaSpace{}%
\AgdaOperator{\AgdaInductiveConstructor{,}}\AgdaSpace{}%
\AgdaBound{A}\AgdaSpace{}%
\AgdaOperator{\AgdaInductiveConstructor{,}}\AgdaSpace{}%
\AgdaFunction{Refl⊑}\AgdaSymbol{)}\<%
\end{code}
\begin{code}[hide]%
\>[0]\AgdaFunction{compatible-blame}\AgdaSymbol{\{}\AgdaBound{Γ}\AgdaSymbol{\}\{}\AgdaBound{A}\AgdaSymbol{\}\{}\AgdaBound{M}\AgdaSymbol{\}}\AgdaSpace{}%
\AgdaBound{⊢M}\AgdaSpace{}%
\AgdaSymbol{=}\AgdaSpace{}%
\AgdaSymbol{(λ}\AgdaSpace{}%
\AgdaBound{γ}\AgdaSpace{}%
\AgdaBound{γ′}\AgdaSpace{}%
\AgdaSymbol{→}\AgdaSpace{}%
\AgdaFunction{LRₜ-blame}\AgdaSymbol{)}\AgdaSpace{}%
\AgdaOperator{\AgdaInductiveConstructor{,}}\AgdaSpace{}%
\AgdaSymbol{(λ}\AgdaSpace{}%
\AgdaBound{γ}\AgdaSpace{}%
\AgdaBound{γ′}\AgdaSpace{}%
\AgdaSymbol{→}\AgdaSpace{}%
\AgdaFunction{LRₜ-blame}\AgdaSymbol{)}\<%
\end{code}

\noindent The proof of compatibility for variables relies on the
following lemma regarding related substitutions.

\begin{code}%
\>[0]\AgdaFunction{lookup-⊑ᴸᴿ}\AgdaSpace{}%
\AgdaSymbol{:}\AgdaSpace{}%
\AgdaSymbol{∀\{}\AgdaBound{dir}\AgdaSymbol{\}}\AgdaSpace{}%
\AgdaSymbol{(}\AgdaBound{Γ}\AgdaSpace{}%
\AgdaSymbol{:}\AgdaSpace{}%
\AgdaDatatype{List}\AgdaSpace{}%
\AgdaFunction{Prec}\AgdaSymbol{)}\AgdaSpace{}%
\AgdaSymbol{→}\AgdaSpace{}%
\AgdaSymbol{(}\AgdaBound{γ}\AgdaSpace{}%
\AgdaBound{γ′}\AgdaSpace{}%
\AgdaSymbol{:}\AgdaSpace{}%
\AgdaFunction{Subst}\AgdaSymbol{)}\AgdaSpace{}%
\AgdaSymbol{→}\AgdaSpace{}%
\AgdaSymbol{∀}\AgdaSpace{}%
\AgdaSymbol{\{}\AgdaBound{A}\AgdaSymbol{\}\{}\AgdaBound{A′}\AgdaSymbol{\}\{}\AgdaBound{A⊑A′}\AgdaSymbol{\}\{}\AgdaBound{x}\AgdaSymbol{\}}\<%
\\
\>[0][@{}l@{\AgdaIndent{0}}]%
\>[2]\AgdaSymbol{→}\AgdaSpace{}%
\AgdaBound{Γ}\AgdaSpace{}%
\AgdaOperator{\AgdaFunction{∋}}\AgdaSpace{}%
\AgdaBound{x}\AgdaSpace{}%
\AgdaOperator{\AgdaFunction{⦂}}\AgdaSpace{}%
\AgdaSymbol{(}\AgdaBound{A}\AgdaSpace{}%
\AgdaOperator{\AgdaInductiveConstructor{,}}\AgdaSpace{}%
\AgdaBound{A′}\AgdaSpace{}%
\AgdaOperator{\AgdaInductiveConstructor{,}}\AgdaSpace{}%
\AgdaBound{A⊑A′}\AgdaSymbol{)}%
\>[29]\AgdaSymbol{→}%
\>[32]\AgdaSymbol{(}\AgdaBound{Γ}\AgdaSpace{}%
\AgdaOperator{\AgdaFunction{∣}}\AgdaSpace{}%
\AgdaBound{dir}\AgdaSpace{}%
\AgdaOperator{\AgdaFunction{⊨}}\AgdaSpace{}%
\AgdaBound{γ}\AgdaSpace{}%
\AgdaOperator{\AgdaFunction{⊑ᴸᴿ}}\AgdaSpace{}%
\AgdaBound{γ′}\AgdaSymbol{)}\AgdaSpace{}%
\AgdaOperator{\AgdaFunction{⊢ᵒ}}\AgdaSpace{}%
\AgdaBound{dir}\AgdaSpace{}%
\AgdaOperator{\AgdaFunction{∣}}\AgdaSpace{}%
\AgdaBound{γ}\AgdaSpace{}%
\AgdaBound{x}\AgdaSpace{}%
\AgdaOperator{\AgdaFunction{⊑ᴸᴿᵥ}}\AgdaSpace{}%
\AgdaBound{γ′}\AgdaSpace{}%
\AgdaBound{x}\AgdaSpace{}%
\AgdaOperator{\AgdaFunction{⦂}}\AgdaSpace{}%
\AgdaBound{A⊑A′}\<%
\end{code}
\begin{code}[hide]%
\>[0]\AgdaFunction{lookup-⊑ᴸᴿ}\AgdaSpace{}%
\AgdaSymbol{\{}\AgdaBound{dir}\AgdaSymbol{\}}\AgdaSpace{}%
\AgdaSymbol{(}\AgdaDottedPattern{\AgdaSymbol{.(}}\AgdaDottedPattern{\AgdaBound{A}}\AgdaSpace{}%
\AgdaDottedPattern{\AgdaOperator{\AgdaInductiveConstructor{,}}}\AgdaSpace{}%
\AgdaDottedPattern{\AgdaBound{A′}}\AgdaSpace{}%
\AgdaDottedPattern{\AgdaOperator{\AgdaInductiveConstructor{,}}}\AgdaSpace{}%
\AgdaDottedPattern{\AgdaBound{A⊑A′}}\AgdaDottedPattern{\AgdaSymbol{)}}\AgdaSpace{}%
\AgdaOperator{\AgdaInductiveConstructor{∷}}\AgdaSpace{}%
\AgdaBound{Γ}\AgdaSymbol{)}\AgdaSpace{}%
\AgdaBound{γ}\AgdaSpace{}%
\AgdaBound{γ′}\AgdaSpace{}%
\AgdaSymbol{\{}\AgdaBound{A}\AgdaSymbol{\}}\AgdaSpace{}%
\AgdaSymbol{\{}\AgdaBound{A′}\AgdaSymbol{\}}\AgdaSpace{}%
\AgdaSymbol{\{}\AgdaBound{A⊑A′}\AgdaSymbol{\}}\AgdaSpace{}%
\AgdaSymbol{\{}\AgdaInductiveConstructor{zero}\AgdaSymbol{\}}\AgdaSpace{}%
\AgdaInductiveConstructor{refl}\AgdaSpace{}%
\AgdaSymbol{=}\AgdaSpace{}%
\AgdaFunction{Zᵒ}\<%
\\
\>[0]\AgdaFunction{lookup-⊑ᴸᴿ}\AgdaSpace{}%
\AgdaSymbol{\{}\AgdaBound{dir}\AgdaSymbol{\}}\AgdaSpace{}%
\AgdaSymbol{(}\AgdaBound{B}\AgdaSpace{}%
\AgdaOperator{\AgdaInductiveConstructor{∷}}\AgdaSpace{}%
\AgdaBound{Γ}\AgdaSymbol{)}\AgdaSpace{}%
\AgdaBound{γ}\AgdaSpace{}%
\AgdaBound{γ′}\AgdaSpace{}%
\AgdaSymbol{\{}\AgdaBound{A}\AgdaSymbol{\}}\AgdaSpace{}%
\AgdaSymbol{\{}\AgdaBound{A′}\AgdaSymbol{\}}\AgdaSpace{}%
\AgdaSymbol{\{}\AgdaBound{A⊑A′}\AgdaSymbol{\}}\AgdaSpace{}%
\AgdaSymbol{\{}\AgdaInductiveConstructor{suc}\AgdaSpace{}%
\AgdaBound{x}\AgdaSymbol{\}}\AgdaSpace{}%
\AgdaBound{∋x}\AgdaSpace{}%
\AgdaSymbol{=}\<%
\\
\>[0][@{}l@{\AgdaIndent{0}}]%
\>[3]\AgdaFunction{Sᵒ}\AgdaSpace{}%
\AgdaSymbol{(}\AgdaFunction{lookup-⊑ᴸᴿ}\AgdaSpace{}%
\AgdaBound{Γ}\AgdaSpace{}%
\AgdaSymbol{(λ}\AgdaSpace{}%
\AgdaBound{z}\AgdaSpace{}%
\AgdaSymbol{→}\AgdaSpace{}%
\AgdaBound{γ}\AgdaSpace{}%
\AgdaSymbol{(}\AgdaInductiveConstructor{suc}\AgdaSpace{}%
\AgdaBound{z}\AgdaSymbol{))}\AgdaSpace{}%
\AgdaSymbol{(λ}\AgdaSpace{}%
\AgdaBound{z}\AgdaSpace{}%
\AgdaSymbol{→}\AgdaSpace{}%
\AgdaBound{γ′}\AgdaSpace{}%
\AgdaSymbol{(}\AgdaInductiveConstructor{suc}\AgdaSpace{}%
\AgdaBound{z}\AgdaSymbol{))}\AgdaSpace{}%
\AgdaBound{∋x}\AgdaSymbol{)}\<%
\end{code}

\begin{code}%
\>[0]\AgdaFunction{compatibility-var}\AgdaSpace{}%
\AgdaSymbol{:}\AgdaSpace{}%
\AgdaSymbol{∀}\AgdaSpace{}%
\AgdaSymbol{\{}\AgdaBound{Γ}\AgdaSpace{}%
\AgdaBound{A}\AgdaSpace{}%
\AgdaBound{A′}\AgdaSpace{}%
\AgdaBound{A⊑A′}\AgdaSpace{}%
\AgdaBound{x}\AgdaSymbol{\}}\AgdaSpace{}%
\AgdaSymbol{→}\AgdaSpace{}%
\AgdaBound{Γ}\AgdaSpace{}%
\AgdaOperator{\AgdaFunction{∋}}\AgdaSpace{}%
\AgdaBound{x}\AgdaSpace{}%
\AgdaOperator{\AgdaFunction{⦂}}\AgdaSpace{}%
\AgdaSymbol{(}\AgdaBound{A}\AgdaSpace{}%
\AgdaOperator{\AgdaInductiveConstructor{,}}\AgdaSpace{}%
\AgdaBound{A′}\AgdaSpace{}%
\AgdaOperator{\AgdaInductiveConstructor{,}}\AgdaSpace{}%
\AgdaBound{A⊑A′}\AgdaSymbol{)}\<%
\\
\>[0][@{}l@{\AgdaIndent{0}}]%
\>[2]\AgdaSymbol{→}\AgdaSpace{}%
\AgdaBound{Γ}\AgdaSpace{}%
\AgdaOperator{\AgdaFunction{⊨}}\AgdaSpace{}%
\AgdaOperator{\AgdaInductiveConstructor{`}}\AgdaSpace{}%
\AgdaBound{x}\AgdaSpace{}%
\AgdaOperator{\AgdaFunction{⊑ᴸᴿ}}\AgdaSpace{}%
\AgdaOperator{\AgdaInductiveConstructor{`}}\AgdaSpace{}%
\AgdaBound{x}\AgdaSpace{}%
\AgdaOperator{\AgdaFunction{⦂}}\AgdaSpace{}%
\AgdaSymbol{(}\AgdaBound{A}\AgdaSpace{}%
\AgdaOperator{\AgdaInductiveConstructor{,}}\AgdaSpace{}%
\AgdaBound{A′}\AgdaSpace{}%
\AgdaOperator{\AgdaInductiveConstructor{,}}\AgdaSpace{}%
\AgdaBound{A⊑A′}\AgdaSymbol{)}\<%
\end{code}
\begin{code}[hide]%
\>[0]\AgdaFunction{compatibility-var}\AgdaSpace{}%
\AgdaSymbol{\{}\AgdaBound{Γ}\AgdaSymbol{\}\{}\AgdaBound{A}\AgdaSymbol{\}\{}\AgdaBound{A′}\AgdaSymbol{\}\{}\AgdaBound{A⊑A′}\AgdaSymbol{\}\{}\AgdaBound{x}\AgdaSymbol{\}}\AgdaSpace{}%
\AgdaBound{∋x}\AgdaSpace{}%
\AgdaSymbol{=}\AgdaSpace{}%
\AgdaFunction{LT}\AgdaSpace{}%
\AgdaOperator{\AgdaInductiveConstructor{,}}\AgdaSpace{}%
\AgdaFunction{GT}\<%
\\
\>[0][@{}l@{\AgdaIndent{0}}]%
\>[2]\AgdaKeyword{where}\<%
\\
\>[2]\AgdaFunction{LT}\AgdaSpace{}%
\AgdaSymbol{:}\AgdaSpace{}%
\AgdaBound{Γ}\AgdaSpace{}%
\AgdaOperator{\AgdaFunction{∣}}\AgdaSpace{}%
\AgdaInductiveConstructor{≼}\AgdaSpace{}%
\AgdaOperator{\AgdaFunction{⊨}}\AgdaSpace{}%
\AgdaOperator{\AgdaInductiveConstructor{`}}\AgdaSpace{}%
\AgdaBound{x}\AgdaSpace{}%
\AgdaOperator{\AgdaFunction{⊑ᴸᴿ}}\AgdaSpace{}%
\AgdaOperator{\AgdaInductiveConstructor{`}}\AgdaSpace{}%
\AgdaBound{x}\AgdaSpace{}%
\AgdaOperator{\AgdaFunction{⦂}}\AgdaSpace{}%
\AgdaSymbol{(}\AgdaBound{A}\AgdaSpace{}%
\AgdaOperator{\AgdaInductiveConstructor{,}}\AgdaSpace{}%
\AgdaBound{A′}\AgdaSpace{}%
\AgdaOperator{\AgdaInductiveConstructor{,}}\AgdaSpace{}%
\AgdaBound{A⊑A′}\AgdaSymbol{)}\<%
\\
\>[2]\AgdaFunction{LT}\AgdaSpace{}%
\AgdaBound{γ}\AgdaSpace{}%
\AgdaBound{γ′}\AgdaSpace{}%
\AgdaKeyword{rewrite}\AgdaSpace{}%
\AgdaFunction{sub-var}\AgdaSpace{}%
\AgdaBound{γ}\AgdaSpace{}%
\AgdaBound{x}\AgdaSpace{}%
\AgdaSymbol{|}\AgdaSpace{}%
\AgdaFunction{sub-var}\AgdaSpace{}%
\AgdaBound{γ′}\AgdaSpace{}%
\AgdaBound{x}\AgdaSpace{}%
\AgdaSymbol{=}\AgdaSpace{}%
\AgdaFunction{LRᵥ⇒LRₜ}\AgdaSpace{}%
\AgdaSymbol{(}\AgdaFunction{lookup-⊑ᴸᴿ}\AgdaSpace{}%
\AgdaBound{Γ}\AgdaSpace{}%
\AgdaBound{γ}\AgdaSpace{}%
\AgdaBound{γ′}\AgdaSpace{}%
\AgdaBound{∋x}\AgdaSymbol{)}\<%
\\
\\[\AgdaEmptyExtraSkip]%
\>[2]\AgdaFunction{GT}\AgdaSpace{}%
\AgdaSymbol{:}\AgdaSpace{}%
\AgdaBound{Γ}\AgdaSpace{}%
\AgdaOperator{\AgdaFunction{∣}}\AgdaSpace{}%
\AgdaInductiveConstructor{≽}\AgdaSpace{}%
\AgdaOperator{\AgdaFunction{⊨}}\AgdaSpace{}%
\AgdaOperator{\AgdaInductiveConstructor{`}}\AgdaSpace{}%
\AgdaBound{x}\AgdaSpace{}%
\AgdaOperator{\AgdaFunction{⊑ᴸᴿ}}\AgdaSpace{}%
\AgdaOperator{\AgdaInductiveConstructor{`}}\AgdaSpace{}%
\AgdaBound{x}\AgdaSpace{}%
\AgdaOperator{\AgdaFunction{⦂}}\AgdaSpace{}%
\AgdaSymbol{(}\AgdaBound{A}\AgdaSpace{}%
\AgdaOperator{\AgdaInductiveConstructor{,}}\AgdaSpace{}%
\AgdaBound{A′}\AgdaSpace{}%
\AgdaOperator{\AgdaInductiveConstructor{,}}\AgdaSpace{}%
\AgdaBound{A⊑A′}\AgdaSymbol{)}\<%
\\
\>[2]\AgdaFunction{GT}\AgdaSpace{}%
\AgdaBound{γ}\AgdaSpace{}%
\AgdaBound{γ′}\AgdaSpace{}%
\AgdaKeyword{rewrite}\AgdaSpace{}%
\AgdaFunction{sub-var}\AgdaSpace{}%
\AgdaBound{γ}\AgdaSpace{}%
\AgdaBound{x}\AgdaSpace{}%
\AgdaSymbol{|}\AgdaSpace{}%
\AgdaFunction{sub-var}\AgdaSpace{}%
\AgdaBound{γ′}\AgdaSpace{}%
\AgdaBound{x}\AgdaSpace{}%
\AgdaSymbol{=}\AgdaSpace{}%
\AgdaFunction{LRᵥ⇒LRₜ}\AgdaSpace{}%
\AgdaSymbol{(}\AgdaFunction{lookup-⊑ᴸᴿ}\AgdaSpace{}%
\AgdaBound{Γ}\AgdaSpace{}%
\AgdaBound{γ}\AgdaSpace{}%
\AgdaBound{γ′}\AgdaSpace{}%
\AgdaBound{∋x}\AgdaSymbol{)}\<%
\end{code}

\paragraph{Compatibility for Lambda}

The proof of compatibility for lambda abstraction has a premise that
says the bodies of the two lambdas are in the logical relation, which
is the induction hypothesis in this case of the fundamental theorem.
The logical relation for lambda requires us to prove
\begin{center}
\textsf{𝒫 ⊢ᵒ (dir ∣ (⟪ ext γ ⟫ N) [ W ] ⊑ᴸᴿₜ (⟪ ext γ′ ⟫ N′) [ W′ ] ⦂ d)}
\end{center}
Using the premise we obtain
\begin{center}
\textsf{𝒫 ⊢ᵒ (dir ∣ ⟪ W • γ ⟫ N ⊑ᴸᴿₜ ⟪ W′ • γ′ ⟫ N′ ⦂ d)}
\end{center}
which is equivalent to what is required thanks to the
\textsf{ext-sub-cons} theorem from the ABT library. As an example of a
proof using SIL, here is the proof in full of compatibility for
lambda.

\begin{code}[hide]%
\>[0]\AgdaFunction{LRᵥ-fun}\AgdaSpace{}%
\AgdaSymbol{:}\AgdaSpace{}%
\AgdaSymbol{∀\{}\AgdaBound{A}\AgdaSpace{}%
\AgdaBound{B}\AgdaSpace{}%
\AgdaBound{A′}\AgdaSpace{}%
\AgdaBound{B′}\AgdaSymbol{\}\{}\AgdaBound{A⊑A′}\AgdaSpace{}%
\AgdaSymbol{:}\AgdaSpace{}%
\AgdaBound{A}\AgdaSpace{}%
\AgdaOperator{\AgdaDatatype{⊑}}\AgdaSpace{}%
\AgdaBound{A′}\AgdaSymbol{\}\{}\AgdaBound{B⊑B′}\AgdaSpace{}%
\AgdaSymbol{:}\AgdaSpace{}%
\AgdaBound{B}\AgdaSpace{}%
\AgdaOperator{\AgdaDatatype{⊑}}\AgdaSpace{}%
\AgdaBound{B′}\AgdaSymbol{\}\{}\AgdaBound{N}\AgdaSymbol{\}\{}\AgdaBound{N′}\AgdaSymbol{\}\{}\AgdaBound{dir}\AgdaSymbol{\}}\<%
\\
\>[0][@{}l@{\AgdaIndent{0}}]%
\>[3]\AgdaSymbol{→}%
\>[473I]\AgdaSymbol{(}\AgdaBound{dir}\AgdaSpace{}%
\AgdaOperator{\AgdaFunction{∣}}\AgdaSpace{}%
\AgdaSymbol{(}\AgdaInductiveConstructor{ƛ}\AgdaSpace{}%
\AgdaBound{N}\AgdaSymbol{)}\AgdaSpace{}%
\AgdaOperator{\AgdaFunction{⊑ᴸᴿᵥ}}\AgdaSpace{}%
\AgdaSymbol{(}\AgdaInductiveConstructor{ƛ}\AgdaSpace{}%
\AgdaBound{N′}\AgdaSymbol{)}\AgdaSpace{}%
\AgdaOperator{\AgdaFunction{⦂}}\AgdaSpace{}%
\AgdaInductiveConstructor{fun⊑}\AgdaSpace{}%
\AgdaBound{A⊑A′}\AgdaSpace{}%
\AgdaBound{B⊑B′}\AgdaSymbol{)}\<%
\\
\>[473I][@{}l@{\AgdaIndent{0}}]%
\>[6]\AgdaOperator{\AgdaFunction{≡ᵒ}}\AgdaSpace{}%
\AgdaSymbol{(}\AgdaFunction{∀ᵒ[}\AgdaSpace{}%
\AgdaBound{W}%
\>[486I]\AgdaFunction{]}\AgdaSpace{}%
\AgdaFunction{∀ᵒ[}\AgdaSpace{}%
\AgdaBound{W′}\AgdaSpace{}%
\AgdaFunction{]}\AgdaSpace{}%
\AgdaSymbol{((}\AgdaOperator{\AgdaFunction{▷ᵒ}}\AgdaSpace{}%
\AgdaSymbol{(}\AgdaBound{dir}\AgdaSpace{}%
\AgdaOperator{\AgdaFunction{∣}}\AgdaSpace{}%
\AgdaBound{W}\AgdaSpace{}%
\AgdaOperator{\AgdaFunction{⊑ᴸᴿᵥ}}\AgdaSpace{}%
\AgdaBound{W′}\AgdaSpace{}%
\AgdaOperator{\AgdaFunction{⦂}}\AgdaSpace{}%
\AgdaBound{A⊑A′}\AgdaSymbol{))}\<%
\\
\>[.][@{}l@{}]\<[486I]%
\>[16]\AgdaOperator{\AgdaFunction{→ᵒ}}\AgdaSpace{}%
\AgdaSymbol{(}\AgdaOperator{\AgdaFunction{▷ᵒ}}\AgdaSpace{}%
\AgdaSymbol{(}\AgdaBound{dir}\AgdaSpace{}%
\AgdaOperator{\AgdaFunction{∣}}\AgdaSpace{}%
\AgdaSymbol{(}\AgdaBound{N}\AgdaSpace{}%
\AgdaOperator{\AgdaFunction{[}}\AgdaSpace{}%
\AgdaBound{W}\AgdaSpace{}%
\AgdaOperator{\AgdaFunction{]}}\AgdaSymbol{)}\AgdaSpace{}%
\AgdaOperator{\AgdaFunction{⊑ᴸᴿₜ}}\AgdaSpace{}%
\AgdaSymbol{(}\AgdaBound{N′}\AgdaSpace{}%
\AgdaOperator{\AgdaFunction{[}}\AgdaSpace{}%
\AgdaBound{W′}\AgdaSpace{}%
\AgdaOperator{\AgdaFunction{]}}\AgdaSymbol{)}\AgdaSpace{}%
\AgdaOperator{\AgdaFunction{⦂}}\AgdaSpace{}%
\AgdaBound{B⊑B′}\AgdaSymbol{))))}\<%
\end{code}
\begin{code}[hide]%
\>[0]\AgdaFunction{LRᵥ-fun}\AgdaSpace{}%
\AgdaSymbol{\{}\AgdaBound{A}\AgdaSymbol{\}\{}\AgdaBound{B}\AgdaSymbol{\}\{}\AgdaBound{A′}\AgdaSymbol{\}\{}\AgdaBound{B′}\AgdaSymbol{\}\{}\AgdaBound{A⊑A′}\AgdaSymbol{\}\{}\AgdaBound{B⊑B′}\AgdaSymbol{\}\{}\AgdaBound{N}\AgdaSymbol{\}\{}\AgdaBound{N′}\AgdaSymbol{\}\{}\AgdaBound{dir}\AgdaSymbol{\}}\AgdaSpace{}%
\AgdaSymbol{=}\<%
\\
\>[0][@{}l@{\AgdaIndent{0}}]%
\>[3]\AgdaKeyword{let}\AgdaSpace{}%
\AgdaBound{X}\AgdaSpace{}%
\AgdaSymbol{=}\AgdaSpace{}%
\AgdaInductiveConstructor{inj₁}\AgdaSpace{}%
\AgdaSymbol{((}\AgdaBound{A}\AgdaSpace{}%
\AgdaOperator{\AgdaInductiveConstructor{⇒}}\AgdaSpace{}%
\AgdaBound{B}\AgdaSpace{}%
\AgdaOperator{\AgdaInductiveConstructor{,}}\AgdaSpace{}%
\AgdaBound{A′}\AgdaSpace{}%
\AgdaOperator{\AgdaInductiveConstructor{⇒}}\AgdaSpace{}%
\AgdaBound{B′}\AgdaSpace{}%
\AgdaOperator{\AgdaInductiveConstructor{,}}\AgdaSpace{}%
\AgdaInductiveConstructor{fun⊑}\AgdaSpace{}%
\AgdaBound{A⊑A′}\AgdaSpace{}%
\AgdaBound{B⊑B′}\AgdaSymbol{)}\AgdaSpace{}%
\AgdaOperator{\AgdaInductiveConstructor{,}}\AgdaSpace{}%
\AgdaBound{dir}\AgdaSpace{}%
\AgdaOperator{\AgdaInductiveConstructor{,}}\AgdaSpace{}%
\AgdaInductiveConstructor{ƛ}\AgdaSpace{}%
\AgdaBound{N}\AgdaSpace{}%
\AgdaOperator{\AgdaInductiveConstructor{,}}\AgdaSpace{}%
\AgdaInductiveConstructor{ƛ}\AgdaSpace{}%
\AgdaBound{N′}\AgdaSymbol{)}\AgdaSpace{}%
\AgdaKeyword{in}\<%
\\
\>[3]\AgdaSymbol{(}\AgdaBound{dir}\AgdaSpace{}%
\AgdaOperator{\AgdaFunction{∣}}\AgdaSpace{}%
\AgdaSymbol{(}\AgdaInductiveConstructor{ƛ}\AgdaSpace{}%
\AgdaBound{N}\AgdaSymbol{)}\AgdaSpace{}%
\AgdaOperator{\AgdaFunction{⊑ᴸᴿᵥ}}\AgdaSpace{}%
\AgdaSymbol{(}\AgdaInductiveConstructor{ƛ}\AgdaSpace{}%
\AgdaBound{N′}\AgdaSymbol{)}\AgdaSpace{}%
\AgdaOperator{\AgdaFunction{⦂}}\AgdaSpace{}%
\AgdaInductiveConstructor{fun⊑}\AgdaSpace{}%
\AgdaBound{A⊑A′}\AgdaSpace{}%
\AgdaBound{B⊑B′}\AgdaSymbol{)}%
\>[47]\AgdaOperator{\AgdaFunction{⩦⟨}}\AgdaSpace{}%
\AgdaFunction{≡ᵒ-refl}\AgdaSpace{}%
\AgdaInductiveConstructor{refl}\AgdaSpace{}%
\AgdaOperator{\AgdaFunction{⟩}}\<%
\\
\>[3]\AgdaFunction{LRₜ⊎LRᵥ}\AgdaSpace{}%
\AgdaBound{X}%
\>[47]\AgdaOperator{\AgdaFunction{⩦⟨}}\AgdaSpace{}%
\AgdaFunction{fixpointᵒ}\AgdaSpace{}%
\AgdaFunction{pre-LRₜ⊎LRᵥ}\AgdaSpace{}%
\AgdaBound{X}\AgdaSpace{}%
\AgdaOperator{\AgdaFunction{⟩}}\<%
\\
\>[3]\AgdaField{\#}\AgdaSpace{}%
\AgdaSymbol{(}\AgdaFunction{pre-LRₜ⊎LRᵥ}\AgdaSpace{}%
\AgdaBound{X}\AgdaSymbol{)}\AgdaSpace{}%
\AgdaSymbol{(}\AgdaFunction{LRₜ⊎LRᵥ}\AgdaSpace{}%
\AgdaOperator{\AgdaInductiveConstructor{,}}\AgdaSpace{}%
\AgdaFunction{ttᵖ}\AgdaSymbol{)}%
\>[47]\AgdaOperator{\AgdaFunction{⩦⟨}}\AgdaSpace{}%
\AgdaFunction{≡ᵒ-refl}\AgdaSpace{}%
\AgdaInductiveConstructor{refl}\AgdaSpace{}%
\AgdaOperator{\AgdaFunction{⟩}}\<%
\\
\>[3]\AgdaSymbol{(}\AgdaFunction{∀ᵒ[}\AgdaSpace{}%
\AgdaBound{W}\AgdaSpace{}%
\AgdaFunction{]}\AgdaSpace{}%
\AgdaFunction{∀ᵒ[}\AgdaSpace{}%
\AgdaBound{W′}%
\>[567I]\AgdaFunction{]}\AgdaSpace{}%
\AgdaSymbol{((}\AgdaOperator{\AgdaFunction{▷ᵒ}}\AgdaSpace{}%
\AgdaSymbol{(}\AgdaBound{dir}\AgdaSpace{}%
\AgdaOperator{\AgdaFunction{∣}}\AgdaSpace{}%
\AgdaBound{W}\AgdaSpace{}%
\AgdaOperator{\AgdaFunction{⊑ᴸᴿᵥ}}\AgdaSpace{}%
\AgdaBound{W′}\AgdaSpace{}%
\AgdaOperator{\AgdaFunction{⦂}}\AgdaSpace{}%
\AgdaBound{A⊑A′}\AgdaSymbol{))}\<%
\\
\>[.][@{}l@{}]\<[567I]%
\>[19]\AgdaOperator{\AgdaFunction{→ᵒ}}\AgdaSpace{}%
\AgdaSymbol{(}\AgdaOperator{\AgdaFunction{▷ᵒ}}\AgdaSpace{}%
\AgdaSymbol{(}\AgdaBound{dir}\AgdaSpace{}%
\AgdaOperator{\AgdaFunction{∣}}\AgdaSpace{}%
\AgdaSymbol{(}\AgdaBound{N}\AgdaSpace{}%
\AgdaOperator{\AgdaFunction{[}}\AgdaSpace{}%
\AgdaBound{W}\AgdaSpace{}%
\AgdaOperator{\AgdaFunction{]}}\AgdaSymbol{)}\AgdaSpace{}%
\AgdaOperator{\AgdaFunction{⊑ᴸᴿₜ}}\AgdaSpace{}%
\AgdaSymbol{(}\AgdaBound{N′}\AgdaSpace{}%
\AgdaOperator{\AgdaFunction{[}}\AgdaSpace{}%
\AgdaBound{W′}\AgdaSpace{}%
\AgdaOperator{\AgdaFunction{]}}\AgdaSymbol{)}\AgdaSpace{}%
\AgdaOperator{\AgdaFunction{⦂}}\AgdaSpace{}%
\AgdaBound{B⊑B′}\AgdaSymbol{))))}\AgdaSpace{}%
\AgdaOperator{\AgdaFunction{∎}}\<%
\end{code}
\begin{code}%
\>[0]\AgdaFunction{compatible-lambda}\AgdaSpace{}%
\AgdaSymbol{:}\AgdaSpace{}%
\AgdaSymbol{∀\{}\AgdaBound{Γ}\AgdaSpace{}%
\AgdaSymbol{:}\AgdaSpace{}%
\AgdaDatatype{List}\AgdaSpace{}%
\AgdaFunction{Prec}\AgdaSymbol{\}\{}\AgdaBound{A}\AgdaSymbol{\}\{}\AgdaBound{B}\AgdaSymbol{\}\{}\AgdaBound{C}\AgdaSymbol{\}\{}\AgdaBound{D}\AgdaSymbol{\}\{}\AgdaBound{N}\AgdaSpace{}%
\AgdaBound{N′}\AgdaSpace{}%
\AgdaSymbol{:}\AgdaSpace{}%
\AgdaDatatype{Term}\AgdaSymbol{\}\{}\AgdaBound{c}\AgdaSpace{}%
\AgdaSymbol{:}\AgdaSpace{}%
\AgdaBound{A}\AgdaSpace{}%
\AgdaOperator{\AgdaDatatype{⊑}}\AgdaSpace{}%
\AgdaBound{C}\AgdaSymbol{\}\{}\AgdaBound{d}\AgdaSpace{}%
\AgdaSymbol{:}\AgdaSpace{}%
\AgdaBound{B}\AgdaSpace{}%
\AgdaOperator{\AgdaDatatype{⊑}}\AgdaSpace{}%
\AgdaBound{D}\AgdaSymbol{\}}\<%
\\
\>[0][@{}l@{\AgdaIndent{0}}]%
\>[3]\AgdaSymbol{→}\AgdaSpace{}%
\AgdaSymbol{((}\AgdaBound{A}\AgdaSpace{}%
\AgdaOperator{\AgdaInductiveConstructor{,}}\AgdaSpace{}%
\AgdaBound{C}\AgdaSpace{}%
\AgdaOperator{\AgdaInductiveConstructor{,}}\AgdaSpace{}%
\AgdaBound{c}\AgdaSymbol{)}\AgdaSpace{}%
\AgdaOperator{\AgdaInductiveConstructor{∷}}\AgdaSpace{}%
\AgdaBound{Γ}\AgdaSymbol{)}\AgdaSpace{}%
\AgdaOperator{\AgdaFunction{⊨}}\AgdaSpace{}%
\AgdaBound{N}\AgdaSpace{}%
\AgdaOperator{\AgdaFunction{⊑ᴸᴿ}}\AgdaSpace{}%
\AgdaBound{N′}\AgdaSpace{}%
\AgdaOperator{\AgdaFunction{⦂}}\AgdaSpace{}%
\AgdaSymbol{(}\AgdaBound{B}\AgdaSpace{}%
\AgdaOperator{\AgdaInductiveConstructor{,}}\AgdaSpace{}%
\AgdaBound{D}\AgdaSpace{}%
\AgdaOperator{\AgdaInductiveConstructor{,}}\AgdaSpace{}%
\AgdaBound{d}\AgdaSymbol{)}\<%
\\
\>[3]\AgdaSymbol{→}\AgdaSpace{}%
\AgdaBound{Γ}\AgdaSpace{}%
\AgdaOperator{\AgdaFunction{⊨}}\AgdaSpace{}%
\AgdaSymbol{(}\AgdaInductiveConstructor{ƛ}\AgdaSpace{}%
\AgdaBound{N}\AgdaSymbol{)}\AgdaSpace{}%
\AgdaOperator{\AgdaFunction{⊑ᴸᴿ}}\AgdaSpace{}%
\AgdaSymbol{(}\AgdaInductiveConstructor{ƛ}\AgdaSpace{}%
\AgdaBound{N′}\AgdaSymbol{)}\AgdaSpace{}%
\AgdaOperator{\AgdaFunction{⦂}}\AgdaSpace{}%
\AgdaSymbol{(}\AgdaBound{A}\AgdaSpace{}%
\AgdaOperator{\AgdaInductiveConstructor{⇒}}\AgdaSpace{}%
\AgdaBound{B}\AgdaSpace{}%
\AgdaOperator{\AgdaInductiveConstructor{,}}\AgdaSpace{}%
\AgdaBound{C}\AgdaSpace{}%
\AgdaOperator{\AgdaInductiveConstructor{⇒}}\AgdaSpace{}%
\AgdaBound{D}\AgdaSpace{}%
\AgdaOperator{\AgdaInductiveConstructor{,}}\AgdaSpace{}%
\AgdaInductiveConstructor{fun⊑}\AgdaSpace{}%
\AgdaBound{c}\AgdaSpace{}%
\AgdaBound{d}\AgdaSymbol{)}\<%
\\
\>[0]\AgdaFunction{compatible-lambda}\AgdaSymbol{\{}\AgdaBound{Γ}\AgdaSymbol{\}\{}\AgdaBound{A}\AgdaSymbol{\}\{}\AgdaBound{B}\AgdaSymbol{\}\{}\AgdaBound{C}\AgdaSymbol{\}\{}\AgdaBound{D}\AgdaSymbol{\}\{}\AgdaBound{N}\AgdaSymbol{\}\{}\AgdaBound{N′}\AgdaSymbol{\}\{}\AgdaBound{c}\AgdaSymbol{\}\{}\AgdaBound{d}\AgdaSymbol{\}}\AgdaSpace{}%
\AgdaBound{⊨N⊑N′}\AgdaSpace{}%
\AgdaSymbol{=}\<%
\\
\>[0][@{}l@{\AgdaIndent{0}}]%
\>[2]\AgdaSymbol{(λ}\AgdaSpace{}%
\AgdaBound{γ}\AgdaSpace{}%
\AgdaBound{γ′}\AgdaSpace{}%
\AgdaSymbol{→}\AgdaSpace{}%
\AgdaFunction{⊢ℰλNλN′}\AgdaSymbol{)}\AgdaSpace{}%
\AgdaOperator{\AgdaInductiveConstructor{,}}\AgdaSpace{}%
\AgdaSymbol{(λ}\AgdaSpace{}%
\AgdaBound{γ}\AgdaSpace{}%
\AgdaBound{γ′}\AgdaSpace{}%
\AgdaSymbol{→}\AgdaSpace{}%
\AgdaFunction{⊢ℰλNλN′}\AgdaSymbol{)}\<%
\\
\>[0][@{}l@{\AgdaIndent{0}}]%
\>[1]\AgdaKeyword{where}%
\>[655I]\AgdaFunction{⊢ℰλNλN′}\AgdaSpace{}%
\AgdaSymbol{:}%
\>[657I]\AgdaSymbol{∀\{}\AgdaBound{dir}\AgdaSymbol{\}\{}\AgdaBound{γ}\AgdaSymbol{\}\{}\AgdaBound{γ′}\AgdaSymbol{\}}\AgdaSpace{}%
\AgdaSymbol{→}\AgdaSpace{}%
\AgdaSymbol{(}\AgdaBound{Γ}\AgdaSpace{}%
\AgdaOperator{\AgdaFunction{∣}}\AgdaSpace{}%
\AgdaBound{dir}\AgdaSpace{}%
\AgdaOperator{\AgdaFunction{⊨}}\AgdaSpace{}%
\AgdaBound{γ}\AgdaSpace{}%
\AgdaOperator{\AgdaFunction{⊑ᴸᴿ}}\AgdaSpace{}%
\AgdaBound{γ′}\AgdaSymbol{)}\<%
\\
\>[.][@{}l@{}]\<[657I]%
\>[17]\AgdaOperator{\AgdaFunction{⊢ᵒ}}\AgdaSpace{}%
\AgdaSymbol{(}\AgdaBound{dir}\AgdaSpace{}%
\AgdaOperator{\AgdaFunction{∣}}\AgdaSpace{}%
\AgdaOperator{\AgdaFunction{⟪}}\AgdaSpace{}%
\AgdaBound{γ}\AgdaSpace{}%
\AgdaOperator{\AgdaFunction{⟫}}\AgdaSpace{}%
\AgdaSymbol{(}\AgdaInductiveConstructor{ƛ}\AgdaSpace{}%
\AgdaBound{N}\AgdaSymbol{)}\AgdaSpace{}%
\AgdaOperator{\AgdaFunction{⊑ᴸᴿₜ}}\AgdaSpace{}%
\AgdaOperator{\AgdaFunction{⟪}}\AgdaSpace{}%
\AgdaBound{γ′}\AgdaSpace{}%
\AgdaOperator{\AgdaFunction{⟫}}\AgdaSpace{}%
\AgdaSymbol{(}\AgdaInductiveConstructor{ƛ}\AgdaSpace{}%
\AgdaBound{N′}\AgdaSymbol{)}\AgdaSpace{}%
\AgdaOperator{\AgdaFunction{⦂}}\AgdaSpace{}%
\AgdaInductiveConstructor{fun⊑}\AgdaSpace{}%
\AgdaBound{c}\AgdaSpace{}%
\AgdaBound{d}\AgdaSymbol{)}\<%
\\
\>[.][@{}l@{}]\<[655I]%
\>[7]\AgdaFunction{⊢ℰλNλN′}\AgdaSpace{}%
\AgdaSymbol{\{}\AgdaBound{dir}\AgdaSymbol{\}\{}\AgdaBound{γ}\AgdaSymbol{\}\{}\AgdaBound{γ′}\AgdaSymbol{\}}\AgdaSpace{}%
\AgdaSymbol{=}\AgdaSpace{}%
\AgdaFunction{LRᵥ⇒LRₜ}\AgdaSpace{}%
\AgdaSymbol{(}\AgdaFunction{substᵒ}\AgdaSpace{}%
\AgdaSymbol{(}\AgdaFunction{≡ᵒ-sym}\AgdaSpace{}%
\AgdaFunction{LRᵥ-fun}\AgdaSymbol{)}\<%
\\
\>[7][@{}l@{\AgdaIndent{0}}]%
\>[9]\AgdaSymbol{(}\AgdaFunction{Λᵒ[}\AgdaSpace{}%
\AgdaBound{W}\AgdaSpace{}%
\AgdaFunction{]}\AgdaSpace{}%
\AgdaFunction{Λᵒ[}\AgdaSpace{}%
\AgdaBound{W′}\AgdaSpace{}%
\AgdaFunction{]}\AgdaSpace{}%
\AgdaFunction{→ᵒI}\AgdaSpace{}%
\AgdaSymbol{\{}\AgdaArgument{P}\AgdaSpace{}%
\AgdaSymbol{=}\AgdaSpace{}%
\AgdaOperator{\AgdaFunction{▷ᵒ}}\AgdaSpace{}%
\AgdaSymbol{(}\AgdaBound{dir}\AgdaSpace{}%
\AgdaOperator{\AgdaFunction{∣}}\AgdaSpace{}%
\AgdaBound{W}\AgdaSpace{}%
\AgdaOperator{\AgdaFunction{⊑ᴸᴿᵥ}}\AgdaSpace{}%
\AgdaBound{W′}\AgdaSpace{}%
\AgdaOperator{\AgdaFunction{⦂}}\AgdaSpace{}%
\AgdaBound{c}\AgdaSymbol{)\}}\<%
\\
\>[9][@{}l@{\AgdaIndent{0}}]%
\>[11]\AgdaKeyword{let}\AgdaSpace{}%
\AgdaBound{IH}\AgdaSpace{}%
\AgdaSymbol{=}\AgdaSpace{}%
\AgdaSymbol{(}\AgdaFunction{proj}\AgdaSpace{}%
\AgdaBound{dir}\AgdaSpace{}%
\AgdaBound{N}\AgdaSpace{}%
\AgdaBound{N′}\AgdaSpace{}%
\AgdaBound{⊨N⊑N′}\AgdaSymbol{)}\AgdaSpace{}%
\AgdaSymbol{(}\AgdaBound{W}\AgdaSpace{}%
\AgdaOperator{\AgdaFunction{•}}\AgdaSpace{}%
\AgdaBound{γ}\AgdaSymbol{)}\AgdaSpace{}%
\AgdaSymbol{(}\AgdaBound{W′}\AgdaSpace{}%
\AgdaOperator{\AgdaFunction{•}}\AgdaSpace{}%
\AgdaBound{γ′}\AgdaSymbol{)}\AgdaSpace{}%
\AgdaKeyword{in}\<%
\\
\>[11]\AgdaSymbol{(}\AgdaFunction{appᵒ}\AgdaSpace{}%
\AgdaSymbol{(}\AgdaFunction{Sᵒ}\AgdaSpace{}%
\AgdaSymbol{(}\AgdaFunction{▷→}\AgdaSpace{}%
\AgdaSymbol{(}\AgdaFunction{monoᵒ}\AgdaSpace{}%
\AgdaSymbol{(}\AgdaFunction{→ᵒI}\AgdaSpace{}%
\AgdaBound{IH}\AgdaSymbol{))))}\AgdaSpace{}%
\AgdaFunction{Zᵒ}\AgdaSymbol{)))}\<%
\end{code}

We note that the use of SIL in the above proof comes with some
tradeoffs. On the one hand, there is no explicit reasoning about step
indices. On the other hand, there is some added verbosity compared to
a proof in raw Agda such as the use of \textsf{appᵒ} for modus-ponens,
the use of de Bruijn indices \textsf{Zᵒ} to refer to premises,
and extra annotations such as \textsf{\{P = ▷ᵒ (dir ∣ W ⊑ᴸᴿᵥ W′ ⦂ c)\}}
that are needed when Agda's type inference fails.

However, there is a bigger problem regarding incremental proof
development in SIL. It is common practice to create a partial proof
with a hole, written \textsf{?}, and one can ask Agda to print what
need to be proved in the hole. For example, instead of \textsf{(→ᵒI
IH)} in the above proof, one might have started with \textsf{(→ᵒI ?)}.
Unfortunately, Agda's message describing what needs to be proved fills
an entire computer screen because Agda normalizes the SIL formulas
into their Agda encodings. We are working on new version of SIL that
uses the \texttt{abstract} feature of Agda to hide the internals of
SIL from its clients, but that also has its challenges. It seems that
the \texttt{prop} extension is necessary so that the fields of
\textsf{Setᵒ} that contain proofs are ignored when proving equations
such as \textsf{fixpointᵒ}.

\paragraph{Anti-reduction and Bind Lemmas}

The remaining compatibility lemmas, for function application and for
injections and projections, require several anti-reduction lemmas
which state that if two terms are in the logical relation, then
walking backwards with one or both of them yields terms that are still
in the logical relation. We formulated these lemmas with explicit step
indices and the meaning function \textsf{\#} because working with the
raw Agda encoding enables the use of Agda's Auto command for automatic
proof search.

\begin{code}%
\>[0]\AgdaFunction{anti-reduction-≼-R-one}\AgdaSpace{}%
\AgdaSymbol{:}\AgdaSpace{}%
\AgdaSymbol{∀\{}\AgdaBound{A}\AgdaSymbol{\}\{}\AgdaBound{A′}\AgdaSymbol{\}\{}\AgdaBound{c}\AgdaSpace{}%
\AgdaSymbol{:}\AgdaSpace{}%
\AgdaBound{A}\AgdaSpace{}%
\AgdaOperator{\AgdaDatatype{⊑}}\AgdaSpace{}%
\AgdaBound{A′}\AgdaSymbol{\}\{}\AgdaBound{M}\AgdaSymbol{\}\{}\AgdaBound{M′}\AgdaSymbol{\}\{}\AgdaBound{N′}\AgdaSymbol{\}\{}\AgdaBound{i}\AgdaSymbol{\}}\<%
\\
\>[0][@{}l@{\AgdaIndent{0}}]%
\>[2]\AgdaSymbol{→}\AgdaSpace{}%
\AgdaField{\#}\AgdaSymbol{(}\AgdaInductiveConstructor{≼}\AgdaSpace{}%
\AgdaOperator{\AgdaFunction{∣}}\AgdaSpace{}%
\AgdaBound{M}\AgdaSpace{}%
\AgdaOperator{\AgdaFunction{⊑ᴸᴿₜ}}\AgdaSpace{}%
\AgdaBound{N′}\AgdaSpace{}%
\AgdaOperator{\AgdaFunction{⦂}}\AgdaSpace{}%
\AgdaBound{c}\AgdaSymbol{)}\AgdaSpace{}%
\AgdaBound{i}%
\>[28]\AgdaSymbol{→}%
\>[31]\AgdaBound{M′}\AgdaSpace{}%
\AgdaOperator{\AgdaDatatype{⟶}}\AgdaSpace{}%
\AgdaBound{N′}%
\>[40]\AgdaSymbol{→}%
\>[43]\AgdaField{\#}\AgdaSymbol{(}\AgdaInductiveConstructor{≼}\AgdaSpace{}%
\AgdaOperator{\AgdaFunction{∣}}\AgdaSpace{}%
\AgdaBound{M}\AgdaSpace{}%
\AgdaOperator{\AgdaFunction{⊑ᴸᴿₜ}}\AgdaSpace{}%
\AgdaBound{M′}\AgdaSpace{}%
\AgdaOperator{\AgdaFunction{⦂}}\AgdaSpace{}%
\AgdaBound{c}\AgdaSymbol{)}\AgdaSpace{}%
\AgdaBound{i}\<%
\end{code}
\begin{code}[hide]%
\>[0]\AgdaFunction{anti-reduction-≼-R-one}\AgdaSpace{}%
\AgdaSymbol{\{}\AgdaArgument{c}\AgdaSpace{}%
\AgdaSymbol{=}\AgdaSpace{}%
\AgdaBound{c}\AgdaSymbol{\}\{}\AgdaBound{M}\AgdaSymbol{\}\{}\AgdaBound{M′}\AgdaSymbol{\}\{}\AgdaBound{N′}\AgdaSymbol{\}\{}\AgdaInductiveConstructor{zero}\AgdaSymbol{\}}\AgdaSpace{}%
\AgdaBound{ℰMN′}\AgdaSpace{}%
\AgdaBound{M′→N′}\AgdaSpace{}%
\AgdaSymbol{=}\<%
\\
\>[0][@{}l@{\AgdaIndent{0}}]%
\>[2]\AgdaField{tz}\AgdaSpace{}%
\AgdaSymbol{(}\AgdaInductiveConstructor{≼}\AgdaSpace{}%
\AgdaOperator{\AgdaFunction{∣}}\AgdaSpace{}%
\AgdaBound{M}\AgdaSpace{}%
\AgdaOperator{\AgdaFunction{⊑ᴸᴿₜ}}\AgdaSpace{}%
\AgdaBound{M′}\AgdaSpace{}%
\AgdaOperator{\AgdaFunction{⦂}}\AgdaSpace{}%
\AgdaBound{c}\AgdaSymbol{)}\<%
\\
\>[0]\AgdaFunction{anti-reduction-≼-R-one}\AgdaSpace{}%
\AgdaSymbol{\{}\AgdaArgument{c}\AgdaSpace{}%
\AgdaSymbol{=}\AgdaSpace{}%
\AgdaBound{c}\AgdaSymbol{\}\{}\AgdaBound{M}\AgdaSymbol{\}\{}\AgdaBound{M′}\AgdaSymbol{\}\{}\AgdaBound{N′}\AgdaSymbol{\}\{}\AgdaInductiveConstructor{suc}\AgdaSpace{}%
\AgdaBound{i}\AgdaSymbol{\}}\AgdaSpace{}%
\AgdaBound{ℰMN′}\AgdaSpace{}%
\AgdaBound{M′→N′}\<%
\\
\>[0][@{}l@{\AgdaIndent{0}}]%
\>[4]\AgdaKeyword{with}\AgdaSpace{}%
\AgdaBound{ℰMN′}\<%
\\
\>[0]\AgdaSymbol{...}\AgdaSpace{}%
\AgdaSymbol{|}%
\>[769I]\AgdaInductiveConstructor{inj₁}\AgdaSpace{}%
\AgdaSymbol{(}\AgdaBound{N}\AgdaSpace{}%
\AgdaOperator{\AgdaInductiveConstructor{,}}\AgdaSpace{}%
\AgdaBound{M→N}\AgdaSpace{}%
\AgdaOperator{\AgdaInductiveConstructor{,}}\AgdaSpace{}%
\AgdaBound{▷ℰNN′}\AgdaSymbol{)}\AgdaSpace{}%
\AgdaSymbol{=}\<%
\\
\>[769I][@{}l@{\AgdaIndent{0}}]%
\>[9]\AgdaKeyword{let}\AgdaSpace{}%
\AgdaBound{ℰNM′si}\AgdaSpace{}%
\AgdaSymbol{=}\AgdaSpace{}%
\AgdaFunction{anti-reduction-≼-R-one}\AgdaSpace{}%
\AgdaBound{▷ℰNN′}\AgdaSpace{}%
\AgdaBound{M′→N′}\AgdaSpace{}%
\AgdaKeyword{in}\<%
\\
\>[9]\AgdaInductiveConstructor{inj₁}\AgdaSpace{}%
\AgdaSymbol{(}\AgdaBound{N}\AgdaSpace{}%
\AgdaOperator{\AgdaInductiveConstructor{,}}\AgdaSpace{}%
\AgdaBound{M→N}\AgdaSpace{}%
\AgdaOperator{\AgdaInductiveConstructor{,}}\AgdaSpace{}%
\AgdaBound{ℰNM′si}\AgdaSymbol{)}\<%
\\
\>[0]\AgdaSymbol{...}\AgdaSpace{}%
\AgdaSymbol{|}\AgdaSpace{}%
\AgdaInductiveConstructor{inj₂}\AgdaSpace{}%
\AgdaSymbol{(}\AgdaInductiveConstructor{inj₁}\AgdaSpace{}%
\AgdaBound{N′→blame}\AgdaSymbol{)}\AgdaSpace{}%
\AgdaSymbol{=}\AgdaSpace{}%
\AgdaInductiveConstructor{inj₂}\AgdaSpace{}%
\AgdaSymbol{(}\AgdaInductiveConstructor{inj₁}\AgdaSpace{}%
\AgdaSymbol{(}\AgdaFunction{unit}\AgdaSpace{}%
\AgdaBound{M′→N′}\AgdaSpace{}%
\AgdaOperator{\AgdaFunction{++}}\AgdaSpace{}%
\AgdaBound{N′→blame}\AgdaSymbol{))}\<%
\\
\>[0]\AgdaSymbol{...}\AgdaSpace{}%
\AgdaSymbol{|}%
\>[799I]\AgdaInductiveConstructor{inj₂}\AgdaSpace{}%
\AgdaSymbol{(}\AgdaInductiveConstructor{inj₂}\AgdaSpace{}%
\AgdaSymbol{(}\AgdaBound{m}\AgdaSpace{}%
\AgdaOperator{\AgdaInductiveConstructor{,}}\AgdaSpace{}%
\AgdaSymbol{(}\AgdaBound{V′}\AgdaSpace{}%
\AgdaOperator{\AgdaInductiveConstructor{,}}\AgdaSpace{}%
\AgdaBound{N′→V′}\AgdaSpace{}%
\AgdaOperator{\AgdaInductiveConstructor{,}}\AgdaSpace{}%
\AgdaBound{v′}\AgdaSpace{}%
\AgdaOperator{\AgdaInductiveConstructor{,}}\AgdaSpace{}%
\AgdaBound{𝒱MV′}\AgdaSymbol{)))}\AgdaSpace{}%
\AgdaSymbol{=}\<%
\\
\>[.][@{}l@{}]\<[799I]%
\>[6]\AgdaInductiveConstructor{inj₂}\AgdaSpace{}%
\AgdaSymbol{(}\AgdaInductiveConstructor{inj₂}\AgdaSpace{}%
\AgdaSymbol{(}\AgdaBound{m}\AgdaSpace{}%
\AgdaOperator{\AgdaInductiveConstructor{,}}\AgdaSpace{}%
\AgdaSymbol{(}\AgdaBound{V′}\AgdaSpace{}%
\AgdaOperator{\AgdaInductiveConstructor{,}}\AgdaSpace{}%
\AgdaSymbol{(}\AgdaFunction{unit}\AgdaSpace{}%
\AgdaBound{M′→N′}\AgdaSpace{}%
\AgdaOperator{\AgdaFunction{++}}\AgdaSpace{}%
\AgdaBound{N′→V′}\AgdaSymbol{)}\AgdaSpace{}%
\AgdaOperator{\AgdaInductiveConstructor{,}}\AgdaSpace{}%
\AgdaBound{v′}\AgdaSpace{}%
\AgdaOperator{\AgdaInductiveConstructor{,}}\AgdaSpace{}%
\AgdaBound{𝒱MV′}\AgdaSymbol{)))}\<%
\end{code}
\begin{code}%
\>[0]\AgdaFunction{anti-reduction-≼-R}\AgdaSpace{}%
\AgdaSymbol{:}\AgdaSpace{}%
\AgdaSymbol{∀\{}\AgdaBound{A}\AgdaSymbol{\}\{}\AgdaBound{A′}\AgdaSymbol{\}\{}\AgdaBound{c}\AgdaSpace{}%
\AgdaSymbol{:}\AgdaSpace{}%
\AgdaBound{A}\AgdaSpace{}%
\AgdaOperator{\AgdaDatatype{⊑}}\AgdaSpace{}%
\AgdaBound{A′}\AgdaSymbol{\}\{}\AgdaBound{M}\AgdaSymbol{\}\{}\AgdaBound{M′}\AgdaSymbol{\}\{}\AgdaBound{N′}\AgdaSymbol{\}\{}\AgdaBound{i}\AgdaSymbol{\}}\<%
\\
\>[0][@{}l@{\AgdaIndent{0}}]%
\>[2]\AgdaSymbol{→}\AgdaSpace{}%
\AgdaField{\#}\AgdaSymbol{(}\AgdaInductiveConstructor{≼}\AgdaSpace{}%
\AgdaOperator{\AgdaFunction{∣}}\AgdaSpace{}%
\AgdaBound{M}\AgdaSpace{}%
\AgdaOperator{\AgdaFunction{⊑ᴸᴿₜ}}\AgdaSpace{}%
\AgdaBound{N′}\AgdaSpace{}%
\AgdaOperator{\AgdaFunction{⦂}}\AgdaSpace{}%
\AgdaBound{c}\AgdaSymbol{)}\AgdaSpace{}%
\AgdaBound{i}%
\>[28]\AgdaSymbol{→}%
\>[31]\AgdaBound{M′}\AgdaSpace{}%
\AgdaOperator{\AgdaDatatype{↠}}\AgdaSpace{}%
\AgdaBound{N′}%
\>[40]\AgdaSymbol{→}%
\>[43]\AgdaField{\#}\AgdaSymbol{(}\AgdaInductiveConstructor{≼}\AgdaSpace{}%
\AgdaOperator{\AgdaFunction{∣}}\AgdaSpace{}%
\AgdaBound{M}\AgdaSpace{}%
\AgdaOperator{\AgdaFunction{⊑ᴸᴿₜ}}\AgdaSpace{}%
\AgdaBound{M′}\AgdaSpace{}%
\AgdaOperator{\AgdaFunction{⦂}}\AgdaSpace{}%
\AgdaBound{c}\AgdaSymbol{)}\AgdaSpace{}%
\AgdaBound{i}\<%
\end{code}
\begin{code}[hide]%
\>[0]\AgdaFunction{anti-reduction-≼-R}\AgdaSpace{}%
\AgdaSymbol{\{}\AgdaArgument{M′}\AgdaSpace{}%
\AgdaSymbol{=}\AgdaSpace{}%
\AgdaBound{M′}\AgdaSymbol{\}}\AgdaSpace{}%
\AgdaBound{ℰMN′}\AgdaSpace{}%
\AgdaSymbol{(}\AgdaDottedPattern{\AgdaSymbol{.}}\AgdaDottedPattern{\AgdaBound{M′}}\AgdaSpace{}%
\AgdaOperator{\AgdaInductiveConstructor{END}}\AgdaSymbol{)}\AgdaSpace{}%
\AgdaSymbol{=}\AgdaSpace{}%
\AgdaBound{ℰMN′}\<%
\\
\>[0]\AgdaFunction{anti-reduction-≼-R}\AgdaSpace{}%
\AgdaSymbol{\{}\AgdaArgument{M′}\AgdaSpace{}%
\AgdaSymbol{=}\AgdaSpace{}%
\AgdaBound{M′}\AgdaSymbol{\}}\AgdaSpace{}%
\AgdaSymbol{\{}\AgdaBound{N′}\AgdaSymbol{\}}\AgdaSpace{}%
\AgdaSymbol{\{}\AgdaBound{i}\AgdaSymbol{\}}\AgdaSpace{}%
\AgdaBound{ℰMN′}\AgdaSpace{}%
\AgdaSymbol{(}\AgdaDottedPattern{\AgdaSymbol{.}}\AgdaDottedPattern{\AgdaBound{M′}}\AgdaSpace{}%
\AgdaOperator{\AgdaInductiveConstructor{⟶⟨}}\AgdaSpace{}%
\AgdaBound{M′→L′}\AgdaSpace{}%
\AgdaOperator{\AgdaInductiveConstructor{⟩}}\AgdaSpace{}%
\AgdaBound{L′→*N′}\AgdaSymbol{)}\AgdaSpace{}%
\AgdaSymbol{=}\<%
\\
\>[0][@{}l@{\AgdaIndent{0}}]%
\>[2]\AgdaFunction{anti-reduction-≼-R-one}\AgdaSpace{}%
\AgdaSymbol{(}\AgdaFunction{anti-reduction-≼-R}\AgdaSpace{}%
\AgdaBound{ℰMN′}\AgdaSpace{}%
\AgdaBound{L′→*N′}\AgdaSymbol{)}\AgdaSpace{}%
\AgdaBound{M′→L′}\<%
\end{code}
\begin{code}%
\>[0]\AgdaFunction{anti-reduction-≽-R-one}\AgdaSpace{}%
\AgdaSymbol{:}\AgdaSpace{}%
\AgdaSymbol{∀\{}\AgdaBound{A}\AgdaSymbol{\}\{}\AgdaBound{A′}\AgdaSymbol{\}\{}\AgdaBound{c}\AgdaSpace{}%
\AgdaSymbol{:}\AgdaSpace{}%
\AgdaBound{A}\AgdaSpace{}%
\AgdaOperator{\AgdaDatatype{⊑}}\AgdaSpace{}%
\AgdaBound{A′}\AgdaSymbol{\}\{}\AgdaBound{M}\AgdaSymbol{\}\{}\AgdaBound{M′}\AgdaSymbol{\}\{}\AgdaBound{N′}\AgdaSymbol{\}\{}\AgdaBound{i}\AgdaSymbol{\}}\<%
\\
\>[0][@{}l@{\AgdaIndent{0}}]%
\>[2]\AgdaSymbol{→}\AgdaSpace{}%
\AgdaField{\#}\AgdaSymbol{(}\AgdaInductiveConstructor{≽}\AgdaSpace{}%
\AgdaOperator{\AgdaFunction{∣}}\AgdaSpace{}%
\AgdaBound{M}\AgdaSpace{}%
\AgdaOperator{\AgdaFunction{⊑ᴸᴿₜ}}\AgdaSpace{}%
\AgdaBound{N′}\AgdaSpace{}%
\AgdaOperator{\AgdaFunction{⦂}}\AgdaSpace{}%
\AgdaBound{c}\AgdaSymbol{)}\AgdaSpace{}%
\AgdaBound{i}%
\>[28]\AgdaSymbol{→}%
\>[31]\AgdaBound{M′}\AgdaSpace{}%
\AgdaOperator{\AgdaDatatype{⟶}}\AgdaSpace{}%
\AgdaBound{N′}%
\>[40]\AgdaSymbol{→}%
\>[43]\AgdaField{\#}\AgdaSymbol{(}\AgdaInductiveConstructor{≽}\AgdaSpace{}%
\AgdaOperator{\AgdaFunction{∣}}\AgdaSpace{}%
\AgdaBound{M}\AgdaSpace{}%
\AgdaOperator{\AgdaFunction{⊑ᴸᴿₜ}}\AgdaSpace{}%
\AgdaBound{M′}\AgdaSpace{}%
\AgdaOperator{\AgdaFunction{⦂}}\AgdaSpace{}%
\AgdaBound{c}\AgdaSymbol{)}\AgdaSpace{}%
\AgdaSymbol{(}\AgdaInductiveConstructor{suc}\AgdaSpace{}%
\AgdaBound{i}\AgdaSymbol{)}\<%
\end{code}
\begin{code}[hide]%
\>[0]\AgdaFunction{anti-reduction-≽-R-one}\AgdaSpace{}%
\AgdaSymbol{\{}\AgdaArgument{c}\AgdaSpace{}%
\AgdaSymbol{=}\AgdaSpace{}%
\AgdaBound{c}\AgdaSymbol{\}}\AgdaSpace{}%
\AgdaSymbol{\{}\AgdaBound{M}\AgdaSymbol{\}}\AgdaSpace{}%
\AgdaSymbol{\{}\AgdaBound{M′}\AgdaSymbol{\}\{}\AgdaBound{N′}\AgdaSymbol{\}}\AgdaSpace{}%
\AgdaSymbol{\{}\AgdaBound{i}\AgdaSymbol{\}}\AgdaSpace{}%
\AgdaBound{ℰ≽MN′}\AgdaSpace{}%
\AgdaBound{M′→N′}\AgdaSpace{}%
\AgdaSymbol{=}\<%
\\
\>[0][@{}l@{\AgdaIndent{0}}]%
\>[2]\AgdaInductiveConstructor{inj₁}\AgdaSpace{}%
\AgdaSymbol{(}\AgdaBound{N′}\AgdaSpace{}%
\AgdaOperator{\AgdaInductiveConstructor{,}}\AgdaSpace{}%
\AgdaBound{M′→N′}\AgdaSpace{}%
\AgdaOperator{\AgdaInductiveConstructor{,}}\AgdaSpace{}%
\AgdaBound{ℰ≽MN′}\AgdaSymbol{)}\<%
\end{code}
\begin{code}%
\>[0]\AgdaFunction{anti-reduction-≽-L-one}\AgdaSpace{}%
\AgdaSymbol{:}\AgdaSpace{}%
\AgdaSymbol{∀\{}\AgdaBound{A}\AgdaSymbol{\}\{}\AgdaBound{A′}\AgdaSymbol{\}\{}\AgdaBound{c}\AgdaSpace{}%
\AgdaSymbol{:}\AgdaSpace{}%
\AgdaBound{A}\AgdaSpace{}%
\AgdaOperator{\AgdaDatatype{⊑}}\AgdaSpace{}%
\AgdaBound{A′}\AgdaSymbol{\}\{}\AgdaBound{M}\AgdaSymbol{\}\{}\AgdaBound{N}\AgdaSymbol{\}\{}\AgdaBound{M′}\AgdaSymbol{\}\{}\AgdaBound{i}\AgdaSymbol{\}}\<%
\\
\>[0][@{}l@{\AgdaIndent{0}}]%
\>[2]\AgdaSymbol{→}\AgdaSpace{}%
\AgdaField{\#}\AgdaSymbol{(}\AgdaInductiveConstructor{≽}\AgdaSpace{}%
\AgdaOperator{\AgdaFunction{∣}}\AgdaSpace{}%
\AgdaBound{N}\AgdaSpace{}%
\AgdaOperator{\AgdaFunction{⊑ᴸᴿₜ}}\AgdaSpace{}%
\AgdaBound{M′}\AgdaSpace{}%
\AgdaOperator{\AgdaFunction{⦂}}\AgdaSpace{}%
\AgdaBound{c}\AgdaSymbol{)}\AgdaSpace{}%
\AgdaBound{i}%
\>[28]\AgdaSymbol{→}%
\>[31]\AgdaBound{M}\AgdaSpace{}%
\AgdaOperator{\AgdaDatatype{⟶}}\AgdaSpace{}%
\AgdaBound{N}%
\>[38]\AgdaSymbol{→}%
\>[41]\AgdaField{\#}\AgdaSymbol{(}\AgdaInductiveConstructor{≽}\AgdaSpace{}%
\AgdaOperator{\AgdaFunction{∣}}\AgdaSpace{}%
\AgdaBound{M}\AgdaSpace{}%
\AgdaOperator{\AgdaFunction{⊑ᴸᴿₜ}}\AgdaSpace{}%
\AgdaBound{M′}\AgdaSpace{}%
\AgdaOperator{\AgdaFunction{⦂}}\AgdaSpace{}%
\AgdaBound{c}\AgdaSymbol{)}\AgdaSpace{}%
\AgdaBound{i}\<%
\end{code}
\begin{code}[hide]%
\>[0]\AgdaFunction{anti-reduction-≽-L-one}\AgdaSpace{}%
\AgdaSymbol{\{}\AgdaArgument{c}\AgdaSpace{}%
\AgdaSymbol{=}\AgdaSpace{}%
\AgdaBound{c}\AgdaSymbol{\}\{}\AgdaBound{M}\AgdaSymbol{\}}\AgdaSpace{}%
\AgdaSymbol{\{}\AgdaBound{N}\AgdaSymbol{\}\{}\AgdaBound{M′}\AgdaSymbol{\}}\AgdaSpace{}%
\AgdaSymbol{\{}\AgdaInductiveConstructor{zero}\AgdaSymbol{\}}\AgdaSpace{}%
\AgdaBound{ℰNM′}\AgdaSpace{}%
\AgdaBound{M→N}\AgdaSpace{}%
\AgdaSymbol{=}\<%
\\
\>[0][@{}l@{\AgdaIndent{0}}]%
\>[4]\AgdaField{tz}\AgdaSpace{}%
\AgdaSymbol{(}\AgdaInductiveConstructor{≽}\AgdaSpace{}%
\AgdaOperator{\AgdaFunction{∣}}\AgdaSpace{}%
\AgdaBound{M}\AgdaSpace{}%
\AgdaOperator{\AgdaFunction{⊑ᴸᴿₜ}}\AgdaSpace{}%
\AgdaBound{M′}\AgdaSpace{}%
\AgdaOperator{\AgdaFunction{⦂}}\AgdaSpace{}%
\AgdaBound{c}\AgdaSymbol{)}\<%
\\
\>[0]\AgdaFunction{anti-reduction-≽-L-one}\AgdaSpace{}%
\AgdaSymbol{\{}\AgdaArgument{M}\AgdaSpace{}%
\AgdaSymbol{=}\AgdaSpace{}%
\AgdaBound{M}\AgdaSymbol{\}}\AgdaSpace{}%
\AgdaSymbol{\{}\AgdaBound{N}\AgdaSymbol{\}\{}\AgdaBound{M′}\AgdaSymbol{\}}%
\>[40]\AgdaSymbol{\{}\AgdaInductiveConstructor{suc}\AgdaSpace{}%
\AgdaBound{i}\AgdaSymbol{\}}\AgdaSpace{}%
\AgdaBound{ℰNM′}\AgdaSpace{}%
\AgdaBound{M→N}\<%
\\
\>[0][@{}l@{\AgdaIndent{0}}]%
\>[4]\AgdaKeyword{with}\AgdaSpace{}%
\AgdaBound{ℰNM′}\<%
\\
\>[0]\AgdaSymbol{...}\AgdaSpace{}%
\AgdaSymbol{|}%
\>[956I]\AgdaInductiveConstructor{inj₁}\AgdaSpace{}%
\AgdaSymbol{(}\AgdaBound{N′}\AgdaSpace{}%
\AgdaOperator{\AgdaInductiveConstructor{,}}\AgdaSpace{}%
\AgdaBound{M′→N′}\AgdaSpace{}%
\AgdaOperator{\AgdaInductiveConstructor{,}}\AgdaSpace{}%
\AgdaBound{▷ℰMN′}\AgdaSymbol{)}\AgdaSpace{}%
\AgdaSymbol{=}\<%
\\
\>[.][@{}l@{}]\<[956I]%
\>[6]\AgdaInductiveConstructor{inj₁}\AgdaSpace{}%
\AgdaSymbol{(}\AgdaBound{N′}\AgdaSpace{}%
\AgdaOperator{\AgdaInductiveConstructor{,}}\AgdaSpace{}%
\AgdaSymbol{(}\AgdaBound{M′→N′}\AgdaSpace{}%
\AgdaOperator{\AgdaInductiveConstructor{,}}\AgdaSpace{}%
\AgdaSymbol{(}\AgdaFunction{anti-reduction-≽-L-one}\AgdaSpace{}%
\AgdaBound{▷ℰMN′}\AgdaSpace{}%
\AgdaBound{M→N}\AgdaSymbol{)))}\<%
\\
\>[0]\AgdaSymbol{...}\AgdaSpace{}%
\AgdaSymbol{|}\AgdaSpace{}%
\AgdaInductiveConstructor{inj₂}\AgdaSpace{}%
\AgdaSymbol{(}\AgdaInductiveConstructor{inj₁}\AgdaSpace{}%
\AgdaInductiveConstructor{isBlame}\AgdaSymbol{)}\AgdaSpace{}%
\AgdaSymbol{=}\AgdaSpace{}%
\AgdaInductiveConstructor{inj₂}\AgdaSpace{}%
\AgdaSymbol{(}\AgdaInductiveConstructor{inj₁}\AgdaSpace{}%
\AgdaInductiveConstructor{isBlame}\AgdaSymbol{)}\<%
\\
\>[0]\AgdaSymbol{...}\AgdaSpace{}%
\AgdaSymbol{|}%
\>[979I]\AgdaInductiveConstructor{inj₂}\AgdaSpace{}%
\AgdaSymbol{(}\AgdaInductiveConstructor{inj₂}\AgdaSpace{}%
\AgdaSymbol{(}\AgdaBound{m′}\AgdaSpace{}%
\AgdaOperator{\AgdaInductiveConstructor{,}}\AgdaSpace{}%
\AgdaBound{V}\AgdaSpace{}%
\AgdaOperator{\AgdaInductiveConstructor{,}}\AgdaSpace{}%
\AgdaBound{N→V}\AgdaSpace{}%
\AgdaOperator{\AgdaInductiveConstructor{,}}\AgdaSpace{}%
\AgdaBound{v}\AgdaSpace{}%
\AgdaOperator{\AgdaInductiveConstructor{,}}\AgdaSpace{}%
\AgdaBound{𝒱VM′}\AgdaSymbol{))}\AgdaSpace{}%
\AgdaSymbol{=}\<%
\\
\>[.][@{}l@{}]\<[979I]%
\>[6]\AgdaInductiveConstructor{inj₂}\AgdaSpace{}%
\AgdaSymbol{(}\AgdaInductiveConstructor{inj₂}\AgdaSpace{}%
\AgdaSymbol{(}\AgdaBound{m′}\AgdaSpace{}%
\AgdaOperator{\AgdaInductiveConstructor{,}}\AgdaSpace{}%
\AgdaBound{V}\AgdaSpace{}%
\AgdaOperator{\AgdaInductiveConstructor{,}}\AgdaSpace{}%
\AgdaSymbol{(}\AgdaFunction{unit}\AgdaSpace{}%
\AgdaBound{M→N}\AgdaSpace{}%
\AgdaOperator{\AgdaFunction{++}}\AgdaSpace{}%
\AgdaBound{N→V}\AgdaSymbol{)}\AgdaSpace{}%
\AgdaOperator{\AgdaInductiveConstructor{,}}\AgdaSpace{}%
\AgdaBound{v}\AgdaSpace{}%
\AgdaOperator{\AgdaInductiveConstructor{,}}\AgdaSpace{}%
\AgdaBound{𝒱VM′}\AgdaSymbol{))}\<%
\end{code}
\begin{code}%
\>[0]\AgdaFunction{anti-reduction-≽-L}\AgdaSpace{}%
\AgdaSymbol{:}\AgdaSpace{}%
\AgdaSymbol{∀\{}\AgdaBound{A}\AgdaSymbol{\}\{}\AgdaBound{A′}\AgdaSymbol{\}\{}\AgdaBound{c}\AgdaSpace{}%
\AgdaSymbol{:}\AgdaSpace{}%
\AgdaBound{A}\AgdaSpace{}%
\AgdaOperator{\AgdaDatatype{⊑}}\AgdaSpace{}%
\AgdaBound{A′}\AgdaSymbol{\}\{}\AgdaBound{M}\AgdaSymbol{\}\{}\AgdaBound{N}\AgdaSymbol{\}\{}\AgdaBound{M′}\AgdaSymbol{\}\{}\AgdaBound{i}\AgdaSymbol{\}}\<%
\\
\>[0][@{}l@{\AgdaIndent{0}}]%
\>[2]\AgdaSymbol{→}\AgdaSpace{}%
\AgdaField{\#}\AgdaSymbol{(}\AgdaInductiveConstructor{≽}\AgdaSpace{}%
\AgdaOperator{\AgdaFunction{∣}}\AgdaSpace{}%
\AgdaBound{N}\AgdaSpace{}%
\AgdaOperator{\AgdaFunction{⊑ᴸᴿₜ}}\AgdaSpace{}%
\AgdaBound{M′}\AgdaSpace{}%
\AgdaOperator{\AgdaFunction{⦂}}\AgdaSpace{}%
\AgdaBound{c}\AgdaSymbol{)}\AgdaSpace{}%
\AgdaBound{i}%
\>[28]\AgdaSymbol{→}%
\>[31]\AgdaBound{M}\AgdaSpace{}%
\AgdaOperator{\AgdaDatatype{↠}}\AgdaSpace{}%
\AgdaBound{N}%
\>[38]\AgdaSymbol{→}%
\>[41]\AgdaField{\#}\AgdaSymbol{(}\AgdaInductiveConstructor{≽}\AgdaSpace{}%
\AgdaOperator{\AgdaFunction{∣}}\AgdaSpace{}%
\AgdaBound{M}\AgdaSpace{}%
\AgdaOperator{\AgdaFunction{⊑ᴸᴿₜ}}\AgdaSpace{}%
\AgdaBound{M′}\AgdaSpace{}%
\AgdaOperator{\AgdaFunction{⦂}}\AgdaSpace{}%
\AgdaBound{c}\AgdaSymbol{)}\AgdaSpace{}%
\AgdaBound{i}\<%
\end{code}
\begin{code}[hide]%
\>[0]\AgdaFunction{anti-reduction-≽-L}\AgdaSpace{}%
\AgdaSymbol{\{}\AgdaArgument{c}\AgdaSpace{}%
\AgdaSymbol{=}\AgdaSpace{}%
\AgdaBound{c}\AgdaSymbol{\}}\AgdaSpace{}%
\AgdaSymbol{\{}\AgdaBound{M}\AgdaSymbol{\}}\AgdaSpace{}%
\AgdaSymbol{\{}\AgdaDottedPattern{\AgdaSymbol{.}}\AgdaDottedPattern{\AgdaBound{M}}\AgdaSymbol{\}}\AgdaSpace{}%
\AgdaSymbol{\{}\AgdaBound{N′}\AgdaSymbol{\}}\AgdaSpace{}%
\AgdaSymbol{\{}\AgdaBound{i}\AgdaSymbol{\}}\AgdaSpace{}%
\AgdaBound{ℰNM′}\AgdaSpace{}%
\AgdaSymbol{(}\AgdaDottedPattern{\AgdaSymbol{.}}\AgdaDottedPattern{\AgdaBound{M}}\AgdaSpace{}%
\AgdaOperator{\AgdaInductiveConstructor{END}}\AgdaSymbol{)}\AgdaSpace{}%
\AgdaSymbol{=}\AgdaSpace{}%
\AgdaBound{ℰNM′}\<%
\\
\>[0]\AgdaFunction{anti-reduction-≽-L}\AgdaSpace{}%
\AgdaSymbol{\{}\AgdaArgument{c}\AgdaSpace{}%
\AgdaSymbol{=}\AgdaSpace{}%
\AgdaBound{c}\AgdaSymbol{\}}\AgdaSpace{}%
\AgdaSymbol{\{}\AgdaBound{M}\AgdaSymbol{\}}\AgdaSpace{}%
\AgdaSymbol{\{}\AgdaBound{M′}\AgdaSymbol{\}}\AgdaSpace{}%
\AgdaSymbol{\{}\AgdaBound{N′}\AgdaSymbol{\}}\AgdaSpace{}%
\AgdaSymbol{\{}\AgdaBound{i}\AgdaSymbol{\}}\AgdaSpace{}%
\AgdaBound{ℰNM′}\AgdaSpace{}%
\AgdaSymbol{(}\AgdaDottedPattern{\AgdaSymbol{.}}\AgdaDottedPattern{\AgdaBound{M}}\AgdaSpace{}%
\AgdaOperator{\AgdaInductiveConstructor{⟶⟨}}\AgdaSpace{}%
\AgdaBound{M→L}\AgdaSpace{}%
\AgdaOperator{\AgdaInductiveConstructor{⟩}}\AgdaSpace{}%
\AgdaBound{L→*N}\AgdaSymbol{)}\AgdaSpace{}%
\AgdaSymbol{=}\<%
\\
\>[0][@{}l@{\AgdaIndent{0}}]%
\>[2]\AgdaFunction{anti-reduction-≽-L-one}\AgdaSpace{}%
\AgdaSymbol{(}\AgdaFunction{anti-reduction-≽-L}\AgdaSpace{}%
\AgdaBound{ℰNM′}\AgdaSpace{}%
\AgdaBound{L→*N}\AgdaSymbol{)}\AgdaSpace{}%
\AgdaBound{M→L}\<%
\end{code}
\begin{code}%
\>[0]\AgdaFunction{anti-reduction-≼-L-one}\AgdaSpace{}%
\AgdaSymbol{:}\AgdaSpace{}%
\AgdaSymbol{∀\{}\AgdaBound{A}\AgdaSymbol{\}\{}\AgdaBound{A′}\AgdaSymbol{\}\{}\AgdaBound{c}\AgdaSpace{}%
\AgdaSymbol{:}\AgdaSpace{}%
\AgdaBound{A}\AgdaSpace{}%
\AgdaOperator{\AgdaDatatype{⊑}}\AgdaSpace{}%
\AgdaBound{A′}\AgdaSymbol{\}\{}\AgdaBound{M}\AgdaSymbol{\}\{}\AgdaBound{N}\AgdaSymbol{\}\{}\AgdaBound{M′}\AgdaSymbol{\}\{}\AgdaBound{i}\AgdaSymbol{\}}\<%
\\
\>[0][@{}l@{\AgdaIndent{0}}]%
\>[2]\AgdaSymbol{→}\AgdaSpace{}%
\AgdaField{\#}\AgdaSymbol{(}\AgdaInductiveConstructor{≼}\AgdaSpace{}%
\AgdaOperator{\AgdaFunction{∣}}\AgdaSpace{}%
\AgdaBound{N}\AgdaSpace{}%
\AgdaOperator{\AgdaFunction{⊑ᴸᴿₜ}}\AgdaSpace{}%
\AgdaBound{M′}\AgdaSpace{}%
\AgdaOperator{\AgdaFunction{⦂}}\AgdaSpace{}%
\AgdaBound{c}\AgdaSymbol{)}\AgdaSpace{}%
\AgdaBound{i}%
\>[28]\AgdaSymbol{→}%
\>[31]\AgdaBound{M}\AgdaSpace{}%
\AgdaOperator{\AgdaDatatype{⟶}}\AgdaSpace{}%
\AgdaBound{N}%
\>[38]\AgdaSymbol{→}%
\>[41]\AgdaField{\#}\AgdaSymbol{(}\AgdaInductiveConstructor{≼}\AgdaSpace{}%
\AgdaOperator{\AgdaFunction{∣}}\AgdaSpace{}%
\AgdaBound{M}\AgdaSpace{}%
\AgdaOperator{\AgdaFunction{⊑ᴸᴿₜ}}\AgdaSpace{}%
\AgdaBound{M′}\AgdaSpace{}%
\AgdaOperator{\AgdaFunction{⦂}}\AgdaSpace{}%
\AgdaBound{c}\AgdaSymbol{)}\AgdaSpace{}%
\AgdaSymbol{(}\AgdaInductiveConstructor{suc}\AgdaSpace{}%
\AgdaBound{i}\AgdaSymbol{)}\<%
\end{code}
\begin{code}[hide]%
\>[0]\AgdaFunction{anti-reduction-≼-L-one}\AgdaSpace{}%
\AgdaSymbol{\{}\AgdaArgument{c}\AgdaSpace{}%
\AgdaSymbol{=}\AgdaSpace{}%
\AgdaBound{c}\AgdaSymbol{\}}\AgdaSpace{}%
\AgdaSymbol{\{}\AgdaBound{M}\AgdaSymbol{\}}\AgdaSpace{}%
\AgdaSymbol{\{}\AgdaBound{N}\AgdaSymbol{\}}\AgdaSpace{}%
\AgdaSymbol{\{}\AgdaBound{M′}\AgdaSymbol{\}}\AgdaSpace{}%
\AgdaSymbol{\{}\AgdaBound{i}\AgdaSymbol{\}}\AgdaSpace{}%
\AgdaBound{ℰ≼NM′i}\AgdaSpace{}%
\AgdaBound{M→N}\AgdaSpace{}%
\AgdaSymbol{=}\AgdaSpace{}%
\AgdaInductiveConstructor{inj₁}\AgdaSpace{}%
\AgdaSymbol{(}\AgdaBound{N}\AgdaSpace{}%
\AgdaOperator{\AgdaInductiveConstructor{,}}\AgdaSpace{}%
\AgdaBound{M→N}\AgdaSpace{}%
\AgdaOperator{\AgdaInductiveConstructor{,}}\AgdaSpace{}%
\AgdaBound{ℰ≼NM′i}\AgdaSymbol{)}\<%
\end{code}
\begin{code}%
\>[0]\AgdaFunction{anti-reduction}\AgdaSpace{}%
\AgdaSymbol{:}\AgdaSpace{}%
\AgdaSymbol{∀\{}\AgdaBound{A}\AgdaSymbol{\}\{}\AgdaBound{A′}\AgdaSymbol{\}\{}\AgdaBound{c}\AgdaSpace{}%
\AgdaSymbol{:}\AgdaSpace{}%
\AgdaBound{A}\AgdaSpace{}%
\AgdaOperator{\AgdaDatatype{⊑}}\AgdaSpace{}%
\AgdaBound{A′}\AgdaSymbol{\}\{}\AgdaBound{M}\AgdaSymbol{\}\{}\AgdaBound{N}\AgdaSymbol{\}\{}\AgdaBound{M′}\AgdaSymbol{\}\{}\AgdaBound{N′}\AgdaSymbol{\}\{}\AgdaBound{i}\AgdaSymbol{\}\{}\AgdaBound{dir}\AgdaSymbol{\}}\<%
\\
\>[0][@{}l@{\AgdaIndent{0}}]%
\>[2]\AgdaSymbol{→}\AgdaSpace{}%
\AgdaField{\#}\AgdaSymbol{(}\AgdaBound{dir}\AgdaSpace{}%
\AgdaOperator{\AgdaFunction{∣}}\AgdaSpace{}%
\AgdaBound{N}\AgdaSpace{}%
\AgdaOperator{\AgdaFunction{⊑ᴸᴿₜ}}\AgdaSpace{}%
\AgdaBound{N′}\AgdaSpace{}%
\AgdaOperator{\AgdaFunction{⦂}}\AgdaSpace{}%
\AgdaBound{c}\AgdaSymbol{)}\AgdaSpace{}%
\AgdaBound{i}%
\>[30]\AgdaSymbol{→}%
\>[33]\AgdaBound{M}\AgdaSpace{}%
\AgdaOperator{\AgdaDatatype{⟶}}\AgdaSpace{}%
\AgdaBound{N}%
\>[40]\AgdaSymbol{→}%
\>[43]\AgdaBound{M′}\AgdaSpace{}%
\AgdaOperator{\AgdaDatatype{⟶}}\AgdaSpace{}%
\AgdaBound{N′}\<%
\\
\>[2]\AgdaSymbol{→}\AgdaSpace{}%
\AgdaField{\#}\AgdaSymbol{(}\AgdaBound{dir}\AgdaSpace{}%
\AgdaOperator{\AgdaFunction{∣}}\AgdaSpace{}%
\AgdaBound{M}\AgdaSpace{}%
\AgdaOperator{\AgdaFunction{⊑ᴸᴿₜ}}\AgdaSpace{}%
\AgdaBound{M′}\AgdaSpace{}%
\AgdaOperator{\AgdaFunction{⦂}}\AgdaSpace{}%
\AgdaBound{c}\AgdaSymbol{)}\AgdaSpace{}%
\AgdaSymbol{(}\AgdaInductiveConstructor{suc}\AgdaSpace{}%
\AgdaBound{i}\AgdaSymbol{)}\<%
\end{code}
\begin{code}[hide]%
\>[0]\AgdaFunction{anti-reduction}\AgdaSpace{}%
\AgdaSymbol{\{}\AgdaArgument{c}\AgdaSpace{}%
\AgdaSymbol{=}\AgdaSpace{}%
\AgdaBound{c}\AgdaSymbol{\}}\AgdaSpace{}%
\AgdaSymbol{\{}\AgdaBound{M}\AgdaSymbol{\}}\AgdaSpace{}%
\AgdaSymbol{\{}\AgdaBound{N}\AgdaSymbol{\}}\AgdaSpace{}%
\AgdaSymbol{\{}\AgdaBound{M′}\AgdaSymbol{\}}\AgdaSpace{}%
\AgdaSymbol{\{}\AgdaBound{N′}\AgdaSymbol{\}}\AgdaSpace{}%
\AgdaSymbol{\{}\AgdaBound{i}\AgdaSymbol{\}}\AgdaSpace{}%
\AgdaSymbol{\{}\AgdaInductiveConstructor{≼}\AgdaSymbol{\}}\AgdaSpace{}%
\AgdaBound{ℰNN′i}\AgdaSpace{}%
\AgdaBound{M→N}\AgdaSpace{}%
\AgdaBound{M′→N′}\AgdaSpace{}%
\AgdaSymbol{=}\<%
\\
\>[0][@{}l@{\AgdaIndent{0}}]%
\>[2]\AgdaKeyword{let}\AgdaSpace{}%
\AgdaBound{ℰMN′si}\AgdaSpace{}%
\AgdaSymbol{=}\AgdaSpace{}%
\AgdaFunction{anti-reduction-≼-L-one}\AgdaSpace{}%
\AgdaBound{ℰNN′i}\AgdaSpace{}%
\AgdaBound{M→N}\AgdaSpace{}%
\AgdaKeyword{in}\<%
\\
\>[2]\AgdaKeyword{let}\AgdaSpace{}%
\AgdaBound{ℰM′N′si}\AgdaSpace{}%
\AgdaSymbol{=}\AgdaSpace{}%
\AgdaFunction{anti-reduction-≼-R-one}\AgdaSpace{}%
\AgdaBound{ℰMN′si}\AgdaSpace{}%
\AgdaBound{M′→N′}\AgdaSpace{}%
\AgdaKeyword{in}\<%
\\
\>[2]\AgdaBound{ℰM′N′si}\<%
\\
\>[0]\AgdaFunction{anti-reduction}\AgdaSpace{}%
\AgdaSymbol{\{}\AgdaArgument{c}\AgdaSpace{}%
\AgdaSymbol{=}\AgdaSpace{}%
\AgdaBound{c}\AgdaSymbol{\}}\AgdaSpace{}%
\AgdaSymbol{\{}\AgdaBound{M}\AgdaSymbol{\}}\AgdaSpace{}%
\AgdaSymbol{\{}\AgdaBound{N}\AgdaSymbol{\}}\AgdaSpace{}%
\AgdaSymbol{\{}\AgdaBound{M′}\AgdaSymbol{\}}\AgdaSpace{}%
\AgdaSymbol{\{}\AgdaBound{N′}\AgdaSymbol{\}}\AgdaSpace{}%
\AgdaSymbol{\{}\AgdaBound{i}\AgdaSymbol{\}}\AgdaSpace{}%
\AgdaSymbol{\{}\AgdaInductiveConstructor{≽}\AgdaSymbol{\}}\AgdaSpace{}%
\AgdaBound{ℰNN′i}\AgdaSpace{}%
\AgdaBound{M→N}\AgdaSpace{}%
\AgdaBound{M′→N′}\AgdaSpace{}%
\AgdaSymbol{=}\<%
\\
\>[0][@{}l@{\AgdaIndent{0}}]%
\>[2]\AgdaKeyword{let}\AgdaSpace{}%
\AgdaBound{ℰM′Nsi}\AgdaSpace{}%
\AgdaSymbol{=}\AgdaSpace{}%
\AgdaFunction{anti-reduction-≽-R-one}\AgdaSpace{}%
\AgdaBound{ℰNN′i}\AgdaSpace{}%
\AgdaBound{M′→N′}\AgdaSpace{}%
\AgdaKeyword{in}\<%
\\
\>[2]\AgdaKeyword{let}\AgdaSpace{}%
\AgdaBound{ℰM′N′si}\AgdaSpace{}%
\AgdaSymbol{=}\AgdaSpace{}%
\AgdaFunction{anti-reduction-≽-L-one}\AgdaSpace{}%
\AgdaBound{ℰM′Nsi}\AgdaSpace{}%
\AgdaBound{M→N}\AgdaSpace{}%
\AgdaKeyword{in}\<%
\\
\>[2]\AgdaBound{ℰM′N′si}\<%
\end{code}

The remaining compatibility lemmas all involve language features with
subexpressions, and one needs to reason about the reduction of those
subexpressions to values. The following \textsf{LRₜ-bind} lemma
performs that reasoning once and for all. It says that if you are
trying to prove that $N$ ⊑ᴸᴿₜ $N′$, if $M$ is a direct subexpression
of $N$ and $M′$ is a direct subexpression of $N′$, so $N = F ⦉ M ⦊$
and $N′ = F′ ⦉ M′ ⦊$ ($F$ and $F′$ are non-empty frames), then one can
replace $M$ and $M′$ with any related values $V$ ⊑ᴸᴿᵥ $V′$ and it
suffices prove $F ⦉ V ⦊$ ⊑ᴸᴿₜ $F′ ⦉ V′ ⦊$.  The proof of the
\textsf{LRₜ-bind} lemma relies on two of the anti-reduction lemmas.

\begin{code}[hide]%
\>[0]\AgdaFunction{bind-premise}\AgdaSpace{}%
\AgdaSymbol{:}\AgdaSpace{}%
\AgdaDatatype{Dir}\AgdaSpace{}%
\AgdaSymbol{→}\AgdaSpace{}%
\AgdaDatatype{PEFrame}\AgdaSpace{}%
\AgdaSymbol{→}\AgdaSpace{}%
\AgdaDatatype{PEFrame}\AgdaSpace{}%
\AgdaSymbol{→}\AgdaSpace{}%
\AgdaDatatype{Term}\AgdaSpace{}%
\AgdaSymbol{→}\AgdaSpace{}%
\AgdaDatatype{Term}\AgdaSpace{}%
\AgdaSymbol{→}\AgdaSpace{}%
\AgdaDatatype{ℕ}\<%
\\
\>[0][@{}l@{\AgdaIndent{0}}]%
\>[3]\AgdaSymbol{→}\AgdaSpace{}%
\AgdaSymbol{∀}\AgdaSpace{}%
\AgdaSymbol{\{}\AgdaBound{B}\AgdaSymbol{\}\{}\AgdaBound{B′}\AgdaSymbol{\}(}\AgdaBound{c}\AgdaSpace{}%
\AgdaSymbol{:}\AgdaSpace{}%
\AgdaBound{B}\AgdaSpace{}%
\AgdaOperator{\AgdaDatatype{⊑}}\AgdaSpace{}%
\AgdaBound{B′}\AgdaSymbol{)}\AgdaSpace{}%
\AgdaSymbol{→}\AgdaSpace{}%
\AgdaSymbol{∀}\AgdaSpace{}%
\AgdaSymbol{\{}\AgdaBound{A}\AgdaSymbol{\}\{}\AgdaBound{A′}\AgdaSymbol{\}}\AgdaSpace{}%
\AgdaSymbol{(}\AgdaBound{d}\AgdaSpace{}%
\AgdaSymbol{:}\AgdaSpace{}%
\AgdaBound{A}\AgdaSpace{}%
\AgdaOperator{\AgdaDatatype{⊑}}\AgdaSpace{}%
\AgdaBound{A′}\AgdaSymbol{)}\AgdaSpace{}%
\AgdaSymbol{→}\AgdaSpace{}%
\AgdaPrimitive{Set}\<%
\\
\>[0]\AgdaFunction{bind-premise}\AgdaSpace{}%
\AgdaBound{dir}\AgdaSpace{}%
\AgdaBound{F}\AgdaSpace{}%
\AgdaBound{F′}\AgdaSpace{}%
\AgdaBound{M}\AgdaSpace{}%
\AgdaBound{M′}\AgdaSpace{}%
\AgdaBound{i}\AgdaSpace{}%
\AgdaBound{c}\AgdaSpace{}%
\AgdaBound{d}\AgdaSpace{}%
\AgdaSymbol{=}\<%
\\
\>[0][@{}l@{\AgdaIndent{0}}]%
\>[4]\AgdaSymbol{(∀}\AgdaSpace{}%
\AgdaBound{j}\AgdaSpace{}%
\AgdaBound{V}\AgdaSpace{}%
\AgdaBound{V′}\AgdaSpace{}%
\AgdaSymbol{→}\AgdaSpace{}%
\AgdaBound{j}\AgdaSpace{}%
\AgdaOperator{\AgdaDatatype{≤}}\AgdaSpace{}%
\AgdaBound{i}\AgdaSpace{}%
\AgdaSymbol{→}\AgdaSpace{}%
\AgdaBound{M}\AgdaSpace{}%
\AgdaOperator{\AgdaDatatype{↠}}\AgdaSpace{}%
\AgdaBound{V}\AgdaSpace{}%
\AgdaSymbol{→}\AgdaSpace{}%
\AgdaDatatype{Value}\AgdaSpace{}%
\AgdaBound{V}\AgdaSpace{}%
\AgdaSymbol{→}\AgdaSpace{}%
\AgdaBound{M′}\AgdaSpace{}%
\AgdaOperator{\AgdaDatatype{↠}}\AgdaSpace{}%
\AgdaBound{V′}\AgdaSpace{}%
\AgdaSymbol{→}\AgdaSpace{}%
\AgdaDatatype{Value}\AgdaSpace{}%
\AgdaBound{V′}\<%
\\
\>[4][@{}l@{\AgdaIndent{0}}]%
\>[5]\AgdaSymbol{→}\AgdaSpace{}%
\AgdaField{\#}\AgdaSpace{}%
\AgdaSymbol{(}\AgdaBound{dir}\AgdaSpace{}%
\AgdaOperator{\AgdaFunction{∣}}\AgdaSpace{}%
\AgdaBound{V}\AgdaSpace{}%
\AgdaOperator{\AgdaFunction{⊑ᴸᴿᵥ}}\AgdaSpace{}%
\AgdaBound{V′}\AgdaSpace{}%
\AgdaOperator{\AgdaFunction{⦂}}\AgdaSpace{}%
\AgdaBound{d}\AgdaSymbol{)}\AgdaSpace{}%
\AgdaBound{j}\<%
\\
\>[5]\AgdaSymbol{→}\AgdaSpace{}%
\AgdaField{\#}\AgdaSpace{}%
\AgdaSymbol{(}\AgdaBound{dir}\AgdaSpace{}%
\AgdaOperator{\AgdaFunction{∣}}\AgdaSpace{}%
\AgdaSymbol{(}\AgdaBound{F}\AgdaSpace{}%
\AgdaOperator{\AgdaFunction{⦉}}\AgdaSpace{}%
\AgdaBound{V}\AgdaSpace{}%
\AgdaOperator{\AgdaFunction{⦊}}\AgdaSymbol{)}\AgdaSpace{}%
\AgdaOperator{\AgdaFunction{⊑ᴸᴿₜ}}\AgdaSpace{}%
\AgdaSymbol{(}\AgdaBound{F′}\AgdaSpace{}%
\AgdaOperator{\AgdaFunction{⦉}}\AgdaSpace{}%
\AgdaBound{V′}\AgdaSpace{}%
\AgdaOperator{\AgdaFunction{⦊}}\AgdaSymbol{)}\AgdaSpace{}%
\AgdaOperator{\AgdaFunction{⦂}}\AgdaSpace{}%
\AgdaBound{c}\AgdaSymbol{)}\AgdaSpace{}%
\AgdaBound{j}\AgdaSymbol{)}\<%
\end{code}
\begin{code}[hide]%
\>[0]\AgdaFunction{LRᵥ→LRₜ-down-one-≼}\AgdaSpace{}%
\AgdaSymbol{:}\AgdaSpace{}%
\AgdaSymbol{∀\{}\AgdaBound{B}\AgdaSymbol{\}\{}\AgdaBound{B′}\AgdaSymbol{\}\{}\AgdaBound{c}\AgdaSpace{}%
\AgdaSymbol{:}\AgdaSpace{}%
\AgdaBound{B}\AgdaSpace{}%
\AgdaOperator{\AgdaDatatype{⊑}}\AgdaSpace{}%
\AgdaBound{B′}\AgdaSymbol{\}\{}\AgdaBound{A}\AgdaSymbol{\}\{}\AgdaBound{A′}\AgdaSymbol{\}\{}\AgdaBound{d}\AgdaSpace{}%
\AgdaSymbol{:}\AgdaSpace{}%
\AgdaBound{A}\AgdaSpace{}%
\AgdaOperator{\AgdaDatatype{⊑}}\AgdaSpace{}%
\AgdaBound{A′}\AgdaSymbol{\}\{}\AgdaBound{F}\AgdaSymbol{\}\{}\AgdaBound{F′}\AgdaSymbol{\}\{}\AgdaBound{i}\AgdaSymbol{\}\{}\AgdaBound{M}\AgdaSymbol{\}\{}\AgdaBound{N}\AgdaSymbol{\}\{}\AgdaBound{M′}\AgdaSymbol{\}}\<%
\\
\>[0][@{}l@{\AgdaIndent{0}}]%
\>[3]\AgdaSymbol{→}\AgdaSpace{}%
\AgdaBound{M}\AgdaSpace{}%
\AgdaOperator{\AgdaDatatype{⟶}}\AgdaSpace{}%
\AgdaBound{N}\<%
\\
\>[3]\AgdaSymbol{→}\AgdaSpace{}%
\AgdaSymbol{(}\AgdaFunction{bind-premise}\AgdaSpace{}%
\AgdaInductiveConstructor{≼}\AgdaSpace{}%
\AgdaBound{F}\AgdaSpace{}%
\AgdaBound{F′}\AgdaSpace{}%
\AgdaBound{M}\AgdaSpace{}%
\AgdaBound{M′}\AgdaSpace{}%
\AgdaSymbol{(}\AgdaInductiveConstructor{suc}\AgdaSpace{}%
\AgdaBound{i}\AgdaSymbol{)}\AgdaSpace{}%
\AgdaBound{c}\AgdaSpace{}%
\AgdaBound{d}\AgdaSymbol{)}\<%
\\
\>[3]\AgdaSymbol{→}\AgdaSpace{}%
\AgdaSymbol{(}\AgdaFunction{bind-premise}\AgdaSpace{}%
\AgdaInductiveConstructor{≼}\AgdaSpace{}%
\AgdaBound{F}\AgdaSpace{}%
\AgdaBound{F′}\AgdaSpace{}%
\AgdaBound{N}\AgdaSpace{}%
\AgdaBound{M′}\AgdaSpace{}%
\AgdaBound{i}\AgdaSpace{}%
\AgdaBound{c}\AgdaSpace{}%
\AgdaBound{d}\AgdaSymbol{)}\<%
\end{code}
\begin{code}[hide]%
\>[0]\AgdaFunction{LRᵥ→LRₜ-down-one-≼}\AgdaSpace{}%
\AgdaSymbol{\{}\AgdaBound{B}\AgdaSymbol{\}\{}\AgdaBound{B′}\AgdaSymbol{\}\{}\AgdaBound{c}\AgdaSymbol{\}\{}\AgdaBound{A}\AgdaSymbol{\}\{}\AgdaBound{A′}\AgdaSymbol{\}\{}\AgdaBound{d}\AgdaSymbol{\}\{}\AgdaBound{F}\AgdaSymbol{\}\{}\AgdaBound{F′}\AgdaSymbol{\}\{}\AgdaBound{i}\AgdaSymbol{\}\{}\AgdaBound{M}\AgdaSymbol{\}\{}\AgdaBound{N}\AgdaSymbol{\}\{}\AgdaBound{M′}\AgdaSymbol{\}}\AgdaSpace{}%
\AgdaBound{M→N}\AgdaSpace{}%
\AgdaBound{LRᵥ→LRₜsi}\<%
\\
\>[0][@{}l@{\AgdaIndent{0}}]%
\>[3]\AgdaBound{j}\AgdaSpace{}%
\AgdaBound{V}\AgdaSpace{}%
\AgdaBound{V′}\AgdaSpace{}%
\AgdaBound{j≤i}\AgdaSpace{}%
\AgdaBound{M→V}\AgdaSpace{}%
\AgdaBound{v}\AgdaSpace{}%
\AgdaBound{M′→V′}\AgdaSpace{}%
\AgdaBound{v′}\AgdaSpace{}%
\AgdaBound{𝒱j}\AgdaSpace{}%
\AgdaSymbol{=}\<%
\\
\>[3]\AgdaBound{LRᵥ→LRₜsi}\AgdaSpace{}%
\AgdaBound{j}\AgdaSpace{}%
\AgdaBound{V}\AgdaSpace{}%
\AgdaBound{V′}\AgdaSpace{}%
\AgdaSymbol{(}\AgdaFunction{≤-trans}\AgdaSpace{}%
\AgdaBound{j≤i}\AgdaSpace{}%
\AgdaSymbol{(}\AgdaFunction{n≤1+n}\AgdaSpace{}%
\AgdaBound{i}\AgdaSymbol{))}\AgdaSpace{}%
\AgdaSymbol{(}\AgdaBound{M}\AgdaSpace{}%
\AgdaOperator{\AgdaInductiveConstructor{⟶⟨}}\AgdaSpace{}%
\AgdaBound{M→N}\AgdaSpace{}%
\AgdaOperator{\AgdaInductiveConstructor{⟩}}\AgdaSpace{}%
\AgdaBound{M→V}\AgdaSymbol{)}\AgdaSpace{}%
\AgdaBound{v}\AgdaSpace{}%
\AgdaBound{M′→V′}\AgdaSpace{}%
\AgdaBound{v′}\AgdaSpace{}%
\AgdaBound{𝒱j}\<%
\end{code}
\begin{code}[hide]%
\>[0]\AgdaFunction{LRᵥ→LRₜ-down-one-≽}\AgdaSpace{}%
\AgdaSymbol{:}\AgdaSpace{}%
\AgdaSymbol{∀\{}\AgdaBound{B}\AgdaSymbol{\}\{}\AgdaBound{B′}\AgdaSymbol{\}\{}\AgdaBound{c}\AgdaSpace{}%
\AgdaSymbol{:}\AgdaSpace{}%
\AgdaBound{B}\AgdaSpace{}%
\AgdaOperator{\AgdaDatatype{⊑}}\AgdaSpace{}%
\AgdaBound{B′}\AgdaSymbol{\}\{}\AgdaBound{A}\AgdaSymbol{\}\{}\AgdaBound{A′}\AgdaSymbol{\}\{}\AgdaBound{d}\AgdaSpace{}%
\AgdaSymbol{:}\AgdaSpace{}%
\AgdaBound{A}\AgdaSpace{}%
\AgdaOperator{\AgdaDatatype{⊑}}\AgdaSpace{}%
\AgdaBound{A′}\AgdaSymbol{\}\{}\AgdaBound{F}\AgdaSymbol{\}\{}\AgdaBound{F′}\AgdaSymbol{\}\{}\AgdaBound{i}\AgdaSymbol{\}\{}\AgdaBound{M}\AgdaSymbol{\}\{}\AgdaBound{M′}\AgdaSymbol{\}\{}\AgdaBound{N′}\AgdaSymbol{\}}\<%
\\
\>[0][@{}l@{\AgdaIndent{0}}]%
\>[3]\AgdaSymbol{→}\AgdaSpace{}%
\AgdaBound{M′}\AgdaSpace{}%
\AgdaOperator{\AgdaDatatype{⟶}}\AgdaSpace{}%
\AgdaBound{N′}\<%
\\
\>[3]\AgdaSymbol{→}\AgdaSpace{}%
\AgdaSymbol{(}\AgdaFunction{bind-premise}\AgdaSpace{}%
\AgdaInductiveConstructor{≽}\AgdaSpace{}%
\AgdaBound{F}\AgdaSpace{}%
\AgdaBound{F′}\AgdaSpace{}%
\AgdaBound{M}\AgdaSpace{}%
\AgdaBound{M′}\AgdaSpace{}%
\AgdaSymbol{(}\AgdaInductiveConstructor{suc}\AgdaSpace{}%
\AgdaBound{i}\AgdaSymbol{)}\AgdaSpace{}%
\AgdaBound{c}\AgdaSpace{}%
\AgdaBound{d}\AgdaSymbol{)}\<%
\\
\>[3]\AgdaSymbol{→}\AgdaSpace{}%
\AgdaSymbol{(}\AgdaFunction{bind-premise}\AgdaSpace{}%
\AgdaInductiveConstructor{≽}\AgdaSpace{}%
\AgdaBound{F}\AgdaSpace{}%
\AgdaBound{F′}\AgdaSpace{}%
\AgdaBound{M}\AgdaSpace{}%
\AgdaBound{N′}\AgdaSpace{}%
\AgdaBound{i}\AgdaSpace{}%
\AgdaBound{c}\AgdaSpace{}%
\AgdaBound{d}\AgdaSymbol{)}\<%
\end{code}
\begin{code}[hide]%
\>[0]\AgdaFunction{LRᵥ→LRₜ-down-one-≽}\AgdaSpace{}%
\AgdaSymbol{\{}\AgdaBound{B}\AgdaSymbol{\}\{}\AgdaBound{B′}\AgdaSymbol{\}\{}\AgdaBound{c}\AgdaSymbol{\}\{}\AgdaBound{A}\AgdaSymbol{\}\{}\AgdaBound{A′}\AgdaSymbol{\}\{}\AgdaBound{d}\AgdaSymbol{\}\{}\AgdaBound{F}\AgdaSymbol{\}\{}\AgdaBound{F′}\AgdaSymbol{\}\{}\AgdaBound{i}\AgdaSymbol{\}\{}\AgdaBound{M}\AgdaSymbol{\}\{}\AgdaBound{N}\AgdaSymbol{\}\{}\AgdaBound{M′}\AgdaSymbol{\}}\AgdaSpace{}%
\AgdaBound{M′→N′}\AgdaSpace{}%
\AgdaBound{LRᵥ→LRₜsi}\<%
\\
\>[0][@{}l@{\AgdaIndent{0}}]%
\>[3]\AgdaBound{j}\AgdaSpace{}%
\AgdaBound{V}\AgdaSpace{}%
\AgdaBound{V′}\AgdaSpace{}%
\AgdaBound{j≤i}\AgdaSpace{}%
\AgdaBound{M→V}\AgdaSpace{}%
\AgdaBound{v}\AgdaSpace{}%
\AgdaBound{M′→V′}\AgdaSpace{}%
\AgdaBound{v′}\AgdaSpace{}%
\AgdaBound{𝒱j}\AgdaSpace{}%
\AgdaSymbol{=}\<%
\\
\>[3]\AgdaBound{LRᵥ→LRₜsi}\AgdaSpace{}%
\AgdaBound{j}\AgdaSpace{}%
\AgdaBound{V}\AgdaSpace{}%
\AgdaBound{V′}\AgdaSpace{}%
\AgdaSymbol{(}\AgdaFunction{≤-trans}\AgdaSpace{}%
\AgdaBound{j≤i}\AgdaSpace{}%
\AgdaSymbol{(}\AgdaFunction{n≤1+n}\AgdaSpace{}%
\AgdaBound{i}\AgdaSymbol{))}\AgdaSpace{}%
\AgdaBound{M→V}\AgdaSpace{}%
\AgdaBound{v}\AgdaSpace{}%
\AgdaSymbol{(}\AgdaBound{N}\AgdaSpace{}%
\AgdaOperator{\AgdaInductiveConstructor{⟶⟨}}\AgdaSpace{}%
\AgdaBound{M′→N′}\AgdaSpace{}%
\AgdaOperator{\AgdaInductiveConstructor{⟩}}\AgdaSpace{}%
\AgdaBound{M′→V′}\AgdaSymbol{)}\AgdaSpace{}%
\AgdaBound{v′}\AgdaSpace{}%
\AgdaBound{𝒱j}\<%
\end{code}
\begin{code}%
\>[0]\AgdaFunction{LRₜ-bind}\AgdaSpace{}%
\AgdaSymbol{:}\AgdaSpace{}%
\AgdaSymbol{∀\{}\AgdaBound{B}\AgdaSymbol{\}\{}\AgdaBound{B′}\AgdaSymbol{\}\{}\AgdaBound{c}\AgdaSpace{}%
\AgdaSymbol{:}\AgdaSpace{}%
\AgdaBound{B}\AgdaSpace{}%
\AgdaOperator{\AgdaDatatype{⊑}}\AgdaSpace{}%
\AgdaBound{B′}\AgdaSymbol{\}\{}\AgdaBound{A}\AgdaSymbol{\}\{}\AgdaBound{A′}\AgdaSymbol{\}\{}\AgdaBound{d}\AgdaSpace{}%
\AgdaSymbol{:}\AgdaSpace{}%
\AgdaBound{A}\AgdaSpace{}%
\AgdaOperator{\AgdaDatatype{⊑}}\AgdaSpace{}%
\AgdaBound{A′}\AgdaSymbol{\}\{}\AgdaBound{F}\AgdaSymbol{\}\{}\AgdaBound{F′}\AgdaSymbol{\}\{}\AgdaBound{M}\AgdaSymbol{\}\{}\AgdaBound{M′}\AgdaSymbol{\}\{}\AgdaBound{i}\AgdaSymbol{\}\{}\AgdaBound{dir}\AgdaSymbol{\}}\<%
\\
\>[0][@{}l@{\AgdaIndent{0}}]%
\>[3]\AgdaSymbol{→}\AgdaSpace{}%
\AgdaField{\#}\AgdaSymbol{(}\AgdaBound{dir}\AgdaSpace{}%
\AgdaOperator{\AgdaFunction{∣}}\AgdaSpace{}%
\AgdaBound{M}\AgdaSpace{}%
\AgdaOperator{\AgdaFunction{⊑ᴸᴿₜ}}\AgdaSpace{}%
\AgdaBound{M′}\AgdaSpace{}%
\AgdaOperator{\AgdaFunction{⦂}}\AgdaSpace{}%
\AgdaBound{d}\AgdaSymbol{)}\AgdaSpace{}%
\AgdaBound{i}\<%
\\
\>[3]\AgdaSymbol{→}\AgdaSpace{}%
\AgdaSymbol{(∀}%
\>[1395I]\AgdaBound{j}\AgdaSpace{}%
\AgdaBound{V}\AgdaSpace{}%
\AgdaBound{V′}\AgdaSpace{}%
\AgdaSymbol{→}\AgdaSpace{}%
\AgdaBound{j}\AgdaSpace{}%
\AgdaOperator{\AgdaDatatype{≤}}\AgdaSpace{}%
\AgdaBound{i}\AgdaSpace{}%
\AgdaSymbol{→}\AgdaSpace{}%
\AgdaBound{M}\AgdaSpace{}%
\AgdaOperator{\AgdaDatatype{↠}}\AgdaSpace{}%
\AgdaBound{V}\AgdaSpace{}%
\AgdaSymbol{→}\AgdaSpace{}%
\AgdaDatatype{Value}\AgdaSpace{}%
\AgdaBound{V}\AgdaSpace{}%
\AgdaSymbol{→}\AgdaSpace{}%
\AgdaBound{M′}\AgdaSpace{}%
\AgdaOperator{\AgdaDatatype{↠}}\AgdaSpace{}%
\AgdaBound{V′}\AgdaSpace{}%
\AgdaSymbol{→}\AgdaSpace{}%
\AgdaDatatype{Value}\AgdaSpace{}%
\AgdaBound{V′}\<%
\\
\>[1395I][@{}l@{\AgdaIndent{0}}]%
\>[9]\AgdaSymbol{→}\AgdaSpace{}%
\AgdaField{\#}\AgdaSymbol{(}\AgdaBound{dir}\AgdaSpace{}%
\AgdaOperator{\AgdaFunction{∣}}\AgdaSpace{}%
\AgdaBound{V}\AgdaSpace{}%
\AgdaOperator{\AgdaFunction{⊑ᴸᴿᵥ}}\AgdaSpace{}%
\AgdaBound{V′}\AgdaSpace{}%
\AgdaOperator{\AgdaFunction{⦂}}\AgdaSpace{}%
\AgdaBound{d}\AgdaSymbol{)}\AgdaSpace{}%
\AgdaBound{j}\AgdaSpace{}%
\AgdaSymbol{→}\AgdaSpace{}%
\AgdaField{\#}\AgdaSymbol{(}\AgdaBound{dir}\AgdaSpace{}%
\AgdaOperator{\AgdaFunction{∣}}\AgdaSpace{}%
\AgdaSymbol{(}\AgdaBound{F}\AgdaSpace{}%
\AgdaOperator{\AgdaFunction{⦉}}\AgdaSpace{}%
\AgdaBound{V}\AgdaSpace{}%
\AgdaOperator{\AgdaFunction{⦊}}\AgdaSymbol{)}\AgdaSpace{}%
\AgdaOperator{\AgdaFunction{⊑ᴸᴿₜ}}\AgdaSpace{}%
\AgdaSymbol{(}\AgdaBound{F′}\AgdaSpace{}%
\AgdaOperator{\AgdaFunction{⦉}}\AgdaSpace{}%
\AgdaBound{V′}\AgdaSpace{}%
\AgdaOperator{\AgdaFunction{⦊}}\AgdaSymbol{)}\AgdaSpace{}%
\AgdaOperator{\AgdaFunction{⦂}}\AgdaSpace{}%
\AgdaBound{c}\AgdaSymbol{)}\AgdaSpace{}%
\AgdaBound{j}\AgdaSymbol{)}\<%
\\
\>[3]\AgdaSymbol{→}\AgdaSpace{}%
\AgdaField{\#}\AgdaSymbol{(}\AgdaBound{dir}\AgdaSpace{}%
\AgdaOperator{\AgdaFunction{∣}}\AgdaSpace{}%
\AgdaSymbol{(}\AgdaBound{F}\AgdaSpace{}%
\AgdaOperator{\AgdaFunction{⦉}}\AgdaSpace{}%
\AgdaBound{M}\AgdaSpace{}%
\AgdaOperator{\AgdaFunction{⦊}}\AgdaSymbol{)}\AgdaSpace{}%
\AgdaOperator{\AgdaFunction{⊑ᴸᴿₜ}}\AgdaSpace{}%
\AgdaSymbol{(}\AgdaBound{F′}\AgdaSpace{}%
\AgdaOperator{\AgdaFunction{⦉}}\AgdaSpace{}%
\AgdaBound{M′}\AgdaSpace{}%
\AgdaOperator{\AgdaFunction{⦊}}\AgdaSymbol{)}\AgdaSpace{}%
\AgdaOperator{\AgdaFunction{⦂}}\AgdaSpace{}%
\AgdaBound{c}\AgdaSymbol{)}\AgdaSpace{}%
\AgdaBound{i}\<%
\end{code}
\begin{code}[hide]%
\>[0]\AgdaFunction{LRₜ-bind}\AgdaSpace{}%
\AgdaSymbol{\{}\AgdaBound{B}\AgdaSymbol{\}\{}\AgdaBound{B′}\AgdaSymbol{\}\{}\AgdaBound{c}\AgdaSymbol{\}\{}\AgdaBound{A}\AgdaSymbol{\}\{}\AgdaBound{A′}\AgdaSymbol{\}\{}\AgdaBound{d}\AgdaSymbol{\}\{}\AgdaBound{F}\AgdaSymbol{\}}\AgdaSpace{}%
\AgdaSymbol{\{}\AgdaBound{F′}\AgdaSymbol{\}}\AgdaSpace{}%
\AgdaSymbol{\{}\AgdaBound{M}\AgdaSymbol{\}}\AgdaSpace{}%
\AgdaSymbol{\{}\AgdaBound{M′}\AgdaSymbol{\}}\AgdaSpace{}%
\AgdaSymbol{\{}\AgdaInductiveConstructor{zero}\AgdaSymbol{\}}\AgdaSpace{}%
\AgdaSymbol{\{}\AgdaBound{dir}\AgdaSymbol{\}}\AgdaSpace{}%
\AgdaBound{ℰMM′sz}\AgdaSpace{}%
\AgdaBound{LRᵥ→LRₜj}\AgdaSpace{}%
\AgdaSymbol{=}\<%
\\
\>[0][@{}l@{\AgdaIndent{0}}]%
\>[4]\AgdaField{tz}\AgdaSpace{}%
\AgdaSymbol{(}\AgdaBound{dir}\AgdaSpace{}%
\AgdaOperator{\AgdaFunction{∣}}\AgdaSpace{}%
\AgdaSymbol{(}\AgdaBound{F}\AgdaSpace{}%
\AgdaOperator{\AgdaFunction{⦉}}\AgdaSpace{}%
\AgdaBound{M}\AgdaSpace{}%
\AgdaOperator{\AgdaFunction{⦊}}\AgdaSymbol{)}\AgdaSpace{}%
\AgdaOperator{\AgdaFunction{⊑ᴸᴿₜ}}\AgdaSpace{}%
\AgdaSymbol{(}\AgdaBound{F′}\AgdaSpace{}%
\AgdaOperator{\AgdaFunction{⦉}}\AgdaSpace{}%
\AgdaBound{M′}\AgdaSpace{}%
\AgdaOperator{\AgdaFunction{⦊}}\AgdaSymbol{)}\AgdaSpace{}%
\AgdaOperator{\AgdaFunction{⦂}}\AgdaSpace{}%
\AgdaBound{c}\AgdaSymbol{)}\<%
\\
\>[0]\AgdaFunction{LRₜ-bind}\AgdaSpace{}%
\AgdaSymbol{\{}\AgdaBound{B}\AgdaSymbol{\}\{}\AgdaBound{B′}\AgdaSymbol{\}\{}\AgdaBound{c}\AgdaSymbol{\}\{}\AgdaBound{A}\AgdaSymbol{\}\{}\AgdaBound{A′}\AgdaSymbol{\}\{}\AgdaBound{d}\AgdaSymbol{\}\{}\AgdaBound{F}\AgdaSymbol{\}\{}\AgdaBound{F′}\AgdaSymbol{\}\{}\AgdaBound{M}\AgdaSymbol{\}\{}\AgdaBound{M′}\AgdaSymbol{\}\{}\AgdaInductiveConstructor{suc}\AgdaSpace{}%
\AgdaBound{i}\AgdaSymbol{\}\{}\AgdaInductiveConstructor{≼}\AgdaSymbol{\}}\AgdaSpace{}%
\AgdaBound{ℰMM′si}\AgdaSpace{}%
\AgdaBound{LRᵥ→LRₜj}\<%
\\
\>[0][@{}l@{\AgdaIndent{0}}]%
\>[4]\AgdaKeyword{with}\AgdaSpace{}%
\AgdaFunction{⇔-to}\AgdaSpace{}%
\AgdaSymbol{(}\AgdaFunction{LRₜ-suc}\AgdaSymbol{\{}\AgdaArgument{dir}\AgdaSpace{}%
\AgdaSymbol{=}\AgdaSpace{}%
\AgdaInductiveConstructor{≼}\AgdaSymbol{\})}\AgdaSpace{}%
\AgdaBound{ℰMM′si}\<%
\\
\>[0]\AgdaSymbol{...}%
\>[1484I]\AgdaSymbol{|}\AgdaSpace{}%
\AgdaInductiveConstructor{inj₁}\AgdaSpace{}%
\AgdaSymbol{(}\AgdaBound{N}\AgdaSpace{}%
\AgdaOperator{\AgdaInductiveConstructor{,}}\AgdaSpace{}%
\AgdaBound{M→N}\AgdaSpace{}%
\AgdaOperator{\AgdaInductiveConstructor{,}}\AgdaSpace{}%
\AgdaBound{▷ℰNM′}\AgdaSymbol{)}\AgdaSpace{}%
\AgdaSymbol{=}\<%
\\
\>[1484I][@{}l@{\AgdaIndent{0}}]%
\>[5]\AgdaKeyword{let}\AgdaSpace{}%
\AgdaBound{IH}\AgdaSpace{}%
\AgdaSymbol{=}%
\>[1494I]\AgdaFunction{LRₜ-bind}\AgdaSymbol{\{}\AgdaArgument{c}\AgdaSpace{}%
\AgdaSymbol{=}\AgdaSpace{}%
\AgdaBound{c}\AgdaSymbol{\}\{}\AgdaArgument{d}\AgdaSpace{}%
\AgdaSymbol{=}\AgdaSpace{}%
\AgdaBound{d}\AgdaSymbol{\}\{}\AgdaBound{F}\AgdaSymbol{\}\{}\AgdaBound{F′}\AgdaSymbol{\}\{}\AgdaBound{N}\AgdaSymbol{\}\{}\AgdaBound{M′}\AgdaSymbol{\}\{}\AgdaBound{i}\AgdaSymbol{\}\{}\AgdaInductiveConstructor{≼}\AgdaSymbol{\}}\AgdaSpace{}%
\AgdaBound{▷ℰNM′}\<%
\\
\>[1494I][@{}l@{\AgdaIndent{0}}]%
\>[16]\AgdaSymbol{(}\AgdaFunction{LRᵥ→LRₜ-down-one-≼}\AgdaSymbol{\{}\AgdaArgument{c}\AgdaSpace{}%
\AgdaSymbol{=}\AgdaSpace{}%
\AgdaBound{c}\AgdaSymbol{\}\{}\AgdaArgument{d}\AgdaSpace{}%
\AgdaSymbol{=}\AgdaSpace{}%
\AgdaBound{d}\AgdaSymbol{\}\{}\AgdaBound{F}\AgdaSymbol{\}\{}\AgdaBound{F′}\AgdaSymbol{\}\{}\AgdaBound{i}\AgdaSymbol{\}\{}\AgdaBound{M}\AgdaSymbol{\}\{}\AgdaBound{N}\AgdaSymbol{\}\{}\AgdaBound{M′}\AgdaSymbol{\}}\<%
\\
\>[16][@{}l@{\AgdaIndent{0}}]%
\>[21]\AgdaBound{M→N}\AgdaSpace{}%
\AgdaBound{LRᵥ→LRₜj}\AgdaSymbol{)}\AgdaSpace{}%
\AgdaKeyword{in}\<%
\\
\>[5][@{}l@{\AgdaIndent{0}}]%
\>[6]\AgdaFunction{⇔-fro}\AgdaSpace{}%
\AgdaSymbol{(}\AgdaFunction{LRₜ-suc}\AgdaSymbol{\{}\AgdaArgument{dir}\AgdaSpace{}%
\AgdaSymbol{=}\AgdaSpace{}%
\AgdaInductiveConstructor{≼}\AgdaSymbol{\})}\AgdaSpace{}%
\AgdaSymbol{(}\AgdaInductiveConstructor{inj₁}\AgdaSpace{}%
\AgdaSymbol{((}\AgdaBound{F}\AgdaSpace{}%
\AgdaOperator{\AgdaFunction{⦉}}\AgdaSpace{}%
\AgdaBound{N}\AgdaSpace{}%
\AgdaOperator{\AgdaFunction{⦊}}\AgdaSymbol{)}\AgdaSpace{}%
\AgdaOperator{\AgdaInductiveConstructor{,}}\AgdaSpace{}%
\AgdaFunction{ξ′}\AgdaSpace{}%
\AgdaBound{F}\AgdaSpace{}%
\AgdaInductiveConstructor{refl}\AgdaSpace{}%
\AgdaInductiveConstructor{refl}\AgdaSpace{}%
\AgdaBound{M→N}\AgdaSpace{}%
\AgdaOperator{\AgdaInductiveConstructor{,}}\AgdaSpace{}%
\AgdaBound{IH}\AgdaSymbol{))}\<%
\\
\>[0]\AgdaFunction{LRₜ-bind}\AgdaSpace{}%
\AgdaSymbol{\{}\AgdaBound{B}\AgdaSymbol{\}\{}\AgdaBound{B′}\AgdaSymbol{\}\{}\AgdaBound{c}\AgdaSymbol{\}\{}\AgdaBound{A}\AgdaSymbol{\}\{}\AgdaBound{A′}\AgdaSymbol{\}\{}\AgdaBound{d}\AgdaSymbol{\}\{}\AgdaBound{F}\AgdaSymbol{\}\{}\AgdaBound{F′}\AgdaSymbol{\}\{}\AgdaBound{M}\AgdaSymbol{\}\{}\AgdaBound{M′}\AgdaSymbol{\}\{}\AgdaInductiveConstructor{suc}\AgdaSpace{}%
\AgdaBound{i}\AgdaSymbol{\}\{}\AgdaInductiveConstructor{≼}\AgdaSymbol{\}}\AgdaSpace{}%
\AgdaBound{ℰMM′si}\AgdaSpace{}%
\AgdaBound{LRᵥ→LRₜj}\<%
\\
\>[0][@{}l@{\AgdaIndent{0}}]%
\>[4]\AgdaSymbol{|}%
\>[1526I]\AgdaInductiveConstructor{inj₂}\AgdaSpace{}%
\AgdaSymbol{(}\AgdaInductiveConstructor{inj₂}\AgdaSpace{}%
\AgdaSymbol{(}\AgdaBound{m}\AgdaSpace{}%
\AgdaOperator{\AgdaInductiveConstructor{,}}\AgdaSpace{}%
\AgdaSymbol{(}\AgdaBound{V′}\AgdaSpace{}%
\AgdaOperator{\AgdaInductiveConstructor{,}}\AgdaSpace{}%
\AgdaBound{M′→V′}\AgdaSpace{}%
\AgdaOperator{\AgdaInductiveConstructor{,}}\AgdaSpace{}%
\AgdaBound{v′}\AgdaSpace{}%
\AgdaOperator{\AgdaInductiveConstructor{,}}\AgdaSpace{}%
\AgdaBound{𝒱MV′}\AgdaSymbol{)))}\AgdaSpace{}%
\AgdaSymbol{=}\<%
\\
\>[.][@{}l@{}]\<[1526I]%
\>[6]\AgdaKeyword{let}\AgdaSpace{}%
\AgdaBound{ℰFMF′V′}\AgdaSpace{}%
\AgdaSymbol{=}\AgdaSpace{}%
\AgdaBound{LRᵥ→LRₜj}\AgdaSpace{}%
\AgdaSymbol{(}\AgdaInductiveConstructor{suc}\AgdaSpace{}%
\AgdaBound{i}\AgdaSymbol{)}\AgdaSpace{}%
\AgdaBound{M}\AgdaSpace{}%
\AgdaBound{V′}\AgdaSpace{}%
\AgdaFunction{≤-refl}\AgdaSpace{}%
\AgdaSymbol{(}\AgdaBound{M}\AgdaSpace{}%
\AgdaOperator{\AgdaInductiveConstructor{END}}\AgdaSymbol{)}\AgdaSpace{}%
\AgdaBound{m}\AgdaSpace{}%
\AgdaBound{M′→V′}\AgdaSpace{}%
\AgdaBound{v′}\AgdaSpace{}%
\AgdaBound{𝒱MV′}\AgdaSpace{}%
\AgdaKeyword{in}\<%
\\
\>[6]\AgdaFunction{anti-reduction-≼-R}\AgdaSpace{}%
\AgdaBound{ℰFMF′V′}\AgdaSpace{}%
\AgdaSymbol{(}\AgdaFunction{ξ′*}\AgdaSpace{}%
\AgdaBound{F′}\AgdaSpace{}%
\AgdaBound{M′→V′}\AgdaSymbol{)}\<%
\\
\>[0]\AgdaFunction{LRₜ-bind}\AgdaSpace{}%
\AgdaSymbol{\{}\AgdaBound{B}\AgdaSymbol{\}\{}\AgdaBound{B′}\AgdaSymbol{\}\{}\AgdaBound{c}\AgdaSymbol{\}\{}\AgdaBound{A}\AgdaSymbol{\}\{}\AgdaBound{A′}\AgdaSymbol{\}\{}\AgdaBound{d}\AgdaSymbol{\}\{}\AgdaBound{F}\AgdaSymbol{\}\{}\AgdaBound{F′}\AgdaSymbol{\}\{}\AgdaBound{M}\AgdaSymbol{\}\{}\AgdaBound{M′}\AgdaSymbol{\}\{}\AgdaInductiveConstructor{suc}\AgdaSpace{}%
\AgdaBound{i}\AgdaSymbol{\}\{}\AgdaInductiveConstructor{≼}\AgdaSymbol{\}}\AgdaSpace{}%
\AgdaBound{ℰMM′si}\AgdaSpace{}%
\AgdaBound{LRᵥ→LRₜj}\<%
\\
\>[0][@{}l@{\AgdaIndent{0}}]%
\>[4]\AgdaSymbol{|}\AgdaSpace{}%
\AgdaInductiveConstructor{inj₂}\AgdaSpace{}%
\AgdaSymbol{(}\AgdaInductiveConstructor{inj₁}\AgdaSpace{}%
\AgdaBound{M′→blame}\AgdaSymbol{)}\AgdaSpace{}%
\AgdaSymbol{=}\AgdaSpace{}%
\AgdaInductiveConstructor{inj₂}\AgdaSpace{}%
\AgdaSymbol{(}\AgdaInductiveConstructor{inj₁}\AgdaSpace{}%
\AgdaSymbol{(}\AgdaFunction{ξ-blame₃}\AgdaSpace{}%
\AgdaBound{F′}\AgdaSpace{}%
\AgdaBound{M′→blame}\AgdaSpace{}%
\AgdaInductiveConstructor{refl}\AgdaSymbol{))}\<%
\\
\>[0]\AgdaFunction{LRₜ-bind}\AgdaSpace{}%
\AgdaSymbol{\{}\AgdaBound{B}\AgdaSymbol{\}\{}\AgdaBound{B′}\AgdaSymbol{\}\{}\AgdaBound{c}\AgdaSymbol{\}\{}\AgdaBound{A}\AgdaSymbol{\}\{}\AgdaBound{A′}\AgdaSymbol{\}\{}\AgdaBound{d}\AgdaSymbol{\}\{}\AgdaBound{F}\AgdaSymbol{\}\{}\AgdaBound{F′}\AgdaSymbol{\}\{}\AgdaBound{M}\AgdaSymbol{\}\{}\AgdaBound{M′}\AgdaSymbol{\}\{}\AgdaInductiveConstructor{suc}\AgdaSpace{}%
\AgdaBound{i}\AgdaSymbol{\}\{}\AgdaInductiveConstructor{≽}\AgdaSymbol{\}}\AgdaSpace{}%
\AgdaBound{ℰMM′si}\AgdaSpace{}%
\AgdaBound{LRᵥ→LRₜj}\<%
\\
\>[0][@{}l@{\AgdaIndent{0}}]%
\>[4]\AgdaKeyword{with}\AgdaSpace{}%
\AgdaFunction{⇔-to}\AgdaSpace{}%
\AgdaSymbol{(}\AgdaFunction{LRₜ-suc}\AgdaSymbol{\{}\AgdaArgument{dir}\AgdaSpace{}%
\AgdaSymbol{=}\AgdaSpace{}%
\AgdaInductiveConstructor{≽}\AgdaSymbol{\})}\AgdaSpace{}%
\AgdaBound{ℰMM′si}\<%
\\
\>[0]\AgdaSymbol{...}\AgdaSpace{}%
\AgdaSymbol{|}%
\>[1581I]\AgdaInductiveConstructor{inj₁}\AgdaSpace{}%
\AgdaSymbol{(}\AgdaBound{N′}\AgdaSpace{}%
\AgdaOperator{\AgdaInductiveConstructor{,}}\AgdaSpace{}%
\AgdaBound{M′→N′}\AgdaSpace{}%
\AgdaOperator{\AgdaInductiveConstructor{,}}\AgdaSpace{}%
\AgdaBound{▷ℰMN′}\AgdaSymbol{)}\AgdaSpace{}%
\AgdaSymbol{=}\<%
\\
\>[.][@{}l@{}]\<[1581I]%
\>[6]\AgdaKeyword{let}%
\>[1588I]\AgdaBound{ℰFMFN′}\AgdaSpace{}%
\AgdaSymbol{:}\AgdaSpace{}%
\AgdaField{\#}\AgdaSpace{}%
\AgdaSymbol{(}\AgdaInductiveConstructor{≽}\AgdaSpace{}%
\AgdaOperator{\AgdaFunction{∣}}\AgdaSpace{}%
\AgdaSymbol{(}\AgdaBound{F}\AgdaSpace{}%
\AgdaOperator{\AgdaFunction{⦉}}\AgdaSpace{}%
\AgdaBound{M}\AgdaSpace{}%
\AgdaOperator{\AgdaFunction{⦊}}\AgdaSymbol{)}\AgdaSpace{}%
\AgdaOperator{\AgdaFunction{⊑ᴸᴿₜ}}\AgdaSpace{}%
\AgdaSymbol{(}\AgdaBound{F′}\AgdaSpace{}%
\AgdaOperator{\AgdaFunction{⦉}}\AgdaSpace{}%
\AgdaBound{N′}\AgdaSpace{}%
\AgdaOperator{\AgdaFunction{⦊}}\AgdaSymbol{)}\AgdaSpace{}%
\AgdaOperator{\AgdaFunction{⦂}}\AgdaSpace{}%
\AgdaBound{c}\AgdaSymbol{)}\AgdaSpace{}%
\AgdaBound{i}\<%
\\
\>[.][@{}l@{}]\<[1588I]%
\>[10]\AgdaBound{ℰFMFN′}\AgdaSpace{}%
\AgdaSymbol{=}%
\>[1606I]\AgdaFunction{LRₜ-bind}\AgdaSymbol{\{}\AgdaArgument{c}\AgdaSpace{}%
\AgdaSymbol{=}\AgdaSpace{}%
\AgdaBound{c}\AgdaSymbol{\}\{}\AgdaArgument{d}\AgdaSpace{}%
\AgdaSymbol{=}\AgdaSpace{}%
\AgdaBound{d}\AgdaSymbol{\}\{}\AgdaBound{F}\AgdaSymbol{\}\{}\AgdaBound{F′}\AgdaSymbol{\}\{}\AgdaBound{M}\AgdaSymbol{\}\{}\AgdaBound{N′}\AgdaSymbol{\}\{}\AgdaBound{i}\AgdaSymbol{\}\{}\AgdaInductiveConstructor{≽}\AgdaSymbol{\}}\AgdaSpace{}%
\AgdaBound{▷ℰMN′}\<%
\\
\>[.][@{}l@{}]\<[1606I]%
\>[19]\AgdaSymbol{(}\AgdaFunction{LRᵥ→LRₜ-down-one-≽}\AgdaSymbol{\{}\AgdaArgument{c}\AgdaSpace{}%
\AgdaSymbol{=}\AgdaSpace{}%
\AgdaBound{c}\AgdaSymbol{\}\{}\AgdaArgument{d}\AgdaSpace{}%
\AgdaSymbol{=}\AgdaSpace{}%
\AgdaBound{d}\AgdaSymbol{\}\{}\AgdaBound{F}\AgdaSymbol{\}\{}\AgdaBound{F′}\AgdaSymbol{\}}\AgdaSpace{}%
\AgdaBound{M′→N′}\AgdaSpace{}%
\AgdaBound{LRᵥ→LRₜj}\AgdaSymbol{)}\AgdaSpace{}%
\AgdaKeyword{in}\<%
\\
\>[6]\AgdaInductiveConstructor{inj₁}\AgdaSpace{}%
\AgdaSymbol{((}\AgdaBound{F′}\AgdaSpace{}%
\AgdaOperator{\AgdaFunction{⦉}}\AgdaSpace{}%
\AgdaBound{N′}\AgdaSpace{}%
\AgdaOperator{\AgdaFunction{⦊}}\AgdaSymbol{)}\AgdaSpace{}%
\AgdaOperator{\AgdaInductiveConstructor{,}}\AgdaSpace{}%
\AgdaSymbol{(}\AgdaFunction{ξ′}\AgdaSpace{}%
\AgdaBound{F′}\AgdaSpace{}%
\AgdaInductiveConstructor{refl}\AgdaSpace{}%
\AgdaInductiveConstructor{refl}\AgdaSpace{}%
\AgdaBound{M′→N′}\AgdaSymbol{)}\AgdaSpace{}%
\AgdaOperator{\AgdaInductiveConstructor{,}}\AgdaSpace{}%
\AgdaBound{ℰFMFN′}\AgdaSymbol{)}\<%
\\
\>[0]\AgdaSymbol{...}%
\>[1631I]\AgdaSymbol{|}\AgdaSpace{}%
\AgdaInductiveConstructor{inj₂}\AgdaSpace{}%
\AgdaSymbol{(}\AgdaInductiveConstructor{inj₁}\AgdaSpace{}%
\AgdaInductiveConstructor{isBlame}\AgdaSymbol{)}\<%
\\
\>[.][@{}l@{}]\<[1631I]%
\>[4]\AgdaKeyword{with}\AgdaSpace{}%
\AgdaBound{F′}\<%
\\
\>[0]\AgdaSymbol{...}\AgdaSpace{}%
\AgdaSymbol{|}\AgdaSpace{}%
\AgdaInductiveConstructor{□}\AgdaSpace{}%
\AgdaSymbol{=}\AgdaSpace{}%
\AgdaInductiveConstructor{inj₂}\AgdaSpace{}%
\AgdaSymbol{(}\AgdaInductiveConstructor{inj₁}\AgdaSpace{}%
\AgdaInductiveConstructor{isBlame}\AgdaSymbol{)}\<%
\\
\>[0]\AgdaSymbol{...}\AgdaSpace{}%
\AgdaSymbol{|}\AgdaSpace{}%
\AgdaOperator{\AgdaInductiveConstructor{`}}\AgdaSpace{}%
\AgdaBound{F″}\AgdaSpace{}%
\AgdaSymbol{=}\AgdaSpace{}%
\AgdaInductiveConstructor{inj₁}\AgdaSpace{}%
\AgdaSymbol{(}\AgdaInductiveConstructor{blame}\AgdaSpace{}%
\AgdaOperator{\AgdaInductiveConstructor{,}}\AgdaSpace{}%
\AgdaInductiveConstructor{ξ-blame}\AgdaSpace{}%
\AgdaBound{F″}\AgdaSpace{}%
\AgdaOperator{\AgdaInductiveConstructor{,}}\AgdaSpace{}%
\AgdaFunction{LRₜ-blame-step}\AgdaSymbol{\{}\AgdaArgument{dir}\AgdaSpace{}%
\AgdaSymbol{=}\AgdaSpace{}%
\AgdaInductiveConstructor{≽}\AgdaSymbol{\})}\<%
\\
\>[0]\AgdaFunction{LRₜ-bind}\AgdaSpace{}%
\AgdaSymbol{\{}\AgdaBound{B}\AgdaSymbol{\}\{}\AgdaBound{B′}\AgdaSymbol{\}\{}\AgdaBound{c}\AgdaSymbol{\}\{}\AgdaBound{A}\AgdaSymbol{\}\{}\AgdaBound{A′}\AgdaSymbol{\}\{}\AgdaBound{d}\AgdaSymbol{\}\{}\AgdaBound{F}\AgdaSymbol{\}\{}\AgdaBound{F′}\AgdaSymbol{\}\{}\AgdaBound{M}\AgdaSymbol{\}\{}\AgdaBound{M′}\AgdaSymbol{\}\{}\AgdaInductiveConstructor{suc}\AgdaSpace{}%
\AgdaBound{i}\AgdaSymbol{\}\{}\AgdaInductiveConstructor{≽}\AgdaSymbol{\}}\AgdaSpace{}%
\AgdaBound{ℰMM′si}\AgdaSpace{}%
\AgdaBound{LRᵥ→LRₜj}\<%
\\
\>[0][@{}l@{\AgdaIndent{0}}]%
\>[4]\AgdaSymbol{|}\AgdaSpace{}%
\AgdaInductiveConstructor{inj₂}\AgdaSpace{}%
\AgdaSymbol{(}\AgdaInductiveConstructor{inj₂}\AgdaSpace{}%
\AgdaSymbol{(}\AgdaBound{m′}\AgdaSpace{}%
\AgdaOperator{\AgdaInductiveConstructor{,}}\AgdaSpace{}%
\AgdaBound{V}\AgdaSpace{}%
\AgdaOperator{\AgdaInductiveConstructor{,}}\AgdaSpace{}%
\AgdaBound{M→V}\AgdaSpace{}%
\AgdaOperator{\AgdaInductiveConstructor{,}}\AgdaSpace{}%
\AgdaBound{v}\AgdaSpace{}%
\AgdaOperator{\AgdaInductiveConstructor{,}}\AgdaSpace{}%
\AgdaBound{𝒱VM′}\AgdaSymbol{))}\AgdaSpace{}%
\AgdaSymbol{=}\<%
\\
\>[4]\AgdaKeyword{let}\AgdaSpace{}%
\AgdaBound{xx}\AgdaSpace{}%
\AgdaSymbol{=}\AgdaSpace{}%
\AgdaBound{LRᵥ→LRₜj}\AgdaSpace{}%
\AgdaSymbol{(}\AgdaInductiveConstructor{suc}\AgdaSpace{}%
\AgdaBound{i}\AgdaSymbol{)}\AgdaSpace{}%
\AgdaBound{V}\AgdaSpace{}%
\AgdaBound{M′}\AgdaSpace{}%
\AgdaFunction{≤-refl}\AgdaSpace{}%
\AgdaBound{M→V}\AgdaSpace{}%
\AgdaBound{v}\AgdaSpace{}%
\AgdaSymbol{(}\AgdaBound{M′}\AgdaSpace{}%
\AgdaOperator{\AgdaInductiveConstructor{END}}\AgdaSymbol{)}\AgdaSpace{}%
\AgdaBound{m′}\AgdaSpace{}%
\AgdaBound{𝒱VM′}\AgdaSpace{}%
\AgdaKeyword{in}\<%
\\
\>[4]\AgdaFunction{anti-reduction-≽-L}\AgdaSpace{}%
\AgdaBound{xx}\AgdaSpace{}%
\AgdaSymbol{(}\AgdaFunction{ξ′*}\AgdaSpace{}%
\AgdaBound{F}\AgdaSpace{}%
\AgdaBound{M→V}\AgdaSymbol{)}\<%
\end{code}

\paragraph{Compatibility for Application}

Here is where the logical relation demonstrates its worth.
Using the \textsf{LRₜ-bind} lemma twice, we go from needing
to prove $L · M$ ⊑ᴸᴿₜ $L′ · M′$ to 
$V · W$ ⊑ᴸᴿₜ $V′ · W′$. Then because $V$ ⊑ᴸᴿᵥ $V′$ at
function type, the logical relation tells us that
$V = λN$, $V′ = λN′$, and \textsf{N[W]} ⊑ᴸᴿₜ \textsf{N′[W′]}
at one step later in time. So we back up one step of β-reduction
using \textsf{anti-reduction} to show that 
\textsf{(λN) · W} ⊑ᴸᴿₜ \textsf{(λN′) · W′}.

\begin{code}[hide]%
\>[0]\AgdaFunction{LRᵥ-fun-elim-step}\AgdaSpace{}%
\AgdaSymbol{:}\AgdaSpace{}%
\AgdaSymbol{∀\{}\AgdaBound{A}\AgdaSymbol{\}\{}\AgdaBound{B}\AgdaSymbol{\}\{}\AgdaBound{A′}\AgdaSymbol{\}\{}\AgdaBound{B′}\AgdaSymbol{\}\{}\AgdaBound{c}\AgdaSpace{}%
\AgdaSymbol{:}\AgdaSpace{}%
\AgdaBound{A}\AgdaSpace{}%
\AgdaOperator{\AgdaDatatype{⊑}}\AgdaSpace{}%
\AgdaBound{A′}\AgdaSymbol{\}\{}\AgdaBound{d}\AgdaSpace{}%
\AgdaSymbol{:}\AgdaSpace{}%
\AgdaBound{B}\AgdaSpace{}%
\AgdaOperator{\AgdaDatatype{⊑}}\AgdaSpace{}%
\AgdaBound{B′}\AgdaSymbol{\}\{}\AgdaBound{V}\AgdaSymbol{\}\{}\AgdaBound{V′}\AgdaSymbol{\}\{}\AgdaBound{dir}\AgdaSymbol{\}\{}\AgdaBound{k}\AgdaSymbol{\}\{}\AgdaBound{j}\AgdaSymbol{\}}\<%
\\
\>[0][@{}l@{\AgdaIndent{0}}]%
\>[2]\AgdaSymbol{→}\AgdaSpace{}%
\AgdaField{\#}\AgdaSymbol{(}\AgdaBound{dir}\AgdaSpace{}%
\AgdaOperator{\AgdaFunction{∣}}\AgdaSpace{}%
\AgdaBound{V}\AgdaSpace{}%
\AgdaOperator{\AgdaFunction{⊑ᴸᴿᵥ}}\AgdaSpace{}%
\AgdaBound{V′}\AgdaSpace{}%
\AgdaOperator{\AgdaFunction{⦂}}\AgdaSpace{}%
\AgdaInductiveConstructor{fun⊑}\AgdaSpace{}%
\AgdaBound{c}\AgdaSpace{}%
\AgdaBound{d}\AgdaSymbol{)}\AgdaSpace{}%
\AgdaSymbol{(}\AgdaInductiveConstructor{suc}\AgdaSpace{}%
\AgdaBound{k}\AgdaSymbol{)}\<%
\\
\>[2]\AgdaSymbol{→}\AgdaSpace{}%
\AgdaBound{j}\AgdaSpace{}%
\AgdaOperator{\AgdaDatatype{≤}}\AgdaSpace{}%
\AgdaBound{k}\<%
\\
\>[2]\AgdaSymbol{→}%
\>[1714I]\AgdaFunction{∃[}\AgdaSpace{}%
\AgdaBound{N}\AgdaSpace{}%
\AgdaFunction{]}\AgdaSpace{}%
\AgdaFunction{∃[}\AgdaSpace{}%
\AgdaBound{N′}\AgdaSpace{}%
\AgdaFunction{]}\AgdaSpace{}%
\AgdaBound{V}\AgdaSpace{}%
\AgdaOperator{\AgdaDatatype{≡}}\AgdaSpace{}%
\AgdaInductiveConstructor{ƛ}\AgdaSpace{}%
\AgdaBound{N}\AgdaSpace{}%
\AgdaOperator{\AgdaFunction{×}}\AgdaSpace{}%
\AgdaBound{V′}\AgdaSpace{}%
\AgdaOperator{\AgdaDatatype{≡}}\AgdaSpace{}%
\AgdaInductiveConstructor{ƛ}\AgdaSpace{}%
\AgdaBound{N′}\<%
\\
\>[1714I][@{}l@{\AgdaIndent{0}}]%
\>[6]\AgdaOperator{\AgdaFunction{×}}\AgdaSpace{}%
\AgdaSymbol{(∀\{}\AgdaBound{W}\AgdaSpace{}%
\AgdaBound{W′}\AgdaSymbol{\}}%
\>[1731I]\AgdaSymbol{→}\AgdaSpace{}%
\AgdaField{\#}\AgdaSpace{}%
\AgdaSymbol{(}\AgdaBound{dir}\AgdaSpace{}%
\AgdaOperator{\AgdaFunction{∣}}\AgdaSpace{}%
\AgdaBound{W}\AgdaSpace{}%
\AgdaOperator{\AgdaFunction{⊑ᴸᴿᵥ}}\AgdaSpace{}%
\AgdaBound{W′}\AgdaSpace{}%
\AgdaOperator{\AgdaFunction{⦂}}\AgdaSpace{}%
\AgdaBound{c}\AgdaSymbol{)}\AgdaSpace{}%
\AgdaBound{j}\<%
\\
\>[.][@{}l@{}]\<[1731I]%
\>[17]\AgdaSymbol{→}\AgdaSpace{}%
\AgdaField{\#}\AgdaSpace{}%
\AgdaSymbol{(}\AgdaBound{dir}\AgdaSpace{}%
\AgdaOperator{\AgdaFunction{∣}}\AgdaSpace{}%
\AgdaSymbol{(}\AgdaBound{N}\AgdaSpace{}%
\AgdaOperator{\AgdaFunction{[}}\AgdaSpace{}%
\AgdaBound{W}\AgdaSpace{}%
\AgdaOperator{\AgdaFunction{]}}\AgdaSymbol{)}\AgdaSpace{}%
\AgdaOperator{\AgdaFunction{⊑ᴸᴿₜ}}\AgdaSpace{}%
\AgdaSymbol{(}\AgdaBound{N′}\AgdaSpace{}%
\AgdaOperator{\AgdaFunction{[}}\AgdaSpace{}%
\AgdaBound{W′}\AgdaSpace{}%
\AgdaOperator{\AgdaFunction{]}}\AgdaSymbol{)}\AgdaSpace{}%
\AgdaOperator{\AgdaFunction{⦂}}\AgdaSpace{}%
\AgdaBound{d}\AgdaSymbol{)}\AgdaSpace{}%
\AgdaBound{j}\AgdaSymbol{)}\<%
\\
\>[0]\AgdaFunction{LRᵥ-fun-elim-step}\AgdaSpace{}%
\AgdaSymbol{\{}\AgdaBound{A}\AgdaSymbol{\}\{}\AgdaBound{B}\AgdaSymbol{\}\{}\AgdaBound{A′}\AgdaSymbol{\}\{}\AgdaBound{B′}\AgdaSymbol{\}\{}\AgdaBound{c}\AgdaSymbol{\}\{}\AgdaBound{d}\AgdaSymbol{\}\{}\AgdaInductiveConstructor{ƛ}\AgdaSpace{}%
\AgdaBound{N}\AgdaSymbol{\}\{}\AgdaInductiveConstructor{ƛ}\AgdaSpace{}%
\AgdaBound{N′}\AgdaSymbol{\}\{}\AgdaBound{dir}\AgdaSymbol{\}\{}\AgdaBound{k}\AgdaSymbol{\}\{}\AgdaBound{j}\AgdaSymbol{\}}\AgdaSpace{}%
\AgdaBound{𝒱VV′}\AgdaSpace{}%
\AgdaBound{j≤k}\AgdaSpace{}%
\AgdaSymbol{=}\<%
\\
\>[0][@{}l@{\AgdaIndent{0}}]%
\>[2]\AgdaBound{N}%
\>[1762I]\AgdaOperator{\AgdaInductiveConstructor{,}}\AgdaSpace{}%
\AgdaBound{N′}\AgdaSpace{}%
\AgdaOperator{\AgdaInductiveConstructor{,}}\AgdaSpace{}%
\AgdaInductiveConstructor{refl}\AgdaSpace{}%
\AgdaOperator{\AgdaInductiveConstructor{,}}\AgdaSpace{}%
\AgdaInductiveConstructor{refl}\AgdaSpace{}%
\AgdaOperator{\AgdaInductiveConstructor{,}}\AgdaSpace{}%
\AgdaSymbol{λ}\AgdaSpace{}%
\AgdaSymbol{\{}\AgdaBound{W}\AgdaSymbol{\}\{}\AgdaBound{W′}\AgdaSymbol{\}}\AgdaSpace{}%
\AgdaBound{𝒱WW′}\AgdaSpace{}%
\AgdaSymbol{→}\<%
\\
\>[.][@{}l@{}]\<[1762I]%
\>[4]\AgdaKeyword{let}\AgdaSpace{}%
\AgdaBound{𝒱λNλN′sj}\AgdaSpace{}%
\AgdaSymbol{=}\AgdaSpace{}%
\AgdaField{down}%
\>[1776I]\AgdaSymbol{(}\AgdaBound{dir}\AgdaSpace{}%
\AgdaOperator{\AgdaFunction{∣}}\AgdaSpace{}%
\AgdaSymbol{(}\AgdaInductiveConstructor{ƛ}\AgdaSpace{}%
\AgdaBound{N}\AgdaSymbol{)}\AgdaSpace{}%
\AgdaOperator{\AgdaFunction{⊑ᴸᴿᵥ}}\AgdaSpace{}%
\AgdaSymbol{(}\AgdaInductiveConstructor{ƛ}\AgdaSpace{}%
\AgdaBound{N′}\AgdaSymbol{)}\AgdaSpace{}%
\AgdaOperator{\AgdaFunction{⦂}}\AgdaSpace{}%
\AgdaInductiveConstructor{fun⊑}\AgdaSpace{}%
\AgdaBound{c}\AgdaSpace{}%
\AgdaBound{d}\AgdaSymbol{)}\<%
\\
\>[.][@{}l@{}]\<[1776I]%
\>[24]\AgdaSymbol{(}\AgdaInductiveConstructor{suc}\AgdaSpace{}%
\AgdaBound{k}\AgdaSymbol{)}\AgdaSpace{}%
\AgdaBound{𝒱VV′}\AgdaSpace{}%
\AgdaSymbol{(}\AgdaInductiveConstructor{suc}\AgdaSpace{}%
\AgdaBound{j}\AgdaSymbol{)}\AgdaSpace{}%
\AgdaSymbol{(}\AgdaInductiveConstructor{s≤s}\AgdaSpace{}%
\AgdaBound{j≤k}\AgdaSymbol{)}\AgdaSpace{}%
\AgdaKeyword{in}\<%
\\
\>[4]\AgdaKeyword{let}\AgdaSpace{}%
\AgdaBound{ℰNWN′W′j}\AgdaSpace{}%
\AgdaSymbol{=}\AgdaSpace{}%
\AgdaBound{𝒱λNλN′sj}\AgdaSpace{}%
\AgdaBound{W}\AgdaSpace{}%
\AgdaBound{W′}\AgdaSpace{}%
\AgdaSymbol{(}\AgdaInductiveConstructor{suc}\AgdaSpace{}%
\AgdaBound{j}\AgdaSymbol{)}\AgdaSpace{}%
\AgdaFunction{≤-refl}\AgdaSpace{}%
\AgdaBound{𝒱WW′}\AgdaSpace{}%
\AgdaKeyword{in}\<%
\\
\>[4]\AgdaBound{ℰNWN′W′j}\<%
\end{code}
\begin{code}%
\>[0]\AgdaFunction{compatible-app}\AgdaSpace{}%
\AgdaSymbol{:}\AgdaSpace{}%
\AgdaSymbol{∀\{}\AgdaBound{Γ}\AgdaSymbol{\}\{}\AgdaBound{A}\AgdaSpace{}%
\AgdaBound{A′}\AgdaSpace{}%
\AgdaBound{B}\AgdaSpace{}%
\AgdaBound{B′}\AgdaSymbol{\}\{}\AgdaBound{c}\AgdaSpace{}%
\AgdaSymbol{:}\AgdaSpace{}%
\AgdaBound{A}\AgdaSpace{}%
\AgdaOperator{\AgdaDatatype{⊑}}\AgdaSpace{}%
\AgdaBound{A′}\AgdaSymbol{\}\{}\AgdaBound{d}\AgdaSpace{}%
\AgdaSymbol{:}\AgdaSpace{}%
\AgdaBound{B}\AgdaSpace{}%
\AgdaOperator{\AgdaDatatype{⊑}}\AgdaSpace{}%
\AgdaBound{B′}\AgdaSymbol{\}\{}\AgdaBound{L}\AgdaSpace{}%
\AgdaBound{L′}\AgdaSpace{}%
\AgdaBound{M}\AgdaSpace{}%
\AgdaBound{M′}\AgdaSymbol{\}}\<%
\\
\>[0][@{}l@{\AgdaIndent{0}}]%
\>[3]\AgdaSymbol{→}\AgdaSpace{}%
\AgdaBound{Γ}\AgdaSpace{}%
\AgdaOperator{\AgdaFunction{⊨}}\AgdaSpace{}%
\AgdaBound{L}\AgdaSpace{}%
\AgdaOperator{\AgdaFunction{⊑ᴸᴿ}}\AgdaSpace{}%
\AgdaBound{L′}\AgdaSpace{}%
\AgdaOperator{\AgdaFunction{⦂}}\AgdaSpace{}%
\AgdaSymbol{(}\AgdaBound{A}\AgdaSpace{}%
\AgdaOperator{\AgdaInductiveConstructor{⇒}}\AgdaSpace{}%
\AgdaBound{B}\AgdaSpace{}%
\AgdaOperator{\AgdaInductiveConstructor{,}}\AgdaSpace{}%
\AgdaBound{A′}\AgdaSpace{}%
\AgdaOperator{\AgdaInductiveConstructor{⇒}}\AgdaSpace{}%
\AgdaBound{B′}\AgdaSpace{}%
\AgdaOperator{\AgdaInductiveConstructor{,}}\AgdaSpace{}%
\AgdaInductiveConstructor{fun⊑}\AgdaSpace{}%
\AgdaBound{c}\AgdaSpace{}%
\AgdaBound{d}\AgdaSymbol{)}\AgdaSpace{}%
\AgdaSymbol{→}\AgdaSpace{}%
\AgdaBound{Γ}\AgdaSpace{}%
\AgdaOperator{\AgdaFunction{⊨}}\AgdaSpace{}%
\AgdaBound{M}\AgdaSpace{}%
\AgdaOperator{\AgdaFunction{⊑ᴸᴿ}}\AgdaSpace{}%
\AgdaBound{M′}\AgdaSpace{}%
\AgdaOperator{\AgdaFunction{⦂}}\AgdaSpace{}%
\AgdaSymbol{(}\AgdaBound{A}\AgdaSpace{}%
\AgdaOperator{\AgdaInductiveConstructor{,}}\AgdaSpace{}%
\AgdaBound{A′}\AgdaSpace{}%
\AgdaOperator{\AgdaInductiveConstructor{,}}\AgdaSpace{}%
\AgdaBound{c}\AgdaSymbol{)}\<%
\\
\>[3]\AgdaSymbol{→}\AgdaSpace{}%
\AgdaBound{Γ}\AgdaSpace{}%
\AgdaOperator{\AgdaFunction{⊨}}\AgdaSpace{}%
\AgdaBound{L}\AgdaSpace{}%
\AgdaOperator{\AgdaInductiveConstructor{·}}\AgdaSpace{}%
\AgdaBound{M}\AgdaSpace{}%
\AgdaOperator{\AgdaFunction{⊑ᴸᴿ}}\AgdaSpace{}%
\AgdaBound{L′}\AgdaSpace{}%
\AgdaOperator{\AgdaInductiveConstructor{·}}\AgdaSpace{}%
\AgdaBound{M′}\AgdaSpace{}%
\AgdaOperator{\AgdaFunction{⦂}}\AgdaSpace{}%
\AgdaSymbol{(}\AgdaBound{B}\AgdaSpace{}%
\AgdaOperator{\AgdaInductiveConstructor{,}}\AgdaSpace{}%
\AgdaBound{B′}\AgdaSpace{}%
\AgdaOperator{\AgdaInductiveConstructor{,}}\AgdaSpace{}%
\AgdaBound{d}\AgdaSymbol{)}\<%
\end{code}
\begin{code}[hide]%
\>[0]\AgdaFunction{compatible-app}\AgdaSpace{}%
\AgdaSymbol{\{}\AgdaBound{Γ}\AgdaSymbol{\}\{}\AgdaBound{A}\AgdaSymbol{\}\{}\AgdaBound{A′}\AgdaSymbol{\}\{}\AgdaBound{B}\AgdaSymbol{\}\{}\AgdaBound{B′}\AgdaSymbol{\}\{}\AgdaBound{c}\AgdaSymbol{\}\{}\AgdaBound{d}\AgdaSymbol{\}\{}\AgdaBound{L}\AgdaSymbol{\}\{}\AgdaBound{L′}\AgdaSymbol{\}\{}\AgdaBound{M}\AgdaSymbol{\}\{}\AgdaBound{M′}\AgdaSymbol{\}}\AgdaSpace{}%
\AgdaBound{⊨L⊑L′}\AgdaSpace{}%
\AgdaBound{⊨M⊑M′}\AgdaSpace{}%
\AgdaSymbol{=}\<%
\\
\>[0][@{}l@{\AgdaIndent{0}}]%
\>[1]\AgdaSymbol{(λ}\AgdaSpace{}%
\AgdaBound{γ}\AgdaSpace{}%
\AgdaBound{γ′}\AgdaSpace{}%
\AgdaSymbol{→}\AgdaSpace{}%
\AgdaFunction{⊢ℰLM⊑LM′}\AgdaSymbol{)}\AgdaSpace{}%
\AgdaOperator{\AgdaInductiveConstructor{,}}\AgdaSpace{}%
\AgdaSymbol{λ}\AgdaSpace{}%
\AgdaBound{γ}\AgdaSpace{}%
\AgdaBound{γ′}\AgdaSpace{}%
\AgdaSymbol{→}\AgdaSpace{}%
\AgdaFunction{⊢ℰLM⊑LM′}\<%
\\
\>[1]\AgdaKeyword{where}\<%
\\
\>[1]\AgdaFunction{⊢ℰLM⊑LM′}\AgdaSpace{}%
\AgdaSymbol{:}\AgdaSpace{}%
\AgdaSymbol{∀\{}\AgdaBound{dir}\AgdaSymbol{\}\{}\AgdaBound{γ}\AgdaSymbol{\}\{}\AgdaBound{γ′}\AgdaSymbol{\}}\AgdaSpace{}%
\AgdaSymbol{→}%
\>[1881I]\AgdaSymbol{(}\AgdaBound{Γ}\AgdaSpace{}%
\AgdaOperator{\AgdaFunction{∣}}\AgdaSpace{}%
\AgdaBound{dir}\AgdaSpace{}%
\AgdaOperator{\AgdaFunction{⊨}}\AgdaSpace{}%
\AgdaBound{γ}\AgdaSpace{}%
\AgdaOperator{\AgdaFunction{⊑ᴸᴿ}}\AgdaSpace{}%
\AgdaBound{γ′}\AgdaSymbol{)}\<%
\\
\>[1881I][@{}l@{\AgdaIndent{0}}]%
\>[29]\AgdaOperator{\AgdaFunction{⊢ᵒ}}\AgdaSpace{}%
\AgdaBound{dir}\AgdaSpace{}%
\AgdaOperator{\AgdaFunction{∣}}\AgdaSpace{}%
\AgdaOperator{\AgdaFunction{⟪}}\AgdaSpace{}%
\AgdaBound{γ}\AgdaSpace{}%
\AgdaOperator{\AgdaFunction{⟫}}\AgdaSpace{}%
\AgdaSymbol{(}\AgdaBound{L}\AgdaSpace{}%
\AgdaOperator{\AgdaInductiveConstructor{·}}\AgdaSpace{}%
\AgdaBound{M}\AgdaSymbol{)}\AgdaSpace{}%
\AgdaOperator{\AgdaFunction{⊑ᴸᴿₜ}}\AgdaSpace{}%
\AgdaOperator{\AgdaFunction{⟪}}\AgdaSpace{}%
\AgdaBound{γ′}\AgdaSpace{}%
\AgdaOperator{\AgdaFunction{⟫}}\AgdaSpace{}%
\AgdaSymbol{(}\AgdaBound{L′}\AgdaSpace{}%
\AgdaOperator{\AgdaInductiveConstructor{·}}\AgdaSpace{}%
\AgdaBound{M′}\AgdaSymbol{)}\AgdaSpace{}%
\AgdaOperator{\AgdaFunction{⦂}}\AgdaSpace{}%
\AgdaBound{d}\<%
\\
\>[1]\AgdaFunction{⊢ℰLM⊑LM′}\AgdaSpace{}%
\AgdaSymbol{\{}\AgdaBound{dir}\AgdaSymbol{\}\{}\AgdaBound{γ}\AgdaSymbol{\}\{}\AgdaBound{γ′}\AgdaSymbol{\}}\AgdaSpace{}%
\AgdaSymbol{=}\AgdaSpace{}%
\AgdaFunction{⊢ᵒ-intro}\AgdaSpace{}%
\AgdaSymbol{λ}\AgdaSpace{}%
\AgdaBound{n}\AgdaSpace{}%
\AgdaBound{𝒫n}\AgdaSpace{}%
\AgdaSymbol{→}\<%
\\
\>[1][@{}l@{\AgdaIndent{0}}]%
\>[2]\AgdaFunction{LRₜ-bind}\AgdaSymbol{\{}\AgdaArgument{c}\AgdaSpace{}%
\AgdaSymbol{=}%
\>[1913I]\AgdaBound{d}\AgdaSymbol{\}\{}\AgdaArgument{d}\AgdaSpace{}%
\AgdaSymbol{=}\AgdaSpace{}%
\AgdaInductiveConstructor{fun⊑}\AgdaSpace{}%
\AgdaBound{c}\AgdaSpace{}%
\AgdaBound{d}\AgdaSymbol{\}}\<%
\\
\>[.][@{}l@{}]\<[1913I]%
\>[15]\AgdaSymbol{\{}\AgdaArgument{F}\AgdaSpace{}%
\AgdaSymbol{=}\AgdaSpace{}%
\AgdaOperator{\AgdaInductiveConstructor{`}}\AgdaSpace{}%
\AgdaSymbol{(}\AgdaOperator{\AgdaInductiveConstructor{□·}}\AgdaSpace{}%
\AgdaSymbol{(}\AgdaOperator{\AgdaFunction{⟪}}\AgdaSpace{}%
\AgdaBound{γ}\AgdaSpace{}%
\AgdaOperator{\AgdaFunction{⟫}}\AgdaSpace{}%
\AgdaBound{M}\AgdaSymbol{))\}\{}\AgdaArgument{F′}\AgdaSpace{}%
\AgdaSymbol{=}\AgdaSpace{}%
\AgdaOperator{\AgdaInductiveConstructor{`}}\AgdaSpace{}%
\AgdaSymbol{(}\AgdaOperator{\AgdaInductiveConstructor{□·}}\AgdaSpace{}%
\AgdaSymbol{(}\AgdaOperator{\AgdaFunction{⟪}}\AgdaSpace{}%
\AgdaBound{γ′}\AgdaSpace{}%
\AgdaOperator{\AgdaFunction{⟫}}\AgdaSpace{}%
\AgdaBound{M′}\AgdaSymbol{))\}}\<%
\\
\>[2]\AgdaSymbol{(}\AgdaFunction{⊢ᵒ-elim}\AgdaSpace{}%
\AgdaSymbol{((}\AgdaFunction{proj}\AgdaSpace{}%
\AgdaBound{dir}\AgdaSpace{}%
\AgdaBound{L}\AgdaSpace{}%
\AgdaBound{L′}\AgdaSpace{}%
\AgdaBound{⊨L⊑L′}\AgdaSymbol{)}\AgdaSpace{}%
\AgdaBound{γ}\AgdaSpace{}%
\AgdaBound{γ′}\AgdaSymbol{)}\AgdaSpace{}%
\AgdaBound{n}\AgdaSpace{}%
\AgdaBound{𝒫n}\AgdaSymbol{)}\<%
\\
\>[2]\AgdaSymbol{λ}\AgdaSpace{}%
\AgdaBound{j}\AgdaSpace{}%
\AgdaBound{V}\AgdaSpace{}%
\AgdaBound{V′}\AgdaSpace{}%
\AgdaBound{j≤n}\AgdaSpace{}%
\AgdaBound{L→V}\AgdaSpace{}%
\AgdaBound{v}\AgdaSpace{}%
\AgdaBound{L′→V′}\AgdaSpace{}%
\AgdaBound{v′}\AgdaSpace{}%
\AgdaBound{𝒱VV′j}\AgdaSpace{}%
\AgdaSymbol{→}\<%
\\
\>[2]\AgdaFunction{LRₜ-bind}\AgdaSymbol{\{}\AgdaArgument{c}\AgdaSpace{}%
\AgdaSymbol{=}\AgdaSpace{}%
\AgdaBound{d}\AgdaSymbol{\}\{}\AgdaArgument{d}\AgdaSpace{}%
\AgdaSymbol{=}\AgdaSpace{}%
\AgdaBound{c}\AgdaSymbol{\}\{}\AgdaArgument{F}\AgdaSpace{}%
\AgdaSymbol{=}\AgdaSpace{}%
\AgdaOperator{\AgdaInductiveConstructor{`}}\AgdaSpace{}%
\AgdaSymbol{(}\AgdaBound{v}\AgdaSpace{}%
\AgdaOperator{\AgdaInductiveConstructor{·□}}\AgdaSymbol{)\}\{}\AgdaArgument{F′}\AgdaSpace{}%
\AgdaSymbol{=}\AgdaSpace{}%
\AgdaOperator{\AgdaInductiveConstructor{`}}\AgdaSpace{}%
\AgdaSymbol{(}\AgdaBound{v′}\AgdaSpace{}%
\AgdaOperator{\AgdaInductiveConstructor{·□}}\AgdaSymbol{)\}}\<%
\\
\>[2][@{}l@{\AgdaIndent{0}}]%
\>[3]\AgdaSymbol{(}\AgdaFunction{⊢ᵒ-elim}\AgdaSpace{}%
\AgdaSymbol{((}\AgdaFunction{proj}\AgdaSpace{}%
\AgdaBound{dir}\AgdaSpace{}%
\AgdaBound{M}\AgdaSpace{}%
\AgdaBound{M′}\AgdaSpace{}%
\AgdaBound{⊨M⊑M′}\AgdaSymbol{)}\AgdaSpace{}%
\AgdaBound{γ}\AgdaSpace{}%
\AgdaBound{γ′}\AgdaSymbol{)}\AgdaSpace{}%
\AgdaBound{j}\<%
\\
\>[3]\AgdaSymbol{(}\AgdaField{down}\AgdaSpace{}%
\AgdaSymbol{(}\AgdaFunction{Πᵒ}\AgdaSpace{}%
\AgdaSymbol{(}\AgdaBound{Γ}\AgdaSpace{}%
\AgdaOperator{\AgdaFunction{∣}}\AgdaSpace{}%
\AgdaBound{dir}\AgdaSpace{}%
\AgdaOperator{\AgdaFunction{⊨}}\AgdaSpace{}%
\AgdaBound{γ}\AgdaSpace{}%
\AgdaOperator{\AgdaFunction{⊑ᴸᴿ}}\AgdaSpace{}%
\AgdaBound{γ′}\AgdaSymbol{))}\AgdaSpace{}%
\AgdaBound{n}\AgdaSpace{}%
\AgdaBound{𝒫n}\AgdaSpace{}%
\AgdaBound{j}\AgdaSpace{}%
\AgdaBound{j≤n}\AgdaSymbol{))}\<%
\\
\>[3]\AgdaSymbol{λ}%
\>[1983I]\AgdaBound{i}\AgdaSpace{}%
\AgdaBound{W}\AgdaSpace{}%
\AgdaBound{W′}\AgdaSpace{}%
\AgdaBound{i≤j}\AgdaSpace{}%
\AgdaBound{M→W}\AgdaSpace{}%
\AgdaBound{w}\AgdaSpace{}%
\AgdaBound{M′→W′}\AgdaSpace{}%
\AgdaBound{w′}\AgdaSpace{}%
\AgdaBound{𝒱WW′i}\AgdaSpace{}%
\AgdaSymbol{→}\<%
\\
\>[.][@{}l@{}]\<[1983I]%
\>[5]\AgdaFunction{Goal}\AgdaSymbol{\{}\AgdaArgument{v}\AgdaSpace{}%
\AgdaSymbol{=}\AgdaSpace{}%
\AgdaBound{v}\AgdaSymbol{\}\{}\AgdaBound{v′}\AgdaSymbol{\}\{}\AgdaArgument{w}\AgdaSpace{}%
\AgdaSymbol{=}\AgdaSpace{}%
\AgdaBound{w}\AgdaSymbol{\}\{}\AgdaBound{w′}\AgdaSymbol{\}}\AgdaSpace{}%
\AgdaBound{i≤j}\AgdaSpace{}%
\AgdaBound{𝒱VV′j}\AgdaSpace{}%
\AgdaBound{𝒱WW′i}\<%
\\
\>[3]\AgdaKeyword{where}\<%
\\
\>[3]\AgdaFunction{Goal}\AgdaSpace{}%
\AgdaSymbol{:}%
\>[2001I]\AgdaSymbol{∀\{}\AgdaBound{V}\AgdaSymbol{\}\{}\AgdaBound{V′}\AgdaSymbol{\}\{}\AgdaBound{v}\AgdaSpace{}%
\AgdaSymbol{:}\AgdaSpace{}%
\AgdaDatatype{Value}\AgdaSpace{}%
\AgdaBound{V}\AgdaSymbol{\}\{}\AgdaBound{v′}\AgdaSpace{}%
\AgdaSymbol{:}\AgdaSpace{}%
\AgdaDatatype{Value}\AgdaSpace{}%
\AgdaBound{V′}\AgdaSymbol{\}}\<%
\\
\>[2001I][@{}l@{\AgdaIndent{0}}]%
\>[11]\AgdaSymbol{\{}\AgdaBound{W}\AgdaSymbol{\}\{}\AgdaBound{W′}\AgdaSymbol{\}\{}\AgdaBound{w}\AgdaSpace{}%
\AgdaSymbol{:}\AgdaSpace{}%
\AgdaDatatype{Value}\AgdaSpace{}%
\AgdaBound{W}\AgdaSymbol{\}\{}\AgdaBound{w′}\AgdaSpace{}%
\AgdaSymbol{:}\AgdaSpace{}%
\AgdaDatatype{Value}\AgdaSpace{}%
\AgdaBound{W′}\AgdaSymbol{\}\{}\AgdaBound{i}\AgdaSymbol{\}\{}\AgdaBound{j}\AgdaSymbol{\}}\<%
\\
\>[3][@{}l@{\AgdaIndent{0}}]%
\>[5]\AgdaSymbol{→}\AgdaSpace{}%
\AgdaBound{i}\AgdaSpace{}%
\AgdaOperator{\AgdaDatatype{≤}}\AgdaSpace{}%
\AgdaBound{j}\<%
\\
\>[5]\AgdaSymbol{→}\AgdaSpace{}%
\AgdaField{\#}\AgdaSpace{}%
\AgdaSymbol{(}\AgdaBound{dir}\AgdaSpace{}%
\AgdaOperator{\AgdaFunction{∣}}\AgdaSpace{}%
\AgdaBound{V}\AgdaSpace{}%
\AgdaOperator{\AgdaFunction{⊑ᴸᴿᵥ}}\AgdaSpace{}%
\AgdaBound{V′}\AgdaSpace{}%
\AgdaOperator{\AgdaFunction{⦂}}\AgdaSpace{}%
\AgdaInductiveConstructor{fun⊑}\AgdaSpace{}%
\AgdaBound{c}\AgdaSpace{}%
\AgdaBound{d}\AgdaSymbol{)}\AgdaSpace{}%
\AgdaBound{j}\<%
\\
\>[5]\AgdaSymbol{→}\AgdaSpace{}%
\AgdaField{\#}\AgdaSpace{}%
\AgdaSymbol{(}\AgdaBound{dir}\AgdaSpace{}%
\AgdaOperator{\AgdaFunction{∣}}\AgdaSpace{}%
\AgdaBound{W}\AgdaSpace{}%
\AgdaOperator{\AgdaFunction{⊑ᴸᴿᵥ}}\AgdaSpace{}%
\AgdaBound{W′}\AgdaSpace{}%
\AgdaOperator{\AgdaFunction{⦂}}\AgdaSpace{}%
\AgdaBound{c}\AgdaSymbol{)}\AgdaSpace{}%
\AgdaBound{i}\<%
\\
\>[5]\AgdaSymbol{→}\AgdaSpace{}%
\AgdaField{\#}\AgdaSpace{}%
\AgdaSymbol{(}\AgdaBound{dir}\AgdaSpace{}%
\AgdaOperator{\AgdaFunction{∣}}\AgdaSpace{}%
\AgdaSymbol{((}\AgdaOperator{\AgdaInductiveConstructor{`}}\AgdaSpace{}%
\AgdaSymbol{(}\AgdaBound{v}\AgdaSpace{}%
\AgdaOperator{\AgdaInductiveConstructor{·□}}\AgdaSymbol{))}\AgdaSpace{}%
\AgdaOperator{\AgdaFunction{⦉}}\AgdaSpace{}%
\AgdaBound{W}\AgdaSpace{}%
\AgdaOperator{\AgdaFunction{⦊}}\AgdaSymbol{)}\AgdaSpace{}%
\AgdaOperator{\AgdaFunction{⊑ᴸᴿₜ}}\AgdaSpace{}%
\AgdaSymbol{((}\AgdaOperator{\AgdaInductiveConstructor{`}}\AgdaSpace{}%
\AgdaSymbol{(}\AgdaBound{v′}\AgdaSpace{}%
\AgdaOperator{\AgdaInductiveConstructor{·□}}\AgdaSymbol{))}\AgdaSpace{}%
\AgdaOperator{\AgdaFunction{⦉}}\AgdaSpace{}%
\AgdaBound{W′}\AgdaSpace{}%
\AgdaOperator{\AgdaFunction{⦊}}\AgdaSymbol{)}\AgdaSpace{}%
\AgdaOperator{\AgdaFunction{⦂}}\AgdaSpace{}%
\AgdaBound{d}\AgdaSymbol{)}\AgdaSpace{}%
\AgdaBound{i}\<%
\\
\>[3]\AgdaFunction{Goal}\AgdaSpace{}%
\AgdaSymbol{\{}\AgdaBound{V}\AgdaSymbol{\}}\AgdaSpace{}%
\AgdaSymbol{\{}\AgdaBound{V′}\AgdaSymbol{\}}\AgdaSpace{}%
\AgdaSymbol{\{}\AgdaBound{v}\AgdaSymbol{\}}\AgdaSpace{}%
\AgdaSymbol{\{}\AgdaBound{v′}\AgdaSymbol{\}}\AgdaSpace{}%
\AgdaSymbol{\{}\AgdaBound{W}\AgdaSymbol{\}}\AgdaSpace{}%
\AgdaSymbol{\{}\AgdaBound{W′}\AgdaSymbol{\}}\AgdaSpace{}%
\AgdaSymbol{\{}\AgdaBound{w}\AgdaSymbol{\}\{}\AgdaBound{w′}\AgdaSymbol{\}\{}\AgdaInductiveConstructor{zero}\AgdaSymbol{\}}\AgdaSpace{}%
\AgdaSymbol{\{}\AgdaBound{j}\AgdaSymbol{\}}\AgdaSpace{}%
\AgdaBound{i≤j}\AgdaSpace{}%
\AgdaBound{𝒱VV′j}\AgdaSpace{}%
\AgdaBound{𝒱WW′i}\AgdaSpace{}%
\AgdaSymbol{=}\<%
\\
\>[3][@{}l@{\AgdaIndent{0}}]%
\>[5]\AgdaField{tz}\AgdaSpace{}%
\AgdaSymbol{(}\AgdaBound{dir}\AgdaSpace{}%
\AgdaOperator{\AgdaFunction{∣}}\AgdaSpace{}%
\AgdaSymbol{(}\AgdaFunction{value}\AgdaSpace{}%
\AgdaBound{v}\AgdaSpace{}%
\AgdaOperator{\AgdaInductiveConstructor{·}}\AgdaSpace{}%
\AgdaBound{W}\AgdaSymbol{)}\AgdaSpace{}%
\AgdaOperator{\AgdaFunction{⊑ᴸᴿₜ}}\AgdaSpace{}%
\AgdaSymbol{(}\AgdaFunction{value}\AgdaSpace{}%
\AgdaBound{v′}\AgdaSpace{}%
\AgdaOperator{\AgdaInductiveConstructor{·}}\AgdaSpace{}%
\AgdaBound{W′}\AgdaSymbol{)}\AgdaSpace{}%
\AgdaOperator{\AgdaFunction{⦂}}\AgdaSpace{}%
\AgdaBound{d}\AgdaSymbol{)}\<%
\\
\>[3]\AgdaFunction{Goal}\AgdaSpace{}%
\AgdaSymbol{\{}\AgdaBound{V}\AgdaSymbol{\}}\AgdaSpace{}%
\AgdaSymbol{\{}\AgdaBound{V′}\AgdaSymbol{\}}\AgdaSpace{}%
\AgdaSymbol{\{}\AgdaBound{v}\AgdaSymbol{\}}\AgdaSpace{}%
\AgdaSymbol{\{}\AgdaBound{v′}\AgdaSymbol{\}}\AgdaSpace{}%
\AgdaSymbol{\{}\AgdaBound{W}\AgdaSymbol{\}}\AgdaSpace{}%
\AgdaSymbol{\{}\AgdaBound{W′}\AgdaSymbol{\}}\AgdaSpace{}%
\AgdaSymbol{\{}\AgdaBound{w}\AgdaSymbol{\}\{}\AgdaBound{w′}\AgdaSymbol{\}\{}\AgdaInductiveConstructor{suc}\AgdaSpace{}%
\AgdaBound{i}\AgdaSymbol{\}}\AgdaSpace{}%
\AgdaSymbol{\{}\AgdaInductiveConstructor{suc}\AgdaSpace{}%
\AgdaBound{j}\AgdaSymbol{\}}\<%
\\
\>[3][@{}l@{\AgdaIndent{0}}]%
\>[7]\AgdaSymbol{(}\AgdaInductiveConstructor{s≤s}\AgdaSpace{}%
\AgdaBound{i≤j}\AgdaSymbol{)}\AgdaSpace{}%
\AgdaBound{𝒱VV′sj}\AgdaSpace{}%
\AgdaBound{𝒱WW′si}\<%
\\
\>[7]\AgdaKeyword{with}\AgdaSpace{}%
\AgdaFunction{LRᵥ-fun-elim-step}\AgdaSymbol{\{}\AgdaBound{A}\AgdaSymbol{\}\{}\AgdaBound{B}\AgdaSymbol{\}\{}\AgdaBound{A′}\AgdaSymbol{\}\{}\AgdaBound{B′}\AgdaSymbol{\}\{}\AgdaBound{c}\AgdaSymbol{\}\{}\AgdaBound{d}\AgdaSymbol{\}\{}\AgdaBound{V}\AgdaSymbol{\}\{}\AgdaBound{V′}\AgdaSymbol{\}\{}\AgdaBound{dir}\AgdaSymbol{\}\{}\AgdaBound{j}\AgdaSymbol{\}\{}\AgdaBound{i}\AgdaSymbol{\}}\AgdaSpace{}%
\AgdaBound{𝒱VV′sj}\AgdaSpace{}%
\AgdaBound{i≤j}\<%
\\
\>[3]\AgdaSymbol{...}%
\>[2097I]\AgdaSymbol{|}\AgdaSpace{}%
\AgdaBound{N}\AgdaSpace{}%
\AgdaOperator{\AgdaInductiveConstructor{,}}\AgdaSpace{}%
\AgdaBound{N′}\AgdaSpace{}%
\AgdaOperator{\AgdaInductiveConstructor{,}}\AgdaSpace{}%
\AgdaInductiveConstructor{refl}\AgdaSpace{}%
\AgdaOperator{\AgdaInductiveConstructor{,}}\AgdaSpace{}%
\AgdaInductiveConstructor{refl}\AgdaSpace{}%
\AgdaOperator{\AgdaInductiveConstructor{,}}\AgdaSpace{}%
\AgdaBound{body}\AgdaSpace{}%
\AgdaSymbol{=}\<%
\\
\>[.][@{}l@{}]\<[2097I]%
\>[7]\AgdaKeyword{let}\AgdaSpace{}%
\AgdaBound{𝒱WW′i}\AgdaSpace{}%
\AgdaSymbol{=}\AgdaSpace{}%
\AgdaField{down}\AgdaSpace{}%
\AgdaSymbol{(}\AgdaBound{dir}\AgdaSpace{}%
\AgdaOperator{\AgdaFunction{∣}}\AgdaSpace{}%
\AgdaBound{W}\AgdaSpace{}%
\AgdaOperator{\AgdaFunction{⊑ᴸᴿᵥ}}\AgdaSpace{}%
\AgdaBound{W′}\AgdaSpace{}%
\AgdaOperator{\AgdaFunction{⦂}}\AgdaSpace{}%
\AgdaBound{c}\AgdaSymbol{)(}\AgdaInductiveConstructor{suc}\AgdaSpace{}%
\AgdaBound{i}\AgdaSymbol{)}\AgdaBound{𝒱WW′si}\AgdaSpace{}%
\AgdaBound{i}\AgdaSpace{}%
\AgdaSymbol{(}\AgdaFunction{n≤1+n}\AgdaSpace{}%
\AgdaBound{i}\AgdaSymbol{)}\AgdaSpace{}%
\AgdaKeyword{in}\<%
\\
\>[7]\AgdaKeyword{let}\AgdaSpace{}%
\AgdaBound{ℰNWNW′i}\AgdaSpace{}%
\AgdaSymbol{=}\AgdaSpace{}%
\AgdaBound{body}\AgdaSymbol{\{}\AgdaBound{W}\AgdaSymbol{\}\{}\AgdaBound{W′}\AgdaSymbol{\}}\AgdaSpace{}%
\AgdaBound{𝒱WW′i}\AgdaSpace{}%
\AgdaKeyword{in}\<%
\\
\>[7]\AgdaFunction{anti-reduction}\AgdaSymbol{\{}\AgdaArgument{c}\AgdaSpace{}%
\AgdaSymbol{=}\AgdaSpace{}%
\AgdaBound{d}\AgdaSymbol{\}\{}\AgdaArgument{i}\AgdaSpace{}%
\AgdaSymbol{=}\AgdaSpace{}%
\AgdaBound{i}\AgdaSymbol{\}\{}\AgdaArgument{dir}\AgdaSpace{}%
\AgdaSymbol{=}\AgdaSpace{}%
\AgdaBound{dir}\AgdaSymbol{\}}\AgdaSpace{}%
\AgdaBound{ℰNWNW′i}\AgdaSpace{}%
\AgdaSymbol{(}\AgdaInductiveConstructor{β}\AgdaSpace{}%
\AgdaBound{w}\AgdaSymbol{)}\AgdaSpace{}%
\AgdaSymbol{(}\AgdaInductiveConstructor{β}\AgdaSpace{}%
\AgdaBound{w′}\AgdaSymbol{)}\<%
\end{code}

\paragraph{Compatibility for Injections}

We have two cases to deal with, the injection can be on the left or
the right. Starting with a projection on the left, \textsf{LRₜ-bind}
takes us from need to prove $M ⟨ G !⟩$ ⊑ᴸᴿ $M′$ to needing $V ⟨ G !⟩$
⊑ᴸᴿ $V′$, assuming $V$ ⊑ᴸᴿ $V′$. We proceed by case analysis on the
direction (≼ or ≽).  For ≼, we need to prove $▷ᵒ (V$ ⊑ᴸᴿᵥ $V′)$, which
follows from $V$ ⊑ᴸᴿ $V′$ by monotonicity. For ≽, we just need to
prove $V$ ⊑ᴸᴿ $V′$, which we have by the assumption.

\begin{code}[hide]%
\>[0]\AgdaFunction{LRᵥ-inject-L-intro-≽}\AgdaSpace{}%
\AgdaSymbol{:}\AgdaSpace{}%
\AgdaSymbol{∀\{}\AgdaBound{G}\AgdaSymbol{\}\{}\AgdaBound{A′}\AgdaSymbol{\}\{}\AgdaBound{c}\AgdaSpace{}%
\AgdaSymbol{:}\AgdaSpace{}%
\AgdaOperator{\AgdaFunction{⌈}}\AgdaSpace{}%
\AgdaBound{G}\AgdaSpace{}%
\AgdaOperator{\AgdaFunction{⌉}}\AgdaSpace{}%
\AgdaOperator{\AgdaDatatype{⊑}}\AgdaSpace{}%
\AgdaBound{A′}\AgdaSymbol{\}\{}\AgdaBound{V}\AgdaSymbol{\}\{}\AgdaBound{V′}\AgdaSymbol{\}\{}\AgdaBound{k}\AgdaSymbol{\}}\<%
\\
\>[0][@{}l@{\AgdaIndent{0}}]%
\>[3]\AgdaSymbol{→}\AgdaSpace{}%
\AgdaField{\#}\AgdaSymbol{(}\AgdaInductiveConstructor{≽}\AgdaSpace{}%
\AgdaOperator{\AgdaFunction{∣}}\AgdaSpace{}%
\AgdaBound{V}\AgdaSpace{}%
\AgdaOperator{\AgdaFunction{⊑ᴸᴿᵥ}}\AgdaSpace{}%
\AgdaBound{V′}\AgdaSpace{}%
\AgdaOperator{\AgdaFunction{⦂}}\AgdaSpace{}%
\AgdaBound{c}\AgdaSymbol{)}\AgdaSpace{}%
\AgdaBound{k}\<%
\\
\>[3]\AgdaSymbol{→}\AgdaSpace{}%
\AgdaField{\#}\AgdaSymbol{(}\AgdaInductiveConstructor{≽}\AgdaSpace{}%
\AgdaOperator{\AgdaFunction{∣}}\AgdaSpace{}%
\AgdaSymbol{(}\AgdaBound{V}\AgdaSpace{}%
\AgdaOperator{\AgdaInductiveConstructor{⟨}}\AgdaSpace{}%
\AgdaBound{G}\AgdaSpace{}%
\AgdaOperator{\AgdaInductiveConstructor{!⟩}}\AgdaSymbol{)}\AgdaSpace{}%
\AgdaOperator{\AgdaFunction{⊑ᴸᴿᵥ}}\AgdaSpace{}%
\AgdaBound{V′}\AgdaSpace{}%
\AgdaOperator{\AgdaFunction{⦂}}\AgdaSpace{}%
\AgdaInductiveConstructor{unk⊑}\AgdaSpace{}%
\AgdaBound{c}\AgdaSymbol{)}\AgdaSpace{}%
\AgdaBound{k}\<%
\end{code}
\begin{code}[hide]%
\>[0]\AgdaFunction{LRᵥ-inject-L-intro-≽}\AgdaSpace{}%
\AgdaSymbol{\{}\AgdaBound{G}\AgdaSymbol{\}\{}\AgdaBound{A′}\AgdaSymbol{\}\{}\AgdaBound{c}\AgdaSymbol{\}\{}\AgdaBound{V}\AgdaSymbol{\}\{}\AgdaBound{V′}\AgdaSymbol{\}\{}\AgdaInductiveConstructor{zero}\AgdaSymbol{\}}\AgdaSpace{}%
\AgdaBound{𝒱VV′k}\AgdaSpace{}%
\AgdaSymbol{=}\<%
\\
\>[0][@{}l@{\AgdaIndent{0}}]%
\>[4]\AgdaField{tz}\AgdaSpace{}%
\AgdaSymbol{(}\AgdaInductiveConstructor{≽}\AgdaSpace{}%
\AgdaOperator{\AgdaFunction{∣}}\AgdaSpace{}%
\AgdaSymbol{(}\AgdaBound{V}\AgdaSpace{}%
\AgdaOperator{\AgdaInductiveConstructor{⟨}}\AgdaSpace{}%
\AgdaBound{G}\AgdaSpace{}%
\AgdaOperator{\AgdaInductiveConstructor{!⟩}}\AgdaSymbol{)}\AgdaSpace{}%
\AgdaOperator{\AgdaFunction{⊑ᴸᴿᵥ}}\AgdaSpace{}%
\AgdaBound{V′}\AgdaSpace{}%
\AgdaOperator{\AgdaFunction{⦂}}\AgdaSpace{}%
\AgdaInductiveConstructor{unk⊑}\AgdaSpace{}%
\AgdaBound{c}\AgdaSymbol{)}\<%
\\
\>[0]\AgdaFunction{LRᵥ-inject-L-intro-≽}\AgdaSpace{}%
\AgdaSymbol{\{}\AgdaBound{G}\AgdaSymbol{\}}\AgdaSpace{}%
\AgdaSymbol{\{}\AgdaBound{A′}\AgdaSymbol{\}}\AgdaSpace{}%
\AgdaSymbol{\{}\AgdaBound{c}\AgdaSymbol{\}}\AgdaSpace{}%
\AgdaSymbol{\{}\AgdaBound{V}\AgdaSymbol{\}}\AgdaSpace{}%
\AgdaSymbol{\{}\AgdaBound{V′}\AgdaSymbol{\}}\AgdaSpace{}%
\AgdaSymbol{\{}\AgdaInductiveConstructor{suc}\AgdaSpace{}%
\AgdaBound{k}\AgdaSymbol{\}}\AgdaSpace{}%
\AgdaBound{𝒱VV′sk}\<%
\\
\>[0][@{}l@{\AgdaIndent{0}}]%
\>[4]\AgdaKeyword{with}\AgdaSpace{}%
\AgdaBound{G}\AgdaSpace{}%
\AgdaOperator{\AgdaFunction{≡ᵍ}}\AgdaSpace{}%
\AgdaBound{G}\<%
\\
\>[0]\AgdaSymbol{...}\AgdaSpace{}%
\AgdaSymbol{|}\AgdaSpace{}%
\AgdaInductiveConstructor{no}\AgdaSpace{}%
\AgdaBound{neq}\AgdaSpace{}%
\AgdaSymbol{=}\AgdaSpace{}%
\AgdaFunction{⊥-elim}\AgdaSpace{}%
\AgdaSymbol{(}\AgdaBound{neq}\AgdaSpace{}%
\AgdaInductiveConstructor{refl}\AgdaSymbol{)}\<%
\\
\>[0]\AgdaSymbol{...}\AgdaSpace{}%
\AgdaSymbol{|}%
\>[2200I]\AgdaInductiveConstructor{yes}\AgdaSpace{}%
\AgdaInductiveConstructor{refl}\AgdaSpace{}%
\AgdaSymbol{=}\<%
\\
\>[.][@{}l@{}]\<[2200I]%
\>[6]\AgdaKeyword{let}\AgdaSpace{}%
\AgdaSymbol{(}\AgdaBound{v}\AgdaSpace{}%
\AgdaOperator{\AgdaInductiveConstructor{,}}\AgdaSpace{}%
\AgdaBound{v′}\AgdaSymbol{)}\AgdaSpace{}%
\AgdaSymbol{=}\AgdaSpace{}%
\AgdaFunction{LRᵥ⇒Value}\AgdaSpace{}%
\AgdaBound{c}\AgdaSpace{}%
\AgdaBound{V}\AgdaSpace{}%
\AgdaBound{V′}\AgdaSpace{}%
\AgdaBound{𝒱VV′sk}\AgdaSpace{}%
\AgdaKeyword{in}\<%
\\
\>[6]\AgdaBound{v}\AgdaSpace{}%
\AgdaOperator{\AgdaInductiveConstructor{,}}\AgdaSpace{}%
\AgdaBound{v′}\AgdaSpace{}%
\AgdaOperator{\AgdaInductiveConstructor{,}}\AgdaSpace{}%
\AgdaBound{𝒱VV′sk}\<%
\end{code}
\begin{code}[hide]%
\>[0]\AgdaFunction{LRᵥ-inject-L-intro}\AgdaSpace{}%
\AgdaSymbol{:}\AgdaSpace{}%
\AgdaSymbol{∀\{}\AgdaBound{G}\AgdaSymbol{\}\{}\AgdaBound{A′}\AgdaSymbol{\}\{}\AgdaBound{c}\AgdaSpace{}%
\AgdaSymbol{:}\AgdaSpace{}%
\AgdaOperator{\AgdaFunction{⌈}}\AgdaSpace{}%
\AgdaBound{G}\AgdaSpace{}%
\AgdaOperator{\AgdaFunction{⌉}}\AgdaSpace{}%
\AgdaOperator{\AgdaDatatype{⊑}}\AgdaSpace{}%
\AgdaBound{A′}\AgdaSymbol{\}\{}\AgdaBound{V}\AgdaSymbol{\}\{}\AgdaBound{V′}\AgdaSymbol{\}\{}\AgdaBound{dir}\AgdaSymbol{\}\{}\AgdaBound{k}\AgdaSymbol{\}}\<%
\\
\>[0][@{}l@{\AgdaIndent{0}}]%
\>[3]\AgdaSymbol{→}\AgdaSpace{}%
\AgdaField{\#}\AgdaSymbol{(}\AgdaBound{dir}\AgdaSpace{}%
\AgdaOperator{\AgdaFunction{∣}}\AgdaSpace{}%
\AgdaBound{V}\AgdaSpace{}%
\AgdaOperator{\AgdaFunction{⊑ᴸᴿᵥ}}\AgdaSpace{}%
\AgdaBound{V′}\AgdaSpace{}%
\AgdaOperator{\AgdaFunction{⦂}}\AgdaSpace{}%
\AgdaBound{c}\AgdaSymbol{)}\AgdaSpace{}%
\AgdaBound{k}\<%
\\
\>[3]\AgdaSymbol{→}\AgdaSpace{}%
\AgdaField{\#}\AgdaSymbol{(}\AgdaBound{dir}\AgdaSpace{}%
\AgdaOperator{\AgdaFunction{∣}}\AgdaSpace{}%
\AgdaSymbol{(}\AgdaBound{V}\AgdaSpace{}%
\AgdaOperator{\AgdaInductiveConstructor{⟨}}\AgdaSpace{}%
\AgdaBound{G}\AgdaSpace{}%
\AgdaOperator{\AgdaInductiveConstructor{!⟩}}\AgdaSymbol{)}\AgdaSpace{}%
\AgdaOperator{\AgdaFunction{⊑ᴸᴿᵥ}}\AgdaSpace{}%
\AgdaBound{V′}\AgdaSpace{}%
\AgdaOperator{\AgdaFunction{⦂}}\AgdaSpace{}%
\AgdaInductiveConstructor{unk⊑}\AgdaSpace{}%
\AgdaBound{c}\AgdaSymbol{)}\AgdaSpace{}%
\AgdaBound{k}\<%
\end{code}
\begin{code}[hide]%
\>[0]\AgdaFunction{LRᵥ-inject-L-intro}\AgdaSpace{}%
\AgdaSymbol{\{}\AgdaBound{G}\AgdaSymbol{\}}\AgdaSpace{}%
\AgdaSymbol{\{}\AgdaBound{A′}\AgdaSymbol{\}}\AgdaSpace{}%
\AgdaSymbol{\{}\AgdaBound{c}\AgdaSymbol{\}}\AgdaSpace{}%
\AgdaSymbol{\{}\AgdaBound{V}\AgdaSymbol{\}}\AgdaSpace{}%
\AgdaSymbol{\{}\AgdaBound{V′}\AgdaSymbol{\}}\AgdaSpace{}%
\AgdaSymbol{\{}\AgdaInductiveConstructor{≼}\AgdaSymbol{\}}\AgdaSpace{}%
\AgdaSymbol{\{}\AgdaInductiveConstructor{zero}\AgdaSymbol{\}}\AgdaSpace{}%
\AgdaBound{𝒱VV′k}\AgdaSpace{}%
\AgdaSymbol{=}\<%
\\
\>[0][@{}l@{\AgdaIndent{0}}]%
\>[4]\AgdaField{tz}\AgdaSpace{}%
\AgdaSymbol{(}\AgdaInductiveConstructor{≼}\AgdaSpace{}%
\AgdaOperator{\AgdaFunction{∣}}\AgdaSpace{}%
\AgdaBound{V}\AgdaSpace{}%
\AgdaOperator{\AgdaInductiveConstructor{⟨}}\AgdaSpace{}%
\AgdaBound{G}\AgdaSpace{}%
\AgdaOperator{\AgdaInductiveConstructor{!⟩}}\AgdaSpace{}%
\AgdaOperator{\AgdaFunction{⊑ᴸᴿᵥ}}\AgdaSpace{}%
\AgdaBound{V′}\AgdaSpace{}%
\AgdaOperator{\AgdaFunction{⦂}}\AgdaSpace{}%
\AgdaInductiveConstructor{unk⊑}\AgdaSpace{}%
\AgdaBound{c}\AgdaSymbol{)}\<%
\\
\>[0]\AgdaFunction{LRᵥ-inject-L-intro}\AgdaSpace{}%
\AgdaSymbol{\{}\AgdaBound{G}\AgdaSymbol{\}}\AgdaSpace{}%
\AgdaSymbol{\{}\AgdaBound{A′}\AgdaSymbol{\}}\AgdaSpace{}%
\AgdaSymbol{\{}\AgdaBound{c}\AgdaSymbol{\}}\AgdaSpace{}%
\AgdaSymbol{\{}\AgdaBound{V}\AgdaSymbol{\}}\AgdaSpace{}%
\AgdaSymbol{\{}\AgdaBound{V′}\AgdaSymbol{\}}\AgdaSpace{}%
\AgdaSymbol{\{}\AgdaInductiveConstructor{≼}\AgdaSymbol{\}}\AgdaSpace{}%
\AgdaSymbol{\{}\AgdaInductiveConstructor{suc}\AgdaSpace{}%
\AgdaBound{k}\AgdaSymbol{\}}\AgdaSpace{}%
\AgdaBound{𝒱VV′sk}\<%
\\
\>[0][@{}l@{\AgdaIndent{0}}]%
\>[4]\AgdaKeyword{with}\AgdaSpace{}%
\AgdaBound{G}\AgdaSpace{}%
\AgdaOperator{\AgdaFunction{≡ᵍ}}\AgdaSpace{}%
\AgdaBound{G}\<%
\\
\>[0]\AgdaSymbol{...}\AgdaSpace{}%
\AgdaSymbol{|}\AgdaSpace{}%
\AgdaInductiveConstructor{no}\AgdaSpace{}%
\AgdaBound{neq}\AgdaSpace{}%
\AgdaSymbol{=}\AgdaSpace{}%
\AgdaFunction{⊥-elim}\AgdaSpace{}%
\AgdaSymbol{(}\AgdaBound{neq}\AgdaSpace{}%
\AgdaInductiveConstructor{refl}\AgdaSymbol{)}\<%
\\
\>[0]\AgdaSymbol{...}%
\>[2284I]\AgdaSymbol{|}\AgdaSpace{}%
\AgdaInductiveConstructor{yes}\AgdaSpace{}%
\AgdaInductiveConstructor{refl}\AgdaSpace{}%
\AgdaSymbol{=}\<%
\\
\>[.][@{}l@{}]\<[2284I]%
\>[4]\AgdaKeyword{let}\AgdaSpace{}%
\AgdaSymbol{(}\AgdaBound{v}\AgdaSpace{}%
\AgdaOperator{\AgdaInductiveConstructor{,}}\AgdaSpace{}%
\AgdaBound{v′}\AgdaSymbol{)}\AgdaSpace{}%
\AgdaSymbol{=}\AgdaSpace{}%
\AgdaFunction{LRᵥ⇒Value}\AgdaSpace{}%
\AgdaBound{c}\AgdaSpace{}%
\AgdaBound{V}\AgdaSpace{}%
\AgdaBound{V′}\AgdaSpace{}%
\AgdaBound{𝒱VV′sk}\AgdaSpace{}%
\AgdaKeyword{in}\<%
\\
\>[4]\AgdaKeyword{let}\AgdaSpace{}%
\AgdaBound{𝒱VV′k}\AgdaSpace{}%
\AgdaSymbol{=}\AgdaSpace{}%
\AgdaField{down}\AgdaSpace{}%
\AgdaSymbol{(}\AgdaInductiveConstructor{≼}\AgdaSpace{}%
\AgdaOperator{\AgdaFunction{∣}}\AgdaSpace{}%
\AgdaBound{V}\AgdaSpace{}%
\AgdaOperator{\AgdaFunction{⊑ᴸᴿᵥ}}\AgdaSpace{}%
\AgdaBound{V′}\AgdaSpace{}%
\AgdaOperator{\AgdaFunction{⦂}}\AgdaSpace{}%
\AgdaBound{c}\AgdaSymbol{)}\AgdaSpace{}%
\AgdaSymbol{(}\AgdaInductiveConstructor{suc}\AgdaSpace{}%
\AgdaBound{k}\AgdaSymbol{)}\AgdaSpace{}%
\AgdaBound{𝒱VV′sk}\AgdaSpace{}%
\AgdaBound{k}\AgdaSpace{}%
\AgdaSymbol{(}\AgdaFunction{n≤1+n}\AgdaSpace{}%
\AgdaBound{k}\AgdaSymbol{)}\AgdaSpace{}%
\AgdaKeyword{in}\<%
\\
\>[4]\AgdaBound{v}\AgdaSpace{}%
\AgdaOperator{\AgdaInductiveConstructor{,}}\AgdaSpace{}%
\AgdaBound{v′}\AgdaSpace{}%
\AgdaOperator{\AgdaInductiveConstructor{,}}\AgdaSpace{}%
\AgdaBound{𝒱VV′k}\<%
\\
\>[0]\AgdaFunction{LRᵥ-inject-L-intro}\AgdaSpace{}%
\AgdaSymbol{\{}\AgdaBound{G}\AgdaSymbol{\}}\AgdaSpace{}%
\AgdaSymbol{\{}\AgdaBound{A′}\AgdaSymbol{\}}\AgdaSpace{}%
\AgdaSymbol{\{}\AgdaBound{c}\AgdaSymbol{\}}\AgdaSpace{}%
\AgdaSymbol{\{}\AgdaBound{V}\AgdaSymbol{\}}\AgdaSpace{}%
\AgdaSymbol{\{}\AgdaBound{V′}\AgdaSymbol{\}}\AgdaSpace{}%
\AgdaSymbol{\{}\AgdaInductiveConstructor{≽}\AgdaSymbol{\}}\AgdaSpace{}%
\AgdaSymbol{\{}\AgdaBound{k}\AgdaSymbol{\}}\AgdaSpace{}%
\AgdaBound{𝒱VV′k}\AgdaSpace{}%
\AgdaSymbol{=}\<%
\\
\>[0][@{}l@{\AgdaIndent{0}}]%
\>[3]\AgdaFunction{LRᵥ-inject-L-intro-≽}\AgdaSymbol{\{}\AgdaBound{G}\AgdaSymbol{\}}\AgdaSpace{}%
\AgdaSymbol{\{}\AgdaBound{A′}\AgdaSymbol{\}}\AgdaSpace{}%
\AgdaSymbol{\{}\AgdaBound{c}\AgdaSymbol{\}}\AgdaSpace{}%
\AgdaSymbol{\{}\AgdaBound{V}\AgdaSymbol{\}}\AgdaSpace{}%
\AgdaSymbol{\{}\AgdaBound{V′}\AgdaSymbol{\}}\AgdaSpace{}%
\AgdaBound{𝒱VV′k}\<%
\end{code}
\begin{code}%
\>[0]\AgdaFunction{compatible-inj-L}\AgdaSpace{}%
\AgdaSymbol{:}\AgdaSpace{}%
\AgdaSymbol{∀\{}\AgdaBound{Γ}\AgdaSymbol{\}\{}\AgdaBound{G}\AgdaSpace{}%
\AgdaBound{A′}\AgdaSymbol{\}\{}\AgdaBound{c}\AgdaSpace{}%
\AgdaSymbol{:}\AgdaSpace{}%
\AgdaOperator{\AgdaFunction{⌈}}\AgdaSpace{}%
\AgdaBound{G}\AgdaSpace{}%
\AgdaOperator{\AgdaFunction{⌉}}\AgdaSpace{}%
\AgdaOperator{\AgdaDatatype{⊑}}\AgdaSpace{}%
\AgdaBound{A′}\AgdaSymbol{\}\{}\AgdaBound{M}\AgdaSpace{}%
\AgdaBound{M′}\AgdaSymbol{\}}\<%
\\
\>[0][@{}l@{\AgdaIndent{0}}]%
\>[3]\AgdaSymbol{→}\AgdaSpace{}%
\AgdaBound{Γ}\AgdaSpace{}%
\AgdaOperator{\AgdaFunction{⊨}}\AgdaSpace{}%
\AgdaBound{M}\AgdaSpace{}%
\AgdaOperator{\AgdaFunction{⊑ᴸᴿ}}\AgdaSpace{}%
\AgdaBound{M′}\AgdaSpace{}%
\AgdaOperator{\AgdaFunction{⦂}}\AgdaSpace{}%
\AgdaSymbol{(}\AgdaOperator{\AgdaFunction{⌈}}\AgdaSpace{}%
\AgdaBound{G}\AgdaSpace{}%
\AgdaOperator{\AgdaFunction{⌉}}\AgdaSpace{}%
\AgdaOperator{\AgdaInductiveConstructor{,}}\AgdaSpace{}%
\AgdaBound{A′}\AgdaSpace{}%
\AgdaOperator{\AgdaInductiveConstructor{,}}\AgdaSpace{}%
\AgdaBound{c}\AgdaSymbol{)}\<%
\\
\>[3]\AgdaSymbol{→}\AgdaSpace{}%
\AgdaBound{Γ}\AgdaSpace{}%
\AgdaOperator{\AgdaFunction{⊨}}\AgdaSpace{}%
\AgdaBound{M}\AgdaSpace{}%
\AgdaOperator{\AgdaInductiveConstructor{⟨}}\AgdaSpace{}%
\AgdaBound{G}\AgdaSpace{}%
\AgdaOperator{\AgdaInductiveConstructor{!⟩}}\AgdaSpace{}%
\AgdaOperator{\AgdaFunction{⊑ᴸᴿ}}\AgdaSpace{}%
\AgdaBound{M′}\AgdaSpace{}%
\AgdaOperator{\AgdaFunction{⦂}}\AgdaSpace{}%
\AgdaSymbol{(}\AgdaInductiveConstructor{★}\AgdaSpace{}%
\AgdaOperator{\AgdaInductiveConstructor{,}}\AgdaSpace{}%
\AgdaBound{A′}\AgdaSpace{}%
\AgdaOperator{\AgdaInductiveConstructor{,}}\AgdaSpace{}%
\AgdaInductiveConstructor{unk⊑}\AgdaSymbol{\{}\AgdaBound{G}\AgdaSymbol{\}\{}\AgdaBound{A′}\AgdaSymbol{\}}\AgdaSpace{}%
\AgdaBound{c}\AgdaSymbol{)}\<%
\end{code}
\begin{code}[hide]%
\>[0]\AgdaFunction{compatible-inj-L}\AgdaSymbol{\{}\AgdaBound{Γ}\AgdaSymbol{\}\{}\AgdaBound{G}\AgdaSymbol{\}\{}\AgdaBound{A′}\AgdaSymbol{\}\{}\AgdaBound{c}\AgdaSymbol{\}\{}\AgdaBound{M}\AgdaSymbol{\}\{}\AgdaBound{M′}\AgdaSymbol{\}}\AgdaSpace{}%
\AgdaBound{⊨M⊑M′}\AgdaSpace{}%
\AgdaSymbol{=}\<%
\\
\>[0][@{}l@{\AgdaIndent{0}}]%
\>[2]\AgdaSymbol{(λ}\AgdaSpace{}%
\AgdaBound{γ}\AgdaSpace{}%
\AgdaBound{γ′}\AgdaSpace{}%
\AgdaSymbol{→}\AgdaSpace{}%
\AgdaFunction{ℰMGM′}\AgdaSymbol{)}\AgdaSpace{}%
\AgdaOperator{\AgdaInductiveConstructor{,}}\AgdaSpace{}%
\AgdaSymbol{(λ}\AgdaSpace{}%
\AgdaBound{γ}\AgdaSpace{}%
\AgdaBound{γ′}\AgdaSpace{}%
\AgdaSymbol{→}\AgdaSpace{}%
\AgdaFunction{ℰMGM′}\AgdaSymbol{)}\<%
\\
\>[2]\AgdaKeyword{where}\<%
\\
\>[2]\AgdaFunction{ℰMGM′}\AgdaSpace{}%
\AgdaSymbol{:}\AgdaSpace{}%
\AgdaSymbol{∀}\AgdaSpace{}%
\AgdaSymbol{\{}\AgdaBound{γ}\AgdaSymbol{\}\{}\AgdaBound{γ′}\AgdaSymbol{\}\{}\AgdaBound{dir}\AgdaSymbol{\}}\<%
\\
\>[2][@{}l@{\AgdaIndent{0}}]%
\>[3]\AgdaSymbol{→}\AgdaSpace{}%
\AgdaSymbol{(}\AgdaBound{Γ}\AgdaSpace{}%
\AgdaOperator{\AgdaFunction{∣}}\AgdaSpace{}%
\AgdaBound{dir}\AgdaSpace{}%
\AgdaOperator{\AgdaFunction{⊨}}\AgdaSpace{}%
\AgdaBound{γ}\AgdaSpace{}%
\AgdaOperator{\AgdaFunction{⊑ᴸᴿ}}\AgdaSpace{}%
\AgdaBound{γ′}\AgdaSymbol{)}\AgdaSpace{}%
\AgdaOperator{\AgdaFunction{⊢ᵒ}}\AgdaSpace{}%
\AgdaSymbol{(}\AgdaBound{dir}\AgdaSpace{}%
\AgdaOperator{\AgdaFunction{∣}}\AgdaSpace{}%
\AgdaSymbol{(}\AgdaOperator{\AgdaFunction{⟪}}\AgdaSpace{}%
\AgdaBound{γ}\AgdaSpace{}%
\AgdaOperator{\AgdaFunction{⟫}}\AgdaSpace{}%
\AgdaBound{M}\AgdaSpace{}%
\AgdaOperator{\AgdaInductiveConstructor{⟨}}\AgdaSpace{}%
\AgdaBound{G}\AgdaSpace{}%
\AgdaOperator{\AgdaInductiveConstructor{!⟩}}\AgdaSymbol{)}\AgdaSpace{}%
\AgdaOperator{\AgdaFunction{⊑ᴸᴿₜ}}\AgdaSpace{}%
\AgdaSymbol{(}\AgdaOperator{\AgdaFunction{⟪}}\AgdaSpace{}%
\AgdaBound{γ′}\AgdaSpace{}%
\AgdaOperator{\AgdaFunction{⟫}}\AgdaSpace{}%
\AgdaBound{M′}\AgdaSymbol{)}\AgdaSpace{}%
\AgdaOperator{\AgdaFunction{⦂}}\AgdaSpace{}%
\AgdaInductiveConstructor{unk⊑}\AgdaSpace{}%
\AgdaBound{c}\AgdaSymbol{)}\<%
\\
\>[2]\AgdaFunction{ℰMGM′}\AgdaSymbol{\{}\AgdaBound{γ}\AgdaSymbol{\}\{}\AgdaBound{γ′}\AgdaSymbol{\}\{}\AgdaBound{dir}\AgdaSymbol{\}}\AgdaSpace{}%
\AgdaSymbol{=}\AgdaSpace{}%
\AgdaFunction{⊢ᵒ-intro}\AgdaSpace{}%
\AgdaSymbol{λ}\AgdaSpace{}%
\AgdaBound{n}\AgdaSpace{}%
\AgdaBound{𝒫n}\AgdaSpace{}%
\AgdaSymbol{→}\<%
\\
\>[2][@{}l@{\AgdaIndent{0}}]%
\>[3]\AgdaFunction{LRₜ-bind}\AgdaSymbol{\{}\AgdaArgument{c}%
\>[2417I]\AgdaSymbol{=}\AgdaSpace{}%
\AgdaInductiveConstructor{unk⊑}\AgdaSpace{}%
\AgdaBound{c}\AgdaSymbol{\}\{}\AgdaArgument{d}\AgdaSpace{}%
\AgdaSymbol{=}\AgdaSpace{}%
\AgdaBound{c}\AgdaSymbol{\}\{}\AgdaArgument{F}\AgdaSpace{}%
\AgdaSymbol{=}\AgdaSpace{}%
\AgdaOperator{\AgdaInductiveConstructor{`}}\AgdaSpace{}%
\AgdaSymbol{(}\AgdaOperator{\AgdaInductiveConstructor{□⟨}}\AgdaSpace{}%
\AgdaBound{G}\AgdaSpace{}%
\AgdaOperator{\AgdaInductiveConstructor{!⟩}}\AgdaSymbol{)\}\{}\AgdaArgument{F′}\AgdaSpace{}%
\AgdaSymbol{=}\AgdaSpace{}%
\AgdaInductiveConstructor{□}\AgdaSymbol{\}}\<%
\\
\>[.][@{}l@{}]\<[2417I]%
\>[14]\AgdaSymbol{\{}\AgdaOperator{\AgdaFunction{⟪}}\AgdaSpace{}%
\AgdaBound{γ}\AgdaSpace{}%
\AgdaOperator{\AgdaFunction{⟫}}\AgdaSpace{}%
\AgdaBound{M}\AgdaSymbol{\}\{}\AgdaOperator{\AgdaFunction{⟪}}\AgdaSpace{}%
\AgdaBound{γ′}\AgdaSpace{}%
\AgdaOperator{\AgdaFunction{⟫}}\AgdaSpace{}%
\AgdaBound{M′}\AgdaSymbol{\}\{}\AgdaBound{n}\AgdaSymbol{\}\{}\AgdaBound{dir}\AgdaSymbol{\}}\<%
\\
\>[3]\AgdaSymbol{(}\AgdaFunction{⊢ᵒ-elim}\AgdaSpace{}%
\AgdaSymbol{((}\AgdaFunction{proj}\AgdaSpace{}%
\AgdaBound{dir}\AgdaSpace{}%
\AgdaBound{M}\AgdaSpace{}%
\AgdaBound{M′}\AgdaSpace{}%
\AgdaBound{⊨M⊑M′}\AgdaSymbol{)}\AgdaSpace{}%
\AgdaBound{γ}\AgdaSpace{}%
\AgdaBound{γ′}\AgdaSymbol{)}\AgdaSpace{}%
\AgdaBound{n}\AgdaSpace{}%
\AgdaBound{𝒫n}\AgdaSymbol{)}\<%
\\
\>[3]\AgdaSymbol{λ}\AgdaSpace{}%
\AgdaBound{j}\AgdaSpace{}%
\AgdaBound{V}\AgdaSpace{}%
\AgdaBound{V′}\AgdaSpace{}%
\AgdaBound{j≤n}\AgdaSpace{}%
\AgdaBound{M→V}\AgdaSpace{}%
\AgdaBound{v}\AgdaSpace{}%
\AgdaBound{M′→V′}\AgdaSpace{}%
\AgdaBound{v′}\AgdaSpace{}%
\AgdaBound{𝒱VV′j}\AgdaSpace{}%
\AgdaSymbol{→}\<%
\\
\>[3]\AgdaFunction{LRᵥ⇒LRₜ-step}\AgdaSymbol{\{}\AgdaInductiveConstructor{★}\AgdaSymbol{\}\{}\AgdaBound{A′}\AgdaSymbol{\}\{}\AgdaInductiveConstructor{unk⊑}\AgdaSpace{}%
\AgdaBound{c}\AgdaSymbol{\}\{}\AgdaBound{V}\AgdaSpace{}%
\AgdaOperator{\AgdaInductiveConstructor{⟨}}\AgdaSpace{}%
\AgdaBound{G}\AgdaSpace{}%
\AgdaOperator{\AgdaInductiveConstructor{!⟩}}\AgdaSymbol{\}\{}\AgdaBound{V′}\AgdaSymbol{\}\{}\AgdaBound{dir}\AgdaSymbol{\}\{}\AgdaBound{j}\AgdaSymbol{\}}\<%
\\
\>[3]\AgdaSymbol{(}\AgdaFunction{LRᵥ-inject-L-intro}\AgdaSymbol{\{}\AgdaBound{G}\AgdaSymbol{\}\{}\AgdaBound{A′}\AgdaSymbol{\}\{}\AgdaBound{c}\AgdaSymbol{\}\{}\AgdaBound{V}\AgdaSymbol{\}\{}\AgdaBound{V′}\AgdaSymbol{\}\{}\AgdaBound{dir}\AgdaSymbol{\}\{}\AgdaBound{j}\AgdaSymbol{\}}\AgdaSpace{}%
\AgdaBound{𝒱VV′j}\AgdaSymbol{)}\<%
\end{code}

Next consider when the injection is on the right.  The
\textsf{LRₜ-bind} lemma takes us from needing to prove $M$ ⊑ᴸᴿ $M′ ⟨ G !⟩$
to needing $V$ ⊑ᴸᴿ $V′ ⟨ G !⟩$ where $V$ ⊑ᴸᴿᵥ $V′$.
We know that $V$ is of type ★ (rule \textsf{⊑-inj-R})
so $V = W ⟨ G !⟩$ and $W$ ⊑ᴸᴿᵥ $V′$ (the \textsf{unk⊑} clause of \textsf{LRᵥ}).
So we conclude that $W ⟨ G !⟩$ ⊑ᴸᴿ $V′ ⟨ G !⟩$
by the \textsf{unk⊑unk} clause of \textsf{LRᵥ}. 

\begin{code}[hide]%
\>[0]\AgdaFunction{LRᵥ-dyn-any-elim-≽}\AgdaSpace{}%
\AgdaSymbol{:}\AgdaSpace{}%
\AgdaSymbol{∀\{}\AgdaBound{V}\AgdaSymbol{\}\{}\AgdaBound{V′}\AgdaSymbol{\}\{}\AgdaBound{k}\AgdaSymbol{\}\{}\AgdaBound{H}\AgdaSymbol{\}\{}\AgdaBound{A′}\AgdaSymbol{\}\{}\AgdaBound{c}\AgdaSpace{}%
\AgdaSymbol{:}\AgdaSpace{}%
\AgdaOperator{\AgdaFunction{⌈}}\AgdaSpace{}%
\AgdaBound{H}\AgdaSpace{}%
\AgdaOperator{\AgdaFunction{⌉}}\AgdaSpace{}%
\AgdaOperator{\AgdaDatatype{⊑}}\AgdaSpace{}%
\AgdaBound{A′}\AgdaSymbol{\}}\<%
\\
\>[0][@{}l@{\AgdaIndent{0}}]%
\>[3]\AgdaSymbol{→}\AgdaSpace{}%
\AgdaField{\#}\AgdaSymbol{(}\AgdaInductiveConstructor{≽}\AgdaSpace{}%
\AgdaOperator{\AgdaFunction{∣}}\AgdaSpace{}%
\AgdaBound{V}\AgdaSpace{}%
\AgdaOperator{\AgdaFunction{⊑ᴸᴿᵥ}}\AgdaSpace{}%
\AgdaBound{V′}\AgdaSpace{}%
\AgdaOperator{\AgdaFunction{⦂}}\AgdaSpace{}%
\AgdaInductiveConstructor{unk⊑}\AgdaSpace{}%
\AgdaBound{c}\AgdaSymbol{)}\AgdaSpace{}%
\AgdaSymbol{(}\AgdaInductiveConstructor{suc}\AgdaSpace{}%
\AgdaBound{k}\AgdaSymbol{)}\<%
\\
\>[3]\AgdaSymbol{→}\AgdaSpace{}%
\AgdaFunction{∃[}\AgdaSpace{}%
\AgdaBound{V₁}\AgdaSpace{}%
\AgdaFunction{]}%
\>[2480I]\AgdaBound{V}\AgdaSpace{}%
\AgdaOperator{\AgdaDatatype{≡}}\AgdaSpace{}%
\AgdaBound{V₁}\AgdaSpace{}%
\AgdaOperator{\AgdaInductiveConstructor{⟨}}\AgdaSpace{}%
\AgdaBound{H}\AgdaSpace{}%
\AgdaOperator{\AgdaInductiveConstructor{!⟩}}\AgdaSpace{}%
\AgdaOperator{\AgdaFunction{×}}\AgdaSpace{}%
\AgdaDatatype{Value}\AgdaSpace{}%
\AgdaBound{V₁}\AgdaSpace{}%
\AgdaOperator{\AgdaFunction{×}}\AgdaSpace{}%
\AgdaDatatype{Value}\AgdaSpace{}%
\AgdaBound{V′}\<%
\\
\>[.][@{}l@{}]\<[2480I]%
\>[13]\AgdaOperator{\AgdaFunction{×}}\AgdaSpace{}%
\AgdaField{\#}\AgdaSymbol{(}\AgdaInductiveConstructor{≽}\AgdaSpace{}%
\AgdaOperator{\AgdaFunction{∣}}\AgdaSpace{}%
\AgdaBound{V₁}\AgdaSpace{}%
\AgdaOperator{\AgdaFunction{⊑ᴸᴿᵥ}}\AgdaSpace{}%
\AgdaBound{V′}\AgdaSpace{}%
\AgdaOperator{\AgdaFunction{⦂}}\AgdaSpace{}%
\AgdaBound{c}\AgdaSymbol{)}\AgdaSpace{}%
\AgdaSymbol{(}\AgdaInductiveConstructor{suc}\AgdaSpace{}%
\AgdaBound{k}\AgdaSymbol{)}\<%
\end{code}
\begin{code}[hide]%
\>[0]\AgdaFunction{LRᵥ-dyn-any-elim-≽}\AgdaSpace{}%
\AgdaSymbol{\{}\AgdaBound{V}\AgdaSpace{}%
\AgdaOperator{\AgdaInductiveConstructor{⟨}}\AgdaSpace{}%
\AgdaBound{G}\AgdaSpace{}%
\AgdaOperator{\AgdaInductiveConstructor{!⟩}}\AgdaSymbol{\}\{}\AgdaBound{V′}\AgdaSymbol{\}\{}\AgdaBound{k}\AgdaSymbol{\}\{}\AgdaBound{H}\AgdaSymbol{\}\{}\AgdaBound{A′}\AgdaSymbol{\}\{}\AgdaBound{c}\AgdaSymbol{\}}\AgdaSpace{}%
\AgdaBound{𝒱VGV′}\<%
\\
\>[0][@{}l@{\AgdaIndent{0}}]%
\>[4]\AgdaKeyword{with}\AgdaSpace{}%
\AgdaBound{G}\AgdaSpace{}%
\AgdaOperator{\AgdaFunction{≡ᵍ}}\AgdaSpace{}%
\AgdaBound{H}\<%
\\
\>[0]\AgdaSymbol{...}\AgdaSpace{}%
\AgdaSymbol{|}\AgdaSpace{}%
\AgdaInductiveConstructor{no}\AgdaSpace{}%
\AgdaBound{neq}\AgdaSpace{}%
\AgdaSymbol{=}\AgdaSpace{}%
\AgdaFunction{⊥-elim}\AgdaSpace{}%
\AgdaBound{𝒱VGV′}\<%
\\
\>[0]\AgdaSymbol{...}%
\>[2515I]\AgdaSymbol{|}\AgdaSpace{}%
\AgdaInductiveConstructor{yes}\AgdaSpace{}%
\AgdaInductiveConstructor{refl}\<%
\\
\>[.][@{}l@{}]\<[2515I]%
\>[4]\AgdaKeyword{with}\AgdaSpace{}%
\AgdaBound{𝒱VGV′}\<%
\\
\>[0]\AgdaSymbol{...}\AgdaSpace{}%
\AgdaSymbol{|}\AgdaSpace{}%
\AgdaBound{v}\AgdaSpace{}%
\AgdaOperator{\AgdaInductiveConstructor{,}}\AgdaSpace{}%
\AgdaBound{v′}\AgdaSpace{}%
\AgdaOperator{\AgdaInductiveConstructor{,}}\AgdaSpace{}%
\AgdaBound{𝒱VV′}\AgdaSpace{}%
\AgdaSymbol{=}\AgdaSpace{}%
\AgdaBound{V}\AgdaSpace{}%
\AgdaOperator{\AgdaInductiveConstructor{,}}\AgdaSpace{}%
\AgdaInductiveConstructor{refl}\AgdaSpace{}%
\AgdaOperator{\AgdaInductiveConstructor{,}}\AgdaSpace{}%
\AgdaBound{v}\AgdaSpace{}%
\AgdaOperator{\AgdaInductiveConstructor{,}}\AgdaSpace{}%
\AgdaBound{v′}\AgdaSpace{}%
\AgdaOperator{\AgdaInductiveConstructor{,}}\AgdaSpace{}%
\AgdaBound{𝒱VV′}\<%
\end{code}
\begin{code}[hide]%
\>[0]\AgdaFunction{LRᵥ-inject-R-intro-≽}\AgdaSpace{}%
\AgdaSymbol{:}\AgdaSpace{}%
\AgdaSymbol{∀\{}\AgdaBound{G}\AgdaSymbol{\}\{}\AgdaBound{c}\AgdaSpace{}%
\AgdaSymbol{:}\AgdaSpace{}%
\AgdaInductiveConstructor{★}\AgdaSpace{}%
\AgdaOperator{\AgdaDatatype{⊑}}\AgdaSpace{}%
\AgdaOperator{\AgdaFunction{⌈}}\AgdaSpace{}%
\AgdaBound{G}\AgdaSpace{}%
\AgdaOperator{\AgdaFunction{⌉}}\AgdaSymbol{\}\{}\AgdaBound{V}\AgdaSymbol{\}\{}\AgdaBound{V′}\AgdaSymbol{\}\{}\AgdaBound{k}\AgdaSymbol{\}}\<%
\\
\>[0][@{}l@{\AgdaIndent{0}}]%
\>[3]\AgdaSymbol{→}\AgdaSpace{}%
\AgdaField{\#}\AgdaSymbol{(}\AgdaInductiveConstructor{≽}\AgdaSpace{}%
\AgdaOperator{\AgdaFunction{∣}}\AgdaSpace{}%
\AgdaBound{V}\AgdaSpace{}%
\AgdaOperator{\AgdaFunction{⊑ᴸᴿᵥ}}\AgdaSpace{}%
\AgdaBound{V′}\AgdaSpace{}%
\AgdaOperator{\AgdaFunction{⦂}}\AgdaSpace{}%
\AgdaBound{c}\AgdaSymbol{)}\AgdaSpace{}%
\AgdaBound{k}\<%
\\
\>[3]\AgdaSymbol{→}\AgdaSpace{}%
\AgdaField{\#}\AgdaSymbol{(}\AgdaInductiveConstructor{≽}\AgdaSpace{}%
\AgdaOperator{\AgdaFunction{∣}}\AgdaSpace{}%
\AgdaBound{V}\AgdaSpace{}%
\AgdaOperator{\AgdaFunction{⊑ᴸᴿᵥ}}\AgdaSpace{}%
\AgdaSymbol{(}\AgdaBound{V′}\AgdaSpace{}%
\AgdaOperator{\AgdaInductiveConstructor{⟨}}\AgdaSpace{}%
\AgdaBound{G}\AgdaSpace{}%
\AgdaOperator{\AgdaInductiveConstructor{!⟩}}\AgdaSymbol{)}\AgdaSpace{}%
\AgdaOperator{\AgdaFunction{⦂}}\AgdaSpace{}%
\AgdaInductiveConstructor{unk⊑unk}\AgdaSymbol{)}\AgdaSpace{}%
\AgdaBound{k}\<%
\end{code}
\begin{code}[hide]%
\>[0]\AgdaFunction{LRᵥ-inject-R-intro-≽}\AgdaSpace{}%
\AgdaSymbol{\{}\AgdaBound{G}\AgdaSymbol{\}}\AgdaSpace{}%
\AgdaSymbol{\{}\AgdaBound{c}\AgdaSymbol{\}}\AgdaSpace{}%
\AgdaSymbol{\{}\AgdaBound{V}\AgdaSymbol{\}}\AgdaSpace{}%
\AgdaSymbol{\{}\AgdaBound{V′}\AgdaSymbol{\}}\AgdaSpace{}%
\AgdaSymbol{\{}\AgdaInductiveConstructor{zero}\AgdaSymbol{\}}\AgdaSpace{}%
\AgdaBound{𝒱VV′}\AgdaSpace{}%
\AgdaSymbol{=}\<%
\\
\>[0][@{}l@{\AgdaIndent{0}}]%
\>[5]\AgdaField{tz}\AgdaSpace{}%
\AgdaSymbol{(}\AgdaInductiveConstructor{≽}\AgdaSpace{}%
\AgdaOperator{\AgdaFunction{∣}}\AgdaSpace{}%
\AgdaBound{V}\AgdaSpace{}%
\AgdaOperator{\AgdaFunction{⊑ᴸᴿᵥ}}\AgdaSpace{}%
\AgdaSymbol{(}\AgdaBound{V′}\AgdaSpace{}%
\AgdaOperator{\AgdaInductiveConstructor{⟨}}\AgdaSpace{}%
\AgdaBound{G}\AgdaSpace{}%
\AgdaOperator{\AgdaInductiveConstructor{!⟩}}\AgdaSymbol{)}\AgdaSpace{}%
\AgdaOperator{\AgdaFunction{⦂}}\AgdaSpace{}%
\AgdaInductiveConstructor{unk⊑unk}\AgdaSymbol{)}\<%
\\
\>[0]\AgdaFunction{LRᵥ-inject-R-intro-≽}\AgdaSpace{}%
\AgdaSymbol{\{}\AgdaBound{G}\AgdaSymbol{\}}\AgdaSpace{}%
\AgdaSymbol{\{}\AgdaBound{c}\AgdaSymbol{\}}\AgdaSpace{}%
\AgdaSymbol{\{}\AgdaBound{V}\AgdaSymbol{\}}\AgdaSpace{}%
\AgdaSymbol{\{}\AgdaBound{V′}\AgdaSymbol{\}}\AgdaSpace{}%
\AgdaSymbol{\{}\AgdaInductiveConstructor{suc}\AgdaSpace{}%
\AgdaBound{k}\AgdaSymbol{\}}\AgdaSpace{}%
\AgdaBound{𝒱VV′sk}\<%
\\
\>[0][@{}l@{\AgdaIndent{0}}]%
\>[4]\AgdaKeyword{with}\AgdaSpace{}%
\AgdaFunction{unk⊑gnd-inv}\AgdaSpace{}%
\AgdaBound{c}\<%
\\
\>[0]\AgdaSymbol{...}%
\>[2588I]\AgdaSymbol{|}\AgdaSpace{}%
\AgdaBound{d}\AgdaSpace{}%
\AgdaOperator{\AgdaInductiveConstructor{,}}\AgdaSpace{}%
\AgdaInductiveConstructor{refl}\<%
\\
\>[.][@{}l@{}]\<[2588I]%
\>[4]\AgdaKeyword{with}\AgdaSpace{}%
\AgdaFunction{LRᵥ-dyn-any-elim-≽}\AgdaSpace{}%
\AgdaSymbol{\{}\AgdaBound{V}\AgdaSymbol{\}\{}\AgdaBound{V′}\AgdaSymbol{\}\{}\AgdaBound{k}\AgdaSymbol{\}\{}\AgdaBound{G}\AgdaSymbol{\}\{\AgdaUnderscore{}\}\{}\AgdaBound{d}\AgdaSymbol{\}}\AgdaSpace{}%
\AgdaBound{𝒱VV′sk}\<%
\\
\>[0]\AgdaSymbol{...}%
\>[2595I]\AgdaSymbol{|}\AgdaSpace{}%
\AgdaBound{V₁}\AgdaSpace{}%
\AgdaOperator{\AgdaInductiveConstructor{,}}\AgdaSpace{}%
\AgdaInductiveConstructor{refl}\AgdaSpace{}%
\AgdaOperator{\AgdaInductiveConstructor{,}}\AgdaSpace{}%
\AgdaBound{v₁}\AgdaSpace{}%
\AgdaOperator{\AgdaInductiveConstructor{,}}\AgdaSpace{}%
\AgdaBound{v′}\AgdaSpace{}%
\AgdaOperator{\AgdaInductiveConstructor{,}}\AgdaSpace{}%
\AgdaBound{𝒱V₁V′sk}\<%
\\
\>[.][@{}l@{}]\<[2595I]%
\>[4]\AgdaKeyword{with}\AgdaSpace{}%
\AgdaBound{G}\AgdaSpace{}%
\AgdaOperator{\AgdaFunction{≡ᵍ}}\AgdaSpace{}%
\AgdaBound{G}\<%
\\
\>[0]\AgdaSymbol{...}\AgdaSpace{}%
\AgdaSymbol{|}\AgdaSpace{}%
\AgdaInductiveConstructor{no}\AgdaSpace{}%
\AgdaBound{neq}\AgdaSpace{}%
\AgdaSymbol{=}\AgdaSpace{}%
\AgdaFunction{⊥-elim}\AgdaSpace{}%
\AgdaBound{𝒱VV′sk}\<%
\\
\>[0]\AgdaSymbol{...}%
\>[2614I]\AgdaSymbol{|}\AgdaSpace{}%
\AgdaInductiveConstructor{yes}\AgdaSpace{}%
\AgdaInductiveConstructor{refl}\<%
\\
\>[.][@{}l@{}]\<[2614I]%
\>[4]\AgdaKeyword{with}\AgdaSpace{}%
\AgdaFunction{gnd-prec-unique}\AgdaSpace{}%
\AgdaBound{d}\AgdaSpace{}%
\AgdaFunction{Refl⊑}\<%
\\
\>[0]\AgdaSymbol{...}%
\>[2620I]\AgdaSymbol{|}\AgdaSpace{}%
\AgdaInductiveConstructor{refl}\AgdaSpace{}%
\AgdaSymbol{=}\<%
\\
\>[.][@{}l@{}]\<[2620I]%
\>[4]\AgdaKeyword{let}\AgdaSpace{}%
\AgdaBound{𝒱V₁V′k}\AgdaSpace{}%
\AgdaSymbol{=}\AgdaSpace{}%
\AgdaField{down}\AgdaSpace{}%
\AgdaSymbol{(}\AgdaInductiveConstructor{≽}\AgdaSpace{}%
\AgdaOperator{\AgdaFunction{∣}}\AgdaSpace{}%
\AgdaBound{V₁}\AgdaSpace{}%
\AgdaOperator{\AgdaFunction{⊑ᴸᴿᵥ}}\AgdaSpace{}%
\AgdaBound{V′}\AgdaSpace{}%
\AgdaOperator{\AgdaFunction{⦂}}\AgdaSpace{}%
\AgdaBound{d}\AgdaSymbol{)}\AgdaSpace{}%
\AgdaSymbol{(}\AgdaInductiveConstructor{suc}\AgdaSpace{}%
\AgdaBound{k}\AgdaSymbol{)}\AgdaSpace{}%
\AgdaBound{𝒱V₁V′sk}\AgdaSpace{}%
\AgdaBound{k}\AgdaSpace{}%
\AgdaSymbol{(}\AgdaFunction{n≤1+n}\AgdaSpace{}%
\AgdaBound{k}\AgdaSymbol{)}\AgdaSpace{}%
\AgdaKeyword{in}\<%
\\
\>[4]\AgdaBound{v₁}\AgdaSpace{}%
\AgdaOperator{\AgdaInductiveConstructor{,}}\AgdaSpace{}%
\AgdaBound{v′}\AgdaSpace{}%
\AgdaOperator{\AgdaInductiveConstructor{,}}\AgdaSpace{}%
\AgdaBound{𝒱V₁V′k}\<%
\end{code}
\begin{code}[hide]%
\>[0]\AgdaFunction{LRᵥ-dyn-any-elim-≼}\AgdaSpace{}%
\AgdaSymbol{:}\AgdaSpace{}%
\AgdaSymbol{∀\{}\AgdaBound{V}\AgdaSymbol{\}\{}\AgdaBound{V′}\AgdaSymbol{\}\{}\AgdaBound{k}\AgdaSymbol{\}\{}\AgdaBound{H}\AgdaSymbol{\}\{}\AgdaBound{A′}\AgdaSymbol{\}\{}\AgdaBound{c}\AgdaSpace{}%
\AgdaSymbol{:}\AgdaSpace{}%
\AgdaOperator{\AgdaFunction{⌈}}\AgdaSpace{}%
\AgdaBound{H}\AgdaSpace{}%
\AgdaOperator{\AgdaFunction{⌉}}\AgdaSpace{}%
\AgdaOperator{\AgdaDatatype{⊑}}\AgdaSpace{}%
\AgdaBound{A′}\AgdaSymbol{\}}\<%
\\
\>[0][@{}l@{\AgdaIndent{0}}]%
\>[3]\AgdaSymbol{→}\AgdaSpace{}%
\AgdaField{\#}\AgdaSymbol{(}\AgdaInductiveConstructor{≼}\AgdaSpace{}%
\AgdaOperator{\AgdaFunction{∣}}\AgdaSpace{}%
\AgdaBound{V}\AgdaSpace{}%
\AgdaOperator{\AgdaFunction{⊑ᴸᴿᵥ}}\AgdaSpace{}%
\AgdaBound{V′}\AgdaSpace{}%
\AgdaOperator{\AgdaFunction{⦂}}\AgdaSpace{}%
\AgdaInductiveConstructor{unk⊑}\AgdaSpace{}%
\AgdaBound{c}\AgdaSymbol{)}\AgdaSpace{}%
\AgdaSymbol{(}\AgdaInductiveConstructor{suc}\AgdaSpace{}%
\AgdaBound{k}\AgdaSymbol{)}\<%
\\
\>[3]\AgdaSymbol{→}\AgdaSpace{}%
\AgdaFunction{∃[}\AgdaSpace{}%
\AgdaBound{V₁}\AgdaSpace{}%
\AgdaFunction{]}\AgdaSpace{}%
\AgdaBound{V}\AgdaSpace{}%
\AgdaOperator{\AgdaDatatype{≡}}\AgdaSpace{}%
\AgdaBound{V₁}\AgdaSpace{}%
\AgdaOperator{\AgdaInductiveConstructor{⟨}}\AgdaSpace{}%
\AgdaBound{H}\AgdaSpace{}%
\AgdaOperator{\AgdaInductiveConstructor{!⟩}}\AgdaSpace{}%
\AgdaOperator{\AgdaFunction{×}}\AgdaSpace{}%
\AgdaDatatype{Value}\AgdaSpace{}%
\AgdaBound{V₁}\AgdaSpace{}%
\AgdaOperator{\AgdaFunction{×}}\AgdaSpace{}%
\AgdaDatatype{Value}\AgdaSpace{}%
\AgdaBound{V′}\AgdaSpace{}%
\AgdaOperator{\AgdaFunction{×}}\AgdaSpace{}%
\AgdaField{\#}\AgdaSymbol{(}\AgdaInductiveConstructor{≼}\AgdaSpace{}%
\AgdaOperator{\AgdaFunction{∣}}\AgdaSpace{}%
\AgdaBound{V₁}\AgdaSpace{}%
\AgdaOperator{\AgdaFunction{⊑ᴸᴿᵥ}}\AgdaSpace{}%
\AgdaBound{V′}\AgdaSpace{}%
\AgdaOperator{\AgdaFunction{⦂}}\AgdaSpace{}%
\AgdaBound{c}\AgdaSymbol{)}\AgdaSpace{}%
\AgdaBound{k}\<%
\end{code}
\begin{code}[hide]%
\>[0]\AgdaFunction{LRᵥ-dyn-any-elim-≼}\AgdaSpace{}%
\AgdaSymbol{\{}\AgdaBound{V}\AgdaSpace{}%
\AgdaOperator{\AgdaInductiveConstructor{⟨}}\AgdaSpace{}%
\AgdaBound{G}\AgdaSpace{}%
\AgdaOperator{\AgdaInductiveConstructor{!⟩}}\AgdaSymbol{\}\{}\AgdaBound{V′}\AgdaSymbol{\}\{}\AgdaBound{k}\AgdaSymbol{\}\{}\AgdaBound{H}\AgdaSymbol{\}\{}\AgdaBound{A′}\AgdaSymbol{\}\{}\AgdaBound{c}\AgdaSymbol{\}}\AgdaSpace{}%
\AgdaBound{𝒱VGV′}\<%
\\
\>[0][@{}l@{\AgdaIndent{0}}]%
\>[4]\AgdaKeyword{with}\AgdaSpace{}%
\AgdaBound{G}\AgdaSpace{}%
\AgdaOperator{\AgdaFunction{≡ᵍ}}\AgdaSpace{}%
\AgdaBound{H}\<%
\\
\>[0]\AgdaSymbol{...}\AgdaSpace{}%
\AgdaSymbol{|}\AgdaSpace{}%
\AgdaInductiveConstructor{no}\AgdaSpace{}%
\AgdaBound{neq}\AgdaSpace{}%
\AgdaSymbol{=}\AgdaSpace{}%
\AgdaFunction{⊥-elim}\AgdaSpace{}%
\AgdaBound{𝒱VGV′}\<%
\\
\>[0]\AgdaSymbol{...}%
\>[2700I]\AgdaSymbol{|}\AgdaSpace{}%
\AgdaInductiveConstructor{yes}\AgdaSpace{}%
\AgdaInductiveConstructor{refl}\<%
\\
\>[.][@{}l@{}]\<[2700I]%
\>[4]\AgdaKeyword{with}\AgdaSpace{}%
\AgdaBound{𝒱VGV′}\<%
\\
\>[0]\AgdaSymbol{...}\AgdaSpace{}%
\AgdaSymbol{|}\AgdaSpace{}%
\AgdaBound{v}\AgdaSpace{}%
\AgdaOperator{\AgdaInductiveConstructor{,}}\AgdaSpace{}%
\AgdaBound{v′}\AgdaSpace{}%
\AgdaOperator{\AgdaInductiveConstructor{,}}\AgdaSpace{}%
\AgdaBound{𝒱VV′}\AgdaSpace{}%
\AgdaSymbol{=}\AgdaSpace{}%
\AgdaBound{V}\AgdaSpace{}%
\AgdaOperator{\AgdaInductiveConstructor{,}}\AgdaSpace{}%
\AgdaInductiveConstructor{refl}\AgdaSpace{}%
\AgdaOperator{\AgdaInductiveConstructor{,}}\AgdaSpace{}%
\AgdaBound{v}\AgdaSpace{}%
\AgdaOperator{\AgdaInductiveConstructor{,}}\AgdaSpace{}%
\AgdaBound{v′}\AgdaSpace{}%
\AgdaOperator{\AgdaInductiveConstructor{,}}\AgdaSpace{}%
\AgdaBound{𝒱VV′}\<%
\end{code}
\begin{code}[hide]%
\>[0]\AgdaFunction{LRᵥ-inject-R-intro-≼}\AgdaSpace{}%
\AgdaSymbol{:}\AgdaSpace{}%
\AgdaSymbol{∀\{}\AgdaBound{G}\AgdaSymbol{\}\{}\AgdaBound{c}\AgdaSpace{}%
\AgdaSymbol{:}\AgdaSpace{}%
\AgdaInductiveConstructor{★}\AgdaSpace{}%
\AgdaOperator{\AgdaDatatype{⊑}}\AgdaSpace{}%
\AgdaOperator{\AgdaFunction{⌈}}\AgdaSpace{}%
\AgdaBound{G}\AgdaSpace{}%
\AgdaOperator{\AgdaFunction{⌉}}\AgdaSymbol{\}\{}\AgdaBound{V}\AgdaSymbol{\}\{}\AgdaBound{V′}\AgdaSymbol{\}\{}\AgdaBound{k}\AgdaSymbol{\}}\<%
\\
\>[0][@{}l@{\AgdaIndent{0}}]%
\>[3]\AgdaSymbol{→}\AgdaSpace{}%
\AgdaField{\#}\AgdaSymbol{(}\AgdaInductiveConstructor{≼}\AgdaSpace{}%
\AgdaOperator{\AgdaFunction{∣}}\AgdaSpace{}%
\AgdaBound{V}\AgdaSpace{}%
\AgdaOperator{\AgdaFunction{⊑ᴸᴿᵥ}}\AgdaSpace{}%
\AgdaBound{V′}\AgdaSpace{}%
\AgdaOperator{\AgdaFunction{⦂}}\AgdaSpace{}%
\AgdaBound{c}\AgdaSymbol{)}\AgdaSpace{}%
\AgdaBound{k}\<%
\\
\>[3]\AgdaSymbol{→}\AgdaSpace{}%
\AgdaField{\#}\AgdaSymbol{(}\AgdaInductiveConstructor{≼}\AgdaSpace{}%
\AgdaOperator{\AgdaFunction{∣}}\AgdaSpace{}%
\AgdaBound{V}\AgdaSpace{}%
\AgdaOperator{\AgdaFunction{⊑ᴸᴿᵥ}}\AgdaSpace{}%
\AgdaSymbol{(}\AgdaBound{V′}\AgdaSpace{}%
\AgdaOperator{\AgdaInductiveConstructor{⟨}}\AgdaSpace{}%
\AgdaBound{G}\AgdaSpace{}%
\AgdaOperator{\AgdaInductiveConstructor{!⟩}}\AgdaSymbol{)}\AgdaSpace{}%
\AgdaOperator{\AgdaFunction{⦂}}\AgdaSpace{}%
\AgdaInductiveConstructor{unk⊑unk}\AgdaSymbol{)}\AgdaSpace{}%
\AgdaBound{k}\<%
\end{code}
\begin{code}[hide]%
\>[0]\AgdaFunction{LRᵥ-inject-R-intro-≼}\AgdaSpace{}%
\AgdaSymbol{\{}\AgdaBound{G}\AgdaSymbol{\}}\AgdaSpace{}%
\AgdaSymbol{\{}\AgdaBound{c}\AgdaSymbol{\}}\AgdaSpace{}%
\AgdaSymbol{\{}\AgdaBound{V}\AgdaSymbol{\}}\AgdaSpace{}%
\AgdaSymbol{\{}\AgdaBound{V′}\AgdaSymbol{\}}\AgdaSpace{}%
\AgdaSymbol{\{}\AgdaInductiveConstructor{zero}\AgdaSymbol{\}}\AgdaSpace{}%
\AgdaBound{𝒱VV′}\AgdaSpace{}%
\AgdaSymbol{=}\<%
\\
\>[0][@{}l@{\AgdaIndent{0}}]%
\>[5]\AgdaField{tz}\AgdaSpace{}%
\AgdaSymbol{(}\AgdaInductiveConstructor{≼}\AgdaSpace{}%
\AgdaOperator{\AgdaFunction{∣}}\AgdaSpace{}%
\AgdaBound{V}\AgdaSpace{}%
\AgdaOperator{\AgdaFunction{⊑ᴸᴿᵥ}}\AgdaSpace{}%
\AgdaSymbol{(}\AgdaBound{V′}\AgdaSpace{}%
\AgdaOperator{\AgdaInductiveConstructor{⟨}}\AgdaSpace{}%
\AgdaBound{G}\AgdaSpace{}%
\AgdaOperator{\AgdaInductiveConstructor{!⟩}}\AgdaSymbol{)}\AgdaSpace{}%
\AgdaOperator{\AgdaFunction{⦂}}\AgdaSpace{}%
\AgdaInductiveConstructor{unk⊑unk}\AgdaSymbol{)}\<%
\\
\>[0]\AgdaFunction{LRᵥ-inject-R-intro-≼}\AgdaSpace{}%
\AgdaSymbol{\{}\AgdaBound{G}\AgdaSymbol{\}}\AgdaSpace{}%
\AgdaSymbol{\{}\AgdaBound{c}\AgdaSymbol{\}}\AgdaSpace{}%
\AgdaSymbol{\{}\AgdaBound{V}\AgdaSymbol{\}}\AgdaSpace{}%
\AgdaSymbol{\{}\AgdaBound{V′}\AgdaSymbol{\}}\AgdaSpace{}%
\AgdaSymbol{\{}\AgdaInductiveConstructor{suc}\AgdaSpace{}%
\AgdaBound{k}\AgdaSymbol{\}}\AgdaSpace{}%
\AgdaBound{𝒱VV′sk}\<%
\\
\>[0][@{}l@{\AgdaIndent{0}}]%
\>[4]\AgdaKeyword{with}\AgdaSpace{}%
\AgdaFunction{unk⊑gnd-inv}\AgdaSpace{}%
\AgdaBound{c}\<%
\\
\>[0]\AgdaSymbol{...}%
\>[2773I]\AgdaSymbol{|}\AgdaSpace{}%
\AgdaBound{d}\AgdaSpace{}%
\AgdaOperator{\AgdaInductiveConstructor{,}}\AgdaSpace{}%
\AgdaInductiveConstructor{refl}\<%
\\
\>[.][@{}l@{}]\<[2773I]%
\>[4]\AgdaKeyword{with}\AgdaSpace{}%
\AgdaFunction{LRᵥ-dyn-any-elim-≼}\AgdaSpace{}%
\AgdaSymbol{\{}\AgdaBound{V}\AgdaSymbol{\}\{}\AgdaBound{V′}\AgdaSymbol{\}\{}\AgdaBound{k}\AgdaSymbol{\}\{}\AgdaBound{G}\AgdaSymbol{\}\{\AgdaUnderscore{}\}\{}\AgdaBound{d}\AgdaSymbol{\}}\AgdaSpace{}%
\AgdaBound{𝒱VV′sk}\<%
\\
\>[0]\AgdaSymbol{...}%
\>[2780I]\AgdaSymbol{|}\AgdaSpace{}%
\AgdaBound{V₁}\AgdaSpace{}%
\AgdaOperator{\AgdaInductiveConstructor{,}}\AgdaSpace{}%
\AgdaInductiveConstructor{refl}\AgdaSpace{}%
\AgdaOperator{\AgdaInductiveConstructor{,}}\AgdaSpace{}%
\AgdaBound{v₁}\AgdaSpace{}%
\AgdaOperator{\AgdaInductiveConstructor{,}}\AgdaSpace{}%
\AgdaBound{v′}\AgdaSpace{}%
\AgdaOperator{\AgdaInductiveConstructor{,}}\AgdaSpace{}%
\AgdaBound{𝒱V₁V′k}\<%
\\
\>[.][@{}l@{}]\<[2780I]%
\>[4]\AgdaKeyword{with}\AgdaSpace{}%
\AgdaBound{G}\AgdaSpace{}%
\AgdaOperator{\AgdaFunction{≡ᵍ}}\AgdaSpace{}%
\AgdaBound{G}\<%
\\
\>[0]\AgdaSymbol{...}\AgdaSpace{}%
\AgdaSymbol{|}\AgdaSpace{}%
\AgdaInductiveConstructor{no}\AgdaSpace{}%
\AgdaBound{neq}\AgdaSpace{}%
\AgdaSymbol{=}\AgdaSpace{}%
\AgdaFunction{⊥-elim}\AgdaSpace{}%
\AgdaBound{𝒱VV′sk}\<%
\\
\>[0]\AgdaSymbol{...}%
\>[2799I]\AgdaSymbol{|}\AgdaSpace{}%
\AgdaInductiveConstructor{yes}\AgdaSpace{}%
\AgdaInductiveConstructor{refl}\<%
\\
\>[.][@{}l@{}]\<[2799I]%
\>[4]\AgdaKeyword{with}\AgdaSpace{}%
\AgdaFunction{gnd-prec-unique}\AgdaSpace{}%
\AgdaBound{d}\AgdaSpace{}%
\AgdaFunction{Refl⊑}\<%
\\
\>[0]\AgdaSymbol{...}\AgdaSpace{}%
\AgdaSymbol{|}\AgdaSpace{}%
\AgdaInductiveConstructor{refl}\AgdaSpace{}%
\AgdaSymbol{=}\AgdaSpace{}%
\AgdaBound{v₁}\AgdaSpace{}%
\AgdaOperator{\AgdaInductiveConstructor{,}}\AgdaSpace{}%
\AgdaBound{v′}\AgdaSpace{}%
\AgdaOperator{\AgdaInductiveConstructor{,}}\AgdaSpace{}%
\AgdaBound{𝒱V₁V′k}\<%
\end{code}
\begin{code}[hide]%
\>[0]\AgdaFunction{LRᵥ-inject-R-intro}\AgdaSpace{}%
\AgdaSymbol{:}\AgdaSpace{}%
\AgdaSymbol{∀\{}\AgdaBound{G}\AgdaSymbol{\}\{}\AgdaBound{c}\AgdaSpace{}%
\AgdaSymbol{:}\AgdaSpace{}%
\AgdaInductiveConstructor{★}\AgdaSpace{}%
\AgdaOperator{\AgdaDatatype{⊑}}\AgdaSpace{}%
\AgdaOperator{\AgdaFunction{⌈}}\AgdaSpace{}%
\AgdaBound{G}\AgdaSpace{}%
\AgdaOperator{\AgdaFunction{⌉}}\AgdaSymbol{\}\{}\AgdaBound{V}\AgdaSymbol{\}\{}\AgdaBound{V′}\AgdaSymbol{\}\{}\AgdaBound{k}\AgdaSymbol{\}\{}\AgdaBound{dir}\AgdaSymbol{\}}\<%
\\
\>[0][@{}l@{\AgdaIndent{0}}]%
\>[3]\AgdaSymbol{→}\AgdaSpace{}%
\AgdaField{\#}\AgdaSymbol{(}\AgdaBound{dir}\AgdaSpace{}%
\AgdaOperator{\AgdaFunction{∣}}\AgdaSpace{}%
\AgdaBound{V}\AgdaSpace{}%
\AgdaOperator{\AgdaFunction{⊑ᴸᴿᵥ}}\AgdaSpace{}%
\AgdaBound{V′}\AgdaSpace{}%
\AgdaOperator{\AgdaFunction{⦂}}\AgdaSpace{}%
\AgdaBound{c}\AgdaSymbol{)}\AgdaSpace{}%
\AgdaBound{k}\<%
\\
\>[3]\AgdaSymbol{→}\AgdaSpace{}%
\AgdaField{\#}\AgdaSymbol{(}\AgdaBound{dir}\AgdaSpace{}%
\AgdaOperator{\AgdaFunction{∣}}\AgdaSpace{}%
\AgdaBound{V}\AgdaSpace{}%
\AgdaOperator{\AgdaFunction{⊑ᴸᴿᵥ}}\AgdaSpace{}%
\AgdaSymbol{(}\AgdaBound{V′}\AgdaSpace{}%
\AgdaOperator{\AgdaInductiveConstructor{⟨}}\AgdaSpace{}%
\AgdaBound{G}\AgdaSpace{}%
\AgdaOperator{\AgdaInductiveConstructor{!⟩}}\AgdaSymbol{)}\AgdaSpace{}%
\AgdaOperator{\AgdaFunction{⦂}}\AgdaSpace{}%
\AgdaInductiveConstructor{unk⊑unk}\AgdaSymbol{)}\AgdaSpace{}%
\AgdaBound{k}\<%
\end{code}
\begin{code}[hide]%
\>[0]\AgdaFunction{LRᵥ-inject-R-intro}\AgdaSpace{}%
\AgdaSymbol{\{}\AgdaBound{G}\AgdaSymbol{\}}\AgdaSpace{}%
\AgdaSymbol{\{}\AgdaBound{c}\AgdaSymbol{\}}\AgdaSpace{}%
\AgdaSymbol{\{}\AgdaBound{V}\AgdaSymbol{\}}\AgdaSpace{}%
\AgdaSymbol{\{}\AgdaBound{V′}\AgdaSymbol{\}}\AgdaSpace{}%
\AgdaSymbol{\{}\AgdaBound{k}\AgdaSymbol{\}}\AgdaSpace{}%
\AgdaSymbol{\{}\AgdaInductiveConstructor{≼}\AgdaSymbol{\}}\AgdaSpace{}%
\AgdaBound{𝒱VV′}\AgdaSpace{}%
\AgdaSymbol{=}\<%
\\
\>[0][@{}l@{\AgdaIndent{0}}]%
\>[3]\AgdaFunction{LRᵥ-inject-R-intro-≼}\AgdaSymbol{\{}\AgdaBound{G}\AgdaSymbol{\}}\AgdaSpace{}%
\AgdaSymbol{\{}\AgdaBound{c}\AgdaSymbol{\}}\AgdaSpace{}%
\AgdaSymbol{\{}\AgdaBound{V}\AgdaSymbol{\}}\AgdaSpace{}%
\AgdaSymbol{\{}\AgdaBound{V′}\AgdaSymbol{\}}\AgdaSpace{}%
\AgdaSymbol{\{}\AgdaBound{k}\AgdaSymbol{\}}\AgdaSpace{}%
\AgdaBound{𝒱VV′}\<%
\\
\>[0]\AgdaFunction{LRᵥ-inject-R-intro}\AgdaSpace{}%
\AgdaSymbol{\{}\AgdaBound{G}\AgdaSymbol{\}}\AgdaSpace{}%
\AgdaSymbol{\{}\AgdaBound{c}\AgdaSymbol{\}}\AgdaSpace{}%
\AgdaSymbol{\{}\AgdaBound{V}\AgdaSymbol{\}}\AgdaSpace{}%
\AgdaSymbol{\{}\AgdaBound{V′}\AgdaSymbol{\}}\AgdaSpace{}%
\AgdaSymbol{\{}\AgdaBound{k}\AgdaSymbol{\}}\AgdaSpace{}%
\AgdaSymbol{\{}\AgdaInductiveConstructor{≽}\AgdaSymbol{\}}\AgdaSpace{}%
\AgdaBound{𝒱VV′}\AgdaSpace{}%
\AgdaSymbol{=}\<%
\\
\>[0][@{}l@{\AgdaIndent{0}}]%
\>[3]\AgdaFunction{LRᵥ-inject-R-intro-≽}\AgdaSymbol{\{}\AgdaBound{G}\AgdaSymbol{\}}\AgdaSpace{}%
\AgdaSymbol{\{}\AgdaBound{c}\AgdaSymbol{\}}\AgdaSpace{}%
\AgdaSymbol{\{}\AgdaBound{V}\AgdaSymbol{\}}\AgdaSpace{}%
\AgdaSymbol{\{}\AgdaBound{V′}\AgdaSymbol{\}}\AgdaSpace{}%
\AgdaSymbol{\{}\AgdaBound{k}\AgdaSymbol{\}}\AgdaSpace{}%
\AgdaBound{𝒱VV′}\<%
\end{code}
\begin{code}%
\>[0]\AgdaFunction{compatible-inj-R}\AgdaSpace{}%
\AgdaSymbol{:}\AgdaSpace{}%
\AgdaSymbol{∀\{}\AgdaBound{Γ}\AgdaSymbol{\}\{}\AgdaBound{G}\AgdaSymbol{\}\{}\AgdaBound{c}\AgdaSpace{}%
\AgdaSymbol{:}\AgdaSpace{}%
\AgdaInductiveConstructor{★}\AgdaSpace{}%
\AgdaOperator{\AgdaDatatype{⊑}}\AgdaSpace{}%
\AgdaOperator{\AgdaFunction{⌈}}\AgdaSpace{}%
\AgdaBound{G}\AgdaSpace{}%
\AgdaOperator{\AgdaFunction{⌉}}\AgdaSpace{}%
\AgdaSymbol{\}\{}\AgdaBound{M}\AgdaSpace{}%
\AgdaBound{M′}\AgdaSymbol{\}}\<%
\\
\>[0][@{}l@{\AgdaIndent{0}}]%
\>[3]\AgdaSymbol{→}\AgdaSpace{}%
\AgdaBound{Γ}\AgdaSpace{}%
\AgdaOperator{\AgdaFunction{⊨}}\AgdaSpace{}%
\AgdaBound{M}\AgdaSpace{}%
\AgdaOperator{\AgdaFunction{⊑ᴸᴿ}}\AgdaSpace{}%
\AgdaBound{M′}\AgdaSpace{}%
\AgdaOperator{\AgdaFunction{⦂}}\AgdaSpace{}%
\AgdaSymbol{(}\AgdaInductiveConstructor{★}\AgdaSpace{}%
\AgdaOperator{\AgdaInductiveConstructor{,}}\AgdaSpace{}%
\AgdaOperator{\AgdaFunction{⌈}}\AgdaSpace{}%
\AgdaBound{G}\AgdaSpace{}%
\AgdaOperator{\AgdaFunction{⌉}}\AgdaSpace{}%
\AgdaOperator{\AgdaInductiveConstructor{,}}\AgdaSpace{}%
\AgdaBound{c}\AgdaSymbol{)}\<%
\\
\>[3]\AgdaSymbol{→}\AgdaSpace{}%
\AgdaBound{Γ}\AgdaSpace{}%
\AgdaOperator{\AgdaFunction{⊨}}\AgdaSpace{}%
\AgdaBound{M}\AgdaSpace{}%
\AgdaOperator{\AgdaFunction{⊑ᴸᴿ}}\AgdaSpace{}%
\AgdaBound{M′}\AgdaSpace{}%
\AgdaOperator{\AgdaInductiveConstructor{⟨}}\AgdaSpace{}%
\AgdaBound{G}\AgdaSpace{}%
\AgdaOperator{\AgdaInductiveConstructor{!⟩}}\AgdaSpace{}%
\AgdaOperator{\AgdaFunction{⦂}}\AgdaSpace{}%
\AgdaSymbol{(}\AgdaInductiveConstructor{★}\AgdaSpace{}%
\AgdaOperator{\AgdaInductiveConstructor{,}}\AgdaSpace{}%
\AgdaInductiveConstructor{★}\AgdaSpace{}%
\AgdaOperator{\AgdaInductiveConstructor{,}}\AgdaSpace{}%
\AgdaInductiveConstructor{unk⊑unk}\AgdaSymbol{)}\<%
\end{code}
\begin{code}[hide]%
\>[0]\AgdaFunction{compatible-inj-R}\AgdaSymbol{\{}\AgdaBound{Γ}\AgdaSymbol{\}\{}\AgdaBound{G}\AgdaSymbol{\}\{}\AgdaBound{c}\AgdaSymbol{\}\{}\AgdaBound{M}\AgdaSymbol{\}\{}\AgdaBound{M′}\AgdaSymbol{\}}\AgdaSpace{}%
\AgdaBound{⊨M⊑M′}\<%
\\
\>[0][@{}l@{\AgdaIndent{0}}]%
\>[4]\AgdaKeyword{with}\AgdaSpace{}%
\AgdaFunction{unk⊑gnd-inv}\AgdaSpace{}%
\AgdaBound{c}\<%
\\
\>[0]\AgdaSymbol{...}\AgdaSpace{}%
\AgdaSymbol{|}\AgdaSpace{}%
\AgdaBound{d}\AgdaSpace{}%
\AgdaOperator{\AgdaInductiveConstructor{,}}\AgdaSpace{}%
\AgdaInductiveConstructor{refl}\AgdaSpace{}%
\AgdaSymbol{=}\AgdaSpace{}%
\AgdaSymbol{(λ}\AgdaSpace{}%
\AgdaBound{γ}\AgdaSpace{}%
\AgdaBound{γ′}\AgdaSpace{}%
\AgdaSymbol{→}\AgdaSpace{}%
\AgdaFunction{ℰMM′G}\AgdaSymbol{)}\AgdaSpace{}%
\AgdaOperator{\AgdaInductiveConstructor{,}}\AgdaSpace{}%
\AgdaSymbol{λ}\AgdaSpace{}%
\AgdaBound{γ}\AgdaSpace{}%
\AgdaBound{γ′}\AgdaSpace{}%
\AgdaSymbol{→}\AgdaSpace{}%
\AgdaFunction{ℰMM′G}\<%
\\
\>[0][@{}l@{\AgdaIndent{0}}]%
\>[2]\AgdaKeyword{where}\<%
\\
\>[2]\AgdaFunction{ℰMM′G}\AgdaSpace{}%
\AgdaSymbol{:}\AgdaSpace{}%
\AgdaSymbol{∀\{}\AgdaBound{γ}\AgdaSymbol{\}\{}\AgdaBound{γ′}\AgdaSymbol{\}\{}\AgdaBound{dir}\AgdaSymbol{\}}\<%
\\
\>[2][@{}l@{\AgdaIndent{0}}]%
\>[4]\AgdaSymbol{→}\AgdaSpace{}%
\AgdaSymbol{(}\AgdaBound{Γ}\AgdaSpace{}%
\AgdaOperator{\AgdaFunction{∣}}\AgdaSpace{}%
\AgdaBound{dir}\AgdaSpace{}%
\AgdaOperator{\AgdaFunction{⊨}}\AgdaSpace{}%
\AgdaBound{γ}\AgdaSpace{}%
\AgdaOperator{\AgdaFunction{⊑ᴸᴿ}}\AgdaSpace{}%
\AgdaBound{γ′}\AgdaSymbol{)}\AgdaSpace{}%
\AgdaOperator{\AgdaFunction{⊢ᵒ}}\AgdaSpace{}%
\AgdaBound{dir}\AgdaSpace{}%
\AgdaOperator{\AgdaFunction{∣}}\AgdaSpace{}%
\AgdaSymbol{(}\AgdaOperator{\AgdaFunction{⟪}}\AgdaSpace{}%
\AgdaBound{γ}\AgdaSpace{}%
\AgdaOperator{\AgdaFunction{⟫}}\AgdaSpace{}%
\AgdaBound{M}\AgdaSymbol{)}\AgdaSpace{}%
\AgdaOperator{\AgdaFunction{⊑ᴸᴿₜ}}\AgdaSpace{}%
\AgdaSymbol{(}\AgdaOperator{\AgdaFunction{⟪}}\AgdaSpace{}%
\AgdaBound{γ′}\AgdaSpace{}%
\AgdaOperator{\AgdaFunction{⟫}}\AgdaSpace{}%
\AgdaBound{M′}\AgdaSpace{}%
\AgdaOperator{\AgdaInductiveConstructor{⟨}}\AgdaSpace{}%
\AgdaBound{G}\AgdaSpace{}%
\AgdaOperator{\AgdaInductiveConstructor{!⟩}}\AgdaSymbol{)}\AgdaSpace{}%
\AgdaOperator{\AgdaFunction{⦂}}\AgdaSpace{}%
\AgdaInductiveConstructor{unk⊑unk}\<%
\\
\>[2]\AgdaFunction{ℰMM′G}\AgdaSpace{}%
\AgdaSymbol{\{}\AgdaBound{γ}\AgdaSymbol{\}\{}\AgdaBound{γ′}\AgdaSymbol{\}\{}\AgdaBound{dir}\AgdaSymbol{\}}\AgdaSpace{}%
\AgdaSymbol{=}\AgdaSpace{}%
\AgdaFunction{⊢ᵒ-intro}\AgdaSpace{}%
\AgdaSymbol{λ}\AgdaSpace{}%
\AgdaBound{n}\AgdaSpace{}%
\AgdaBound{𝒫n}\AgdaSpace{}%
\AgdaSymbol{→}\<%
\\
\>[2][@{}l@{\AgdaIndent{0}}]%
\>[3]\AgdaFunction{LRₜ-bind}\AgdaSymbol{\{}\AgdaArgument{c}%
\>[2955I]\AgdaSymbol{=}\AgdaSpace{}%
\AgdaInductiveConstructor{unk⊑unk}\AgdaSymbol{\}\{}\AgdaArgument{d}\AgdaSpace{}%
\AgdaSymbol{=}\AgdaSpace{}%
\AgdaInductiveConstructor{unk⊑}\AgdaSpace{}%
\AgdaBound{d}\AgdaSymbol{\}\{}\AgdaArgument{F}\AgdaSpace{}%
\AgdaSymbol{=}\AgdaSpace{}%
\AgdaInductiveConstructor{□}\AgdaSymbol{\}\{}\AgdaArgument{F′}\AgdaSpace{}%
\AgdaSymbol{=}\AgdaSpace{}%
\AgdaOperator{\AgdaInductiveConstructor{`}}\AgdaSpace{}%
\AgdaSymbol{(}\AgdaOperator{\AgdaInductiveConstructor{□⟨}}\AgdaSpace{}%
\AgdaBound{G}\AgdaSpace{}%
\AgdaOperator{\AgdaInductiveConstructor{!⟩}}\AgdaSymbol{)\}}\<%
\\
\>[.][@{}l@{}]\<[2955I]%
\>[14]\AgdaSymbol{\{}\AgdaOperator{\AgdaFunction{⟪}}\AgdaSpace{}%
\AgdaBound{γ}\AgdaSpace{}%
\AgdaOperator{\AgdaFunction{⟫}}\AgdaSpace{}%
\AgdaBound{M}\AgdaSymbol{\}\{}\AgdaOperator{\AgdaFunction{⟪}}\AgdaSpace{}%
\AgdaBound{γ′}\AgdaSpace{}%
\AgdaOperator{\AgdaFunction{⟫}}\AgdaSpace{}%
\AgdaBound{M′}\AgdaSymbol{\}\{}\AgdaBound{n}\AgdaSymbol{\}\{}\AgdaBound{dir}\AgdaSymbol{\}}\<%
\\
\>[3]\AgdaSymbol{(}\AgdaFunction{⊢ᵒ-elim}\AgdaSpace{}%
\AgdaSymbol{((}\AgdaFunction{proj}\AgdaSpace{}%
\AgdaBound{dir}\AgdaSpace{}%
\AgdaBound{M}\AgdaSpace{}%
\AgdaBound{M′}\AgdaSpace{}%
\AgdaBound{⊨M⊑M′}\AgdaSymbol{)}\AgdaSpace{}%
\AgdaBound{γ}\AgdaSpace{}%
\AgdaBound{γ′}\AgdaSymbol{)}\AgdaSpace{}%
\AgdaBound{n}\AgdaSpace{}%
\AgdaBound{𝒫n}\AgdaSymbol{)}\<%
\\
\>[3]\AgdaSymbol{λ}\AgdaSpace{}%
\AgdaBound{j}\AgdaSpace{}%
\AgdaBound{V}\AgdaSpace{}%
\AgdaBound{V′}\AgdaSpace{}%
\AgdaBound{j≤n}\AgdaSpace{}%
\AgdaBound{M→V}\AgdaSpace{}%
\AgdaBound{v}\AgdaSpace{}%
\AgdaBound{M′→V′}\AgdaSpace{}%
\AgdaBound{v′}\AgdaSpace{}%
\AgdaBound{𝒱VV′j}\AgdaSpace{}%
\AgdaSymbol{→}\<%
\\
\>[3]\AgdaFunction{LRᵥ⇒LRₜ-step}\AgdaSymbol{\{}\AgdaInductiveConstructor{★}\AgdaSymbol{\}\{}\AgdaInductiveConstructor{★}\AgdaSymbol{\}\{}\AgdaInductiveConstructor{unk⊑unk}\AgdaSymbol{\}\{}\AgdaBound{V}\AgdaSymbol{\}\{}\AgdaBound{V′}\AgdaSpace{}%
\AgdaOperator{\AgdaInductiveConstructor{⟨}}\AgdaSpace{}%
\AgdaBound{G}\AgdaSpace{}%
\AgdaOperator{\AgdaInductiveConstructor{!⟩}}\AgdaSymbol{\}\{}\AgdaBound{dir}\AgdaSymbol{\}\{}\AgdaBound{j}\AgdaSymbol{\}}\<%
\\
\>[3]\AgdaSymbol{(}\AgdaFunction{LRᵥ-inject-R-intro}\AgdaSymbol{\{}\AgdaBound{G}\AgdaSymbol{\}\{}\AgdaInductiveConstructor{unk⊑}\AgdaSpace{}%
\AgdaBound{d}\AgdaSymbol{\}\{}\AgdaBound{V}\AgdaSymbol{\}\{}\AgdaBound{V′}\AgdaSymbol{\}\{}\AgdaBound{j}\AgdaSymbol{\}}\AgdaSpace{}%
\AgdaBound{𝒱VV′j}\AgdaSpace{}%
\AgdaSymbol{)}\<%
\end{code}

\paragraph{Compatibility for Projections}

We can have a projection on the left (rule \textsf{⊑-proj-L}) or the
right (rule \textsf{⊑-proj-R}).
Starting on the left, 
\textsf{LRₜ-bind} changes the goal to $V ⟨ H ?⟩$ ⊑ᴸᴿₜ $V′$
assuming that $V$ ⊑ᴸᴿ $V′$.
We need to ensure that $V ⟨ H ?⟩$ reduces to a value without error.
Thankfully, \textsf{⊑-proj-L} says the types of $V$ and $V′$ are related
by \textsf{unk⊑ c} with $c : H ⊑ A′$, and that clause of \textsf{LRᵥ}
tells us that $V = W ⟨ H !⟩$ and
$W$ ⊑ᴸᴿᵥ $V′$. So $W ⟨ H !⟩ ⟨ H ?⟩ \longrightarrow W$
and by anti-reduction we conclude that $W ⟨ H !⟩ ⟨ H ?⟩$ ⊑ᴸᴿₜ $V′$.

\begin{code}%
\>[0]\AgdaFunction{compatible-proj-L}\AgdaSpace{}%
\AgdaSymbol{:}\AgdaSpace{}%
\AgdaSymbol{∀\{}\AgdaBound{Γ}\AgdaSymbol{\}\{}\AgdaBound{H}\AgdaSymbol{\}\{}\AgdaBound{A′}\AgdaSymbol{\}\{}\AgdaBound{c}\AgdaSpace{}%
\AgdaSymbol{:}\AgdaSpace{}%
\AgdaOperator{\AgdaFunction{⌈}}\AgdaSpace{}%
\AgdaBound{H}\AgdaSpace{}%
\AgdaOperator{\AgdaFunction{⌉}}\AgdaSpace{}%
\AgdaOperator{\AgdaDatatype{⊑}}\AgdaSpace{}%
\AgdaBound{A′}\AgdaSymbol{\}\{}\AgdaBound{M}\AgdaSymbol{\}\{}\AgdaBound{M′}\AgdaSymbol{\}}\<%
\\
\>[0][@{}l@{\AgdaIndent{0}}]%
\>[3]\AgdaSymbol{→}\AgdaSpace{}%
\AgdaBound{Γ}\AgdaSpace{}%
\AgdaOperator{\AgdaFunction{⊨}}\AgdaSpace{}%
\AgdaBound{M}\AgdaSpace{}%
\AgdaOperator{\AgdaFunction{⊑ᴸᴿ}}\AgdaSpace{}%
\AgdaBound{M′}\AgdaSpace{}%
\AgdaOperator{\AgdaFunction{⦂}}\AgdaSpace{}%
\AgdaSymbol{(}\AgdaInductiveConstructor{★}\AgdaSpace{}%
\AgdaOperator{\AgdaInductiveConstructor{,}}\AgdaSpace{}%
\AgdaBound{A′}\AgdaSpace{}%
\AgdaOperator{\AgdaInductiveConstructor{,}}%
\>[31]\AgdaInductiveConstructor{unk⊑}\AgdaSpace{}%
\AgdaBound{c}\AgdaSymbol{)}\<%
\\
\>[3]\AgdaSymbol{→}\AgdaSpace{}%
\AgdaBound{Γ}\AgdaSpace{}%
\AgdaOperator{\AgdaFunction{⊨}}\AgdaSpace{}%
\AgdaBound{M}\AgdaSpace{}%
\AgdaOperator{\AgdaInductiveConstructor{⟨}}\AgdaSpace{}%
\AgdaBound{H}\AgdaSpace{}%
\AgdaOperator{\AgdaInductiveConstructor{?⟩}}\AgdaSpace{}%
\AgdaOperator{\AgdaFunction{⊑ᴸᴿ}}\AgdaSpace{}%
\AgdaBound{M′}\AgdaSpace{}%
\AgdaOperator{\AgdaFunction{⦂}}\AgdaSpace{}%
\AgdaSymbol{(}\AgdaOperator{\AgdaFunction{⌈}}\AgdaSpace{}%
\AgdaBound{H}\AgdaSpace{}%
\AgdaOperator{\AgdaFunction{⌉}}\AgdaSpace{}%
\AgdaOperator{\AgdaInductiveConstructor{,}}\AgdaSpace{}%
\AgdaBound{A′}\AgdaSpace{}%
\AgdaOperator{\AgdaInductiveConstructor{,}}\AgdaSpace{}%
\AgdaBound{c}\AgdaSymbol{)}\<%
\end{code}
\begin{code}[hide]%
\>[0]\AgdaFunction{compatible-proj-L}\AgdaSpace{}%
\AgdaSymbol{\{}\AgdaBound{Γ}\AgdaSymbol{\}\{}\AgdaBound{H}\AgdaSymbol{\}\{}\AgdaBound{A′}\AgdaSymbol{\}\{}\AgdaBound{c}\AgdaSymbol{\}\{}\AgdaBound{M}\AgdaSymbol{\}\{}\AgdaBound{M′}\AgdaSymbol{\}}\AgdaSpace{}%
\AgdaBound{⊨M⊑M′}\AgdaSpace{}%
\AgdaSymbol{=}\<%
\\
\>[0][@{}l@{\AgdaIndent{0}}]%
\>[2]\AgdaSymbol{(λ}\AgdaSpace{}%
\AgdaBound{γ}\AgdaSpace{}%
\AgdaBound{γ′}\AgdaSpace{}%
\AgdaSymbol{→}\AgdaSpace{}%
\AgdaFunction{ℰMHM′}\AgdaSymbol{)}\AgdaSpace{}%
\AgdaOperator{\AgdaInductiveConstructor{,}}\AgdaSpace{}%
\AgdaSymbol{λ}\AgdaSpace{}%
\AgdaBound{γ}\AgdaSpace{}%
\AgdaBound{γ′}\AgdaSpace{}%
\AgdaSymbol{→}\AgdaSpace{}%
\AgdaFunction{ℰMHM′}\<%
\\
\>[2]\AgdaKeyword{where}\<%
\\
\>[2]\AgdaFunction{ℰMHM′}\AgdaSpace{}%
\AgdaSymbol{:}\AgdaSpace{}%
\AgdaSymbol{∀\{}\AgdaBound{γ}\AgdaSymbol{\}\{}\AgdaBound{γ′}\AgdaSymbol{\}\{}\AgdaBound{dir}\AgdaSymbol{\}}\AgdaSpace{}%
\AgdaSymbol{→}\AgdaSpace{}%
\AgdaSymbol{(}\AgdaBound{Γ}\AgdaSpace{}%
\AgdaOperator{\AgdaFunction{∣}}\AgdaSpace{}%
\AgdaBound{dir}\AgdaSpace{}%
\AgdaOperator{\AgdaFunction{⊨}}\AgdaSpace{}%
\AgdaBound{γ}\AgdaSpace{}%
\AgdaOperator{\AgdaFunction{⊑ᴸᴿ}}\AgdaSpace{}%
\AgdaBound{γ′}\AgdaSymbol{)}\<%
\\
\>[2][@{}l@{\AgdaIndent{0}}]%
\>[7]\AgdaOperator{\AgdaFunction{⊢ᵒ}}\AgdaSpace{}%
\AgdaBound{dir}\AgdaSpace{}%
\AgdaOperator{\AgdaFunction{∣}}\AgdaSpace{}%
\AgdaSymbol{(}\AgdaOperator{\AgdaFunction{⟪}}\AgdaSpace{}%
\AgdaBound{γ}\AgdaSpace{}%
\AgdaOperator{\AgdaFunction{⟫}}\AgdaSpace{}%
\AgdaBound{M}\AgdaSpace{}%
\AgdaOperator{\AgdaInductiveConstructor{⟨}}\AgdaSpace{}%
\AgdaBound{H}\AgdaSpace{}%
\AgdaOperator{\AgdaInductiveConstructor{?⟩}}\AgdaSymbol{)}\AgdaSpace{}%
\AgdaOperator{\AgdaFunction{⊑ᴸᴿₜ}}\AgdaSpace{}%
\AgdaSymbol{(}\AgdaOperator{\AgdaFunction{⟪}}\AgdaSpace{}%
\AgdaBound{γ′}\AgdaSpace{}%
\AgdaOperator{\AgdaFunction{⟫}}\AgdaSpace{}%
\AgdaBound{M′}\AgdaSymbol{)}\AgdaSpace{}%
\AgdaOperator{\AgdaFunction{⦂}}\AgdaSpace{}%
\AgdaBound{c}\<%
\\
\>[2]\AgdaFunction{ℰMHM′}\AgdaSpace{}%
\AgdaSymbol{\{}\AgdaBound{γ}\AgdaSymbol{\}\{}\AgdaBound{γ′}\AgdaSymbol{\}\{}\AgdaBound{dir}\AgdaSymbol{\}}\AgdaSpace{}%
\AgdaSymbol{=}\AgdaSpace{}%
\AgdaFunction{⊢ᵒ-intro}\AgdaSpace{}%
\AgdaSymbol{λ}\AgdaSpace{}%
\AgdaBound{n}\AgdaSpace{}%
\AgdaBound{𝒫n}\AgdaSpace{}%
\AgdaSymbol{→}\<%
\\
\>[2][@{}l@{\AgdaIndent{0}}]%
\>[3]\AgdaFunction{LRₜ-bind}\AgdaSymbol{\{}\AgdaArgument{c}%
\>[3079I]\AgdaSymbol{=}\AgdaSpace{}%
\AgdaBound{c}\AgdaSymbol{\}\{}\AgdaArgument{d}\AgdaSpace{}%
\AgdaSymbol{=}\AgdaSpace{}%
\AgdaInductiveConstructor{unk⊑}\AgdaSpace{}%
\AgdaBound{c}\AgdaSymbol{\}\{}\AgdaArgument{F}\AgdaSpace{}%
\AgdaSymbol{=}\AgdaSpace{}%
\AgdaOperator{\AgdaInductiveConstructor{`}}\AgdaSpace{}%
\AgdaSymbol{(}\AgdaOperator{\AgdaInductiveConstructor{□⟨}}\AgdaSpace{}%
\AgdaBound{H}\AgdaSpace{}%
\AgdaOperator{\AgdaInductiveConstructor{?⟩}}\AgdaSymbol{)\}\{}\AgdaArgument{F′}\AgdaSpace{}%
\AgdaSymbol{=}\AgdaSpace{}%
\AgdaInductiveConstructor{□}\AgdaSymbol{\}}\<%
\\
\>[.][@{}l@{}]\<[3079I]%
\>[14]\AgdaSymbol{\{}\AgdaOperator{\AgdaFunction{⟪}}\AgdaSpace{}%
\AgdaBound{γ}\AgdaSpace{}%
\AgdaOperator{\AgdaFunction{⟫}}\AgdaSpace{}%
\AgdaBound{M}\AgdaSymbol{\}\{}\AgdaOperator{\AgdaFunction{⟪}}\AgdaSpace{}%
\AgdaBound{γ′}\AgdaSpace{}%
\AgdaOperator{\AgdaFunction{⟫}}\AgdaSpace{}%
\AgdaBound{M′}\AgdaSymbol{\}\{}\AgdaBound{n}\AgdaSymbol{\}\{}\AgdaBound{dir}\AgdaSymbol{\}}\<%
\\
\>[3]\AgdaSymbol{(}\AgdaFunction{⊢ᵒ-elim}\AgdaSpace{}%
\AgdaSymbol{((}\AgdaFunction{proj}\AgdaSpace{}%
\AgdaBound{dir}\AgdaSpace{}%
\AgdaBound{M}\AgdaSpace{}%
\AgdaBound{M′}\AgdaSpace{}%
\AgdaBound{⊨M⊑M′}\AgdaSymbol{)}\AgdaSpace{}%
\AgdaBound{γ}\AgdaSpace{}%
\AgdaBound{γ′}\AgdaSymbol{)}\AgdaSpace{}%
\AgdaBound{n}\AgdaSpace{}%
\AgdaBound{𝒫n}\AgdaSymbol{)}\<%
\\
\>[3]\AgdaSymbol{λ}\AgdaSpace{}%
\AgdaBound{j}\AgdaSpace{}%
\AgdaBound{V}\AgdaSpace{}%
\AgdaBound{V′}\AgdaSpace{}%
\AgdaBound{j≤n}\AgdaSpace{}%
\AgdaBound{M→V}\AgdaSpace{}%
\AgdaBound{v}\AgdaSpace{}%
\AgdaBound{M′→V′}\AgdaSpace{}%
\AgdaBound{v′}\AgdaSpace{}%
\AgdaBound{𝒱VV′j}\AgdaSpace{}%
\AgdaSymbol{→}\AgdaSpace{}%
\AgdaFunction{Goal}\AgdaSymbol{\{}\AgdaBound{j}\AgdaSymbol{\}\{}\AgdaBound{V}\AgdaSymbol{\}\{}\AgdaBound{V′}\AgdaSymbol{\}\{}\AgdaBound{dir}\AgdaSymbol{\}}\AgdaSpace{}%
\AgdaBound{𝒱VV′j}\<%
\\
\>[3]\AgdaKeyword{where}\<%
\\
\>[3]\AgdaFunction{Goal}\AgdaSpace{}%
\AgdaSymbol{:}\AgdaSpace{}%
\AgdaSymbol{∀\{}\AgdaBound{j}\AgdaSymbol{\}\{}\AgdaBound{V}\AgdaSymbol{\}\{}\AgdaBound{V′}\AgdaSymbol{\}\{}\AgdaBound{dir}\AgdaSymbol{\}}\<%
\\
\>[3][@{}l@{\AgdaIndent{0}}]%
\>[7]\AgdaSymbol{→}\AgdaSpace{}%
\AgdaField{\#}\AgdaSymbol{(}\AgdaBound{dir}\AgdaSpace{}%
\AgdaOperator{\AgdaFunction{∣}}\AgdaSpace{}%
\AgdaBound{V}\AgdaSpace{}%
\AgdaOperator{\AgdaFunction{⊑ᴸᴿᵥ}}\AgdaSpace{}%
\AgdaBound{V′}\AgdaSpace{}%
\AgdaOperator{\AgdaFunction{⦂}}\AgdaSpace{}%
\AgdaInductiveConstructor{unk⊑}\AgdaSpace{}%
\AgdaBound{c}\AgdaSymbol{)}\AgdaSpace{}%
\AgdaBound{j}\<%
\\
\>[7]\AgdaSymbol{→}\AgdaSpace{}%
\AgdaField{\#}\AgdaSymbol{(}\AgdaBound{dir}\AgdaSpace{}%
\AgdaOperator{\AgdaFunction{∣}}\AgdaSpace{}%
\AgdaSymbol{(}\AgdaBound{V}\AgdaSpace{}%
\AgdaOperator{\AgdaInductiveConstructor{⟨}}\AgdaSpace{}%
\AgdaBound{H}\AgdaSpace{}%
\AgdaOperator{\AgdaInductiveConstructor{?⟩}}\AgdaSymbol{)}\AgdaSpace{}%
\AgdaOperator{\AgdaFunction{⊑ᴸᴿₜ}}\AgdaSpace{}%
\AgdaBound{V′}\AgdaSpace{}%
\AgdaOperator{\AgdaFunction{⦂}}\AgdaSpace{}%
\AgdaBound{c}\AgdaSymbol{)}\AgdaSpace{}%
\AgdaBound{j}\<%
\\
\>[3]\AgdaFunction{Goal}\AgdaSpace{}%
\AgdaSymbol{\{}\AgdaInductiveConstructor{zero}\AgdaSymbol{\}}\AgdaSpace{}%
\AgdaSymbol{\{}\AgdaBound{V}\AgdaSymbol{\}}\AgdaSpace{}%
\AgdaSymbol{\{}\AgdaBound{V′}\AgdaSymbol{\}\{}\AgdaBound{dir}\AgdaSymbol{\}}\AgdaSpace{}%
\AgdaBound{𝒱VV′j}\AgdaSpace{}%
\AgdaSymbol{=}\<%
\\
\>[3][@{}l@{\AgdaIndent{0}}]%
\>[7]\AgdaField{tz}\AgdaSpace{}%
\AgdaSymbol{(}\AgdaBound{dir}\AgdaSpace{}%
\AgdaOperator{\AgdaFunction{∣}}\AgdaSpace{}%
\AgdaSymbol{(}\AgdaBound{V}\AgdaSpace{}%
\AgdaOperator{\AgdaInductiveConstructor{⟨}}\AgdaSpace{}%
\AgdaBound{H}\AgdaSpace{}%
\AgdaOperator{\AgdaInductiveConstructor{?⟩}}\AgdaSymbol{)}\AgdaSpace{}%
\AgdaOperator{\AgdaFunction{⊑ᴸᴿₜ}}\AgdaSpace{}%
\AgdaBound{V′}\AgdaSpace{}%
\AgdaOperator{\AgdaFunction{⦂}}\AgdaSpace{}%
\AgdaBound{c}\AgdaSymbol{)}\<%
\\
\>[3]\AgdaFunction{Goal}\AgdaSpace{}%
\AgdaSymbol{\{}\AgdaInductiveConstructor{suc}\AgdaSpace{}%
\AgdaBound{j}\AgdaSymbol{\}}\AgdaSpace{}%
\AgdaSymbol{\{}\AgdaBound{V}\AgdaSymbol{\}}\AgdaSpace{}%
\AgdaSymbol{\{}\AgdaBound{V′}\AgdaSymbol{\}\{}\AgdaInductiveConstructor{≼}\AgdaSymbol{\}}\AgdaSpace{}%
\AgdaBound{𝒱VV′sj}\<%
\\
\>[3][@{}l@{\AgdaIndent{0}}]%
\>[7]\AgdaKeyword{with}\AgdaSpace{}%
\AgdaFunction{LRᵥ-dyn-any-elim-≼}\AgdaSymbol{\{}\AgdaBound{V}\AgdaSymbol{\}\{}\AgdaBound{V′}\AgdaSymbol{\}\{}\AgdaBound{j}\AgdaSymbol{\}\{}\AgdaBound{H}\AgdaSymbol{\}\{}\AgdaBound{A′}\AgdaSymbol{\}\{}\AgdaBound{c}\AgdaSymbol{\}}\AgdaSpace{}%
\AgdaBound{𝒱VV′sj}\<%
\\
\>[3]\AgdaSymbol{...}%
\>[3162I]\AgdaSymbol{|}\AgdaSpace{}%
\AgdaBound{V₁}\AgdaSpace{}%
\AgdaOperator{\AgdaInductiveConstructor{,}}\AgdaSpace{}%
\AgdaInductiveConstructor{refl}\AgdaSpace{}%
\AgdaOperator{\AgdaInductiveConstructor{,}}\AgdaSpace{}%
\AgdaBound{v₁}\AgdaSpace{}%
\AgdaOperator{\AgdaInductiveConstructor{,}}\AgdaSpace{}%
\AgdaBound{v′}\AgdaSpace{}%
\AgdaOperator{\AgdaInductiveConstructor{,}}\AgdaSpace{}%
\AgdaBound{𝒱V₁V′j}\AgdaSpace{}%
\AgdaSymbol{=}\<%
\\
\>[.][@{}l@{}]\<[3162I]%
\>[7]\AgdaKeyword{let}\AgdaSpace{}%
\AgdaBound{V₁HH→V₁}\AgdaSpace{}%
\AgdaSymbol{=}\AgdaSpace{}%
\AgdaInductiveConstructor{collapse}\AgdaSymbol{\{}\AgdaBound{H}\AgdaSymbol{\}\{}\AgdaArgument{V}\AgdaSpace{}%
\AgdaSymbol{=}\AgdaSpace{}%
\AgdaBound{V₁}\AgdaSymbol{\}}\AgdaSpace{}%
\AgdaBound{v₁}\AgdaSpace{}%
\AgdaInductiveConstructor{refl}\AgdaSpace{}%
\AgdaKeyword{in}\<%
\\
\>[7]\AgdaKeyword{let}\AgdaSpace{}%
\AgdaBound{ℰV₁V′j}\AgdaSpace{}%
\AgdaSymbol{=}\AgdaSpace{}%
\AgdaFunction{LRᵥ⇒LRₜ-step}\AgdaSymbol{\{}\AgdaOperator{\AgdaFunction{⌈}}\AgdaSpace{}%
\AgdaBound{H}\AgdaSpace{}%
\AgdaOperator{\AgdaFunction{⌉}}\AgdaSymbol{\}\{}\AgdaBound{A′}\AgdaSymbol{\}\{}\AgdaBound{c}\AgdaSymbol{\}\{}\AgdaBound{V₁}\AgdaSymbol{\}\{}\AgdaBound{V′}\AgdaSymbol{\}\{}\AgdaInductiveConstructor{≼}\AgdaSymbol{\}\{}\AgdaBound{j}\AgdaSymbol{\}}\AgdaSpace{}%
\AgdaBound{𝒱V₁V′j}\AgdaSpace{}%
\AgdaKeyword{in}\<%
\\
\>[7]\AgdaFunction{anti-reduction-≼-L-one}\AgdaSpace{}%
\AgdaBound{ℰV₁V′j}\AgdaSpace{}%
\AgdaBound{V₁HH→V₁}\<%
\\
\>[3]\AgdaFunction{Goal}\AgdaSpace{}%
\AgdaSymbol{\{}\AgdaInductiveConstructor{suc}\AgdaSpace{}%
\AgdaBound{j}\AgdaSymbol{\}}\AgdaSpace{}%
\AgdaSymbol{\{}\AgdaBound{V}\AgdaSymbol{\}}\AgdaSpace{}%
\AgdaSymbol{\{}\AgdaBound{V′}\AgdaSymbol{\}\{}\AgdaInductiveConstructor{≽}\AgdaSymbol{\}}\AgdaSpace{}%
\AgdaBound{𝒱VV′sj}\<%
\\
\>[3][@{}l@{\AgdaIndent{0}}]%
\>[7]\AgdaKeyword{with}\AgdaSpace{}%
\AgdaFunction{LRᵥ-dyn-any-elim-≽}\AgdaSymbol{\{}\AgdaBound{V}\AgdaSymbol{\}\{}\AgdaBound{V′}\AgdaSymbol{\}\{}\AgdaBound{j}\AgdaSymbol{\}\{}\AgdaBound{H}\AgdaSymbol{\}\{}\AgdaBound{A′}\AgdaSymbol{\}\{}\AgdaBound{c}\AgdaSymbol{\}}\AgdaSpace{}%
\AgdaBound{𝒱VV′sj}\<%
\\
\>[3]\AgdaSymbol{...}%
\>[3197I]\AgdaSymbol{|}\AgdaSpace{}%
\AgdaBound{V₁}\AgdaSpace{}%
\AgdaOperator{\AgdaInductiveConstructor{,}}\AgdaSpace{}%
\AgdaInductiveConstructor{refl}\AgdaSpace{}%
\AgdaOperator{\AgdaInductiveConstructor{,}}\AgdaSpace{}%
\AgdaBound{v₁}\AgdaSpace{}%
\AgdaOperator{\AgdaInductiveConstructor{,}}\AgdaSpace{}%
\AgdaBound{v′}\AgdaSpace{}%
\AgdaOperator{\AgdaInductiveConstructor{,}}\AgdaSpace{}%
\AgdaBound{𝒱V₁V′sj}\AgdaSpace{}%
\AgdaSymbol{=}\<%
\\
\>[.][@{}l@{}]\<[3197I]%
\>[7]\AgdaKeyword{let}\AgdaSpace{}%
\AgdaBound{V₁HH→V₁}\AgdaSpace{}%
\AgdaSymbol{=}\AgdaSpace{}%
\AgdaInductiveConstructor{collapse}\AgdaSymbol{\{}\AgdaBound{H}\AgdaSymbol{\}\{}\AgdaArgument{V}\AgdaSpace{}%
\AgdaSymbol{=}\AgdaSpace{}%
\AgdaBound{V₁}\AgdaSymbol{\}}\AgdaSpace{}%
\AgdaBound{v₁}\AgdaSpace{}%
\AgdaInductiveConstructor{refl}\AgdaSpace{}%
\AgdaKeyword{in}\<%
\\
\>[7]\AgdaInductiveConstructor{inj₂}\AgdaSpace{}%
\AgdaSymbol{(}\AgdaInductiveConstructor{inj₂}\AgdaSpace{}%
\AgdaSymbol{(}\AgdaBound{v′}\AgdaSpace{}%
\AgdaOperator{\AgdaInductiveConstructor{,}}\AgdaSpace{}%
\AgdaBound{V₁}\AgdaSpace{}%
\AgdaOperator{\AgdaInductiveConstructor{,}}\AgdaSpace{}%
\AgdaFunction{unit}\AgdaSpace{}%
\AgdaBound{V₁HH→V₁}\AgdaSpace{}%
\AgdaOperator{\AgdaInductiveConstructor{,}}\AgdaSpace{}%
\AgdaBound{v₁}\AgdaSpace{}%
\AgdaOperator{\AgdaInductiveConstructor{,}}\AgdaSpace{}%
\AgdaBound{𝒱V₁V′sj}\AgdaSymbol{))}\<%
\end{code}

When the projection is on the right, there is less to worry about.
\textsf{LRₜ-bind} changes the goal to $V$ ⊑ᴸᴿₜ $V′ ⟨ H ?⟩$
assuming that $V$ ⊑ᴸᴿᵥ $V′$. We have $V′ = W′ ⟨ G !⟩$
and $V$ ⊑ᴸᴿᵥ $W′$.
If $G = H$ then $W′ ⟨ G !⟩⟨H ?⟩ \longrightarrow W′$
and by anti-reduction, $V$ ⊑ᴸᴿₜ $W′ ⟨ G !⟩ ⟨ H ?⟩$.
If $G ≠ H$, then $W′ ⟨ G !⟩⟨H ?⟩ \longrightarrow \textsf{blame}$.
Since $V$ ⊑ᴸᴿₜ \textsf{blame}, we use anti-reduction
to conclude $V$ ⊑ᴸᴿₜ $W′ ⟨ G !⟩⟨H ?⟩$.

\begin{AgdaSuppressSpace}
\begin{code}%
\>[0]\AgdaFunction{compatible-proj-R}\AgdaSpace{}%
\AgdaSymbol{:}\AgdaSpace{}%
\AgdaSymbol{∀\{}\AgdaBound{Γ}\AgdaSymbol{\}\{}\AgdaBound{H}\AgdaSymbol{\}\{}\AgdaBound{c}\AgdaSpace{}%
\AgdaSymbol{:}\AgdaSpace{}%
\AgdaInductiveConstructor{★}\AgdaSpace{}%
\AgdaOperator{\AgdaDatatype{⊑}}\AgdaSpace{}%
\AgdaOperator{\AgdaFunction{⌈}}\AgdaSpace{}%
\AgdaBound{H}\AgdaSpace{}%
\AgdaOperator{\AgdaFunction{⌉}}\AgdaSymbol{\}\{}\AgdaBound{M}\AgdaSymbol{\}\{}\AgdaBound{M′}\AgdaSymbol{\}}\<%
\\
\>[0][@{}l@{\AgdaIndent{0}}]%
\>[3]\AgdaSymbol{→}\AgdaSpace{}%
\AgdaBound{Γ}\AgdaSpace{}%
\AgdaOperator{\AgdaFunction{⊨}}\AgdaSpace{}%
\AgdaBound{M}\AgdaSpace{}%
\AgdaOperator{\AgdaFunction{⊑ᴸᴿ}}\AgdaSpace{}%
\AgdaBound{M′}\AgdaSpace{}%
\AgdaOperator{\AgdaFunction{⦂}}\AgdaSpace{}%
\AgdaSymbol{(}\AgdaInductiveConstructor{★}\AgdaSpace{}%
\AgdaOperator{\AgdaInductiveConstructor{,}}\AgdaSpace{}%
\AgdaInductiveConstructor{★}\AgdaSpace{}%
\AgdaOperator{\AgdaInductiveConstructor{,}}\AgdaSpace{}%
\AgdaInductiveConstructor{unk⊑unk}\AgdaSymbol{)}\<%
\\
\>[3]\AgdaSymbol{→}\AgdaSpace{}%
\AgdaBound{Γ}\AgdaSpace{}%
\AgdaOperator{\AgdaFunction{⊨}}\AgdaSpace{}%
\AgdaBound{M}\AgdaSpace{}%
\AgdaOperator{\AgdaFunction{⊑ᴸᴿ}}\AgdaSpace{}%
\AgdaBound{M′}\AgdaSpace{}%
\AgdaOperator{\AgdaInductiveConstructor{⟨}}\AgdaSpace{}%
\AgdaBound{H}\AgdaSpace{}%
\AgdaOperator{\AgdaInductiveConstructor{?⟩}}\AgdaSpace{}%
\AgdaOperator{\AgdaFunction{⦂}}\AgdaSpace{}%
\AgdaSymbol{(}\AgdaInductiveConstructor{★}\AgdaSpace{}%
\AgdaOperator{\AgdaInductiveConstructor{,}}\AgdaSpace{}%
\AgdaOperator{\AgdaFunction{⌈}}\AgdaSpace{}%
\AgdaBound{H}\AgdaSpace{}%
\AgdaOperator{\AgdaFunction{⌉}}\AgdaSpace{}%
\AgdaOperator{\AgdaInductiveConstructor{,}}\AgdaSpace{}%
\AgdaBound{c}\AgdaSymbol{)}\<%
\end{code}
\begin{code}[hide]%
\>[0]\AgdaFunction{compatible-proj-R}\AgdaSpace{}%
\AgdaSymbol{\{}\AgdaBound{Γ}\AgdaSymbol{\}\{}\AgdaBound{H}\AgdaSymbol{\}\{}\AgdaBound{c}\AgdaSymbol{\}\{}\AgdaBound{M}\AgdaSymbol{\}\{}\AgdaBound{M′}\AgdaSymbol{\}}\AgdaSpace{}%
\AgdaBound{⊨M⊑M′}\<%
\\
\>[0][@{}l@{\AgdaIndent{0}}]%
\>[4]\AgdaKeyword{with}\AgdaSpace{}%
\AgdaFunction{unk⊑gnd-inv}\AgdaSpace{}%
\AgdaBound{c}\<%
\\
\>[0]\AgdaSymbol{...}%
\>[3266I]\AgdaSymbol{|}\AgdaSpace{}%
\AgdaBound{d}\AgdaSpace{}%
\AgdaOperator{\AgdaInductiveConstructor{,}}\AgdaSpace{}%
\AgdaInductiveConstructor{refl}\AgdaSpace{}%
\AgdaSymbol{=}\AgdaSpace{}%
\AgdaSymbol{(λ}\AgdaSpace{}%
\AgdaBound{γ}\AgdaSpace{}%
\AgdaBound{γ′}\AgdaSpace{}%
\AgdaSymbol{→}\AgdaSpace{}%
\AgdaFunction{ℰMM′H}\AgdaSymbol{)}\AgdaSpace{}%
\AgdaOperator{\AgdaInductiveConstructor{,}}\AgdaSpace{}%
\AgdaSymbol{λ}\AgdaSpace{}%
\AgdaBound{γ}\AgdaSpace{}%
\AgdaBound{γ′}\AgdaSpace{}%
\AgdaSymbol{→}\AgdaSpace{}%
\AgdaFunction{ℰMM′H}\<%
\\
\>[.][@{}l@{}]\<[3266I]%
\>[4]\AgdaKeyword{where}\<%
\\
\>[4]\AgdaFunction{ℰMM′H}\AgdaSpace{}%
\AgdaSymbol{:}%
\>[3283I]\AgdaSymbol{∀\{}\AgdaBound{γ}\AgdaSymbol{\}\{}\AgdaBound{γ′}\AgdaSymbol{\}\{}\AgdaBound{dir}\AgdaSymbol{\}}\AgdaSpace{}%
\AgdaSymbol{→}\AgdaSpace{}%
\AgdaSymbol{(}\AgdaBound{Γ}\AgdaSpace{}%
\AgdaOperator{\AgdaFunction{∣}}\AgdaSpace{}%
\AgdaBound{dir}\AgdaSpace{}%
\AgdaOperator{\AgdaFunction{⊨}}\AgdaSpace{}%
\AgdaBound{γ}\AgdaSpace{}%
\AgdaOperator{\AgdaFunction{⊑ᴸᴿ}}\AgdaSpace{}%
\AgdaBound{γ′}\AgdaSymbol{)}\<%
\\
\>[3283I][@{}l@{\AgdaIndent{0}}]%
\>[13]\AgdaOperator{\AgdaFunction{⊢ᵒ}}\AgdaSpace{}%
\AgdaBound{dir}\AgdaSpace{}%
\AgdaOperator{\AgdaFunction{∣}}\AgdaSpace{}%
\AgdaSymbol{(}\AgdaOperator{\AgdaFunction{⟪}}\AgdaSpace{}%
\AgdaBound{γ}\AgdaSpace{}%
\AgdaOperator{\AgdaFunction{⟫}}\AgdaSpace{}%
\AgdaBound{M}\AgdaSymbol{)}\AgdaSpace{}%
\AgdaOperator{\AgdaFunction{⊑ᴸᴿₜ}}\AgdaSpace{}%
\AgdaSymbol{(}\AgdaOperator{\AgdaFunction{⟪}}\AgdaSpace{}%
\AgdaBound{γ′}\AgdaSpace{}%
\AgdaOperator{\AgdaFunction{⟫}}\AgdaSpace{}%
\AgdaBound{M′}\AgdaSpace{}%
\AgdaOperator{\AgdaInductiveConstructor{⟨}}\AgdaSpace{}%
\AgdaBound{H}\AgdaSpace{}%
\AgdaOperator{\AgdaInductiveConstructor{?⟩}}\AgdaSymbol{)}\AgdaSpace{}%
\AgdaOperator{\AgdaFunction{⦂}}\AgdaSpace{}%
\AgdaInductiveConstructor{unk⊑}\AgdaSpace{}%
\AgdaBound{d}\<%
\\
\>[4]\AgdaFunction{ℰMM′H}\AgdaSpace{}%
\AgdaSymbol{\{}\AgdaBound{γ}\AgdaSymbol{\}\{}\AgdaBound{γ′}\AgdaSymbol{\}\{}\AgdaBound{dir}\AgdaSymbol{\}}\AgdaSpace{}%
\AgdaSymbol{=}\AgdaSpace{}%
\AgdaFunction{⊢ᵒ-intro}\AgdaSpace{}%
\AgdaSymbol{λ}\AgdaSpace{}%
\AgdaBound{n}\AgdaSpace{}%
\AgdaBound{𝒫n}\AgdaSpace{}%
\AgdaSymbol{→}\<%
\\
\>[4][@{}l@{\AgdaIndent{0}}]%
\>[5]\AgdaFunction{LRₜ-bind}\AgdaSymbol{\{}\AgdaArgument{c}%
\>[3316I]\AgdaSymbol{=}\AgdaSpace{}%
\AgdaBound{c}\AgdaSymbol{\}\{}\AgdaArgument{d}\AgdaSpace{}%
\AgdaSymbol{=}\AgdaSpace{}%
\AgdaInductiveConstructor{unk⊑unk}\AgdaSymbol{\}\{}\AgdaArgument{F}\AgdaSpace{}%
\AgdaSymbol{=}\AgdaSpace{}%
\AgdaInductiveConstructor{□}\AgdaSymbol{\}\{}\AgdaArgument{F′}\AgdaSpace{}%
\AgdaSymbol{=}\AgdaSpace{}%
\AgdaOperator{\AgdaInductiveConstructor{`}}\AgdaSpace{}%
\AgdaOperator{\AgdaInductiveConstructor{□⟨}}\AgdaSpace{}%
\AgdaBound{H}\AgdaSpace{}%
\AgdaOperator{\AgdaInductiveConstructor{?⟩}}\AgdaSymbol{\}}\<%
\\
\>[.][@{}l@{}]\<[3316I]%
\>[16]\AgdaSymbol{\{}\AgdaOperator{\AgdaFunction{⟪}}\AgdaSpace{}%
\AgdaBound{γ}\AgdaSpace{}%
\AgdaOperator{\AgdaFunction{⟫}}\AgdaSpace{}%
\AgdaBound{M}\AgdaSymbol{\}\{}\AgdaOperator{\AgdaFunction{⟪}}\AgdaSpace{}%
\AgdaBound{γ′}\AgdaSpace{}%
\AgdaOperator{\AgdaFunction{⟫}}\AgdaSpace{}%
\AgdaBound{M′}\AgdaSymbol{\}\{}\AgdaBound{n}\AgdaSymbol{\}\{}\AgdaBound{dir}\AgdaSymbol{\}}\<%
\\
\>[5]\AgdaSymbol{(}\AgdaFunction{⊢ᵒ-elim}\AgdaSpace{}%
\AgdaSymbol{((}\AgdaFunction{proj}\AgdaSpace{}%
\AgdaBound{dir}\AgdaSpace{}%
\AgdaBound{M}\AgdaSpace{}%
\AgdaBound{M′}\AgdaSpace{}%
\AgdaBound{⊨M⊑M′}\AgdaSymbol{)}\AgdaSpace{}%
\AgdaBound{γ}\AgdaSpace{}%
\AgdaBound{γ′}\AgdaSymbol{)}\AgdaSpace{}%
\AgdaBound{n}\AgdaSpace{}%
\AgdaBound{𝒫n}\AgdaSymbol{)}\<%
\\
\>[5]\AgdaSymbol{λ}\AgdaSpace{}%
\AgdaBound{j}\AgdaSpace{}%
\AgdaBound{V}\AgdaSpace{}%
\AgdaBound{V′}\AgdaSpace{}%
\AgdaBound{j≤n}\AgdaSpace{}%
\AgdaBound{M→V}\AgdaSpace{}%
\AgdaBound{v}\AgdaSpace{}%
\AgdaBound{M′→V′}\AgdaSpace{}%
\AgdaBound{v′}\AgdaSpace{}%
\AgdaBound{𝒱VV′j}\AgdaSpace{}%
\AgdaSymbol{→}\<%
\\
\>[5]\AgdaFunction{Goal}\AgdaSpace{}%
\AgdaSymbol{\{}\AgdaBound{j}\AgdaSymbol{\}\{}\AgdaBound{V}\AgdaSymbol{\}\{}\AgdaBound{V′}\AgdaSymbol{\}\{}\AgdaBound{dir}\AgdaSymbol{\}}\AgdaSpace{}%
\AgdaBound{𝒱VV′j}\<%
\\
\>[5]\AgdaKeyword{where}\<%
\\
\>[5]\AgdaFunction{Goal}\AgdaSpace{}%
\AgdaSymbol{:}\AgdaSpace{}%
\AgdaSymbol{∀\{}\AgdaBound{j}\AgdaSymbol{\}\{}\AgdaBound{V}\AgdaSymbol{\}\{}\AgdaBound{V′}\AgdaSymbol{\}\{}\AgdaBound{dir}\AgdaSymbol{\}}\<%
\\
\>[5][@{}l@{\AgdaIndent{0}}]%
\>[8]\AgdaSymbol{→}\AgdaSpace{}%
\AgdaField{\#}\AgdaSpace{}%
\AgdaSymbol{(}\AgdaBound{dir}\AgdaSpace{}%
\AgdaOperator{\AgdaFunction{∣}}\AgdaSpace{}%
\AgdaBound{V}\AgdaSpace{}%
\AgdaOperator{\AgdaFunction{⊑ᴸᴿᵥ}}\AgdaSpace{}%
\AgdaBound{V′}\AgdaSpace{}%
\AgdaOperator{\AgdaFunction{⦂}}\AgdaSpace{}%
\AgdaInductiveConstructor{unk⊑unk}\AgdaSymbol{)}\AgdaSpace{}%
\AgdaBound{j}\<%
\\
\>[8]\AgdaSymbol{→}\AgdaSpace{}%
\AgdaField{\#}\AgdaSpace{}%
\AgdaSymbol{(}\AgdaBound{dir}\AgdaSpace{}%
\AgdaOperator{\AgdaFunction{∣}}\AgdaSpace{}%
\AgdaBound{V}\AgdaSpace{}%
\AgdaOperator{\AgdaFunction{⊑ᴸᴿₜ}}\AgdaSpace{}%
\AgdaSymbol{(}\AgdaBound{V′}\AgdaSpace{}%
\AgdaOperator{\AgdaInductiveConstructor{⟨}}\AgdaSpace{}%
\AgdaBound{H}\AgdaSpace{}%
\AgdaOperator{\AgdaInductiveConstructor{?⟩}}\AgdaSymbol{)}\AgdaSpace{}%
\AgdaOperator{\AgdaFunction{⦂}}\AgdaSpace{}%
\AgdaInductiveConstructor{unk⊑}\AgdaSpace{}%
\AgdaBound{d}\AgdaSymbol{)}\AgdaSpace{}%
\AgdaBound{j}\<%
\\
\>[5]\AgdaFunction{Goal}\AgdaSpace{}%
\AgdaSymbol{\{}\AgdaInductiveConstructor{zero}\AgdaSymbol{\}}\AgdaSpace{}%
\AgdaSymbol{\{}\AgdaBound{V}\AgdaSymbol{\}}\AgdaSpace{}%
\AgdaSymbol{\{}\AgdaBound{V′}\AgdaSymbol{\}\{}\AgdaBound{dir}\AgdaSymbol{\}}\AgdaSpace{}%
\AgdaBound{𝒱VV′j}\AgdaSpace{}%
\AgdaSymbol{=}\<%
\\
\>[5][@{}l@{\AgdaIndent{0}}]%
\>[9]\AgdaField{tz}\AgdaSpace{}%
\AgdaSymbol{(}\AgdaBound{dir}\AgdaSpace{}%
\AgdaOperator{\AgdaFunction{∣}}\AgdaSpace{}%
\AgdaBound{V}\AgdaSpace{}%
\AgdaOperator{\AgdaFunction{⊑ᴸᴿₜ}}\AgdaSpace{}%
\AgdaSymbol{(}\AgdaBound{V′}\AgdaSpace{}%
\AgdaOperator{\AgdaInductiveConstructor{⟨}}\AgdaSpace{}%
\AgdaBound{H}\AgdaSpace{}%
\AgdaOperator{\AgdaInductiveConstructor{?⟩}}\AgdaSymbol{)}\AgdaSpace{}%
\AgdaOperator{\AgdaFunction{⦂}}\AgdaSpace{}%
\AgdaInductiveConstructor{unk⊑}\AgdaSpace{}%
\AgdaBound{d}\AgdaSymbol{)}\<%
\\
\>[5]\AgdaFunction{Goal}\AgdaSpace{}%
\AgdaSymbol{\{}\AgdaInductiveConstructor{suc}\AgdaSpace{}%
\AgdaBound{j}\AgdaSymbol{\}}\AgdaSpace{}%
\AgdaSymbol{\{}\AgdaBound{V₁}\AgdaSpace{}%
\AgdaOperator{\AgdaInductiveConstructor{⟨}}\AgdaSpace{}%
\AgdaBound{G}\AgdaSpace{}%
\AgdaOperator{\AgdaInductiveConstructor{!⟩}}\AgdaSymbol{\}}\AgdaSpace{}%
\AgdaSymbol{\{}\AgdaBound{V′₁}\AgdaSpace{}%
\AgdaOperator{\AgdaInductiveConstructor{⟨}}\AgdaSpace{}%
\AgdaBound{H₂}\AgdaSpace{}%
\AgdaOperator{\AgdaInductiveConstructor{!⟩}}\AgdaSymbol{\}\{}\AgdaBound{dir}\AgdaSymbol{\}}\AgdaSpace{}%
\AgdaBound{𝒱VV′sj}\<%
\\
\>[5][@{}l@{\AgdaIndent{0}}]%
\>[9]\AgdaKeyword{with}\AgdaSpace{}%
\AgdaBound{G}\AgdaSpace{}%
\AgdaOperator{\AgdaFunction{≡ᵍ}}\AgdaSpace{}%
\AgdaBound{H₂}\AgdaSpace{}%
\AgdaSymbol{|}\AgdaSpace{}%
\AgdaBound{𝒱VV′sj}\<%
\\
\>[5]\AgdaSymbol{...}\AgdaSpace{}%
\AgdaSymbol{|}\AgdaSpace{}%
\AgdaInductiveConstructor{no}\AgdaSpace{}%
\AgdaBound{neq}\AgdaSpace{}%
\AgdaSymbol{|}\AgdaSpace{}%
\AgdaSymbol{()}\<%
\\
\>[5]\AgdaSymbol{...}%
\>[3415I]\AgdaSymbol{|}\AgdaSpace{}%
\AgdaInductiveConstructor{yes}\AgdaSpace{}%
\AgdaInductiveConstructor{refl}\AgdaSpace{}%
\AgdaSymbol{|}\AgdaSpace{}%
\AgdaBound{v₁}\AgdaSpace{}%
\AgdaOperator{\AgdaInductiveConstructor{,}}\AgdaSpace{}%
\AgdaBound{v′}\AgdaSpace{}%
\AgdaOperator{\AgdaInductiveConstructor{,}}\AgdaSpace{}%
\AgdaBound{𝒱V₁V′₁j}\<%
\\
\>[.][@{}l@{}]\<[3415I]%
\>[9]\AgdaKeyword{with}\AgdaSpace{}%
\AgdaBound{G}\AgdaSpace{}%
\AgdaOperator{\AgdaFunction{≡ᵍ}}\AgdaSpace{}%
\AgdaBound{G}\<%
\\
\>[5]\AgdaSymbol{...}\AgdaSpace{}%
\AgdaSymbol{|}\AgdaSpace{}%
\AgdaInductiveConstructor{no}\AgdaSpace{}%
\AgdaBound{neq}\AgdaSpace{}%
\AgdaSymbol{=}\AgdaSpace{}%
\AgdaFunction{⊥-elim}\AgdaSpace{}%
\AgdaSymbol{(}\AgdaBound{neq}\AgdaSpace{}%
\AgdaInductiveConstructor{refl}\AgdaSymbol{)}\<%
\\
\>[5]\AgdaSymbol{...}%
\>[3434I]\AgdaSymbol{|}\AgdaSpace{}%
\AgdaInductiveConstructor{yes}\AgdaSpace{}%
\AgdaInductiveConstructor{refl}\<%
\\
\>[.][@{}l@{}]\<[3434I]%
\>[9]\AgdaKeyword{with}\AgdaSpace{}%
\AgdaBound{G}\AgdaSpace{}%
\AgdaOperator{\AgdaFunction{≡ᵍ}}\AgdaSpace{}%
\AgdaBound{H}\<%
\\
\>[9]\AgdaComment{\{--------\ Case\ G\ ≢\ H\ ---------\}}\<%
\\
\>[5]\AgdaSymbol{...}%
\>[3440I]\AgdaSymbol{|}\AgdaSpace{}%
\AgdaInductiveConstructor{no}\AgdaSpace{}%
\AgdaBound{neq}\<%
\\
\>[.][@{}l@{}]\<[3440I]%
\>[9]\AgdaKeyword{with}\AgdaSpace{}%
\AgdaBound{dir}\<%
\\
\>[9]\AgdaComment{\{--------\ Subcase\ ≼\ ---------\}}\<%
\\
\>[5]\AgdaSymbol{...}%
\>[3444I]\AgdaSymbol{|}\AgdaSpace{}%
\AgdaInductiveConstructor{≼}\AgdaSpace{}%
\AgdaSymbol{=}\AgdaSpace{}%
\AgdaInductiveConstructor{inj₂}\AgdaSpace{}%
\AgdaSymbol{(}\AgdaInductiveConstructor{inj₁}\AgdaSpace{}%
\AgdaSymbol{(}\AgdaFunction{unit}\AgdaSpace{}%
\AgdaSymbol{(}\AgdaInductiveConstructor{collide}\AgdaSpace{}%
\AgdaBound{v′}\AgdaSpace{}%
\AgdaBound{neq}\AgdaSpace{}%
\AgdaInductiveConstructor{refl}\AgdaSymbol{)))}\<%
\\
\>[.][@{}l@{}]\<[3444I]%
\>[9]\AgdaComment{\{--------\ Subcase\ ≽\ ---------\}}\<%
\\
\>[5]\AgdaSymbol{...}\AgdaSpace{}%
\AgdaSymbol{|}\AgdaSpace{}%
\AgdaInductiveConstructor{≽}\AgdaSpace{}%
\AgdaSymbol{=}\AgdaSpace{}%
\AgdaFunction{anti-reduction-≽-R-one}%
\>[3458I]\AgdaSymbol{(}\AgdaFunction{LRₜ-blame-step}\AgdaSymbol{\{}\AgdaInductiveConstructor{★}\AgdaSymbol{\}\{}\AgdaOperator{\AgdaFunction{⌈}}\AgdaSpace{}%
\AgdaBound{H}\AgdaSpace{}%
\AgdaOperator{\AgdaFunction{⌉}}\AgdaSymbol{\}\{}\AgdaInductiveConstructor{unk⊑}\AgdaSpace{}%
\AgdaBound{d}\AgdaSymbol{\}\{}\AgdaInductiveConstructor{≽}\AgdaSymbol{\})}\<%
\\
\>[.][@{}l@{}]\<[3458I]%
\>[38]\AgdaSymbol{(}\AgdaInductiveConstructor{collide}\AgdaSpace{}%
\AgdaBound{v′}\AgdaSpace{}%
\AgdaBound{neq}\AgdaSpace{}%
\AgdaInductiveConstructor{refl}\AgdaSymbol{)}\<%
\\
\>[5]\AgdaFunction{Goal}\AgdaSpace{}%
\AgdaSymbol{\{}\AgdaInductiveConstructor{suc}\AgdaSpace{}%
\AgdaBound{j}\AgdaSymbol{\}}\AgdaSpace{}%
\AgdaSymbol{\{}\AgdaBound{V₁}\AgdaSpace{}%
\AgdaOperator{\AgdaInductiveConstructor{⟨}}\AgdaSpace{}%
\AgdaBound{G}\AgdaSpace{}%
\AgdaOperator{\AgdaInductiveConstructor{!⟩}}\AgdaSymbol{\}}\AgdaSpace{}%
\AgdaSymbol{\{}\AgdaBound{V′₁}\AgdaSpace{}%
\AgdaOperator{\AgdaInductiveConstructor{⟨}}\AgdaSpace{}%
\AgdaBound{H₂}\AgdaSpace{}%
\AgdaOperator{\AgdaInductiveConstructor{!⟩}}\AgdaSymbol{\}\{}\AgdaBound{dir}\AgdaSymbol{\}}\AgdaSpace{}%
\AgdaBound{𝒱VV′sj}\<%
\\
\>[5][@{}l@{\AgdaIndent{0}}]%
\>[9]\AgdaSymbol{|}\AgdaSpace{}%
\AgdaInductiveConstructor{yes}\AgdaSpace{}%
\AgdaInductiveConstructor{refl}\AgdaSpace{}%
\AgdaSymbol{|}\AgdaSpace{}%
\AgdaBound{v₁}\AgdaSpace{}%
\AgdaOperator{\AgdaInductiveConstructor{,}}\AgdaSpace{}%
\AgdaBound{v′}\AgdaSpace{}%
\AgdaOperator{\AgdaInductiveConstructor{,}}\AgdaSpace{}%
\AgdaBound{𝒱V₁V′₁j}\AgdaSpace{}%
\AgdaSymbol{|}\AgdaSpace{}%
\AgdaInductiveConstructor{yes}\AgdaSpace{}%
\AgdaInductiveConstructor{refl}\<%
\\
\>[9]\AgdaComment{\{--------\ Case\ G\ ≡\ H\ ---------\}}\<%
\\
\>[9]\AgdaSymbol{|}\AgdaSpace{}%
\AgdaInductiveConstructor{yes}\AgdaSpace{}%
\AgdaInductiveConstructor{refl}\<%
\\
\>[9]\AgdaKeyword{with}\AgdaSpace{}%
\AgdaBound{dir}\<%
\\
\>[9]\AgdaComment{\{--------\ Subcase\ ≼\ ---------\}}\<%
\\
\>[5]\AgdaSymbol{...}%
\>[3490I]\AgdaSymbol{|}\AgdaSpace{}%
\AgdaInductiveConstructor{≼}\<%
\\
\>[.][@{}l@{}]\<[3490I]%
\>[9]\AgdaKeyword{with}\AgdaSpace{}%
\AgdaBound{G}\AgdaSpace{}%
\AgdaOperator{\AgdaFunction{≡ᵍ}}\AgdaSpace{}%
\AgdaBound{G}\<%
\\
\>[5]\AgdaSymbol{...}\AgdaSpace{}%
\AgdaSymbol{|}\AgdaSpace{}%
\AgdaInductiveConstructor{no}\AgdaSpace{}%
\AgdaBound{neq}\AgdaSpace{}%
\AgdaSymbol{=}\AgdaSpace{}%
\AgdaFunction{⊥-elim}\AgdaSpace{}%
\AgdaSymbol{(}\AgdaBound{neq}\AgdaSpace{}%
\AgdaInductiveConstructor{refl}\AgdaSymbol{)}\<%
\\
\>[5]\AgdaSymbol{...}%
\>[3502I]\AgdaSymbol{|}\AgdaSpace{}%
\AgdaInductiveConstructor{yes}\AgdaSpace{}%
\AgdaInductiveConstructor{refl}\<%
\\
\>[.][@{}l@{}]\<[3502I]%
\>[9]\AgdaKeyword{with}\AgdaSpace{}%
\AgdaFunction{gnd-prec-unique}\AgdaSpace{}%
\AgdaBound{d}\AgdaSpace{}%
\AgdaFunction{Refl⊑}\<%
\\
\>[5]\AgdaSymbol{...}\AgdaSpace{}%
\AgdaSymbol{|}%
\>[3509I]\AgdaInductiveConstructor{refl}\AgdaSpace{}%
\AgdaSymbol{=}\<%
\\
\>[.][@{}l@{}]\<[3509I]%
\>[11]\AgdaKeyword{let}\AgdaSpace{}%
\AgdaBound{V₁G⊑V′₁sj}\AgdaSpace{}%
\AgdaSymbol{=}\AgdaSpace{}%
\AgdaBound{v₁}\AgdaSpace{}%
\AgdaOperator{\AgdaInductiveConstructor{,}}\AgdaSpace{}%
\AgdaBound{v′}\AgdaSpace{}%
\AgdaOperator{\AgdaInductiveConstructor{,}}\AgdaSpace{}%
\AgdaBound{𝒱V₁V′₁j}\AgdaSpace{}%
\AgdaKeyword{in}\<%
\\
\>[11]\AgdaInductiveConstructor{inj₂}\AgdaSpace{}%
\AgdaSymbol{(}\AgdaInductiveConstructor{inj₂}%
\>[3520I]\AgdaSymbol{(}\AgdaBound{v₁}\AgdaSpace{}%
\AgdaOperator{\AgdaInductiveConstructor{〈}}\AgdaSpace{}%
\AgdaBound{G}\AgdaSpace{}%
\AgdaOperator{\AgdaInductiveConstructor{〉}}\AgdaSpace{}%
\AgdaOperator{\AgdaInductiveConstructor{,}}\<%
\\
\>[3520I][@{}l@{\AgdaIndent{0}}]%
\>[23]\AgdaSymbol{(}\AgdaBound{V′₁}\AgdaSpace{}%
\AgdaOperator{\AgdaInductiveConstructor{,}}\AgdaSpace{}%
\AgdaFunction{unit}\AgdaSpace{}%
\AgdaSymbol{(}\AgdaInductiveConstructor{collapse}\AgdaSpace{}%
\AgdaBound{v′}\AgdaSpace{}%
\AgdaInductiveConstructor{refl}\AgdaSymbol{)}\AgdaSpace{}%
\AgdaOperator{\AgdaInductiveConstructor{,}}\AgdaSpace{}%
\AgdaBound{v′}\AgdaSpace{}%
\AgdaOperator{\AgdaInductiveConstructor{,}}\AgdaSpace{}%
\AgdaBound{V₁G⊑V′₁sj}\AgdaSymbol{)))}\<%
\\
\>[5]\AgdaFunction{Goal}\AgdaSpace{}%
\AgdaSymbol{\{}\AgdaInductiveConstructor{suc}\AgdaSpace{}%
\AgdaBound{j}\AgdaSymbol{\}}\AgdaSpace{}%
\AgdaSymbol{\{}\AgdaBound{V₁}\AgdaSpace{}%
\AgdaOperator{\AgdaInductiveConstructor{⟨}}\AgdaSpace{}%
\AgdaBound{G}\AgdaSpace{}%
\AgdaOperator{\AgdaInductiveConstructor{!⟩}}\AgdaSymbol{\}}\AgdaSpace{}%
\AgdaSymbol{\{}\AgdaBound{V′₁}\AgdaSpace{}%
\AgdaOperator{\AgdaInductiveConstructor{⟨}}\AgdaSpace{}%
\AgdaBound{H₂}\AgdaSpace{}%
\AgdaOperator{\AgdaInductiveConstructor{!⟩}}\AgdaSymbol{\}\{}\AgdaBound{dir}\AgdaSymbol{\}}\AgdaSpace{}%
\AgdaBound{𝒱VV′sj}\<%
\\
\>[5][@{}l@{\AgdaIndent{0}}]%
\>[9]\AgdaSymbol{|}\AgdaSpace{}%
\AgdaInductiveConstructor{yes}\AgdaSpace{}%
\AgdaInductiveConstructor{refl}\AgdaSpace{}%
\AgdaSymbol{|}\AgdaSpace{}%
\AgdaBound{v₁}\AgdaSpace{}%
\AgdaOperator{\AgdaInductiveConstructor{,}}\AgdaSpace{}%
\AgdaBound{v′}\AgdaSpace{}%
\AgdaOperator{\AgdaInductiveConstructor{,}}\AgdaSpace{}%
\AgdaBound{𝒱V₁V′₁j}\AgdaSpace{}%
\AgdaSymbol{|}\AgdaSpace{}%
\AgdaInductiveConstructor{yes}\AgdaSpace{}%
\AgdaInductiveConstructor{refl}\<%
\\
\>[9]\AgdaSymbol{|}\AgdaSpace{}%
\AgdaInductiveConstructor{yes}\AgdaSpace{}%
\AgdaInductiveConstructor{refl}\<%
\\
\>[9]\AgdaComment{\{--------\ Subcase\ ≽\ ---------\}}\<%
\\
\>[9]\AgdaSymbol{|}\AgdaSpace{}%
\AgdaInductiveConstructor{≽}\<%
\\
\>[9]\AgdaKeyword{with}\AgdaSpace{}%
\AgdaFunction{gnd-prec-unique}\AgdaSpace{}%
\AgdaBound{d}\AgdaSpace{}%
\AgdaFunction{Refl⊑}\<%
\\
\>[5]\AgdaSymbol{...}%
\>[3562I]\AgdaSymbol{|}\AgdaSpace{}%
\AgdaInductiveConstructor{refl}\AgdaSpace{}%
\AgdaSymbol{=}\<%
\\
\>[.][@{}l@{}]\<[3562I]%
\>[9]\AgdaKeyword{let}\AgdaSpace{}%
\AgdaBound{𝒱VGV′j}\AgdaSpace{}%
\AgdaSymbol{=}\AgdaSpace{}%
\AgdaFunction{LRᵥ-inject-L-intro-≽}\AgdaSpace{}%
\AgdaSymbol{\{}\AgdaBound{G}\AgdaSymbol{\}\{}\AgdaOperator{\AgdaFunction{⌈}}\AgdaSpace{}%
\AgdaBound{G}\AgdaSpace{}%
\AgdaOperator{\AgdaFunction{⌉}}\AgdaSymbol{\}\{}\AgdaBound{d}\AgdaSymbol{\}}\AgdaSpace{}%
\AgdaBound{𝒱V₁V′₁j}\AgdaSpace{}%
\AgdaKeyword{in}\<%
\\
\>[9]\AgdaKeyword{let}\AgdaSpace{}%
\AgdaBound{ℰVGV′j}\AgdaSpace{}%
\AgdaSymbol{=}\AgdaSpace{}%
\AgdaFunction{LRᵥ⇒LRₜ-step}\AgdaSymbol{\{}\AgdaArgument{V}\AgdaSpace{}%
\AgdaSymbol{=}\AgdaSpace{}%
\AgdaBound{V₁}\AgdaSpace{}%
\AgdaOperator{\AgdaInductiveConstructor{⟨}}\AgdaSpace{}%
\AgdaBound{G}\AgdaSpace{}%
\AgdaOperator{\AgdaInductiveConstructor{!⟩}}\AgdaSymbol{\}\{}\AgdaBound{V′₁}\AgdaSymbol{\}\{}\AgdaInductiveConstructor{≽}\AgdaSymbol{\}}\AgdaSpace{}%
\AgdaBound{𝒱VGV′j}\AgdaSpace{}%
\AgdaKeyword{in}\<%
\\
\>[9]\AgdaFunction{anti-reduction-≽-R-one}\AgdaSpace{}%
\AgdaBound{ℰVGV′j}\AgdaSpace{}%
\AgdaSymbol{(}\AgdaInductiveConstructor{collapse}\AgdaSpace{}%
\AgdaBound{v′}\AgdaSpace{}%
\AgdaInductiveConstructor{refl}\AgdaSymbol{)}\<%
\end{code}
\end{AgdaSuppressSpace}

\paragraph{Fundamental Theorem}

The proof is by induction on $M ⊑ M′$, using the appropriate
Compatibility Lemma in each case.

\begin{code}%
\>[0]\AgdaFunction{fundamental}\AgdaSpace{}%
\AgdaSymbol{:}\AgdaSpace{}%
\AgdaSymbol{∀}\AgdaSpace{}%
\AgdaSymbol{\{}\AgdaBound{Γ}\AgdaSymbol{\}\{}\AgdaBound{A}\AgdaSymbol{\}\{}\AgdaBound{A′}\AgdaSymbol{\}\{}\AgdaBound{A⊑A′}\AgdaSpace{}%
\AgdaSymbol{:}\AgdaSpace{}%
\AgdaBound{A}\AgdaSpace{}%
\AgdaOperator{\AgdaDatatype{⊑}}\AgdaSpace{}%
\AgdaBound{A′}\AgdaSymbol{\}}\AgdaSpace{}%
\AgdaSymbol{→}\AgdaSpace{}%
\AgdaSymbol{(}\AgdaBound{M}\AgdaSpace{}%
\AgdaBound{M′}\AgdaSpace{}%
\AgdaSymbol{:}\AgdaSpace{}%
\AgdaDatatype{Term}\AgdaSymbol{)}\<%
\\
\>[0][@{}l@{\AgdaIndent{0}}]%
\>[2]\AgdaSymbol{→}\AgdaSpace{}%
\AgdaBound{Γ}\AgdaSpace{}%
\AgdaOperator{\AgdaDatatype{⊩}}\AgdaSpace{}%
\AgdaBound{M}\AgdaSpace{}%
\AgdaOperator{\AgdaDatatype{⊑}}\AgdaSpace{}%
\AgdaBound{M′}\AgdaSpace{}%
\AgdaOperator{\AgdaDatatype{⦂}}\AgdaSpace{}%
\AgdaBound{A⊑A′}%
\>[23]\AgdaSymbol{→}%
\>[26]\AgdaBound{Γ}\AgdaSpace{}%
\AgdaOperator{\AgdaFunction{⊨}}\AgdaSpace{}%
\AgdaBound{M}\AgdaSpace{}%
\AgdaOperator{\AgdaFunction{⊑ᴸᴿ}}\AgdaSpace{}%
\AgdaBound{M′}\AgdaSpace{}%
\AgdaOperator{\AgdaFunction{⦂}}\AgdaSpace{}%
\AgdaSymbol{(}\AgdaBound{A}\AgdaSpace{}%
\AgdaOperator{\AgdaInductiveConstructor{,}}\AgdaSpace{}%
\AgdaBound{A′}\AgdaSpace{}%
\AgdaOperator{\AgdaInductiveConstructor{,}}\AgdaSpace{}%
\AgdaBound{A⊑A′}\AgdaSymbol{)}\<%
\end{code}
\begin{code}[hide]%
\>[0]\AgdaFunction{fundamental}\AgdaSpace{}%
\AgdaSymbol{\{}\AgdaBound{Γ}\AgdaSymbol{\}}\AgdaSpace{}%
\AgdaSymbol{\{}\AgdaBound{A}\AgdaSymbol{\}}\AgdaSpace{}%
\AgdaSymbol{\{}\AgdaBound{A′}\AgdaSymbol{\}}\AgdaSpace{}%
\AgdaSymbol{\{}\AgdaBound{A⊑A′}\AgdaSymbol{\}}\AgdaSpace{}%
\AgdaDottedPattern{\AgdaSymbol{.(}}\AgdaDottedPattern{\AgdaOperator{\AgdaInductiveConstructor{`}}}\AgdaSpace{}%
\AgdaDottedPattern{\AgdaSymbol{\AgdaUnderscore{})}}\AgdaSpace{}%
\AgdaDottedPattern{\AgdaSymbol{.(}}\AgdaDottedPattern{\AgdaOperator{\AgdaInductiveConstructor{`}}}\AgdaSpace{}%
\AgdaDottedPattern{\AgdaSymbol{\AgdaUnderscore{})}}\AgdaSpace{}%
\AgdaSymbol{(}\AgdaInductiveConstructor{⊑-var}\AgdaSpace{}%
\AgdaBound{∋x}\AgdaSymbol{)}\AgdaSpace{}%
\AgdaSymbol{=}\<%
\\
\>[0][@{}l@{\AgdaIndent{0}}]%
\>[4]\AgdaFunction{compatibility-var}\AgdaSpace{}%
\AgdaBound{∋x}\<%
\\
\>[0]\AgdaFunction{fundamental}\AgdaSpace{}%
\AgdaSymbol{\{}\AgdaBound{Γ}\AgdaSymbol{\}}\AgdaSpace{}%
\AgdaSymbol{\{\AgdaUnderscore{}\}}\AgdaSpace{}%
\AgdaSymbol{\{\AgdaUnderscore{}\}}\AgdaSpace{}%
\AgdaSymbol{\{}\AgdaInductiveConstructor{base⊑}\AgdaSymbol{\}}\AgdaSpace{}%
\AgdaSymbol{(}\AgdaInductiveConstructor{\$}\AgdaSpace{}%
\AgdaBound{c}\AgdaSymbol{)}\AgdaSpace{}%
\AgdaSymbol{(}\AgdaInductiveConstructor{\$}\AgdaSpace{}%
\AgdaBound{c}\AgdaSymbol{)}\AgdaSpace{}%
\AgdaInductiveConstructor{⊑-lit}\AgdaSpace{}%
\AgdaSymbol{=}\<%
\\
\>[0][@{}l@{\AgdaIndent{0}}]%
\>[4]\AgdaFunction{compatible-literal}\<%
\\
\>[0]\AgdaFunction{fundamental}\AgdaSpace{}%
\AgdaSymbol{\{}\AgdaBound{Γ}\AgdaSymbol{\}}\AgdaSpace{}%
\AgdaSymbol{\{}\AgdaBound{A}\AgdaSymbol{\}}\AgdaSpace{}%
\AgdaSymbol{\{}\AgdaBound{A′}\AgdaSymbol{\}}\AgdaSpace{}%
\AgdaSymbol{\{}\AgdaBound{A⊑A′}\AgdaSymbol{\}}\AgdaSpace{}%
\AgdaSymbol{(}\AgdaBound{L}\AgdaSpace{}%
\AgdaOperator{\AgdaInductiveConstructor{·}}\AgdaSpace{}%
\AgdaBound{M}\AgdaSymbol{)}\AgdaSpace{}%
\AgdaSymbol{(}\AgdaBound{L′}\AgdaSpace{}%
\AgdaOperator{\AgdaInductiveConstructor{·}}\AgdaSpace{}%
\AgdaBound{M′}\AgdaSymbol{)}\AgdaSpace{}%
\AgdaSymbol{(}\AgdaInductiveConstructor{⊑-app}\AgdaSpace{}%
\AgdaBound{⊢L⊑L′}\AgdaSpace{}%
\AgdaBound{⊢M⊑M′}\AgdaSymbol{)}\AgdaSpace{}%
\AgdaSymbol{=}\<%
\\
\>[0][@{}l@{\AgdaIndent{0}}]%
\>[4]\AgdaFunction{compatible-app}\AgdaSymbol{\{}\AgdaArgument{L}\AgdaSpace{}%
\AgdaSymbol{=}\AgdaSpace{}%
\AgdaBound{L}\AgdaSymbol{\}\{}\AgdaBound{L′}\AgdaSymbol{\}\{}\AgdaBound{M}\AgdaSymbol{\}\{}\AgdaBound{M′}\AgdaSymbol{\}}\AgdaSpace{}%
\AgdaSymbol{(}\AgdaFunction{fundamental}\AgdaSpace{}%
\AgdaBound{L}\AgdaSpace{}%
\AgdaBound{L′}\AgdaSpace{}%
\AgdaBound{⊢L⊑L′}\AgdaSymbol{)}\AgdaSpace{}%
\AgdaSymbol{(}\AgdaFunction{fundamental}\AgdaSpace{}%
\AgdaBound{M}\AgdaSpace{}%
\AgdaBound{M′}\AgdaSpace{}%
\AgdaBound{⊢M⊑M′}\AgdaSymbol{)}\<%
\\
\>[0]\AgdaFunction{fundamental}\AgdaSpace{}%
\AgdaSymbol{\{}\AgdaBound{Γ}\AgdaSymbol{\}}\AgdaSpace{}%
\AgdaSymbol{\{}\AgdaDottedPattern{\AgdaSymbol{.(\AgdaUnderscore{}}}\AgdaSpace{}%
\AgdaDottedPattern{\AgdaOperator{\AgdaInductiveConstructor{⇒}}}\AgdaSpace{}%
\AgdaDottedPattern{\AgdaSymbol{\AgdaUnderscore{})}}\AgdaSymbol{\}}\AgdaSpace{}%
\AgdaSymbol{\{}\AgdaDottedPattern{\AgdaSymbol{.(\AgdaUnderscore{}}}\AgdaSpace{}%
\AgdaDottedPattern{\AgdaOperator{\AgdaInductiveConstructor{⇒}}}\AgdaSpace{}%
\AgdaDottedPattern{\AgdaSymbol{\AgdaUnderscore{})}}\AgdaSymbol{\}}\AgdaSpace{}%
\AgdaSymbol{\{}\AgdaDottedPattern{\AgdaSymbol{.(}}\AgdaDottedPattern{\AgdaInductiveConstructor{fun⊑}}\AgdaSpace{}%
\AgdaDottedPattern{\AgdaSymbol{\AgdaUnderscore{}}}\AgdaSpace{}%
\AgdaDottedPattern{\AgdaSymbol{\AgdaUnderscore{})}}\AgdaSymbol{\}}\AgdaSpace{}%
\AgdaSymbol{(}\AgdaInductiveConstructor{ƛ}\AgdaSpace{}%
\AgdaBound{N}\AgdaSymbol{)(}\AgdaInductiveConstructor{ƛ}\AgdaSpace{}%
\AgdaBound{N′}\AgdaSymbol{)}\AgdaSpace{}%
\AgdaSymbol{(}\AgdaInductiveConstructor{⊑-lam}\AgdaSpace{}%
\AgdaBound{⊢N⊑N′}\AgdaSymbol{)}\AgdaSpace{}%
\AgdaSymbol{=}\<%
\\
\>[0][@{}l@{\AgdaIndent{0}}]%
\>[4]\AgdaFunction{compatible-lambda}\AgdaSymbol{\{}\AgdaArgument{N}\AgdaSpace{}%
\AgdaSymbol{=}\AgdaSpace{}%
\AgdaBound{N}\AgdaSymbol{\}\{}\AgdaBound{N′}\AgdaSymbol{\}}\AgdaSpace{}%
\AgdaSymbol{(}\AgdaFunction{fundamental}\AgdaSpace{}%
\AgdaBound{N}\AgdaSpace{}%
\AgdaBound{N′}\AgdaSpace{}%
\AgdaBound{⊢N⊑N′}\AgdaSymbol{)}\<%
\\
\>[0]\AgdaFunction{fundamental}\AgdaSpace{}%
\AgdaSymbol{\{}\AgdaBound{Γ}\AgdaSymbol{\}}\AgdaSpace{}%
\AgdaSymbol{\{}\AgdaInductiveConstructor{★}\AgdaSymbol{\}}\AgdaSpace{}%
\AgdaSymbol{\{}\AgdaBound{A′}\AgdaSymbol{\}}\AgdaSpace{}%
\AgdaSymbol{\{}\AgdaInductiveConstructor{unk⊑}\AgdaSpace{}%
\AgdaBound{c}\AgdaSymbol{\}}\AgdaSpace{}%
\AgdaSymbol{(}\AgdaBound{M}\AgdaSpace{}%
\AgdaOperator{\AgdaInductiveConstructor{⟨}}\AgdaSpace{}%
\AgdaBound{G}\AgdaSpace{}%
\AgdaOperator{\AgdaInductiveConstructor{!⟩}}\AgdaSymbol{)}\AgdaSpace{}%
\AgdaBound{M′}\AgdaSpace{}%
\AgdaSymbol{(}\AgdaInductiveConstructor{⊑-inj-L}\AgdaSpace{}%
\AgdaBound{⊢M⊑M′}\AgdaSymbol{)}\AgdaSpace{}%
\AgdaSymbol{=}\<%
\\
\>[0][@{}l@{\AgdaIndent{0}}]%
\>[4]\AgdaFunction{compatible-inj-L}\AgdaSymbol{\{}\AgdaArgument{G}\AgdaSpace{}%
\AgdaSymbol{=}%
\>[26]\AgdaBound{G}\AgdaSymbol{\}\{}\AgdaArgument{M}\AgdaSpace{}%
\AgdaSymbol{=}\AgdaSpace{}%
\AgdaBound{M}\AgdaSymbol{\}\{}\AgdaBound{M′}\AgdaSymbol{\}}\AgdaSpace{}%
\AgdaSymbol{(}\AgdaFunction{fundamental}\AgdaSpace{}%
\AgdaBound{M}\AgdaSpace{}%
\AgdaBound{M′}\AgdaSpace{}%
\AgdaBound{⊢M⊑M′}\AgdaSymbol{)}\<%
\\
\>[0]\AgdaFunction{fundamental}\AgdaSpace{}%
\AgdaSymbol{\{}\AgdaBound{Γ}\AgdaSymbol{\}}\AgdaSpace{}%
\AgdaSymbol{\{}\AgdaInductiveConstructor{★}\AgdaSymbol{\}}\AgdaSpace{}%
\AgdaSymbol{\{}\AgdaInductiveConstructor{★}\AgdaSymbol{\}}\AgdaSpace{}%
\AgdaSymbol{\{}\AgdaDottedPattern{\AgdaSymbol{.}}\AgdaDottedPattern{\AgdaInductiveConstructor{unk⊑unk}}\AgdaSymbol{\}}\AgdaSpace{}%
\AgdaBound{M}\AgdaSpace{}%
\AgdaSymbol{(}\AgdaBound{M′}\AgdaSpace{}%
\AgdaOperator{\AgdaInductiveConstructor{⟨}}\AgdaSpace{}%
\AgdaBound{G}\AgdaSpace{}%
\AgdaOperator{\AgdaInductiveConstructor{!⟩}}\AgdaSymbol{)}\AgdaSpace{}%
\AgdaSymbol{(}\AgdaInductiveConstructor{⊑-inj-R}\AgdaSpace{}%
\AgdaBound{⊢M⊑M′}\AgdaSymbol{)}\AgdaSpace{}%
\AgdaSymbol{=}\<%
\\
\>[0][@{}l@{\AgdaIndent{0}}]%
\>[4]\AgdaFunction{compatible-inj-R}\AgdaSymbol{\{}\AgdaBound{Γ}\AgdaSymbol{\}\{}\AgdaArgument{G}\AgdaSpace{}%
\AgdaSymbol{=}\AgdaSpace{}%
\AgdaBound{G}\AgdaSymbol{\}\{}\AgdaArgument{M}\AgdaSpace{}%
\AgdaSymbol{=}\AgdaSpace{}%
\AgdaBound{M}\AgdaSymbol{\}\{}\AgdaBound{M′}\AgdaSymbol{\}}\AgdaSpace{}%
\AgdaSymbol{(}\AgdaFunction{fundamental}\AgdaSpace{}%
\AgdaBound{M}\AgdaSpace{}%
\AgdaBound{M′}\AgdaSpace{}%
\AgdaBound{⊢M⊑M′}\AgdaSymbol{)}\<%
\\
\>[0]\AgdaFunction{fundamental}\AgdaSpace{}%
\AgdaSymbol{\{}\AgdaBound{Γ}\AgdaSymbol{\}}\AgdaSpace{}%
\AgdaSymbol{\{\AgdaUnderscore{}\}}\AgdaSpace{}%
\AgdaSymbol{\{}\AgdaBound{A′}\AgdaSymbol{\}}\AgdaSpace{}%
\AgdaSymbol{\{}\AgdaBound{A⊑A′}\AgdaSymbol{\}}\AgdaSpace{}%
\AgdaSymbol{(}\AgdaBound{M}\AgdaSpace{}%
\AgdaOperator{\AgdaInductiveConstructor{⟨}}\AgdaSpace{}%
\AgdaBound{H}\AgdaSpace{}%
\AgdaOperator{\AgdaInductiveConstructor{?⟩}}\AgdaSymbol{)}\AgdaSpace{}%
\AgdaBound{M′}\AgdaSpace{}%
\AgdaSymbol{(}\AgdaInductiveConstructor{⊑-proj-L}\AgdaSpace{}%
\AgdaBound{⊢M⊑M′}\AgdaSymbol{)}\AgdaSpace{}%
\AgdaSymbol{=}\<%
\\
\>[0][@{}l@{\AgdaIndent{0}}]%
\>[4]\AgdaFunction{compatible-proj-L}\AgdaSymbol{\{}\AgdaBound{Γ}\AgdaSymbol{\}\{}\AgdaBound{H}\AgdaSymbol{\}\{}\AgdaBound{A′}\AgdaSymbol{\}\{}\AgdaArgument{M}\AgdaSpace{}%
\AgdaSymbol{=}\AgdaSpace{}%
\AgdaBound{M}\AgdaSymbol{\}\{}\AgdaBound{M′}\AgdaSymbol{\}}\AgdaSpace{}%
\AgdaSymbol{(}\AgdaFunction{fundamental}\AgdaSpace{}%
\AgdaBound{M}\AgdaSpace{}%
\AgdaBound{M′}\AgdaSpace{}%
\AgdaBound{⊢M⊑M′}\AgdaSymbol{)}\<%
\\
\>[0]\AgdaFunction{fundamental}\AgdaSpace{}%
\AgdaSymbol{\{}\AgdaBound{Γ}\AgdaSymbol{\}}\AgdaSpace{}%
\AgdaSymbol{\{}\AgdaBound{A}\AgdaSymbol{\}}\AgdaSpace{}%
\AgdaSymbol{\{}\AgdaDottedPattern{\AgdaSymbol{.(}}\AgdaDottedPattern{\AgdaOperator{\AgdaFunction{⌈}}}\AgdaSpace{}%
\AgdaDottedPattern{\AgdaSymbol{\AgdaUnderscore{}}}\AgdaSpace{}%
\AgdaDottedPattern{\AgdaOperator{\AgdaFunction{⌉}}}\AgdaDottedPattern{\AgdaSymbol{)}}\AgdaSymbol{\}}\AgdaSpace{}%
\AgdaSymbol{\{}\AgdaBound{A⊑A′}\AgdaSymbol{\}}\AgdaSpace{}%
\AgdaBound{M}\AgdaSpace{}%
\AgdaSymbol{(}\AgdaBound{M′}\AgdaSpace{}%
\AgdaOperator{\AgdaInductiveConstructor{⟨}}\AgdaSpace{}%
\AgdaBound{H′}\AgdaSpace{}%
\AgdaOperator{\AgdaInductiveConstructor{?⟩}}\AgdaSymbol{)}\AgdaSpace{}%
\AgdaSymbol{(}\AgdaInductiveConstructor{⊑-proj-R}\AgdaSpace{}%
\AgdaBound{⊢M⊑M′}\AgdaSymbol{)}\AgdaSpace{}%
\AgdaSymbol{=}\<%
\\
\>[0][@{}l@{\AgdaIndent{0}}]%
\>[4]\AgdaFunction{compatible-proj-R}\AgdaSymbol{\{}\AgdaArgument{M}\AgdaSpace{}%
\AgdaSymbol{=}\AgdaSpace{}%
\AgdaBound{M}\AgdaSymbol{\}\{}\AgdaBound{M′}\AgdaSymbol{\}}\AgdaSpace{}%
\AgdaSymbol{(}\AgdaFunction{fundamental}\AgdaSpace{}%
\AgdaBound{M}\AgdaSpace{}%
\AgdaBound{M′}\AgdaSpace{}%
\AgdaBound{⊢M⊑M′}\AgdaSymbol{)}\<%
\\
\>[0]\AgdaFunction{fundamental}\AgdaSpace{}%
\AgdaSymbol{\{}\AgdaBound{Γ}\AgdaSymbol{\}}\AgdaSpace{}%
\AgdaSymbol{\{}\AgdaBound{A}\AgdaSymbol{\}}\AgdaSpace{}%
\AgdaSymbol{\{}\AgdaDottedPattern{\AgdaSymbol{.}}\AgdaDottedPattern{\AgdaBound{A}}\AgdaSymbol{\}}\AgdaSpace{}%
\AgdaSymbol{\{}\AgdaDottedPattern{\AgdaSymbol{.}}\AgdaDottedPattern{\AgdaFunction{Refl⊑}}\AgdaSymbol{\}}\AgdaSpace{}%
\AgdaBound{M}\AgdaSpace{}%
\AgdaDottedPattern{\AgdaSymbol{.}}\AgdaDottedPattern{\AgdaInductiveConstructor{blame}}\AgdaSpace{}%
\AgdaSymbol{(}\AgdaInductiveConstructor{⊑-blame}\AgdaSpace{}%
\AgdaBound{⊢M∶A}\AgdaSymbol{)}\AgdaSpace{}%
\AgdaSymbol{=}\<%
\\
\>[0][@{}l@{\AgdaIndent{0}}]%
\>[4]\AgdaFunction{compatible-blame}\AgdaSpace{}%
\AgdaBound{⊢M∶A}\<%
\end{code}


\begin{code}[hide]%
\>[0]\AgdaSymbol{\{-\#}\AgdaSpace{}%
\AgdaKeyword{OPTIONS}\AgdaSpace{}%
\AgdaPragma{--rewriting}\AgdaSpace{}%
\AgdaSymbol{\#-\}}\<%
\\
\>[0]\AgdaKeyword{module}\AgdaSpace{}%
\AgdaModule{LogRel.PeterGG}\AgdaSpace{}%
\AgdaKeyword{where}\<%
\\
\\[\AgdaEmptyExtraSkip]%
\>[0]\AgdaKeyword{open}\AgdaSpace{}%
\AgdaKeyword{import}\AgdaSpace{}%
\AgdaModule{Data.Empty}\AgdaSpace{}%
\AgdaKeyword{using}\AgdaSpace{}%
\AgdaSymbol{(}\AgdaDatatype{⊥}\AgdaSymbol{;}\AgdaSpace{}%
\AgdaFunction{⊥-elim}\AgdaSymbol{)}\<%
\\
\>[0]\AgdaKeyword{open}\AgdaSpace{}%
\AgdaKeyword{import}\AgdaSpace{}%
\AgdaModule{Data.List}\AgdaSpace{}%
\AgdaKeyword{using}\AgdaSpace{}%
\AgdaSymbol{(}\AgdaDatatype{List}\AgdaSymbol{;}\AgdaSpace{}%
\AgdaInductiveConstructor{[]}\AgdaSymbol{;}\AgdaSpace{}%
\AgdaOperator{\AgdaInductiveConstructor{\AgdaUnderscore{}∷\AgdaUnderscore{}}}\AgdaSymbol{;}\AgdaSpace{}%
\AgdaFunction{map}\AgdaSymbol{;}\AgdaSpace{}%
\AgdaFunction{length}\AgdaSymbol{)}\<%
\\
\>[0]\AgdaKeyword{open}\AgdaSpace{}%
\AgdaKeyword{import}\AgdaSpace{}%
\AgdaModule{Data.Nat}\<%
\\
\>[0]\AgdaKeyword{open}\AgdaSpace{}%
\AgdaKeyword{import}\AgdaSpace{}%
\AgdaModule{Data.Product}\AgdaSpace{}%
\AgdaKeyword{using}\AgdaSpace{}%
\AgdaSymbol{(}\AgdaOperator{\AgdaInductiveConstructor{\AgdaUnderscore{},\AgdaUnderscore{}}}\AgdaSymbol{;}\AgdaOperator{\AgdaFunction{\AgdaUnderscore{}×\AgdaUnderscore{}}}\AgdaSymbol{;}\AgdaSpace{}%
\AgdaField{proj₁}\AgdaSymbol{;}\AgdaSpace{}%
\AgdaField{proj₂}\AgdaSymbol{;}\AgdaSpace{}%
\AgdaFunction{Σ-syntax}\AgdaSymbol{;}\AgdaSpace{}%
\AgdaFunction{∃-syntax}\AgdaSymbol{)}\<%
\\
\>[0]\AgdaKeyword{open}\AgdaSpace{}%
\AgdaKeyword{import}\AgdaSpace{}%
\AgdaModule{Data.Sum}\AgdaSpace{}%
\AgdaKeyword{using}\AgdaSpace{}%
\AgdaSymbol{(}\AgdaOperator{\AgdaDatatype{\AgdaUnderscore{}⊎\AgdaUnderscore{}}}\AgdaSymbol{;}\AgdaSpace{}%
\AgdaInductiveConstructor{inj₁}\AgdaSymbol{;}\AgdaSpace{}%
\AgdaInductiveConstructor{inj₂}\AgdaSymbol{)}\<%
\\
\>[0]\AgdaKeyword{open}\AgdaSpace{}%
\AgdaKeyword{import}\AgdaSpace{}%
\AgdaModule{Data.Unit}\AgdaSpace{}%
\AgdaKeyword{using}\AgdaSpace{}%
\AgdaSymbol{(}\AgdaRecord{⊤}\AgdaSymbol{;}\AgdaSpace{}%
\AgdaInductiveConstructor{tt}\AgdaSymbol{)}\<%
\\
\>[0]\AgdaKeyword{open}\AgdaSpace{}%
\AgdaKeyword{import}\AgdaSpace{}%
\AgdaModule{Relation.Binary.PropositionalEquality}\AgdaSpace{}%
\AgdaSymbol{as}\AgdaSpace{}%
\AgdaModule{Eq}\<%
\\
\>[0][@{}l@{\AgdaIndent{0}}]%
\>[2]\AgdaKeyword{using}\AgdaSpace{}%
\AgdaSymbol{(}\AgdaOperator{\AgdaDatatype{\AgdaUnderscore{}≡\AgdaUnderscore{}}}\AgdaSymbol{;}\AgdaSpace{}%
\AgdaOperator{\AgdaFunction{\AgdaUnderscore{}≢\AgdaUnderscore{}}}\AgdaSymbol{;}\AgdaSpace{}%
\AgdaInductiveConstructor{refl}\AgdaSymbol{;}\AgdaSpace{}%
\AgdaFunction{sym}\AgdaSymbol{;}\AgdaSpace{}%
\AgdaFunction{cong}\AgdaSymbol{;}\AgdaSpace{}%
\AgdaFunction{subst}\AgdaSymbol{;}\AgdaSpace{}%
\AgdaFunction{trans}\AgdaSymbol{)}\<%
\\
\\[\AgdaEmptyExtraSkip]%
\>[0]\AgdaKeyword{open}\AgdaSpace{}%
\AgdaKeyword{import}\AgdaSpace{}%
\AgdaModule{LogRel.PeterCastCalculus}\<%
\\
\>[0]\AgdaKeyword{open}\AgdaSpace{}%
\AgdaKeyword{import}\AgdaSpace{}%
\AgdaModule{LogRel.PeterPrecision}\<%
\\
\>[0]\AgdaKeyword{open}\AgdaSpace{}%
\AgdaKeyword{import}\AgdaSpace{}%
\AgdaModule{LogRel.PeterLogRel}\<%
\\
\>[0]\AgdaKeyword{open}\AgdaSpace{}%
\AgdaKeyword{import}\AgdaSpace{}%
\AgdaModule{LogRel.PeterFundamental}\<%
\\
\>[0]\AgdaKeyword{open}\AgdaSpace{}%
\AgdaKeyword{import}\AgdaSpace{}%
\AgdaModule{StepIndexedLogic}\<%
\end{code}

\section{Proof of the Gradual Guarantee}
\label{sec:gradual-guarantee}

The next step in the proof of the gradual guarantee is to connect the
logical relation to the behavior that's required by the gradual
guarantee. (Recall the \textsf{gradual} predicate defined in
Section~\ref{sec:precision}.) The proof goes through an intermediate
step that uses the following notion of semantic approximation for a
fixed number of reduction steps.

\begin{code}%
\>[0]\AgdaOperator{\AgdaFunction{\AgdaUnderscore{}⊨\AgdaUnderscore{}⊑\AgdaUnderscore{}for\AgdaUnderscore{}}}\AgdaSpace{}%
\AgdaSymbol{:}\AgdaSpace{}%
\AgdaDatatype{Dir}\AgdaSpace{}%
\AgdaSymbol{→}\AgdaSpace{}%
\AgdaDatatype{Term}\AgdaSpace{}%
\AgdaSymbol{→}\AgdaSpace{}%
\AgdaDatatype{Term}\AgdaSpace{}%
\AgdaSymbol{→}\AgdaSpace{}%
\AgdaDatatype{ℕ}\AgdaSpace{}%
\AgdaSymbol{→}\AgdaSpace{}%
\AgdaPrimitive{Set}\<%
\\
\>[0]\AgdaInductiveConstructor{≼}\AgdaSpace{}%
\AgdaOperator{\AgdaFunction{⊨}}\AgdaSpace{}%
\AgdaBound{M}\AgdaSpace{}%
\AgdaOperator{\AgdaFunction{⊑}}\AgdaSpace{}%
\AgdaBound{M′}\AgdaSpace{}%
\AgdaOperator{\AgdaFunction{for}}\AgdaSpace{}%
\AgdaBound{k}\AgdaSpace{}%
\AgdaSymbol{=}\AgdaSpace{}%
\AgdaSymbol{(}\AgdaBound{M}\AgdaSpace{}%
\AgdaOperator{\AgdaFunction{⇓}}\AgdaSpace{}%
\AgdaOperator{\AgdaFunction{×}}\AgdaSpace{}%
\AgdaBound{M′}\AgdaSpace{}%
\AgdaOperator{\AgdaFunction{⇓}}\AgdaSymbol{)}\AgdaSpace{}%
\AgdaOperator{\AgdaDatatype{⊎}}\AgdaSpace{}%
\AgdaSymbol{(}\AgdaBound{M′}\AgdaSpace{}%
\AgdaOperator{\AgdaDatatype{↠}}\AgdaSpace{}%
\AgdaInductiveConstructor{blame}\AgdaSymbol{)}\AgdaSpace{}%
\AgdaOperator{\AgdaDatatype{⊎}}\AgdaSpace{}%
\AgdaFunction{∃[}\AgdaSpace{}%
\AgdaBound{N}\AgdaSpace{}%
\AgdaFunction{]}\AgdaSpace{}%
\AgdaFunction{Σ[}\AgdaSpace{}%
\AgdaBound{r}\AgdaSpace{}%
\AgdaFunction{∈}\AgdaSpace{}%
\AgdaBound{M}\AgdaSpace{}%
\AgdaOperator{\AgdaDatatype{↠}}\AgdaSpace{}%
\AgdaBound{N}\AgdaSpace{}%
\AgdaFunction{]}\AgdaSpace{}%
\AgdaFunction{len}\AgdaSpace{}%
\AgdaBound{r}\AgdaSpace{}%
\AgdaOperator{\AgdaDatatype{≡}}\AgdaSpace{}%
\AgdaBound{k}\<%
\\
\>[0]\AgdaInductiveConstructor{≽}\AgdaSpace{}%
\AgdaOperator{\AgdaFunction{⊨}}\AgdaSpace{}%
\AgdaBound{M}\AgdaSpace{}%
\AgdaOperator{\AgdaFunction{⊑}}\AgdaSpace{}%
\AgdaBound{M′}\AgdaSpace{}%
\AgdaOperator{\AgdaFunction{for}}\AgdaSpace{}%
\AgdaBound{k}\AgdaSpace{}%
\AgdaSymbol{=}\AgdaSpace{}%
\AgdaSymbol{(}\AgdaBound{M}\AgdaSpace{}%
\AgdaOperator{\AgdaFunction{⇓}}\AgdaSpace{}%
\AgdaOperator{\AgdaFunction{×}}\AgdaSpace{}%
\AgdaBound{M′}\AgdaSpace{}%
\AgdaOperator{\AgdaFunction{⇓}}\AgdaSymbol{)}\AgdaSpace{}%
\AgdaOperator{\AgdaDatatype{⊎}}\AgdaSpace{}%
\AgdaSymbol{(}\AgdaBound{M′}\AgdaSpace{}%
\AgdaOperator{\AgdaDatatype{↠}}\AgdaSpace{}%
\AgdaInductiveConstructor{blame}\AgdaSymbol{)}\AgdaSpace{}%
\AgdaOperator{\AgdaDatatype{⊎}}\AgdaSpace{}%
\AgdaFunction{∃[}\AgdaSpace{}%
\AgdaBound{N′}\AgdaSpace{}%
\AgdaFunction{]}\AgdaSpace{}%
\AgdaFunction{Σ[}\AgdaSpace{}%
\AgdaBound{r}\AgdaSpace{}%
\AgdaFunction{∈}\AgdaSpace{}%
\AgdaBound{M′}\AgdaSpace{}%
\AgdaOperator{\AgdaDatatype{↠}}\AgdaSpace{}%
\AgdaBound{N′}\AgdaSpace{}%
\AgdaFunction{]}\AgdaSpace{}%
\AgdaFunction{len}\AgdaSpace{}%
\AgdaBound{r}\AgdaSpace{}%
\AgdaOperator{\AgdaDatatype{≡}}\AgdaSpace{}%
\AgdaBound{k}\<%
\\
\\[\AgdaEmptyExtraSkip]%
\>[0]\AgdaOperator{\AgdaFunction{⊨\AgdaUnderscore{}⊑\AgdaUnderscore{}for\AgdaUnderscore{}}}\AgdaSpace{}%
\AgdaSymbol{:}\AgdaSpace{}%
\AgdaDatatype{Term}\AgdaSpace{}%
\AgdaSymbol{→}\AgdaSpace{}%
\AgdaDatatype{Term}\AgdaSpace{}%
\AgdaSymbol{→}\AgdaSpace{}%
\AgdaDatatype{ℕ}\AgdaSpace{}%
\AgdaSymbol{→}\AgdaSpace{}%
\AgdaPrimitive{Set}\<%
\\
\>[0]\AgdaOperator{\AgdaFunction{⊨}}\AgdaSpace{}%
\AgdaBound{M}\AgdaSpace{}%
\AgdaOperator{\AgdaFunction{⊑}}\AgdaSpace{}%
\AgdaBound{M′}\AgdaSpace{}%
\AgdaOperator{\AgdaFunction{for}}\AgdaSpace{}%
\AgdaBound{k}\AgdaSpace{}%
\AgdaSymbol{=}\AgdaSpace{}%
\AgdaSymbol{(}\AgdaInductiveConstructor{≼}\AgdaSpace{}%
\AgdaOperator{\AgdaFunction{⊨}}\AgdaSpace{}%
\AgdaBound{M}\AgdaSpace{}%
\AgdaOperator{\AgdaFunction{⊑}}\AgdaSpace{}%
\AgdaBound{M′}\AgdaSpace{}%
\AgdaOperator{\AgdaFunction{for}}\AgdaSpace{}%
\AgdaBound{k}\AgdaSymbol{)}\AgdaSpace{}%
\AgdaOperator{\AgdaFunction{×}}\AgdaSpace{}%
\AgdaSymbol{(}\AgdaInductiveConstructor{≽}\AgdaSpace{}%
\AgdaOperator{\AgdaFunction{⊨}}\AgdaSpace{}%
\AgdaBound{M}\AgdaSpace{}%
\AgdaOperator{\AgdaFunction{⊑}}\AgdaSpace{}%
\AgdaBound{M′}\AgdaSpace{}%
\AgdaOperator{\AgdaFunction{for}}\AgdaSpace{}%
\AgdaBound{k}\AgdaSymbol{)}\<%
\end{code}

\noindent The proof that the logical relation implies semantic
approximation is a straightforward induction on the step index $k$.

\begin{code}%
\>[0]\AgdaFunction{LR⇒sem-approx}\AgdaSpace{}%
\AgdaSymbol{:}\AgdaSpace{}%
\AgdaSymbol{∀\{}\AgdaBound{A}\AgdaSymbol{\}\{}\AgdaBound{A′}\AgdaSymbol{\}\{}\AgdaBound{A⊑A′}\AgdaSpace{}%
\AgdaSymbol{:}\AgdaSpace{}%
\AgdaBound{A}\AgdaSpace{}%
\AgdaOperator{\AgdaDatatype{⊑}}\AgdaSpace{}%
\AgdaBound{A′}\AgdaSymbol{\}\{}\AgdaBound{M}\AgdaSymbol{\}\{}\AgdaBound{M′}\AgdaSymbol{\}\{}\AgdaBound{k}\AgdaSymbol{\}\{}\AgdaBound{dir}\AgdaSymbol{\}}\<%
\\
\>[0][@{}l@{\AgdaIndent{0}}]%
\>[2]\AgdaSymbol{→}\AgdaSpace{}%
\AgdaField{\#}\AgdaSymbol{(}\AgdaBound{dir}\AgdaSpace{}%
\AgdaOperator{\AgdaFunction{∣}}\AgdaSpace{}%
\AgdaBound{M}\AgdaSpace{}%
\AgdaOperator{\AgdaFunction{⊑ᴸᴿₜ}}\AgdaSpace{}%
\AgdaBound{M′}\AgdaSpace{}%
\AgdaOperator{\AgdaFunction{⦂}}\AgdaSpace{}%
\AgdaBound{A⊑A′}\AgdaSymbol{)}\AgdaSpace{}%
\AgdaSymbol{(}\AgdaInductiveConstructor{suc}\AgdaSpace{}%
\AgdaBound{k}\AgdaSymbol{)}%
\>[39]\AgdaSymbol{→}%
\>[42]\AgdaBound{dir}\AgdaSpace{}%
\AgdaOperator{\AgdaFunction{⊨}}\AgdaSpace{}%
\AgdaBound{M}\AgdaSpace{}%
\AgdaOperator{\AgdaFunction{⊑}}\AgdaSpace{}%
\AgdaBound{M′}\AgdaSpace{}%
\AgdaOperator{\AgdaFunction{for}}\AgdaSpace{}%
\AgdaBound{k}\<%
\end{code}
\begin{code}[hide]%
\>[0]\AgdaFunction{LR⇒sem-approx}\AgdaSpace{}%
\AgdaSymbol{\{}\AgdaBound{A}\AgdaSymbol{\}}\AgdaSpace{}%
\AgdaSymbol{\{}\AgdaBound{A′}\AgdaSymbol{\}}\AgdaSpace{}%
\AgdaSymbol{\{}\AgdaBound{A⊑A′}\AgdaSymbol{\}}\AgdaSpace{}%
\AgdaSymbol{\{}\AgdaBound{M}\AgdaSymbol{\}}\AgdaSpace{}%
\AgdaSymbol{\{}\AgdaBound{M′}\AgdaSymbol{\}}\AgdaSpace{}%
\AgdaSymbol{\{}\AgdaInductiveConstructor{zero}\AgdaSymbol{\}}\AgdaSpace{}%
\AgdaSymbol{\{}\AgdaInductiveConstructor{≼}\AgdaSymbol{\}}\AgdaSpace{}%
\AgdaBound{M⊑M′sk}\AgdaSpace{}%
\AgdaSymbol{=}\<%
\\
\>[0][@{}l@{\AgdaIndent{0}}]%
\>[4]\AgdaInductiveConstructor{inj₂}\AgdaSpace{}%
\AgdaSymbol{(}\AgdaInductiveConstructor{inj₂}\AgdaSpace{}%
\AgdaSymbol{(}\AgdaBound{M}\AgdaSpace{}%
\AgdaOperator{\AgdaInductiveConstructor{,}}\AgdaSpace{}%
\AgdaSymbol{(}\AgdaBound{M}\AgdaSpace{}%
\AgdaOperator{\AgdaInductiveConstructor{END}}\AgdaSymbol{)}\AgdaSpace{}%
\AgdaOperator{\AgdaInductiveConstructor{,}}\AgdaSpace{}%
\AgdaInductiveConstructor{refl}\AgdaSymbol{))}\<%
\\
\>[0]\AgdaFunction{LR⇒sem-approx}\AgdaSpace{}%
\AgdaSymbol{\{}\AgdaBound{A}\AgdaSymbol{\}}\AgdaSpace{}%
\AgdaSymbol{\{}\AgdaBound{A′}\AgdaSymbol{\}}\AgdaSpace{}%
\AgdaSymbol{\{}\AgdaBound{A⊑A′}\AgdaSymbol{\}}\AgdaSpace{}%
\AgdaSymbol{\{}\AgdaBound{M}\AgdaSymbol{\}}\AgdaSpace{}%
\AgdaSymbol{\{}\AgdaBound{M′}\AgdaSymbol{\}}\AgdaSpace{}%
\AgdaSymbol{\{}\AgdaInductiveConstructor{suc}\AgdaSpace{}%
\AgdaBound{k}\AgdaSymbol{\}}\AgdaSpace{}%
\AgdaSymbol{\{}\AgdaInductiveConstructor{≼}\AgdaSymbol{\}}\AgdaSpace{}%
\AgdaBound{M⊑M′sk}\<%
\\
\>[0][@{}l@{\AgdaIndent{0}}]%
\>[4]\AgdaKeyword{with}\AgdaSpace{}%
\AgdaFunction{⇔-to}\AgdaSpace{}%
\AgdaSymbol{(}\AgdaFunction{LRₜ-suc}\AgdaSymbol{\{}\AgdaArgument{dir}\AgdaSpace{}%
\AgdaSymbol{=}\AgdaSpace{}%
\AgdaInductiveConstructor{≼}\AgdaSymbol{\})}\AgdaSpace{}%
\AgdaBound{M⊑M′sk}\<%
\\
\>[0]\AgdaSymbol{...}\AgdaSpace{}%
\AgdaSymbol{|}%
\>[213I]\AgdaInductiveConstructor{inj₂}\AgdaSpace{}%
\AgdaSymbol{(}\AgdaInductiveConstructor{inj₁}\AgdaSpace{}%
\AgdaBound{M′→blame}\AgdaSymbol{)}\AgdaSpace{}%
\AgdaSymbol{=}\<%
\\
\>[.][@{}l@{}]\<[213I]%
\>[6]\AgdaInductiveConstructor{inj₂}\AgdaSpace{}%
\AgdaSymbol{(}\AgdaInductiveConstructor{inj₁}\AgdaSpace{}%
\AgdaBound{M′→blame}\AgdaSymbol{)}\<%
\\
\>[0]\AgdaSymbol{...}\AgdaSpace{}%
\AgdaSymbol{|}%
\>[220I]\AgdaInductiveConstructor{inj₂}\AgdaSpace{}%
\AgdaSymbol{(}\AgdaInductiveConstructor{inj₂}\AgdaSpace{}%
\AgdaSymbol{(}\AgdaBound{m}\AgdaSpace{}%
\AgdaOperator{\AgdaInductiveConstructor{,}}\AgdaSpace{}%
\AgdaSymbol{(}\AgdaBound{V′}\AgdaSpace{}%
\AgdaOperator{\AgdaInductiveConstructor{,}}\AgdaSpace{}%
\AgdaBound{M′→V′}\AgdaSpace{}%
\AgdaOperator{\AgdaInductiveConstructor{,}}\AgdaSpace{}%
\AgdaBound{v′}\AgdaSpace{}%
\AgdaOperator{\AgdaInductiveConstructor{,}}\AgdaSpace{}%
\AgdaBound{𝒱≼V′M}\AgdaSymbol{)))}\AgdaSpace{}%
\AgdaSymbol{=}\<%
\\
\>[.][@{}l@{}]\<[220I]%
\>[6]\AgdaInductiveConstructor{inj₁}\AgdaSpace{}%
\AgdaSymbol{((}\AgdaBound{M}\AgdaSpace{}%
\AgdaOperator{\AgdaInductiveConstructor{,}}\AgdaSpace{}%
\AgdaSymbol{(}\AgdaBound{M}\AgdaSpace{}%
\AgdaOperator{\AgdaInductiveConstructor{END}}\AgdaSymbol{)}\AgdaSpace{}%
\AgdaOperator{\AgdaInductiveConstructor{,}}\AgdaSpace{}%
\AgdaBound{m}\AgdaSymbol{)}\AgdaSpace{}%
\AgdaOperator{\AgdaInductiveConstructor{,}}\AgdaSpace{}%
\AgdaSymbol{(}\AgdaBound{V′}\AgdaSpace{}%
\AgdaOperator{\AgdaInductiveConstructor{,}}\AgdaSpace{}%
\AgdaBound{M′→V′}\AgdaSpace{}%
\AgdaOperator{\AgdaInductiveConstructor{,}}\AgdaSpace{}%
\AgdaBound{v′}\AgdaSymbol{))}\<%
\\
\>[0]\AgdaSymbol{...}%
\>[244I]\AgdaSymbol{|}\AgdaSpace{}%
\AgdaInductiveConstructor{inj₁}\AgdaSpace{}%
\AgdaSymbol{(}\AgdaBound{N}\AgdaSpace{}%
\AgdaOperator{\AgdaInductiveConstructor{,}}\AgdaSpace{}%
\AgdaBound{M→N}\AgdaSpace{}%
\AgdaOperator{\AgdaInductiveConstructor{,}}\AgdaSpace{}%
\AgdaBound{▷N⊑M′}\AgdaSymbol{)}\<%
\\
\>[.][@{}l@{}]\<[244I]%
\>[4]\AgdaKeyword{with}\AgdaSpace{}%
\AgdaFunction{LR⇒sem-approx}\AgdaSymbol{\{}\AgdaArgument{k}\AgdaSpace{}%
\AgdaSymbol{=}\AgdaSpace{}%
\AgdaBound{k}\AgdaSymbol{\}\{}\AgdaArgument{dir}\AgdaSpace{}%
\AgdaSymbol{=}\AgdaSpace{}%
\AgdaInductiveConstructor{≼}\AgdaSymbol{\}}\AgdaSpace{}%
\AgdaBound{▷N⊑M′}\<%
\\
\>[0]\AgdaSymbol{...}\AgdaSpace{}%
\AgdaSymbol{|}%
\>[258I]\AgdaInductiveConstructor{inj₁}\AgdaSpace{}%
\AgdaSymbol{((}\AgdaBound{V}\AgdaSpace{}%
\AgdaOperator{\AgdaInductiveConstructor{,}}\AgdaSpace{}%
\AgdaBound{M→V}\AgdaSpace{}%
\AgdaOperator{\AgdaInductiveConstructor{,}}\AgdaSpace{}%
\AgdaBound{v}\AgdaSymbol{)}\AgdaSpace{}%
\AgdaOperator{\AgdaInductiveConstructor{,}}\AgdaSpace{}%
\AgdaSymbol{(}\AgdaBound{V′}\AgdaSpace{}%
\AgdaOperator{\AgdaInductiveConstructor{,}}\AgdaSpace{}%
\AgdaBound{M′→V′}\AgdaSpace{}%
\AgdaOperator{\AgdaInductiveConstructor{,}}\AgdaSpace{}%
\AgdaBound{v′}\AgdaSymbol{))}\AgdaSpace{}%
\AgdaSymbol{=}\<%
\\
\>[.][@{}l@{}]\<[258I]%
\>[6]\AgdaInductiveConstructor{inj₁}\AgdaSpace{}%
\AgdaSymbol{((}\AgdaBound{V}\AgdaSpace{}%
\AgdaOperator{\AgdaInductiveConstructor{,}}\AgdaSpace{}%
\AgdaSymbol{(}\AgdaBound{M}\AgdaSpace{}%
\AgdaOperator{\AgdaInductiveConstructor{⟶⟨}}\AgdaSpace{}%
\AgdaBound{M→N}\AgdaSpace{}%
\AgdaOperator{\AgdaInductiveConstructor{⟩}}\AgdaSpace{}%
\AgdaBound{M→V}\AgdaSymbol{)}\AgdaSpace{}%
\AgdaOperator{\AgdaInductiveConstructor{,}}\AgdaSpace{}%
\AgdaBound{v}\AgdaSymbol{)}\AgdaSpace{}%
\AgdaOperator{\AgdaInductiveConstructor{,}}\AgdaSpace{}%
\AgdaSymbol{(}\AgdaBound{V′}\AgdaSpace{}%
\AgdaOperator{\AgdaInductiveConstructor{,}}\AgdaSpace{}%
\AgdaBound{M′→V′}\AgdaSpace{}%
\AgdaOperator{\AgdaInductiveConstructor{,}}\AgdaSpace{}%
\AgdaBound{v′}\AgdaSymbol{))}\<%
\\
\>[0]\AgdaSymbol{...}\AgdaSpace{}%
\AgdaSymbol{|}%
\>[287I]\AgdaInductiveConstructor{inj₂}\AgdaSpace{}%
\AgdaSymbol{(}\AgdaInductiveConstructor{inj₁}\AgdaSpace{}%
\AgdaBound{M′→blame}\AgdaSymbol{)}\AgdaSpace{}%
\AgdaSymbol{=}\<%
\\
\>[.][@{}l@{}]\<[287I]%
\>[6]\AgdaInductiveConstructor{inj₂}\AgdaSpace{}%
\AgdaSymbol{(}\AgdaInductiveConstructor{inj₁}\AgdaSpace{}%
\AgdaBound{M′→blame}\AgdaSymbol{)}\<%
\\
\>[0]\AgdaSymbol{...}\AgdaSpace{}%
\AgdaSymbol{|}%
\>[294I]\AgdaInductiveConstructor{inj₂}\AgdaSpace{}%
\AgdaSymbol{(}\AgdaInductiveConstructor{inj₂}\AgdaSpace{}%
\AgdaSymbol{(}\AgdaBound{L}\AgdaSpace{}%
\AgdaOperator{\AgdaInductiveConstructor{,}}\AgdaSpace{}%
\AgdaBound{N→L}\AgdaSpace{}%
\AgdaOperator{\AgdaInductiveConstructor{,}}\AgdaSpace{}%
\AgdaBound{eq}\AgdaSymbol{))}\AgdaSpace{}%
\AgdaSymbol{=}\<%
\\
\>[.][@{}l@{}]\<[294I]%
\>[6]\AgdaInductiveConstructor{inj₂}\AgdaSpace{}%
\AgdaSymbol{(}\AgdaInductiveConstructor{inj₂}\AgdaSpace{}%
\AgdaSymbol{(}\AgdaBound{L}\AgdaSpace{}%
\AgdaOperator{\AgdaInductiveConstructor{,}}\AgdaSpace{}%
\AgdaSymbol{(}\AgdaBound{M}\AgdaSpace{}%
\AgdaOperator{\AgdaInductiveConstructor{⟶⟨}}\AgdaSpace{}%
\AgdaBound{M→N}\AgdaSpace{}%
\AgdaOperator{\AgdaInductiveConstructor{⟩}}\AgdaSpace{}%
\AgdaBound{N→L}\AgdaSymbol{)}\AgdaSpace{}%
\AgdaOperator{\AgdaInductiveConstructor{,}}\AgdaSpace{}%
\AgdaFunction{cong}\AgdaSpace{}%
\AgdaInductiveConstructor{suc}\AgdaSpace{}%
\AgdaBound{eq}\AgdaSymbol{))}\<%
\\
\>[0]\AgdaFunction{LR⇒sem-approx}\AgdaSpace{}%
\AgdaSymbol{\{}\AgdaBound{A}\AgdaSymbol{\}}\AgdaSpace{}%
\AgdaSymbol{\{}\AgdaBound{A′}\AgdaSymbol{\}}\AgdaSpace{}%
\AgdaSymbol{\{}\AgdaBound{A⊑A′}\AgdaSymbol{\}}\AgdaSpace{}%
\AgdaSymbol{\{}\AgdaBound{M}\AgdaSymbol{\}}\AgdaSpace{}%
\AgdaSymbol{\{}\AgdaBound{M′}\AgdaSymbol{\}}\AgdaSpace{}%
\AgdaSymbol{\{}\AgdaInductiveConstructor{zero}\AgdaSymbol{\}}\AgdaSpace{}%
\AgdaSymbol{\{}\AgdaInductiveConstructor{≽}\AgdaSymbol{\}}\AgdaSpace{}%
\AgdaBound{M⊑M′sk}\AgdaSpace{}%
\AgdaSymbol{=}\<%
\\
\>[0][@{}l@{\AgdaIndent{0}}]%
\>[4]\AgdaInductiveConstructor{inj₂}\AgdaSpace{}%
\AgdaSymbol{(}\AgdaInductiveConstructor{inj₂}\AgdaSpace{}%
\AgdaSymbol{(}\AgdaBound{M′}\AgdaSpace{}%
\AgdaOperator{\AgdaInductiveConstructor{,}}\AgdaSpace{}%
\AgdaSymbol{(}\AgdaBound{M′}\AgdaSpace{}%
\AgdaOperator{\AgdaInductiveConstructor{END}}\AgdaSymbol{)}\AgdaSpace{}%
\AgdaOperator{\AgdaInductiveConstructor{,}}\AgdaSpace{}%
\AgdaInductiveConstructor{refl}\AgdaSymbol{))}\<%
\\
\>[0]\AgdaFunction{LR⇒sem-approx}\AgdaSpace{}%
\AgdaSymbol{\{}\AgdaBound{A}\AgdaSymbol{\}}\AgdaSpace{}%
\AgdaSymbol{\{}\AgdaBound{A′}\AgdaSymbol{\}}\AgdaSpace{}%
\AgdaSymbol{\{}\AgdaBound{A⊑A′}\AgdaSymbol{\}}\AgdaSpace{}%
\AgdaSymbol{\{}\AgdaBound{M}\AgdaSymbol{\}}\AgdaSpace{}%
\AgdaSymbol{\{}\AgdaBound{M′}\AgdaSymbol{\}}\AgdaSpace{}%
\AgdaSymbol{\{}\AgdaInductiveConstructor{suc}\AgdaSpace{}%
\AgdaBound{k}\AgdaSymbol{\}}\AgdaSpace{}%
\AgdaSymbol{\{}\AgdaInductiveConstructor{≽}\AgdaSymbol{\}}\AgdaSpace{}%
\AgdaBound{M⊑M′sk}\<%
\\
\>[0][@{}l@{\AgdaIndent{0}}]%
\>[4]\AgdaKeyword{with}\AgdaSpace{}%
\AgdaFunction{⇔-to}\AgdaSpace{}%
\AgdaSymbol{(}\AgdaFunction{LRₜ-suc}\AgdaSymbol{\{}\AgdaArgument{dir}\AgdaSpace{}%
\AgdaSymbol{=}\AgdaSpace{}%
\AgdaInductiveConstructor{≽}\AgdaSymbol{\})}\AgdaSpace{}%
\AgdaBound{M⊑M′sk}\<%
\\
\>[0]\AgdaSymbol{...}\AgdaSpace{}%
\AgdaSymbol{|}%
\>[345I]\AgdaInductiveConstructor{inj₂}\AgdaSpace{}%
\AgdaSymbol{(}\AgdaInductiveConstructor{inj₁}\AgdaSpace{}%
\AgdaInductiveConstructor{isBlame}\AgdaSymbol{)}\AgdaSpace{}%
\AgdaSymbol{=}\<%
\\
\>[.][@{}l@{}]\<[345I]%
\>[6]\AgdaInductiveConstructor{inj₂}\AgdaSpace{}%
\AgdaSymbol{(}\AgdaInductiveConstructor{inj₁}\AgdaSpace{}%
\AgdaSymbol{(}\AgdaInductiveConstructor{blame}\AgdaSpace{}%
\AgdaOperator{\AgdaInductiveConstructor{END}}\AgdaSymbol{))}\<%
\\
\>[0]\AgdaSymbol{...}\AgdaSpace{}%
\AgdaSymbol{|}%
\>[353I]\AgdaInductiveConstructor{inj₂}\AgdaSpace{}%
\AgdaSymbol{(}\AgdaInductiveConstructor{inj₂}\AgdaSpace{}%
\AgdaSymbol{(}\AgdaBound{m′}\AgdaSpace{}%
\AgdaOperator{\AgdaInductiveConstructor{,}}\AgdaSpace{}%
\AgdaBound{V}\AgdaSpace{}%
\AgdaOperator{\AgdaInductiveConstructor{,}}\AgdaSpace{}%
\AgdaBound{M→V}\AgdaSpace{}%
\AgdaOperator{\AgdaInductiveConstructor{,}}\AgdaSpace{}%
\AgdaBound{v}\AgdaSpace{}%
\AgdaOperator{\AgdaInductiveConstructor{,}}\AgdaSpace{}%
\AgdaBound{𝒱≽VM′}\AgdaSymbol{))}\AgdaSpace{}%
\AgdaSymbol{=}\<%
\\
\>[.][@{}l@{}]\<[353I]%
\>[6]\AgdaInductiveConstructor{inj₁}\AgdaSpace{}%
\AgdaSymbol{((}\AgdaBound{V}\AgdaSpace{}%
\AgdaOperator{\AgdaInductiveConstructor{,}}\AgdaSpace{}%
\AgdaBound{M→V}\AgdaSpace{}%
\AgdaOperator{\AgdaInductiveConstructor{,}}\AgdaSpace{}%
\AgdaBound{v}\AgdaSymbol{)}\AgdaSpace{}%
\AgdaOperator{\AgdaInductiveConstructor{,}}\AgdaSpace{}%
\AgdaBound{M′}\AgdaSpace{}%
\AgdaOperator{\AgdaInductiveConstructor{,}}\AgdaSpace{}%
\AgdaSymbol{(}\AgdaBound{M′}\AgdaSpace{}%
\AgdaOperator{\AgdaInductiveConstructor{END}}\AgdaSymbol{)}\AgdaSpace{}%
\AgdaOperator{\AgdaInductiveConstructor{,}}\AgdaSpace{}%
\AgdaBound{m′}\AgdaSymbol{)}\<%
\\
\>[0]\AgdaSymbol{...}%
\>[377I]\AgdaSymbol{|}\AgdaSpace{}%
\AgdaInductiveConstructor{inj₁}\AgdaSpace{}%
\AgdaSymbol{(}\AgdaBound{N′}\AgdaSpace{}%
\AgdaOperator{\AgdaInductiveConstructor{,}}\AgdaSpace{}%
\AgdaBound{M′→N′}\AgdaSpace{}%
\AgdaOperator{\AgdaInductiveConstructor{,}}\AgdaSpace{}%
\AgdaBound{▷M⊑N′}\AgdaSymbol{)}\<%
\\
\>[.][@{}l@{}]\<[377I]%
\>[4]\AgdaKeyword{with}\AgdaSpace{}%
\AgdaFunction{LR⇒sem-approx}\AgdaSymbol{\{}\AgdaArgument{k}\AgdaSpace{}%
\AgdaSymbol{=}\AgdaSpace{}%
\AgdaBound{k}\AgdaSymbol{\}\{}\AgdaArgument{dir}\AgdaSpace{}%
\AgdaSymbol{=}\AgdaSpace{}%
\AgdaInductiveConstructor{≽}\AgdaSymbol{\}}\AgdaSpace{}%
\AgdaBound{▷M⊑N′}\<%
\\
\>[0]\AgdaSymbol{...}\AgdaSpace{}%
\AgdaSymbol{|}%
\>[391I]\AgdaInductiveConstructor{inj₁}\AgdaSpace{}%
\AgdaSymbol{((}\AgdaBound{V}\AgdaSpace{}%
\AgdaOperator{\AgdaInductiveConstructor{,}}\AgdaSpace{}%
\AgdaBound{M→V}\AgdaSpace{}%
\AgdaOperator{\AgdaInductiveConstructor{,}}\AgdaSpace{}%
\AgdaBound{v}\AgdaSymbol{)}\AgdaSpace{}%
\AgdaOperator{\AgdaInductiveConstructor{,}}\AgdaSpace{}%
\AgdaSymbol{(}\AgdaBound{V′}\AgdaSpace{}%
\AgdaOperator{\AgdaInductiveConstructor{,}}\AgdaSpace{}%
\AgdaBound{N′→V′}\AgdaSpace{}%
\AgdaOperator{\AgdaInductiveConstructor{,}}\AgdaSpace{}%
\AgdaBound{v′}\AgdaSymbol{))}\AgdaSpace{}%
\AgdaSymbol{=}\<%
\\
\>[.][@{}l@{}]\<[391I]%
\>[6]\AgdaInductiveConstructor{inj₁}\AgdaSpace{}%
\AgdaSymbol{((}\AgdaBound{V}\AgdaSpace{}%
\AgdaOperator{\AgdaInductiveConstructor{,}}\AgdaSpace{}%
\AgdaBound{M→V}\AgdaSpace{}%
\AgdaOperator{\AgdaInductiveConstructor{,}}\AgdaSpace{}%
\AgdaBound{v}\AgdaSymbol{)}\AgdaSpace{}%
\AgdaOperator{\AgdaInductiveConstructor{,}}\AgdaSpace{}%
\AgdaBound{V′}\AgdaSpace{}%
\AgdaOperator{\AgdaInductiveConstructor{,}}\AgdaSpace{}%
\AgdaSymbol{(}\AgdaBound{M′}\AgdaSpace{}%
\AgdaOperator{\AgdaInductiveConstructor{⟶⟨}}\AgdaSpace{}%
\AgdaBound{M′→N′}\AgdaSpace{}%
\AgdaOperator{\AgdaInductiveConstructor{⟩}}\AgdaSpace{}%
\AgdaBound{N′→V′}\AgdaSymbol{)}\AgdaSpace{}%
\AgdaOperator{\AgdaInductiveConstructor{,}}\AgdaSpace{}%
\AgdaBound{v′}\AgdaSymbol{)}\<%
\\
\>[0]\AgdaSymbol{...}\AgdaSpace{}%
\AgdaSymbol{|}\AgdaSpace{}%
\AgdaInductiveConstructor{inj₂}\AgdaSpace{}%
\AgdaSymbol{(}\AgdaInductiveConstructor{inj₁}\AgdaSpace{}%
\AgdaBound{N′→blame}\AgdaSymbol{)}\AgdaSpace{}%
\AgdaSymbol{=}\AgdaSpace{}%
\AgdaInductiveConstructor{inj₂}\AgdaSpace{}%
\AgdaSymbol{(}\AgdaInductiveConstructor{inj₁}\AgdaSpace{}%
\AgdaSymbol{(}\AgdaBound{M′}\AgdaSpace{}%
\AgdaOperator{\AgdaInductiveConstructor{⟶⟨}}\AgdaSpace{}%
\AgdaBound{M′→N′}\AgdaSpace{}%
\AgdaOperator{\AgdaInductiveConstructor{⟩}}\AgdaSpace{}%
\AgdaBound{N′→blame}\AgdaSymbol{))}\<%
\\
\>[0]\AgdaSymbol{...}\AgdaSpace{}%
\AgdaSymbol{|}%
\>[432I]\AgdaInductiveConstructor{inj₂}\AgdaSpace{}%
\AgdaSymbol{(}\AgdaInductiveConstructor{inj₂}\AgdaSpace{}%
\AgdaSymbol{(}\AgdaBound{L′}\AgdaSpace{}%
\AgdaOperator{\AgdaInductiveConstructor{,}}\AgdaSpace{}%
\AgdaBound{N′→L′}\AgdaSpace{}%
\AgdaOperator{\AgdaInductiveConstructor{,}}\AgdaSpace{}%
\AgdaBound{eq}\AgdaSymbol{))}\AgdaSpace{}%
\AgdaSymbol{=}\<%
\\
\>[.][@{}l@{}]\<[432I]%
\>[6]\AgdaInductiveConstructor{inj₂}\AgdaSpace{}%
\AgdaSymbol{(}\AgdaInductiveConstructor{inj₂}\AgdaSpace{}%
\AgdaSymbol{(}\AgdaBound{L′}\AgdaSpace{}%
\AgdaOperator{\AgdaInductiveConstructor{,}}\AgdaSpace{}%
\AgdaSymbol{(}\AgdaBound{M′}\AgdaSpace{}%
\AgdaOperator{\AgdaInductiveConstructor{⟶⟨}}\AgdaSpace{}%
\AgdaBound{M′→N′}\AgdaSpace{}%
\AgdaOperator{\AgdaInductiveConstructor{⟩}}\AgdaSpace{}%
\AgdaBound{N′→L′}\AgdaSymbol{)}\AgdaSpace{}%
\AgdaOperator{\AgdaInductiveConstructor{,}}\AgdaSpace{}%
\AgdaFunction{cong}\AgdaSpace{}%
\AgdaInductiveConstructor{suc}\AgdaSpace{}%
\AgdaBound{eq}\AgdaSymbol{))}\<%
\end{code}

\noindent The proof that semantic approximation implies the gradual
guarantee relies on a proof of determinism for the Cast Calculus.

\begin{code}%
\>[0]\AgdaFunction{sem-approx⇒GG}\AgdaSpace{}%
\AgdaSymbol{:}\AgdaSpace{}%
\AgdaSymbol{∀\{}\AgdaBound{A}\AgdaSymbol{\}\{}\AgdaBound{A′}\AgdaSymbol{\}\{}\AgdaBound{A⊑A′}\AgdaSpace{}%
\AgdaSymbol{:}\AgdaSpace{}%
\AgdaBound{A}\AgdaSpace{}%
\AgdaOperator{\AgdaDatatype{⊑}}\AgdaSpace{}%
\AgdaBound{A′}\AgdaSymbol{\}\{}\AgdaBound{M}\AgdaSymbol{\}\{}\AgdaBound{M′}\AgdaSymbol{\}}\<%
\\
\>[0][@{}l@{\AgdaIndent{0}}]%
\>[3]\AgdaSymbol{→}\AgdaSpace{}%
\AgdaSymbol{(∀}\AgdaSpace{}%
\AgdaBound{k}\AgdaSpace{}%
\AgdaSymbol{→}\AgdaSpace{}%
\AgdaOperator{\AgdaFunction{⊨}}\AgdaSpace{}%
\AgdaBound{M}\AgdaSpace{}%
\AgdaOperator{\AgdaFunction{⊑}}\AgdaSpace{}%
\AgdaBound{M′}\AgdaSpace{}%
\AgdaOperator{\AgdaFunction{for}}\AgdaSpace{}%
\AgdaBound{k}\AgdaSymbol{)}%
\>[29]\AgdaSymbol{→}%
\>[32]\AgdaFunction{gradual}\AgdaSpace{}%
\AgdaBound{M}\AgdaSpace{}%
\AgdaBound{M′}\<%
\end{code}
\begin{code}[hide]%
\>[0]\AgdaFunction{sem-approx⇒GG}\AgdaSpace{}%
\AgdaSymbol{\{}\AgdaBound{A}\AgdaSymbol{\}\{}\AgdaBound{A′}\AgdaSymbol{\}\{}\AgdaBound{A⊑A′}\AgdaSymbol{\}\{}\AgdaBound{M}\AgdaSymbol{\}\{}\AgdaBound{M′}\AgdaSymbol{\}}\AgdaSpace{}%
\AgdaBound{⊨M⊑M′}\AgdaSpace{}%
\AgdaSymbol{=}\<%
\\
\>[0][@{}l@{\AgdaIndent{0}}]%
\>[2]\AgdaFunction{to-value-right}\AgdaSpace{}%
\AgdaOperator{\AgdaInductiveConstructor{,}}\AgdaSpace{}%
\AgdaFunction{diverge-right}\AgdaSpace{}%
\AgdaOperator{\AgdaInductiveConstructor{,}}\AgdaSpace{}%
\AgdaFunction{to-value-left}\AgdaSpace{}%
\AgdaOperator{\AgdaInductiveConstructor{,}}\AgdaSpace{}%
\AgdaFunction{diverge-left}\AgdaSpace{}%
\AgdaOperator{\AgdaInductiveConstructor{,}}\AgdaSpace{}%
\AgdaFunction{blame-blame}\<%
\\
\>[2]\AgdaKeyword{where}\<%
\\
\>[2]\AgdaFunction{to-value-right}\AgdaSpace{}%
\AgdaSymbol{:}\AgdaSpace{}%
\AgdaBound{M′}\AgdaSpace{}%
\AgdaOperator{\AgdaFunction{⇓}}\AgdaSpace{}%
\AgdaSymbol{→}\AgdaSpace{}%
\AgdaBound{M}\AgdaSpace{}%
\AgdaOperator{\AgdaFunction{⇓}}\<%
\\
\>[2]\AgdaFunction{to-value-right}\AgdaSpace{}%
\AgdaSymbol{(}\AgdaBound{V′}\AgdaSpace{}%
\AgdaOperator{\AgdaInductiveConstructor{,}}\AgdaSpace{}%
\AgdaBound{M′→V′}\AgdaSpace{}%
\AgdaOperator{\AgdaInductiveConstructor{,}}\AgdaSpace{}%
\AgdaBound{v′}\AgdaSymbol{)}\<%
\\
\>[2][@{}l@{\AgdaIndent{0}}]%
\>[6]\AgdaKeyword{with}\AgdaSpace{}%
\AgdaField{proj₂}\AgdaSpace{}%
\AgdaSymbol{(}\AgdaBound{⊨M⊑M′}\AgdaSpace{}%
\AgdaSymbol{(}\AgdaInductiveConstructor{suc}\AgdaSpace{}%
\AgdaSymbol{(}\AgdaFunction{len}\AgdaSpace{}%
\AgdaBound{M′→V′}\AgdaSymbol{)))}\<%
\\
\>[2]\AgdaSymbol{...}\AgdaSpace{}%
\AgdaSymbol{|}\AgdaSpace{}%
\AgdaInductiveConstructor{inj₁}\AgdaSpace{}%
\AgdaSymbol{((}\AgdaBound{V}\AgdaSpace{}%
\AgdaOperator{\AgdaInductiveConstructor{,}}\AgdaSpace{}%
\AgdaBound{M→V}\AgdaSpace{}%
\AgdaOperator{\AgdaInductiveConstructor{,}}\AgdaSpace{}%
\AgdaBound{v}\AgdaSymbol{)}\AgdaSpace{}%
\AgdaOperator{\AgdaInductiveConstructor{,}}\AgdaSpace{}%
\AgdaSymbol{\AgdaUnderscore{})}\AgdaSpace{}%
\AgdaSymbol{=}\AgdaSpace{}%
\AgdaBound{V}\AgdaSpace{}%
\AgdaOperator{\AgdaInductiveConstructor{,}}\AgdaSpace{}%
\AgdaBound{M→V}\AgdaSpace{}%
\AgdaOperator{\AgdaInductiveConstructor{,}}\AgdaSpace{}%
\AgdaBound{v}\<%
\\
\>[2]\AgdaSymbol{...}\AgdaSpace{}%
\AgdaSymbol{|}%
\>[512I]\AgdaInductiveConstructor{inj₂}\AgdaSpace{}%
\AgdaSymbol{(}\AgdaInductiveConstructor{inj₁}\AgdaSpace{}%
\AgdaBound{M′→blame}\AgdaSymbol{)}\AgdaSpace{}%
\AgdaSymbol{=}\<%
\\
\>[.][@{}l@{}]\<[512I]%
\>[8]\AgdaFunction{⊥-elim}\AgdaSpace{}%
\AgdaSymbol{(}\AgdaFunction{cant-reduce-value-and-blame}\AgdaSpace{}%
\AgdaBound{v′}\AgdaSpace{}%
\AgdaBound{M′→V′}\AgdaSpace{}%
\AgdaBound{M′→blame}\AgdaSymbol{)}\<%
\\
\>[2]\AgdaSymbol{...}\AgdaSpace{}%
\AgdaSymbol{|}%
\>[521I]\AgdaInductiveConstructor{inj₂}\AgdaSpace{}%
\AgdaSymbol{(}\AgdaInductiveConstructor{inj₂}\AgdaSpace{}%
\AgdaSymbol{(}\AgdaBound{N′}\AgdaSpace{}%
\AgdaOperator{\AgdaInductiveConstructor{,}}\AgdaSpace{}%
\AgdaBound{M′→N′}\AgdaSpace{}%
\AgdaOperator{\AgdaInductiveConstructor{,}}\AgdaSpace{}%
\AgdaBound{eq}\AgdaSymbol{))}\AgdaSpace{}%
\AgdaSymbol{=}\<%
\\
\>[.][@{}l@{}]\<[521I]%
\>[8]\AgdaFunction{⊥-elim}\AgdaSpace{}%
\AgdaSymbol{(}\AgdaFunction{step-value-plus-one}\AgdaSpace{}%
\AgdaBound{M′→N′}\AgdaSpace{}%
\AgdaBound{M′→V′}\AgdaSpace{}%
\AgdaBound{v′}\AgdaSpace{}%
\AgdaBound{eq}\AgdaSymbol{)}\<%
\\
\>[0]\<%
\\
\>[2]\AgdaFunction{diverge-right}\AgdaSpace{}%
\AgdaSymbol{:}\AgdaSpace{}%
\AgdaBound{M′}\AgdaSpace{}%
\AgdaOperator{\AgdaFunction{⇑}}\AgdaSpace{}%
\AgdaSymbol{→}\AgdaSpace{}%
\AgdaBound{M}\AgdaSpace{}%
\AgdaOperator{\AgdaFunction{⇑}}\<%
\\
\>[2]\AgdaFunction{diverge-right}\AgdaSpace{}%
\AgdaBound{divM′}\AgdaSpace{}%
\AgdaBound{k}\<%
\\
\>[2][@{}l@{\AgdaIndent{0}}]%
\>[6]\AgdaKeyword{with}\AgdaSpace{}%
\AgdaField{proj₁}\AgdaSpace{}%
\AgdaSymbol{(}\AgdaBound{⊨M⊑M′}\AgdaSpace{}%
\AgdaBound{k}\AgdaSymbol{)}\<%
\\
\>[2]\AgdaSymbol{...}\AgdaSpace{}%
\AgdaSymbol{|}%
\>[546I]\AgdaInductiveConstructor{inj₁}\AgdaSpace{}%
\AgdaSymbol{((}\AgdaBound{V}\AgdaSpace{}%
\AgdaOperator{\AgdaInductiveConstructor{,}}\AgdaSpace{}%
\AgdaBound{M→V}\AgdaSpace{}%
\AgdaOperator{\AgdaInductiveConstructor{,}}\AgdaSpace{}%
\AgdaBound{v}\AgdaSymbol{)}\AgdaSpace{}%
\AgdaOperator{\AgdaInductiveConstructor{,}}\AgdaSpace{}%
\AgdaSymbol{(}\AgdaBound{V′}\AgdaSpace{}%
\AgdaOperator{\AgdaInductiveConstructor{,}}\AgdaSpace{}%
\AgdaBound{M′→V′}\AgdaSpace{}%
\AgdaOperator{\AgdaInductiveConstructor{,}}\AgdaSpace{}%
\AgdaBound{v′}\AgdaSymbol{))}\AgdaSpace{}%
\AgdaSymbol{=}\<%
\\
\>[.][@{}l@{}]\<[546I]%
\>[8]\AgdaFunction{⊥-elim}\AgdaSpace{}%
\AgdaSymbol{(}\AgdaFunction{diverge-not-halt}\AgdaSpace{}%
\AgdaBound{divM′}\AgdaSpace{}%
\AgdaSymbol{(}\AgdaInductiveConstructor{inj₂}\AgdaSpace{}%
\AgdaSymbol{(}\AgdaBound{V′}\AgdaSpace{}%
\AgdaOperator{\AgdaInductiveConstructor{,}}\AgdaSpace{}%
\AgdaBound{M′→V′}\AgdaSpace{}%
\AgdaOperator{\AgdaInductiveConstructor{,}}\AgdaSpace{}%
\AgdaBound{v′}\AgdaSymbol{)))}\<%
\\
\>[2]\AgdaSymbol{...}\AgdaSpace{}%
\AgdaSymbol{|}%
\>[568I]\AgdaInductiveConstructor{inj₂}\AgdaSpace{}%
\AgdaSymbol{(}\AgdaInductiveConstructor{inj₁}\AgdaSpace{}%
\AgdaBound{M′→blame}\AgdaSymbol{)}\AgdaSpace{}%
\AgdaSymbol{=}\<%
\\
\>[.][@{}l@{}]\<[568I]%
\>[8]\AgdaFunction{⊥-elim}\AgdaSpace{}%
\AgdaSymbol{(}\AgdaFunction{diverge-not-halt}\AgdaSpace{}%
\AgdaBound{divM′}\AgdaSpace{}%
\AgdaSymbol{(}\AgdaInductiveConstructor{inj₁}\AgdaSpace{}%
\AgdaBound{M′→blame}\AgdaSymbol{))}\<%
\\
\>[2]\AgdaSymbol{...}\AgdaSpace{}%
\AgdaSymbol{|}\AgdaSpace{}%
\AgdaInductiveConstructor{inj₂}\AgdaSpace{}%
\AgdaSymbol{(}\AgdaInductiveConstructor{inj₂}\AgdaSpace{}%
\AgdaSymbol{(}\AgdaBound{N}\AgdaSpace{}%
\AgdaOperator{\AgdaInductiveConstructor{,}}\AgdaSpace{}%
\AgdaBound{M→N}\AgdaSpace{}%
\AgdaOperator{\AgdaInductiveConstructor{,}}\AgdaSpace{}%
\AgdaBound{eq}\AgdaSymbol{))}\AgdaSpace{}%
\AgdaSymbol{=}\AgdaSpace{}%
\AgdaBound{N}\AgdaSpace{}%
\AgdaOperator{\AgdaInductiveConstructor{,}}\AgdaSpace{}%
\AgdaBound{M→N}\AgdaSpace{}%
\AgdaOperator{\AgdaInductiveConstructor{,}}\AgdaSpace{}%
\AgdaFunction{sym}\AgdaSpace{}%
\AgdaBound{eq}\<%
\\
\\[\AgdaEmptyExtraSkip]%
\>[2]\AgdaFunction{to-value-left}\AgdaSpace{}%
\AgdaSymbol{:}\AgdaSpace{}%
\AgdaBound{M}\AgdaSpace{}%
\AgdaOperator{\AgdaFunction{⇓}}\AgdaSpace{}%
\AgdaSymbol{→}\AgdaSpace{}%
\AgdaBound{M′}\AgdaSpace{}%
\AgdaOperator{\AgdaFunction{⇓}}\AgdaSpace{}%
\AgdaOperator{\AgdaDatatype{⊎}}\AgdaSpace{}%
\AgdaBound{M′}\AgdaSpace{}%
\AgdaOperator{\AgdaDatatype{↠}}\AgdaSpace{}%
\AgdaInductiveConstructor{blame}\<%
\\
\>[2]\AgdaFunction{to-value-left}\AgdaSpace{}%
\AgdaSymbol{(}\AgdaBound{V}\AgdaSpace{}%
\AgdaOperator{\AgdaInductiveConstructor{,}}\AgdaSpace{}%
\AgdaBound{M→V}\AgdaSpace{}%
\AgdaOperator{\AgdaInductiveConstructor{,}}\AgdaSpace{}%
\AgdaBound{v}\AgdaSymbol{)}\<%
\\
\>[2][@{}l@{\AgdaIndent{0}}]%
\>[6]\AgdaKeyword{with}\AgdaSpace{}%
\AgdaField{proj₁}\AgdaSpace{}%
\AgdaSymbol{(}\AgdaBound{⊨M⊑M′}\AgdaSpace{}%
\AgdaSymbol{(}\AgdaInductiveConstructor{suc}\AgdaSpace{}%
\AgdaSymbol{(}\AgdaFunction{len}\AgdaSpace{}%
\AgdaBound{M→V}\AgdaSymbol{)))}\<%
\\
\>[2]\AgdaSymbol{...}\AgdaSpace{}%
\AgdaSymbol{|}\AgdaSpace{}%
\AgdaInductiveConstructor{inj₁}\AgdaSpace{}%
\AgdaSymbol{((}\AgdaBound{V}\AgdaSpace{}%
\AgdaOperator{\AgdaInductiveConstructor{,}}\AgdaSpace{}%
\AgdaBound{M→V}\AgdaSpace{}%
\AgdaOperator{\AgdaInductiveConstructor{,}}\AgdaSpace{}%
\AgdaBound{v}\AgdaSymbol{)}\AgdaSpace{}%
\AgdaOperator{\AgdaInductiveConstructor{,}}\AgdaSpace{}%
\AgdaSymbol{(}\AgdaBound{V′}\AgdaSpace{}%
\AgdaOperator{\AgdaInductiveConstructor{,}}\AgdaSpace{}%
\AgdaBound{M′→V′}\AgdaSpace{}%
\AgdaOperator{\AgdaInductiveConstructor{,}}\AgdaSpace{}%
\AgdaBound{v′}\AgdaSymbol{))}\AgdaSpace{}%
\AgdaSymbol{=}\AgdaSpace{}%
\AgdaInductiveConstructor{inj₁}\AgdaSpace{}%
\AgdaSymbol{(}\AgdaBound{V′}\AgdaSpace{}%
\AgdaOperator{\AgdaInductiveConstructor{,}}\AgdaSpace{}%
\AgdaBound{M′→V′}\AgdaSpace{}%
\AgdaOperator{\AgdaInductiveConstructor{,}}\AgdaSpace{}%
\AgdaBound{v′}\AgdaSymbol{)}\<%
\\
\>[2]\AgdaSymbol{...}\AgdaSpace{}%
\AgdaSymbol{|}\AgdaSpace{}%
\AgdaInductiveConstructor{inj₂}\AgdaSpace{}%
\AgdaSymbol{(}\AgdaInductiveConstructor{inj₁}\AgdaSpace{}%
\AgdaBound{M′→blame}\AgdaSymbol{)}\AgdaSpace{}%
\AgdaSymbol{=}\AgdaSpace{}%
\AgdaInductiveConstructor{inj₂}\AgdaSpace{}%
\AgdaBound{M′→blame}\<%
\\
\>[2]\AgdaSymbol{...}\AgdaSpace{}%
\AgdaSymbol{|}%
\>[639I]\AgdaInductiveConstructor{inj₂}\AgdaSpace{}%
\AgdaSymbol{(}\AgdaInductiveConstructor{inj₂}\AgdaSpace{}%
\AgdaSymbol{(}\AgdaBound{N}\AgdaSpace{}%
\AgdaOperator{\AgdaInductiveConstructor{,}}\AgdaSpace{}%
\AgdaBound{M→N}\AgdaSpace{}%
\AgdaOperator{\AgdaInductiveConstructor{,}}\AgdaSpace{}%
\AgdaBound{eq}\AgdaSymbol{))}\AgdaSpace{}%
\AgdaSymbol{=}\<%
\\
\>[.][@{}l@{}]\<[639I]%
\>[8]\AgdaFunction{⊥-elim}\AgdaSpace{}%
\AgdaSymbol{(}\AgdaFunction{step-value-plus-one}\AgdaSpace{}%
\AgdaBound{M→N}\AgdaSpace{}%
\AgdaBound{M→V}\AgdaSpace{}%
\AgdaBound{v}\AgdaSpace{}%
\AgdaBound{eq}\AgdaSymbol{)}\<%
\\
\\[\AgdaEmptyExtraSkip]%
\>[2]\AgdaFunction{diverge-left}\AgdaSpace{}%
\AgdaSymbol{:}\AgdaSpace{}%
\AgdaBound{M}\AgdaSpace{}%
\AgdaOperator{\AgdaFunction{⇑}}\AgdaSpace{}%
\AgdaSymbol{→}\AgdaSpace{}%
\AgdaBound{M′}\AgdaSpace{}%
\AgdaOperator{\AgdaFunction{⇑⊎blame}}\<%
\\
\>[2]\AgdaFunction{diverge-left}\AgdaSpace{}%
\AgdaBound{divM}\AgdaSpace{}%
\AgdaBound{k}\<%
\\
\>[2][@{}l@{\AgdaIndent{0}}]%
\>[6]\AgdaKeyword{with}\AgdaSpace{}%
\AgdaField{proj₂}\AgdaSpace{}%
\AgdaSymbol{(}\AgdaBound{⊨M⊑M′}\AgdaSpace{}%
\AgdaBound{k}\AgdaSymbol{)}\<%
\\
\>[2]\AgdaSymbol{...}\AgdaSpace{}%
\AgdaSymbol{|}%
\>[664I]\AgdaInductiveConstructor{inj₁}\AgdaSpace{}%
\AgdaSymbol{((}\AgdaBound{V}\AgdaSpace{}%
\AgdaOperator{\AgdaInductiveConstructor{,}}\AgdaSpace{}%
\AgdaBound{M→V}\AgdaSpace{}%
\AgdaOperator{\AgdaInductiveConstructor{,}}\AgdaSpace{}%
\AgdaBound{v}\AgdaSymbol{)}\AgdaSpace{}%
\AgdaOperator{\AgdaInductiveConstructor{,}}\AgdaSpace{}%
\AgdaSymbol{\AgdaUnderscore{})}\AgdaSpace{}%
\AgdaSymbol{=}\<%
\\
\>[.][@{}l@{}]\<[664I]%
\>[8]\AgdaFunction{⊥-elim}\AgdaSpace{}%
\AgdaSymbol{(}\AgdaFunction{diverge-not-halt}\AgdaSpace{}%
\AgdaBound{divM}\AgdaSpace{}%
\AgdaSymbol{(}\AgdaInductiveConstructor{inj₂}\AgdaSpace{}%
\AgdaSymbol{(}\AgdaBound{V}\AgdaSpace{}%
\AgdaOperator{\AgdaInductiveConstructor{,}}\AgdaSpace{}%
\AgdaBound{M→V}\AgdaSpace{}%
\AgdaOperator{\AgdaInductiveConstructor{,}}\AgdaSpace{}%
\AgdaBound{v}\AgdaSymbol{)))}\<%
\\
\>[2]\AgdaSymbol{...}\AgdaSpace{}%
\AgdaSymbol{|}\AgdaSpace{}%
\AgdaInductiveConstructor{inj₂}\AgdaSpace{}%
\AgdaSymbol{(}\AgdaInductiveConstructor{inj₁}\AgdaSpace{}%
\AgdaBound{M′→blame}\AgdaSymbol{)}\AgdaSpace{}%
\AgdaSymbol{=}\AgdaSpace{}%
\AgdaInductiveConstructor{blame}\AgdaSpace{}%
\AgdaOperator{\AgdaInductiveConstructor{,}}\AgdaSpace{}%
\AgdaSymbol{(}\AgdaBound{M′→blame}\AgdaSpace{}%
\AgdaOperator{\AgdaInductiveConstructor{,}}\AgdaSpace{}%
\AgdaSymbol{(}\AgdaInductiveConstructor{inj₂}\AgdaSpace{}%
\AgdaInductiveConstructor{refl}\AgdaSymbol{))}\<%
\\
\>[2]\AgdaSymbol{...}\AgdaSpace{}%
\AgdaSymbol{|}\AgdaSpace{}%
\AgdaInductiveConstructor{inj₂}\AgdaSpace{}%
\AgdaSymbol{(}\AgdaInductiveConstructor{inj₂}\AgdaSpace{}%
\AgdaSymbol{(}\AgdaBound{N′}\AgdaSpace{}%
\AgdaOperator{\AgdaInductiveConstructor{,}}\AgdaSpace{}%
\AgdaBound{M′→N′}\AgdaSpace{}%
\AgdaOperator{\AgdaInductiveConstructor{,}}\AgdaSpace{}%
\AgdaBound{eq}\AgdaSymbol{))}\AgdaSpace{}%
\AgdaSymbol{=}\AgdaSpace{}%
\AgdaBound{N′}\AgdaSpace{}%
\AgdaOperator{\AgdaInductiveConstructor{,}}\AgdaSpace{}%
\AgdaSymbol{(}\AgdaBound{M′→N′}\AgdaSpace{}%
\AgdaOperator{\AgdaInductiveConstructor{,}}\AgdaSpace{}%
\AgdaSymbol{(}\AgdaInductiveConstructor{inj₁}\AgdaSpace{}%
\AgdaSymbol{(}\AgdaFunction{sym}\AgdaSpace{}%
\AgdaBound{eq}\AgdaSymbol{)))}\<%
\\
\\[\AgdaEmptyExtraSkip]%
\>[2]\AgdaFunction{blame-blame}\AgdaSpace{}%
\AgdaSymbol{:}\AgdaSpace{}%
\AgdaSymbol{(}\AgdaBound{M}\AgdaSpace{}%
\AgdaOperator{\AgdaDatatype{↠}}\AgdaSpace{}%
\AgdaInductiveConstructor{blame}\AgdaSpace{}%
\AgdaSymbol{→}\AgdaSpace{}%
\AgdaBound{M′}\AgdaSpace{}%
\AgdaOperator{\AgdaDatatype{↠}}\AgdaSpace{}%
\AgdaInductiveConstructor{blame}\AgdaSymbol{)}\<%
\\
\>[2]\AgdaFunction{blame-blame}\AgdaSpace{}%
\AgdaBound{M→blame}\<%
\\
\>[2][@{}l@{\AgdaIndent{0}}]%
\>[6]\AgdaKeyword{with}\AgdaSpace{}%
\AgdaField{proj₁}\AgdaSpace{}%
\AgdaSymbol{(}\AgdaBound{⊨M⊑M′}\AgdaSpace{}%
\AgdaSymbol{(}\AgdaInductiveConstructor{suc}\AgdaSpace{}%
\AgdaSymbol{(}\AgdaFunction{len}\AgdaSpace{}%
\AgdaBound{M→blame}\AgdaSymbol{)))}\<%
\\
\>[2]\AgdaSymbol{...}\AgdaSpace{}%
\AgdaSymbol{|}%
\>[723I]\AgdaInductiveConstructor{inj₁}\AgdaSpace{}%
\AgdaSymbol{((}\AgdaBound{V}\AgdaSpace{}%
\AgdaOperator{\AgdaInductiveConstructor{,}}\AgdaSpace{}%
\AgdaBound{M→V}\AgdaSpace{}%
\AgdaOperator{\AgdaInductiveConstructor{,}}\AgdaSpace{}%
\AgdaBound{v}\AgdaSymbol{)}\AgdaSpace{}%
\AgdaOperator{\AgdaInductiveConstructor{,}}\AgdaSpace{}%
\AgdaSymbol{(}\AgdaBound{V′}\AgdaSpace{}%
\AgdaOperator{\AgdaInductiveConstructor{,}}\AgdaSpace{}%
\AgdaBound{M′→V′}\AgdaSpace{}%
\AgdaOperator{\AgdaInductiveConstructor{,}}\AgdaSpace{}%
\AgdaBound{v′}\AgdaSymbol{))}\AgdaSpace{}%
\AgdaSymbol{=}\<%
\\
\>[.][@{}l@{}]\<[723I]%
\>[8]\AgdaFunction{⊥-elim}\AgdaSpace{}%
\AgdaSymbol{(}\AgdaFunction{cant-reduce-value-and-blame}\AgdaSpace{}%
\AgdaBound{v}\AgdaSpace{}%
\AgdaBound{M→V}\AgdaSpace{}%
\AgdaBound{M→blame}\AgdaSymbol{)}\<%
\\
\>[2]\AgdaSymbol{...}\AgdaSpace{}%
\AgdaSymbol{|}\AgdaSpace{}%
\AgdaInductiveConstructor{inj₂}\AgdaSpace{}%
\AgdaSymbol{(}\AgdaInductiveConstructor{inj₁}\AgdaSpace{}%
\AgdaBound{M′→blame}\AgdaSymbol{)}\AgdaSpace{}%
\AgdaSymbol{=}\AgdaSpace{}%
\AgdaBound{M′→blame}\<%
\\
\>[2]\AgdaSymbol{...}\AgdaSpace{}%
\AgdaSymbol{|}%
\>[747I]\AgdaInductiveConstructor{inj₂}\AgdaSpace{}%
\AgdaSymbol{(}\AgdaInductiveConstructor{inj₂}\AgdaSpace{}%
\AgdaSymbol{(}\AgdaBound{N}\AgdaSpace{}%
\AgdaOperator{\AgdaInductiveConstructor{,}}\AgdaSpace{}%
\AgdaBound{M→N}\AgdaSpace{}%
\AgdaOperator{\AgdaInductiveConstructor{,}}\AgdaSpace{}%
\AgdaBound{eq}\AgdaSymbol{))}\AgdaSpace{}%
\AgdaSymbol{=}\<%
\\
\>[.][@{}l@{}]\<[747I]%
\>[8]\AgdaFunction{⊥-elim}\AgdaSpace{}%
\AgdaSymbol{(}\AgdaFunction{step-blame-plus-one}\AgdaSpace{}%
\AgdaBound{M→N}\AgdaSpace{}%
\AgdaBound{M→blame}\AgdaSpace{}%
\AgdaBound{eq}\AgdaSymbol{)}\<%
\end{code}

\noindent We put these two proofs together to show that the logical
relation implies the gradual guarantee.

\begin{code}%
\>[0]\AgdaFunction{LR⇒GG}\AgdaSpace{}%
\AgdaSymbol{:}\AgdaSpace{}%
\AgdaSymbol{∀\{}\AgdaBound{A}\AgdaSymbol{\}\{}\AgdaBound{A′}\AgdaSymbol{\}\{}\AgdaBound{A⊑A′}\AgdaSpace{}%
\AgdaSymbol{:}\AgdaSpace{}%
\AgdaBound{A}\AgdaSpace{}%
\AgdaOperator{\AgdaDatatype{⊑}}\AgdaSpace{}%
\AgdaBound{A′}\AgdaSymbol{\}\{}\AgdaBound{M}\AgdaSymbol{\}\{}\AgdaBound{M′}\AgdaSymbol{\}}%
\>[40]\AgdaSymbol{→}\AgdaSpace{}%
\AgdaInductiveConstructor{[]}\AgdaSpace{}%
\AgdaOperator{\AgdaFunction{⊢ᵒ}}\AgdaSpace{}%
\AgdaBound{M}\AgdaSpace{}%
\AgdaOperator{\AgdaFunction{⊑ᴸᴿₜ}}\AgdaSpace{}%
\AgdaBound{M′}\AgdaSpace{}%
\AgdaOperator{\AgdaFunction{⦂}}\AgdaSpace{}%
\AgdaBound{A⊑A′}%
\>[66]\AgdaSymbol{→}%
\>[69]\AgdaFunction{gradual}\AgdaSpace{}%
\AgdaBound{M}\AgdaSpace{}%
\AgdaBound{M′}\<%
\end{code}
\begin{code}[hide]%
\>[0]\AgdaFunction{LR⇒GG}\AgdaSpace{}%
\AgdaSymbol{\{}\AgdaBound{A}\AgdaSymbol{\}\{}\AgdaBound{A′}\AgdaSymbol{\}\{}\AgdaBound{A⊑A′}\AgdaSymbol{\}\{}\AgdaBound{M}\AgdaSymbol{\}\{}\AgdaBound{M′}\AgdaSymbol{\}}\AgdaSpace{}%
\AgdaBound{⊨M⊑M′}\AgdaSpace{}%
\AgdaSymbol{=}\<%
\\
\>[0][@{}l@{\AgdaIndent{0}}]%
\>[2]\AgdaFunction{sem-approx⇒GG}\AgdaSymbol{\{}\AgdaArgument{A⊑A′}\AgdaSpace{}%
\AgdaSymbol{=}\AgdaSpace{}%
\AgdaBound{A⊑A′}\AgdaSymbol{\}}\AgdaSpace{}%
\AgdaSymbol{(λ}\AgdaSpace{}%
\AgdaBound{k}\AgdaSpace{}%
\AgdaSymbol{→}\AgdaSpace{}%
\AgdaFunction{≼⊨M⊑M′}\AgdaSpace{}%
\AgdaOperator{\AgdaInductiveConstructor{,}}\AgdaSpace{}%
\AgdaFunction{≽⊨M⊑M′}\AgdaSymbol{)}\<%
\\
\>[2]\AgdaKeyword{where}\<%
\\
\>[2]\AgdaFunction{≼⊨M⊑M′}\AgdaSpace{}%
\AgdaSymbol{:}\AgdaSpace{}%
\AgdaSymbol{∀\{}\AgdaBound{k}\AgdaSymbol{\}}\AgdaSpace{}%
\AgdaSymbol{→}\AgdaSpace{}%
\AgdaInductiveConstructor{≼}\AgdaSpace{}%
\AgdaOperator{\AgdaFunction{⊨}}\AgdaSpace{}%
\AgdaBound{M}\AgdaSpace{}%
\AgdaOperator{\AgdaFunction{⊑}}\AgdaSpace{}%
\AgdaBound{M′}\AgdaSpace{}%
\AgdaOperator{\AgdaFunction{for}}\AgdaSpace{}%
\AgdaBound{k}\<%
\\
\>[2]\AgdaFunction{≼⊨M⊑M′}\AgdaSpace{}%
\AgdaSymbol{\{}\AgdaBound{k}\AgdaSymbol{\}}\AgdaSpace{}%
\AgdaSymbol{=}%
\>[797I]\AgdaFunction{LR⇒sem-approx}\AgdaSpace{}%
\AgdaSymbol{\{}\AgdaArgument{k}\AgdaSpace{}%
\AgdaSymbol{=}\AgdaSpace{}%
\AgdaBound{k}\AgdaSymbol{\}\{}\AgdaArgument{dir}\AgdaSpace{}%
\AgdaSymbol{=}\AgdaSpace{}%
\AgdaInductiveConstructor{≼}\AgdaSymbol{\}}\<%
\\
\>[797I][@{}l@{\AgdaIndent{0}}]%
\>[19]\AgdaSymbol{(}\AgdaFunction{⊢ᵒ-elim}\AgdaSpace{}%
\AgdaSymbol{(}\AgdaFunction{proj₁ᵒ}\AgdaSpace{}%
\AgdaBound{⊨M⊑M′}\AgdaSymbol{)}\AgdaSpace{}%
\AgdaSymbol{(}\AgdaInductiveConstructor{suc}\AgdaSpace{}%
\AgdaBound{k}\AgdaSymbol{)}\AgdaSpace{}%
\AgdaInductiveConstructor{tt}\AgdaSymbol{)}\<%
\\
\>[2]\AgdaFunction{≽⊨M⊑M′}\AgdaSpace{}%
\AgdaSymbol{:}\AgdaSpace{}%
\AgdaSymbol{∀\{}\AgdaBound{k}\AgdaSymbol{\}}\AgdaSpace{}%
\AgdaSymbol{→}\AgdaSpace{}%
\AgdaInductiveConstructor{≽}\AgdaSpace{}%
\AgdaOperator{\AgdaFunction{⊨}}\AgdaSpace{}%
\AgdaBound{M}\AgdaSpace{}%
\AgdaOperator{\AgdaFunction{⊑}}\AgdaSpace{}%
\AgdaBound{M′}\AgdaSpace{}%
\AgdaOperator{\AgdaFunction{for}}\AgdaSpace{}%
\AgdaBound{k}\<%
\\
\>[2]\AgdaFunction{≽⊨M⊑M′}\AgdaSpace{}%
\AgdaSymbol{\{}\AgdaBound{k}\AgdaSymbol{\}}\AgdaSpace{}%
\AgdaSymbol{=}%
\>[820I]\AgdaFunction{LR⇒sem-approx}\AgdaSpace{}%
\AgdaSymbol{\{}\AgdaArgument{k}\AgdaSpace{}%
\AgdaSymbol{=}\AgdaSpace{}%
\AgdaBound{k}\AgdaSymbol{\}\{}\AgdaArgument{dir}\AgdaSpace{}%
\AgdaSymbol{=}\AgdaSpace{}%
\AgdaInductiveConstructor{≽}\AgdaSymbol{\}}\<%
\\
\>[820I][@{}l@{\AgdaIndent{0}}]%
\>[19]\AgdaSymbol{(}\AgdaFunction{⊢ᵒ-elim}\AgdaSpace{}%
\AgdaSymbol{(}\AgdaFunction{proj₂ᵒ}\AgdaSpace{}%
\AgdaBound{⊨M⊑M′}\AgdaSymbol{)}\AgdaSpace{}%
\AgdaSymbol{(}\AgdaInductiveConstructor{suc}\AgdaSpace{}%
\AgdaBound{k}\AgdaSymbol{)}\AgdaSpace{}%
\AgdaInductiveConstructor{tt}\AgdaSymbol{)}\<%
\end{code}

\noindent The gradual guarantee then follows by use of the Fundamental
Lemma to obtain $M$ ⊑ᴸᴿₜ $M′$ and then \textsf{LR⇒GG} to
conclude that \textsf{gradual M M′}.

\begin{code}%
\>[0]\AgdaFunction{gradual-guarantee}\AgdaSpace{}%
\AgdaSymbol{:}\AgdaSpace{}%
\AgdaSymbol{∀}\AgdaSpace{}%
\AgdaSymbol{\{}\AgdaBound{A}\AgdaSymbol{\}\{}\AgdaBound{A′}\AgdaSymbol{\}\{}\AgdaBound{A⊑A′}\AgdaSpace{}%
\AgdaSymbol{:}\AgdaSpace{}%
\AgdaBound{A}\AgdaSpace{}%
\AgdaOperator{\AgdaDatatype{⊑}}\AgdaSpace{}%
\AgdaBound{A′}\AgdaSymbol{\}}\AgdaSpace{}%
\AgdaSymbol{→}\AgdaSpace{}%
\AgdaSymbol{(}\AgdaBound{M}\AgdaSpace{}%
\AgdaBound{M′}\AgdaSpace{}%
\AgdaSymbol{:}\AgdaSpace{}%
\AgdaDatatype{Term}\AgdaSymbol{)}\<%
\\
\>[0][@{}l@{\AgdaIndent{0}}]%
\>[3]\AgdaSymbol{→}\AgdaSpace{}%
\AgdaInductiveConstructor{[]}\AgdaSpace{}%
\AgdaOperator{\AgdaDatatype{⊩}}\AgdaSpace{}%
\AgdaBound{M}\AgdaSpace{}%
\AgdaOperator{\AgdaDatatype{⊑}}\AgdaSpace{}%
\AgdaBound{M′}\AgdaSpace{}%
\AgdaOperator{\AgdaDatatype{⦂}}\AgdaSpace{}%
\AgdaBound{A⊑A′}%
\>[25]\AgdaSymbol{→}%
\>[28]\AgdaFunction{gradual}\AgdaSpace{}%
\AgdaBound{M}\AgdaSpace{}%
\AgdaBound{M′}\<%
\\
\>[0]\AgdaFunction{gradual-guarantee}\AgdaSpace{}%
\AgdaSymbol{\{}\AgdaBound{A}\AgdaSymbol{\}\{}\AgdaBound{A′}\AgdaSymbol{\}\{}\AgdaBound{A⊑A′}\AgdaSymbol{\}}\AgdaSpace{}%
\AgdaBound{M}\AgdaSpace{}%
\AgdaBound{M′}\AgdaSpace{}%
\AgdaBound{M⊑M′}\AgdaSpace{}%
\AgdaSymbol{=}\<%
\\
\>[0][@{}l@{\AgdaIndent{0}}]%
\>[2]\AgdaKeyword{let}\AgdaSpace{}%
\AgdaSymbol{(}\AgdaBound{⊨≼M⊑ᴸᴿM′}\AgdaSpace{}%
\AgdaOperator{\AgdaInductiveConstructor{,}}\AgdaSpace{}%
\AgdaBound{⊨≽M⊑ᴸᴿM′}\AgdaSymbol{)}\AgdaSpace{}%
\AgdaSymbol{=}\AgdaSpace{}%
\AgdaFunction{fundamental}\AgdaSpace{}%
\AgdaBound{M}\AgdaSpace{}%
\AgdaBound{M′}\AgdaSpace{}%
\AgdaBound{M⊑M′}\AgdaSpace{}%
\AgdaKeyword{in}\<%
\\
\>[2]\AgdaFunction{LR⇒GG}\AgdaSpace{}%
\AgdaSymbol{(}\AgdaBound{⊨≼M⊑ᴸᴿM′}\AgdaSpace{}%
\AgdaFunction{id}\AgdaSpace{}%
\AgdaFunction{id}\AgdaSpace{}%
\AgdaOperator{\AgdaFunction{,ᵒ}}\AgdaSpace{}%
\AgdaBound{⊨≽M⊑ᴸᴿM′}\AgdaSpace{}%
\AgdaFunction{id}\AgdaSpace{}%
\AgdaFunction{id}\AgdaSymbol{)}\<%
\end{code}


\section{Conclusion and Acknowledgments}
\label{sec:conclusion}

This paper presented the first mechanized proof of the gradual
guarantee using step-indexed logical relations. One naturally wonders
how using step-indexed logical relations compares to a
simulation-based proof. One rough comparison is the number of lines of
code in Agda. Wadler, Thiemann, and I developed a simulation-based
proof of the gradual guarantee for a similar cast calculus in Agda,
which came in at 3,200 LOC of which 232 lines are proofs about substitution
(equivalent to what is provided in the ABT library).
The logical-relations proof presented here
was somewhat shorter, at 2,300 LOC, though it makes use of the
Abstract Binding Tree Library (900 LOC) and the Step-Indexed Logic
Library (2100 LOC). These LOC numbers confirm my feeling that the
total effort to create the SIL and ABT libraries and prove the gradual
guarantee via logical relations was higher than to prove the gradual
guarantee via simulation. However, if one discounts the SIL and ABT
libraries because they are reusable and language independent, then the
remaining effort to prove the gradual guarantee via logical relations
was lower than via simulation.

As mentioned at various points in this paper, there are some rough
edges to the SIL and ABT libraries, primarily due to challenges
regarding leaky abstractions. For SIL, we mentioned how Agda's
output shows normalized versions of the SIL formulas, which exposes
the underlying encodings and are too large to be readable. We have
created a new version of SIL that uses Agda's \texttt{abstract}
feature and look forward to updating the proof of the gradual
guarantee to use the new version of SIL. Regarding the ABT library,
there are also challenges regarding (1) Agda output not always using
the concise pattern syntax and (2) Agda's automated case splitting
does not work for ABT-generated languages.

This work was conducted in the context of a collaboration with Peter
Thiemann and Philip Wadler where we have been exploring how to
mechanize blame calculi in Agda and study the polymorphic blame
calculus. My understanding of step-indexed logical relations was
improved by reading Peter Thiemann's proof in Agda of type safety for
a typed $\lambda$-calculus with \textsf{fix} (for defining recursive
function) using step-indexed logical relations. The initiative to
build an Agda version of the LSLR logic came from Philip Wadler.

\bibliographystyle{eptcs}
\bibliography{all}
\end{document}